\documentclass[11pt,twoside]{cernrepp}
\usepackage{epsf}
\usepackage{axodraw}
\usepackage{epsfig}                            
\usepackage{graphicx}
\usepackage{rotate}
\usepackage{latexsym}
\usepackage{here}
\usepackage{amssymb}
\renewcommand{\cal}{\mathcal}
\begin{document}
\marginparwidth 3cm
\newcommand{\vj}[4]{{\sl #1~}{\bf #2 }\ifnum#3<100 (19#3) \else (#3) \fi #4}
\newcommand{\ej}[3]{{\bf #1~}\ifnum#2<100 (19#2) \else (#2) \fi #3}
\newcommand{\ap}[3]{\vj{Ann.~Phys.}{#1}{#2}{#3}}
\newcommand{\app}[3]{\vj{Acta~Phys.~Pol.}{#1}{#2}{#3}}
\newcommand{\cmp}[3]{\vj{Commun. Math. Phys.}{#1}{#2}{#3}}
\newcommand{\cnpp}[3]{\vj{Comments Nucl. Part. Phys.}{#1}{#2}{#3}}
\newcommand{\cpc}[3]{\vj{Comp. Phys. Commun.}{#1}{#2}{#3}}
\newcommand{\epj}[3]{\vj{Eur. Phys. J.}{#1}{#2}{#3}}
\newcommand{\hpa}[3]{\vj{Helv. Phys.~Acta}{#1}{#2}{#3}} 
\newcommand{\ijmp}[3]{\vj{Int. J. Mod. Phys.}{#1}{#2}{#3}}
\newcommand{\jetp}[3]{\vj{JETP}{#1}{#2}{#3}}
\newcommand{\jetpl}[3]{\vj{JETP Lett.}{#1}{#2}{#3}}
\newcommand{\jmp}[3]{\vj{J. Math. Phys.}{#1}{#2}{#3}}
\newcommand{\jp}[3]{\vj{J. Phys.}{#1}{#2}{#3}}
\newcommand{\lnc}[3]{\vj{Lett. Nuovo Cimento}{#1}{#2}{#3}}
\newcommand{\mpl}[3]{\vj{Mod. Phys. Lett.}{#1}{#2}{#3}}
\newcommand{\nc}[3]{\vj{Nuovo Cimento}{#1}{#2}{#3}}
\newcommand{\nim}[3]{\vj{Nucl. Instr. Meth.}{#1}{#2}{#3}}
\newcommand{\np}[3]{\vj{Nucl. Phys.}{#1}{#2}{#3}}
\newcommand{\pl}[3]{\vj{Phys. Lett.}{#1}{#2}{#3}}
\newcommand{\prp}[3]{\vj{Phys. Rep.}{#1}{#2}{#3}}
\newcommand{\pr}[3]{\vj{Phys. Rev.}{#1}{#2}{#3}}
\newcommand{\prl}[3]{\vj{Phys. Rev. Lett.}{#1}{#2}{#3}}
\newcommand{\ptp}[3]{\vj{Prog. Theor. Phys.}{#1}{#2}{#3}}
\newcommand{\rpp}[3]{\vj{Rep. Prog. Phys.}{#1}{#2}{#3}}
\newcommand{\rmp}[3]{\vj{Rev. Mod. Phys.}{#1}{#2}{#3}}
\newcommand{\rnc}[3]{\vj{Rivista del Nuovo Cim.}{#1}{#2}{#3}}
\newcommand{\sjnp}[3]{\vj{Sov. J. Nucl. Phys.}{#1}{#2}{#3}}
\newcommand{\sptp}[3]{\vj{Suppl. Prog. Theor. Phys.}{#1}{#2}{#3}}
\newcommand{\zp}[3]{\vj{Z. Phys.}{#1}{#2}{#3}}
\newcommand{\jop}[3]{\vj{Journal of Physics} {\bf #1} (#2) #3}
\newcommand{\ibid}[3]{\vj{ibid.} {\bf #1} (#2) #3}
\newcommand{\hep}[1]{{\sl hep--ph/}{#1}}
%
%
\newcommand{\rmf}{{\rm f}}
\newcommand{\vb}{V}
\newcommand{\hq}{{\hat Q}}
\newcommand{\ben}{\begin{enumerate}}
\newcommand{\een}{\end{enumerate}}
\newcommand{\asums}[1]{\sum_{#1}}
\newcommand{\bei}{\begin{itemize}}
\newcommand{\eei}{\end{itemize}}
\newcommand{\dr}{\Delta r}
\newcommand{\ft}{t}
\newcommand{\tabn}[1]{Tab.(\ref{#1})}
\newcommand{\tabns}[2]{Tabs.(\ref{#1}--\ref{#2})}
\newcommand{\fnul}{\nu_l}
\newcommand{\fnue}{\nu_e}
\newcommand{\fnum}{\nu_{\mu}}
\newcommand{\fnut}{\nu_{\tau}}
\newcommand{\gb}{g} 
\newcommand{\ib}{i}
\newcommand{\fe}{e}
\newcommand{\ff}{f}
\newcommand{\fep}{e^{+}}
\newcommand{\fem}{e^{-}}
\newcommand{\fepm}{e^{\pm}}
\newcommand{\fp}{f^{+}}
\newcommand{\fm}{f^{-}}
\newcommand{\bqas}{\begin{eqnarray*}}
\newcommand{\eqas}{\end{eqnarray*}}
\newcommand{\scff}[1]{C_{#1}}                    
\newcommand{\TeV}{\;\mathrm{TeV}}                     
\newcommand{\gz}{\Gamma_{_{\zb}}}
\newcommand{\gw}{\Gamma_{_{\wb}}}
\newcommand{\sW}{p_{_W}}
\newcommand{\sZ}{p_{_Z}}
\newcommand{\ssp}{s_p}
\newcommand{\fW}{f_{_W}}
\newcommand{\fZ}{f_{_Z}}
\newcommand{\bzm}{M_{_0}}
\newcommand{\stw}{s_{\theta}}             
\newcommand{\ctw}{c_{\theta}}
\newcommand{\stws}{s_{\theta}^2}
\newcommand{\stwc}{s_{\theta}^3}
\newcommand{\stwf}{s_{\theta}^4}
\newcommand{\stwx}{s_{\theta}^6}
\newcommand{\ctws}{c_{\theta}^2}
\newcommand{\ctwc}{c_{\theta}^3}
\newcommand{\ctwf}{c_{\theta}^4}
\newcommand{\ctwx}{c_{\theta}^6}
\newcommand{\stwfiv}{s_{\theta}^5}
\newcommand{\ctwfiv}{c_{\theta}^5}
\newcommand{\stwsix}{s_{\theta}^6}
\newcommand{\ctwsix}{c_{\theta}^6}
\newcommand{\drii}[2]{\delta_{#1#2}}                    
\newcommand{\mlones}{m^2}
\newcommand{\Smat}{{\cal{S}}}
\newcommand{\me}{m_e}
\newcommand{\mm}{m_{\mu}}
\newcommand{\mtau}{m_{\tau}}
\newcommand{\muq}{m_u}
\newcommand{\md}{m_d}
\newcommand{\muqp}{m'_u}
\newcommand{\mdqp}{m'_d}
\newcommand{\mc}{m_c}
\newcommand{\ms}{m_s}
\newcommand{\mb}{m_b}
\newcommand{\eqn}[1]{Eq.(\ref{#1})}
\newcommand{\eqns}[2]{Eqs.(\ref{#1}--\ref{#2})}
\newcommand{\eqnss}[1]{Eqs.(\ref{#1})}
\newcommand{\eqnsc}[2]{Eqs.(\ref{#1},~\ref{#2})}
\newcommand{\tbn}[1]{Tab.~\ref{#1}}
\newcommand{\tbns}[2]{Tabs.~\ref{#1}--\ref{#2}}
\newcommand{\tbnsc}[2]{Tabs.~\ref{#1},~\ref{#2}}
\newcommand{\fig}[1]{Fig.~\ref{#1}}
\newcommand{\figs}[2]{Figs.~\ref{#1}--\ref{#2}}
\newcommand{\figsc}[2]{Figs.~\ref{#1},~\ref{#2}}
\newcommand{\sect}[1]{Sect.~\ref{#1}}
\newcommand{\subsect}[1]{Sub-Sect.~\ref{#1}}
\newcommand{\subsects}[2]{Sub-Sects.~\ref{#1},~\ref{#2}}
\newcommand{\pmom}{p}
\newcommand{\pmomp}{p'}
\newcommand{\pmoms}{p^2}
\newcommand{\pmomq}{p^4}
\newcommand{\pmomx}{p^6}
\newcommand{\pmomi}[1]{p_{#1}}
\newcommand{\pmomis}[1]{p^2_{#1}}
\newcommand{\Glone}{\Gamma}
\newcommand{\gbs}{g^2}
\newcommand{\gbc}{g^3}
\newcommand{\gbf}{g^4}
\newcommand{\sgh}{{\hat\Sigma}}
\newcommand{\Pgg}{\Pi_{\ph\ph}}
\newcommand{\Ptg}{\Pi_{_{3Q}}}
\newcommand{\Ptt}{\Pi_{_{33}}}
\newcommand{\Pzg}{\Pi_{_{\zb\ab}}}
\newcommand{\Pzga}[2]{\Pi^{#1}_{_{\zb\ab}}\lpar#2\rpar}
\newcommand{\Pf}{\Pi_{_F}}
\newcommand{\Sgg}{\Sigma_{_{\ab\ab}}}
\newcommand{\Szg}{\Sigma_{_{\zb\ab}}}
\newcommand{\SVV}{\Sigma_{_{\vb\vb}}}
\newcommand{\USvv}{{\hat\Sigma}_{_{\vb\vb}}}
\newcommand{\Sww}{\Sigma_{_{\wb\wb}}}
\newcommand{\Swwg}{\Sigma^{_G}_{_{\wb\wb}}}
\newcommand{\Szz}{\Sigma_{_{\zb\zb}}}
\newcommand{\Shh}{\Sigma_{_{\hb\hb}}}
\newcommand{\Spzz}{\Sigma'_{_{\zb\zb}}}
\newcommand{\Stg}{\Sigma_{_{3Q}}}
\newcommand{\Stt}{\Sigma_{_{33}}}
\newcommand{\bSww}{{\overline\Sigma}_{_{WW}}}
\newcommand{\bStg}{{\overline\Sigma}_{_{3Q}}}
\newcommand{\bStt}{{\overline\Sigma}_{_{33}}}
\newcommand{\sman}{s}
\newcommand{\tman}{t}
\newcommand{\uman}{u}
\newcommand{\smani}[1]{s_{#1}}
\newcommand{\bsmani}[1]{{\bar{s}}_{#1}}
\newcommand{\smans}{s^2}
\newcommand{\tmans}{t^2}
\newcommand{\umans}{u^2}
\newcommand{\ec}{e}
\newcommand{\ecs}{e^2}
\newcommand{\ect}{e^3}
\newcommand{\ecq}{e^4}
\newcommand{\Reb}{{\rm{Re}}}
\newcommand{\Imb}{{\rm{Im}}}
\newcommand{\sany}{s}
\newcommand{\cany}{c}
\newcommand{\sanys}{s^2}
\newcommand{\canys}{c^2}
\newcommand{\scats}{s}
\newcommand{\scatss}{s^2}
\newcommand{\scatsi}[1]{s_{#1}}
\newcommand{\scatsis}[1]{s^2_{#1}}
\newcommand{\scatst}[2]{s_{#1}^{#2}}
\newcommand{\scatc}{c}
\newcommand{\scatcs}{c^2}
\newcommand{\scatci}[1]{c_{#1}}
\newcommand{\scatcis}[1]{c^2_{#1}}
\newcommand{\scatct}[2]{c_{#1}^{#2}}
\newcommand{\bq}{\begin{equation}}                   
\newcommand{\eq}{\end{equation}}
\newcommand{\bqa}{\begin{eqnarray}}
\newcommand{\eqa}{\end{eqnarray}}
\newcommand{\ba}[1]{\begin{array}{#1}}
\newcommand{\ea}{\end{array}}
\newcommand{\lpar}{\left(}                            
\newcommand{\rpar}{\right)} 
\newcommand{\lrbr}{\left[}
\newcommand{\rrbr}{\right]}
\newcommand{\lcbr}{\left\{}
\newcommand{\rcbr}{\right\}} 
\newcommand{\ph}{\gamma}
\newcommand{\zb}{Z}
\newcommand{\wb}{W}            
\newcommand{\wbp}{W^{+}}
\newcommand{\wbm}{W^{-}}
\newcommand{\wbpm}{W^{\pm}}
\newcommand{\barf}{\overline f}                
\newcommand{\barl}{\overline l}
\newcommand{\barq}{\overline q}
\newcommand{\barqp}{\overline{q}'}
\newcommand{\barb}{\overline b}
\newcommand{\bart}{\overline t}
\newcommand{\barc}{\overline c}
\newcommand{\baru}{\overline u}
\newcommand{\bard}{\overline d}
\newcommand{\bars}{\overline s}
\newcommand{\barv}{\overline v}
\newcommand{\barnu}{\overline{\nu}}
\newcommand{\barne}{\overline{\nu}_{\fe}}
\newcommand{\barnm}{\overline{\nu}_{\flm}}
\newcommand{\barnt}{\overline{\nu}_{\flt}}
\newcommand{\mh}{M_{_H}}
\newcommand{\mls}{m^2_l}
\newcommand{\mVs}{M^2_{_V}}
\newcommand{\mw}{M_{_W}}
\newcommand{\mws}{M^2_{_W}}
\newcommand{\mwc}{M^3_{_W}}
\newcommand{\LMs}{M^2}
\newcommand{\LMc}{M^3}
\newcommand{\mz}{M_{_Z}}
\newcommand{\mzs}{M^2_{_Z}}
\newcommand{\mzc}{M^3_{_Z}}
\newcommand{\bzms}{M^2_{_0}}
\newcommand{\bzmc}{M^3_{_0}}
\newcommand{\bhms}{M^2_{_{0H}}}
\newcommand{\mhs}{M^2_{_H}}
\newcommand{\mfs}{m^2_f}
\newcommand{\mfc}{m^3_f}
\newcommand{\mfps}{m^2_{f'}}
\newcommand{\mfhs}{m^2_{h}}
\newcommand{\mfpc}{m^3_{f'}}
\newcommand{\mts}{m^2_t}
\newcommand{\mes}{m^2_e}
\newcommand{\mms}{m^2_{\mu}}
\newcommand{\mmc}{m^3_{\mu}}
\newcommand{\mmfour}{m^4_{\mu}}
\newcommand{\mmf}{m^5_{\mu}}
\newcommand{\mmfive}{m^5_{\mu}}
\newcommand{\mmsix}{m^6_{\mu}}
\newcommand{\mminv}{\frac{1}{m_{\mu}}}
\newcommand{\mtaus}{m^2_{\tau}}
\newcommand{\mus}{m^2_u}
\newcommand{\mds}{m^2_d}
\newcommand{\muqps}{m'^2_u}
\newcommand{\mdqps}{m'^2_d}
\newcommand{\mcs}{m^2_c}
\newcommand{\mss}{m^2_s}
\newcommand{\mbs}{m^2_b}
\newcommand{\mups}{M^2_u}
\newcommand{\mdps}{M^2_d}
\newcommand{\mcps}{M^2_c}
\newcommand{\msps}{M^2_s}
\newcommand{\mbps}{M^2_b}
\newcommand{\mt}{m_t}
\newcommand{\gf}{G_{\ssF}}
\newcommand{\ssZ}{{\scriptscriptstyle{\zb}}}
\newcommand{\ssW}{{\scriptscriptstyle{\wb}}}
\newcommand{\ssH}{{\scriptscriptstyle{\hb}}}
\newcommand{\ssV}{{\scriptscriptstyle{\vb}}}
\newcommand{\ssA}{{\scriptscriptstyle{A}}}
\newcommand{\ssB}{{\scriptscriptstyle{B}}}
\newcommand{\ssC}{{\scriptscriptstyle{C}}}
\newcommand{\ssD}{{\scriptscriptstyle{D}}}
\newcommand{\ssF}{{\scriptscriptstyle{F}}}
\newcommand{\ssG}{{\scriptscriptstyle{G}}}
\newcommand{\ssL}{{\scriptscriptstyle{L}}}
\newcommand{\ssM}{{\scriptscriptstyle{M}}}
\newcommand{\ssN}{{\scriptscriptstyle{N}}}
\newcommand{\ssP}{{\scriptscriptstyle{P}}}
\newcommand{\ssQ}{{\scriptscriptstyle{Q}}}
\newcommand{\ssR}{{\scriptscriptstyle{R}}}
\newcommand{\ssS}{{\scriptscriptstyle{S}}}
\newcommand{\ssT}{{\scriptscriptstyle{T}}}
\newcommand{\ssU}{{\scriptscriptstyle{U}}}
\newcommand{\ssX}{{\scriptscriptstyle{X}}}
\newcommand{\ssY}{{\scriptscriptstyle{Y}}}
\newcommand{\ssI}{{\scriptscriptstyle{I}}}
\newcommand{\ord}[1]{{\cal O}\lpar#1\rpar}
\newcommand{\als}{\alpha_{_S}}
%
%
\def\beq{\begin{equation}}
\def\eeq{\end{equation}}
\def\beqar{\begin{eqnarray}}
\def\eeqar{\end{eqnarray}}
\def\barr#1{\begin{array}{#1}}
\def\earr{\end{array}}
\def\bfi{\begin{figure}}
\def\efi{\end{figure}}
\def\btab{\begin{table}}
\def\etab{\end{table}}
\def\bce{\begin{center}}
\def\ece{\end{center}}
\def\nn{\nonumber}
\def\nl{\nonumber\\}

\def\al{\alpha}
\def\be{\beta}
\def\ga{\gamma}

\def\refeq#1{\mbox{(\ref{#1})}}
\def\reffi#1{\mbox{Figure~\ref{#1}}}
\def\reffis#1{\mbox{Figures~\ref{#1}}}
\def\refta#1{\mbox{Table~\ref{#1}}}
\def\reftas#1{\mbox{Tables~\ref{#1}}}
\def\citere#1{\mbox{Ref.~\cite{#1}}}
\def\citeres#1{\mbox{Refs.~\cite{#1}}}

\newcommand{\GeV}{\unskip\,\mathrm{GeV}}
\newcommand{\MeV}{\unskip\,\mathrm{MeV}}

\newcommand{\ri}{{\mathrm{i}}}

\renewcommand{\O}{{\cal O}}
\newcommand{\M}{{\cal{M}}}

\def\mathswitchr#1{\relax\ifmmode{\mathrm{#1}}\else$\mathrm{#1}$\fi}
\newcommand{\PW}{\mathswitchr W}
\newcommand{\PZ}{\mathswitchr Z}
\newcommand{\PH}{\mathswitchr H}
\newcommand{\Pe}{\mathswitchr e}
\newcommand{\Pd}{\mathswitchr d}
\newcommand{\Pu}{\mathswitchr u}
\newcommand{\Ps}{\mathswitchr s}
\newcommand{\Pc}{\mathswitchr c}
\newcommand{\Pb}{\mathswitchr b}
\newcommand{\Pt}{\mathswitchr t}
\newcommand{\Pep}{\mathswitchr {e^+}}
\newcommand{\Pem}{\mathswitchr {e^-}}
\newcommand{\PWp}{\mathswitchr {W^+}}
\newcommand{\PWm}{\mathswitchr {W^-}}

\def\mathswitch#1{\relax\ifmmode#1\else$#1$\fi}
\newcommand{\MW}{\mathswitch {M_\PW}}
\newcommand{\MZ}{\mathswitch {M_\PZ}}
\newcommand{\MH}{\mathswitch {M_\PH}}
\newcommand{\Me}{\mathswitch {m_\Pe}}
\newcommand{\GW}{\Gamma_{\PW}}
\newcommand{\GZ}{\Gamma_{\PZ}}

\newcommand{\cw}{\mathswitch {c_\mathrm{w}}}
\newcommand{\sw}{\mathswitch {s_\mathrm{w}}}
\newcommand{\GF}{\mathswitch {G_\mu}}

\newcommand{\lsim}
{\mathrel{\raisebox{-.3em}{$\stackrel{\displaystyle <}{\sim}$}}}
\newcommand{\gsim}
{\mathrel{\raisebox{-.3em}{$\stackrel{\displaystyle >}{\sim}$}}}

\def\ie{i.e.\ }
\def\eg{e.g.\ }
\def\cf{cf.\ }

\newcommand{\DPA}{{\mathrm{DPA}}}
\newcommand{\virt}{{\mathrm{virt}}}
\newcommand{\fact}{{\mathrm{fact}}}
\newcommand{\nonfact}{{\mathrm{nfact}}}

\newcommand{\eeWW}{{\Pe^+ \Pe^-\to \PW^+ \PW^-}}
\newcommand{\Wpff}{{\PW^+ \to f_1\bar f_2}}
\newcommand{\Wmff}{{\PW^- \to f_3\bar f_4}}
\newcommand{\eeWWffff}{\Pep\Pem\to\PW\PW\to 4f}
\newcommand{\eeffff}{\Pep\Pem\to 4f}
\newcommand{\eeffffg}{\eeffff\ga}

\newcommand{\mpar}[1]{{\marginpar{\hbadness10000%
                      \sloppy\hfuzz10pt\boldmath\bf#1}}%
                      \typeout{marginpar: #1}\ignorespaces}
\def\draftdate{\relax}
\def\mua{\marginpar[\boldmath\hfil$\uparrow$]%
                   {\boldmath$\uparrow$\hfil}%
                    \typeout{marginpar: $\uparrow$}\ignorespaces}
\def\mda{\marginpar[\boldmath\hfil$\downarrow$]%
                   {\boldmath$\downarrow$\hfil}%
                    \typeout{marginpar: $\downarrow$}\ignorespaces}
\def\mla{\marginpar[\boldmath\hfil$\rightarrow$]%
                   {\boldmath$\leftarrow $\hfil}%
                    \typeout{marginpar: $\leftrightarrow$}\ignorespaces}
\def\Mua{\marginpar[\boldmath\hfil$\Uparrow$]%
                   {\boldmath$\Uparrow$\hfil}%
                    \typeout{marginpar: $\Uparrow$}\ignorespaces}
\def\Mda{\marginpar[\boldmath\hfil$\Downarrow$]%
                   {\boldmath$\Downarrow$\hfil}%
                    \typeout{marginpar: $\Downarrow$}\ignorespaces}
\def\Mla{\marginpar[\boldmath\hfil$\Rightarrow$]%
                   {\boldmath$\Leftarrow $\hfil}%
                    \typeout{marginpar: $\Leftrightarrow$}\ignorespaces}
%
%
\newcommand{\sss}[1]{\mbox{\scriptsize #1}}
\newcommand{\real}{{\rm{Re}}}
\newcommand{\imag}{{\rm{Im}}}
\newcommand{\J}{{\cal J}}
\newcommand{\I}{{\cal I}}
\newcommand{\OO}{{\cal O}}
\newcommand{\PP}{{\cal P}}
\newcommand{\D}{{\cal D}}
\newcommand{\T}{\mbox{T}}
\newcommand{\Tr}{\mbox{Tr}\,}
\newcommand{\F}{\mbox{F}}
\newcommand{\BBCG}{\mbox{G}}
\newcommand{\TF}{\mbox{\boldmath $F$}}
\newcommand{\TA}{\mbox{\boldmath $A$}}
\newcommand{\LL}{{\cal L}}
\newcommand{\LNL}{{\cal L}_{\sss{NL}}}
\newcommand{\SNL}{{\cal S}_{\sss{NL}}}
\newcommand{\SL}{{\cal S}_{\sss{L}}}
\newcommand{\Ds}{D\hspace{-0.63em}/\hspace{0.1em}}
\newcommand{\TAs}{\TA\hspace{-0.65em}/\hspace{0.1em}}
\newcommand{\partials}{\partial\hspace{-0.50em}/\hspace{0.1em}}
%
%
\newcommand{\bfig}{\begin{center}\begin{picture}}
\newcommand{\efig}[1]{\end{picture}\\{\small #1}\end{center}}
\newcommand{\flin}[2]{\ArrowLine(#1)(#2)}
\newcommand{\wlin}[2]{\DashLine(#1)(#2){2.5}}
\newcommand{\zlin}[2]{\DashLine(#1)(#2){5}}
\newcommand{\glin}[3]{\Photon(#1)(#2){2}{#3}}
\newcommand{\lin}[2]{\Line(#1)(#2)}
\newcommand{\sof}{\SetOffset}
\newcommand{\bmip}[2]{\begin{minipage}[t]{#1pt}\bfig(#1,#2)}
\newcommand{\emip}[1]{\efig{#1}\end{minipage}}
\newcommand{\putk}[2]{\Text(#1)[r]{$p_{#2}$}}
\newcommand{\putp}[2]{\Text(#1)[l]{$p_{#2}$}}
\newcommand{\ibidem}{{\it ibidem\/},}
\newcommand{\vpb}{}
\newcommand{\p}[1]{{\scriptstyle{\,(#1)}}}
%
%
\newcommand{\parent}[1]{\lpar#1\rpar}
\newcommand{\rbrak}[1]{\lrbr#1\rrbr}
\newcommand{\ra}{\rightarrow}
\newcommand{\beanon}{\begin{eqnarray*}}
\newcommand{\eeanon}{\end{eqnarray*}}
\newcommand{\ul}{\underline}
\newcommand{\ol}{\overline}
\newcommand{\dotp}{\!\cdot\!}
\newcommand{\thet}[1]{\theta\lpar#1\rpar}
\newcommand{\delt}[1]{\delta\lpar#1\rpar}
\newcommand{\gtap}{\stackrel{\displaystyle >}{\,_{\! \,_{\displaystyle
\sim}}}}  
\newcommand{\ltap}{\stackrel{\displaystyle <}{\,_{\! \,_{\displaystyle
\sim}}}}  
\newcommand{\umu}{^{\mu}}
\newcommand{\lmu}{_{\mu}}
\newcommand{\ub}{\bar{u}}
\newcommand{\vbar}{\bar{v}}
\newcommand{\gp}{(1+\gamma^5)}
\newcommand{\ep}{\epsilon}
\newcommand{\sla}[1]{/\!\!\!#1}
\newcommand{\suml}{\sum\limits}
\renewcommand{\to}{\rightarrow}
\newcommand{\unity}{1\!\!1}
\renewcommand{\iff}{\;\;\Longleftrightarrow\;\;}
\newcommand{\gsovermu}{\kappa}
\newcommand{\mut}{m_t^2}
\newcommand{\gmw}{\Gamma_{_W}}
\newcommand{\gmz}{\Gamma_{_Z}}
\newcommand\pb{\;[\mathrm{pb}]}
\newcommand{\processccten}{$e^-e^+\to \mu^-\bar{\nu}_\mu u\bar{d}$}
\newcommand{\processcceleven}{$e^-e^+\to s\bar{c} u\bar{d}$}
\newcommand{\processcctwenty}{$e^-e^+\to e^-\bar{\nu}_eu\bar{d}$}
\newcommand{\processccemu}{$e^-e^+\to e^-\bar{\nu}_e\mu^+\nu_\mu$}
\newcommand{\processmixeevv}{$e^-e^+\to e^-e^+\nu_e\bar{\nu}_e$}
\newcommand{\processnceevv}{$e^-e^+\to e^-e^+\nu_\mu\bar{\nu}_\mu$}
\newcommand{\wph}{{\tt WPHACT}}
\newcommand{\wto}{{\tt WTO}}
\newcommand{\com}[1] {{(\bf #1})}
%
%
\def\Was{W\c as}
\def\Order#1{${\cal O}(#1$)}
\def\Ordpr#1{${\cal O}(#1)_{prag}$}
\def\bbe{\bar{\beta}}
\def\tbe{\tilde{\beta}}
\def\tal{\tilde{\alpha}}
\def\tom{\tilde{\omega}}
\def\half{ {1\over 2} }
\def\alf1{ {\alpha\over\pi} }

%
%
\title{\bf Four-Fermion Production in Electron-Positron Collisions}

\author{
Martin W.~Gr\"unewald$^{1}$ and Giampiero Passarino$^{2,3}$\\[0.5cm]
   E.~Accomando$^{4}$,
   A.~Ballestrero$^{3}$,
   P.~Bambade$^{5,6}$,
   D.~Bardin$^{7}$,
   W.~Beenakker$^{8}$,
   F.~Berends$^{9}$,
   E.~Boos$^{10}$,
   A.~Chapovsky$^{8}$,
   A.~Denner$^{4}$,
   S.~Dittmaier$^{11}$,
   M.~Dubinin$^{10}$,
   J.~B.~Hansen$^{6}$,
   V.~Ilyin$^{10}$,
   S.~Jadach$^{12,25,26}$,
   Y.~Kurihara$^{13}$,
   M.~Kuroda$^{14}$,
   E.~Maina$^{2,3}$,
   G.~Montagna$^{15,16}$,
   M.~Moretti$^{17}$,
   O.~Nicrosini$^{16,15}$,
   A.~Olshevsky$^{7,6}$, 
   M.~Osmo$^{15,16}$,
   A.~Pallavicini$^{15,16}$,
   C.~G.~Papadopoulos$^{18}$,
   H.~T.~Phillips$^{19,\dagger}$,
   F.~Piccinini$^{16,15}$,
   R.~Pittau$^{2,3}$,
   W.~Placzek$^{26,27}$,
   T.~Riemann$^{20}$,
   M.~Roth$^{21}$,
   A.~S.~Schmidt-Kaerst$^{22}$,
   Y.~Shimizu$^{13}$,
   M.~Skrzypek$^{25,26}$,
   R.~Tanaka$^{23}$,
   M.~Verzocchi$^{6}$,
   D.~Wackeroth$^{24}$,
   B.~F.~L.~Ward$^{26,28,29}$,
   Z.~W\c{a}s$^{25,26}$}
\maketitle

\begin{itemize}

\item[$^1$]
             Institut f\"ur Physik, Humboldt-Universit\"at zu Berlin, Germany
\item[$^2$]
             Dipartimento di Fisica Teorica,
             Universit\`a di Torino, Torino, Italy
\item[$^3$]
             INFN, Sezione di Torino, Torino, Italy

\item[$^4$]
             Paul Scherrer Institut, Villigen, Switzerland

\item[$^5$]
             LAL, B.P. 34, F-91898 Orsay Cedex

\item[$^6$]
             CERN, EP Division, CH-1211 Geneva 23, Switzerland

\item[$^7$]
             LNP, JINR, RU-141980 Dubna, Russia

\item[$^8$]
             Physics Department, University of Durham, Durham, England

\item[$^9$]
             Instituut-Lorentz, University of Leiden, The Netherlands

\item[$^{10}$]
             Institute of Nuclear Physics, Moscow State University,
             Moscow, Russia

\item[$^{11}$] Theoretische Physik, Universit\"at Bielefeld,
               Bielefeld, Germany

\item[$^{12}$] DESY, Theory Division, D-22603 Hamburg, Germany

\item[$^{13}$]
             High Energy Accelerator Research Organization,
             Tsukuba, Japan

\item[$^{14}$]
             Institute of Physics, Meiji-Gakuin University, Yokomama, Japan

\item[$^{15}$]
             Dipartimento di Fisica Teorica e Nucleare,
             Universit\`a di Pavia, Italy

\item[$^{16}$]
             INFN, Sezione di Pavia, Pavia, Italy

\item[$^{17}$]
             Dipartimento di Fisica, Universit\`a di Ferrara,
             INFN, sezione di Ferrara

\item[$^{18}$]
             Insitute of Nuclear Physics, NCSR `Democritos', 
             15310 Athens, Greece

\item[$^{19}$]
             University College, London, England

\item[$^{20}$]
             Theory Group, DESY, D-15738 Zeuthen, Germany
                                                                           
\item[$^{21}$]
             Institut f\"ur Theoretische Physik, Universit\"at Leipzig,
             Leipzig, Germany

\item[$^{22}$]
             RWTH Aachen, Germany

\item[$^{23}$] LPNHE, Ecole Polytechnique, F-91128 Palaiseau CEDEX, FRANCE

\item[$^{24}$]
             Department of Physics and Astronomy, University of Rochester,
             Rochester NY, USA

\item[$^{25}$] Institute of Nuclear Physics, ul. Kawiory 26a, 30-055 
               Cracow, Poland

\item[$^{26}$] CERN, Theory Division, CH-1211 Geneva 23, Switzerland

\item[$^{27}$] Institute of Computer Science, Jagellonian University,\\
               ul. Nawojki 11, 30-072 Cracow, Poland

\item[$^{28}$] Department of Physics and Astronomy,\\
               The University of Tennessee, Knoxville, Tennessee 
               37996-1200, USA

\item[$^{29}$] SLAC, Stanford University, Stanford, California 94309, USA

\item[$\dagger$] {\em previously at} Rutherford Appleton Lab,
                  United Kingdom
 
\end{itemize}

\clearpage

$ $
\vfill

\centerline{\bf\large The LEP-2 Monte Carlo Workshop 1999/2000}

\vskip 1cm

\centerline{\bf\LARGE Four-Fermion Production in Electron-Positron Collisions}

\vskip 1cm

\centerline{\bf\large Four-Fermion Working Group Report} 

\vskip 6cm

\centerline{\bf\LARGE Abstract} 

\vskip 2cm 

This report summarises the results of the four-fermion working group
of the LEP2-MC workshop, held at CERN from 1999 to 2000.  Recent
developments in the calculation of four-fermion processes in
electron-positron collisions at LEP-2 centre-of-mass energies are
presented, concentrating on predictions for four main reactions:
W-pair production, visible photons in four-fermion events, single-W
production and Z-pair production.  Based on a comparison of results
derived within different approaches, theoretical uncertainties on
these predictions are established.

\vfill

$ $


%
%
\tableofcontents 
%
%
%
\section{Introduction}

During the year 1999 an informal workshop on Monte Carlo (MC)
generators and programs took place at CERN, concentrating on processes
in $e^+e^-$ interactions at LEP~2 
centre-of-mass energies (161 GeV to 210 GeV).  One of the
goals was to summarize and review critically the progress made in
theoretical calculations and their implementation in computer programs
since the 1995 workshop on {\it Physics at LEP2}.  One of the reasons
for this report was the need of having an official statement on
various physics processes and the accuracy of their predictions,
before deciding on LEP~2 activities in the year 2000.

This part of the workshop report summarizes the findings in the area
of {\em Four-Fermion} final states.  At the beginning of the workshop
the following goals were identified for the {\em Four-Fermion}
sub-group:
\begin{itemize}
\item[a)] Describe the new calculations and improvements in the
  theoretical understanding and in the upgraded MC implementations.
\item[b)] Indicate where new contributions have changed previous
  predictions in the MC adopted by the collaborations, and specify
  why, how and by how much.
\item[c)] In those cases where a substantial discrepancy has been
  registered and the physical origin has been understood,
  recommendations should be made on what to use.
\item[d)] In those cases where we have found incompleteness of the
  existing MC, but no complete improvement is available, we should be
  able to indicate a sound estimate of the theoretical uncertainty,
  and possibly way and time scale for the solution.
\end{itemize}
Our strategy is determined by the physics issues arising in the
experimental analyses performed at LEP~2.  Therefore, the four LEP
Collaborations have been asked to provide a list of relevant processes
together with the level of theoretical accuracy needed.

Clearly, the LEP experiments investigate many different processes.
For theoretical predictions we thus have to manage with lots of
different sets of cuts.  At the beginning of our activities the four
experiments have presented us with lists that reflect rather diverse
styles and different approaches: The complexity of the observables
varied greatly, ranging from those defined by simple phase-space cuts
on four-fermion (+ photon) level to complete event-selection
procedures requiring parton shower and hadronization of quark systems.

An effort was made to settle as much as possible on a set of
quasi-realistic but simple cuts for each process.  We have collected
processes and/or phase space regions where improved theoretical
predictions are desirable.  A weight has been assigned to each process
according to its relevance and urgency.

The focus of activity has been on improving the theoretical
predictions for the relevant processes and/or phase space regions.
Also, all contributors have been asked to give an estimate for the
remaining theoretical uncertainty.  As a consequence, the output of
the whole operation should not be a mere collection of comparison
tables but a coherent attempt in assessing the theoretical uncertainty
to be associated to any specific process.

The realm of theoretical uncertainty is ill defined and in order to
reach a general consensus one cannot be satisfied with just some
statement on the overall agreement among different programs.  Whenever
differences are found, one has to make sure that they are due to
physics, and not to some different input.  So our project had to
foresee a preliminary phase with more of a technical benchmark.  Once
trivial discrepancies are understood and sorted out, one can start
digging into inevitable differences arising from different
implementations of common theoretical wisdom.

In a vast majority of cases the main theoretical problem is
represented by the inclusion of QED radiation. Therefore, one of the
main questions was: can we improve upon our treatment of QED radiation
and/or give some safe estimate of the theoretical uncertainty
associated with it?

Below we will present our reference table of four-fermion processes.
It is an idealised common ground where, in principle, all theoretical
predictions should be compared. More advanced setups would be
accessible only to a more limited number of generators, built for that
specific purpose.

It is useful to recall that the ultimate, perfect program does not
exist and, most likely, will never exist. Roughly speaking, programs
belong to two quite distinct classes.  On one side there are event
generators, usually interfaced with parton shower and hadronization
packages.  They may miss some fine points of the theoretical knowledge
but represent an essential ingredient in the experimental analyses
concerning the evaluation of signal efficiencies and backgrounds.
Thus they create the necessary bridge between the raw data recorded by
the detectors and the background-subtracted efficiency-corrected
results published.  At the other end of this cosmos we have
semi-analytical programs that are not meant to generate events.
Rather, they show their power in dealing with the signal, furnishing
the implementation of (almost) everything available in the literature
concerning the calculation of specific processes. In either case, we
want to know about the theoretical uncertainty, process by process, to
make clear which program is able to achieve that level of accuracy
under which configuration.
For $\wb$-pair production, however, the scenario is slightly changed: 
We have now MC event generators that, at the same time, represent
a state-of-the-art calculation. Nevertheless, we do not have yet
the ultimate MC: the one with radiative corrections,
virtual/soft/hard photons, DPA, complete phase-space including single-$\wb$,
single-$\zb$, $\zb\ph^*$ and able to produce weight-1 events in finite time.

The results presented in this report are based on several different
approaches and on comparisons of their numerical predictions.  They
are calculated with the following computer codes: {\tt BBC}, {\tt
  Comp\-HEP}, {\tt GENTLE}, {\tt grc4f}, {\tt KORALW/YFSWW/YFSZZ}, {\tt
  NEXTCALIBUR}, {\tt PHEGAS/HELAC}, {\tt RacoonWW}, {\tt SWAP/WRAP},
{\tt WPHACT} and {\tt WTO/ZZTO}.

This article is organised as follows. In \sect{ffp} we
present the four-fermion processes looked at in detail. Then we review
the most recent theoretical developments in four-fermion physics in
$e^+e^-$ interactions.  In \sect{sectWW} we discuss the CC03
$\sigma_{\ssW\ssW}$ cross-section and predictions based on the DPA.
Here different approaches are compared. In
\sect{sect4fg} we discuss the radiative process with $4\rmf + \ph$
final states. In \sect{sectsw} the single-$\wb$ production is
critically discussed. Finally the NC02 cross-section,
$\sigma_{\ssZ\ssZ}$ is analysed in \sect{sectZZ}
Conclusions and outlook are presented in \sect{conco}

\clearpage

\section{Four-fermion processes}
\label{ffp}

Here we present our basic reference table and specify the
calculational setup.  One should read it as summarizing our original
manifest. After reading the following sections, it will be instructive
to come back here with a critical eye: not all the items and questions
listed below have found a satisfactory answer.  This was, somehow,
foreseeable. If one thinks carefully one will easily discover some
important message also for those issues that remain unsolved: they
cannot be solved in any reasonable time scale and the associated
effect is a real source of uncertainty.

\subsection{List of processes}

The following list provides the observables together with precision
tags in $\%$, as requested by the experimental Collaborations.  
The accuracy of MC simulations should be better than
the requested precision tag, i.e.  the {\em physics} uncertainty
should be smaller and at worst the one indicated. How much better is
left to the contributors. For benchmarking it is certainly advisable
to use the maximum available precision. 

In general, radiative corrections and radiative photons in the final
state should be considered for all processes, including the discussion
of photon energy and polar-angle spectra.  Typical minimal
requirements on real photons are: energy $E_{\ph} > 1\,$GeV; polar
angle $|\cos\theta_{\ph}| < 0.985, 0.997, 0.9995$ depending on
channel; and minimal angle between photon and any charged final-state
fermion $\xi>5^\circ$.

\begin{itemize}
\item $\wb\wb$ and $\zb\zb$ type signal:
\end{itemize}

\begin{enumerate}
  
\item $e^+e^- \to \wb\wb \to \,$ all (CC03). The full phase space is
  needed and the inclusive cross-section accuracy is $0.2\%$, which is
  $1/3$ of experimental accuracy combining all LEP~2 energies,

  The spectrum for the photon energy and the polar angle is needed
  ($|\cos\theta_{\ph}| < 0.985~(0.997)$).

\item $e^+e^- \to \zb\zb \to \,$ all (NC02).  The full phase space is
  needed and the inclusive cross-section accuracy is $1\%$. The
  spectrum for the photon energy and the polar angle is needed
  ($|\cos\theta_{\ph}| < 0.985~(0.997)$).

\item $e^+e^- \to l\nu l\nu(\ph)$ where all $\{e/\mu/\tau\} \,\otimes
  \{e/\mu/\tau\}$ combinations are requested with the following
  conditions: ($|\cos\theta_{l_1/l_2}| < 0.985$, $E_{l_1/l_2} >
  5\,$GeV, $M(l^+l^-) > 10~(45)\,$GeV (full and high-mass region).
  The inclusive cross-section accuracy is $4\%$ for individual
  combination; the inclusive cross-section accuracy is $1\%$ for the
  summed one; photon energy and polar angle spectrum
  ($|\cos\theta_{\ph}| < 0.985~(0.997)$).
  
\item $e^+e^- \to \barq q e \nu(\ph)$ (CC20), $q$-flavour blind,
  $|\cos\theta_e| < 0.985$, $E_e > 5\,$GeV, $M(q\barq ) >
  10~(45)\,$GeV (full and high-mass region); inclusive cross-section
  accuracy is $1\%$ (5\% for low-mass region); photon energy and
  polar angle spectrum ($|\cos\theta_{\ph}| < 0.985~(0.997)$).
  
\item $e^+e^- \to \barq q \mu\nu(\ph)$ and $e^+e^- \to \barq q
  \tau\nu(\ph)$ (incl. tau polarization in tau decay) (CC10),
  $|\cos\theta_{\mu/\tau}| < 0.985, E_{\mu/\tau} > 5\,$GeV, $M(q\barq
  ) > 10~(45)\,$GeV (full and high-mass region), inclusive
  cross-section accuracy $1\%$.  Photon energy and polar angle
  ($|\cos\theta_{\ph}| < 0.985~(0.997)$) spectrum.
  
\item $e^+e^- \to q\barq q\barq (\ph)$, flavour blind, $bb q\barq $,
  $bbbb$.  At least two pairs with $M(q_i,q_j) > 10~(45)\,$GeV (full
  and high-mass region), inclusive cross-section accuracy $1\%$.
  photon energy and polar angle ($|\cos\theta_{\ph}| < 0.985~(0.997)$)
  spectrum.
  
\item $e^+e^- \to q\barq l^+l^-(\ph)$, $q$-flavour blind, heavy
  $q$-flavors, $l=e/\mu/\tau$, $|\cos\theta_{l_1}| < 0.985$, no cut
  on 2nd lepton (only one lepton tagged), $M(q\barq ) > 10~(45)\,$GeV
  (full and high-mass region), inclusive cross-section accuracy $2\%$.
  Photon energy and polar angle ($|\cos\theta_{\ph}| < 0.985~(0.997)$)
  spectrum.
  
\item $e^+e^- \to q\barq l^+l^-(\ph)$, $q$-flavour blind, heavy
  $q$-flavors, $|\cos\theta_{l_1}|,|\cos\theta_{l_2}| < 0.985$ (both
  leptons tagged), full and high-mass regions: $M(l^+l^-) >
  10~(45)\,$GeV, $M(q\barq ) > 10~(45)\,$GeV, inclusive cross-section
  accuracy $2\%$.  Photon energy and polar angle ($|\cos\theta_{\ph}|
  < 0.985~(0.997)$) spectrum.
  
\item $e^+e^- \to q\barq e^+e^-(\ph)$, $q$-flavour blind, heavy
  $q$-flavors, with one electron in the beam pipe, $|\cos\theta_e| >
  0.997$, and one electron tagged, $|\cos\theta_e| < 0.985$, $M(q\barq
  ) > 10~(45)\,$GeV (full and high-mass region) .  Photon energy and
  polar angle ($|\cos\theta_{\ph}| < 0.985~(0.997)$) spectrum.
  
\item $e^+e^- \to q\barq \nu\barnu(\ph)$, $q$-flavour blind, heavy
  $q$-flavors, $M(q\barq ) > 10~(45)\,$GeV, inclusive cross-section
  accuracy $4\%$ (10\% for low-mass region).  Photon energy and polar
  angle ($|\cos\theta_{\ph}| < 0.985~(0.997)$) spectrum.
  
\item $e^+e^- \to l^+l^-L^+L^-(\ph)$ and $e^+e^- \to
  l^+l^-l^+l^-(\ph)$ (all possible charged lepton flavour
  combinations): 3 or 4 leptons within acceptance $|\cos\theta| <
  0.985$, $M(l^+l^-)$ and $M(L^+L^-) > 10~(45)\,$GeV (full and
  high-mass region).  Photon energy and polar angle
  ($|\cos\theta_{\ph}| < 0.985~(0.997)$) spectrum.

\end{enumerate}

\begin{itemize}
\item Single-$\wb$ type signal:
\end{itemize}

\begin{enumerate}
  
\item $e^+e^- \to q\barq e\nu(\ph)$, $|\cos\theta_e| > 0.997$, either
  $M(q\barq ) > 45\,$GeV or $E_{q_1}, E_{q_2} > 15\,$GeV, inclusive
  cross-section accuracy $3\%$, photon energy and 
polar angle ($|\cos\theta_{\ph}| < 0.997~(0.9995)$) spectrum.
  
\item $e^+e^- \to e\nu e\nu(\ph)$, $|\cos\theta_e| > 0.997$, $E_e >
  15\,$GeV, $|\cos\theta_e| < 0.7~(0.95)$, inclusive cross-section
  accuracy $5\%$, photon energy and polar angle ($|\cos\theta_{\ph}| <
  0.997~(0.9995)$) spectrum.
  
\item $e^+e^- \to e\nu\mu\nu(\ph)$ and $e^+e^- \to e\nu\tau\nu(\ph)$,
  $|\cos\theta_e| > 0.997$, $E_{\mu/\tau} > 15\,$GeV,
  $|\cos\theta_{\mu/\tau}| < 0.95$, inclusive cross-section accuracy
  $5\%$, photon energy and polar angle ($|\cos\theta_{\ph}| <
  0.997~(0.9995)$) spectrum.

\end{enumerate}
This list deserves already few words of comment.

For hadronic systems (CC or NC), there is usually a requirement of at
least $45$~GeV invariant mass ($\wb$ and $\zb$ signal) or at least
$10$~GeV (background for other processes).  Even lower invariant
masses, say down to $1\,$GeV, should be handled by the dedicated
$\ph\ph$ subgroup. For leptons, there should be no problem to go down
to lower invariant masses or energies than listed above.

We consider as radiative events those events with real photons where
at least one photon passes the photon requirements listed above, and
as non-radiative events those with no photon or only photons below the
minimal photon requirements. In case of non-radiative and radiative
events, the cross section and its accuracy is needed. In case of
non-radiative events, this amounts to adding up virtual and soft
radiative corrections. In case of radiative events, some distributions
are needed in addition, in particular photon energy and polar angle,
and photon angle with respect to the nearest charged final-state
fermion.

\subsection{Questions to theory}

We now elaborate in more detail on specific questions
associated to specific processes.

\begin{itemize}
\item {$\O(\al)$ electroweak corrections to $e^+e^- \to \wb\wb \to 4\,\rmf$.}
\end{itemize}

Until 1999, the LEP experiments were using a $2\%$ theoretical
uncertainty on the calculation of the CC03 $\wb$-pair cross section, not
changed since the 1995 LEP~2 workshop. Although no complete one-loop
$\ord{\alpha}$ EW calculation exist yet 
for off-shell $e^+e^- \to \wb\wb \to 4\,$f production, we wish the
theoretical uncertainty to be below $1\%$ ($0.5\%$ if possible) with
justification.  Also the uncertainties in CC03 vs. $4\rmf$
corrections when measuring the $\wb\wb$ cross section should be
understood.

\begin{itemize}
\item {Photon radiation (ISR) with $p_t$ in $\wb\wb$ and
    $\zb\zb$-dominated channels.}
\end{itemize}

The principle effects will be on the selection efficiency and on the
differential distributions used for W mass and triple gauge boson
coupling (TGC) studies.  The interest in photons is twofold: photons
explicitly identified as such - usually at larger polar angles - and
photons which simply create noticeable activity in the detector.  The
latter is, for example, also important in single-$\wb$ type analysis,
therefore the photon angular range is extended to very low polar
angles.

\begin{itemize}
\item {Single $\wb$ channels.}
\end{itemize}

For the single-$\wb$ process there are several issues to be addressed.
In the region of high invariant masses of the $\wb$~boson (above $45\,$GeV) 
this process is important for
both searches and TGC measurements.  
One topic of investigation should
be ISR: this process is dominated by $t$-channel diagrams, whereas the
current MC program implement ISR assuming s-channel reactions.  A
second issue is the treatment of the $\alpha_{\rm QED}$ scale, not only
for single-$\wb$ but also for single-$\zb$ and for $\zb\ph^*$. Is it
better to re-weight on a event by event basis or on a diagram basis?

One of the outcomes of the workshop should be a recommendation on the
mass cut which distinguishes the high mass region (more reliable) from
the low mass region, i.e. the lower value to which the $5\%$ (or
better) precision tag applies.

The importance of ISR in this channel is threefold: (a) change in
total cross-section due to normal radiative corrections, (b) change in
event distributions used to make cuts which changes the fraction of
the total that fall inside our cuts, (c) fraction of events with
identified photons - this forms a background to some of the chargino
searches where a detected gamma is required.

Since the single-$\wb$ topology is defined as the one where a
high mass object is found in the detector and the electron is not
observed, we would like to know how the presence of $p_t$ ISR changes
the fraction of events where the electron gets {\em kicked} out of the
beam-pipe, how the differential distributions are distorted for TGC
studies and what the explicit hard photon rate is.

The low mass region (below $45$~GeV) is mostly important for searches
and studied within the $\ph\ph$ sub-group.  One would like to trust
the MC predictions down to $5-10\,$GeV invariant mass for the hadronic
system.  The required precision should also be around $5$ to $10\%$.

\subsection{Input parameter set}

A set of parameters must be specified for the calculation
of $\ord{\alpha}$ predictions (CC03 and to some extent also NC02).
Once radiative corrections are included, the question of
Renormalization Scheme (RS) and of Input Parameter Set (IPS) becomes
relevant.  For calculation, the following input parameters are used:
\bqa
\mz &=& 91.1867\,\GeV, \quad 1/\alpha(0) = 137.0359895, \nl
\gf &=& 1.16637\times 10^{-5}\,\GeV^{-2}.
\eqa
As far as masses are concerned one should use:
\begin{description}
\item[Leptons:] PDG values, i.e.
\bqa
\me &=& 0.51099907\,\MeV, \quad \mm = 105.658389\,\MeV,\nl
\mtau &=& 1.77705\,\GeV.
\eqa
\item[Quarks:] for light quarks one should make a distinction; for
phase space:
\bq
\muq = 5\,\MeV, \qquad \md= 10\,\MeV, \qquad
\mbox{only relevant for single-}\wb,
\eq
while, in principle, these masses should {\em not} be used in deriving 
$\alpha_{\rm QED}(s)$ from $\alpha_{\rm QED}(0)$. 
\end{description}
Here the recommendation follows the agreement in our community on
using the following strategy for the evaluation of $\alpha_{\rm
  QED}$ at the mass of the $\zb$. Define:
\bq
\alpha(\mz) = \frac{\alpha(0)}{1-\Delta\alpha^{(5)}(\mz) 
-\Delta_{\rm{top}}(\mz)-\Delta^{\alpha\als}_{\rm{top}}(\mz)}\;, 
\eq
where one has $\Delta\alpha^{(5)}(\mz) = \Delta\alpha_{\rm{lept}} +
\Delta\alpha^{(5)}_{\rm{had}}$. 

The input parameter should be $\Delta\alpha^{(5)}_{\rm{had}}$, as it
is the contribution with the largest uncertainty, while the
calculation of the top contributions to $\Delta\alpha$ is left for the
code. This should become common to all codes.  Codes should include,
for $\Delta\alpha_{\rm{lept}}$, the recently computed $\ord{\alpha^3}$
terms of \cite{ste} and use as default $\Delta\alpha^{(5)}_{\rm{had}}
= 0.0280398$, taken from \cite{ej}.  Using the default one obtains
$1/\alpha^{(5)}(\mz) = 128.877$, to which one must add the $\ft\bart$
contribution and the $\ord{\alpha\als}$ correction induced by the
$\ft\bart$ loop with gluon exchange, \cite{ak}.  Therefore, light
quark masses should {\em not} appear in the evaluation of $\alpha_{\rm
  QED}(\mz)$ and one should end up with:
\bqa
1/\alpha(\mz) &=& 128.887, \nl
\hbox{for} &{}& \mz= 91.1867\,GeV, \quad \mt = 175\,\GeV, \nl
&{}& \mh = 150\,\GeV, \quad \als(\mz) = 0.119.
\eqa
Furthermore, one should use:
\bq
\als(\mz)= 0.119, \quad \mh = 150\,\GeV, \quad \mw = 80.350\,\GeV.
\eq
The quantities $\gz,\gw$ should be understood as computed in the
minimal standard model, e.g. $\gz = 2.49471\,\GeV$ and $\gw = 2.08699\GeV$
for our IPS.

Now we come to the most important point, what to do with IPS in the
presence of radiative corrections. In principle, all RS and all IPS
are equally good and accepted, and differences are true estimates of
some component of the theoretical uncertainty. However, we want to
make sure that differences are not due to technical precision. The IPS
that we want to specify is over-complete, let us repeat,
\bqa
\gf &=& 1.16637\times 10^{-5}\,\GeV^{-2}, \qquad 1/\alpha(\mz) = 128.887, \nl
\mz &=& 91.1867\,\GeV, \qquad \mw = 80.350\,\GeV, \nl 
\als(\mz) &=& 0.119, \qquad \mh = 150\,\GeV.
\eqa

Clearly, once radiative correction are on, $\stw = \stw^{\rm
  xxx-scheme}$ and we don't care anymore since enough radiative
corrections should be included to make all schemes equivalent to
$\ord{\alpha}$.  Thus, for $\ord{\alpha}$ numbers $\stw$ drops out.
Perhaps we should give the highest marks to schemes where $\mw$ is in
the IPS; after all, experiments measure $\mw$ at LEP~2 and any scheme
where $\mw$ is not a primary quantity in the IPS is as bad as a scheme
for LEP~1 where $\mz$ is a derived quantity.

Nevertheless, we can use the over-completeness of the present IPS to
set some internal consistency: it is a good idea to have an
over-complete set of IPS, nevertheless consistent, so that everybody
can make his favourite choice of the RS.  Since we include values for
$\alpha(\mz)$ and for $\gf$ we can, as well, fine-tune the numbers so
that the internal relations hold, to the best of our knowledge. The
recommendation, in this case, is as follows: 
\bei
\item write down your favorite equation
\bq
f\lpar \mz,\mw,\mt,\mh,\als(\mz),\alpha(0),\gf\rpar = 0,
\eq
\item keep everything fixed but $\mt$ which, in turn, is derived as a
solution of the consistency equation (for OMS this involves typically
$\dr$). 
\eei
Even this solution is RS-dependent but variation should be minimal,
sort of irrelevant.  For instance, one could use the following result
(derived from {\tt TOPAZ0}~\cite{topaz0}):
\bq
\mt = 174.17\,\GeV \quad \mbox{Default for CC03}\,\,\,\ord{\alpha}. 
\eq
With $\mw = 80.350\,GeV$ and $\mh = 150\,$GeV we are in a lucky
situation, $\mt$ doesn't change too much.  For more solutions, we
refer to \tabn{moresol}.

\begin{table}[htbp]\centering
\begin{tabular}{|c|c|}
\hline
$\mh\,[\GeV]$   & $\mt\,[\GeV]$  \\
\hline
100 & 170.03 \\
150 & 174.17 \\
180 & 176.14 \\
250 & 179.90 \\
\hline
\end{tabular}
\caption[]{$\mt$ as a function of $\mh$.\label{moresol}}
\end{table}

\subsection{Comparisons for $4\,\rmf$ results}

There is an old tradition in LEP physics, new theoretical ideas and
improvements should always be cross-checked before being adopted in the
analysis of the experimental data. In this Report we present
accurate and detailed comparisons between different generators.
In most cases the authors have agreed to coordinate their action in
understanding the features of the generators, their intrinsic differences and
the goodness of their agreement or disagreement for the predictions.
In so doing, and for the attuned comparisons, they can exclude that eventual 
disagreement may originate from trivial sources, like different input 
parameters.

Before entering into a detailed study of the numerical results it is important
to underline how an estimate of the theoretical uncertainty emerges from
the many sets of numbers obtained with the available generators.
First of all one may distinguish between {\em intrinsic} and {\em parametric}
uncertainties.
The latter are normally associated with a variation of the input parameters
according to the precision with which they are known. These uncertainties 
will eventually shrink when more accurate measurements will become available.

In this Report we are mainly devoted to a discussion of the intrinsic
uncertainties associated with the choice of one scheme versus another.
With one generator alone one cannot simulate the shift of a given quantity 
due to a change in the renormalization scheme. Thus the corresponding 
theoretical band in that quantity should be obtained from the differences in 
the prediction of the generators.
On top of that we should also take into account the possibility of having
different implementations of radiative corrections within one code.
Many implementations of radiative corrections and of DPA are equally plausible 
and differ by non-leading higher order contributions, which however may 
become relevant in view of the achieved or projected experimental precision. 
This sort of intrinsic theoretical uncertainty can very well be estimated by 
staying within each single generator. However, since there are no reasons 
to expect that these will be the same in different generators, only the full 
collection of different sources will, in the end, give a reliable information 
on how accurate an observable may be considered from a theoretical point of 
view.

\section{Phenomenology of unstable particles}

In order to extract the $\wb\wb$ signal from the full set of $\eeffff$
processes, the CC03 cross-section was introduced and discussed in~\cite{EGWWP}.
In lowest order, this cross-section is simply based on the three
$\wb\wb$~signal diagrams with the full four-particle kinematics with
off-shell $\wb$~bosons. Compared to the full set of diagrams, the CC03
subset depends only trivially on the final state and allows to combine
all channels easily.  However, since the CC03 cross-section is based
on a subset of diagrams, it is gauge-dependent and usually defined in
the 't~Hooft--Feynman gauge.  While the CC03 cross-section is not an
observable, it is nevertheless a useful quantity at LEP~2 energies where it can
be classified as a pseudo-observable. It contains the interesting physics, 
such as the non-abelian couplings and the sensitivity of the total cross 
section to $\mw$ near the $\wb$-pair threshold. 
The goal of this common definition is to be able to combine the
different final state measurements from different experiments so that the 
new theoretical calculations can be checked with data at a level better than 
$1\%$.
Note, however, that the CC03 
cross-section will become very problematic at linear-collider energies, where 
the background diagrams and the gauge dependences are much larger.

It is worth summarizing the status of the $\wb\wb$ cross-section prior to
the 2000 Winter Conferences. Nominally, any calculation for
$\eeWWffff$ was a tree level
calculation and one could try the standard procedure of
including, in a reasonable way, as much as possible of the universal
corrections by constructing an improved Born approximation (hereafter IBA).
This is the way the data have been analyzed so far, mostly with the help
of {\tt GENTLE}.
Different programs have been compared for CC03, see Ref.~\cite{EGWWP}:
when one puts the same input parameters, renormalization scheme, etc, a 
technical agreement at the $0.1\%$ level is found. The universal
corrections are not enough, since we wish 
the theoretical uncertainty to be below $1\%$ ($0.5\%$ seems possible) with 
justification.   

Indeed, we have clear indications that non-universal electroweak corrections
for $\wb\wb$(CC03) cross-section are not small and even larger than the
experimental LEP accuracy.
{\tt GENTLE} will produce a CC03 cross-section, typically in the $\gf$-scheme,
with universal ISR QED and non-universal ISR/FSR QED corrections, implemented
with the so-called current-splitting technique. The corresponding curve has 
been used for the definition of the Standard Model prediction with a 
$\pm 2\%$ systematic error assigned to it.
This error estimate \cite{bo92,LEP2WWreport} is based on the knowledge of 
both leading and full $\ord{\alpha}$ corrections to on-shell $\wb$-pair 
production.
Note that, in {\tt GENTLE}, the non-universal ISR correction with
current-splitting technique reads as $+0.4\%$ effect at LEP~2 energies.
               
Recently, a new electroweak $\ord{\alpha}$ CC03 cross-section has become
available, in the framework of DPA, showing a result that is
$2.5 \div 3\%$
smaller than the CC03 cross-section from {\tt GENTLE}.   
This is a big effect since the combined experimental accuracy of LEP
experiments is even smaller.
It is, therefore, of the upmost importance to understand the structure
of a DPA-corrected CC03 cross-section.  

The double-pole approximation (DPA)
of the lowest-order cross-section emerges from the CC03 diagrams upon
projecting the $\wb$-boson momenta in the matrix element to their on-shell
values. This means that the DPA is based on the residue of the double
resonance, which is a gauge-invariant quantity, because it is directly
related to the sub-processes of on-shell $\wb$-pair production and on-shell
$\wb$~decay.  In contrast to the CC03 cross-section, the DPA is
theoretically well-defined. The price to be paid for this is the
exclusion of the threshold region, where the DPA is not valid. On the
other hand, the DPA provides a convenient framework
for the inclusion of radiative corrections.

\subsection{Dealing with unstable particles}
\label{sec:intro}

Most of our {\em technical} problems originate from the 
complications naturally pertaining to the gauge structure of the theory
and to the presence of unstable particles. As an interlude, we would like
to summarize the nominal essence of the theoretical basis of all generators.
In this respect one should remember that several, new, theoretical ideas were
fully developed also as a consequence of the previous workshop on 
$\wb\wb$-physics (Physics at LEP2, Yellow report CERN/96-01, February 1996)
and, in turn, many generators have profited from the 
most recent theoretical development.
Furthermore, this Section will be a natural place where to add some
consideration about the fine points in the DPA-procedure.

Four-fermion production processes, with or without radiative corrections, all 
involve fermions in the initial and final state and unstable gauge bosons as 
intermediate particles. Sometimes a photon is also present in the final 
state. If complete sets of graphs contributing to a given process are taken 
into account, the associated matrix elements are in principle gauge-invariant, 
i.e.~they are independent of gauge fixing and respect Ward identities.
This is, however, not guaranteed for incomplete sets of graphs like the ones 
corresponding to the off-shell $\wb$-pair production process (CC03). 
Indeed this process has been found to violate the $SU(2)$ Ward 
identities~\cite{ww-review}. 

In addition, the unstable gauge bosons that appear as intermediate 
particles can give rise to poles $1/(p^2-M^2)$ in physical observables if they 
are treated as stable particles. In view of the high precision of the LEP~2
experiments, the proper treatment of these unstable particles has become a 
demanding exercise, since on-shell approximations are simply not good enough 
anymore. A proper treatment of unstable particles requires the re-summation of 
the corresponding self-energies to all orders. In this way the singularities 
originating from the poles in the on-shell propagators are regularized by the 
imaginary parts contained in the self-energies, which are closely related to 
the decay widths ($\Gamma$) of the unstable particles. The perturbative 
re-summation itself involves a simple geometric series and is therefore easy 
to perform. However, this simple procedure harbours the serious risk of 
breaking 
gauge invariance. Gauge invariance is guaranteed order by order in perturbation
theory. Unfortunately one takes into account only part of the higher-order 
terms by re-summing the self-energies. This results in a mixing of different 
orders of perturbation theory and thereby jeopardizes gauge invariance, even 
if the self-energies themselves are extracted in a gauge-invariant way. Apart 
from being theoretically unacceptable, gauge-breaking effects can also lead to 
large errors in the MC predictions. At LEP~2 energies this problem
occurs for instance
in the reactions $\,e^+e^- \to e^-\barnu_e u\bard,\,e^+\nu_e \baru d\,$ 
for forward-scattered beam particles~\cite{BHF1}.

Based on this observation, it is clear that a gauge-invariant scheme is 
required for the treatment of unstable particles. It should be stressed, 
however, that any such scheme is arbitrary to a greater or lesser extent: 
since the Dyson summation must necessarily be taken to all orders of 
perturbation theory, and we are not able to compute the complete set of 
{\em all} Feynman diagrams to {\em all} orders, the various 
schemes differ even if they lead to formally gauge-invariant results. Bearing 
this in mind, we need besides gauge invariance some physical motivation for 
choosing a particular scheme. In this context two options can be mentioned.
Either one can try to {\em subtract} gauge-violating terms or one can try to 
{\em add} gauge-restoring terms to the calculation.

The first option is the so-called {\em pole scheme}~\cite{pole-scheme}.
In this scheme one decomposes the complete amplitude by expanding around the 
poles. As the physically observable residues of the poles are gauge-invariant, 
gauge invariance is not broken if the finite width is taken into account in 
the pole terms $\propto 1/(p^2-M^2)$. In reactions with multiple 
unstable-particle resonances it is rather awkward to perform the complete 
pole-scheme expansion with all its subtleties in the treatment of the mapping
of the off-shell phase space on the on-shell phase space.

Therefore one usually approximates the expansion by 
retaining only the terms with the highest degree of resonance. 
This approximation is called the leading-pole approximation and is closely 
related to on-shell production and decay of the unstable particles. 
The accuracy of the approximation is typically $\ord{\Gamma/M}$, making it a 
suitable tool for calculating {\em radiative corrections}, since in that case 
the errors are further suppressed by powers of the coupling constant. 
Since diagrams with a lower degree of resonance do not
feature in the leading-pole approximation, it is not an adequate approach for 
describing lowest-order reactions. So, for lowest-order reactions one needs an 
alternative approach.

The second option is based on the fundamentally different philosophy of trying 
to determine and include the minimal set of Feynman diagrams that is necessary 
for compensating the gauge violation caused by the self-energy graphs. 
This is obviously a theoretically very satisfying solution, but it may cause 
an increase in the complexity of the matrix elements and consequently a slowing
down of the numerical calculations. Two methods have been developed along these
lines. 

First of all, for the gauge bosons we are guided by the observation 
that the {\em lowest-order} decay widths are exclusively given by the imaginary
parts of the fermion loops in the one-loop self-energies. It is therefore 
natural to perform a Dyson summation of these fermionic one-loop self-energies 
and to include the other possible one-particle-irreducible fermionic one-loop 
corrections ({\em fermion-loop scheme})~\cite{BHF1,fls-baur,BHF2,fls-mf}. 
For the {\em lowest-order} LEP~2 process $e^+e^- \to 4\rmf$ this amounts 
to adding the fermionic corrections to the triple gauge-boson vertex. 
The complete set of fermionic contributions forms a manifestly gauge-invariant 
subset, since it involves the closed subset of all $\ord{[N_c^f\alpha/\pi]^n}$ 
contributions (with $N_c^f$ denoting the colour degeneracy of fermion $f$). 
Moreover, it obeys all Ward identities exactly, even with re-summed 
propagators, as shown in Ref.~\cite{BHF2} for two- and four-fermion production.
For any particle reaction this can be deduced from the
fact that the Ward identities of the underlying gauge symmetry, which
are obeyed by the fermion loops, survive such a consistent Dyson summation, in 
contrast to the Slavnov--Taylor identities of the BRS symmetry, as shown in
Ref.~\cite{de96} in the framework of the background-field formalism 
\cite{de95}.
The limitation of the fermion-loop scheme is due to the
fact that it does not apply to particles with bosonic decay modes
and that on resonance one perturbative order is lost.
This in turn disqualifies it as a candidate for handling radiative 
corrections. Moreover, the inclusion of a full-fledged set of one-loop 
corrections in a lowest-order amplitude tends to over-complicate things for
reactions like $e^+e^- \to 4\rmf\gamma$.

Recently a novel non-diagrammatic technique has been proposed for arbitrary 
{\em tree-level} reactions, involving all possible unstable particles and an 
unspecified amount of stable external particles~\cite{non-local}. By using 
gauge-invariant non-local effective Lagrangians, it is possible to generate 
the self-energy effects in the propagators as well as the required 
gauge-restoring terms in the multi-particle (3-point, 4-point, etc.) 
interactions. In this way the full set of Ward identities can be solved, while 
keeping the gauge-restoring terms to a minimum. 

A simplified version of this non-diagrammatic technique is the
{\em complex-mass scheme}, which was introduced in Ref.~\cite{racoonww_ee4fa}
for the reactions $e^+e^- \to 4\rmf$ and $e^+e^- \to 4\rmf\gamma$. 
In this scheme, the modifications of the
vertices that are necessary to compensate the width effects of the
propagators are obtained by analytically continuing the corresponding
mass parameters in all Feynman rules consistently, leading to complex
couplings. The {\em complex-mass scheme} preserves all Ward identities
and works for arbitrary lowest-order predictions. As a small drawback we note, 
that for 
space-like gauge-boson momenta the propagators are complex
in the {\em complex-mass scheme}, whereas perturbation theory in fact predicts
the absence of any imaginary contribution to the propagator.
This leads to complex couplings through
gauge restoration and it will change, potentially, the CP structure of the 
theoretical predictions, whenever imaginary parts are redistributed between 
vertex functions.
  
We must admit that the effect on the CP structure has not been investigated 
in any scheme. However, for the Fermion-Loop scheme one does not see any
problem with CP and for the non-local approach the modifications of the 
vertices have the feature that no imaginary parts are generated
for space-like particles.
One can also use the non-local approach starting from
proper imaginary parts for time-like and unproper ones for
space-like propagators and then look for a solution. One
finds the complex mass scheme. As such it is confirmed
by the non-local method, but only when one starts with an
ad-hoc ansatz.

\subsection{The leading-pole approximation}
\label{sec:LPA}

As mentioned above, the pole scheme consists in decomposing the complete 
amplitude by expanding around the poles of the unstable particles. 
The residues in this expansion are physically observable and therefore 
gauge-invariant. The pole-scheme expansion can be viewed as a gauge-invariant 
prescription for performing an expansion in powers of $\Gamma/M$. It should
be noted that there is no unique definition of the residues. Their calculation
involves a mapping of off-shell matrix elements with off-shell kinematics on 
on-resonance matrix elements with restricted kinematics. This mapping,
however, is not unambiguously fixed. After all, it involves more 
than just the invariant masses of the unstable particles and one thus has to 
specify the variables that have to be kept fixed in the mapping.
The resulting implementation dependence manifests itself in differences of 
sub-leading nature, \eg~$\ord{\Gamma/M}$ suppressed deviations in the 
leading pole-scheme residue. In special regions of phase space, where the
matrix elements vary rapidly, the implementation dependence can take 
noticeable proportions. This happens in particular near phase-space
boundaries, like thresholds.

In order to make these statements a bit more transparent, we sketch the 
pole-scheme method for a single unstable particle.
In this case the Dyson re-summed lowest-order matrix element is given by
\begin{eqnarray}
  \label{pole-scheme}
  \M^\infty 
     &=& \frac{W(p^2,\omega)}{p^2-\tilde{M}^2}\,\sum_{n=0}^{\infty}
         \,\Biggl( \frac{-\tilde{\Sigma}(p^2)}{p^2-\tilde{M}^2} 
           \Biggr)^n 
      =\ \frac{W(p^2,\omega)}{p^2-\tilde{M}^2+\tilde{\Sigma}(p^2)}
         \nonumber \\[1mm]
     &=& \frac{W(M^2,\omega)}{p^2-M^2}\,\frac{1}{Z(M^2)} + \Biggl[  
         \frac{W(p^2,\omega)}{p^2-\tilde{M}^2+\tilde{\Sigma}(p^2)}
         - \frac{W(M^2,\omega)}{p^2-M^2}\,\frac{1}{Z(M^2)} \Biggr],
\end{eqnarray}
where $\tilde{\Sigma}(p^2)$ is the unrenormalized self-energy of the 
unstable particle with momentum $p$ and unrenormalized mass 
$\tilde{M}$. The renormalized quantity $M^2$ is the pole in the complex 
$p^2$ plane, whereas $Z(M^2)$ denotes the wave-function factor:
\beq
  \label{pole-scheme/def}
  M^2-\tilde{M}^2+\tilde{\Sigma}(M^2) = 0, \quad\quad
  Z(M^2) = 1+\tilde{\Sigma}'(M^2). 
\eeq
The first term in the last expression of Eq.~(\ref{pole-scheme}) represents the
single-pole residue, which is closely related to on-shell production and 
decay of the unstable particle. The second term between the square brackets
has no pole and can be
expanded in powers of $\,p^2-M^2$. The argument $\omega$ denotes the
dependence on the other variables, i.e.~the implementation dependence.
After all, the unstable particle is always accompanied by other
particles in the production and decay stages. 

For instance, consider the
LEP1 reaction $e^+e^- \to \bar{f}f$. In the mapping $p_{\ssZ}^2 \to M^2$ one 
can either keep $\,t=(p_{e^-}-p_{f})^2=-p_{\ssZ}^2(1-\cos\theta)/2\,$ fixed or
$\cos\theta$. In the former mapping $\cos\theta_{\sss{pole}}$ is obtained 
from the on-shell relation $\,\cos\theta_{\sss{pole}} = 1+2t/M^2$, whereas in 
the latter mapping $\,t_{\sss{pole}} = -M^2(1-\cos\theta)/2$. It may
be that a particular mapping leads to an unphysical point in the
on-shell phase space. In the present example $t_{\sss{pole}}$ will
always be physical when $\cos\theta$ is kept fixed in the mapping.
However, since $|\cos\theta_{\sss{pole}}| > 1$ for $t<-\Reb M^2$, it
is clear that mappings with fixed Mandelstam variables harbour the potential
risk of producing such unphysical phase-space points.%
\footnote{In the resonance region, $|p_{\ssZ}^2-M^2| \ll |M^2|$, the unphysical
          on-shell phase-space points occur near the edge of the off-shell
          phase space, since $t<-\Reb M^2$ requires $\cos\theta \approx -1$.}

This can have repercussions on the convergence of the pole-scheme expansion.
Therefore it is recommended to use implementations that are free of unphysical 
on-shell phase-space points. 

The issue of taking angles instead of Mandelstam variables was raised
in Ref.~\cite{LEP2WWreport} (see text after Eq.(58) there) and in the
second reference of~\cite{pole-scheme} (see paragraph after Eq.(2)).
For the DPA presented in Ref.~\cite{dpa-ww}, in discussing
the treatment of the mapping of the off-shell phase space on the on-shell 
phase space, angles and completely decoupled off-shell invariant masses 
for the $\wb$ bosons were used.
Finally, in Ref.~\cite{nonfactrcs} the numerical effects coming from 
different phase-space treatments was considered also numerically. 
Specifically, the non-factorizable corrections were considered for different
choices of Mandelstam variables used in the DPA.

The at present only workable approach for evaluating the radiative corrections 
to resonance-pair-production processes, like $\wb$-pair 
production, involves the so-called leading-pole approximation (LPA). This 
approximation restricts the complete pole-scheme expansion to the term with the
highest degree of resonance. In the case of $\wb$-pair production only the 
double-pole residues are hence considered. This is usually referred to as the
DPA. The intrinsic error associated with this 
procedure is $\alpha\Gamma_{\ssW}/(\pi \mw)\times\ln(\dots) \lsim
0.5\%$,
except far off resonance, where the pole-scheme expansion cannot be viewed as 
an effective expansion in powers of $\Gamma/M$, and close to phase-space 
boundaries, where the DPA cannot be trusted to produce the dominant 
contributions. 
In the above error estimate, the $\ln(\dots)$ represents leading
logarithms or other possible enhancement factors in the corrections.
In the latter situations also the implementation
dependence of the double-pole residues can lead to enhanced errors. 
Close to the nominal (on-shell) $\wb$-pair threshold, for instance, the 
intrinsic error is effectively enhanced by a factor 
$\mw/(\sqrt{s}-2\mw)\equiv \mw/\Delta E$. 
In view of this it is wise to apply the DPA only if the energy is several 
$\Gamma_{\ssW}$ above the threshold. 

\begin{figure}[htbp]
  \begin{center}
  \begin{picture}(200,130)(0,-10)
    \ArrowLine(43,58)(25,40)        \Text(7,40)[lc]{$e^+$}
    \ArrowLine(25,90)(43,72)        \Text(7,92)[lc]{$e^-$}
    \Photon(50,65)(150,95){1}{12}   \Text(100,105)[]{$\wb$}  
    \Photon(50,65)(150,35){1}{12}   \Text(100,25)[]{$\wb$}
    \ArrowLine(155,98)(175,110)     \Text(190,110)[rc]{$f_1'$}
    \ArrowLine(175,80)(155,92)      \Text(190,80)[rc]{$\bar{f}_1$}
    \ArrowLine(175,50)(155,38)      \Text(190,50)[rc]{$\bar{f}_2'$}
    \ArrowLine(155,32)(175,20)      \Text(190,20)[rc]{$f_2$}
    \DashLine(75,115)(75,15){5}     \Text(40,5)[]{production}
    \DashLine(125,115)(125,15){5}   \Text(150,5)[]{decays}
    \GCirc(50,65){10}{1}
    \GCirc(100,80){10}{0.8}
    \GCirc(100,50){10}{0.8}
    \GCirc(150,95){10}{1}
    \GCirc(150,35){10}{1}
  \end{picture}
  \end{center}
  \caption[]{The generic structure of the virtual factorizable $\wb$-pair 
             contributions. The shaded circles indicate the Breit--Wigner 
             resonances, whereas the open circles denote the Green functions 
             for the production and decay sub-processes up to $\ord{\alpha}$ 
             precision.}
  \label{fig:WWfact}
\end{figure}
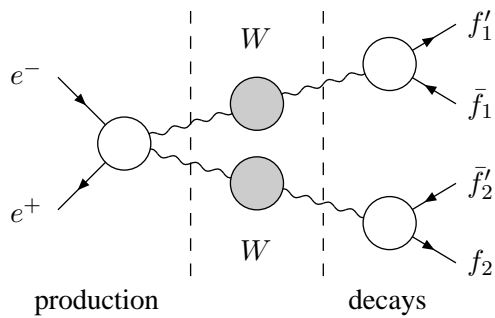

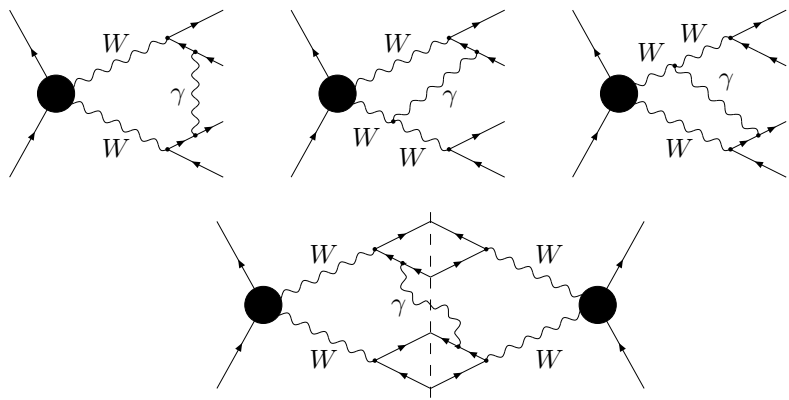
\begin{figure}[htbp]
  \begin{center}
  \unitlength .7pt\small\SetScale{0.7}
  \begin{picture}(120,100)(0,0)
  \ArrowLine(30,50)( 5, 95)
  \ArrowLine( 5, 5)(30, 50)
  \Photon(30,50)(90,80){2}{6}
  \Photon(30,50)(90,20){2}{6}
  \GCirc(30,50){10}{0}
  \Vertex(90,80){1.2}
  \Vertex(90,20){1.2}
  \ArrowLine(90,80)(120, 95)
  \ArrowLine(120,65)(105,72.5)
  \ArrowLine(105,72.5)(90,80)
  \Vertex(105,72.5){1.2}
  \ArrowLine(120, 5)( 90,20)
  \ArrowLine( 90,20)(105,27.5)
  \ArrowLine(105,27.5)(120,35)
  \Vertex(105,27.5){1.2}
  \Photon(105,27.5)(105,72.5){2}{4.5}
  \put(92,47){$\gamma$}
  \put(55,73){$\wb$}
  \put(55,16){$\wb$}
  \end{picture}
  \quad\quad
  \begin{picture}(120,100)(0,0)
  \ArrowLine(30,50)( 5, 95)
  \ArrowLine( 5, 5)(30, 50)
  \Photon(30,50)(90,80){2}{6}
  \Photon(30,50)(90,20){2}{6}
  \Vertex(60,35){1.2}
  \GCirc(30,50){10}{0}
  \Vertex(90,80){1.2}
  \Vertex(90,20){1.2}
  \ArrowLine(90,80)(120, 95)
  \ArrowLine(120,65)(105,72.5)
  \ArrowLine(105,72.5)(90,80)
  \Vertex(105,72.5){1.2}
  \ArrowLine(120, 5)(90,20)
  \ArrowLine(90,20)(120,35)
  \Photon(60,35)(105,72.5){2}{5}
  \put(87,46){$\gamma$}
  \put(63,11){$\wb$}
  \put(38,22){$\wb$}
  \put(55,73){$\wb$}
  \end{picture}
  \quad\quad 
  \begin{picture}(160,100)(0,0)
  \ArrowLine(30,50)( 5, 95)
  \ArrowLine( 5, 5)(30, 50)
  \Photon(30,50)(90,80){-2}{6}
  \Photon(30,50)(90,20){2}{6}
  \Vertex(60,65){1.2}
  \GCirc(30,50){10}{0}
  \Vertex(90,80){1.2}
  \Vertex(90,20){1.2}
  \ArrowLine(90,80)(120, 95)
  \ArrowLine(120,65)(105,72.5)
  \ArrowLine(105,72.5)(90,80)
  \Vertex(105,27.5){1.2}
  \ArrowLine(120, 5)(90,20)
  \ArrowLine(105,27.5)(120,35)
  \ArrowLine(90,20)(105,27.5)
  \Photon(60,65)(105,27.5){-2}{5}
  \put(84,55){$\gamma$}
  \put(63,78){$\wb$}
  \put(40,68){$\wb$}
  \put(55,16){$\wb$}
  \end{picture}
  \\[2ex]
  \unitlength .7pt\small\SetScale{0.7}
  \begin{picture}(240,100)(0,0)
  \ArrowLine(30,50)( 5, 95)
  \ArrowLine( 5, 5)(30, 50)
  \Photon(30,50)(90,80){2}{6}
  \Photon(30,50)(90,20){2}{6}
  \GCirc(30,50){10}{0}
  \Vertex(90,80){1.2}
  \Vertex(90,20){1.2}
  \ArrowLine(90,80)(120, 95)
  \ArrowLine(120,65)(105,72.5)
  \ArrowLine(105,72.5)(90,80)
  \ArrowLine(120, 5)( 90,20)
  \ArrowLine( 90,20)(120,35)
  \Vertex(105,72.5){1.2}
  \PhotonArc(120,65)(15,150,270){2}{3}
  \put(55,73){$\wb$}
  \put(55,16){$\wb$}
  \put(99,47){$\gamma$}
  \DashLine(120,0)(120,100){6}
  \PhotonArc(120,35)(15,-30,90){2}{3}
  \Vertex(135,27.5){1.2}
  \ArrowLine(150,80)(120,95)
  \ArrowLine(120,65)(150,80)
  \ArrowLine(120, 5)(150,20)
  \ArrowLine(150,20)(135,27.5)
  \ArrowLine(135,27.5)(120,35)
  \Vertex(150,80){1.2}
  \Vertex(150,20){1.2}
  \Photon(210,50)(150,80){2}{6}
  \Photon(210,50)(150,20){2}{6}
  \ArrowLine(210,50)(235,95)
  \ArrowLine(235, 5)(210,50)
  \GCirc(210,50){10}{0}
  \put(177,73){$\wb$}
  \put(177,16){$\wb$}
  \end{picture}
  \end{center}
  \caption[]{Examples for virtual (top) and real (bottom) non-factorizable 
             corrections to $\wb$-pair production. The black circles denote 
             the lowest-order Green functions for the production of the
             virtual $\wb$-boson pair.}
  \label{fig:WWnf}
\end{figure}

In the DPA one can identify two types of contributions. One type comprises all 
diagrams that are strictly reducible at both unstable $\wb$-boson lines
(see Fig.~\ref{fig:WWfact}).
These corrections are therefore called factorizable and can be attributed 
unambiguously either to the production of the $\wb$-boson pair or to one of 
the subsequent decays. The second type consists of all diagrams in which the 
production and/or decay sub-processes are not independent and which therefore
do not seem to have two overall $\wb$ propagators as factors 
(see Fig.~\ref{fig:WWnf}). 
We refer to these effects as non-factorizable corrections.%
\footnote{It should be noted that the exact split-up between factorizable and
          non-factorizable radiative corrections requires a precise 
          (gauge-invariant) definition. We will come back to this point.}

In the DPA the non-factorizable corrections arise 
exclusively from the exchange or emission of photons with 
$E_\gamma \lsim \Gamma_{\ssW}$ \cite{nf-cancellations}. 
Hard photons as well as massive-particle exchanges do not lead to 
double-resonant contributions. The physical picture behind all of this is that 
in the DPA the $\wb$-pair process can be viewed as consisting of several 
sub-processes: the production of the $\wb$-boson pair, the propagation of the 
$\wb$ bosons, and the subsequent decay of the unstable $\wb$ bosons. 
The production and decay are hard sub-processes, which occur on a relatively 
short time interval, $\ord{1/\mw}$. They are in general distinguishable as they 
are well separated by a relatively big propagation interval, 
$\ord{1/\Gamma_{\ssW}}$. Consequently, the corresponding amplitudes have 
certain factorization properties. The same holds for the radiative corrections
to the sub-processes. The only way the various stages can be interconnected
is via the radiation of soft photons with energy of $\ord{\Gamma_{\ssW}}$. 

As is clear from the above-given discussion of the DPA, a specific prescription
has to be given for the calculation of the DPA residues. Or, in other words, we
have to fix the implementation of the mapping of the full off-shell phase space
on the kinematically restricted (on-resonance) one. Two strategies have been
adopted in the literature~\cite{dpa-ww,racoonww_res}. One can opt to always 
extract pure double-pole residues~\cite{dpa-ww}. This means in particular that 
after the integration over decay kinematics and invariant masses has been
performed the on-shell cross-section should be recovered. 
Alternatively, one 
can decide to exclude the off-shell phase space from the mapping and apply the 
residue only to the matrix elements~\cite{racoonww_res,ja97}. 
We will come back to 
the conceptual and numerical differences between these two implementation 
strategies in the detailed discussion of the DPA programs. At this point we
merely note that the numerical differences can serve as an estimate of the
theoretical uncertainty of the DPA procedure.
Ref.~\cite{ja97} also used the approach
in which the full off-shell phase space is maintained and the residue
is only applied to the matrix elements.

In the rest of this section we will explain those aspects of the DPA procedure
that are common to both implementation methods. To this end we focus on the 
lowest-order reaction
\beq
\label{ee->ww->4f}
        e^{+}(q_1)\,e^{-}(q_2)\to \wbp(p_1)\,\wbm(p_2)
        \to
        \bar{f}_1(k_1)f_1'(k_1')\,f_2(k_2)\bar{f}_2'(k_2'),
\eeq
involving only those diagrams that contain as factors the Breit--Wigner 
propagators for the $\wbp$ and $\wbm$ bosons. Here $\bar{f}_1$ and $f_1'$ 
are the decay products of the $\wbp$ boson, and $f_2$ and $\bar{f}_2'$ those 
of the $\wbm$ boson. It should be noted that a large part 
of the radiative corrections in DPA to this reaction can be treated in a way 
similar to the lowest-order case, which is therefore a good starting point.
The amplitude for process (\ref{ee->ww->4f}) takes the form
\beq
\label{pole/A}
        \M =
        \sum_{\lambda_{1},\lambda_{2}}
        \Pi_{\lambda_{1}  \lambda_{2}}(M_{1}, M_{2}) \
        \frac{\Delta^{(+)}_{\lambda_{1}}(M_{1})}{D_{1}} \
        \frac{\Delta^{(-)}_{\lambda_{2}}(M_{2})}{D_{2}} \ ,
\eeq
where any dependence on the helicities of the initial- and final-state
fermions has been suppressed, and 
\beq
        D_{i} = M_{i}^{2}-\mws+i \mw\Gamma_{\ssW},\quad\quad
        M_{i}^{2} = (k_{i}+k_{i}')^{2}.
\eeq
The quantities $\Delta^{(+)}_{\lambda_{1}}(M_{1})$ and  
$\Delta^{(-)}_{\lambda_{2}}(M_{2})$ are the off-shell $\wb$-decay amplitudes 
for specific spin-polari\-zation states $\lambda_{1}$ (for the $\wbp$) and
$\lambda_{2}$ (for the $\wbm$), with $\lambda_i=(-1,0,+1)$. 
The off-shell $\wb$-pair production amplitude
$\Pi_{\lambda_{1}\lambda_{2}}(M_{1}, M_{2})$ depends on the invariant
fermion-pair masses $M_{i}$ and on the polarizations $\lambda_{i}$
of the virtual $\wb$ bosons. In the limit $M_{i} \to \mw$ the amplitudes 
$\Pi$ and $\Delta^{(\pm)}$ go over into the on-shell production and decay 
amplitudes.

In the cross-section the above factorization leads to
\beq
\label{pole/|A|^2}
        \sum\limits_{\sss{fermion helicities}} |\M|^{2}\ \ 
        =
        \sum\limits_{\lambda_1,\lambda_2,\lambda_1',\lambda_2'}
        \PP_{[\lambda_{1}\lambda_{2}][\lambda_{1}'\lambda_{2}']}(M_{1},M_{2}) \
        \frac{\D_{\lambda_{1}\lambda_{1}'}(M_{1})}{|D_{1}|^{2}} \
        \frac{\D_{\lambda_{2}\lambda_{2}'}(M_{2})}{|D_{2}|^{2}}.
\eeq
In Eq.~(\ref{pole/|A|^2}) the production part is given by a $9\times9$~density 
matrix
\beq
\label{pole/prod}
\PP_{[\lambda_{1}\lambda_{2}][\lambda_{1}'\lambda_{2}']}(M_{1},M_{2})\ \ 
         =
         \sum\limits_{\sss{$e^{\pm}$ helicities}}
         \Pi_{\lambda_{1}\lambda_{2}}(M_{1},M_{2}) \
         \Pi_{\lambda_{1}'\lambda_{2}'}^{*}(M_{1},M_{2}).
\eeq
Similarly the decay part is governed by $3\times3$~density matrices
\beq
\label{pole/decay}
         \D_{\lambda_{i}\lambda_{i}'}(M_{i})\ \ 
         =
         \sum\limits_{\sss{fermion helicities}}
         \Delta_{\lambda_{i}}(M_{i}) \
         \Delta_{\lambda_{i}'}^{\!*}(M_{i}),
\eeq
where the summation is performed over the helicities of the final-state
fermions.
 
It is clear that Eq.~(\ref{pole/prod}) is closely related to the absolute 
square of the matrix element for stable unpolarized $\wb$-pair production. 
In that case the cross-section contains the trace of the above density matrix
\beq
\label{pole/prod_on}
        \mbox{\bf Tr} \ \PP(\mw,\mw) 
        =
        \sum\limits_{\lambda_1,\lambda_2}
        \PP_{[\lambda_{1}\lambda_{2}][\lambda_{1}\lambda_{2}]}(\mw,\mw)\ \ 
        = 
        \sum\limits_{\sss{all polarizations}}
        |\Pi_{\lambda_{1} \lambda_{2}}(\mw, \mw)|^2.
\eeq
The decay of an unpolarized on-shell $\wb$ boson is determined by
\beq
\label{pole/dec_on}
        \mbox{\bf Tr} \ \D(\mw) 
        =
        \sum\limits_{\lambda_{i}} \D_{\lambda_{i}\lambda_{i}}(\mw)\ \ 
        =
        \sum\limits_{\sss{all polarizations}}
        |\Delta_{\lambda_{i}}(\mw)|^{2}.
\eeq
Note, however, that also the off-diagonal elements of $\PP(\mw,\mw)$ and 
$\D(\mw)$ are required for determining Eq.~(\ref{pole/|A|^2}) in the limit
$M_{i} \to \mw$. 

As a next step it is useful to describe the kinematics of process
(\ref{ee->ww->4f}) in a factorized way, i.e.~using the invariant masses 
$M_{1}$ and $M_{2}$ of the fermion pairs. The differential cross-section 
takes the form
\beq
        d\sigma
        =
        \frac{1}{2s} \sum |\M|^{2}\, d\Gamma_{4\rmf}
        =
        \frac{1}{2s} \sum |\M|^{2}\, d\Gamma_{\sss{pr}}\cdot 
        d\Gamma_{\sss{dec}}^{+} \cdot d\Gamma_{\sss{dec}}^{-}   
        \cdot \frac{dM_{1}^{2}}{2\pi} \cdot \frac{dM_{2}^{2}}{2\pi}, 
\eeq
where $d\Gamma_{4\rmf}$ indicates the complete four-fermion phase-space 
factor
and $s=(q_1+q_2)^2$ the centre-of-mass energy squared. The phase-space factors
for the production and decay sub-processes, $d\Gamma_{\sss{pr}}$ and 
$d\Gamma_{\sss{dec}}^{\pm}$, read 
\begin{eqnarray}
     d\Gamma_{\sss{pr}}\
     &=& \frac{1}{(2\pi)^{2}}\, \delta(q_{1}+q_{2}-p_{1}-p_{2})\,
         \frac{d \vec{p}_{1}}{2p_{10}}\,
         \frac{d \vec{p}_{2}}{2p_{20}}, 
         \nonumber \\
     d\Gamma_{\sss{dec}}^{+}
     &=& \frac{1}{(2\pi)^{2}}\, \delta(p_{1}-k_{1}-k_{1}')\,
         \frac{d \vec{k}_{1}}{2k_{10}}\,
         \frac{d \vec{k}_{1}'}{2k_{10}'}, 
         \nonumber \\      
     d\Gamma_{\sss{dec}}^{-}
     &=& \frac{1}{(2\pi)^{2}}\, \delta(p_{2}-k_{2}-k_{2}')\,
         \frac{d \vec{k}_{2}}{2k_{20}}\,
         \frac{d \vec{k}_{2}'}{2k_{20}'}.
\end{eqnarray}
When the factorized form for $\sum|\M|^{2}$ is inserted one obtains
\begin{eqnarray}
\label{pole/15}
  d\sigma
  &=& \frac{1}{2s}\sum\limits_{\lambda_1,\lambda_2,\lambda_1',\lambda_2'}
      \PP_{[\lambda_{1}\lambda_{2}][\lambda_{1}'\lambda_{2}']}(M_{1},M_{2})\,
      d\Gamma_{\sss{pr}}
      \times
      \D_{\lambda_{1}\lambda_{1}'}(M_{1})\, d\Gamma_{\sss{dec}}^{+}
      \times
      \D_{\lambda_{2}\lambda_{2}'}(M_{2})\, d\Gamma_{\sss{dec}}^{-}
      \times
      \nonumber \\
  & & \times\, 
      \frac{1}{2\pi}\,\frac{d M_{1}^{2}}{|D_{1}|^{2}}
      \times
      \frac{1}{2\pi}\,\frac{d M_{2}^{2}}{|D_{2}|^{2}},
\end{eqnarray}
which is the common starting point for any of the DPA implementations.


\subsection{Radiative corrections in double-pole approximation}
\label{sec:LPA/RC}

A full calculation of the complete electroweak $\ord{\alpha}$
corrections to $\eeffff(+\gamma)$ for all four-fermion final states is
beyond present possibilities. While the real bremsstrahlung
corrections induced by $\eeffffg$ are known exactly
\cite{fu94,racoonww_ee4fa,pv4fg}, there are severe technical
and conceptual problems with the virtual corrections to four-fermion
production. Fortunately, the full account of the $\ord{\alpha}$
corrections is not needed at the level of accuracy demanded by LEP~2.
For $\wb$-pair-mediated processes, $\eeWWffff$, the required accuracy of
predictions is of the order of $0.5\%$ for integrated quantities.  At
this level, the corrections to $\wb$-pair production can be treated in the
DPA.
In regions of phase space where two resonant
$\wb$~bosons do not dominate the cross-sections, such as in the
$\wb\wb$-threshold region or in the single-$\wb$ domain, the DPA is, of course,
not valid and one should resort to other approximations as the 
Weizs\"acker-Williams for single-$\wb$~\cite{swc}. 

Since only diagrams with two nearly resonant $\wb$~bosons are relevant for 
the DPA, the number of graphs is reduced considerably, and a generic treatment
of all four-fermion final states is possible. 
Obviously all diagrams
that appear for the pair production and the decay of on-shell $\wb$~bosons
are also relevant for the pole expansion in the DPA. 
Since such
contributions involve a product of two independent Breit--Wigner
factors for the $\wb$~resonances, they are called {\em factorizable}
corrections. However, there exist
also doubly-resonant corrections in which the production and decay
sub-processes do not proceed independently. Power counting reveals that
such corrections are only doubly-resonant if the particle that is
exchanged by the sub-processes is a low-energetic photon. Owing to the
complicated off-shell behaviour of these corrections, they are called
{\em non-factorizable}.

While the definition of the DPA is straightforward for the virtual
corrections, it is problematic for the real corrections. 
The problem is due to the momentum carried away by photon radiation.
The invariant masses of the $\wb$~bosons in contributions in which the
photon is emitted in the $\wb$-pair production subprocess differ from
those where the photon is emitted in the $\wb$-decay sub-processes.
The corresponding Breit--Wigner resonances overlap if
the energy of the emitted photon is of the order of $\GW$.
It is not obvious how to define the DPA for such photons.
Therefore, the results based on a DPA for the real
corrections have to be treated with some caution.

According to the above classification, there are four categories of
contributions to $\ord{\alpha}$ corrections in DPA: factorizable and
non-factorizable ones both for virtual and real corrections.
In the following the salient features of those four parts are described.


\subsubsection{Virtual corrections}
\label{sec:LPA/RC/virt}

As a first step we discuss how to separate the virtual corrections into a sum 
of factorizable and non-factorizable virtual corrections. The diagrammatic 
split-up according to reducible and irreducible $\wb$-boson lines is an 
illustrative way of understanding the different nature of the two classes of
corrections, but since the double-resonant diagrams are not gauge-invariant 
by themselves the precise split-up needs to be defined properly. 

We can make use of the fact that there are effectively two scales in the 
problem: $\mw$ and $\Gamma_{\ssW}$. Let us now consider virtual corrections 
coming from photons with different energies:
\begin{itemize}
  \item  soft photons, $E_{\gamma} \ll \Gamma_{\ssW}$, 
  \item  semi-soft photons, $E_{\gamma} = \ord{\Gamma_{\ssW}}$,
  \item  hard photons, $\Gamma_{\ssW} \ll E_{\gamma} = \ord{\mw}$.
\end{itemize}
Only soft and semi-soft photons contribute to both factorizable and 
non-factorizable corrections. The latter being defined to describe interactions
between different stages of the off-shell process. The reason for this is that
only these photons can induce relatively long-range interactions and thereby 
allow the various sub-processes, which are separated by a propagation interval 
of $\ord{1/\Gamma_{\ssW}}$, to communicate with each other. Virtual corrections 
involving the exchange of hard photons or 
of massive particles contribute 
exclusively to the factorizable corrections. In view of the short range of the 
interactions induced by these particles, their contribution to the 
non-factorizable corrections are suppressed by at least
$\ord{\Gamma_{\ssW}/\mw}$.

As hard photons contribute to the factorizable corrections only, we merely need
to define a split-up for soft and semi-soft photons. It is impossible to do 
this in a consistent gauge-invariant way on the basis of diagrams. 
In Refs.~\cite{nonfactrcs,dpa-ww}
it was shown that only part of particular diagrams should
be attributed to the non-factorizable corrections, the rest being of 
factorizable nature. The complete set of non-factorizable corrections was 
obtained by collecting all terms that contain the ratios 
$D_{i}/[D_{i}\pm 2kp_{i}]$, where $k$ denotes the momentum of the (semi-)soft 
photon. The so-defined non-factorizable corrections
read~\cite{dpa-ww}
\beq
\label{corr/virt/nf_mtrx}
 \M_{\sss{nf}}^{\sss{virt}}
 =
 i\M_{0}^{\sss{DPA}}
 \int\frac{d^{4}k}{(2\pi)^{4}[k^{2}+io]}
 \biggl[
 \bigl(\J^{\mu}_{0}+\J^{\mu}_{\oplus}\bigr)\J_{+,\,\mu}+
 \bigl(\J^{\mu}_{0}+\J^{\mu}_{\ominus}\bigr)\J_{-,\,\mu}+
 \J^{\mu}_{+}\J_{-,\,\mu}
 \biggr],
\eeq
which contains the gauge-invariant currents
$$
 \J_{0}^{\mu}
 =
 e\Biggl[
  \frac{p_{1}^{\mu}}{kp_{1}+io}
 +\frac{p_{2}^{\mu}}{-kp_{2}+io}
 \Biggr],
$$
\beq
 \J_{\oplus}^{\mu}
 = -\,
 e\Biggl[
  \frac{q_{1}^{\mu}}{kq_{1}+io}
 -\frac{q_{2}^{\mu}}{kq_{2}+io}
 \Biggr],
 \ \ \
 \J_{\ominus}^{\mu}
 = +\,
 e\Biggl[
  \frac{q_{1}^{\mu}}{-kq_{1}+io}
 -\frac{q_{2}^{\mu}}{-kq_{2}+io}
 \Biggr]
\eeq
for photon emission from the production stage of the process, and 
\begin{eqnarray}
 \J_{+}^{\mu}
 &=& -\,e\Biggl[ \frac{p_{1}^{\mu}}{kp_{1}+io}
                + Q_{f_1}\frac{k_{1}^{\mu}}{kk_{1}+io}
                - Q_{f_1'}\frac{{k_{1}'}^{\mu}}{kk_{1}'+io}
        \Biggr]\frac{D_{1}}{D_{1}+2kp_{1}},
        \nonumber \\
 \J_{-}^{\mu}
 &=& -\,e\Biggl[ \frac{p_{2}^{\mu}}{-kp_{2}+io}
                + Q_{f_2}\frac{k_{2}^{\mu}}{-kk_{2}+io}
                - Q_{f_2'}\frac{{k_{2}'}^{\mu}}{-kk_{2}'+io}
        \Biggr]\frac{D_{2}}{D_{2}-2kp_{2}}
\end{eqnarray}
for photon emission from the decay stages of the process.
Here $\M_{0}^{\sss{DPA}}$ is the lowest-order matrix element in DPA and 
$Q_{f}$ stands for the charge of fermion $f$ in units of $e$.
Since Eq.~(\ref{corr/virt/nf_mtrx}) contains (at least) all contributions from 
diagrams with irreducible $\wb$-boson lines, it can be viewed as a 
gauge-invariant extension of the set of $\wb$-irreducible diagrams.
In general one has to calculate all of the integrals appearing in the above
expressions.
The complete set of integrals has been given in
Ref.~\cite{be97} and explicit expressions for the full set of
virtual factorizable corrections can be found in \cite{de00}.
However, if one is interested in the sum of virtual corrections and
real-photon radiation, then some simplifications occur depending on
the treatment of the photon\footnote{Note that Eq.~(\ref{corr/virt/nf_mtrx}) 
is UV-finite and contains
$4$- and $5$-point integrals. In fact it was observed that certain
combinations of these $4$- and $5$-point integrals are equal to a simple
(Coulomb-like) $3$-point integral plus a constant. This simple $3$-point
integral has an artificial UV divergence, which cancels against the
constant and can be regulated by either a cut-off (BBC) or by keeping
the DPA-subleading $k^2$ contributions in the denominators (RACOONWW).
The final answer of course does not depend on this.}.

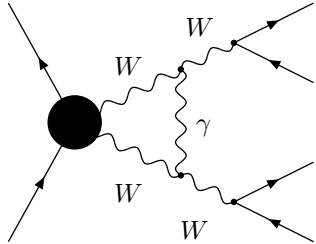
\begin{figure}[htbp]
  \begin{center} 
  \unitlength   1pt\small\SetScale{1.0}
  \begin{picture}(170,100)(0,0)
  \ArrowLine(30,50)( 5, 95)
  \ArrowLine( 5, 5)(30, 50)
  \Photon(30,50)(90,80){2}{6}
  \Photon(30,50)(90,20){2}{6}
  \Photon(70,30)(70,70){2}{4}
  \Vertex(70,30){1.2}
  \Vertex(70,70){1.2}
  \GCirc(30,50){10}{0}
  \Vertex(90,80){1.2}
  \Vertex(90,20){1.2}
  \ArrowLine(90,80)(120, 95)
  \ArrowLine(120,65)(90,80)
  \ArrowLine(120, 5)( 90,20)
  \ArrowLine( 90,20)(120,35)
  \put(76,47){$\gamma$}
  \put(45,68){$\wb$}
  \put(45,20){$\wb$}
  \put(72,83){$\wb$}
  \put(70,6){$\wb$}
  \end{picture} 
  \end{center}
  \caption[]{The Coulomb graph, contributing to both factorizable and 
             non-factorizable corrections.}
  \label{fig:coulomb}
\end{figure}

If the radiated (real) photon is treated inclusively,
then many of the terms in Eq.~(\ref{corr/virt/nf_mtrx}) cancel \cite{nf-cancellations}.
In this context the difference in the signs of the $io$ parts appearing 
in the currents $\J_{\ominus}$ and $\J_{\oplus}$ are crucial. These signs 
actually determine which interference terms give rise to a non-vanishing 
non-factorizable contribution after virtual and real-photon corrections
have been added. As a result of such considerations only a very limited subset
of `final-state' interferences survives for inclusive photons: the virtual 
corrections corresponding to Figs.~\ref{fig:WWnf} and \ref{fig:coulomb} as 
well as the associated real-photon corrections. 

The sum of virtual and real 
non-factorizable corrections has been calculated,
\citeres{me96, be97, nonfactrcs, nonfactrcslett}%
\footnote{The original result of the older calculation \cite{me96}
  does not agree with the two more recent results \cite{be97,nonfactrcs},
  which are in mutual agreement. As known from the authors of
  \citere{me96}, their corrected results also agree with the ones of
  \citeres{be97,nonfactrcs}.}.  
It has been shown in \citere{nf-cancellations} that this
sum vanishes if the invariant masses of both $\wb$~bosons are integrated
over, i.e.\ in particular that the full non-factorizable correction to
the total cross-section is zero in DPA.  

In \citeres{be97,nonfactrcs,nonfactrcslett} the
full non-factorizable corrections have also been discussed
numerically. They vanish on top of the double resonance and are of the
order of $1\%$ in its vicinity. The shift in the $\wb$~invariant-mass
distributions is only of the order of a few MeV.  These results can be
reproduced by a simple approximation \cite{ch99} based on the
so-called screened Coulomb ansatz. However, it is important to note that
all these numerical results on non-factorizable corrections
are based on the DPA for real corrections and have been obtained in
idealized treatment of phase space, namely the assumption that the
$\wb$-boson momenta can be reconstructed from the fermion momenta alone,
i.e.\ without photon recombination. It is not clear 
how these results change in physical situations with photon recombination.

The virtual factorizable corrections consist of all hard contributions and the
left-over part of the semi-soft ones. The so-defined factorizable corrections 
have the nice feature that they can be expressed in terms of corrections to 
on-shell sub-processes, i.e.~the production of two on-shell W bosons and their 
subsequent on-shell decays. The corresponding matrix element can be expressed 
in the same way as described at lowest-order:
\begin{eqnarray}
  \label{mvirt}
  \M_{\sss{fact}}^{\sss{virt}} = \sum_{\lambda_{1},\lambda_{2}}
        \Pi_{\lambda_{1}  \lambda_{2}}(M_{1}, M_{2}) \
        \frac{\Delta^{(+)}_{\lambda_{1}}(M_{1})}{D_{1}} \
        \frac{\Delta^{(-)}_{\lambda_{2}}(M_{2})}{D_{2}}.
\end{eqnarray}
Here two of the amplitudes are taken at lowest order, whereas the remaining
one contains all possible one-loop contributions, including the $\wb$
wave-function factors that appear in Eq.~(\ref{pole-scheme}). In this way the 
well-known on-shell radiative corrections to the production and decay of pairs 
of $\wb$ 
bosons~\cite{prod-corr,decay-corr} appear 
as basic building blocks of the factorizable corrections.%
\footnote{Note that the complete density matrix is 
          required in this case, in contrast to the pure on-shell calculation
          which involves the 
          diagonal elements of the density matrix only.} 
In the semi-soft limit the photonic
virtual factorizable corrections to the 
production stage, contained in $\Pi$, cancel against the corresponding 
real-photon corrections. Non-vanishing contributions from $\Pi$ occur as 
soon as the $k^2$ terms in the propagators
cannot be neglected anymore. An example of this is the factorizable
correction from the Coulomb graph Fig.~\ref{fig:coulomb}.   
For the on-shell (factorizable) part of the Coulomb effect photons with
momenta $k_0=\ord{\Delta E}$ and $\,|\vec{k}|=\ord{\sqrt{\mw\Delta E}\,}$
are important~\cite{coulomb}, i.e.~$k^{2}$ cannot be neglected
in the propagators of the unstable particles. Since we stay well away from
the $\wb$-pair threshold ($\Delta E \equiv \sqrt{s}-2 \mw \gg \Gamma_{\ssW}$), 
this situation occurs outside the realm of the semi-soft photons. 
This fits nicely into the picture of the production stage being a hard 
subprocess, governed by relatively short time scales as compared with the 
much longer time scales required for the non-factorizable corrections, which 
interconnect the different sub-processes.


\subsubsection{Real-photon radiation}
\label{sec:LPA/RC/real}

In this subsection we discuss the aspects of real-photon radiation in the DPA
as used in \cite{dpa-ww}. To this end we consider the process
\beq
\label{ee->ww->4fa}
      e^{+}(q_1)\,e^{-}(q_2) \to \wbp(p_1)\,\wbm(p_2)\,\bigl[\gamma(k)\bigr]
      \to
      \bar{f}_1(k_1)f_1'(k_1')\,f_2(k_2)\bar{f}_2'(k_2')\,\gamma(k),
\eeq  
where in the intermediate state there may or may not be a photon. We will 
show how to extract the gauge-invariant double-pole residues in 
different situations. The exact cross-section for process (\ref{ee->ww->4fa}) 
can be written in the following form
\beq
        \label{sig_rad}
        d\sigma
        =
        \frac{1}{2s}
        |\M_{\gamma}|^{2} 
        d\Gamma_{4\rmf\gamma} 
        =
        \frac{1}{2s} 
        \Biggl[
        2\Reb\biggl( \M_{0}\M_{+}^{*} + \M_{0}\M_{-}^{*} 
                    + \M_{+}\M_{-}^{*} \biggr)
        + |\M_{0}|^2 + |\M_+|^2 + |\M_{-}|^2
        \Biggr] 
        d\Gamma_{4\rmf\gamma},
\eeq
where $d\Gamma_{4\rmf\gamma}$ indicates the complete five-particle 
phase-space factor, and the matrix elements $\M_{0}$ and $\M_{\pm}$ correspond
to the diagrams where the photon is attached to the 
production or decay stage of the three $\wb$-pair diagrams, respectively.
This split-up can be achieved with the help of the partial-fraction 
decomposition \cite{Be85}
\beq
  \label{decomp}
  \frac{1}{D_i (D_i+2pk)} = \frac{1}{2pk}\,\Biggl(\frac{1}{D_i}
                                                  - \frac{1}{D_i+2pk} \Biggr).
\eeq
Each contribution to the cross-section can be written in terms of polarization 
density matrices, which originate from the amplitudes
\beq
\label{prodrad}
        \M_{0} =
        \Pi_{\gamma}(M_{1}, M_{2}) \
        \frac{\Delta^{(+)}(M_{1})}{D_{1}} \
        \frac{\Delta^{(-)}(M_{2})}{D_{2}} \ ,
\eeq
\beq
\label{dec+rad}
        \M_{+} =
        \Pi(M_{1\gamma}, M_{2}) \
        \frac{\Delta^{(+)}_{\gamma}(M_{1\gamma})}{D_{1\gamma}} \
        \frac{\Delta^{(-)}(M_{2})}{D_{2}} \ ,
\eeq
\beq
\label{dec-rad}
        \M_{-} =
        \Pi(M_{1}, M_{2\gamma}) \
        \frac{\Delta^{(+)}(M_{1})}{D_{1}} \
        \frac{\Delta^{(-)}_{\gamma}(M_{2\gamma})}{D_{2\gamma}} \ ,
\eeq
where all polarization indices for the W bosons and the photon have been 
suppressed, and 
\beq
        D_{i\gamma}=D_{i}+2kk_{i}+2kk_{i}', 
        \ \ \ 
        M_{i\gamma}^{2}=M_{i}^{2}+2kk_{i}+2kk_{i}',
        \ \ \ 
        M_{i}^{2} = (k_{i}+k_{i}')^{2}.
\eeq
The matrix elements $\Pi_{\gamma}$ and $\Delta^{(\pm)}_{\gamma}$ describe the
production and decay of the $\wb$ bosons accompanied by the radiation of a 
photon. The matrix elements without subscript $\gamma$ have been 
introduced in Eq.~(\ref{pole/A}).

In the calculation of the Born matrix element and virtual corrections only two 
poles could be identified in the amplitudes, originating from the 
Breit--Wigner propagators $1/D_{i}$. The pole-scheme expansion was performed 
around these two poles. In contrast, the bremsstrahlung matrix element has 
four in general different poles, originating from the four Breit--Wigner 
propagators $1/D_{i}$ and $1/D_{i\gamma}$. As mentioned above, the matrix 
element can be rewritten as a sum of three matrix elements ($\M_0,\M_+,\M_-$),
each of which only contain two Breit--Wigner propagators. For these three
individual matrix elements the pole-scheme expansion is fixed, as before, to 
an expansion around the corresponding two poles. However, when calculating 
cross-sections [see Eq.~(\ref{sig_rad})] the mapping of the five-particle 
phase space introduces a new type of ambiguity. The interference terms in 
Eq.~(\ref{sig_rad}) involve two different double-pole expansions 
simultaneously. One might think this will pose a problem, since there is no 
natural choice for the phase-space mapping in those cases. As we will see
later, however, only photons with $E_{\gamma} \lsim \Gamma_W \ll M_W$
give noticeable contributions to these interference terms. This means
that one can apply a soft-photon-like (semi-soft) approximation (see below). 

In Ref.~\cite{dpa-ww} it was argued that the resulting ambiguity in
the phase-space mapping will not have significant repercussions on the
quality of the DPA calculation, in the same way as stable-particle
calculations are not significantly affected by the photon momentum in
the soft-photon regime. We note, however, that there is still some
controversy on this issue.

Let us return now to the three earlier-defined regimes for the photon energy:
\begin{itemize}
  \item for hard photons [$E_{\gamma}\gg \Gamma_{\ssW}$] the Breit--Wigner 
        poles of the W-boson resonances before and after photon radiation are 
        well separated in phase space (see $M_{i\gamma}^{2}$ and $M_{i}^{2}$ 
        defined above). As a result, the interference terms in
        Eq.~(\ref{sig_rad}) can be neglected. 
        This leads to three {\em distinct}
        regions of on-shell contributions, where the photon can be assigned 
        unambiguously to the W-pair-production subprocess or to one of the 
        two decays. This assignment is determined by the pair of invariant 
        masses (out of $M_{i}^{2}$ and $M_{i\gamma}^{2}$) that is in the 
        $\mws$ region. Therefore, the double-pole residue
        can be expressed as the sum of the three on-shell contributions 
        without increasing the intrinsic error of the DPA. Note that in the 
        same way it is also possible to assign the photon to 
        one of the sub-processes, since misassignment errors are suppressed,
        assuming for convenience that all final-state momenta can
        ideally be measured.
  \item for semi-soft photons [$E_{\gamma} = {\cal O}(\Gamma_{\ssW})$] the 
        Breit--Wigner poles are relatively close together in phase space, 
        resulting in a substantial overlap of the line shapes. The assignment 
        of the photon is now subject to larger errors. Moreover, since the 
        interference terms in Eq.~(\ref{sig_rad}) cannot be neglected,
        a proper prescription for calculating the DPA residues (i.e.~the 
        phase-space mapping) is required~\cite{be97,nonfactrcs,dpa-ww}.
  \item for soft photons [$E_{\gamma}\ll \Gamma_{\ssW}$] the Breit--Wigner 
        poles are on top of each other, resulting in a pole-scheme expansion 
        that is identical to the one without the photon.
\end{itemize}

Let us first consider the hard-photon regime in more detail. Due to the fact 
that the poles are well separated in the hard-photon regime, it is clear that
the interference terms are suppressed and can be neglected:
\beq
        d\sigma
        =
        \frac{1}{2s}
        \biggl[ |\M_{0}|^{2}+|\M_{+}|^{2}+|\M_{-}|^{2}\biggr]
        d\Gamma_{4\rmf\gamma}.
\eeq
Note that each of the three terms has two poles, originating from two 
resonant propagators. However, these poles are different for different 
terms. The phase-space factor can be rewritten in three equivalent ways.
The first is
\beq
        d\Gamma_{4\rmf\gamma} = d\Gamma_{\sss{0}}^{\gamma} = 
        d\Gamma_{\sss{pr}}^{\gamma}\cdot 
        d\Gamma_{\sss{dec}}^{+} \cdot d\Gamma_{\sss{dec}}^{-}   
        \cdot   
        \frac{dM_{1}^{2}}{2\pi} \cdot \frac{dM_{2}^{2}}{2\pi},
\eeq
with
\beq
        d\Gamma_{\sss{pr}}^{\gamma}
        =
        \frac{1}{(2\pi)^{2}}\,
        \delta(q_{1}+q_{2}-p_{1}-p_{2}-k)\, \frac{d\vec{p}_{1}}{2p_{10}}\, 
        \frac{d\vec{p}_{2}}{2p_{20}}\,    \frac{d\vec{k}}{(2\pi)^{3}2k_{0}}.
\eeq
The two others are 
\beq
        d\Gamma_{4\rmf\gamma} = d\Gamma_{+}^{\gamma} = 
        d\Gamma_{\sss{pr}}\cdot 
        d\Gamma_{\sss{dec}}^{+\gamma} \cdot d\Gamma_{\sss{dec}}^{-}     
        \cdot   
        \frac{dM_{1\gamma}^{2}}{2\pi} \cdot \frac{dM_{2}^{2}}{2\pi},
\eeq
with
\beq
        d\Gamma_{\sss{dec}}^{+\gamma}
        =
        \frac{1}{(2\pi)^{2}}\,
        \delta(p_{1}-k_{1}-k_{1}'-k)\, \frac{d\vec{k}_{1}}{2k_{10}}\, 
        \frac{d\vec{k}_{1}'}{2k_{10}'}\,  \frac{d\vec{k}}{(2\pi)^{3}2k_{0}},
\eeq
and a similar expression for $d\Gamma_{-}^{\gamma}$.
The phase-space factors $d\Gamma_{\sss{pr}}$ and $d\Gamma_{\sss{dec}}^{\pm}$
are just the lowest-order ones. The cross-section can then be written in the 
following equivalent form
\beq
\label{corr/hard}
        d\sigma
        =
        \frac{1}{2s}
        \biggl[ 
         |\M_{0}|^{2}\,d\Gamma_{0}^{\gamma}
        +|\M_{+}|^{2}\,d\Gamma_{+}^{\gamma}
        +|\M_{-}|^{2}\,d\Gamma_{-}^{\gamma}
        \biggr].
\eeq
In order to extract gauge-invariant quantities, the DPA limit should be 
taken. This amounts to taking the limit $p_{1,2}^{2}\to \mws$, using a
particular prescription for mapping the full off-shell phase space on the 
kinematically restricted on-resonance one.
Note however that $p_{1,2}$ can be different according to the 
$\delta$-functions in the decay parts of the different phase-space factors. 
To be specific, the production term $|\M_{0}|^{2}$ has poles at 
$p_{i}^{2}=M_{i}^{2}=\mws$, $|\M_{+}|^{2}$ has poles at 
$p_{1}^{2}=M_{1\gamma}^{2}=\mws$ and $p_{2}^{2}=M_{2}^{2}=\mws$, and 
$|\M_{-}|^{2}$ has poles at $p_{1}^{2}=M_{1}^{2}=\mws$ and
$p_{2}^{2}=M_{2\gamma}^{2}=\mws$.

We complete our survey of the different photon-energy regimes by
considering semi-soft and soft photons.
The split-up of factorizable and non-factorizable real-photon corrections
proceeds in the same way as described in the previous subsection for
virtual corrections. The result reads in semi-soft approximation
\beq
\label{corr/real/mtrx}
        d\sigma
        =
        \frac{1}{2s}|\M_{\gamma}|^{2} d\Gamma_{\sss{4f}\gamma}
        \approx 
        -\, d\sigma_{\sss{DPA}}^{0}
        \frac{d\vec{k}}{(2\pi)^{3}2k_{0}}
        \Biggl[
        2\Reb\biggl(\I_{0}^{\mu}\I_{+,\,\mu}^{*}+\I_{0}^{\mu}\I_{-,\,\mu}^{*}
              +\I_{+}^{\mu}\I_{-,\,\mu}^{*}\biggr)
         +|\I_{0}^2|+|\I_{+}^2|+|\I_{-}^2|
         \Biggr].
\eeq
The gauge-invariant currents $\I_{0}$ and $\I_{\pm}$ are given by
\begin{eqnarray}
\label{corr/real/currents}
 \I_{0}^{\mu}
 &=& e\Biggl[ \frac{p_{1}^{\mu}}{kp_{1}}-\frac{p_{2}^{\mu}}{kp_{2}}
              -\frac{q_{1}^{\mu}}{kq_{1}} + \frac{q_{2}^{\mu}}{kq_{2}}
      \Biggr],
      \nonumber \\
 \I_{+}^{\mu} 
 &=& -\,e\Biggl[ \frac{p_{1}^{\mu}}{kp_{1}}
               +Q_{f_1}\frac{k_{1}^{\mu}}{kk_{1}}
               -Q_{f_1'}\frac{{k_{1}'}^{\mu}}{kk_{1}'}
       \Biggr]\frac{D_{1}}{D_{1}+2kp_{1}},
      \nonumber \\ 
 \I_{-}^{\mu} 
 &=& +\,e\Biggl[ \frac{p_{2}^{\mu}}{kp_{2}}
               + Q_{f_2}\frac{k_{2}^{\mu}}{kk_{2}}
               - Q_{f_2'}\frac{{k_{2}'}^{\mu}}{kk_{2}'}
       \Biggr]\frac{D_{2}}{D_{2}+2kp_{2}}.
\end{eqnarray}
The first three interference terms in Eq.~(\ref{corr/real/mtrx})
correspond to the real non-factorizable corrections. 
The last three squared terms in Eq.~(\ref{corr/real/mtrx}) belong to the 
factorizable real-photon corrections. They constitute the semi-soft limit of 
Eq.~(\ref{corr/hard}). 

\subsection{A hybrid scheme -- virtual corrections in DPA and real
corrections from full matrix elements}

The reliability of the error estimate of
$(\alpha/\pi)\times(\GW/\mw)\times\ln(\cdots)\lsim 0.5\%$ for the
accuracy of the DPA can, of course, only be controlled by a comparison
to calculations that are based on the full matrix elements. While 
for the virtual corrections such results do not exist yet, the situation
for the real corrections is much better, since full matrix-element
calculations for the processes $\eeffffg$ are available
\cite{fu94,racoonww_ee4fa,pv4fg}.
The latter results seem to be of particular importance, because the
above error estimate for {\em real} corrections in DPA is subject of
some controversy. 

Although it deserves some care, it is possible to combine the virtual
$\ord{\alpha}$ corrections in DPA with real corrections from the
full $\eeffffg$ lowest-order matrix elements. The non-trivial point in
this combination lies in the relations of IR and mass singularities in
virtual and real corrections.  The singularities have the form of a
universal radiator function multiplied or convoluted with the
respective lowest-order matrix element $\M_0$ of the non-radiative
process.  Since $\M_0$ appears in DPA for the virtual correction
($\M_0^{\DPA}$), but as full matrix element for the real ones, a
simple summation of virtual and real corrections would lead to a
mismatch in the singularity structure and eventually to totally wrong
results.  A solution of this problem is to extract those singular
parts from the real photon contribution that exactly match the
singular parts of the virtual photon contribution, then to replace the
full $|\M_0|^2$ 
by $|\M_0^{\DPA}|^2$ in these terms and finally to add this
modified part to the virtual corrections.  This modification is
allowed in the range of validity of the DPA and leads to a proper
matching of all IR and mass singularities.  The described approach for
such a hybrid DPA scheme is followed in the {\tt RacoonWW} program
\cite{racoonww_res,de00}. More details of this approach can also be
found in \sect{racoonww}.

A particular advantage of this method is due to the fact that the
leading ISR logarithms, which are part of the extracted singularities of
the real corrections, can be easily kept with the full matrix element 
$\M_0$ (see \cite{de00} for details). In this way, the logarithmic
enhancement factor $\ln(\dots)$ 
does not involve large contributions
from the electron mass,
i.e.\ corrections like $\ln(\Me^2/s)$.
In the hybrid scheme, also the non-factorizable corrections have to be
treated carefully. If the full matrix elements for photon radiation is
employed, one cannot exploit any cancellations between real and
virtual non-factorizable corrections, as it is done in the
calculations of \cite{me96,be97,nonfactrcs,nonfactrcslett}. Instead, one needs 
the full set
of non-factorizable virtual corrections, which includes also photons
coupling to the initial state. Such results can be derived from
Eq.~(\ref{corr/virt/nf_mtrx}) and Ref.~\cite{be97}, and are explicitly
given in Ref.~\cite{de00}.

\subsection{Intrinsic ambiguities and reliability of the double-pole
approximation}

The theoretical accuracy of theoretical predictions is indeed at the core of 
the workshop. For this reason it has already been discussed extensively in 
a purely theoretical context. 
Although only the numerical comparisons can tell us where the present
theoretical uncertainty really stands, it is not superfluous that the relevant 
facts are summarized in one place.

An improved assessment of the theoretical uncertainty can be obtained
by varying predictions within the intrinsic freedom of the followed
approach for the DPA.  For instance, any kind of DPA makes use of an
on-shell projection of the off-shell four-fermion phase space to the
phase space with on-shell $\wb$~bosons. The difference between
different on-shell projections is part of the theoretical uncertainty
of the DPA approach and should be considered in predictions (see
\sect{se:tuwithracoonww} for a numerical discussion).

It is a fact of life that questions of principle are sometimes
of scarce practical relevance. CC03 contains gauge-invariance-breaking
terms but what is their numerical impact at LEP~2 energies?
It is quite a known fact that, when computed in the 't Hooft-Feynman gauge,
they are unimportant. At least they are for the $\wb\wb$ total cross-section
-- the signal -- and we can verify this statement by comparing the 
gauge-dependent CC03 with the full gauge-invariant cross-section (CC11 for
instance) including background diagrams. There is a general agreement, dating
from the '95 workshop that the difference is less than $0.2\%$ at LEP~2
energies.

It is bizarre that one can render the Born CC03 diagrams gauge-invariant
at the prize of large numerical variations; it is enough to project 
the kinematics in the matrix elements 
onto the on-shell phase space, while keeping the off-shellness in 
the Breit-Wigner propagators. 
However, this changes the cross-section by several per cent!
Therefore, the use of DPA at Born level (CC03) is numerically 
not recommendable. Once more, for lowest-order reactions one needs an 
alternative approach and for predictions that have a DPA Born and a DPA 
$\ord{\alpha}$ and nothing else the expected accuracy is no more than 
$\Gamma_{\ssW}/\mw \approx 2.5\%$. The difference between Born CC03 and Born DPA
should not enter in the discussion of the theoretical uncertainty.

At the Born level one can accept a non-gauge-invariant CC03 
cross-section (at least in the 't Hooft-Feynman gauge) as a reasonable
quantity at LEP~2 energies. For higher energies one should be more careful.

The same phenomenon will occur when we include radiative corrections and
we would like to add some comment on the DPA procedure, in particular
on the choice of projecting the kinematics.

For high enough energies, any process $e^+e^- \to VV$ will be a dominant
source of four-fermion final states
due to the double resonant enhancement and hence
CC03(NC02) will be a good approximation to the total cross-section for
four-fermion production in a situation where we exclude certain regions
of the phase space, e.g., a small scattering angle of the outgoing electron
in single-$\wb$ production. 

Thus, for example, to calculate the cross-section
$e^+e^- \to VV$ one proceeds as described above; one calculates the
matrix element for $e^+e^- \to VV \to 4\rmf$ and extracts the part resonant
in the invariant masses of the pairs, $k^2_+, k^2_-$. The general matrix
element takes the form
\bq
{\cal M}\lpar \dots,k_+,k_-,\dots\rpar = \asums{i} 
M_i\lpar \dots,k_+,k_-,\dots\rpar\,A_i\lpar\dots,k^2_+,k^2_-\rpar,
\eq
where the $M_i$ contain the spinor and Lorentz tensor structure of the
matrix element, \eg they have the external fermionic wave-functions attached.
The $A_i$ are Lorentz scalars that depend
on the invariants of the problem
and become non-trivial and difficult to compute when higher order corrections
are included. One way of looking at the DPA-procedure is to say that the
resonant part is extracted from the $A_i$, by Laurent expansion.
The external particle wave functions, 
and hence the $M_i$,
should not be affected by the process
hence the kinematics of the problem should be left unchanged because the
final state integrations involve only the fermions, stable on-shell particles.
The gauge nature of the theory is intimately connected with the $A_i$ not
with kinematics.

Whenever we have processes with external, {\em unstable}, vector-bosons, like 
in $\wb\wb \to \wb\wb$ or $\zb\zb \to \zb\zb$, the Higgs resonance will
appear in the $s$-channel and by shifting \eg a factor $s$ from the $M_i$ 
to the $A_i$ one gets factors $s/M^2$ which violate unitarity at high 
energies~\cite{zzzz}.
This can be avoided by making the splitting between the $A_i$ and 
$M_i$ with some care. Here, for $e^+e^- \to \wb\wb,\zb\zb$, the corresponding 
factors do not directly violate unitarity. Nevertheless, one could expect 
that Ward identities are violated by the splitting by terms of the order
$k_{\pm}^2/M^2 -1$, i.e. non double-resonant terms negligible in the DPA
approach. If, on the other hand, one includes 
the $M_i$ in the DPA, as commonly done, one has on-shell matrix elements and 
the WI are fulfilled, at the price of expanding kinematics.

We do not necessarily expect an improvement of the accuracy when taking the 
$M_i$ exactly, but comparing results with DPA applied to $M_i$ or not could 
give an additional estimate on the theoretical uncertainty, of the order of 
$\alpha/\pi\lpar{\rm CC03 Born/CC03 DPA-1}\rpar$. 
We expect that, well above threshold, this will not exceed the quoted $0.5\%$
DPA precision, which involves logarithmic enhancement factors.
                                                        
Another questionable point in DPA is connected to the fact that a
particular mapping may lead to an unphysical point in the on-shell 
phase-space (c.f.~\sect{sec:LPA}).
Even if we do not expand the kinematics in the $M_i$
there are Landau singularities in the 
$A_i$ at the edge of the off-shell phase space. If one performs a DPA 
projection in the $A_i$, these Landau singularities move into the on-shell 
phase space, although only at a distance $(k^2-M^2)/M^2$ from the boundary
\cite{nonfactrcs}.
This might happen when the $A_i$ are parametrized in
terms of invariants. If on the other hand, one parametrizes the $A_i$ in
terms of angles and energies, this can be more easily avoided.

Note that the formulation of a DPA where the on-shell projection is
not applied to the $M_i$ has been implemented 
the formulation of the LPA of Ref.~\cite{ja97} (eqs.(1) and (2)).

\subsection{Remarks on DPA corrections to distributions inclusive
               w.r.t.\ photons}
\label{sec:LPA/RC/distr}

The DPA corrections to distributions that are inclusive w.r.t.\ photons depend
in a very sensitive way on how the four-particle phase space is parametrized, 
or, in other words, on the way the distributions are defined after the photon 
has been integrated out. This statement sounds obvious, but nevertheless 
deserves some special attention.

In particular the invariant-mass distributions ($\wb$ line shapes) are 
affected.
In reactions with two resonances the invariant masses have to be defined from 
the decay products. Depending on the precise definition of the invariant 
masses different sources of large Breit--Wigner distortions can be 
identified~\cite{fsr,YFS-fsr,racoonww_res}, 
in contrast to the situation at LEP1 where only 
initial-state radiation (ISR) can cause such distortions. 

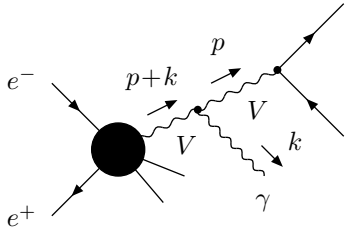
\begin{figure}[htbp]
  \begin{center}
  \unitlength  1pt\small\SetScale{1.0}
  \begin{picture}(150,90)(0,15)
    \ArrowLine(43,43)(25,25)          \Text(8,25)[lc]{$e^+$}
    \ArrowLine(25,75)(43,57)          \Text(8,77)[lc]{$e^-$}
    \Photon(50,50)(110,80){1}{9}
    \LongArrow(61,63)(71,68)          \Text(63,77)[]{$p\!+\!k$}
                                      \Text(76,52)[]{$V$} 
    \LongArrow(85,75)(95,80)          \Text(88,89)[]{$p$} 
                                      \Text(103,65)[]{$V$}    
    \Line(50,50)(75,40)
    \Line(50,50)(67,30)
    \ArrowLine(110,80)(135,105)
    \ArrowLine(135,55)(110,80)
    \Vertex(110,80){1.5}
    \Photon(80,65)(105,40){-1}{6}      \Text(105,30)[]{$\gamma$}
    \Vertex(80,65){1.5}
    \LongArrow(104,51.5)(111,44.5)    \Text(117,53.5)[]{$k$}
    \GCirc(50,50){10}{0}
  \end{picture}
  \end{center} 
  \caption[]{Photon radiation from an unstable particle $V$. Virtualities:
             $s_{\ssV} = p^2$ and $s'_{\ssV} = (p+k)^2$.}
  \label{FSRfig}
\end{figure}

In Ref.~\cite{fsr} 
it has been shown that also final-state radiation (FSR) can induce distortions.
This is a general property of resonance-pair reactions, irrespective of the 
adopted scheme for implementing the finite-width effects. The only decisive 
factor for the distortion to take place is whether the virtuality of the 
unstable particle is defined with ($s'_{\ssV}$) or without ($s_{\ssV}$) the 
radiated photon (see Fig.~\ref{FSRfig}). 

Upon integration over the photon momentum, the former definition 
(cf $M_{i\gamma}^2$ defined in Sec.~\ref{sec:LPA/RC/real}) is free of large FSR
effects from the $V$-decay system. It can only receive large corrections from 
the other (production or {\em decay}) stages of the process. The latter 
definition (cf $M_i^2$ defined in Sec.~\ref{sec:LPA/RC/real}), 
however, does give rise to large FSR effects from the {\em V-decay system}. 
In contrast to the LEP1 case, where the ISR-corrected line shape receives 
contributions from effectively {\em lower} $\zb$-boson virtualities, the 
$s_{\ssV}$ line shape receives contributions from effectively {\em higher} 
virtualities $s'_{\ssV}$ of the unstable particle. As was argued above, only 
sufficiently hard photons ($E_{\gamma}\gg \Gamma_{\ssV}$) can be properly 
assigned to one of the on-shell production or decay stages of the process in 
the DPA. For semi-soft photons [$E_{\gamma} = {\cal O}(\Gamma_{\ssV})$],
however, the assignment is not so clear-cut and will be determined by the experimental
event-selection procedure.

Event selection procedures that involve an  invariant-mass 
definition in terms of the decay products without the photon
give rise to large  FSR-induced distortion effects~\cite{fsr}. 
These are caused by semi-soft photons, since 
hard FSR photons move the virtuality $s'_{\ssV}$ of the unstable particle far off 
resonance for near-resonance $s_{\ssV}$ values, resulting in a suppressed 
contribution to the $s_{\ssV}$ line shape. This picture fits in nicely with the 
negligible overlap of the three on-shell double-pole contributions for hard 
photons, discussed above. The reason why the FSR distortions can be rather 
large lies in the fact that the final-state collinear singularities 
[$\propto\frac{\alpha}{\pi}\,Q_{f}^{2}\ln(m_f^2/M_{\ssV}^2)
         \ln(\Gamma_{\ssV}/M_{\ssV})$] do not vanish, even not for fully 
inclusive photons. After all, a fixed value of $s_{\ssV}$ 
makes it impossible to sum over all degenerate final states by a mere 
integration over the photon momentum. So the KLN theorem does not apply in 
this case. These FSR distortion effects result in shifts in the 
measurement of the $\wb$-boson mass of the order of 40\,MeV, as has
been qualitatively confirmed in Ref.~\cite{YFS-fsr}.

This situation changes for event-selection procedures in which not all photons 
can be separated from the charged fermions.
If photon recombination has to be taken into account,
i.e.\ if photons within a finite cone around the charged fermions have to be 
combined with the corresponding fermion into a single particle, the
mentioned mass singularities connected to final-state fermions
disappear. The KLN theorem applies and the large fermion-mass 
logarithms are effectively replaced by logarithms depending on the cone 
size~\cite{fsr}.
In \citere{YFS-fsr} this expectation has been confirmed numerically,
showing that the large negative shifts in the peak position of the 
$\wb$~invariant-mass distribution obtained without photon recombination are
reduced. In \citere{racoonww_res} it has been shown that the effect of
photon recombination can even overcompensate the momentum loss from FSR
if the recombination is very inclusive. This is due to the recombination
of photons that are radiated off the initial state or off particles
belonging to the other decaying $\wb$~boson. The resulting positive peak
shifts can amount to several $10\MeV$. Explicit numerical results on
$\wb$~invariant-mass distributions can also be found in 
\sect{YFSWWvsRacoonWW}.

Finally we mention a special property of the non-factorizable corrections.
When considering pure angular distributions with an inclusive treatment of the 
photons, one should 
integrate over the photon phase space {\em and} the invariant masses 
$M_{i}^{2}$. After integrating out both invariant masses the non-factorizable 
corrections will vanish, which is a typical feature of the non-factorizable 
interference effects~\cite{nf-cancellations}.

\subsection{Double-pole approximations in practice}

For LEP~2 energies three different groups%
\footnote{Another DPA has been discussed in \citere{ku99} for
linear-collider energies.}  have formulated versions of a DPA 
for $\eeWWffff(+\gamma)$.  While Beenakker, Berends and Chapovsky
\cite{dpa-ww}, called BBC in the following, formulated a semi-analytic
DPA, the other two groups implemented variants of the DPA 
in the event generators {\tt YFSWW} \cite{ja97,ja99} and {\tt RacoonWW}
\cite{de00,racoonww_res}.
The basic features of these different implementations are summarized in
the following.

\subsubsection{The {\tt YFSWW} approach}

{\tt YFSWW}: $\ord{\alpha}$ correction to $\eeWW$ in LPA, 
using the results of Ref.~\cite{ew1}, leading-log
corrections to leptonic $\wb$~decays via {\tt PHOTOS} (up to two
radiative photons with finite $p_t$ according to the exact
$\ord{\alpha}$ soft limit), $\wb$ decays 
normalized to branching ratios, quark
hadronization with {\tt JETSET} and $\tau$ decays 
with {\tt TAUOLA} (including radiative corrections), YFS exponentiation for 
ISR and photon emission from $\wb$-bosons,
off-shell Coulomb singularity, no full non-factorizable corrections -- only
an approximation in terms of the {\em screened} Coulomb ansatz of
Ref.~\cite{ch99}, approximate $\wb$~spin correlations (incomplete correlation
beyond Born) -- they are
missing only in a non-IR non-LL part of EW virtual corrections.

\subsubsection{The {\tt BBC} approach}

BBC: semi-analytical calculation of complete $\ord{\alpha}$ corrections
in DPA (with both factorizable and non-factorizable corrections and
$\wb$~spin correlations), no background. Since the DPA is only valid well 
above threshold, the on-shell part of the Coulomb singularity is automatically
included as part of the factorizable corrections and the off-shell part is 
contained in the non-factorizable corrections, as discussed in 
Ref.~\cite{be97}.

\subsubsection{The {\tt RacoonWW} approach}

{\tt RacoonWW} treats the virtual $\ord{\alpha}$ corrections to
$\eeWWffff$ in DPA. No further approximations beyond the pole
expansion of the matrix element are made, i.e.\ non-factorizable
corrections are included, and $\wb$-spin correlations are respected.
The Coulomb singularity is part of the virtual corrections,
and the corresponding part that goes beyond DPA 
has been added as discussed in \citere{nonfactrcs}. The real $\ord{\alpha}$
corrections are based on the full $4\rmf+\gamma$ matrix element (of the
CC11 class), so that the full kinematics is supported also for photon
radiation. All matrix elements are based on massless fermions, and fermion
masses are introduced only for collinear photon emission that is
inclusive within a (small) finite cone for each fermion.
Thus, a photon collinear to an outgoing fermion has to be recombined
with the corresponding fermion, and a photon close to the beams has to
be considered as invisible.
Initial-state radiation beyond 
$\ord{\alpha}$ is treated in the structure-function approach, 
including soft-photon exponentiation and leading-log contributions 
up to $\ord{\alpha^3}$.


\subsection{The fermion-loop and non-local approaches}
\label{sec:resum}

As was mentioned above, the alternative to {\em subtracting} sub-leading
gauge-violating terms is to {\em add} gauge-restoring terms to the calculation.
In order to do this, one has to add to the amplitude those terms that are 
needed for satisfying the Ward identities. This is not easy to do in general. 
The following observation helps. The very fact that the perturbative amplitudes
require re-summation of the self-energies indicates that either the 
perturbative expansion parameter (coupling constant) is not the proper one, or 
alternatively that the quantity that is expanded (i.e.~the lowest-order 
Lagrangian of the Standard Model) is not the best choice. This observation 
leads one to consider first the one-loop corrected effective potential of the 
Standard Model before doing Born calculations, in order to avoid Dyson 
re-summation of the self-energies.

For the discussion of the fermion-loop and non-local approaches it is therefore
worthwhile to first have a closer look at the origin of the gauge-invariance 
problem associated with the re-summation of self-energies. To this end we 
consider the simple example of an unbroken non-abelian $SU(N)$ gauge theory 
with fermions and subsequently integrate out these fermions~\cite{non-local}. 

First we fix the notations and introduce some conventions. The $SU(N)$ 
generators in the fundamental representation
are denoted by $\T^a$ with $\,a=1,\cdots,N^2\!-\!1$. They are normalized 
according to $\Tr(\T^a\T^b)=\delta^{ab}/2$ and obey the commutation relation
$\Bigl[ \T^a,\T^b \Bigr] = if^{abc}\,\T^c$. In the adjoint representation the 
generators $\F^a$ are given by $(\F^a)^{bc} = -if^{abc}$. The Lagrangian of 
the unbroken $SU(N)$ gauge theory with fermions can be written as 
\beq
\label{lagr}
  \LL(x) 
  = 
  -\,\frac{1}{2}\,\Tr\Bigl[ \TF_{\mu\nu}(x)\,\TF^{\mu\nu}(x) \Bigr] 
  + \bar{\psi}(x)\, (i\,\Ds - m)\, \psi(x),
\eeq
with
\beq
  \TF_{\mu\nu} \equiv \T^a F_{\mu\nu}^a = \frac{i}{g}\,[D_{\mu},D_{\nu}], 
  \qquad 
  D_{\mu} = \partial_{\mu} - ig\,\T^a A_{\mu}^a 
            \equiv \partial_{\mu} - ig\TA_{\mu}.
\eeq
Here $\psi$ is a fermionic $N$-plet in the fundamental representation of
$SU(N)$ and $A_{\mu}^a$ are the ($N^2\!-\!1$) non-abelian $SU(N)$ gauge fields,
which form a multiplet in the adjoint representation. 
The Lagrangian (\ref{lagr}) is invariant under the 
$SU(N)$ gauge transformations
$$
  \psi(x) \to \psi'(x) = \BBCG(x)\,\psi(x), 
$$
\beq
\label{gauge_trans}
  \TA_{\mu}(x) \to \TA'_{\mu}(x) = \BBCG(x)\,\TA_{\mu}(x)\,\BBCG^{-1}(x)
           + \frac{i}{g}\,\BBCG(x)\Bigl[ \partial_{\mu}\BBCG^{-1}(x) \Bigr],
\eeq
with the $SU(N)$ group element defined as $\BBCG(x)=\exp[ig\,\T^a\theta^a(x)]$. 
The covariant derivative $D_{\mu}$ and field strength $\TF_{\mu\nu}$ both 
transform in the adjoint representation
\beq
  D_{\mu} \to \BBCG(x)\,D_{\mu}\,\BBCG^{-1}(x), 
  \qquad 
  \TF_{\mu\nu}(x) \to \BBCG(x)\,\TF_{\mu\nu}(x)\,\BBCG^{-1}(x).
\eeq
Since the Lagrangian is quadratic in the fermion fields, one can integrate 
them out exactly in the functional integral. The resulting effective action is 
then given by
\beq
  i\,S_{\sss{eff}}[J] 
  = 
  i\int d^4x\,\biggl\{ 
            -\,\frac{1}{2}\,\Tr\Bigl[ \TF_{\mu\nu}(x)\,\TF^{\mu\nu}(x) \Bigr]
            + J_{\mu}^a(x)\,A^{a,\,\mu}(x) \biggr\}
  + \Tr\Bigl[ \ln(-\,\Ds - i\,m) \Bigr],            
\eeq
with $J_{\mu}^a(x)$ denoting the gauge-field sources. The trace on the 
right-hand side has to be taken in group, spinor, and coordinate space.
As a next step one can expand the effective action in terms of the coupling 
constant
\begin{eqnarray}
\label{fls/expansion}
  \Tr\Bigl[ \ln(-\,\Ds - i\,m) \Bigr] &=& 
            \Tr\Bigl[ \ln(-\,\partials - i\,m) \Bigr] 
          + \Tr\biggl[ \ln\Bigl( 1+\frac{g}{i\,\partials-m}\,\TAs 
                          \Bigr) \biggr]
  \nonumber \\[1mm]
                                             &=& 
            \Tr\Bigl[ \ln(-\,\partials - i\,m) \Bigr]
          + \sum\limits_{n=1}^{\infty} \frac{(-1)^{n-1}}{n}\,
            \Tr\Biggl[ \biggl( \frac{g}{i\,\partials-m}\,\TAs
                       \biggr)^n \Biggr].
\end{eqnarray}
Note that the left-hand side of Eq.~(\ref{fls/expansion}) is gauge-invariant as
a result of the trace-log operation. In contrast, the separate terms of the 
expansion on the right-hand side are not gauge-invariant. This is due to the 
fact that, unlike in the abelian case, the non-abelian gauge transformation 
(\ref{gauge_trans}) mixes different powers of the gauge field $A_{\mu}$ in 
Eq.~(\ref{fls/expansion}). 
Thus, if one truncates the series on the 
right-hand side of Eq.~(\ref{fls/expansion}) one will in general break gauge 
invariance. From Eq.~(\ref{fls/expansion}) it is also clear that the fermionic 
part of the effective action induces higher-order interactions between the 
gauge bosons. 

What are these higher-order interactions? Let us consider the quadratic
gauge-field contribution 
\beq
\label{fls/AA} 
  -\,\frac{1}{2}\,\Tr\Biggl[ \biggl( \frac{g}{i\,\partials-m}\,\TAs
                             \biggr)^2 \Biggr]
  =
  -\,\frac{1}{2}\,\int d^4x\,d^4y\,\Tr\Bigl[ O(x,y)\,O(y,x) \Bigr],
\eeq
where
\beq 
  O(x,y) = g\,S^{(0)}_{\sss{F}}(x-y)\,\TAs(y)
\eeq
and $i\,S^{(0)}_{\sss{F}}(x-y)=\,<\!0\,|\,T(\psi(x)\,\bar{\psi}(y))\,
|\,0\!>_{\sss{free}}\,$ is the free fermion propagator. The trace on the 
right-hand side of Eq.~(\ref{fls/AA})
has to be taken in group and spinor space. A quick glance at this quadratic 
gauge-field contribution reveals that it is just the one-loop self-energy of 
the gauge boson induced by a fermion loop. In the same way, the higher-order 
terms $\sim g^n A^n$ in Eq.~(\ref{fls/expansion}) are just the fermion-loop 
contributions to the $n$-point gauge-boson vertices.

One can truncate the expansion in Eq.~(\ref{fls/expansion}) at $n=2$, thus 
taking into account only the gauge-boson self-energy term and neglecting the 
fermion-loop contributions to the higher-point gauge-boson vertices. This is
evidently the simplest procedure for performing the Dyson re-summation of the 
fermion-loop self-energies. However, as was pointed out above, truncation of
Eq.~(\ref{fls/expansion}) at any finite order in $g$ in general breaks 
gauge invariance. This leads to the important observation that, {\em although 
the re-summed fermion-loop self-energies are gauge-independent by themselves, 
the re-summation is nevertheless responsible for gauge-breaking effects in the
higher-point gauge-boson interactions through its inherent mixed-order nature}.
Another way of understanding this is provided by the gauge-boson Ward
identities. Since the once-contracted $n$-point gauge-boson vertex can be 
expressed in terms of $(n\!-\!1)$-point vertices, it is clear that gauge 
invariance is violated if the self-energies are re-summed without adding the 
necessary compensating terms to the higher-point vertices.

An alternative is to keep all the terms in Eq.~(\ref{fls/expansion}). Then the 
matrix elements derived from the effective action will be gauge-invariant. 
Keeping all the terms means that we will have to take into account not only 
the fermion-loop self-energy in the propagator, but also all the possible 
fermion-loop contributions to the higher-point gauge-boson vertices. This is 
exactly the prescription of the fermion-loop scheme 
(FLS)~\cite{BHF1,fls-baur,BHF2,fls-mf}. 
Although the FLS guarantees gauge invariance of the matrix elements, it has 
disadvantages as well. Its general applicability is limited to those situations
where non-fermionic particles can effectively be discarded in the 
self-energies, as is for instance the case for $\Gamma_{\ssW}$ and $\Gamma_{\ssZ}$ at 
lowest order. Another disadvantage is that in the FLS one is forced to do the 
loop calculations, even when calculating lowest-order quantities. For example, 
the calculation of the tree-level matrix element for the process 
$e^+e^- \to 4\rmf\gamma\,$ involves a four-point gauge-boson interaction, which 
has to be corrected by fermion loops in the FLS. This over-complicates an 
otherwise lowest-order calculation.

It is clear that the FLS provides more than we actually need. It does not only 
provide gauge invariance for the Dyson re-summed matrix elements at a given 
order in the coupling constant, but it also takes into account all
fermion-loop corrections at that given order. In the vicinity of unstable
particle resonances the imaginary parts of the fermion-loop self-energies are 
effectively enhanced by ${\cal O}(1/g^2)$ with respect to the other 
fermion-loop corrections. Therefore, what is really needed is only a minimal 
subset of the non-enhanced contributions such that gauge invariance
is restored. In a sense one is looking for a minimal solution of a system of 
Ward identities. The FLS provides a solution, but this solution is far from 
minimal and is only practical for particles that decay exclusively into 
fermions. Since the decay of unstable particles is a physical phenomenon,
it seems likely that there exists a simpler and more natural method for 
constructing a solution to a system of Ward identities, without an explicit
reference to fermions. This is precisely the philosophy behind the non-local
approach~\cite{non-local}. This approach consists in using gauge-invariant 
non-local effective Lagrangians for generating both the self-energy effects in 
the propagators and the required gauge-restoring terms in the higher-point 
interactions. In this way the full set of Ward identities can be solved, while 
keeping the gauge-restoring terms to a minimum.


\subsubsection{The fermion-loop scheme}
\label{sec:resum/fls}

The Fermion-Loop scheme developed in~\cite{BHF1} and refined in
\cite{BHF2} makes the approximation of neglecting all masses for
the incoming and outgoing fermions in the processes $e^+e^- \to n\,$fermions.
It is possible, however, to go beyond this approximation~\cite{abm,tfl} 
and give the construction of an exact Fermion-Loop scheme (EFL)~\cite{tfl}, 
\ie, a scheme for incorporating the finite-width effects in the theoretical 
predictions for tree-level, LEP~2 and beyond, processes.

One can work in the 't Hooft-Feynman gauge and create all relevant
building blocks, namely the vector-vector~\cite{book}, vector-scalar and 
scalar-scalar~\cite{tfl} transitions of the theory, all of them one-loop 
re-summed. 
The loops, entering the scheme, contain fermions and, as done before 
in~\cite{BHF2}, one allows for a non-zero top quark mass inside loops. 
There is a very simple relation between re-summed transitions and running 
parameters, since Dyson re-summation is most easily expressed in terms of 
running couplings and running mixing angles.

In the EFL generalization, it is particularly convenient to introduce
additional running quantities. They are the running masses of the vector
bosons, $\bzms(p^2) = M^2(p^2)/c^2(p^2)$, formally connected to the location
of the $\wb$ and $\zb$ complex poles.
After introducing these running masses, it is straightforward to prove that
all $\Smat$-matrix elements of the theory assume a very simple structure.
Coupling constants, mixing angles and masses are promoted to running quantities
and the $\Smat$-matrix elements retain their Born-like structure, with
running parameters instead of bare ones, and vector-scalar or scalar-scalar
transitions disappear if we employ unitary-gauge--like vector boson propagators
where the masses appearing in the denominator of propagators are the running 
ones. If the $\wb-\wb$ and $\phi-\phi$ transitions are denoted by 
$S^{\mu\nu}_{\ssW}$ and by $S_{\phi}$ with,
\bqa
S^{\mu\nu}_{\ssW} &=& \frac{g^2}{16\,\pi^2}\,\Sigma^{\mu\nu}_{\ssW},  \quad
\Sigma^{\mu\nu}_{\ssW} = \Sigma^0_{\ssW}\,\delta^{\mu\nu} + 
\Sigma^1_{\ssW}\,p^{\mu}p^{\nu},  \nl
S_{\phi} &=& \frac{g^2}{16\,\pi^2}\,\Sigma_{\phi},
\eqa
then, the $\wb$-boson running mass is defined by the following 
equation (note the metric):
\bq
\frac{1}{M^2(p^2)} = \frac{1}{M^2}\,{{p^2-S^0_{\ssW}+\frac{M^2}{p^2}\,S_{\phi}}
\over {p^2-S_{\phi}}},
\eq
The whole amplitude can be written in terms of a $\wb$-boson 
exchange diagram, if we make use of the following effective propagator:
\bq
\Delta^{\mu\nu}_{\rm eff} = \frac{1}{p^2+M^2-S^0_{\ssW}}\,\Big[
\delta^{\mu\nu} + \frac{p^{\mu}p^{\nu}}{M^2(p^2)}\Big].
\eq
For the vertices we need that all vector-boson lines be off mass-shell and 
non-conserved and, moreover, a Ward identity has to be computed and not only 
the corresponding amplitude. Therefore, the number of terms increases 
considerably with respect to the standard formulation of the FL-scheme and we 
refer to Ref.~\cite{tfl} for all details.  

The renormalization of ultraviolet divergences can be easily extended to the 
EFL-scheme by showing that all ultraviolet divergent parts of
the one-loop vertices, $\ph\wb\wb, \ph\wb\phi, \ph\phi\wb$ and
$\ph\phi\phi$ for instance, are proportional to the lowest order part.
Therefore, the only combinations that appear are of the form $1/g^2 + 
\vb\vb\vb\,$vertex or $M^2/g^2 + \vb\vb\phi\,$ vertex etc.
All of them are, by construction, ultraviolet finite.

Equipped with this generalization of the Fermion-Loop scheme, one can
prove the fully-massive $U(1)$ Ward identity which is required for a
correct treatment of the single-$\wb$ processes. As a 
by-product of the method, the cross-section for single-$\wb$ production
automatically evaluates all channels at the right scale, without having
to use ad hoc re-scalings and avoiding the approximation of a unique scale
for all terms contributing to the cross-section. 

The generalization of the Fermion-Loop scheme goes beyond its, most
obvious, application to single-$\wb$ processes and allows for a gauge
invariant treatment of all $e^+e^- \to n\,$fermion processes with a
correct evaluation of the relevant scales. Therefore, the EFL-scheme can
be applied to several other processes like $e^+e^- \to \zb\ph^*$ and,
in general to $e^+e^- \to 6\,$fermion processes, with the inclusion
of a stable, external, top quark, but it does not apply to reactions involving
the physical Higgs boson. 
Furthermore, the scheme misses those corrections to the total decay
width in the propagator denominators that are induced by
two-loop contributions.

\subsubsection{The non-local approach}
\label{sec:resum/NL}

The main idea of the non-local approach is to rearrange the series on the 
right-hand side of Eq.~(\ref{fls/expansion}) in such a way that each term 
becomes gauge-invariant by itself. Subsequent truncation of the series at a 
given term is then allowed. It is possible to approximate 
Eq.~(\ref{fls/expansion}) by means of an effective Lagrangian in such a way 
that the resulting effective action has the following properties:
\begin{itemize}
  \item  it generates the Dyson re-summed transverse gauge-boson self-energy
         in the propagator. This means that it contributes to the gauge-boson 
         two-point function. Hence, the effective lagrangian should depend at 
         least on two gauge-boson fields.
  \item  the Dyson re-summed self-energy is in general not a constant, but 
         rather a function dependent on the interaction between the 
         gauge-bosons and the fermions. This means that the effective 
         Lagrangian should in general be non-local (bi-local) in the gauge 
         fields. Thus the gauge fields should be taken at two different 
         space--time points.
  \item  it is gauge-invariant. As such the effective Lagrangian should have 
         the form of an infinite tower of gauge fields. 
\end{itemize}
For the gauging procedure of the non-local Lagrangians we 
will need a special ingredient, the {\em path-ordered exponential}, which is 
defined as 
\beq
\label{Pexp}
  U(x,y) = U^{\dagger}(y,x) \ =\ \mbox{P}\exp\Biggl[ 
             -\,ig\int\limits_x^y\TA_{\mu}(\omega)\,d\omega^{\mu} \Biggr]
\eeq
Here $d\omega^{\mu}$ is the element of integration along some path 
$\Omega(x,y)$ that connects the points $x$ and $y$.%
\footnote{In principle we are free to choose this particular path. This freedom
          is just one out of the many freedoms that characterize the treatment 
          of unstable particles (as mentioned earlier). It just reflects the 
          fact that in a perturbative expansion one is free to pick up 
          additional higher-order contributions, since the answer at any given
          (truncated) order will not be changed by such additional terms.}
The so-defined path-ordered exponential transforms as 
\beq
  U(x,y) \to \BBCG(x)\,U(x,y)\,\BBCG^{-1}(y)
\eeq
under the $SU(N)$ gauge transformations. It hence carries the gauge 
transformation from one space-time point to the other. 

For a $SU(N)$ Yang--Mills theory the non-local action with the above-described
properties takes the form
\beq
\label{ww/action}
  \SNL 
  = 
  -\,\frac{1}{2} \int d^4x\,d^4y\,\Sigma_{\sss{NL}}(x-y)\, 
  \Tr\Bigl[ U(y,x)\,\TF_{\mu\nu}(x)\,U(x,y)\,\TF^{\mu\nu}(y) \Bigr]
  \equiv
   \int d^4x\,d^4y\,\LNL(x,y),
\eeq
with $\LNL(x,y)$ the non-local effective Lagrangian.
As required, the action contains bilinear gauge-boson interactions. The induced
infinite tower of higher-point gauge-boson interactions, which are also of
progressively higher order in the coupling constant $g$, is needed for 
restoring gauge invariance.

It is important to stress at this point that this term in the effective action
should not be understood as a new fundamental interaction. It is generated by 
radiative corrections. From the point 
of view of general properties of non-local Lagrangians, the non-local 
coefficient $\Sigma_{\sss{NL}}(x-y)$ is arbitrary. In practice, however, it is 
fixed by the explicit interaction between the gauge-bosons and the fermions
in the underlying fundamental theory. In our simple example this connection 
is given by Eq.~(\ref{fls/expansion}).

Let us now derive the two-point function as an example of the Feynman rules 
generated by Eq.~(\ref{ww/action}): 
\beq
  \begin{picture}(80,30)(0,-2)
    \Photon(0,0)(80,0){2}{10}
    \ArrowLine(25,-12)(26,-12)
    \Line(15,-12)(25,-12)
    \ArrowLine(55,-12)(54,-12)
    \Line(65,-12)(55,-12)
    \Text(0,8)[lb]{$a_1,\mu_1$}
    \Text(0,-8)[lt]{$q_1$}
    \Text(80,8)[rb]{$a_2,\mu_2$}
    \Text(80,-8)[rt]{$q_2$}
    \GCirc(40,0){3}{0}
  \end{picture} 
  \hspace*{5ex}
  :\
  i\,\Sigma^{a_1a_2,\,\mu_1\mu_2}(x_1,x_2) 
  = 
  \frac{i\,\delta^2 (\SL+\SNL)}{\delta A^{a_1}_{\mu_1}(x_1)\,
        \delta A^{a_2}_{\mu_2}(x_2)}\Biggl.\Biggr|_{A=0}, 
  \vspace*{2ex}
\eeq
where the local action $\SL$ follows from the gauge-boson term in 
Eq.~(\ref{lagr}). The Fourier transform of this two-point function can be 
calculated in a straightforward way, since the path-ordered exponentials are
effectively unity. The result reads 
\beq
\label{ww/self-energy}
  i\,\tilde{\Sigma}^{a_1a_2,\,\mu_1\mu_2}(q_1,q_2)
  =
  i\,\delta^{a_1a_2} \Bigl( q_1^{\mu}q_1^{\nu} - q_1^2 g^{\mu\nu} \Bigr)\,
  \Bigl[ 1 + \tilde{\Sigma}_{\sss{NL}}(q_1^2) \Bigr]
  \,(2\pi)^4\,\delta^{(4)}(q_1+q_2).
\eeq
Note that this two-point interaction is transverse, as it should be for an 
unbroken theory. The non-local coefficient acts as a (dimensionless) 
correction to the transverse free gauge-boson propagator, exactly what is
needed for the Dyson re-summation of the gauge-boson self-energies. The infinite
tower of gauge-restoring higher-point gauge-boson interactions are provided by
the gauge-boson fields present in both $\TF_{\mu\nu}$ and the path-ordered
exponential occurring in Eq.~(\ref{ww/action}). For explicit Feynman rules we
refer to Ref.~\cite{non-local}. 

Although the above-described non-local procedure provides a gauge-invariant 
framework for performing the Dyson re-summation of the gauge-boson 
self-energies, we want to stress that it is not unique. We have seen above 
that the FLS provides a different solution of the system 
of gauge-boson Ward identities. In the context of non-local effective 
Lagrangians it is always possible to add additional towers of gauge-boson 
interactions that start with three-point interactions and therefore do not 
influence the Dyson re-summation of the gauge-boson self-energies.

In the light of the discussion presented in \sect{sec:resum}, we rearrange
the series on the right-hand side of Eq.~(\ref{fls/expansion}) according to 
gauge-invariant towers of gauge-boson interactions labelled by the minimum 
number of gauge bosons that are involved in the non-local interaction. 
Effectively this constitutes an expansion in powers of the coupling constant 
$g$, since a higher minimum number of particles in the interaction is 
equivalent to a higher minimum order in $g$. In order to achieve minimality we 
have truncated this series at the lowest effective order. This should not
be viewed as some {\em ad hoc} recipe, but rather as a systematic expansion 
of the effective potential.

Up to now we have seen how the non-local effective Lagrangian method works for 
unstable gauge bosons in a simple $SU(N)$ gauge theory with fermions.
In the Standard Model there are different types of unstable particles: the
top-quark, the massive gauge-bosons, the Higgs boson. In Ref.~\cite{non-local}
it was shown how to extend the above-described method in such a way that it
allows the description of all the unstable particles in terms of bi-local 
effective Lagrangians. 

\section{The CC03 cross-section, $\sigma_{\ssW\ssW}$}
\label{sectWW}

As mentioned before, a new electroweak $\ord{\alpha}$ CC03 cross-section is
available, showing a result that is between $2.5\%$ and $3\%$ smaller than 
the old 1995  CC03 cross-section predicted with {\tt GENTLE}.   
This is a big effect since the combined experimental accuracy of LEP
experiments is even smaller.

In the '95 workshop~\cite{EGWWP} predictions for CC03 were produced with 
variations in the IPS which agreed 
at the level of $1\%$, and then a $2\%$ theoretical
error was quoted, to be conservative. How does this estimate compare with the
present shift of $2.5 \div 3\%$ downwards?
This is a $1.25$ to $1.5$ sigma difference, totally acceptable
within the area of statistics. Certainly, this is more of a systematic 
theoretical uncertainty which is hard to quantify, but still: it is
compatible and in agreement. 
However, a comment is needed here. In '95 several groups produced tuned
comparisons for CC03 agreeing at the level of one part in $10^4$.
Then they moved to the Best-You-Can approach, defined by
switching on all flags to get the best physics description according
to the flag description of individual codes. The program {\tt GENTLE}, 
in its BYC-mode, 
was selected to represent the Standard Model. However, if we take other codes,
noticeably {\tt WPHACT} and {\tt WTO}, we easily discover CC03, Born-like,
predictions that have a maximal $+1.6\%$ shift with respect to {\tt RacoonWW}
($+1.3\%$ with respect to {\tt YFSWW}) 
at the highest energy. Therefore,
the old estimate of $2\%$ in theoretical accuracy was not underestimated.

It is important to discuss the numerical
predictions for the DPA-corrected CC03 cross-section.  
Therefore, in this Section, we present numerical results and also an accurate
description of the comparisons between different approaches, {\tt YFSWW},
{\tt BBC} and {\tt RacoonWW}.
In principle, one would like to understand the effect of DPA and, therefore, is
interested in the ratio (with DPA)/(without DPA), both with ISR, (naive) QCD
etc. for each of the programs.
For this Report, however, this was not done and we have to take the old 
results (\eg {\tt GENTLE}) for a comparison {\em new} -- {\em old}.
By comparing different calculations one can numerically check the
quality of the DPA for CC03.

\subsection{Description of the programs and their results}

\subsubsection*{CC03 with {\tt RacoonWW}}
\label{racoonww}

\subsection*{Authors}

\begin{tabular}{l}
A.Denner, S.Dittmaier, M.Roth and D.Wackeroth
\end{tabular}

\subsubsection*{General description}

The program {\tt RacoonWW} \cite{de00} evaluates cross-sections
and differential distributions for the reactions $\Pep\Pem\to 4\rmf$ and
$\Pep\Pem\to 4\rmf+\gamma$ for all four-fermion final states.  For the
W-pair mediated channels $\Pep\Pem\to\PW\PW\to 4\rmf(+\gamma)$ the full
virtual $\ord{\alpha}$ corrections are taken into account in
DPA, while for the corresponding real
corrections the full  $4\rmf+\gamma$ matrix elements are used.

\subsubsection*{Features of the program}
\vskip 0.3cm
\begin{itemize}
\item {\bf Lowest order:} the full matrix elements for all $4\rmf$ final
  states are included, and the contribution of the CC03 matrix
  elements or of other subsets of diagrams is provided as an option.
  All external fermions are assumed to be massless.
\item
{\bf Virtual \boldmath{$\ord{\alpha}$} corrections:} the full one-loop
corrections are included in DPA, \ie all
factorizable corrections \cite{prod-corr} and
non-factorizable corrections \cite{nonfactrcs}. In this way, full 
$\wb$-spin correlations are taken into account.
\item
  {\bf Real corrections --- \boldmath{$4\rmf+\gamma$} production}: the
  cross-sections are based on the full matrix-element calculation
  \cite{racoonww_ee4fa} for all $4\rmf+\gamma$ final states with massless
  fermions.  If the process $\eeffff+\gamma$ is investigated with a
  separable photon, i.e.\ if the photon is neither soft nor collinear
  to a charged fermion, all $4\rmf+\gamma$ final states are possible, and
  subsets of diagrams can be chosen as options (e.g.\ boson-pair
  production diagrams, QCD background diagrams).  If the real
  corrections to $\eeWWffff$ are calculated, the full
$4\rmf+\gamma$ matrix elements for the CC11 class%
\footnote{The CC11 class is the smallest gauge-invariant subset of
  diagrams for $\eeffff$ that contains all graphs with two resonant
  $\wb$~bosons; in this class only those background diagrams are
  missing that are peculiar to $\Pe^\pm$, $\nu_\Pe$, $\barnu_\Pe$, or
  $f\barf$ pairs in the final state.}  are taken, i.e.\ photon
radiation from background diagrams is partially included. 

  Depending on the choice of the user, the cancellation of collinear
  and infrared singularities is performed within the phase-space slicing
  method or within the subtraction formalism of \citere{subtract}.
  In both cases, care is taken in avoiding mismatch between the singularities
  of the virtual and the real corrections, which is non-trivial owing to
  the application of the DPA to the virtual corrections only.
The treatment of fermion-mass singularities is described below in more
detail.
\item {\bf ISR:} higher-order ISR is implemented via structure
  functions for the incoming ${e^+}$ and ${e^-}$. The
  structure functions used are those of \citere{structurefunctions}
  with the `BETA' choice, \ie the collinear-soft leading logarithms
  are exponentiated.  If the $\ord{\alpha}$ corrections to $\eeWWffff$
  are included, the $\ord{\alpha}$ contributions already contained in
  the structure functions are subtracted, in order to avoid double
  counting, and the full CC11 Born matrix elements are used in the
  convolution.
\item{\bf Treatment of collinear photons:} the program is only
  applicable to observables that involve no mass-singular
  contributions from the final state.  
  These mass singularities cancel
  if all photons collinear to a charged final-state fermion are
  combined with this fermion%
\footnote{Note that without photon recombination, only the total cross 
section (without any cuts) fulfills this requirement, whereas distributions 
or cuts that make use of fermion momenta in general involve mass-singular
corrections.}.
  The recombination procedure is
  controlled by recombination cuts, i.e.\ photon emission angles and
  photon energies, or invariant masses of photon--fermion pairs.
  Specifically, first the charged fermion that is closest to the
  photon according to these criteria (emission angle or invariant
  mass) is selected, and secondly the photon is recombined with this
  fermion if it is within the recombination cuts for a final-state
  fermion and discarded for an initial-state fermion.  The mass
  singularities that remain from collinear photon emission off
  initial-state electron or positron [i.e.\ the $(\alpha\ln\Me)^n$
  terms] are included in the structure functions.

\item
{\bf Coulomb singularity:} within DPA it is fully included in the 
  $\ord{\alpha}$ corrections.  The full off-shell 
  behaviour of the singularity as described in  \citere{nonfactrcs}
  can be switched on as an option. 
\item
{\bf Finite gauge-boson widths:}
  in the tree-level processes $\Pep\Pem\to 4\rmf,4\rmf+\gamma$ several options
  are included, such as fixed-width, running-width, and complex-mass
  scheme \cite{racoonww_ee4fa}.
  If $\ord{\alpha}$ corrections are taken into account the
  fixed width is automatically used.
  
\item {\bf Cuts:} since each event is completely specified, in
  principle any conceivable phase-space cut can be implemented.
  However, since all fermions are taken to be massless, singularities
  can occur in photon-exchange channels, rendering cuts unavoidable.
  In particular, if a charged fermion--anti-fermion pair is produced,
  a lower cut on its invariant mass has to be specified, or if a
  final-state electron or positron is present, cuts on its minimal
  angle to the beam and its minimal energy are required. For
  calculations based on restricted sets of diagrams, not all cuts are
  necessary; in particular, no cut at all is needed for the CC03
  diagrams.

\item {\bf QCD contributions:} gluon-exchange contributions can be
  switched on in the tree-level processes $\Pep\Pem\to 4\rmf,4\rmf+\gamma$.
  Gluon-emission processes $\Pep\Pem\to 4\rmf+g$ can be calculated for the
  CC11 class of $4\rmf$ final states.

  For the QCD corrections to $\eeWWffff$, one can choose between the
  naive QCD factors of $(1+\alpha_{\mathrm{s}}/\pi)$ per hadronically
  decaying W~boson and the full $\ord{\alpha_{\mathrm{s}}}$ corrections 
  in DPA. The full calculation is performed in the 
  same way as the photonic parts of the $\ord{\alpha}$ corrections.
\item {\bf IBA:} the program includes, 
  as an option, an improved Born
  approximation (IBA) \cite{bo92}, which involves the leading ISR
  logarithms, the
  running of the electromagnetic coupling, corrections associated with
  the $\rho$ parameter, and the Coulomb singularity.
\item {\bf Subsets of diagrams:} 
  For lowest-order predictions of $\eeffff,4\rmf+\gamma$ there is
  the possibility to select subsets of diagrams, such as those
  including the pair production of $\PW$, $\PZ$, $\PZ/\gamma^*$, or
  $\PW/\PZ/\gamma^*$ bosons.  Furthermore, all diagrams corresponding
  to the CC11 process class can be selected.
\item
{\bf Intrinsic ambiguities:}
  the accuracy of the DPA can be studied by changing the DPA within its
  intrinsic ambiguities. This is described in \sect{se:tuwithracoonww}.
\end{itemize}

\subsubsection*{Program layout}

{\tt RacoonWW} consists of two nearly independent 
Monte Carlo programs: one uses phase-space slicing and the other the 
subtraction method of \citere{subtract}. 
Only the main control program, the routines for photon 
recombination and phase-space cuts, and the calculation of the 
matrix elements are commonly used. 
The numerical integration is performed with the multi-channel Monte
Carlo technique \cite{Multichannel} and adaptive weight optimization
\cite{Kl94}. The generator produces weighted events.

\subsubsection*{Input parameters/schemes}

{\tt RacoonWW} needs the following input parameters:
\beqar
&& \alpha(0), \alpha(\mz), \gf, \alpha_{\mathrm{s}}, \qquad
\mw, \mz, \mh, \Gamma_{\ssW}, \Gamma_{\ssZ}, \nn\\
&& m_f, \quad f = \Pe, \mu, \tau, \Pu, \Pc, \Pt, \Pd, \Ps, \Pb. 
\label{eq:racoonww_input}
\eeqar
The weak mixing angle is fixed by $\cw^2=1-\sw^2=\mws/\mzs$, and the
quark-mixing matrix is set to unity. The masses of external fermions
are consistently set to zero where possible. While the masses of the
final-state fermions appear only as regulators, the mass singular
logarithms of ISR depend on $\Me$.  The user can choose between the
externally fixed $\wb$~width $\Gamma_\PW$ and an internally calculated
value including electroweak and/or QCD one-loop radiative corrections.

\begin{sloppypar}
The parameter set \refeq{eq:racoonww_input} is over-complete.
The program supports three different input schemes, fixing the
independent parameters. We recommend to use the $\gf$ scheme where
the tree level is fixed by $\gf$, $\MW$, and $\MZ$ and the relative
$\O(\alpha)$ corrections are calculated with $\al(0)$.
\end{sloppypar}
The code is available from the authors upon request.

\subsubsection*{Numerical results}

In \tabn{xsrac} we list the predictions of {\tt RacoonWW} for the total
CC03 cross-section including radiative corrections (best-with-CC03-Born as
defined below). We give the results for one leptonic channel, for one
semi-leptonic channel, for one hadronic channel, and for the sum of all
channels separately. Note that for CC03 and negligible
fermion masses the results are independent of the
final state within these channels. No cuts are applied.
While in all other {\tt RacoonWW} results in this report LL
$\ord{\al^3}$ corrections according to \citere{structurefunctions} are included, in this table only the LL
$\ord{\al^2}$ terms are taken into account. The LL
$\ord{\al^3}$ contributions reduce the cross-sections by only about $0.02\%$
The given errors are purely statistical. The error for the total cross
section were obtained by adding the (statistically correlated) errors
of the various channels linearly.

\clearpage

\begin{table}[htbp]\centering
\def~{\phantom{0}}
\begin{tabular}{|c|c|c|c|c|}
\hline
&&&&\\
$\sqrt{s}\,$[GeV] & lept. [fb] & semi-lept. [fb] & hadr. [fb] & total [pb] \\
&&&&\\
\hline
&&&&\\
172.086  &   142.088(71)~ &   442.50(36) &   1376.14(67)~ &   12.0934(76)~\\
176.000  &   160.076(78)~ &   498.03(25) &   1550.04(75)~ &   13.6171(67)~\\
182.655  &   180.697(89)~ &   562.22(28) &   1749.48(86)~ &   15.3708(76)~\\
188.628  &   190.882(96)~ &   594.31(55) &   1848.07(92)~ &   16.2420(111)\\
191.583  &   194.271(118) &   604.12(31) &   1880.19(94)~ &   16.5187(85)~\\
195.519  &   197.320(123) &   614.11(31) &   1911.45(97)~ &   16.7910(88)~\\
199.516  &   199.497(103) &   620.53(33) &   1931.28(99)~ &   16.9670(89)~\\
201.624  &   200.200(104) &   622.65(33) &   1937.94(100) &   17.0254(89)~\\
210.000  &   200.910(107) &   624.95(33) &   1945.00(103) &   17.0876(91)~\\
&&&&\\
\hline
\end{tabular}
\vspace*{3mm}
\caption[]{Cross-sections for $e^+e^- \to \wbp\wbm \to 4\rmf$ 
from {\tt RacoonWW}.}
\label{xsrac}
\end{table}

In the following we show the predictions from {\tt RacoonWW} for
the $M(\wbm)$ invariant mass distributions in four different configurations:

\begin{tabular}{ll}
4f-Born:           &  full $\eeffff$ Born without radiative corrections; \\
best-with-4f-Born: &  full $\eeffff$ Born plus radiative corrections \\
                     &   including ISR beyond $\ord{\alpha}$,  \\
                     &   soft photon exponentiation, \\
                     &   LL $\ord{\alpha^3}$, and naive QCD ; \\
CC03-Born:           &   CC03 Born without radiative corrections; \\
best-with-CC03-Born: &   CC03 Born plus radiative corrections \\
                     &    including ISR beyond $\ord{\alpha}$, \\
                     &    soft photon exponentiation, \\
                     &    LL $\ord{\alpha^3}$, and naive QCD, \\
\end{tabular}

for the three final states, $\mu^+\nu_{\mu}\tau^-
\barnu_{\tau}, u \bard \mu^- \barnu_{\mu}$ and $u \bard s \barc$ 
at $\sqrt{s}=200\GeV$. 

As explained in the text, DPA sits only in the virtual correction in
the {\tt RacoonWW} approach. Everything else is (or can be) calculated
from full $4f(\gamma)$ matrix elements. This means that
best-with-4f-Born and best-with-CC03-Born contain the same DPA part
(the virtual correction).

All distributions have been obtained with the following cut
and photon recombination procedure:
\newcommand{\recomb}{{\mathrm{rec}}}%
\begin{enumerate}
\item[--]  All photons within a cone of $5^\circ$ around the beams are
  treated as invisible, \ie their momenta are disregarded when
  calculating angles, energies, and invariant masses.
\item[--]  Next, the invariant masses of the photon with
  each of the charged final-state fermions are calculated. If the
  smallest one is smaller than $M_\recomb$ or if the photon energy is
  smaller than $1\,\GeV$, the photon is combined with
  the corresponding fermion, \ie the momenta of the photon and the
  fermion are added and associated with the momentum of the fermion,
  and the photon is discarded.
\item[--] Finally, all events are discarded in which one of the
  final-state charged fermions is within a cone of $10^\circ$ around
  the beams. No other cuts are applied.
\end{enumerate}
We consider the cases of a tight recombination cut $M_\recomb= 5\,\GeV$
({\em bare}) and of a loose recombination cut $M_\recomb= 25\,\GeV$
({\em calo}). Born predictions are independent of the recombination
cut. The $\wbm$ invariant-mass is always defined via
the four-momenta (after eventual recombination with the photon)
of the $\wbm$ decay fermions.

\clearpage

In \fig{rac_dpa_12} (left) we show the CC03-Born predictions for 
the $M(\wbm)$ distributions for all three final states. 
The r.h.s.\ of \fig{rac_dpa_12} shows {\em best-with-CC03-Born} with the
{\em bare} recombination cut , i.e.\ the corrections are included in
DPA.
In \fig{rac_dpa_34} we show the effect of the radiative corrections by
computing the ratio of the invariant-mass distributions including
radiative corrections and the Born distributions both for {\em bare}
  and {\em calo} recombination.
In the peak region, i.e. $|M(W^-)-\mw| < \Gamma_{\ssW}/2$, the effects of
radiative corrections lower the line-shape by approximately $3\%$
$(5\%)$ ($u \bard s \barc$), $7\%$ $(7\%)$ ($u \bard \mu^-
\barnu_{\mu}$), and $11\%$ $(12\%)$ ($\mu^+\nu_{\mu} \tau^-
\barnu_{\tau}$) for bare (calo) distributions. The differences
between the final states originate mainly from the (naive) QCD corrections.

The shape of the relative corrections to the invariant-mass
distributions can be understood as follows. For small recombination
cuts (bare), in most of the events the \PWm~bosons are defined from the
decay fermions only. If a photon is emitted from the decay fermions
and not recombined, the invariant mass of the fermions is smaller than
the one of the decaying \PWm~boson. This leads to an enhancement of the
distribution for invariant masses below the \PW~resonance. This effect
becomes smaller with increasing recombination 
cut $M_\recomb$. On the other hand, if the recombination cut gets large, the
probability increases that the recombined fermion momenta receive
contributions from photons that are radiated during the \PW-pair production
subprocess or from the decay fermions of the \PWp~boson.
This leads to positive corrections above the considered \PWm~resonance.
The effect is larger for the hadronic invariant mass since in this
case, two decay fermions (the two quarks) can be combined with the
photon. The effect of the squared charges of the final-state fermions
is marginal in this case because the contribution of initial-state
fermions dominates.

In \fig{rac_dpa_56} (left) we show the 4f-Born predictions
for the $M(\wbm)$ distributions, without radiative corrections, 
i.e.\ the invariant mass distributions are constructed from
all diagrams without the restriction to the CC03 diagrams.  
The ratio 4f-Born/CC03-Born, shown in \fig{rac_dpa_56} (right) for the 
$udsc$ final state, 
confirms the goodness of the CC03 approximation for final states
involving no electrons in
describing the $\wb\wb$ cross-section at LEP~2 energies, especially in the
peak region.
The ratio best-with-4F-Born/best-with-CC03-Born
is nearly the same as the one shown on the r.h.s. of \fig{rac_dpa_56}, since
the corrections contained in the numerator and the denominator are
the same.
In \fig{rac_dpa_78} we show the ratio best-with-4f-Born/CC03-Born 
for both the bare (right) and the calo (left) $\wbm$ invariant-mass
distributions, exhibiting the
combined effect of including radiative corrections and background diagrams.

Further numerical results from {\tt RacoonWW} can be found in
Ref.~\cite{racoonww_res} and, for the same set-up as here, in
\sect{YFSWWvsRacoonWW}.

\begin{figure}[p]
{\centerline{
\setlength{\unitlength}{1cm}
\begin{picture}(14,7.3)
\put(-3.8,-13.2){\includegraphics{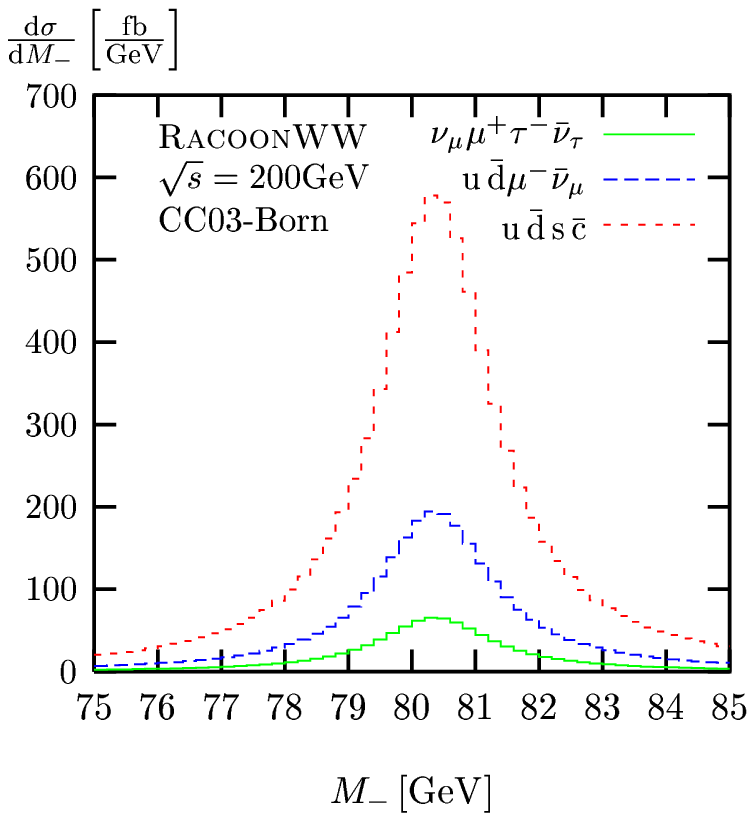}}
\put( 4.0,-13.2){\includegraphics{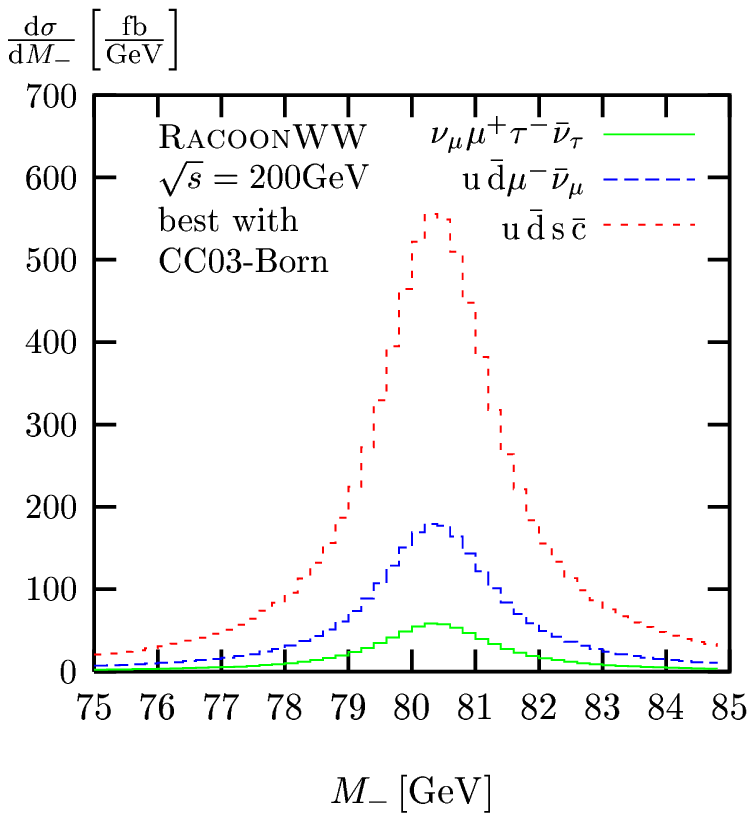}}
\end{picture} }}
\caption{The $\wbm$ invariant-mass distributions for CC03-Born (left) 
and best-with-CC03-Born (right) with bare recombination 
cuts from {\tt RacoonWW}.}
\label{rac_dpa_12}
\efi

\begin{figure}[p]
{\centerline{
\setlength{\unitlength}{1cm}
\begin{picture}(14,7.3)
\put(-3.8,-13.2){\includegraphics{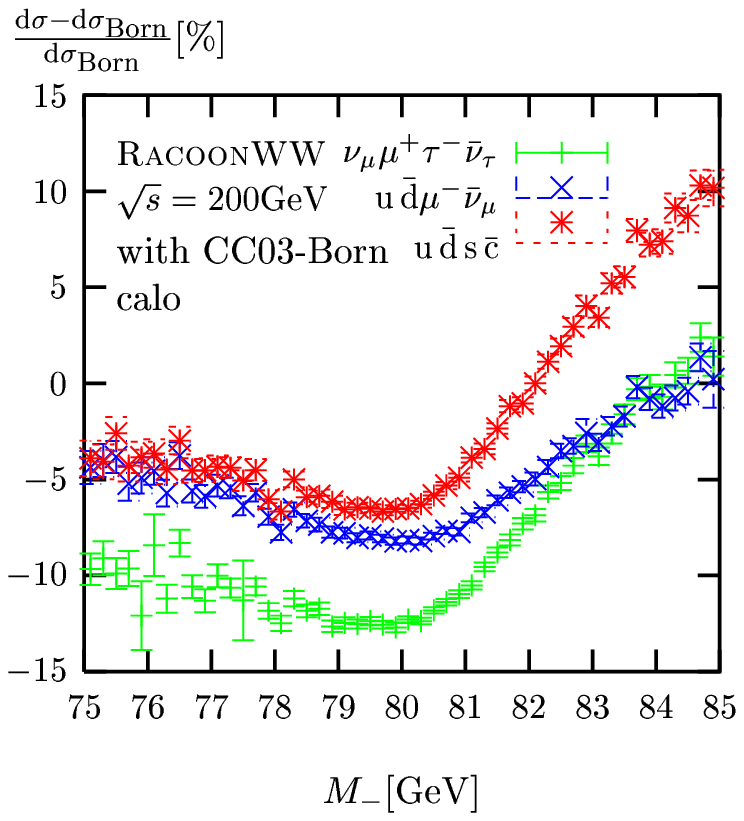}}
\put( 4.0,-13.2){\includegraphics{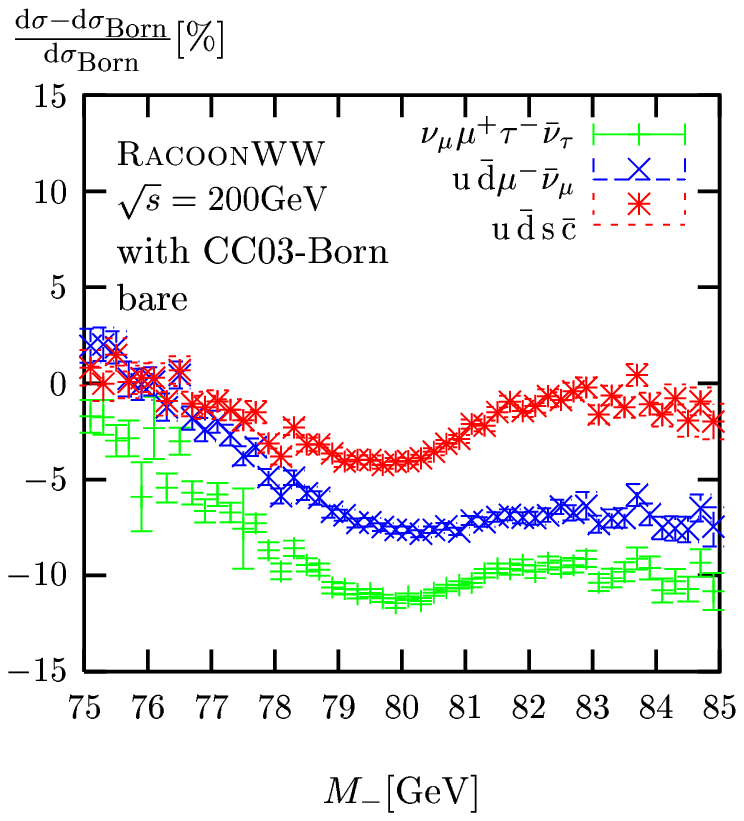}}
\end{picture} }}
\caption{The relative corrections (best-with-CC03-Born/CC03-Born - 1)
for the bare (right) and calo (left) 
$\wbm$ invariant-mass distributions from {\tt RacoonWW}.}
\label{rac_dpa_34}
\efi

\begin{figure}[p]
{\centerline{
\setlength{\unitlength}{1cm}
\begin{picture}(14,7.3)
\put(-3.8,-13.2){\includegraphics{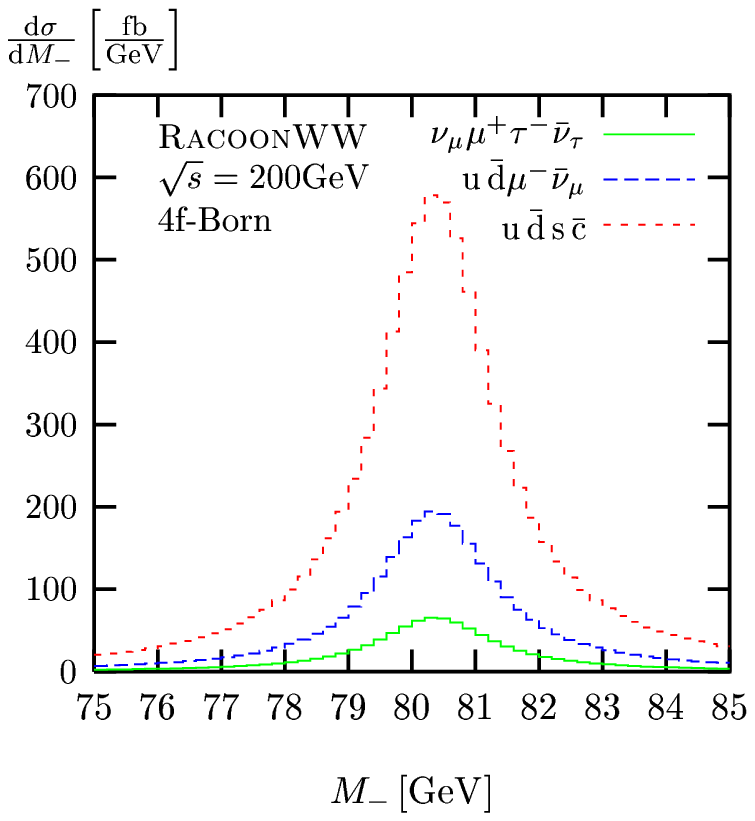}}
\put( 4.0,-13.2){\includegraphics{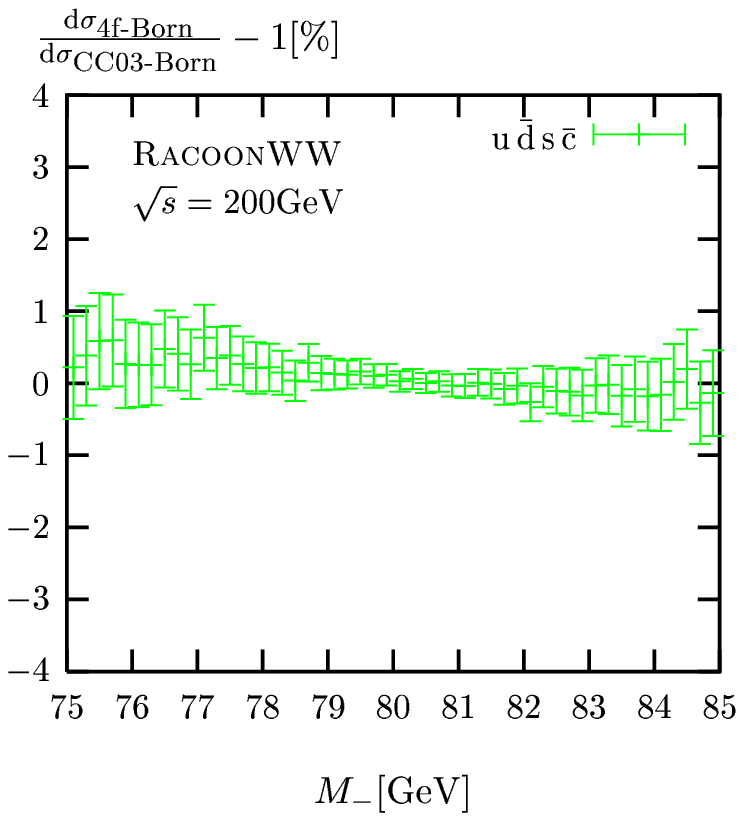}}
\end{picture} }}
\caption{$\wbm$ invariant-mass distributions for the 4f-Born (left) 
from {\tt RacoonWW}.
The ratio (4f-Born/CC03-Born - 1) (right) 
is also shown for the process $e^+ e^- \to u \bard s \barc$.}
\label{rac_dpa_56}
\efi

\begin{figure}[p]
{\centerline{
\setlength{\unitlength}{1cm}
\begin{picture}(14,7.3)
\put(-3.8,-13.2){\includegraphics{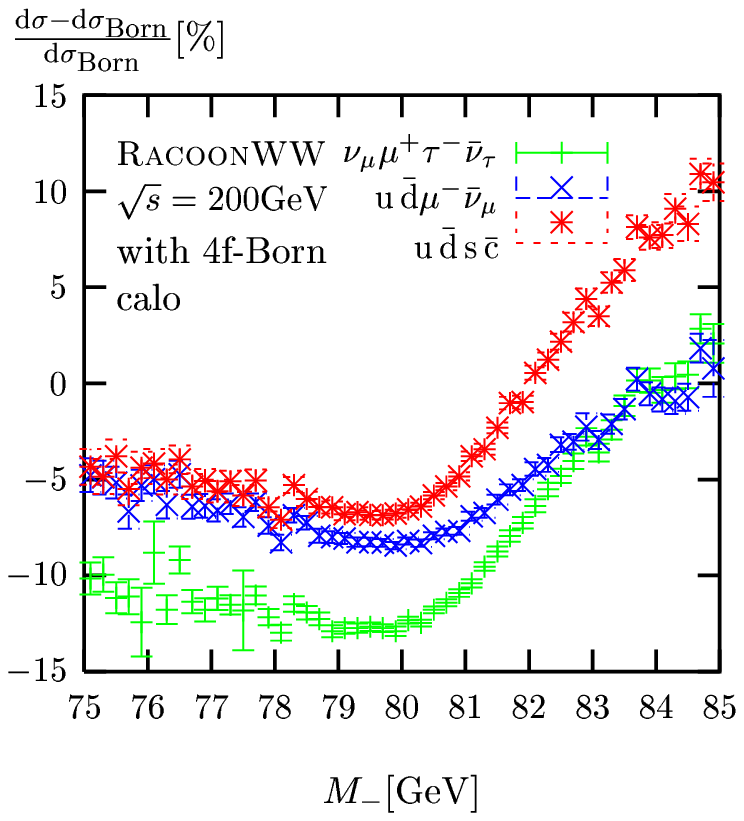}}
\put( 4.0,-13.2){\includegraphics{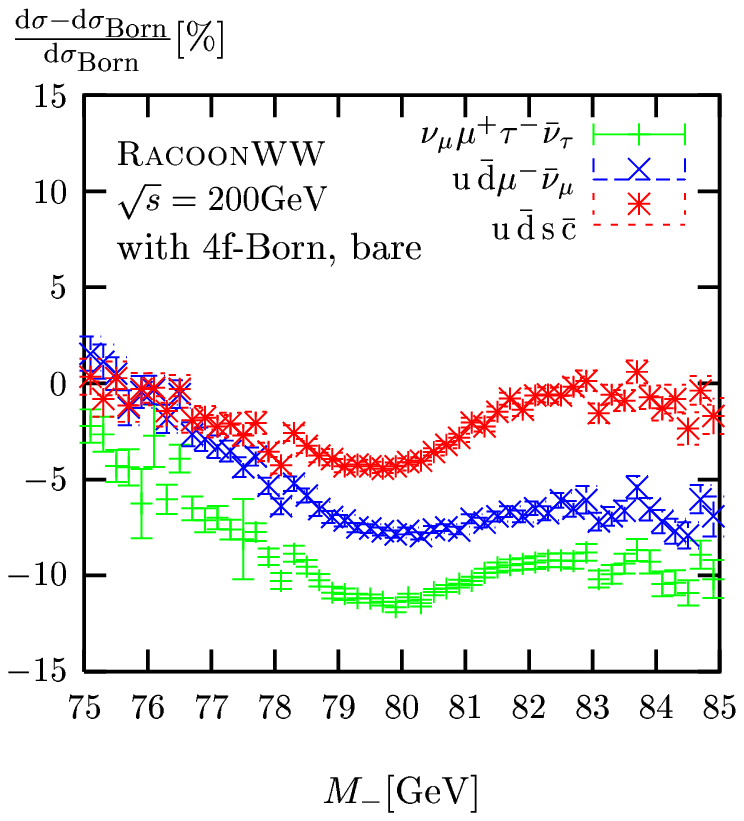}}
\end{picture} }}
\caption{The relative corrections (best-with-4f-Born / CC03-Born -1) 
for bare (right) and calo (left)
$\wbm$ invariant-mass distributions from {\tt RacoonWW} for all three 
final states.}
\label{rac_dpa_78}
\efi

\clearpage

\subsubsection*{CC03 with {\tt KORALW/YFSWW}}

\subsubsection*{Authors}
\begin{tabular}{l}
S.~Jadach, W.~Placzek, M.~Skrzypek, B.~Ward and Z.~Was  \\  
\end{tabular}

\subsubsection*{General Description}

The program {\tt KORALW1.42} has been fully documented and published
in Ref.~\cite{koralw:1999a,koralw:1999b}. Here one can find
the differences between {\tt YFSWW3} and {\tt KORALW} 
in terms of radiative corrections.

Thus, here we describe {\tt YFSWW3} first. This latter program
evaluates the the double resonant process
$ e^+e^-\rightarrow \wbp\wbm\rightarrow 4f$
in the presence of multiple photon radiation using Monte Carlo
event generator techniques. The theoretical formulation is based, 
in the leading pole approximation (LPA), on
the exact $\ord{\alpha}_{\rm prod}$ YFS exponentiation, 
with $\ord{\alpha}$
corrections (both weak and QED) to the production 
process taken from Ref.~\cite{ew1},
combined with $\ord{\alpha^3}$ LL ISR corrections in the YFS scheme
and with FSR implemented
in the $\ord{\alpha^2}$ LL approximation using {\tt PHOTOS}~\cite{photos:1994}.
Anomalous $\wb\wb\vb$ couplings are supported.
The Monte Carlo algorithm used to realize the YFS exponentiation
is based on the YFS3 algorithm presented in Ref.~\cite{yfs3:1992}
and in Ref.~\cite{koralz4:1994}. This algorithm is now described
in detail in Ref.~\cite{ceex2:1999}. 
In this way, one achieves an event-by-event
realization of our calculation in which arbitrary detector
cuts are possible and in which infrared singularities
are cancelled to all orders in $\alpha$. A detailed description of
this work can be found in 
Refs.~\cite{yfsww2:1996,yfsww3:1998,yfsww3:2000a,yfsww3:2000b}.
The program {\tt KoralW}~1.42 evaluates all four-fermion processes in $e^+e^-$
annihilation by means of the Monte Carlo techniques. It generates all
four-fermion final states with multi-branch dedicated Monte Carlo
pre-samplers and complete, massive, Born matrix elements. The
pre-samplers cover the entire phase space. Multi-photon
bremsstrahlung is
implemented in the ISR approximation within the YFS formulation
with the
$\ord{\alpha^3}$ leading-log matrix element. The anomalous
$\wb\wb\vb$ couplings are implemented in CC03 approximation. The standard
decay libraries ({\tt JETSET, PHOTOS, TAUOLA}) are interfaced. The
semi-analytical
CC03-type code {\tt KorWan} for differential and total cross-sections is
included. It operates both in weighted (integrator) and unweighted
(event generator) modes. The detailed description of this work can be
found in Refs.\ \cite{koralw:1995a, koralw:1995b, koralw:1997,
koralw:1999b, koralw:2000} and the long write-up of the program
in Ref.\ \cite{koralw:1999a} .

\subsubsection*{Features of the Program}

As the program {\tt KORALW1.42} is already published in 
Ref.~\cite{koralw:1999a,koralw:1999b},
we again start with the features of the {\tt YFSWW3} program.
The latter code is a complete Monte Carlo event generator and gives for each
event the final particle four-momenta for the entire 
$4f+n\gamma$
final state over the entire phase space for each final state particle.
The events may be weighted or unweighted, as it is more or less convenient
for the user accordingly. The code features two realizations of the LPA,
which are described in Refs.~\cite{yfsww3:1998,yfsww3:2000a,yfsww3:2000b}
wherein we also discuss their respective relative merits.

The operation of the code is entirely analogous to that of the MC's
{\tt YFS3 and YFS2} in Refs.~\cite{yfs3:1992,yfs2:1990}. A crude distribution
based on the primitive Born level distribution and the most dominant
part of the YFS form factors that can be treated analytically is used to
generate a background population of events. The weight for these
events is then computed by standard rejection techniques involving the
ratio of the complete distribution and the crude distribution. As the
user wishes, these weights may be either used directly with the events,
which have the four-momenta of all final state particles available, or they
may be accepted/rejected against a maximal weight WTMAX to produce unweighted
events via again standard MC methods. Standard final statistics
of the run are provided, such as statistical error analysis, 
total cross-sections, etc. The total phase space for the process is always
active in the code.

The program prints certain control outputs. The most important
output of the program is the series of Monte Carlo events. The
total cross-section in $pb$ is available for arbitrary cuts 
in the same standard way as it is for {\tt YFS3} and {\tt YFS2}, i.e. the user
may impose arbitrary detector cuts by the usual rejection methods.
The program is available from the authors via e-mail.
The program is currently posted on {\sl WWW} at
{\sl http://enigma.phys.utk.edu} as well as on {\sl anonymous ftp} at
{\sl enigma.phys.utk.edu} in the form of a {\sl tar.gz} file 
in the {\sl /pub/YFSWW/} directory
together with all relevant papers and documentation in postscript.

As far as the $\wb$-pair physics is concerned the {\tt KoralW} is
optimized to operate together with the {\tt YFSWW} program: {\tt
KoralW}
provides the complete background (beyond CC03) simulation by
including
{\em all} the Born level Feynman diagrams of a given process,
whereas
the signal process (CC03) is simulated by {\tt YFSWW} including
first
order corrections to $\wb$ production. The final prediction is then
obtained by adding and subtracting appropriate results. 

In order to facilitate this {\em add and subtract} procedure both
programs
have been re-organized in the following way: (1) The CC03
anomalous Born
matrix element and corresponding phase-space generator, covering
the
entire phase-space, are the same in both codes. (2) The ISR, based on
YFS principle, with $\ord{\alpha^3}$ leading-log matrix
element and
finite transverse photon momenta is also the same in both codes (in
the
case of {\tt YFSWW} it requires switching off the bremsstrahlung off
$\wb$-pair). (3) The FSR is realized in both codes in the same way
with
the help of {\tt PHOTOS} library. (4) The input data cards are in the same
format for both codes and can be stored in one data file with
common
data base of parameters along with keys specific for both programs.

The features (1) -- (3) guarantee that the common for both programs
Born+ISR+FSR CC03 part can be defined and conveniently
subtracted.
This is a non-trivial feature, as for instance there are a number of
different implementations of photonic cascades available amongst
four-fermion Monte Carlo codes. The feature (4) is a matter of
convenience as it allows for coherent and safe handling of the input
parameters.
For CC03, we note for clarity that {\tt YFSWW3} and {\tt KoralW}~1.42
differ in that {\tt YFSWW3} has the YFS exponentiated exact NL $\ord{\alpha}$
correction to the production process whereas {\tt KoralW}~1.42 does not.
             
\subsubsection*{Numerical results}

We start with predictions for the total cross-section, shown in 
\tabns{xsyfs183}{xsyfs200}, where the Born approximation and the {\em best}
results are shown.
These results in \tabns{xsyfs183}{xsyfs200} already show the
size of the NL $\ord{\alpha}$ correction, $\sim 1.5-2.0\%$, when compared
to the analogous results from programs such as {\tt GENTLE},
see for example Ref.~\cite{tan2000}.
In the sub-section below on the comparison between 
{\tt RacoonWW} and {\tt YFSWW3}, results such as those in 
\tabns{xsyfs183}{xsyfs200} are used to arrive at the current precision 
on the total $\wb\wb$ signal cross-section at LEP~2 energies.

\begin{table}[p]\centering
\begin{tabular}{|c|c|c|}
\hline
&&\\
Channel & Born & Best \\
&&\\
\hline
&&\\
$u \bard s \barc$     &   1.96325(37) &  1.59365(86) \\
$u \bard d \baru$     &   1.96369(41) &  1.59572(71) \\
$u \bard \mu^- \barnu_{\mu}$  &  0.65441(14) &  0.53901(22) \\
$u \bard e^-  \barnu_e$       & 0.65458(12) & 0.53899(23)  \\
$\mu^- \barnu_{\mu} \tau^+ \nu_{\tau}$   & 0.21809(4) & 0.18193(7) \\
all $\wb\wb$ channels   &  17.66681(351) &  15.49161(618)  \\
&&\\
\hline
\end{tabular}
\vspace*{3mm}
\caption[]{Cross-sections [fb] for $e^+e^- \to \wbp\wbm$ from {\tt YFSWW}
at $\sqrt{s} = 183\,$GeV.}
\label{xsyfs183}
\end{table}
\begin{table}[p]\centering
\begin{tabular}{|c|c|c|}
\hline
&&\\
Channel & Born & Best \\
&&\\
\hline
&&\\
$u \bard s \barc$     & 2.03231(39) &  1.68293(93) \\ 
$u \bard d \baru$     & 2.03285(40) &  1.68565(76) \\
$u \bard \mu^- \barnu_{\mu}$  & 0.67756(14) & 0.56931(24) \\
$u \bard e^-  \barnu_e$       & 0.67756(14) & 0.56931(24) \\
$\mu^- \barnu_{\mu} \tau^+ \nu_{\tau}$   & 0.22573(4) & 0.19220(8)\\
all $\wb\wb$ channels   & 18.29266(354) & 16.36329(694) \\
&&\\
\hline
\end{tabular}
\vspace*{3mm}
\caption[]{Cross-sections [fb] for $e^+e^- \to \wbp\wbm$ from {\tt YFSWW}
at $\sqrt{s} = 189\,$GeV.}
\label{xsyfs189}
\end{table}
\begin{table}[p]\centering
\begin{tabular}{|c|c|c|}
\hline
&&\\
Channel & Born & Best \\
&&\\
\hline
&&\\
$u \bard s \barc$     &  2.06691(40) & 1.75725(96) \\ 
$u \bard d \baru$     &  2.06737(41) & 1.76065(82) \\
$u \bard \mu^- \barnu_{\mu}$  & 0.68899(16) &  0.59440(26) \\
$u \bard e^-  \barnu_e$       & 0.68913(13) &  0.59444(27) \\
$\mu^- \barnu_{\mu} \tau^+ \nu_{\tau}$   &  0.22957(5) & 0.20065(9) \\
all $\wb\wb$ channels   & 18.59649(383) &  17.09010(771) \\
&&\\
\hline
\end{tabular}
\vspace*{3mm}
\caption[]{Cross-sections [fb] for $e^+e^- \to \wbp\wbm$ from {\tt YFSWW}
at $\sqrt{s} = 200\,$GeV.}
\label{xsyfs200}
\end{table}

Turning now to {\tt KORALW}, we note that it has multiple-options in the
presence multi-photonic events. It can define distributions for 
\ben
\item visible $\ph$ (radiative/hardest); 
\item all photons, i.e. no cuts, in which case one can take only 
a) the most energetic photon to determine energy and angles 
({\em all/hardest}), b) the sum ({\em all/sum}). 
\een

A sample of results is shown in \figsc{koral1}{koral3} where
we present various differential
distributions for $e^+e^- \to \baru d \barl \nu_l$ including all
background graphs and emission of multiple photons with finite
transverse momenta from initial and final states generated by {\tt
KoralW}.
The following general cuts have been used for all plots:
$M_{ud}\geq 10$ GeV, $E_l \geq 5$ GeV and $|\cos\theta_l| \leq
0.985$.

In the first plot of \fig{koral1} the photon energy distributions
are shown for: the hardest of all photons, the hardest of visible (radiative)
photons and the sum of all photons. A visible photon is defined
as having energy of at least $1\,$GeV, separated by at least $5^\circ$
from all charged fermions and having $|\cos\theta_{\gamma}|\leq 0.985$.
Apart from the natural big difference between visible and invisible
photons one can also see a substantial effect due to emission of more
than one photon ({\em hardest} vs. {\em sum}). A similar pattern for
the electron final state is shown in \fig{koral2}.
In the second plot of \fig{koral1} the angular distributions of
the hardest and hardest visible photon are shown.

\begin{figure}[p]
\vskip -2cm
\epsfig{file=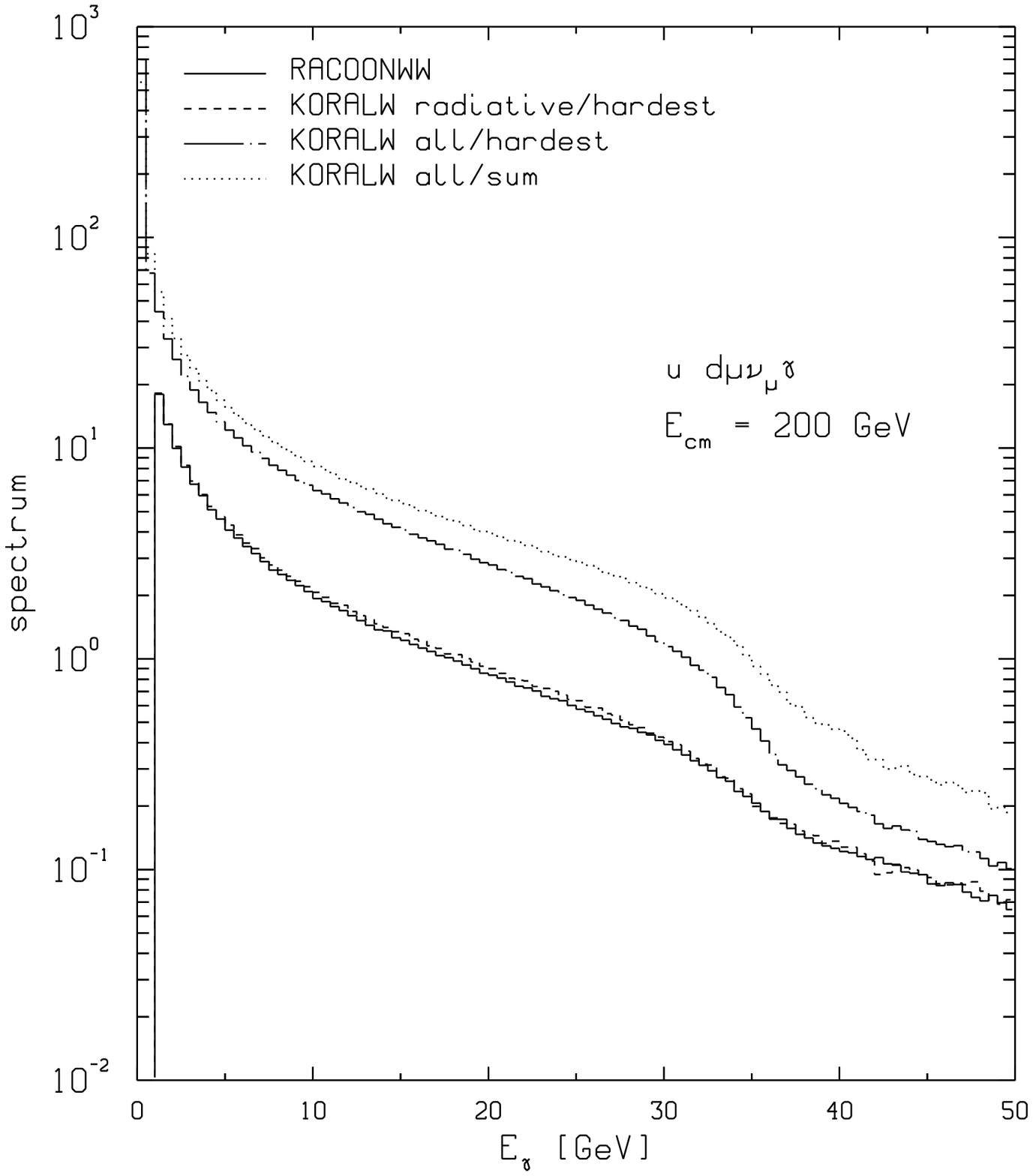,width=0.49\linewidth}
\hfill
\epsfig{file=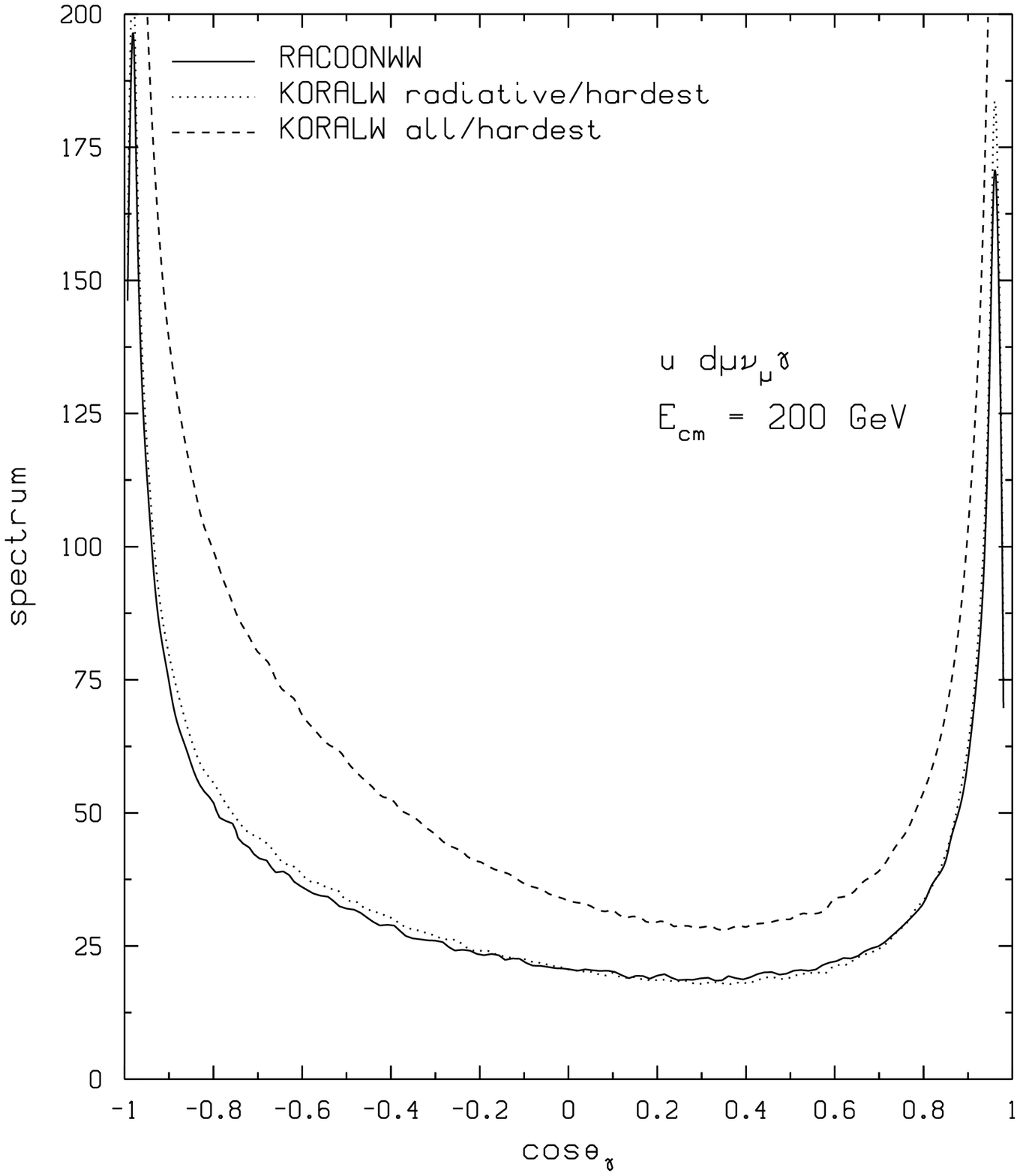,width=0.49\linewidth}
\vskip -2cm
\caption[]{{\tt KORALW} $E_{\ph}$ and $\cos\theta_{\ph}$ spectra for 
$u \bard \mu^- \barnu_{\mu}\ph$.}
\label{koral1}
\efi

\begin{figure}[p]
\vskip -2cm
\epsfig{file=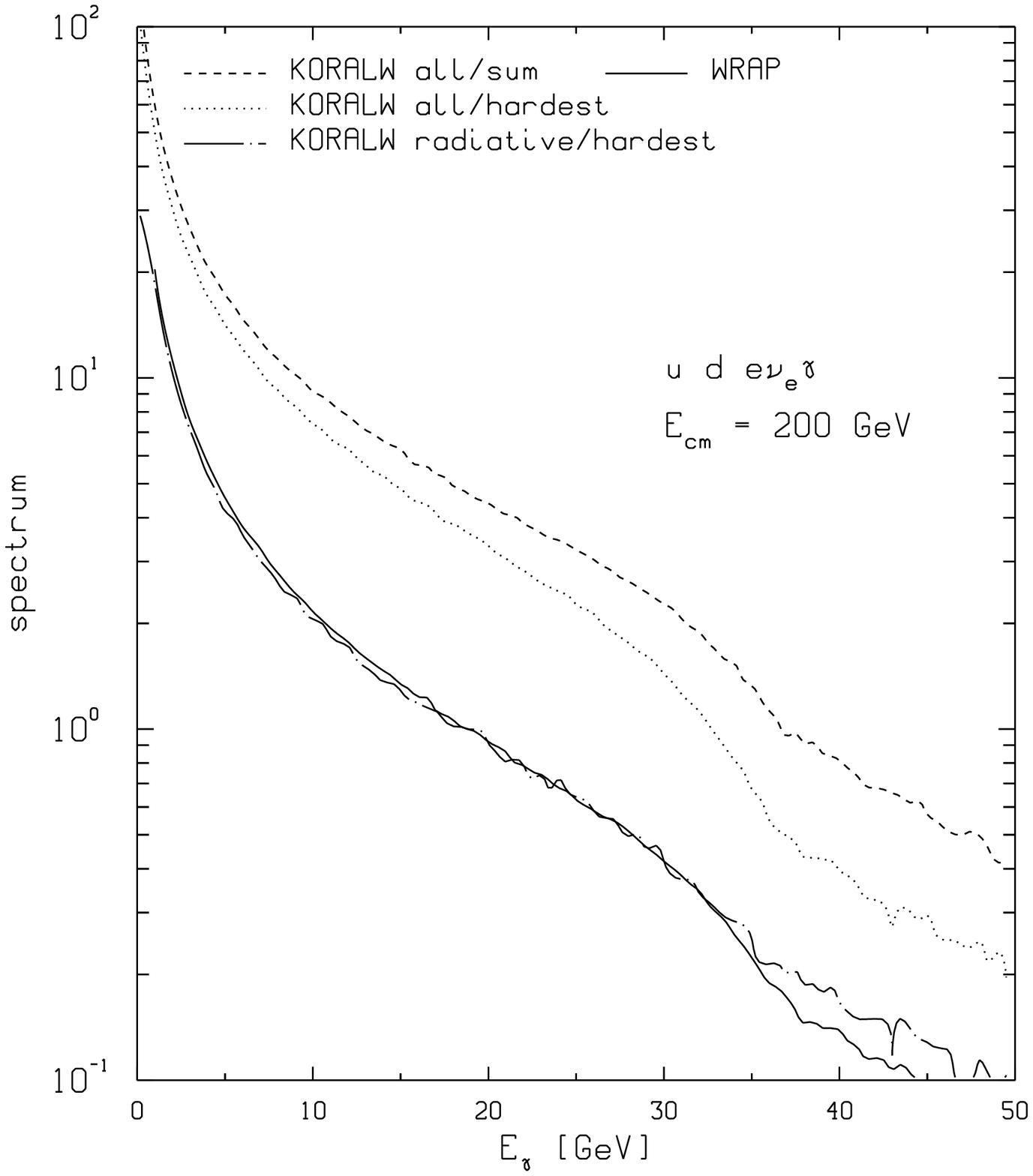,width=0.49\linewidth}
\vskip -2cm
\caption[]{{\tt KORALW} $E_{\ph}$ spectra for $u \bard e^- 
\barnu_e\ph$.}
\label{koral2}
\efi

\clearpage

In \fig{koral3} the invariant mass distributions are
shown. {\em mass}$_1$ denotes the $ud$-system invariant mass and
{\em mass}$_2$ the $\mu\nu_\mu$ mass. {\em Calo} mass includes all
photons that
have either energy smaller than $1\,$GeV or their angle to any final
state
charged particle less than $10^\circ$ for leptons or $25^\circ$ for
quarks.
In the case of leptons one can see the familiar pattern of reduction of
the cross-section below the peak (and weak change above)
due to FSR when going from the {\em Bare} to {\em Calo} mass definition
(cf.\ eg.\ Ref.\ \cite{YFS-fsr}).
In the case of hadrons the FSR is not generated.

\begin{figure}[htbp]
\vskip -1cm
\epsfig{file=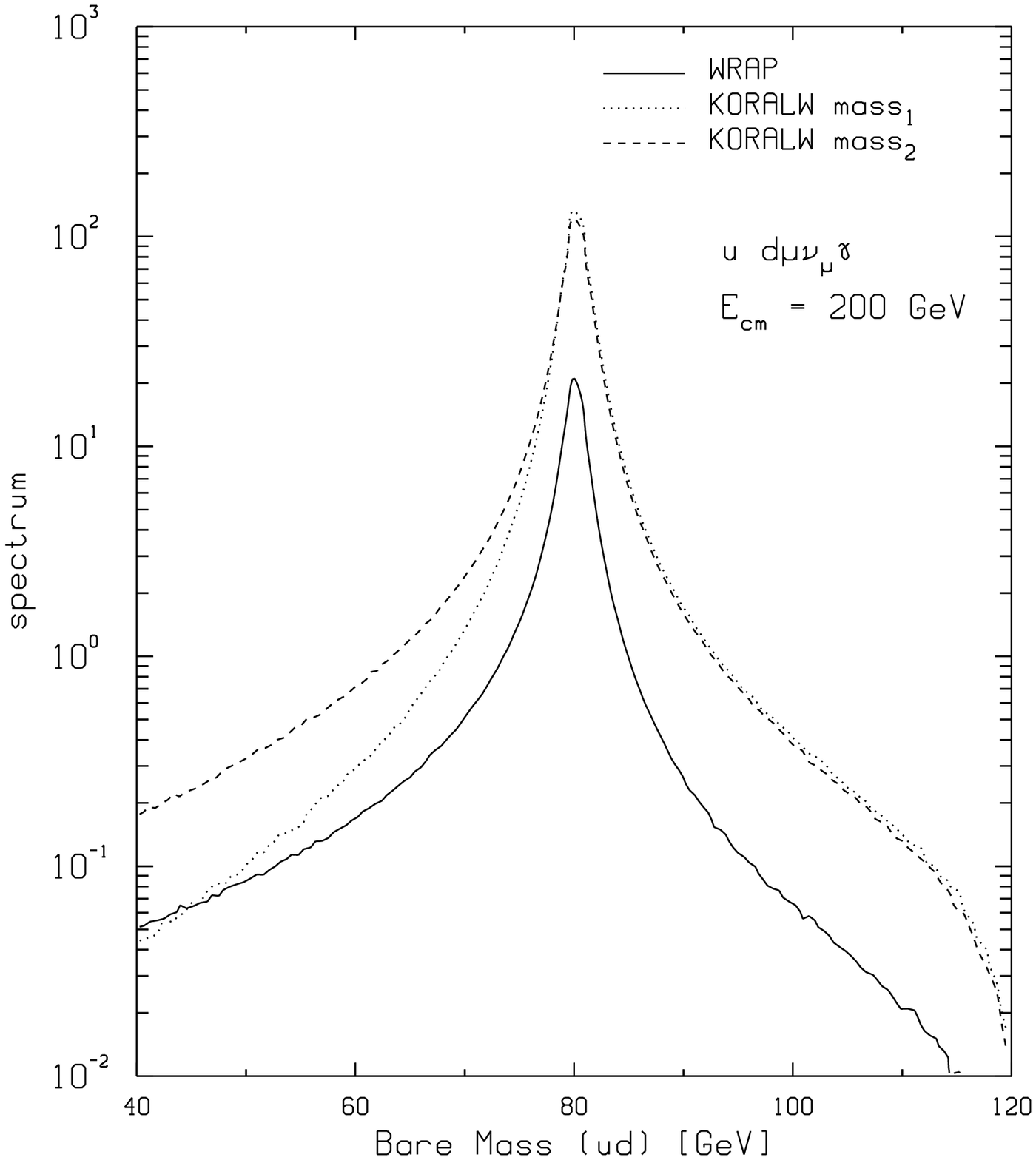,width=0.49\linewidth}
\hfill
\epsfig{file=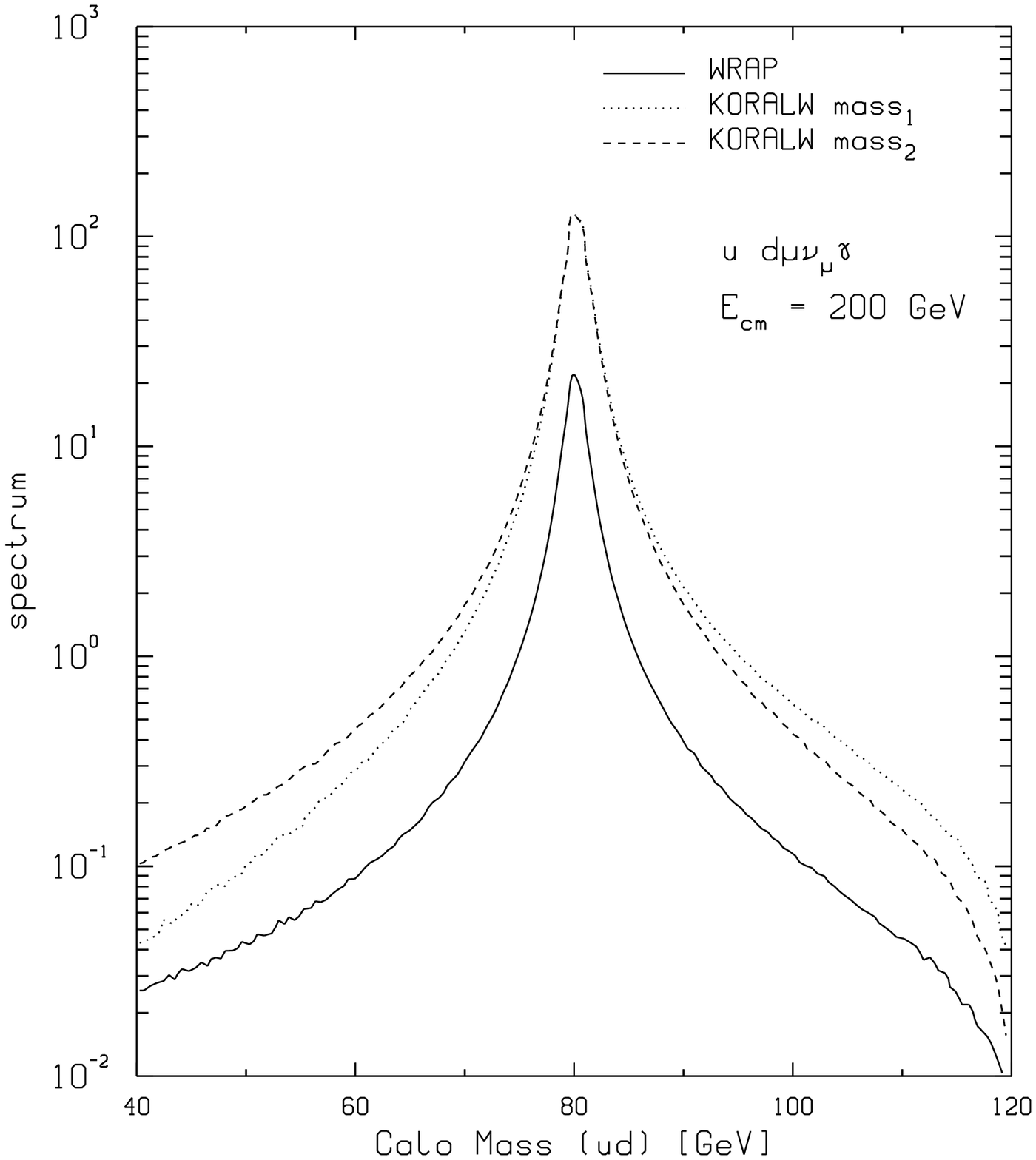,width=0.49\linewidth}
\vskip -1cm
\caption[]{{\tt KORALW} {\em bare, calo} $M$ spectra for $u \bard \mu^- 
\barnu_{\mu}\ph$.}
\label{koral3}
\efi



\subsubsection*{CC03 with {\tt GENTLE}}
\label{GENTLEcomp}

\subsubsection*{Authors}

\begin{tabular}{l}
D.~Bardin, A.~Olchevski and T.~Riemann \\
\end{tabular}

\newcommand{\gn}{{\tt GENTLE}}
\newcommand{\ee}{$e^+e^-$}

We describe shortly the {\tt GENTLE} development after v.2.00 (1996).
{\tt GENTLE~~ v.2.10} ~~(March 2000)~\cite{Bardin:1996zz,EGWWP}, 
with authors
D.~Bardin, J. Biebel, D.~Lehner, A. Leike, A. Olchevski and T. Riemann
can be obtained from:  
{\tt{http://www.ifh.de/$\sim$riemann/doc/Gentle/gentle.html}},\\
{\tt /afs/cern.ch/user/b/bardindy/public/Gentle2\_10}

Program developments since {\tt GENTLE} v.2.00 (used in the 1996 LEP~2 
workshop): 
\\
\gn\ {\tt v.2.01} (14 March 1998) compared to  v.2.00: 
\\
Angular distribution (with anomalous couplings) extended from 
CC03 class to CC11 class \cite{Biebel:1998ww,Biebel:1997id1}.
\\
\gn\ {\tt v.2.02} (11 Sept 1998) compared to v.2.01:
\\
For CC cross-sections, also a constant $\wb$ width may be chosen;
minor bugs eliminated. 
\\
{\tt ZAC v.0.9.4} (12.02.1999) \cite{Biebel:1998pi}: new package, 
includes anomalous couplings
and calculates the angular distribution for polarized $Z$ pair production 
in the NC08 class. 
\\
\gn\ {\tt v.2.10} (March 2000) differs from v.2.02 by the following features:
\\
-- for the CC cross-sections, above threshold the Coulomb correction was 
   modified.
\\
-- the NC cross-sections in package {\tt 4fan} include now besides the NC32 
class also the NC02 process;
also some new options introduced, see flag descriptions below.

As is well-known, recent comparisons for the total CC03 cross-section showed
that {\tt GENTLE} v.2.00 overestimated it by about $2\%$. 
The reason was
understood in a study made by the {\tt RacoonWW} collaboration
\cite{Dittmaier:2000pr}. It was found that the Coulomb correction
as computed in references \cite{coulomb}
overestimates the FSR QED correction above the $2\wb$ threshold. 
Such a behaviour was not excluded, of course, because old calculations
control only the leading term at threshold
$\ord{1/\beta_{\ssW}}$, where $\beta^2_{\ssW}=1-4\mws/s$. 
Only more complete calculations, using e.g. the DPA, may check how precisely 
the $1/\beta_{\ssW}$ approximation works.

An introduction of a simple suppression factor 
\begin{eqnarray}
\mbox{max}
\left(1-\frac{\beta_{\ssW}}{\beta_{\ssW}|_{\sqrt{s}=200\,GeV}},\;0\right)
\label{fudge_CC}
\end{eqnarray}
switching off the Coulomb correction smoothly between $\sqrt{s} = 2\,\mw$ 
and $200\,$GeV improves the numerical agreement with {\tt RacoonWW}
considerably.
In this sense, the introduction of such a {\em fudge} factor is
justified by a more complete calculation based on DPA.

Compared to {\tt GENTLE v.2.00}, new or extended flag regimes in 
\gn\ {\tt v.2.10} allow for:
\\
(a) {\tt IFUDGF}=0,1: switching the Coulomb suppression factor off/on  
                      (for {\tt IPROC}=1, ie, CC);
\\
(b) {\tt IIQCD}=0,1: without/with inclusive (naive) treatment of QCD 
                     corrections (for {\tt IPROC}=2, ie, NC); 
\\
(c) {\tt IIFSR}=0,1,2: choice of final state QED corrections 
                       [none, at scale $s$=$\mzs$, or at scale $s_i$] 
                        (for {\tt IPROC}=2);
\\
(d) {\tt ICHNNL}=0,1: switching between NC02 and NC32 classes  
                      (for {\tt IPROC}=2);
\\
(e) {\tt IGAMWS}=0,1: switching between constant and $s$-dependent 
                $\wb$ width (for {\tt IPROC}=1);
\\
(f) {\tt IINPT}=2: use of the $\gf$ 
             input scheme (for {\tt IPROC}=2)
             See section 2.13, eq. (8) of \cite{EGWWP}: 
$s_{\theta}^2 \equiv (1-\mws/\mzs)$, 
$g^2 \equiv 4\sqrt{2} \gf \mws$.
\\
Further, by calling subroutine {\tt WUFLAG}, one may redefine the numerical 
value of $\alpha_{em}(\mzs)$={\tt ALPHFS} (for {\tt IIFSR}=1).

Remaining electroweak corrections, genuine weak corrections in particular, are 
not included in {\tt GENTLE}.
Although, we have several choices of input parameters. 
We may recall here that {\tt GENTLE} v.2.00 had two options: 
$\alpha$-scheme and $\gf$-scheme,
as defined by Eqs.(71). In {\tt GENTLE} v.2.10 this is extended to the 
NC32 family.
A sample of the numerical results is shown in \tabn{tab_gentle_cc}.
The CC table is produced with the following {\tt GENTLE} flag settings:
\\
{\tt IPROC,IINPT,IONSHL,IBORNF,IBCKGR,ICHNNL} = 1 1 1 1 0 0
\\
{\tt IGAMZS,IGAMWS,IGAMW,IDCS,IANO,IBIN} = 0 0 1 0 0 0
\\
{\tt ICONVL,IZERO,IQEDHS,ITNONU,IZETTA} = x x x 0 1
\\
{\tt ICOLMB,IFUDGF,IIFSR,IIQCD} = 2 1 0 1
\\
{\tt IMAP,IRMAX,IRSTP,IMMIN,IMMAX} = 1 0 1 1 1 
\\

\begin{table}[t]\centering
\begin{tabular}{|c|l|c|c|c|}
\hline
&&&&\\
$\sqrt{s}~$[GeV] & {\tt RacoonWW} & {\tt GENTLE} $-$ & {\tt GENTLE}~2.10 & 
{\tt GENTLE} $+$  \\
&&&&\\
\hline
&&&&\\
172.086  & 12.0934\,(76)  &  12.0366  &  12.0457  &  12.1289  \\
176.000  & 13.6171\,(67)  &  13.5651  &  13.5723  &  13.6655  \\
182.655  & 15.3708\,(76)  &  15.2628  &  15.2731  &  15.3771  \\
188.628  & 16.2420\,(111) &  16.1723  &  16.1839  &  16.2935  \\
191.583  & 16.5187\,(85)  &  16.4749  &  16.4869  &  16.5983  \\
195.519  & 16.7910\,(88)  &  16.7674  &  16.7797  &  16.8927  \\
199.516  & 16.9670\,(89)  &  16.9590  &  16.9723  &  17.0864  \\
201.624  & 17.0254\,(89)  &  17.0309  &  17.0435  &  17.1579  \\
210.000  & 17.0876\,(91)  &  17.1419  &  17.1539  &  17.2687  \\
&&&&\\
\hline
\end{tabular}
\vspace*{3mm}
\caption[]{Cross-sections [pb] for $e^+e^- \to \wbp\wbm \to 4\rmf$;
first column: {\tt RacoonWW} \cite{racoonww_res,Dittmaier:2000pr},
second column: {\tt GENTLE 2.10}, third and fourth columns estimate variations due 
to theoretical uncertainties. Flags: {\tt ICONVL,IZERO,IQEDHS}=001,100,013.
\label{tab_gentle_cc}
}
\end{table}

As seen from the Table, there is a very good agreement between {\tt GENTLE} 
v.2.10 and {\tt RacoonWW}. It is important to emphasize, that the introduction 
of a suppression factor, Eq.(\ref{fudge_CC}),
is the only modification as compared to v.2.00 which overestimated 
the total cross-section by about $2\%$.
In this respect one could say that, following {\tt GENTLE}'s example,
all programs that do not include DPA may, nevertheless, give an {\em effective}
description of CC03 that emulates the results of DPA, \eg {\tt RacoonWW}. 
Nevertheless, only programs including DPA represent a state-of-the-art
calculation.
Indeed, the Coulomb correction is just part of the full $O(\alpha)$ correction 
and cannot be split from the rest unambiguously at energies well above
threshold. However, an improved Born approximation (IBA) comes significantly 
closer to the $\ord{\alpha}$-corrected result if the Coulomb singularity is 
switched off above threshold with some weight function $f(\beta_{\ssW})$. 
This was already done in the IBA of 
\citere{bo92}, where $f(\beta_{\ssW})$ reduced the Coulomb part from $2\%$ to
about $1\%$ at $\sqrt{s}=200\GeV$. The more radical $f(\beta_{\ssW})$ of 
(\ref{fudge_CC}) reduces the $2\%$ to zero at $\sqrt{s}=200\GeV$.

Concerning the theoretical uncertainties given in Table 
\ref{tab_gentle_cc}, one should understand that they are exclusively due to 
ISR as it is implemented within the {\tt GENTLE} approach. 
As seen, they are of the order of $0.75\%$. 
Again, a complete approach, like the DPA, is better suited 
to provide a safe estimate of theoretical uncertainties.


\subsubsection*{Comparison between {\tt RacoonWW} and {\tt BBC} results}

\subsubsection*{Authors}

\begin{tabular}{ll}
{\tt RacoonWW} &  A. Denner, S. Dittmaier, M. Roth and D. Wackeroth  \\
{\tt BBC}      &  F. Berends, W. Beenakker and A. Chapovsky \\
\end{tabular}

In this section we compare the Monte Carlo generator {\tt RacoonWW}
\cite{de00,racoonww_res} with the semi-analytical benchmark 
program 
\cite{dpa-ww} of Berends, Beenakker and Chapovsky, called {\tt BBC} in the 
following. The numerical comparison has been done for the leptonic 
channel $\Pep\Pem\to\nu_\mu\mu^+\tau^-\barnu_\tau$ and the input 
parameters of \citere{dpa-ww}.
As explained in more detail below, in this section the {\tt RacoonWW} 
results are not calculated with the preferred options, but rather in a
setup as close as possible to the {\tt BBC} approach.

\begin{sloppypar}
The two programs include the complete electroweak $\O(\al)$
corrections to $\eeWWffff(+\gamma)$, both including the
non-factorizable corrections and $\wb$-spin correlations, which at present
is only possible within the DPA formalism.  Although both programs use
the DPA, nevertheless there are differences between these two
calculations. One is technical, the usual difference between a
flexible Monte Carlo calculation, which is 
also meant for experimental use,
and a more rigid semi-analytical one, which was constructed as a
benchmark for future calculations.  The other difference is in the
implementation of the DPA. The {\tt BBC} calculation adheres strictly to DPA
definitions, so also the phase space and photon emission are taken in
DPA. In {\tt RacoonWW} the matrix elements for virtual corrections
are calculated
in the DPA, but the exact off-shell phase space is used.  For real
photon radiation the DPA is not used. Instead all Born diagrams for
$\eeffffg$ (including the background) are taken into account
and the finite width is introduced in the fixed-width scheme.
Formally this procedure is not gauge-invariant, but it has been checked
numerically with a gauge-invariant calculation (complex-mass scheme).
The matching between the virtual and real corrections, which is
necessary in order to cancel the IR and mass singularities, is done in
such a way that the leading-logarithmic corrections arising from ISR
are taken into account exactly, \ie not in DPA. 
By comparing the two calculations one can numerically check the
quality of the DPA for real-photon radiation.  
The expected differences in the relative corrections between both
approaches are formally of $\ord{\alpha/\pi \times \GW/\Delta E}$,
with $\Delta E=\sqrt{s} -2\MW$ near the $\wb$-pair production threshold.
\end{sloppypar}

The differences in the approaches have important consequences. With
{\tt RacoonWW} predictions can be obtained for general cuts and
physically relevant situations.  The fact that the masses of the
final-state fermions are neglected restricts the applicability of the
program to those observables that are free of mass singularities 
connected to the final state.
This means, in particular, that collinear photons have to be combined
with the corresponding fermions. This combination depends on the
experimental situation, which in turn depends on the type of final
state. The semi-analytical approach is of course less
flexible for implementing the experimental cuts.  In the benchmark {\tt BBC}
calculation some of the integrations were performed analytically in
order to speed up the numerical evaluations. For instance, the
invariant-mass distributions were treated differently from observables
where the invariant masses have been integrated over.
This is not a requirement in general in the DPA if one is prepared 
to do more of the integrations numerically. 
On the other hand, a treatment of mass-singular observables, i.e.\ 
ones without photon recombination, can be easily performed in the
semi-analytical approach.

For the total cross-section, the differences between the two
approaches should be of the naively expected DPA accuracy of
$\ord{\Gamma_{\ssW}/\Delta E}$ relative to the $\ord{\alpha}$ correction.  
In \reffi{fi:totcs}
we show the prediction of {\tt BBC} as points with error-bars and the
prediction of {\tt RacoonWW} as a curve together with error-bars for
some points. All error-bars are purely statistical. 

\vfill

\begin{figure}[htb]
\centerline{\epsfig{file=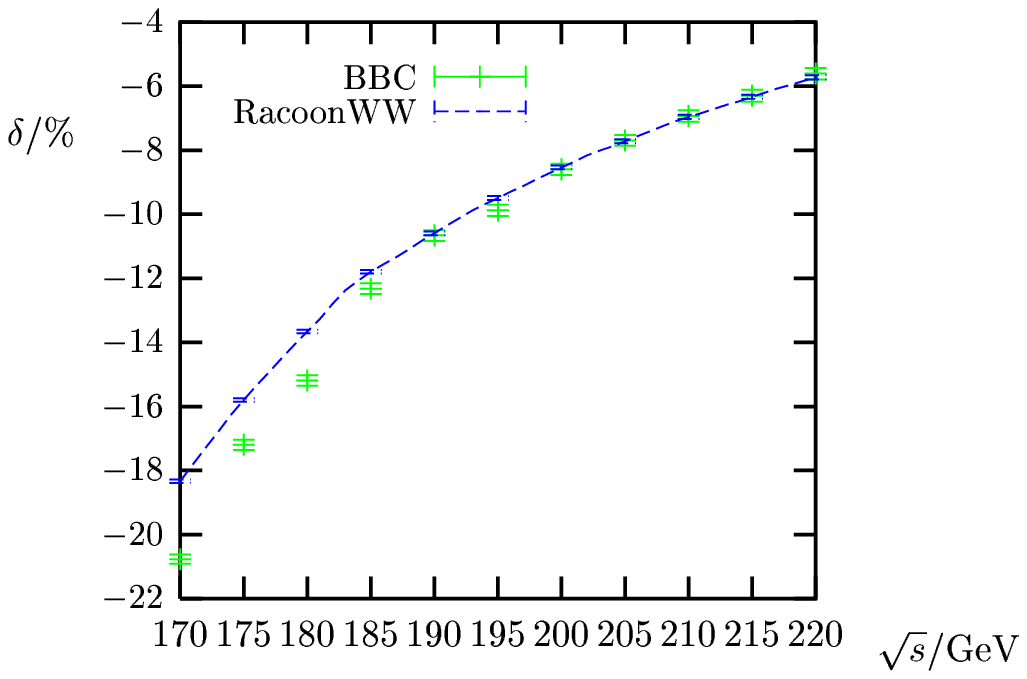,width=\linewidth}}
\caption{Relative $\ord{\alpha}$ corrections to the total cross-section
of $\protect\Pep\protect\Pem\to\nu_\mu\mu^+\tau^-\barnu_\tau$}
\label{fi:totcs}
\end{figure}

\clearpage

As shown in
\reffi{fi:totcs}, both calculations agree very well above $185\,\GeV$.
Below this energy the differences in the implementation of the DPA
become visible, in agreement with the expected relative error of
$\ord{\Gamma_{\ssW}/\Delta E}$.
The main effect originates probably from the different treatment
of the $\O(\al)$ ISR and the phase space. 
While {\tt BBC} treat the complete $\O(\al)$ correction (including ISR) in DPA
and use the on-shell phase space consistently, in {\tt RacoonWW}%
\footnote{The exponentiation of ISR has been switched off in 
{\tt RacoonWW} for this comparison, the on-shell Coulomb
singularity has been used and no naive QCD corrections are included. 
Moreover, the lowest-order cross-section used for normalization is
calculated in DPA with on-shell phase space. This allows to compare
directly the relative corrections of both approaches.}
the universal leading-log part of the $\O(\al)$ ISR correction is applied 
to the {\em full} CC11 cross-section, and the off-shell phase space is used
throughout.
Below about $170\,\GeV$ the DPA cannot be trusted any more for both
virtual corrections and real-photon radiation, since the
kinetic energy of the $\wb$~bosons becomes of the order of the $\wb$~width.
The large deviations of up to $2\%$ in the energy range between $170$ and 
$180\GeV$ can be partially attributed to the fact that {\tt BBC} treats also 
the leading logarithmic ISR corrections in DPA which is not done in 
{\tt RacoonWW}.
Therefore this difference cannot be viewed automatically as a theoretical 
uncertainty of the Monte Carlo programs.

For angular and energy distributions unavoidable differences arise
from the definition of the phase-space variables in the presence of
photon recombination.  
When defining the momenta of the $\wb$~bosons for angular distributions,
{\tt BBC} chooses to assign the photon to one of the production/decay
sub-processes. If the detected photon is hard, $E_{\gamma}\gg\Gamma_{\ssW}$,
then this is theoretically possible.  The error in the assignment is
suppressed by $\O(\Gamma_{\ssW}/\Delta E)$.  If the detected photon is
semi-soft, $E_{\gamma}\sim\Gamma_{\ssW}$, then it is
impossible to assign it to any of the sub-processes, but
as the photon momentum is much smaller than the $\wb$-boson momentum,
the error associated with this procedure is suppressed by the same
relative $\O(\Gamma_{\ssW}/\Delta E)$.  The angles are then determined from
the resulting $\wb$-momenta and the original fermion momenta.  In {\tt
RacoonWW}, all angles are defined from the fermion momenta after
eventual photon recombination.
To this end, the invariant masses of the photon with each of the
charged initial- or final-state fermions are calculated. If the
smallest of these invariant masses is smaller than $M_{\mathrm{rec}}$
and the fermion corresponding to this invariant mass is a final-state
particle, the photon is recombined with this fermion.
The two different angle definitions lead to a redistribution of events
in the angular distributions, which arises, in particular, from hard
photon emission.

The relative corrections to the distributions in the 
cosines of the 
polar production angle, $\theta_{\mathrm{W}}=\angle(\Pe^+,\PW^+)$, and the 
decay angle, $\theta_{\mu\mathrm{W}}=\angle(\mu^+,\PW^+)$, are compared 
for $\sqrt{s}=184\,\GeV$ in 
\reffi{fi:angle}.  

\begin{figure}[htb]
\centerline{
\epsfig{file=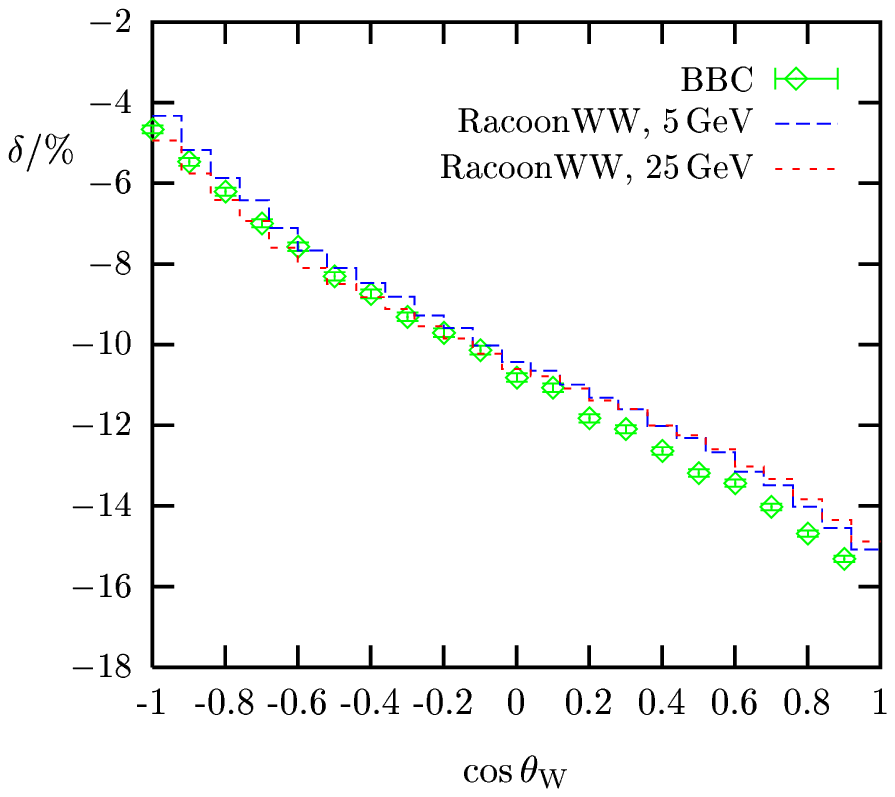,width=0.49\linewidth}
\hfill
\epsfig{file=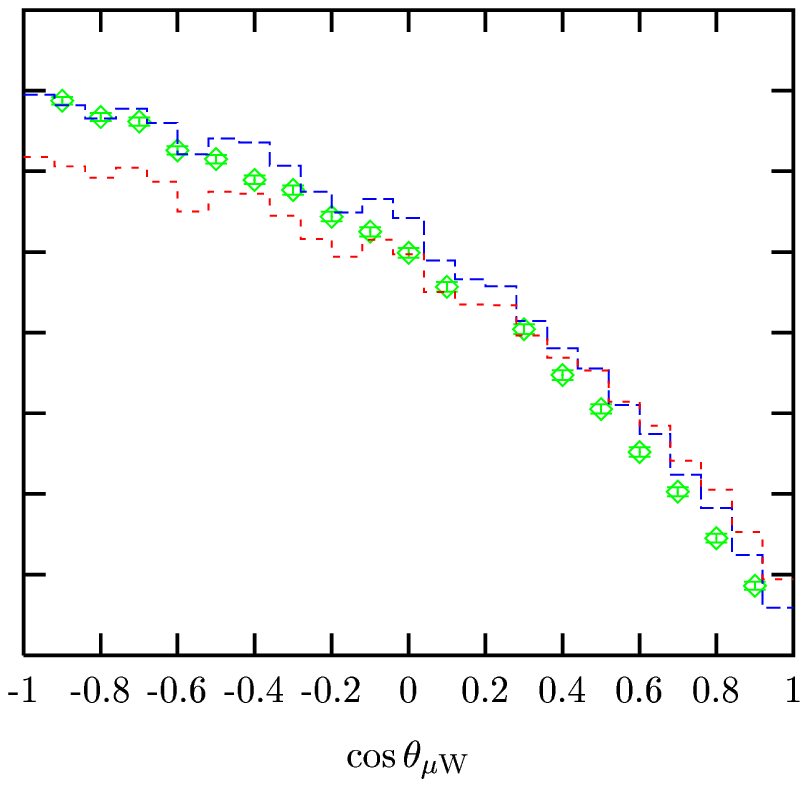,width=0.49\linewidth}}
\caption[]{Relative $\ord{\alpha}$ corrections to the $\wb$-production and 
  decay angle distributions for
  $\protect\Pep\protect\Pem\to\nu_\mu\mu^+\tau^-\barnu_\tau$ at
  $\sqrt{s}=184\,\GeV$ for the two different values of
  $M_{\mathrm{rec}}=5\,\GeV$ and $25\,\GeV$}
\label{fi:angle}
\end{figure}

The results of {\tt BBC} are again shown as points with
error-bars.  The results of {\tt RacoonWW} are plotted as histograms
for two different photon recombination cuts $M_{\mathrm{rec}}=5\,\GeV$
or $25\,\GeV$.
The relative corrections in the two recombination schemes differ at
the level of $0.5 \div 1\%$, with the largest differences   
for large angles where the cross-section is
small.  The deviations between {\tt BBC} and {\tt RacoonWW} are 
somewhat larger than this and also
larger than in the case of the total cross-section, but of the same order of 
magnitude.
A repetition of the analysis at $\sqrt{s}=250\,\GeV$ has shown that
the deviations at large angles grow with increasing centre-of-mass
energy, since also the hard-photon redistribution effects grow with
energy.

Invariant-mass distributions depend crucially on the treatment of the
real photons. Since this is fundamentally different in {\tt RacoonWW}
and {\tt BBC}, it does not make sense to compare these distributions between
the two programs. Specifically, {\tt BBC} define the $\wb$~invariant masses 
from the fermion momenta only ({\em bare} or muon-like)
which make them sensitive to the collinear mass singularities. 
In {\tt RacoonWW}, the photons are always recombined with the fermions
({\em calorimetric} or electron-like).
The actual mass shifts crucially depend on the experimental setup.
They are of the order of several $10 \MeV$ and negative for
the bare procedure. In the calorimetric treatment these
mass shifts are reduced and can even become positive depending on the
recombination procedure.

As was already mentioned earlier, the most important difference between
the two approaches is the treatment of real-photon radiation.
Therefore, it is important to compare distributions that are exclusive in 
the photon variables. As an example of such a distribution we present 
in \reffi{fi:photon_spectrum} a comparison of the photon spectrum, 
$E_\gamma \, d\sigma/d E_\gamma$, as a function of photon energy at the 
CM energy $184\GeV$. 

\begin{figure}[htb]
\centerline{
\epsfig{file=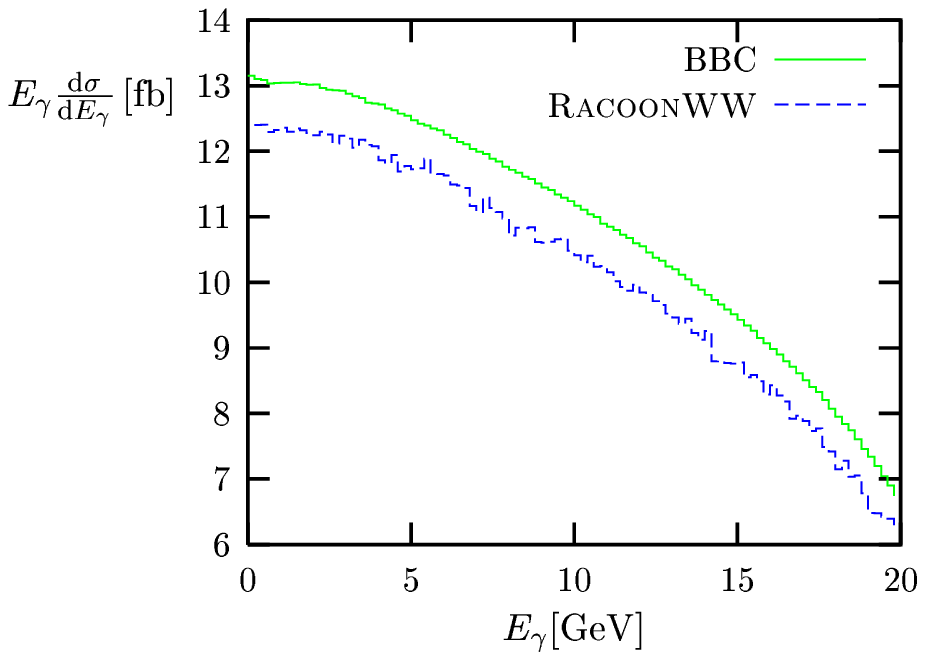,height=6.0cm,angle=0}
\epsfig{file=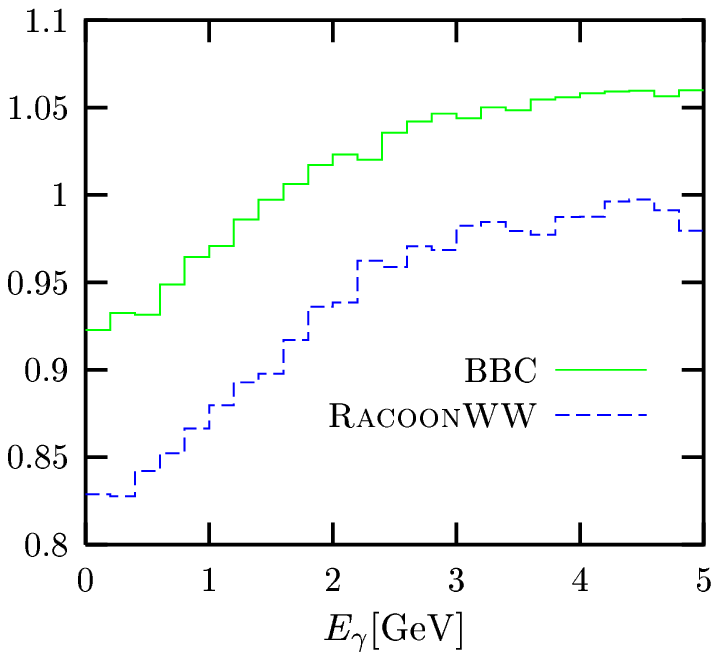,height=6.0cm,angle=0}}
\caption[]{Photon energy spectrum, $E_\gamma \, d\sigma/d E_\gamma$, at 
the CM energy $184\GeV$ for the two different sets of angular cuts, as 
described in the text.}
\label{fi:photon_spectrum}
\end{figure}

The spectrum is shown for two different sets of 
angular cuts, which restrict the angles between the photon and the beam 
momenta, $\angle(\Pe^{\pm},\gamma)$, the photon and final-state lepton 
momenta, $\angle(\ell^{\pm},\gamma)$, and the beam and final-state lepton 
momenta, $\angle(\ell^{\pm},\Pe^{\pm})$:
\begin{enumerate}
\item $\angle(\Pe^{\pm},\gamma)>1\deg$, $\angle(\ell^{\pm},\gamma)>5\deg$ and
        $\angle(\ell^{\pm},\Pe^{\pm})>10\deg$,
\item  $\angle(\Pe^{\pm},\gamma)>50\deg$, $\angle(\ell^{\pm},\gamma)>50\deg$ 
        and $\angle(\ell^{\pm},\Pe^{\pm})>10\deg$.
\end{enumerate}
The first set of cuts is closer to experiment, but the second 
suppresses the dominant contribution of ISR in the real-photonic factorizable
corrections. Since the second set of cuts removes a large part of the 
phase space, statistics in the first case is about ten times bigger than 
in the second case. However, the second set of cuts renders non-factorizable 
and factorizable radiation of a comparable order, thus checking the former.
Figure~\ref{fi:photon_spectrum} reveals an agreement between the two
approaches within $\sim 10\%$ for both sets of cuts, which is of the order 
of the naive expectation 
for the DPA error of $\ord{\Gamma_{\ssW}/\Delta E}$.
Note a peculiar decrease of the photon energy spectrum at lower photon 
energies for the second set of cuts. It was numerically checked in the BBC
approach that this decrease is due to non-factorizable contributions 
(interference between various stages of the process). More precisely, 
the non-factorizable part amounts to roughly 20\% of the complete
contribution and is negative for $E_\gamma\ll\GW$; it 
tends to zero above $E_\gamma\sim\GW$.


\hyphenation{YFSWW}
\hyphenation{RacoonWW}

\subsubsection*{Comparison of {\tt RacoonWW} and {\tt YFSWW3} results}
\label{YFSWWvsRacoonWW}

\subsubsection*{Authors}

\begin{tabular}{ll}
{\tt RacoonWW} &  A.Denner, S.Dittmaier, M.Roth and D.Wackeroth \\[0.3cm]
{\tt YFSWW3}   &  S.~Jadach, W.~Placzek, M.~Skrzypek, B.~Ward and Z.~Was  \\  
\end{tabular}

In this section we compare results obtained with the Monte Carlo 
generators {\tt RacoonWW} \cite{de00,racoonww_res} and {\tt YFSWW3} 
\cite{YFSWW3}. The numerical comparison has been done for the LEP~2 input 
parameter set. This comparison is restricted 
to the CC03 contributions for $\Pep\Pem\to\PW\PW\to 4f$, i.e.\ 
background diagrams have been omitted%
\footnote{Note that the real corrections in {\tt RacoonWW} include the
  background diagrams of the CC11 class, and the ISR is convoluted with
  this class of diagrams. For LEP~2 energies, however, the difference
  induced by these background diagrams with respect to the Born should be 
  at the per mille level.}.

First we recall that
{\tt RacoonWW} contains the complete electroweak $\O(\al)$ corrections
to $\eeWWffff(+\gamma)$ within the DPA, including the non-factorizable
corrections and $\wb$-spin correlations. Real-photon
emission is based on the full $\eeffffg$ matrix element (of the CC11
class), and ISR beyond $\O(\al)$ is treated in the structure-function
approach with soft-photon exponentiation and leading-logarithmic
contributions in $\O(\al^3)$. 
To be more precise, for $4f$ and $4f\gamma$ (with a hard non-collinear
$\gamma$) at tree level all final states are supported, i.e.\ also Mix43,
\ie $\baru d \bard u$. 
If, however, soft and collinear photons are allowed, the virtual correction 
to $e^+e^-\to \wb\wb\to 4\rmf$ is required. In this case, {\tt RacoonWW} takes 
photon radiation from the CC11 class into account\footnote{To do this, in 
any program, for Mix43 would require virtual corrections to $\zb$-pair 
production, which are not implemented.}.
The singular Coulomb correction is included with its full off-shell
behaviour.QCD corrections are taken
into account by the naive QCD factors $(1+\alpha_{\mathrm{s}}/\pi)$
for hadronically decaying $\wb$~bosons.  

In {\tt YFSWW3} the exact
$\O(\al)$ electroweak corrections to $\eeWW$ are implemented together
with YFS exponentiation of the corresponding soft-photon effects for
the production process as defined in the DPA, which is equivalent to
the LPA as defined in Ref.~\cite{yfsww3:2000} for this process.  ISR
beyond $\O(\al)$ is taken into account up to $\O(\al^3)$ in
leading-logarithmic approximation. The full off-shell behaviour of the
singular Coulomb correction is included. The corrections to the
$\wb$~decay, including naive QCD corrections, are implemented by using the
corrected branching ratios. In this way, the total cross-section
receives the full $\O(\al)$ corrections in DPA. Taking this cross
section as normalization, final-state radiation with up to two photons
is generated by {\tt PHOTOS}, which is based on a leading-logarithmic
(LL) approximation in which finite $p_{\mathrm{T}}$ effects are taken into
account in such a way that the soft limit of the respective exact
$\ord{\alpha}$ $p_{\mathrm{T}}$ spectrum is reproduced.

For observables where the decay of the $\wb$~bosons and their off-shellness
are integrated out, the expected differences between the two calculations 
are of the order of the accuracy of the DPA, \ie of the relative order 
$\ord{\al\GW/\Delta E}$, modulo possible enhancement factors.
Here $\Delta E$ is a typical energy scale for the considered observable, 
\ie $\Delta E\sim \sqrt{s} -2\MW$ for
the total cross-section near the $\wb$-pair production threshold.  For
observables that depend on the momenta of the decay products larger
differences can be expected. This holds, in particular, for observables
involving a real photon. 
While such observables are based on the full lowest-order matrix element
for $\eeffffg$ in {\tt RacoonWW},
in {\tt YFSWW3} the multi-photon radiation in the $\wb\wb$ production
(within the YFS scheme) is combined with $\O(\al^2)$ LL radiation in 
$\wb$-decays (done by {\tt PHOTOS}), 
\ie the real photon radiation is treated in DPA and
some finite $\O(\al)$ terms from FSR are neglected, but the
treatment of the leading logarithms goes beyond strict $\O(\al)$.
%

For the total cross-section, the differences between the two
approaches should be of the naively expected DPA accuracy,
\ie below 0.5\% for $\sqrt{s}>180\,\GeV$.
In \refta{ta:totcsYFSRacnocuts} 
we compare the results from both generators for the total cross-section
without any cuts.
The {\em best} numbers correspond to the inclusion of all corrections
implemented in the programs.
Independently of the channel both programs differ by $0.2\div0.3\%$,
which is of the order of the intrinsic ambiguity of any DPA
implementation, i.e.\ the numbers are consistent with each other.

\begin{table}[t]
\bce
\begin{tabular}{|c|c|c|c|}
\hline
\multicolumn{2}{|c|}{\bf no cuts}&
\multicolumn{2}{|c|}{\bf$\sigma_{\mathrm{tot}}[\mathrm{fb}]$}\\
\hline
final state & program & Born & best \nl
\hline\hline
& {\tt YFSWW3} & 219.770(23) & 199.995(62) \nl
$\nu_\mu\mu^+\tau^-\bar\nu_\tau$
& {\tt RacoonWW} & 219.836(40) & 199.551(46) \nl
\cline{2-4}
& (Y--R)/Y & $-0.03(2)$\% &  0.22(4)\% \nl
\hline\hline
& {\tt YFSWW3} & 659.64(07) & 622.71(19) \nl
$\Pu\bar\Pd\mu^-\bar\nu_\mu$
& {\tt RacoonWW} & 659.51(12) & 621.06(14) \nl
\cline{2-4}
& (Y--R)/Y & $0.02(2)$\% &  0.27(4)\% \nl
\hline\hline
& {\tt YFSWW3} & 1978.18(21) & 1937.40(61) \nl
$\Pu\bar\Pd\Ps\bar\Pc$
& {\tt RacoonWW} & 1978.53(36) & 1932.20(44) \nl
\cline{2-4}
& (Y--R)/Y & $-0.02(2)$\% &  0.27(4)\% \nl
\hline
\end{tabular}
\ece
\caption[]{Total cross-sections for CC03 from {\tt RacoonWW} and {\tt
  YFSWW3} at $\sqrt{s}=200\,\GeV$ without cuts. The numbers in
  parentheses are statistical errors corresponding to the last digits.}
\label{ta:totcsYFSRacnocuts}
\end{table}

The results of {\tt YFSWW3} presented here differ from the ones presented at the
winter conferences, where still a difference of $0.7\%$ between the
programs was reported. The main point is that the results
in Table~\ref{ta:totcsYFSRacnocuts} are obtained with version 1.14
whereas those presented at the winter conferences were obtained with
version 1.13. Version 1.14, which has benefitted from the detailed comparison
between the {\tt RacoonWW} and {\tt YFSWW3} virtual corrections, represents,
according to renormalization group improved YFS theory~\cite{bflw:1987},
an improved re-summation of the higher order corrections as compared
to version 1.13. We stress that we (the {\tt RacoonWW} and {\tt YFSWW3} 
groups) have 
also checked that, when we use the same couplings, our 
$\ord{\alpha}$ virtual plus
soft corrections in the W-pair production building block 
agree differentially at the sub-per mille level and agree
for the total cross section at $<0.01\%$. This is an important cross check
on both programs. However, as a by-product of this detailed comparison,
we have realized that the $\gf$ scheme of Refs.~\cite{gmusch}
has only the IR divergent part of the virtual photonic corrections
with coupling $\alpha(0)$ whereas the renormalization group equation
implies that any photon of 4-momentum $q$ should couple completely with
$\alpha(0)$ when $q^2\rightarrow 0$, where $\alpha(q^2)$ is the running
renormalized QED coupling.
In version 1.14 of {\tt YFSWW3}, we have
made this improvement as implied by the renormalization
group equation~\cite{bflw:1987}. The generic size of the resulting shift
in the {\tt YFSWW3} prediction can be understood by isolating the well-known
soft plus virtual LL ISR correction to the process at hand, which
has in $\ord{\alpha}$ the expression~\cite{gmusch}
\begin{equation}
\delta^{v+s}_{ISR,LL}=\beta\ln k_0 + 
\frac{\alpha}{\pi}\left(\frac{3}{2}L+\frac{\pi^2}{3}-2\right),
\label{eq:s+vISRLL}
\end{equation}
where $\beta\equiv \frac{2\alpha}{\pi}(L-1)$,~$L=\ln(s/m_e^2)$, and
$k_0$ is a dummy soft cut-off which cancels out of the cross section as
usual. In the $\gf$ scheme of Refs.~\cite{gmusch} which is used in
{\tt YFSWW3}-1.13, only the part $\beta \ln k_0+(\alpha/\pi)(\pi^2/3)$ of
$\delta^{v+s}_{ISR,LL}$ has the coupling $\alpha(0)$ and
the remaining part of $\delta^{\rm v+s}_{ISR,LL}$ has the coupling
$\alpha_{\gf}\cong\alpha(0)/(1-0.0371)$. The renormalization group
improved YFS theory implies, however, that $\alpha(0)$ should be used
for all the terms in $\delta^{v+s}_{ISR,LL}$. This is done in
{\tt YFSWW3}-1.14 and results in the normalization shift
$\left((\alpha(0)-\alpha_{\gf})/\pi\right)(1.5L-2)$,
which at 200GeV is $\sim -0.33\%$. This explains most of the
change in the normalization of {\tt YFSWW3}-1.14 vs that of {\tt YFSWW3}-1.13.
Moreover, it does not contradict the expected total precision tag of
either version of {\tt YFSWW3} at their respective stages of testing.
We stress that, according to the renormalization group equation,
version 1.14 is an improvement over version 1.13 -- it better
represents the true effect of the respective higher order corrections.
More details of the actual scheme of renormalization and re-summation
used in {\tt YFSWW3}-1.14 will appear elsewhere~\cite{yfsww3:2000b}.

In {\tt RacoonWW}, the coupling $\alpha(0)$ is used everywhere in the
relative $\ord{\alpha}$ corrections, even in the $\gf$ scheme, in order
to include the appropriate coupling for the (dominant) photonic
corrections. A switch in {\tt YFSWW3}-1.14 to this
scheme shifts the maximal differences between the programs to
$0.34\%$, somewhat larger than the $0.27\%$ shown. This confirms the
expectation that the effects of unknown higher-order corrections are
at the level of $0.1\%$.

It should be noted that the results in Table~\ref{ta:totcsYFSRacnocuts}
lie by $2\div3\%$ below the
LL-type predictions given by {\tt GENTLE}~\cite{Bardin:1996zz}
(see also Section~\ref{GENTLEcomp}).
As stated above, however, this consideration
only applies to {\tt GENTLE} in some {\em special} setup. The disagreement
with all other codes active in the `95 workshop~\cite{EGWWP} is within
$1.5\%$.
The fact that two independent Monte Carlo calculations with physical
precision at the level of 
 ${\cal O}(\frac{\alpha}{\pi}\frac{\Gamma_{\ssW}}{\mw})$
now agree to $0.2\div 0.3\%$
at $200\,$GeV for this total cross section is truly an important
improvement over the situation in the '95 workshop~\cite{EGWWP}.

In the following we consider observables obtained with the cut and
photon recombination procedure as given in the description of
numerical results of {\tt RacoonWW} in \sect{racoonww}.  We
again consider the cases of a tight recombination cut $M_\recomb=
5\,\GeV$ ({\em bare}) and of a loose recombination cut $M_\recomb=
25\,\GeV$ ({\em calo}).

Table~\ref{ta:totcsYFSRaccuts} shows the analogous cross-sections to
\refta{ta:totcsYFSRacnocuts} but now with the described {\em bare} cuts
applied. The difference of $0.2\div0.3\%$
between the two compared programs does not change by the applied cuts.
When turning from {\em bare} to {\em calo} cuts the results for the
cross-sections do not change significantly; of course, the lowest-order
results do not change at all.

In the following relative corrections to various distributions for the
semi-leptonic channel $\Pep\Pem\to\Pu\bar\Pd\mu^-\barnu_\mu$ are compared 
at $\sqrt{s}=200\,\GeV$. All these distributions have been calculated
using the above set of separation and recombination cuts.

The corrections to the cosine of the production angle for the $\PWp$
and $\PWm$ bosons are shown in \reffis{fi:costh_Wp-YFSrac} and
\ref{fi:costh_Wm-YFSrac}, respectively, for the {\em bare} (left) and the
{\em calo} (right) recombination schemes.
The distributions are compatible with each other to better than $1\%$.
The largest differences are of the order of $1\%$ and appear
in general for large scattering angles.

\clearpage

\begin{figure}[p]
{\centerline{
\setlength{\unitlength}{1cm}
\begin{picture}(14,7.8)
\put(-2.9,-13.2){\includegraphics{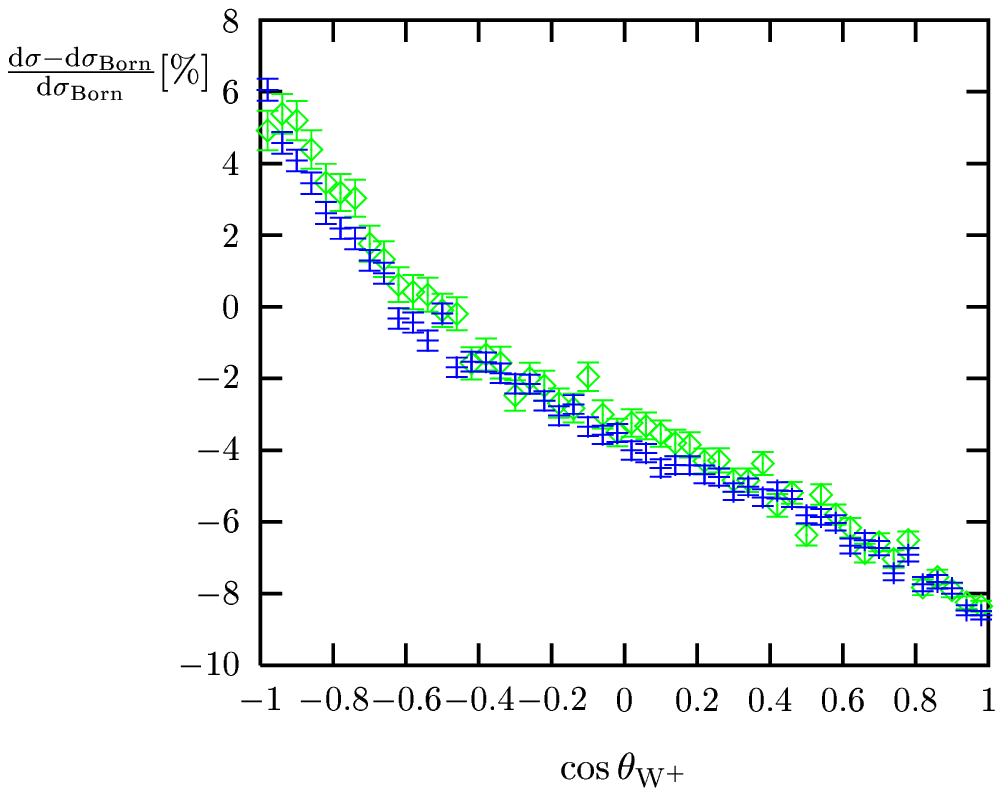}}
\put( 4.2,-13.2){\includegraphics{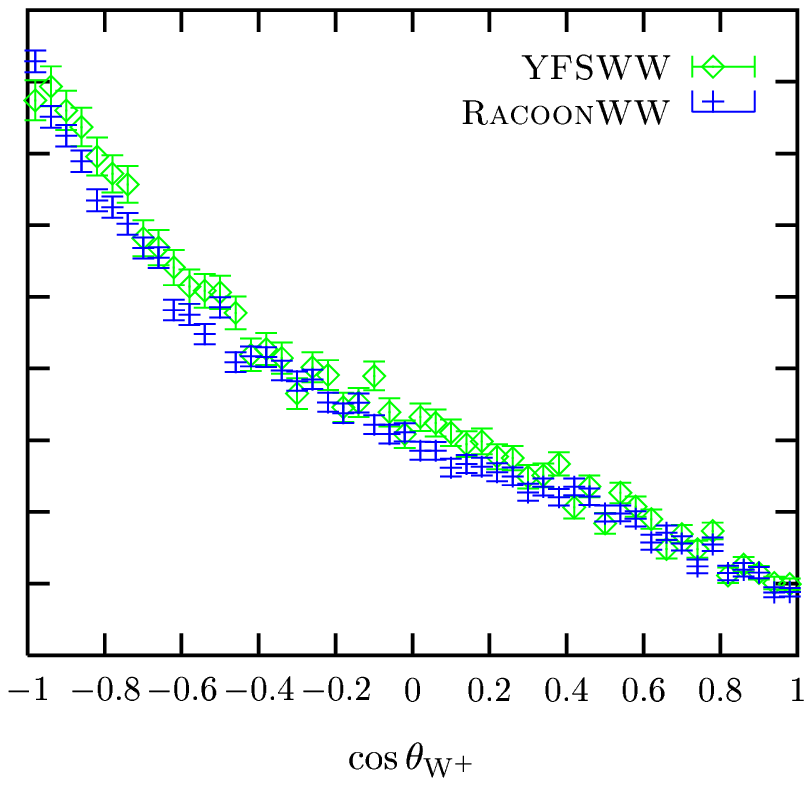}}
\end{picture} }}
\caption[]{Distribution in the cosine of the $\protect\PWp$ production angle
  with respect to the $\protect\Pep$ beam for the {\em bare} (left)
  and {\em calo} (right) setup at $\protect\sqrt{s}=200\,\protect\GeV$
  for \protect$\Pu\bar\Pd\mu^-\barnu_\mu$ final state.}
\label{fi:costh_Wp-YFSrac}
\efi

\begin{figure}[p]
{\centerline{
\setlength{\unitlength}{1cm}
\begin{picture}(14,7.8)
\put(-2.9,-13.2){\includegraphics{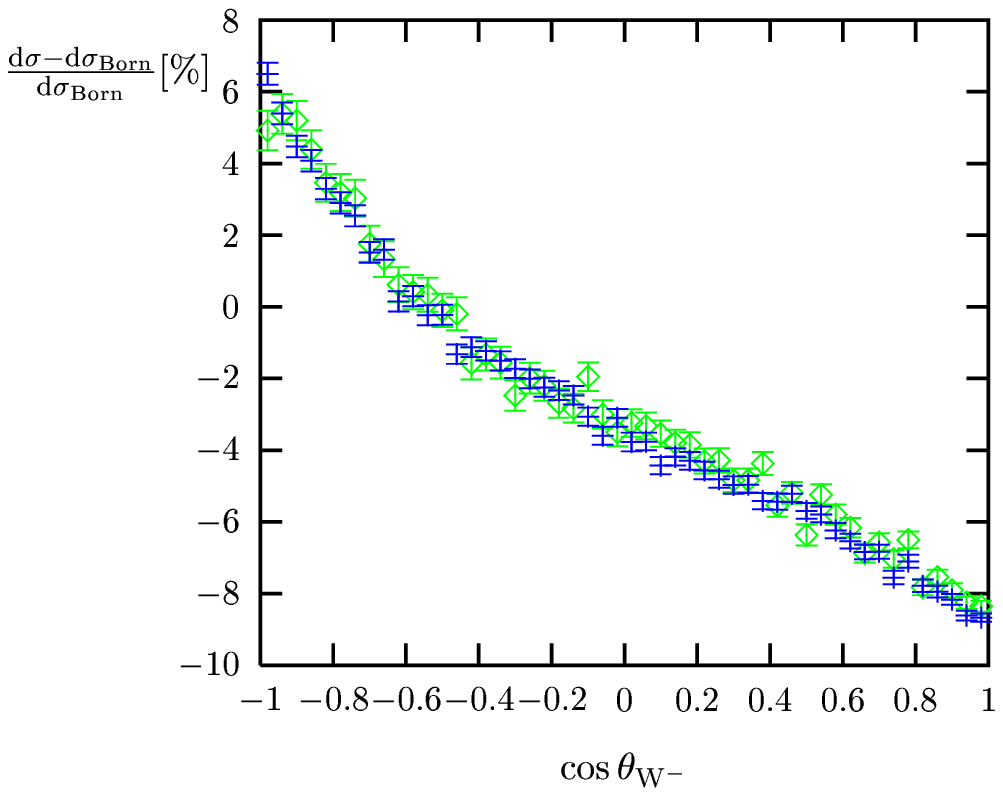}}
\put( 4.2,-13.2){\includegraphics{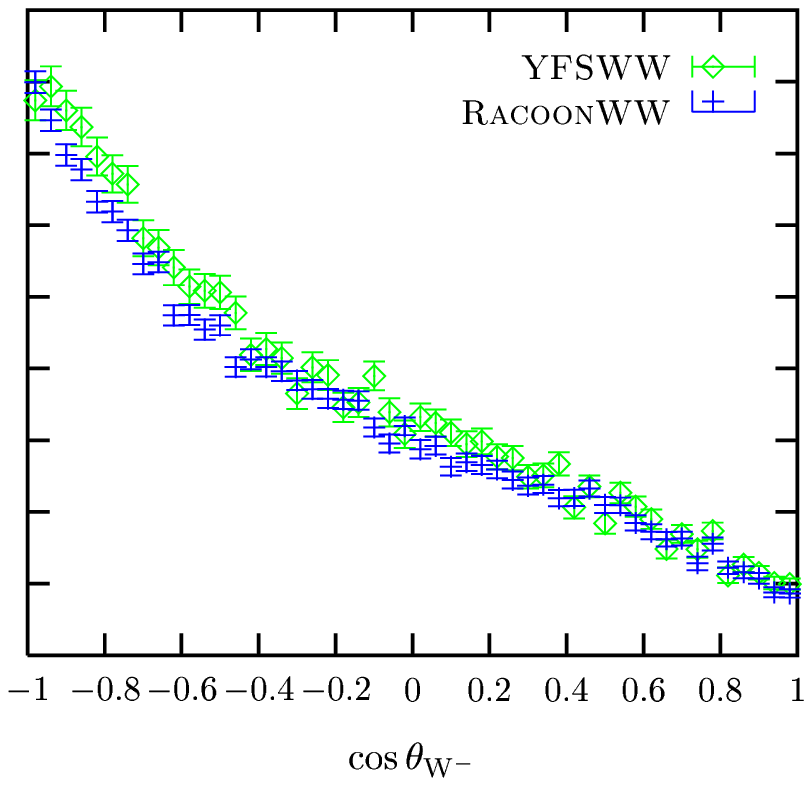}}
\end{picture} }}
\caption[]{Distribution in the cosine of the $\protect\PWm$ production angle
  with respect to the $\protect\Pem$ beam for the {\em bare} (left) and
  {\em calo} (right) setup at $\protect\sqrt{s}=200\,\protect\GeV$
for \protect$\Pu\bar\Pd\mu^-\barnu_\mu$ final state.}
\label{fi:costh_Wm-YFSrac}
\efi

\clearpage

\begin{table}[ht]
\bce
\begin{tabular}{|c|c|c|c|}
\hline
\multicolumn{2}{|c|}{\bf with {\em bare} cuts}&
\multicolumn{2}{|c|}{\bf$\sigma_{\mathrm{tot}}[\mathrm{fb}]$}\\
\hline
final state & program & Born & best \nl
\hline\hline
& {\tt YFSWW3} & 210.918(23) & 192.147(63) \nl
$\nu_\mu\mu^+\tau^-\bar\nu_\tau$
& {\tt RacoonWW} & 211.034(39) & 191.686(46) \nl
\cline{2-4}
& (Y--R)/Y & $-0.05(2)$\% &  0.24(4)\% \nl
\hline\hline
& {\tt YFSWW3} & 627.18(07) & 592.68(19) \nl
$\Pu\bar\Pd\mu^-\bar\nu_\mu$
& {\tt RacoonWW} & 627.22(12) & 590.94(14) \nl
\cline{2-4}
& (Y--R)/Y & $-0.01(2)$\% &  0.29(4)\% \nl
\hline\hline
& {\tt YFSWW3} & 1863.40(21) & 1826.80(62) \nl
$\Pu\bar\Pd\Ps\bar\Pc$
& {\tt RacoonWW} & 1864.28(35) & 1821.16(43) \nl
\cline{2-4}
& (Y--R)/Y & $-0.05(2)$\% &  0.31(4)\% \nl
\hline
\end{tabular}
\ece
\caption[]{Total cross-sections for CC03 from {\tt  YFSWW3} and {\tt
    RacoonWW} at $\sqrt{s}=200\protect\GeV$ with {\em bare} cuts (see
  text). The numbers in parentheses are statistical errors
  corresponding to the last digits.}
\label{ta:totcsYFSRaccuts}
\end{table}

The corrections to the invariant mass distributions for the $\PWp$ and
$\PWm$ bosons are shown in \reffis{fi:mwpYFSrac} and
\ref{fi:mwmYFSrac} for the {\em bare} (left) and the {\em calo}
(right) recombination scheme.  The distributions are statistically
compatible with each other everywhere and agree within 1\%.  It should
be noted that the distortion of the distributions is mainly due to
radiation off the final state and the $\wb$~bosons.  It may seem
remarkable that the LL approach of {\tt PHOTOS} properly accounts for
these distortion effects.  But one should remember that {\tt PHOTOS}
was fine-tuned to describe the exact $\ord{\alpha^1}$ FSR for the
radiative $\zb$ and $\tau$ decays, like $\zb \to \mu^-\mu^+(\gamma)$
and $\tau\to \mu\nu\bar{\nu}(\gamma)$.  {\tt PHOTOS} was also
cross-checked against the exact matrix element for the
$\wb\to\mu\nu\gamma$ process.

\figsc{fi:EgamYFSrac}{fi:cthgamYFSrac}, and \fig{fi:cthgfYFSrac} 
show the distributions in the photon energy
$E_\ga$, in the cosine of the polar angle of the photon (w.r.t.\ the
$\Pep$ axis), and in the
angle between the photon and the nearest final-state charged fermion
from the two programs and in the two recombination schemes. 

The differences are of the order of $15 \div 20\%$. 
Differences of this order may be expected, since photonic observables
are no corrections anymore, but belong to the class of $\eeffffg$ processes,
since $\eeffff$ does not contribute here.
Whether or not the observed differences are consistent with the
differences in the treatments of the real photon emission
in the two programs is under investigation.

\clearpage

\begin{figure}[ht]
{\centerline{
\setlength{\unitlength}{1cm}
\begin{picture}(14,7.8)
\put(-2.9,-13.2){\includegraphics{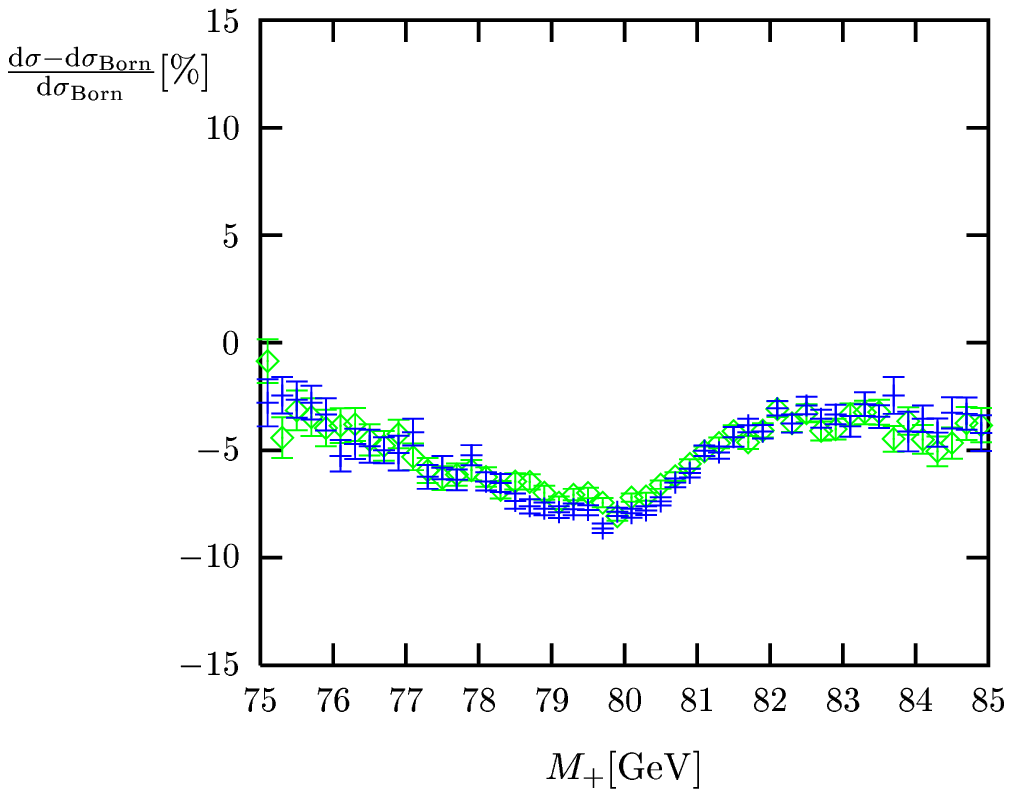}}
\put( 4.2,-13.2){\includegraphics{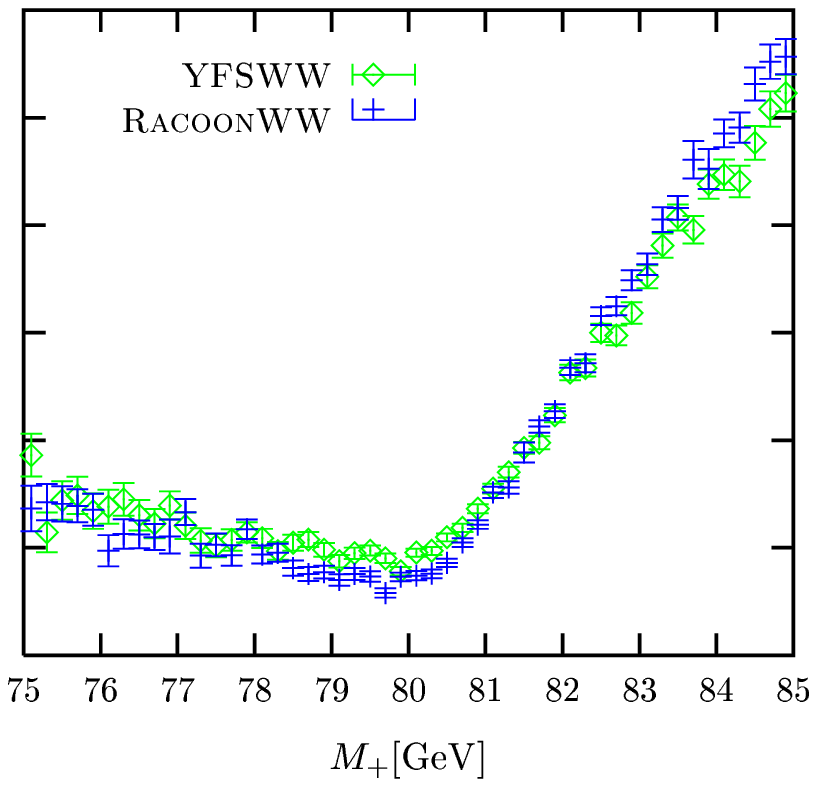}}
\end{picture} }}
\caption[]{Distribution in the $\protect\PWp$ invariant mass for {\em bare} (left) and
  {\em calo} (right) setup at $\protect\sqrt{s}=200\,\protect\GeV$
for \protect$\Pu\bar\Pd\mu^-\barnu_\mu$ final state.}
\label{fi:mwpYFSrac}
\efi

\begin{figure}[hb]
{\centerline{
\setlength{\unitlength}{1cm}
\begin{picture}(14,7.8)
\put(-2.9,-13.2){\includegraphics{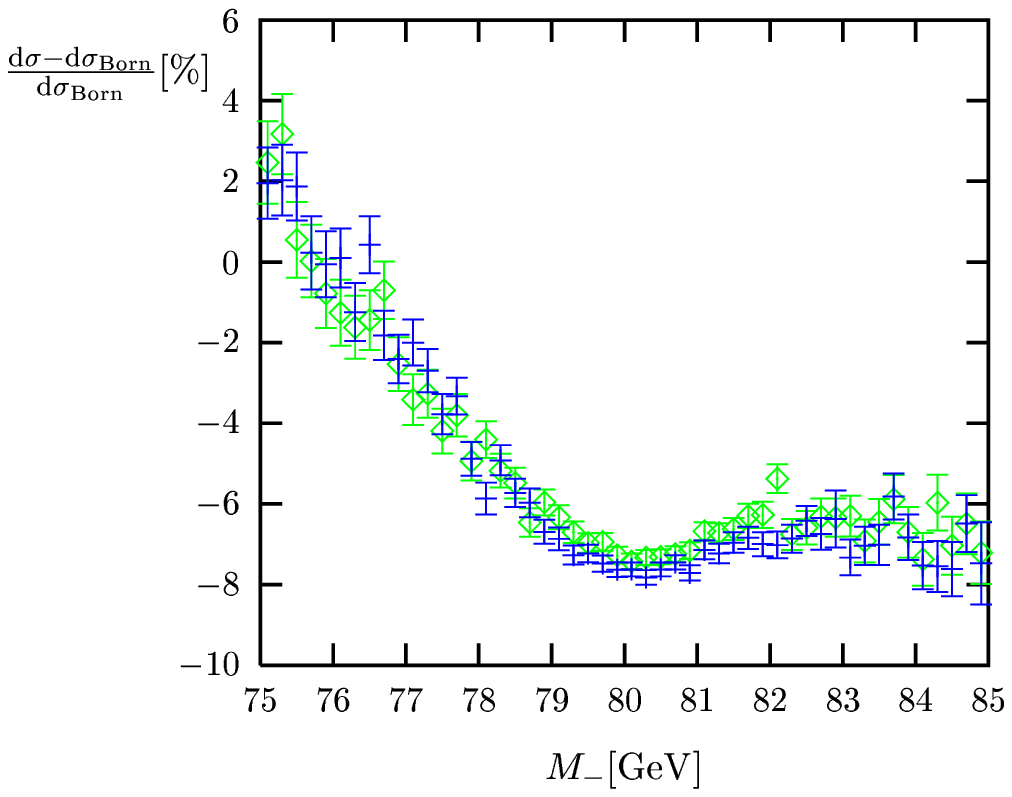}}
\put( 4.2,-13.2){\includegraphics{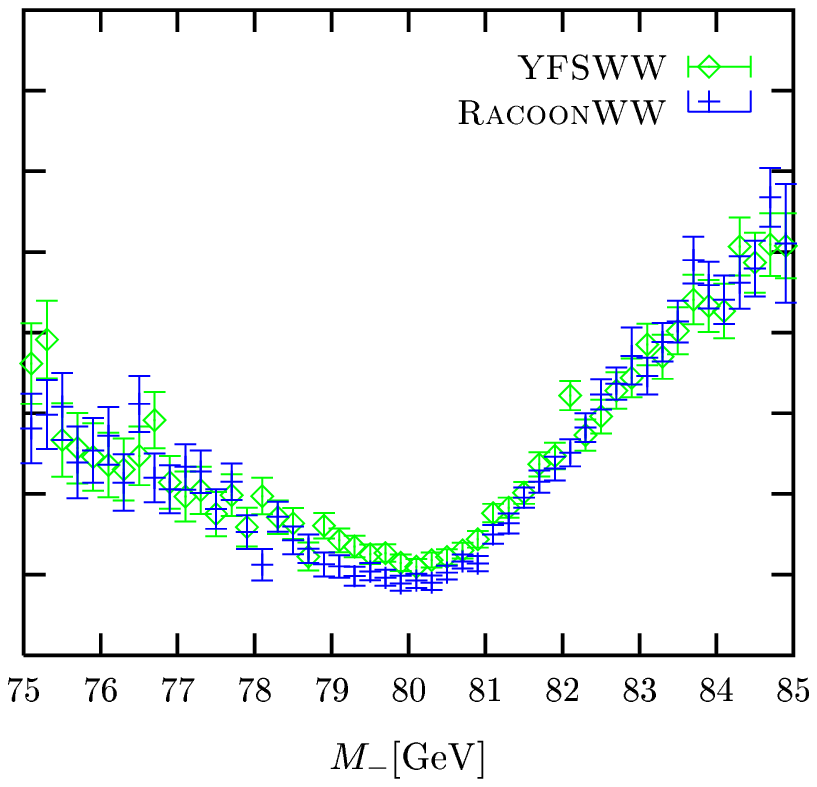}}
\end{picture} }}
\caption[]{Distribution in the $\protect\PWm$ invariant mass for {\em bare} (left) and
  {\em calo} (right) setup, $\protect\sqrt{s}=200\,\protect\GeV$,
for \protect$\Pu\bar\Pd\mu^-\barnu_\mu$ final state.}
\label{fi:mwmYFSrac}
\efi

\clearpage

\begin{figure}[ht]
{\centerline{
\setlength{\unitlength}{1cm}
\begin{picture}(14,7)
{
\put(-2.9,-12.2){\includegraphics{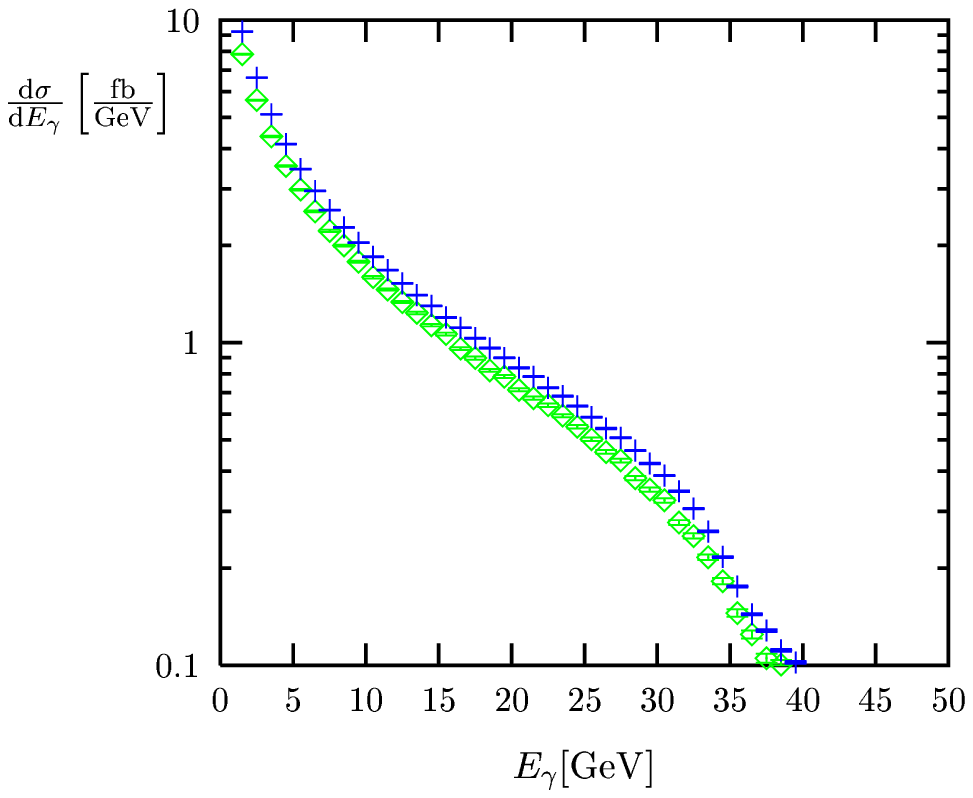}}
\put( 4.3,-12.2){\includegraphics{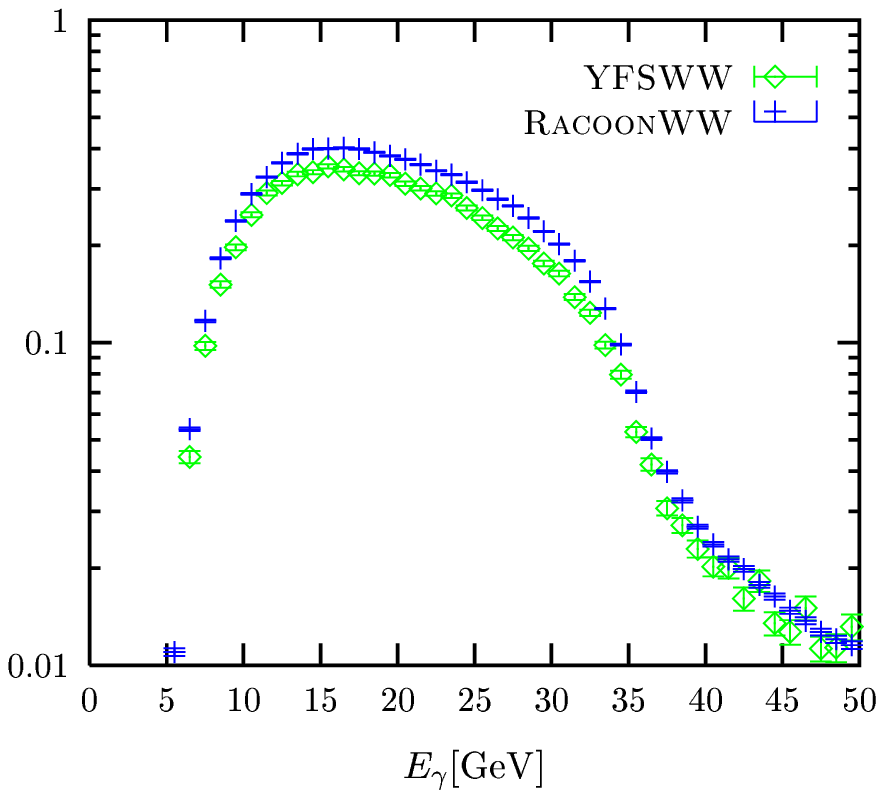}}
}
\end{picture} }}
\caption[]{Distribution in photon energy $E_\ga$ (from {\tt YFSWW3} $E_\ga$ of
  the hardest photon) for the {\em bare} (left) and {\em calo} (right)
  setup at $\protect\sqrt{s}=200\,\protect\GeV$ for
  \protect$\Pu\bar\Pd\mu^-\barnu_\mu$ final state.}
\label{fi:EgamYFSrac}
\efi
\vskip -1cm
\begin{figure}[hb]
{\centerline{
\setlength{\unitlength}{1cm}
\begin{picture}(14,7)
{
\put(-2.9,-12.2){\includegraphics{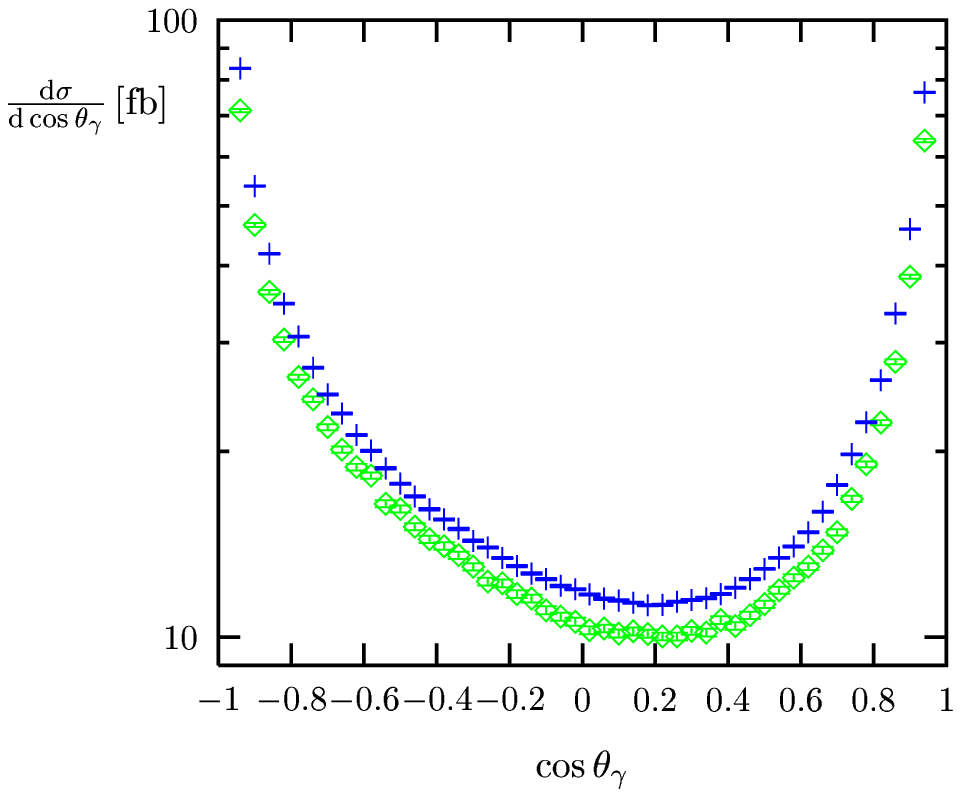}}
\put( 4.3,-12.2){\includegraphics{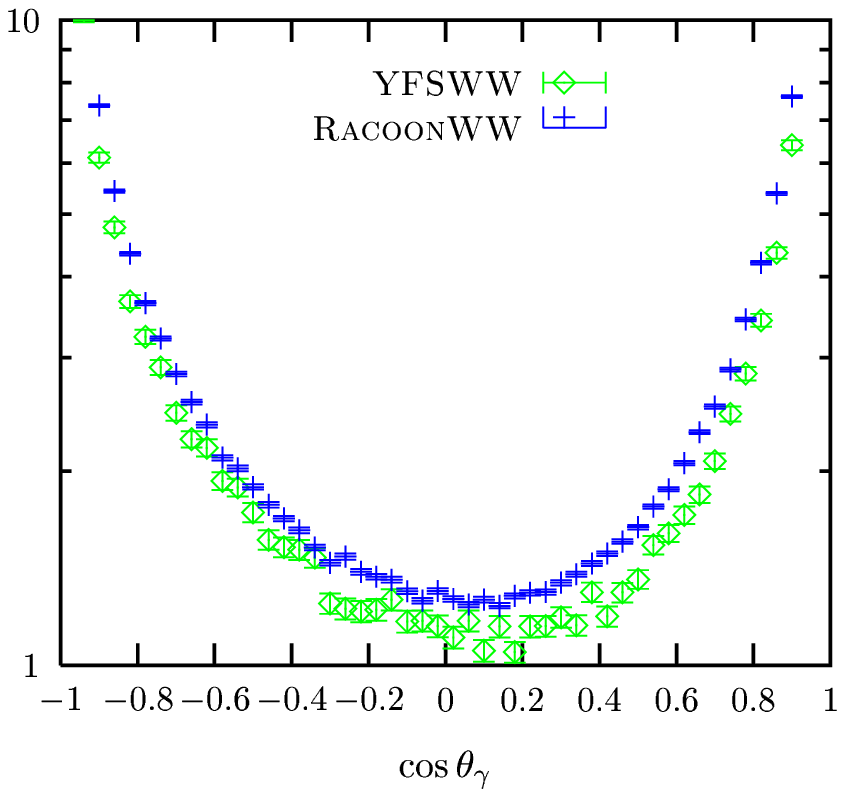}}
}
\end{picture} }}
\caption[]{Distribution in the cosine of the polar angle of the (in
  {\tt YFSWW3} hardest) photon w.r.t. the $\protect\Pep$ beam for {\em
    bare} (left) and {\em calo} (right) setup at
  $\protect\sqrt{s}=200\,\protect\GeV$ for
  \protect$\Pu\bar\Pd\mu^-\barnu_\mu$ final state.}
\label{fi:cthgamYFSrac}
\efi

\clearpage

\begin{figure}[ht]
{\centerline{
\setlength{\unitlength}{1cm}
\begin{picture}(14,7.8)
\put(-2.9,-12.2){\includegraphics{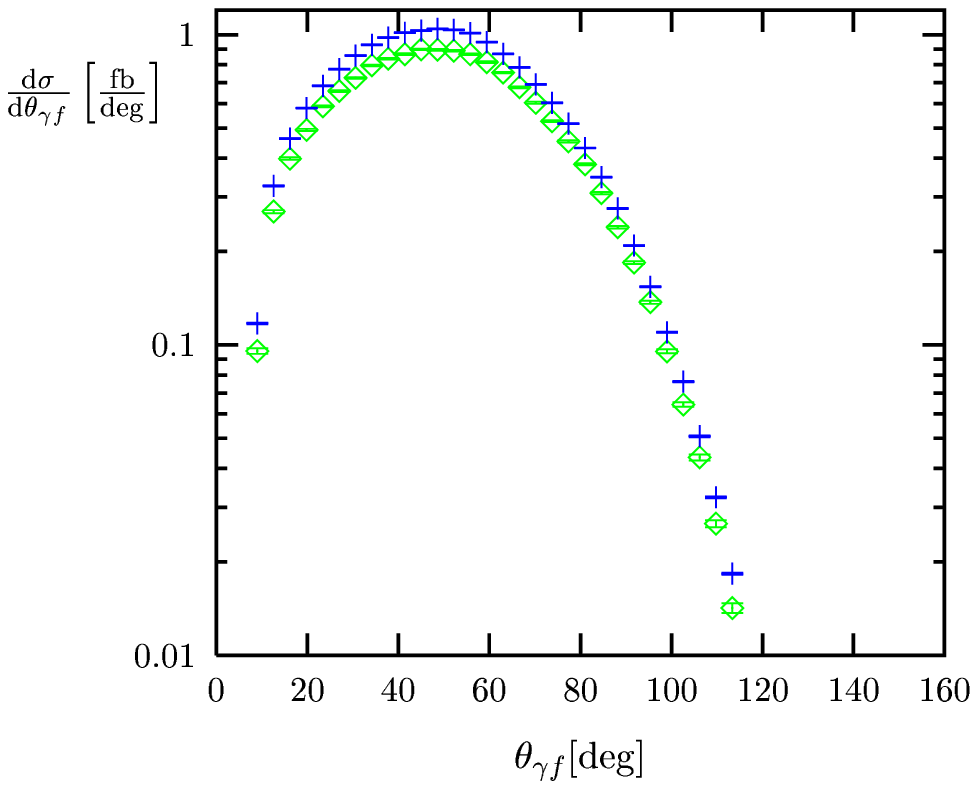}}
\put( 4.3,-12.2){\includegraphics{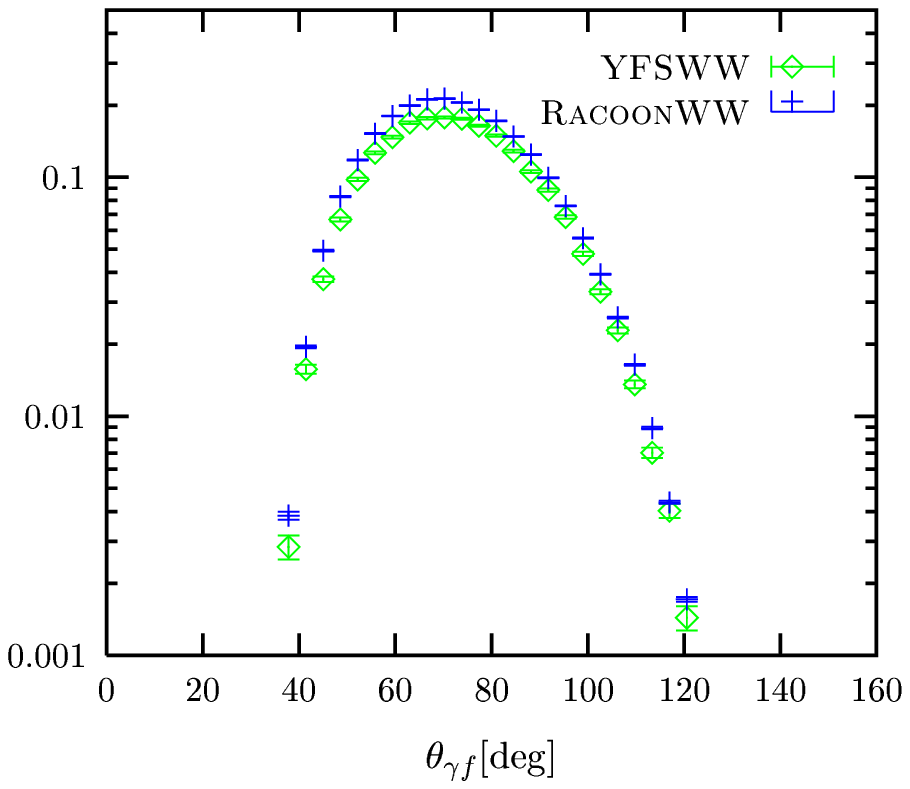}}
\end{picture} }}
\caption[]{Distribution in the angle between the (for {\tt YFSWW3} hardest)
  photon and the nearest final-state charged fermion for {\em bare} (left) and
  {\em calo} (right) setup at $\protect\sqrt{s}=200\,\protect\GeV$}
\label{fi:cthgfYFSrac}
\efi

\subsection{Internal estimate of theoretical uncertainty for CC03}

Here we give a quantitative statement on the theoretical precision for
DPA-approximation.

\subsubsection*{Estimating the theoretical uncertainty of the DPA with
{\tt RacoonWW}}
\label{se:tuwithracoonww}

\subsection*{Authors}

\begin{tabular}{l}
A.Denner, S. Dittmaier, M. Roth and D.Wackeroth
\end{tabular}

All existing calculations of electroweak corrections to $\eeWWffff$
are based on DPA. A naive estimate of
the accuracy of this approach yields
$(\al/\pi)\times\ln(\ldots)\times\GW/\MW$, where $\GW/\MW$ is the
generic accuracy of the DPA, $\al/\pi$ results from considering
one-loop corrections, and $\ln(\ldots)$ represents leading logarithms
or other possible enhancement factors in the corrections. This naive estimate
suggests that the DPA has an uncertainty of some $0.1\%$.
Note, however, that this estimate can fail whenever small scales
become relevant. In particular near the $\wb$-pair threshold, the estimate
should be replaced by 
$(\al/\pi)\times\ln(\ldots)\times\GW/(\sqrt{s}-2\MW)$.

In order to investigate the accuracy of the DPA quantitatively,
a number of tests have been performed with {\tt RacoonWW}. 
The implementation of the DPA has been modified
within the formal level of $\alpha\Gamma_\PW/\MW$,
and the obtained results have been compared.
Note that in {\tt RacoonWW} only the virtual corrections are 
treated in DPA, while real photon emission is based on the full
$\eeffffg$ matrix element with the exact five-particle phase space.
Thus, only the $2\to 4$ part is effected by the following modifications. 
Specifically, three types of uncertainties have been considered 
(see \citere{de00} for details):
\begin{itemize}
\item Different on-shell projections: \\
In order to define a DPA one has to specify a projection of the
physical momenta to a set of momenta for on-shell $\wb$-pair production
and decay%
\footnote{This option only illustrates the effect of different on-shell
projections in the four-particle phase space; if real photonic
corrections are treated in DPA the impact of different projections may
be larger.}.
This can be done in an obvious way by fixing the direction
of one of the \PW~bosons and of one of the final-state fermions
originating from either \PW~boson in the CM frame of the incoming
$\Pep\Pem$ pair. The default in {\tt RacoonWW} is to fix the
directions of the momenta of the fermions (not of the anti-fermions)
resulting from the $\PWp$ and $\PWm$ decays ({\em def}). A different 
projection is obtained by fixing the direction of the anti-fermion from 
the $\PWp$ decay ({\em proj}) instead of the fermion direction.
\item Treatment of soft photons: \\
In {\tt RacoonWW}, the virtual photon contribution
is treated in DPA, while real photon radiation is 
fully taken into account. These two contributions have to be matched in
such a way that IR and mass singularities cancel. This requirement only 
fixes the universal, singular parts, but leaves some freedom to treat 
non-universal, non-singular contributions in DPA or not. 
For instance, in the branch of {\tt RacoonWW} that employs the
subtraction method of \citere{subtract}, the endpoint contributions of
the subtraction functions are calculated in DPA and added to the virtual 
photon contribution as default.
As an option, {\tt RacoonWW} allows to treat also the universal
(IR-sensitive) part of the virtual photon 
contribution off-shell by extracting an U(1)-invariant 
factor \`a la YFS~\cite{ye61} from the virtual photon contribution and
adding it to the real photon contribution, \ie
this soft+virtual part of the photonic correction is treated off shell 
({\em eik}). The two described treatments only differ by terms of the form 
$(\alpha/\pi)\times\pi^2\times \ord{1}$
which are either multiplied with the DPA ({\em def}) or with the full 
off-shell Born cross-sections ({\em eik}). 
\item On-shell versus off-shell Coulomb singularity: \\
The Coulomb singularity is (up to higher orders) 
fully contained in the virtual
$\ord{\alpha}$ correction in DPA. Performing the on-shell projection
to the full virtual correction leads to the on-shell Coulomb
singularity.
However, since the Coulomb singularity is an important
correction in the LEP~2 energy range and is also known beyond DPA,
{\tt RacoonWW} includes this extra off-shell Coulomb correction as default.
Switching the extra off-shell 
parts of the Coulomb correction off ({\em Coul}), yields an
effect of the order of the accuracy of the DPA.
\end{itemize}

In the following table and figures the total cross-section
and various distributions have been compared for the different versions 
of the DPA defined above.
The results have been obtained using the LEP~2 input parameter set
and the set of separation and recombination cuts
as given in the description of
numerical results of {\tt RacoonWW} in \sect{racoonww}.  
The recombination cut is chosen to be $M_\recomb= 25\,\GeV$.
As default, we take the {\tt RacoonWW} results (best-with-4f-Born) 
of \sect{YFSWWvsRacoonWW}
for the process $\Pep\Pem\to\Pu\bar\Pd\mu^-\barnu_\mu(\gamma)$ at
$\sqrt{s}=200\,\GeV$, which are based on the above input. The only
differences are that the naive QCD factors and
ISR corrections beyond $\ord{\alpha}$ are not included in the
results of this section.
The results for the total cross-section are shown in
\refta{tab:sigma_DPA_unc}.

\begin{table}[htbp]
\centerline{
\begin{tabular}{|c||c|c|c|c|}
\hline
& def & proj & eik & Coul \\
\hline \hline
$\sigma/\mathrm{pb}$ & 570.53(46) & 570.37(46) & 570.47(46) & 571.28(46) \\
\hline
$\delta/\%$ & 0 & $-0.03$ & $-0.01$ & 0.13 \\
\hline
\end{tabular}
}
\caption{{\tt RacoonWW} predictions for the total cross-section of
$\protect\Pep\protect\Pem\to\protect\Pu\bar\protect\Pd\mu^-\barnu_\mu(\gamma)$
at $\protect\sqrt{s}=200\,\protect\GeV$ 
in various versions of the DPA and relative differences
$\delta=\sigma/\sigma_{\mathrm{def}}-1$}
\label{tab:sigma_DPA_unc}
\end{table}

Note that these cross-sections are calculated with the above cuts.
We find relative differences at the level of $0.1\%$. 
As expected, the prediction that is based on the on-shell Coulomb
correction is somewhat higher than the exact off-shell treatment, since
off-shell effects screen the positive Coulomb singularity.
The results in \refta{tab:sigma_DPA_unc} have been obtained using 
phase-space slicing for the treatment of the IR and collinear
singularities. If the subtraction method is used instead, the
resulting cross-section is about 0.01\% smaller.

In \reffis{fi:tu_thwp_thwpmu} and \ref{fi:tu_emu_mm} we show the
differences of the {\em proj}, {\em eik}, and {\em Coul} modifications to
the default version of the DPA for some distributions. For the
distribution in the cosine of 
the W-production angle $\theta_{\PW^+}$ and in the
W-decay angle $\theta_{\PW^-\mu^-}$ (see \reffi{fi:tu_thwp_thwpmu})
the relative differences are of the order of $0.1 \div 0.2\%$ 
for all angles, which is of the expected order for the intrinsic DPA
uncertainty. 

\begin{figure}[p]
{\centerline{
\setlength{\unitlength}{1cm}
\begin{picture}(14,7.8)
\put(-2.9,-13.2){\includegraphics{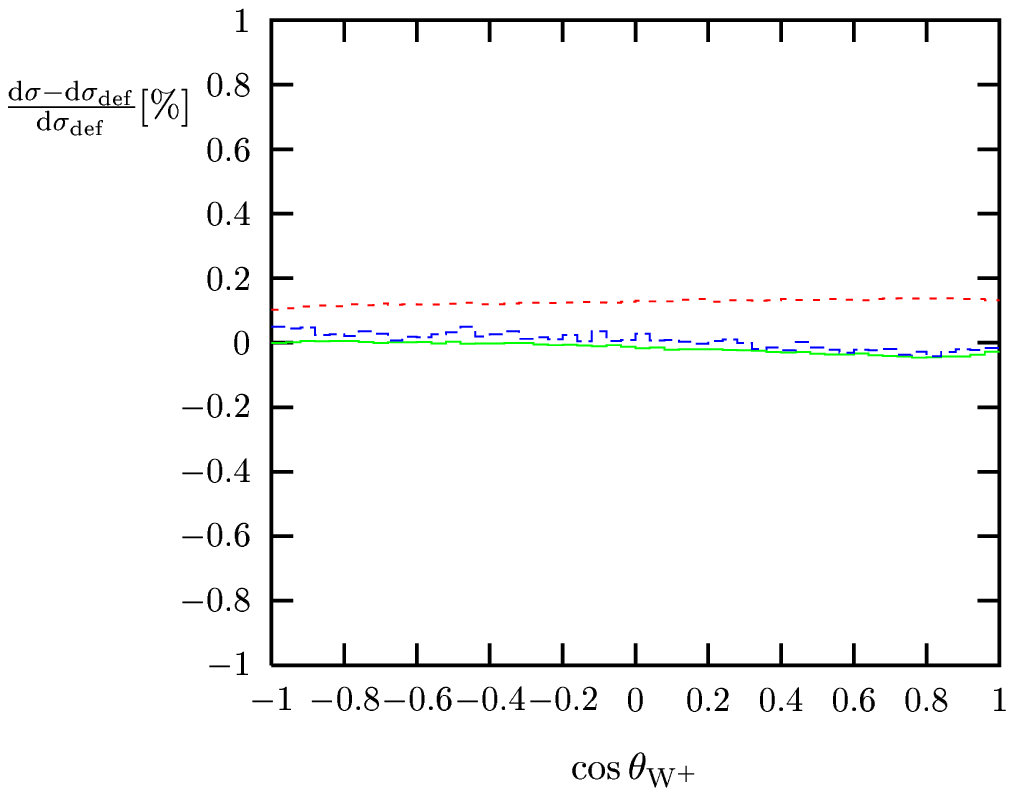}}
\put( 4,-13.2){\includegraphics{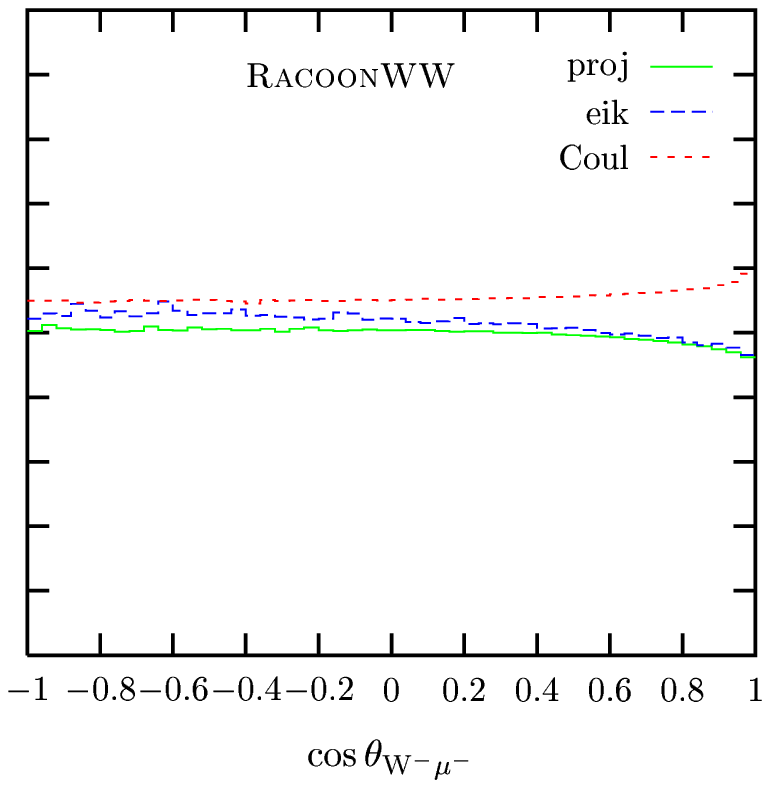}}
\end{picture} }}
\caption{Theoretical uncertainty within the DPA for distributions in the
W-production and W-decay angles
for $\protect\Pep\protect\Pem\to\protect\Pu\bar\protect\Pd\mu^-\barnu_\mu(\gamma)$ at $\protect\sqrt{s}=200\,\protect\GeV$} 
\label{fi:tu_thwp_thwpmu}
\efi

\begin{figure}[p]
{\centerline{
\setlength{\unitlength}{1cm}
\begin{picture}(14,7.8)
\put(-2.9,-13.2){\includegraphics{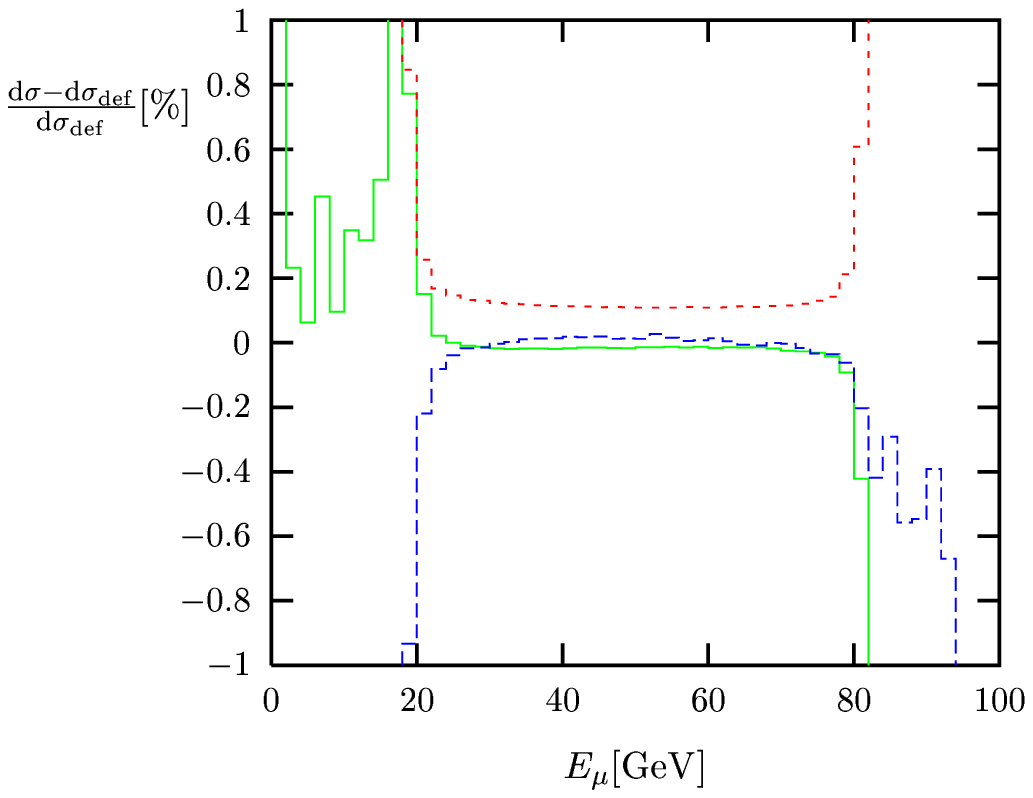}}
\put( 4,-13.2){\includegraphics{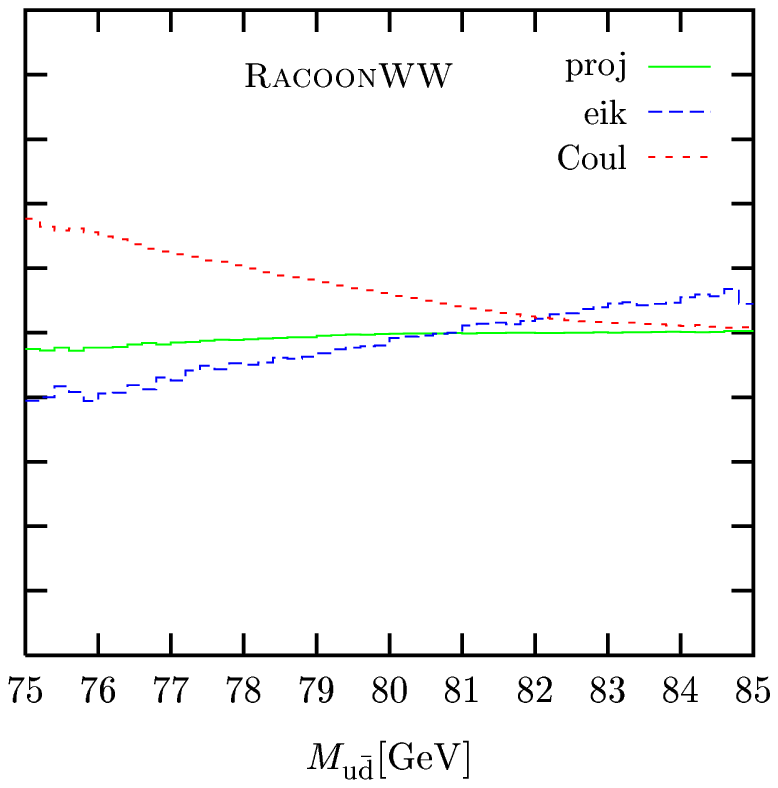}}
\end{picture} }}
\caption[]{Theoretical uncertainty within the DPA for distributions in
the $\mu$ energy and in the $\protect\Pu\bar\protect\Pd$ invariant mass for
$\protect\Pep\protect\Pem\to\protect\Pu\bar\protect\Pd\mu^-\barnu_\mu(\gamma)$ at $\sqrt{s}=200\,\protect\GeV$}
\label{fi:tu_emu_mm}
\efi

For the $\mu$-energy distribution, shown in the l.h.s.\ 
of \reffi{fi:tu_emu_mm}, the differences are typically of
the same order, as long as $E_\mu$ is in the range for $\wb$-pair
production, which is $20.2\,\GeV<E_\mu<79.8\,\GeV$ at $\sqrt{s}=200\,\GeV$.
Outside this region, the four-fermion process is not dominated by the
$\wb$-pair diagrams, and the DPA is not reliable anymore, which is also
indicated by large intrinsic ambiguities.  The r.h.s.\ of
\reffi{fi:tu_emu_mm} shows the DPA uncertainties for the 
$\Pu\bar\Pd$
invariant-mass distribution. Within a window of $2\Gamma_\PW$ around
the W~resonance the relative differences between the considered
modifications are also at the level of $0.1 \div 0.2\%$.  
The differences grow with the distance from the resonance point.

The discussed results illustrate that the intrinsic ambiguities of the
DPA, as applied in {\tt RacoonWW}, are at the level of a few per mil,
whenever resonant $\wb$-pair production dominates the considered observable.

\subsubsection*{Estimating the theoretical uncertainty of the DPA with
{\tt YFSWW3-KoralW}}

The accuracy of the combined result from {\tt YFSWW3}~1.13 and
our all 4-fermion process MC
{\tt KoralW}~1.42 ~\cite{koralw:1999}
as presented in Ref.~\cite{yfsww3:2000b} is expected to be
below $0.5\%$ for the total cross-section when all tests are finished.
These tests are currently in progress.

\subsection{Summary and conclusions}

In this Section we have compared different theoretical predictions for the
CC03 cross-section that have been used to analyze the data in terms of all
$\wb$-pair final states, $4q(qqqq)$ and non-$4q(qq{\rm l}\nu, 
{\rm l}\nu{\rm l}\nu)$. 
The major achievement in this area is represented by inclusion of radiative
corrections in DPA for the $\wb\wb$ cross-section.

Data are collected from $161\,$GeV up to $210\,$GeV.  
One should remember that below
some threshold ($\approx 170\,\GeV$) the DPA cannot be trusted any more for 
both virtual corrections and real-photon radiation, since the kinetic energy 
of the $\wb$~bosons becomes of the order of the $\wb$~width.
{\tt RacoonWW} has shown that the intrinsic ambiguities of its implementation
of the DPA are at the level of a few per mille.

For the total CC03 cross-section, the differences between {\tt RacoonWW}
and {\tt YFSWW} should be of the naively expected DPA accuracy,
\ie below 0.5\% for $\sqrt{s} > 180\,\GeV$.
And, indeed, independently of the channel, the two MC differ by $0.2\div0.3\%$
in the results presented herein and this increases to 0.4\% if uncertainties
from unknown higher-order corrections are taken into account.
Note that, with {\em bare} cuts applied, the difference of $0.2 \div 0.3\%$
shown here between the two compared programs does not change.

The corrections to the distribution in the cosine of the production
angle for the $\PWp$ and $\PWm$ bosons have also been analyzed for the
{\em bare} and the {\em calo} recombination algorithms. They are
compatible with each other at a level below 1\%.
Although 
compatible with the statistical accuracy, the deviations seem to become
somewhat larger for large scattering angles.
The corrections to the invariant mass distributions for the $\PWp$ and
$\PWm$ bosons, again with {\em bare} and {\em calo} recombinations 
are statistically compatible between the two Monte Carlo programs
everywhere  and agree within 1\%.

Another comparison, shown in \reffi{fi:totcs}, indicates that {\tt RacoonWW}
and {\tt BBC} calculations agree very well for the total $\wb$-pair production
cross-section above $185\,\GeV$.
Below this energy the differences in the implementation of the DPA
become visible, in agreement with the expected relative error of
$\ord{\alpha/\pi\times\Gamma_{\ssW}/\Delta E}$.
However, for angular and energy distributions unavoidable differences 
at the level of $1\div 2\%$ arise
between the two predictions, as a consequence of the definition of the 
phase-space variables in the presence of photon recombination.  
Although the {\tt BBC}-calculation has not been implemented in a MonteCarlo 
it can be used for obtaining a relative $\ord{\alpha}$ correction factor
where one has an estimated internal accuracy ranging from $1.5\%$ at
lower energies to $0.3\%$ at $210\,$GeV.

In conclusion, from the direct comparisons of {\tt RacoonWW} and {\tt YFSWW3},
supported by {\tt BBC},
we can estimate an overall theoretical uncertainty of the current predictions
for the total $\wb\wb$ cross-section at $0.4\%$ at $200\,\,\GeV$. 
The $0.4\%$ precision tag is an important conclusion of this Workshop.

For other energies no 
complete investigations of the theoretical
uncertainty have been performed. However, based on the error estimate
of $0.4\%$ for $200\GeV$, the intrinsic uncertainty of the DPA of $0.2\%$
at $200\GeV$ and the generic energy dependence of this uncertainty 
given by $\gw/(E_{\rm CMS}-2\MW)$ we estimate an uncertainty of the
predictions of {\tt RacoonWW} and {\tt YFSWW3} of 
$0.5\%$ for $180\GeV$ and $0.7\%$ for $170\GeV$.
This could be somewhat further reduced, if the sources of the
differences between the different programs are found.

Results for the $\wb\wb$ cross-section at $\ord{\alpha}$ are also 
available from {\tt GRACE} but a comparison with the other codes is not yet 
at the level of those already presented where a considerable amount of time 
was invested to try to understand differences towards a safe estimate of 
theoretical uncertainty.

\section{Four fermions plus a visible photon}
\label{sect4fg}

The class of processes that are investigated at LEP~2 are
$e^+e^- \to \wbp\wbm \to 4\,$f, single-$\wb$ production,
$\zb$-boson-pair production, single-$\zb$ production. LEP~2 and also future
linear colliders will allow us to study a new class of processes, $e^+e^- \to
4\rmf + \ph$.

The physical interest of the latter is twofold. They can be used to obtain
informations on the quartic gauge-boson couplings and include the production
processes of three gauge-bosons, $\wbp\wbm\ph$, $\zb\zb\ph$ and $\zb\ph\ph$.
In this case the photon is visible by definition and we term the corresponding
process {\em radiative}, i.e. we consider as radiative events those events 
with photons where at least one photon passes the experimental photon 
requirements, for instance $E_{\ph} > 1\,$GeV, $\cos\theta_{\ph} < 
0.985(0.997)$ and $\theta_{f - \ph} > 5^\circ$.

Note that for all final states, the invariant mass needs a more precise
definition in case radiative photons are present in the event.  
From a calculational point of view, there is always a minimal invariant mass
(energy and separation angle) below which photons are not
{\em resolved}. Thus we need to specify fermion-photon invariant mass or
fermion-photon energies and separation angles, 
below which the the photon are combined with the fermion and 
above which the photons are not included in the mass calculation. 
A {\em bare} mass would set these cuts rather tight, excluding photons from 
the $f\barf$ mass, a {\em calo} mass would set the separation cuts looser.
Theorists like cuts on $M(\ph - {\rm nearest f})$.
Experimentalists like cuts on energies and angles.
In the following we list both TH(eory)-cuts and EXP(erimental)-cuts.

\begin{description}
\item[TH cuts]:
\bei
\item[{\em bare}] $M(\barf f + (\ph))$ including photons if $M(f+\ph) < 
5\,$GeV,
\item[{\em calo}] $M(\barf f + (\ph))$ including photons if $M(f+\ph) < 25\,
$GeV.
\eei

\item[EXP-cuts]:

\bei
\item[{\em bare}]: $M(\barf_1 f_2 + (\ph))$, photons less than 1 GeV or less 
than $1^\circ$ away from $f_1$ or $f_2$ are included;

\item[{\em calo}]: $M(\barl_1 l_2 + (\ph))$, photons less than 1 GeV 
or less than $10^\circ$ away from charged leptons are included,  \\
$M(\barq_1 q_2 + (\ph))$, photons less than 1 GeV or less than $25^\circ$ away 
from either quark $q_1, q_2$ 
are included, which takes at least the major difference 
between fermions - quarks versus leptons - into account.

\eei
\end{description}

These definitions serve for benchmarking distributions, not so much to
mimic an actual experimental strategy, which is of course fermion
dependent. 
In other words this is an approximation to the experimental side: if the 
fermion is a muon, even $0^\circ$ opening angles can be separated 
experimentally. In 
addition, for identified photons one still may or may not choose to recombine 
the photon with the fermion. 

Furthermore, $e^+e^- \to 4\rmf + \ph$ is
an important building block for
the radiative corrections to the Born process $e^+e^- \to 4\,$f, hence
non-radiative events are those with no photon or only photons below the
minimal photon requirements. In case of non-radiative events, this amounts 
to adding up virtual and soft radiative corrections. 
The effect of $\ord{\alpha}$ QED corrections very often amounts to several
percent, mostly originating from collinear photon radiation off highly
energetic particles and from virtual photon exchange. For initial state
radiation, for instance, we have three types of corrections,
a) $\ord{\alpha/\pi\,\ln(\me/Q)}$ with $Q \gg \me$ being the typical scale at
which the process occur, b) $\ord{\alpha/\pi}$ from hard photons that must, 
nevertheless, be included for a $1\%$ precision tag, c) leading 
$\ord{\alpha^2}$, or higher corrections that becomes relevant for a precision 
tag below the $1\%$ thresholds.

Owing to the fact that a theoretical prediction with a typical accuracy of 
some fraction of a percent must include all QED corrections, we face the
complexity of it. Handling the singularities of the squared matrix element 
represents a formidable task; in any bremsstrahlung process the integrand
blows up for arbitrary small photon energies and similar problems arise from
collinear emission off the charged particles.

A general comment about this section is that some of the programs, but not 
all, implement $4\rmf + \ph$ at the level of (exact) matrix elements.
Few programs have only an effective treatment of photons via
structure functions, with or without $p_t$.
Furthermore we also have to distinguish between massless vs. massive
calculations.

\subsection{Description of the programs and their results}

\subsubsection*{$4\rmf + \ph$ with {\tt RacoonWW}}

\subsubsection*{Authors}

\begin{tabular}{l}
A.Denner, S.Dittmaier, M.Roth and D.Wackeroth
\end{tabular}

\subsubsection*{General description}

The program {\tt RacoonWW} \cite{de00} evaluates cross-sections
and differential distributions for the reactions $\Pep\Pem\to 4\rmf$ and
$\Pep\Pem\to 4\rmf+\gamma$ for all four-fermion final states. The long
write-up has already been presented in \sect{racoonww},
so that we only stress the features that are peculiar to $4\rmf+\gamma$
production with a separated hard photon.
The calculation is based on full $4\rmf+\gamma$ matrix elements for
all final possible states. Since fermion masses are neglected, lower cuts
on the invariant mass of $\rmf\bar\rmf$ pairs and on $\Pe^\pm$
emission angles have to be imposed, in addition to the angular and
energy cuts for the hard photon. {\tt RacoonWW} supports different ways
to treat finite gauge-boson widths (fixed and running widths,
complex-mass scheme) and allows to select subsets of graphs ($VV\gamma$
signal diagrams, QCD background). Detailed numerical results on 
$4\rmf+\gamma$ production with {\tt RacoonWW} can be found in 
Ref.~\cite{racoonww_ee4fa} and in \sect{ac4fg}.

\subsubsection*{$4\rmf + \ph$ with {\tt PHEGAS/HELAC}}

\subsubsection*{Author}

\begin{tabular}{l}
C.~G.~Papadopoulos\\
\end{tabular}

This section refers to a novel Monte Carlo program
that is capable to deal with any tree-order process involving 
any particle and interaction described by the
Standard Model, including QCD.

The program consists of two modules:
\begin{enumerate}
\item {\tt HELAC} which is a matrix element computation-tool~\cite{ref1}
based on Dyson-Schwinger equations, and
\item {\tt PHEGAS} an automatic phase-space generator~\cite{ref2} capable to
simulate all peaking structures of the amplitude.
\end{enumerate}
 
The over all code is using a Monte Carlo integration based on
multichannel optimization~\cite{ref3}.
 
\subsubsection*{{\tt HELAC}}

The matrix element is evaluated using a recursive approach
based on Dyson-Schwinger equations.
The computational cost exhibits an exponential growth~($\simeq 3^n$)
as a function of the number of external particles ($n$)  which for
multi-particle processes results to a very important increase in the
efficiency as compared with the traditional Feynman-graph approach
whose computational cost grows factorially~($\simeq n!$).
In order to optimize code's efficiency the computational strategy 
consists of two phases. In the first
phase a solution to the recursive equations is established
in terms of an integer array
containing all relevant information for the process under consideration.
This is the {\it initialization} phase and is performed once at the
beginning of the execution of the program.
In the second phase, using the already generated information,
the actual computation is performed resulting to the numerical
evaluation of the amplitude for each specific phase-space point provided.

In order to consistently describe unstable particles the fixed width
as well as the complex width schemes have been included. ISR and running
couplings are also an option and work is in progress to implement 
higher order corrections within the approach of reference~\cite{non-local}.

In order to deal with numerical stability problems, besides the
{\em double precision}, a {\em quadruple} as well as
a {\em multi-precision}~\cite{ref4} version is available.
This makes {\tt HELAC} able to deal with processes exhibiting
strong collinear singularities, like $e^-e^+ \to e^-e^+\mu^-\mu^+$
at zero scattering angles. 
Moreover all particle masses and vertices of the Standard Model, 
including QCD, in both the Feynman and unitary gauges are incorporated. 

\subsubsection*{{\tt PHEGAS}}

Although several matrix element computational tools were available
in the past that can deal with arbitrary processes, 
to the one or to the other extent~\cite{ref5},
phase-space generators were always developed according to a specific
process or a class of processes~\cite{ref7}. 
{\tt PHEGAS} is a phase space generator that 
incorporates in an automatic way all possible kinematical mappings
for any given process, using the relevant information provided
by {\tt HELAC}. To this end each Feynman graph contributing to the process
under consideration gives rise to a kinematical mapping. The integration
is performed via a Monte Carlo multichannel approach and during the 
computation, weight  optimization selects automatically 
those kinematical mappings that are relevant for the process under
consideration. 

As a first highly non-trivial test {\tt PHEGAS/HELAC} has been used to 
produce results for four-fermion plus a visible photon within
the current study. Nevertheless, it is worthwhile to emphasize 
that {\tt PHEGAS/HELAC} is able
to deal with any process involving any Standard Model particle and 
is by no means restricted to $e^+e^-\to \mbox{4 f}
+\gamma$ reactions. A detailed presentation of the code, the
implemented algorithms as well as the incorporated physics effects 
will be available in the near future~\cite{ref2}.

\subsubsection*{$4\rmf + \ph$ with {\tt WRAP}}

\subsubsection*{Authors}

\begin{tabular}{l}
G. Montagna, M. Moretti, O. Nicrosini, M. Osmo and F. Piccinini \\
\end{tabular}

\subsubsection*{Description of the Method.}
\label{method4fg}

Contributions of the Pavia/{\tt ALPHA} group to the subject of 
four fermions plus gamma final states are summarized.

\subsubsection*{Hard-scattering matrix element}
\label{me4fg}

The exact tree-level matrix elements for the processes with four fermions 
plus a visible photon in the final state are computed by means of the 
{\tt ALPHA} algorithm~\cite{alpha}. 
At present, the processes which can be mediated by 
two $\wb$-bosons ( CC processes) or by two $\zb$-bosons 
( NC processes) are accounted for. 
The effect of finite fermion masses is taken into 
account exactly both in the kinematics and dynamics. 
The contribution of anomalous trilinear gauge couplings
can be also simulated, after having implemented in {\tt ALPHA} 
and cross-checked the parameterization in terms 
of $\Delta k_{\gamma}$, $\lambda_{\gamma}$, 
$\delta_Z$, $\Delta k_Z$ and $\lambda_Z$ 
of refs.~\cite{gg,hhpz}.
The genuinely anomalous quartic gauge boson couplings, involving 
at least one photon and relevant for 
this process at tree-level, are also included, 
according to the parameterization of Ref.~\cite{bbkpgs}.
Final cross-checks on anomalous 
quartic couplings are in progress.
The fixed-width scheme is adopted as gauge-restoring approach, as 
motivated in comparison with other gauge-invariance-preserving 
schemes in Ref.~\cite{racoonww_ee4fa}.

\subsubsection*{Radiative corrections}
\label{rc4fg}

The phenomenologically relevant Leading Log (LL) QED radiative corrections, 
due to initial-state radiation (ISR), are implemented via the Structure 
Function (SF) formalism~\cite{sf}, according to the two following options:
\begin{itemize}

\item[--]  collinear SF $D(x,s)$;

\item[--]  $p_t$-dependent SF 
$\tilde{D} (x, \cos\theta_{\gamma}; s)$, i.e. a combination 
of the collinear SF $D(x,s)$ with an angular factor for photon
radiation inspired by the leading behaviour 
$1/(p\cdot k)$~\cite{mnpcpc,mmnp}.

\end{itemize}
In fact, as discussed in detail in refs.~\cite{pv4fg,mmnp}, 
due to the presence of an observed photon in the final state, 
the treatment of ISR in terms of collinear SF turns out to be 
inadequate because affected by double counting
between the pre-emission photons 
(described by the SF) and the observed one (described by the hard-scattering 
matrix element).\footnote{In the tuned comparison with {\tt RacoonWW} the
effect of ISR SF was switched off.}
By keeping under control also 
the transverse degrees of freedom of ISR, as 
allowed by $p_t$-dependent SF, it is possible 
to remove the double-counting effects, following the 
procedure for the calculation of the QED corrected cross-section 
discussed in Ref.~\cite{mmnp,pv4fg}, i.e.
\bq
\sigma_{QED}^{4\rmf + 1\gamma} = \int d x_1 d x_2 
d c_{\gamma}^{(1)} d c_{\gamma}^{(2)} \,
\tilde{D} (x_1, c_{\gamma}^{(1)}; s) \tilde{D} (x_2, c_{\gamma}^{(2)}; s)
\Theta({\rm cuts}) d\sigma^{4\rmf + 1\gamma} \, ,
\label{pt4fg}
\eq
where $c_{\gamma}^{(i)} \equiv \cos\theta_{\gamma}^{(i)}$, $i=1,2$. 
According to eq.~(\ref{pt4fg}), an {\em equivalent} photon is generated for 
each colliding lepton and accepted as a higher-order ISR contribution if:
\begin{itemize}
\item[--]  the energy of the equivalent photon is below the
 threshold for the observed photon $E_{\gamma}^{\rm min}$, for arbitrary angles;  or
\item[--]  the angle of the  equivalent photon is outside the angular 
acceptance for the observed photons, for arbitrary energies. 
\end{itemize} 
Within the angular acceptance of the detected photon, 
the cross-section is evaluated by means of the exact matrix 
element for the processes $e^+ e^- \to 4\rmf + \gamma$.
Therefore, eq.~(\ref{pt4fg}) applies to the signature of four fermions plus 
exactly one photon in the final state, corrected by the effects of 
undetected soft and/or collinear ISR.
The $Q^2$-scale entering the QED SF is fixed to be $Q^2 = s$.

\subsubsection*{Computational tool and obtained results}
\label{rot4fg}

The theoretical features sketched above have been implemented into a massive
MonteCarlo (MC) program, named {\tt WRAP} ({\tt W} {\tt R}adiative process with 
{\tt A}lpha \& {\tt P}avia). The multi-channel importance sampling technique is 
employed to perform the phase-space integration, 
paying particular attention to the infrared and collinear peaking structures 
due to photon emission.  
The code supports realistic event selections and can be employed either as
a cross-section calculator or as a true event generator.
Results obtained in the present study can be summarized as follows:
We have performed a critical analysis of the effect of ISR 
(see \figs{fig1_4fg}{fig3_4fg}) and a study of the impact of finite fermion 
masses (see \tabn{fig4_4fg}),
Finally, we have tuned comparisons with the predictions of other codes, 
especially with {\tt RacoonWW} (see Sec. \ref{ac4fg}).

The impact of ISR via collinear SF on the $4\rmf+\gamma$ integrated 
cross-section of the CC10 final state $\mu^- \; \barnu_{\mu}\; u\;
 \bard\; \gamma$ is shown in \figs{fig1_4fg}{fig2_4fg}, as a function of the 
LEP~2 c.m.s. energy (\fig{fig1_4fg}) and of the photon energy threshold 
at $\sqrt{s} = 192$~GeV (\fig{fig2_4fg}). \fig{fig1_4fg} shows that 
ISR in the collinear approximation reduces the Born cross-section 
between $16-12\%$ in the c.m.s. range $180-190\,$GeV and 
at the $10\%$ level close to $200\,$GeV, for 
 the considered photon separation cuts. In particular, 
 at $\sqrt{s} =$~192~GeV the reduction factor as due to ISR 
 is $12-13\%$, almost independent of the photon detection 
 threshold, as shown in \fig{fig2_4fg}.

Note that collinear SF contradicts photon detection criteria, as discussed
before. However, in order to get a first estimate of the correction 
due to ISR, collinear SF can be used, 
since the error introduced by this treatment (double-counting effects) 
is estimated in \fig{fig3_4fg}, by comparing collinear and $p_t$ structure
functions.

\clearpage

\begin{figure}[p]
\begin{center}
\includegraphics[width=12cm]{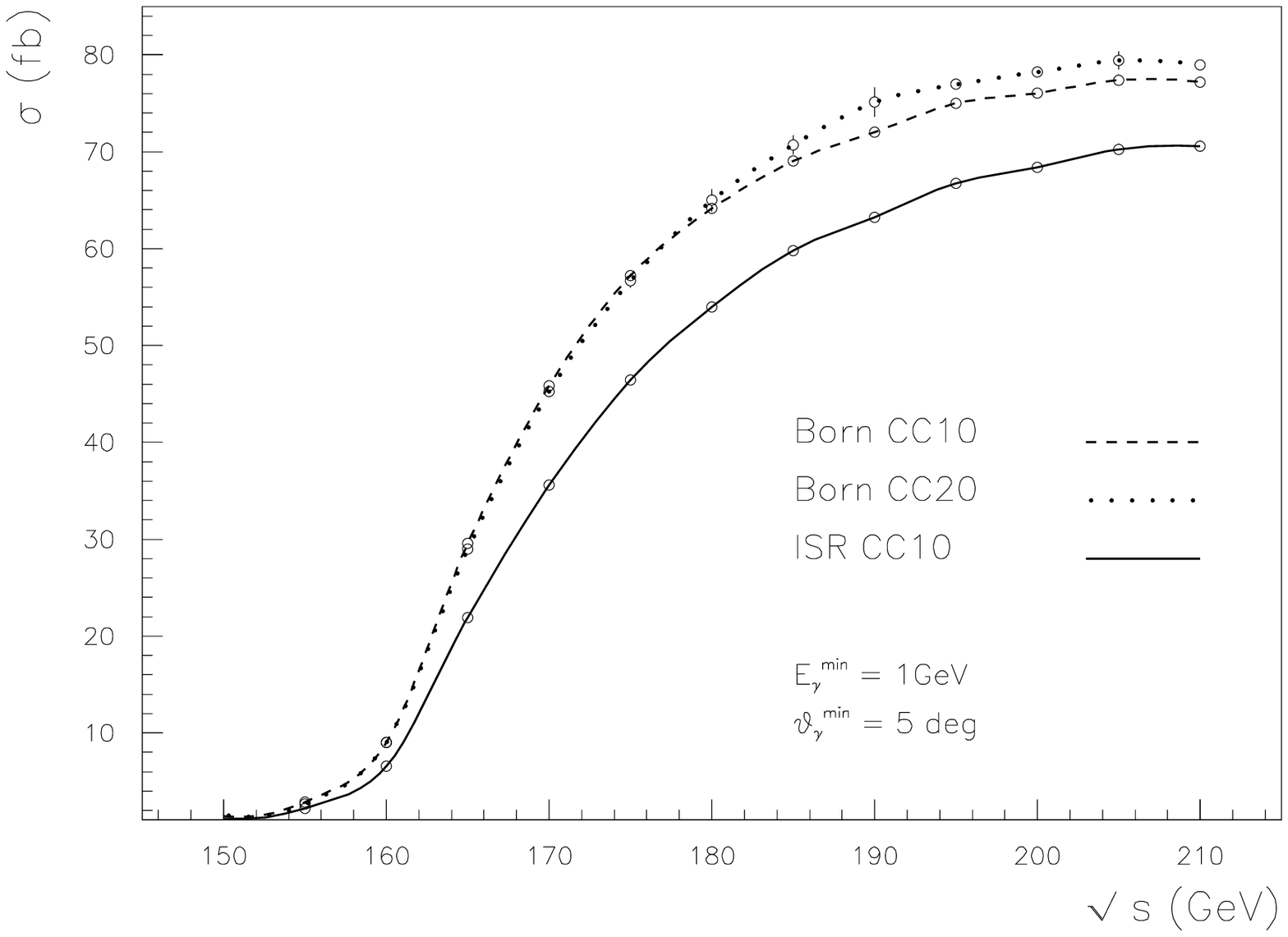}
\caption[]{The effect of ISR, simulated 
by collinear SF, on the integrated cross 
section of the  CC10 final state  
$\mu^- \; \barnu_{\mu}\; u\; \bard\; \gamma$ as a 
function of the LEP~2 c.m.s. energy. The Born cross-section 
for the CC20 final state 
$e^- \;\barnu_e\; u\; \bard\; \gamma$ is also shown.}
\label{fig1_4fg}
\end{center}
\end{figure}
\begin{figure}[p]
\begin{center}
\includegraphics[width=12cm]{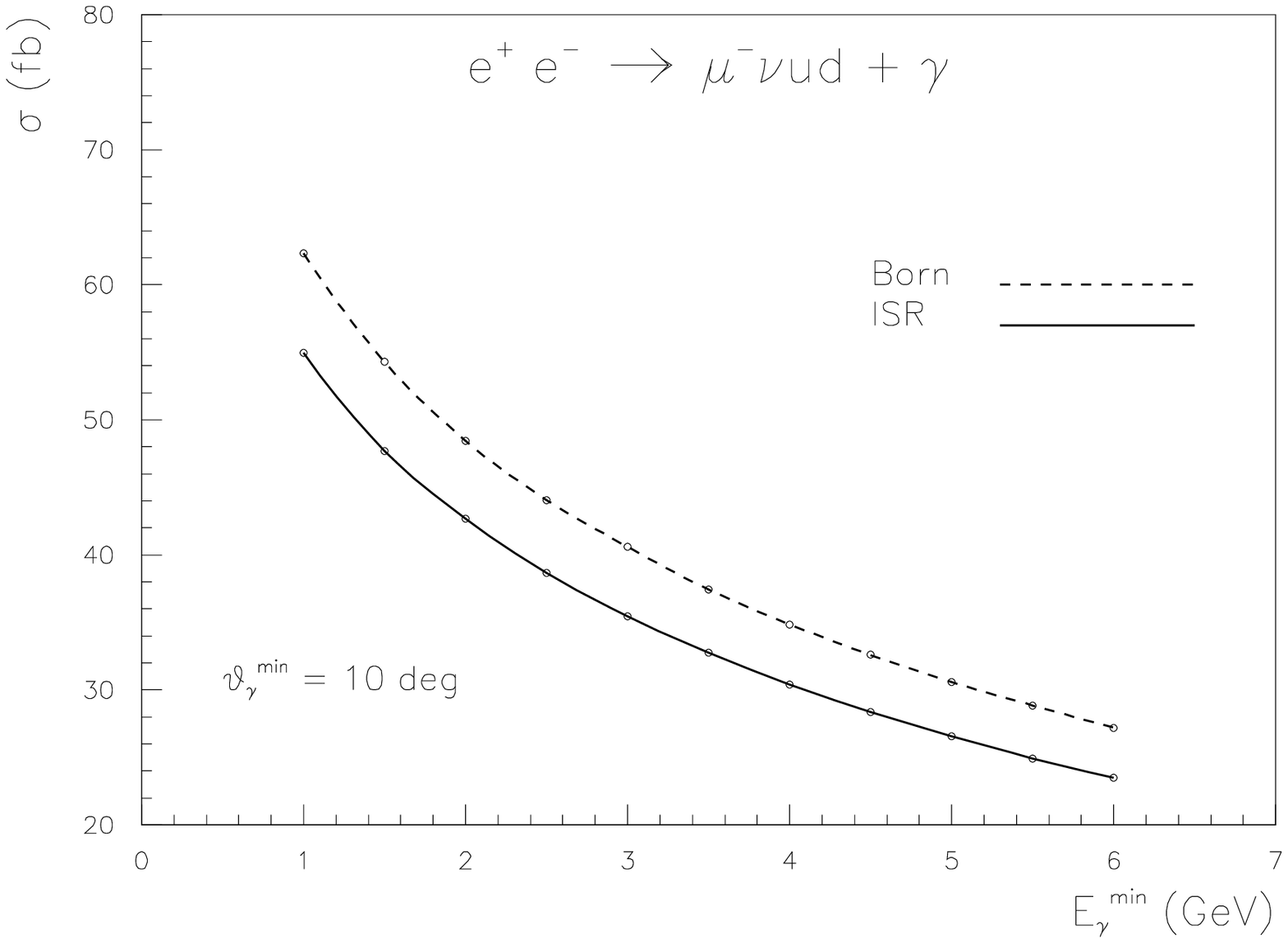}
\caption{The effect of ISR via collinear SF 
to the cross-section of the CC10 final state  
$\mu^- \; \barnu_{\mu}\; u\; \bard\; \gamma$, as a 
function of the minimum energy of the observed photon, 
at 192 GeV.}
\label{fig2_4fg}
\end{center}
\end{figure}

\clearpage

\begin{figure}[p]
\begin{center}
\includegraphics[width=12cm]{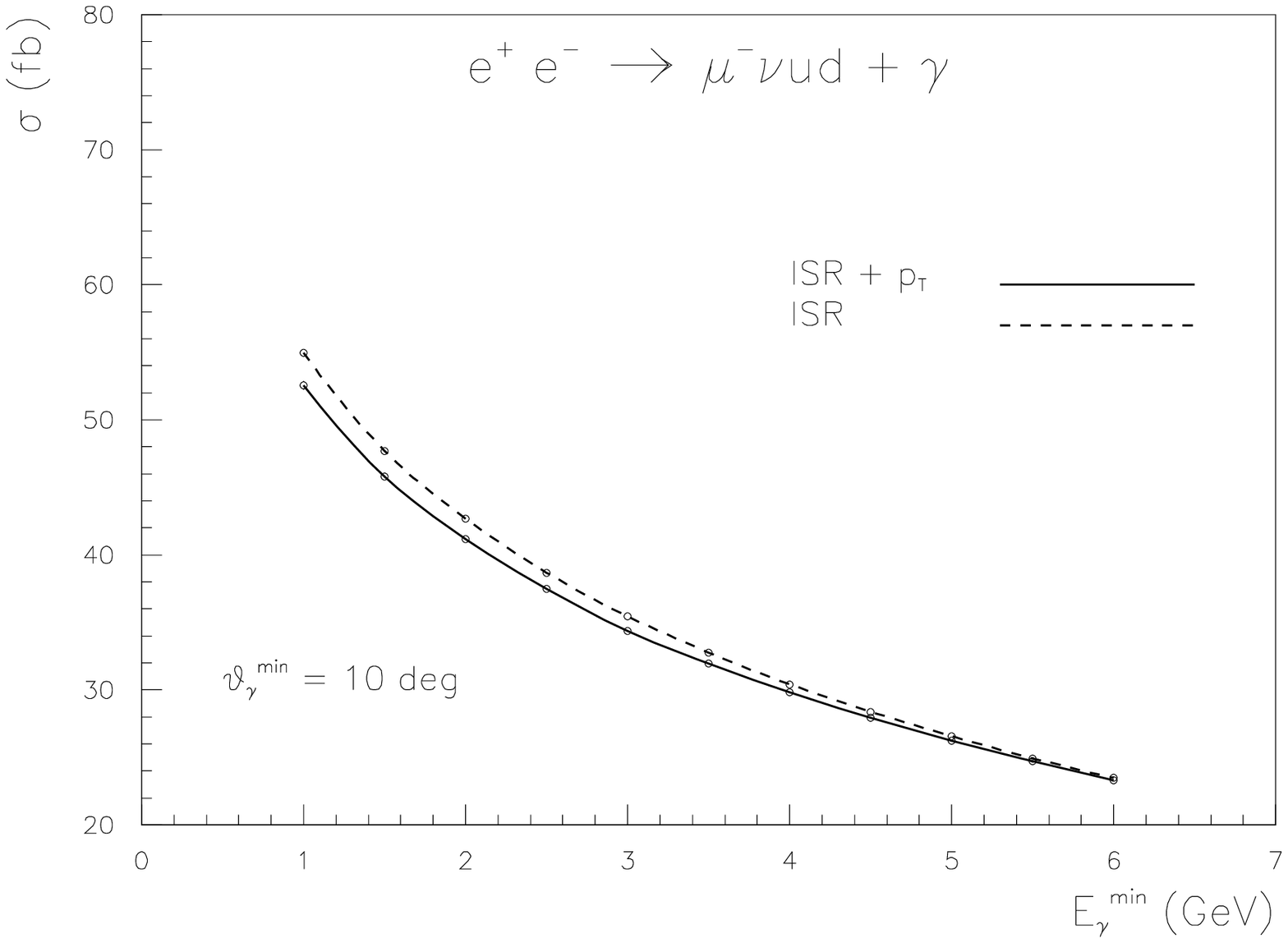}
\caption{Comparison between the effects of 
ISR via collinear SF (dashed line) and $p_t$-dependent SF 
(solid line), respectively,
for the cross 
section of the the CC10 final state  
$\mu^- \; \barnu_{\mu}\; u\; \bard\; \gamma$, as a 
function of the minimum energy of the observed photon, 
at 192 GeV.}
\label{fig3_4fg}
\end{center}
\end{figure}

\begin{table}[p]
\begin{center}
\begin{tabular}{||l|l|l|l||}\hline  
$\vartheta_{\gamma -q}$ (deg) & $\vartheta_{\gamma -\mu}$ [deg]   & 
cross-section [fb] & \, \,  $\delta$ (\%) \\ \hline\hline
& & &\\ 
$5^{\circ}$     & $1.0^{\circ}$      & $90.157 \pm 0.036$  & $1.92 \pm 0.08$ \\ 

                &                    & $91.903 \pm 0.035$  & \\  
& & &\\ \hline
& & &\\
$5^{\circ}$     & $0.1^{\circ}$      & $104.777 \pm 0.046$ & $9.31 \pm 0.09$ \\ 

                &                    & $115.004 \pm 0.044$ &   \\  
 
&& &\\ \hline
& & &\\
 $5^{\circ}$    & $0.0^{\circ}$      & $105.438 \pm 0.045$ & \\ 

                &                    &   & \\  
 
&& &\\ \hline

\end{tabular}            
\end{center}
\caption[]{Comparison between massive and massless Born cross sections for the 
process $\mu^- \barnu_{\mu} c \bars + \gamma$ at $\sqrt{s}=200$~GeV, 
as obtained by means of WRAP.
$\theta_{\gamma -f}$, with $f = q, \mu$ is the minimum separation angle 
between the photon and 
final state charged fermions. In the third column, the first result refers to
the
massive case, and the second one to the massless case. Relative difference 
is shown in the last column.}
\label{fig4_4fg}
\end{table} 

\clearpage

As far as fermion masses are concerned we show in \tabn{fig4_4fg} a 
comparison between the cross-section for the final state
$\mu^- \; \barnu_{\mu}\; c\; \bars\; \gamma$ 
in the massless approximation is compared with the 
same cross-section in the presence of finite masses
for the final state fermions. The mass values and cuts used are: 
$m_\mu = 0.105$~GeV, $m_s = 0.3$~GeV, 
$m_c = 1.55$~GeV, with $M_{cs} \geq 3$~GeV. 
In the considered channel with a muon in the final 
state, the minimum separation angle between the quarks and the photon is 
maintained fixed at $5^\circ$, while the separation angle between the 
muon and the photon is varied from $1^\circ$ down to zero. 
It can be seen that the mass effects on the 
the integrated cross section are of the order of $1\%$ for not
too small separation angles, but it may reach, not 
surprisingly, the $10\%$ level in more stringent conditions, where 
only a massive $4\rmf + \gamma$ calculation can provide a reliable 
prediction in the presence of muons in the final state.


\subsubsection*{$4{\rm f}+\ph$ with {\tt CompHEP}}

\subsubsection*{Authors}

\begin{tabular}{l}
E.~Boos, M.~Dubinin and V.~Ilyin \\
\end{tabular}

\subsubsection*{General description}

The program {\tt CompHEP}~\cite{CompHEP} calculates cross-sections
and distributions for all channels $e^+ e^- \to 4f$ and
$e^+ e^- \to 4f \, + \gamma$. The calculation is based on a
tree-level matrix element for the complete set of diagrams.
Finite fermion masses are taken into account both in the
matrix element and in the four or five particle phase space
parameterization. The fixed-width prescription is used for
the gauge boson propagators.
In so far as {\tt CompHEP} uses the squared diagrams technique,
the calculation for the five particle states with radiative gamma is
CPU time consuming and in the following only the results
for the channel $e^+ e^- \to \gamma \mu \barnu\barnu_{\mu} u \bard$
(2556 squared diagrams) are presented (\fig{dubinin1}, \fig{dubinin2},
where the factor
$\alpha(0)/\alpha_{\gf}$ is not accounted for). We
used the standard set of cuts including EXP-cuts
for the distributions in the {\em bare} and {\em calo} mass. 

\subsubsection*{On-shell $\wb$ boson approximation for 
             $e^+ e^- \to \gamma \mu \barnu_{\mu} u \bard$}

In the $2 \to 4$ approximation of the on-shell $\wb$ boson 
$e^+ e^- \to \gamma \mu \barnu_{\mu} \wbp$ for the $2 \to 5$ process
$e^+ e^- \to \gamma \mu \barnu_{\mu} u \bard$ the number of
diagrams is much smaller ($31$ for the $4$-body and $71$ for the $5$-body
final state). It is interesting to find out if a simpler on-shell
$\wb$ approximation reproduces with enough likelihood the total
rate and distributions given by the exact $2 \to 5$ tree level
amplitude. The possibility to describe quantitatively
the $5$-body distributions of radiative events by some trivial change of
the normalization in the $4$-body results could be attractive.

We calculated the cross section of the process
$e^+ e^- \to \gamma \mu \barnu_{\mu} \wbp$ multiplied by a factor given by 
the following on-shell $\wb$ isotropic decay to $u \bard$. 
Vectors of the $u$, $\bard$ quarks momenta generated randomly
in the rest frame of the $\wb$ were boosted to the $e^+ e^-$ 
c.m.s., where the standard kinematical cuts were introduced:
$E_{\gamma} \geq$ 1 GeV, $E_{\mu} \geq 5\,$GeV,  
$|\cos\theta(\gamma e)| \leq 0.985$. Furthermore,  
$|\cos\theta(\mu e)| \leq 0.985$, 
$\theta(\gamma,\mu)$, $\theta(\gamma,u)$, and $\theta(\gamma,\bard)$
$\geq$ 5$^{\circ}$.
Such a scheme of calculation is based on the well-known
approximation of infinitely small $\wb$ width $\mw \Gamma_{\rm tot}/[(M^2_{u
\bard}-\mws)^2 + \mws\Gamma_{tot}^2] \Rightarrow 
\pi \, \delta (M^2_{u \bard}-\mws)$ 
and have been widely used for the simulation of the $3$- and $4$-body final
states in many generators. The simulation by {\tt PYTHIA} generator
\cite{PYTHIA}
follows slightly better scheme, when the $\wb$ decay products 
invariant mass is distributed according to the Breit-Wigner and gamma
radiation from quarks can be switched on in the approximation of
final state shower. 

\begin{figure}[p]
\epsfig{file=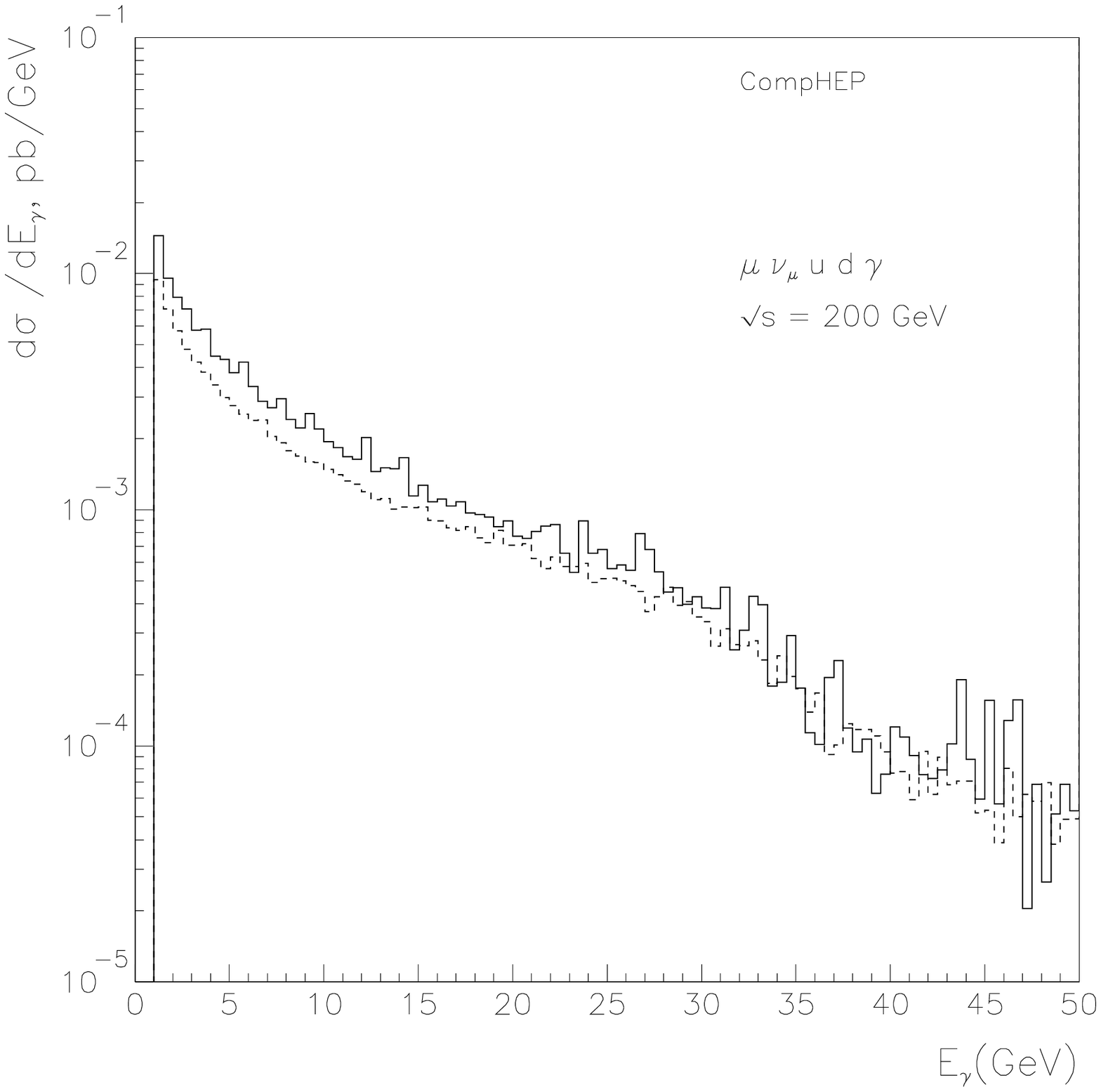,width=0.49\linewidth}
\hfill
\epsfig{file=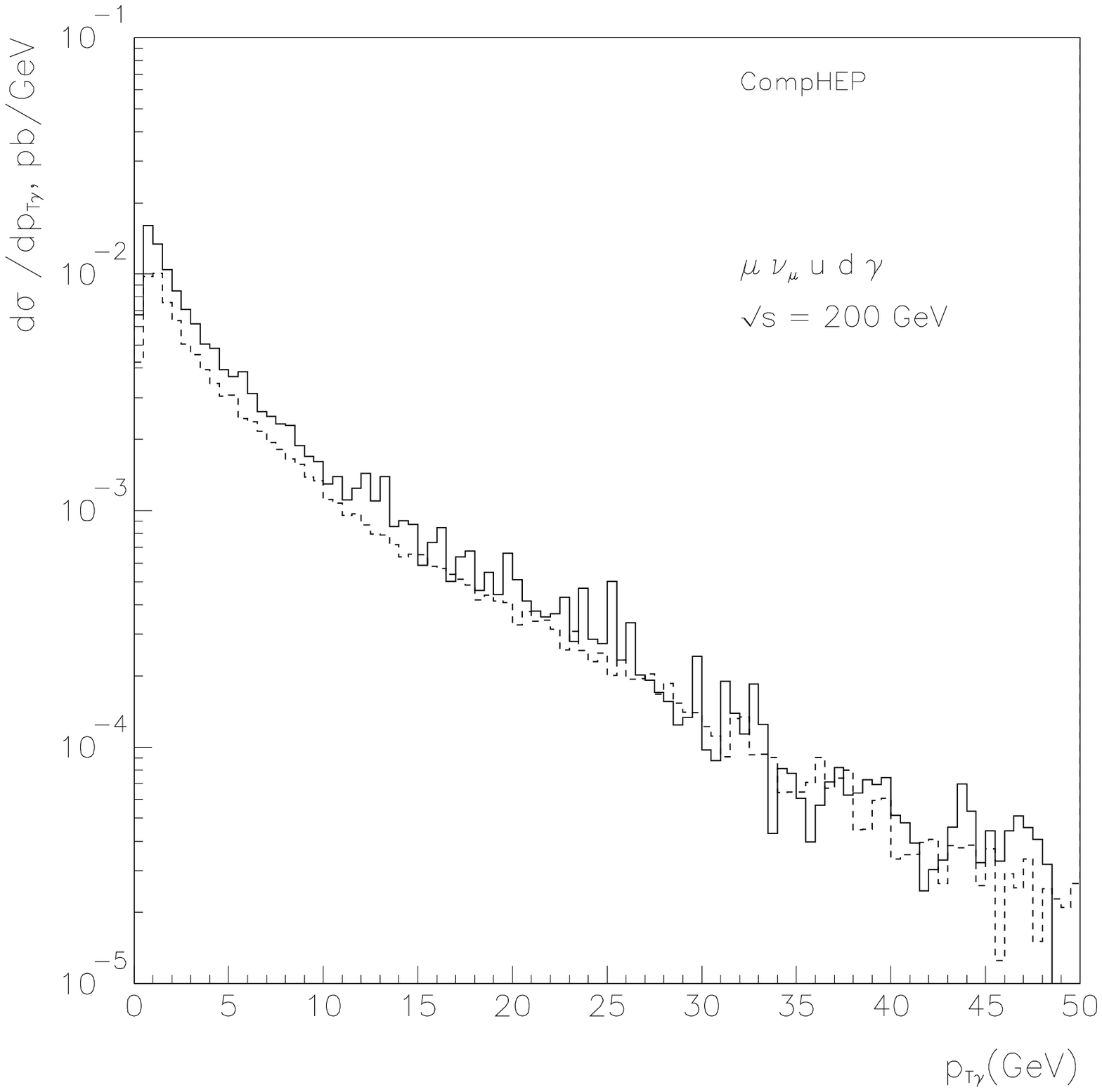,width=0.49\linewidth}
\vskip 1cm
\epsfig{file=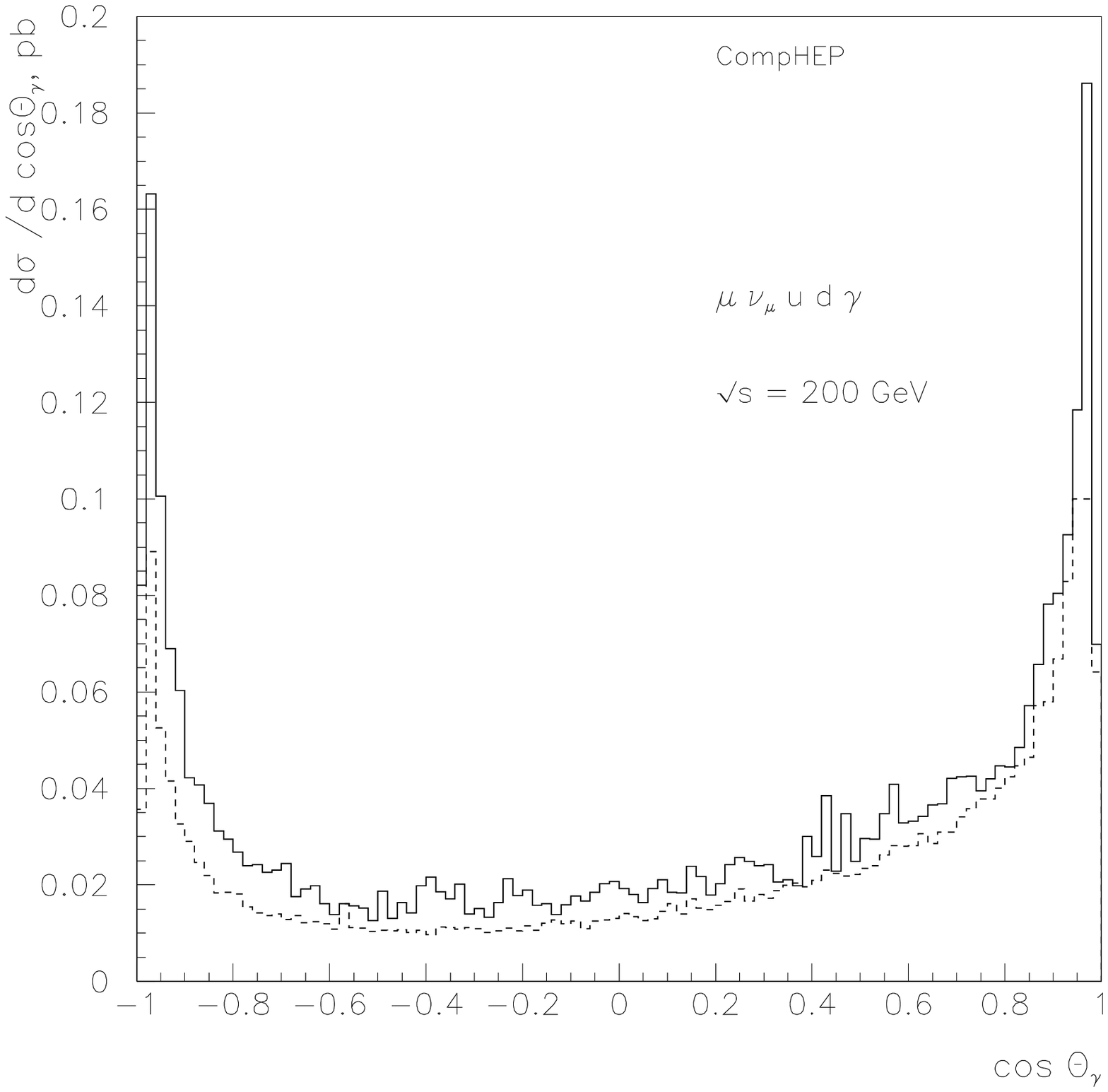,width=0.49\linewidth}
\hfill
\epsfig{file=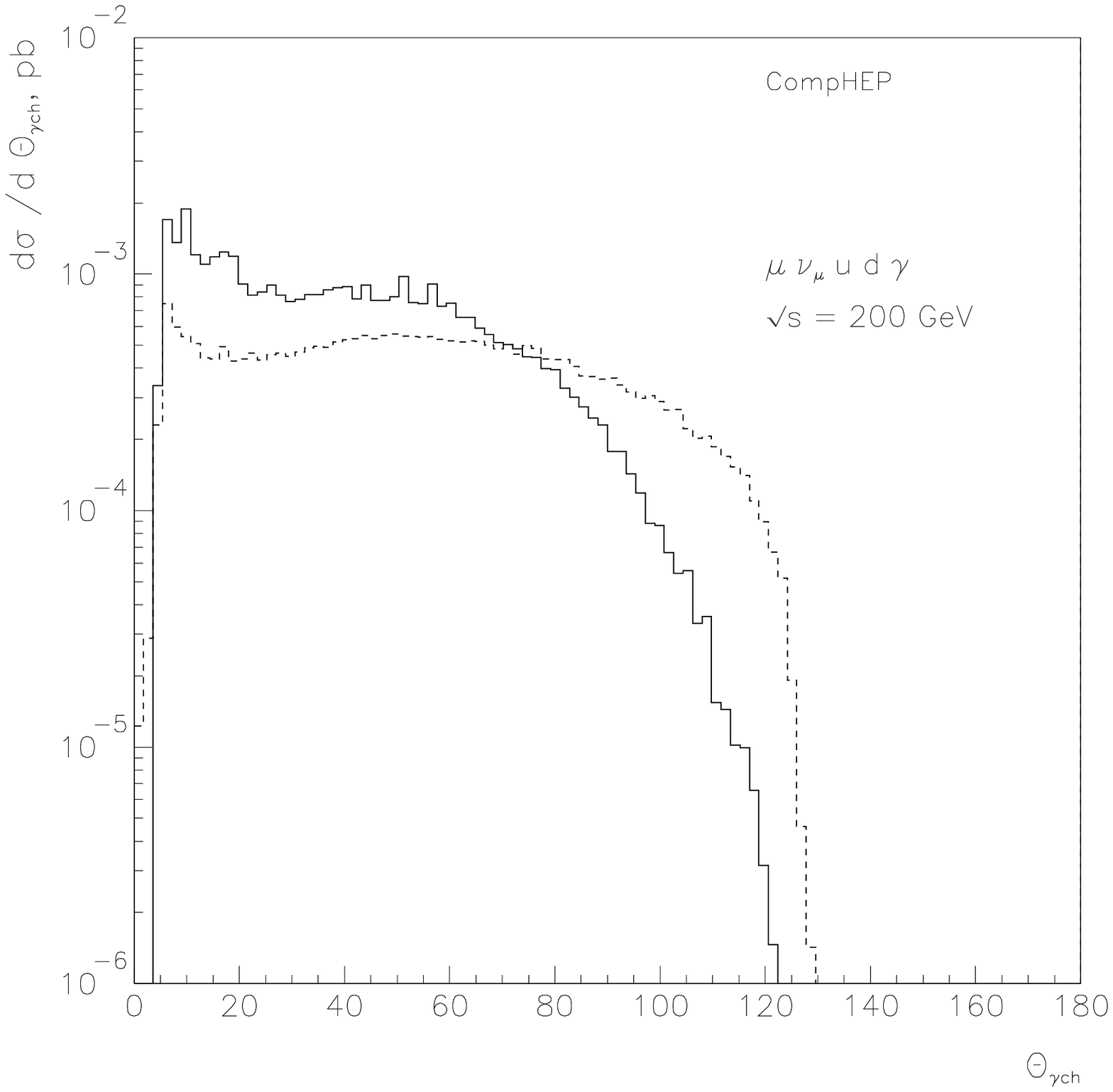,width=0.49\linewidth}
\caption[]{Distributions in the gamma energy, gamma transverse momentum,
  gamma angle with the beam, and in the opening angle between the
  gamma and the nearest charged fermion. The distributions for the
  $e^+ e^- \to \gamma \mu \bar \fnum u \bard$ are shown by the solid
  line and the distributions for the $e^+ e^- \to \gamma \mu \bar
  \fnum \wbp$ with the following $\wb$ isotropic decay are shown by
  the dashed line.}
\label{dubinin1}
\efi
\begin{figure}[p]
\epsfig{file=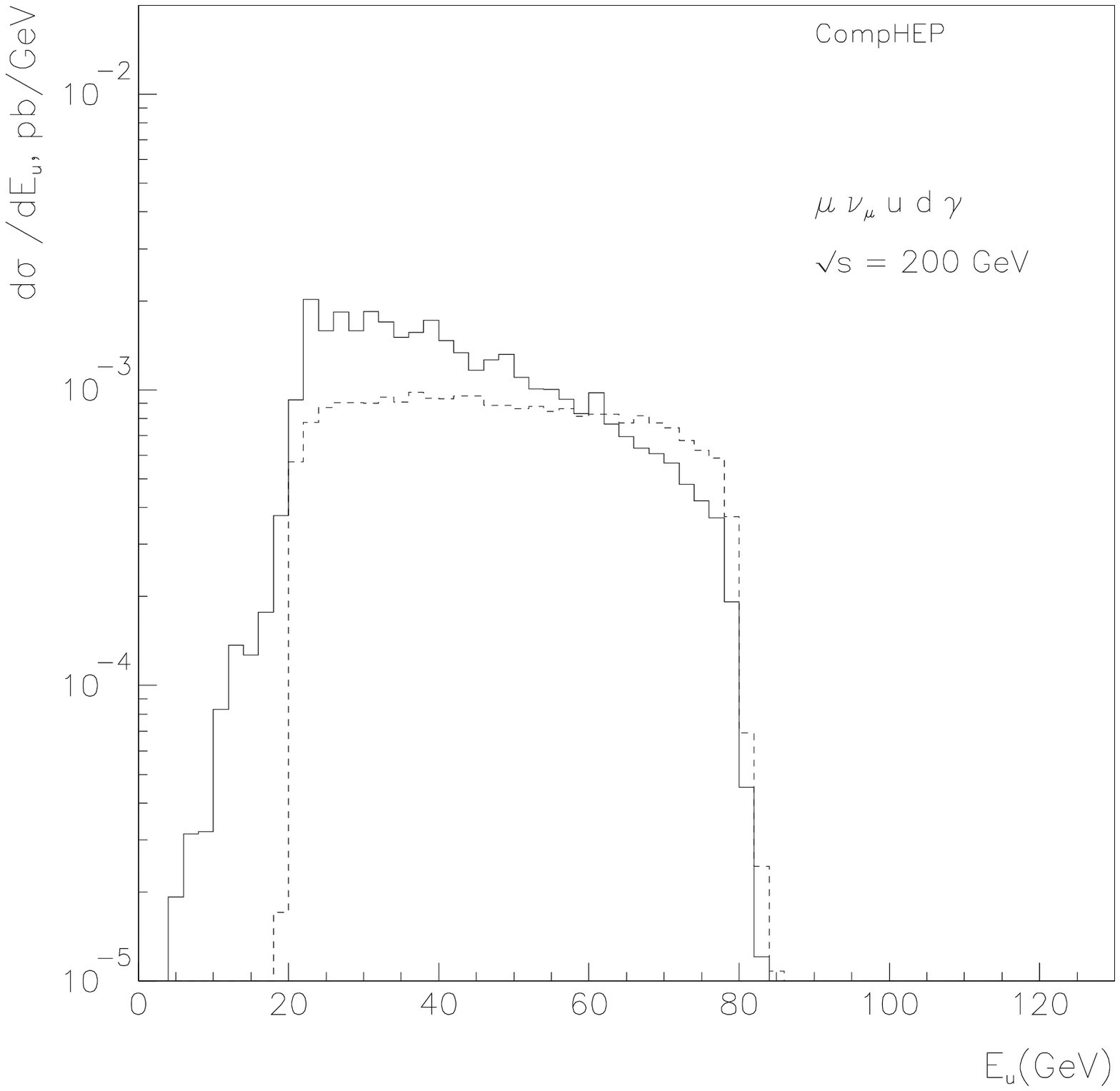,width=0.49\linewidth}
\hfill
\epsfig{file=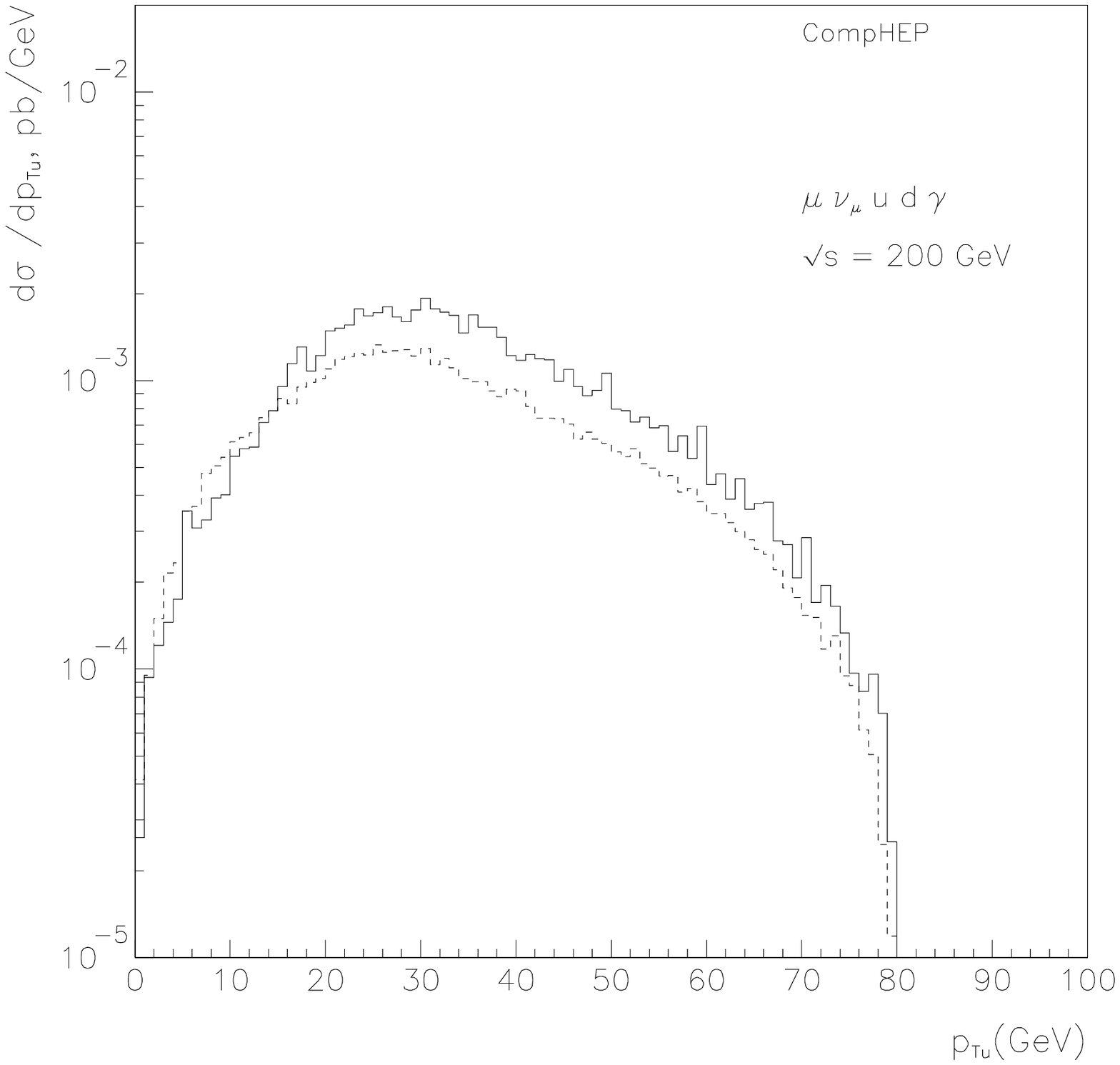,width=0.49\linewidth}
\vskip 1cm
\epsfig{file=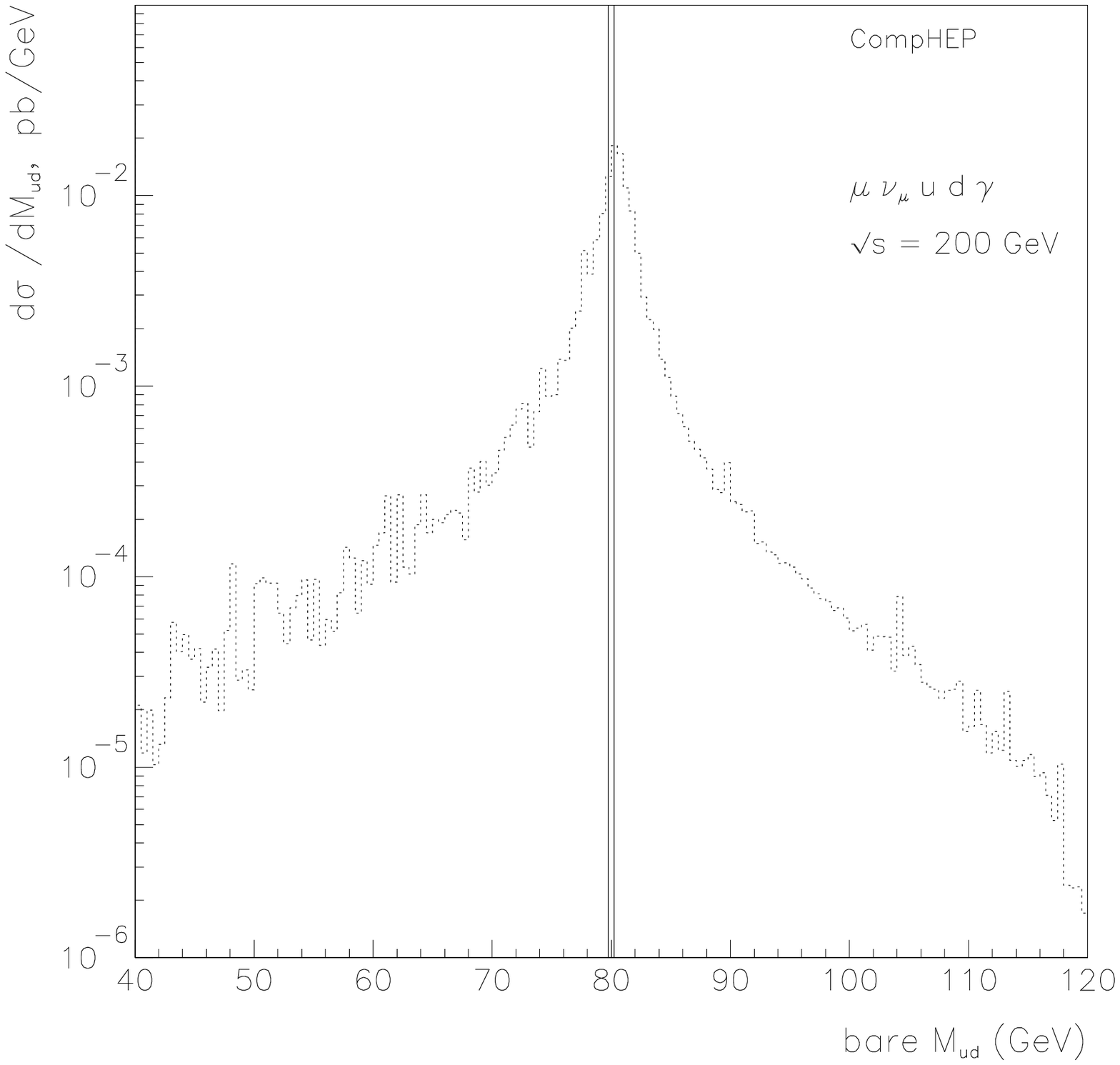,width=0.49\linewidth}
\hfill
\epsfig{file=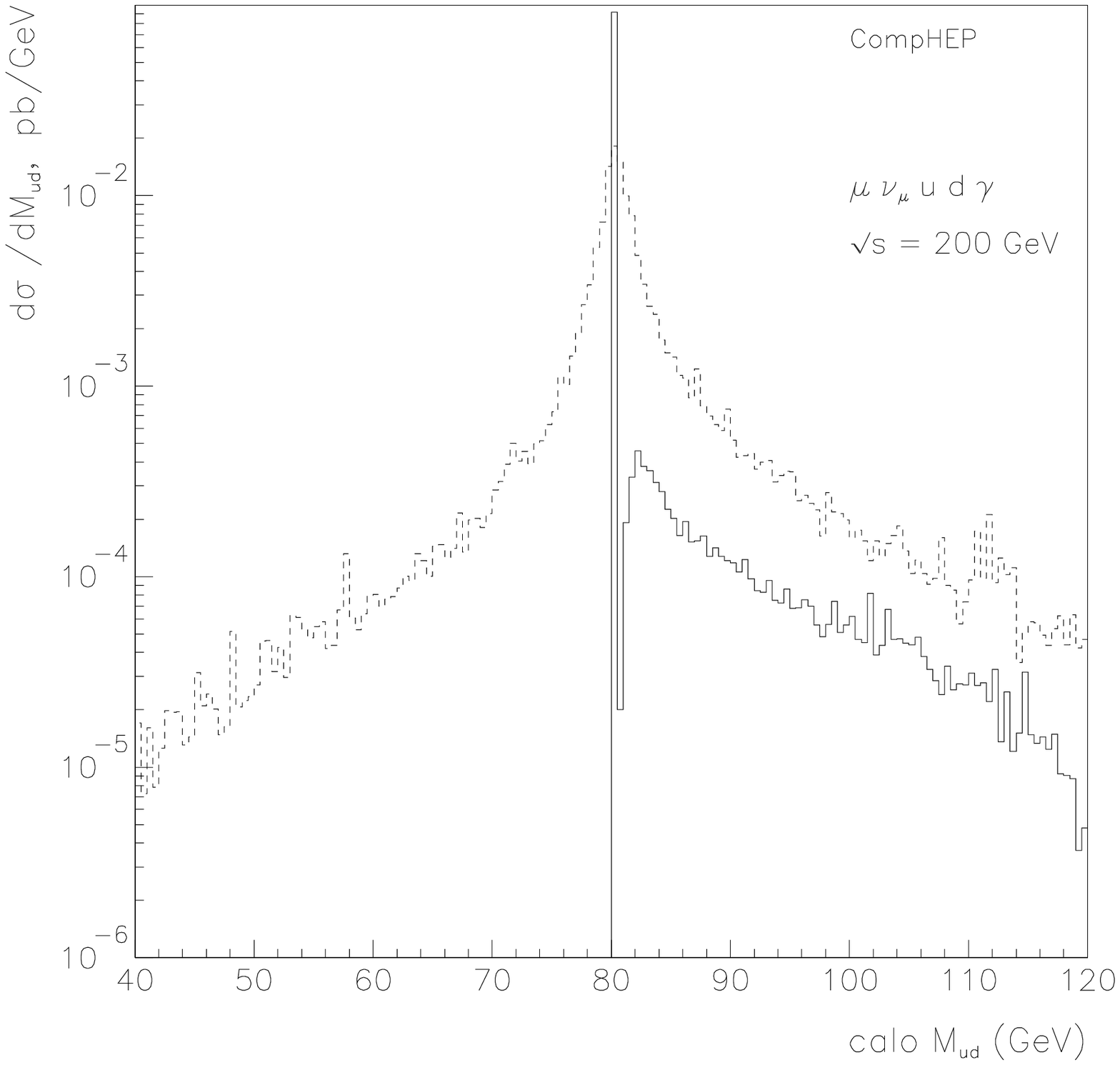,width=0.49\linewidth}
\caption[]{Upper row of plots - distributions in the quark energy and
  the quark transverse momentum for the channel $e^+ e^- \to \gamma
  \mu \bar \fnum u \bard$ (solid) and the approximation $e^+ e^- \to
  \gamma \mu \bar \fnum \wbp$ (dashed).  Lower row of plots -
  distributions in the 'bare' and 'calo' $M_{ud}$ invariant mass for
  the approximation $e^+ e^- \to \gamma \mu \bar \fnum \wbp$ (solid)
  and the exact $2 \to 5$ process $e^+ e^- \to \gamma \mu \bar \fnum u
  \bard$ (dashed/dotted).}
\label{dubinin2}
\efi

The total rate of the $\sigma(e^+ e^- \to \gamma \mu \barnu_{\mu} \wbp)$ 
Br$(\wbp \to u \bard)$
is equal to $49.4(2)\,$fb to be compared with the exact $2 \to 5$ result
$69.1(9)\,$ fb. Missing contribution of the omitted diagrams, especially from
the phase
space regions near the collinear and infrared poles of the photons
radiated from the initial state and the $u$, $d$ quarks leads to
substantial underestimate of the rate. 
Peaks of the forward and back-scattered photons (\fig{dubinin1}), radiated 
from the initial $e^+$, $e^-$, are much stronger underestimated than the 
photon distribution in the central
rapidity region. Distributions in the quark energy and transverse
momentum (upper plots in \fig{dubinin2}) are rather different in the exact and
approximate calculation. For the exact calculation the quark energy
spectrum more rapidly decreases than for the approximation where the
photon radiation from quarks is not accounted for. 
In the exact $5$-body consideration the $\wb$ boson is
created in a rather well defined polarization state, so the approximation
of an isotropic on-shell $\wb$ decay could be unsatisfactory for angular
variables.
Large difference of the distributions in the photon-fermion (muon or
quark) angle (lower plot in \fig{dubinin1}) is caused by a simple
combinatorial
reason. {\em Calo} jet-jet mass (lower plot in \fig{dubinin2}) contains the
unresolved photon radiated from the initial state or from the muon, so
only $M_{u \bard \gamma} \geq \mw$ is possible. 
   
It follows that in the case of four fermion events with radiative photon
the approximation of the on-shell $\wb$ isotropic decay does not, generally
speaking, satisfactorily describe both the total rate and the full set of
final particle distributions.

\subsubsection*{$4\rmf + \ph$ via Structure Functions 
with {\tt NEXTCALIBUR}}

\subsubsection*{Authors}

\begin{tabular}{l}
F.A.Berends, C.~G.~Papadopoulos and R.Pittau \\
\end{tabular}

In this Section we show illustrative results for the processes
$e^+ e^- \to \mu^- \mu^+ u \bar u (\gamma)$ ($\zb\zb$ signal) and
$e^+ e^- \to \mu^- \bar \nu_\mu u \bar d (\gamma)$ ($\wb\wb$ signal).
Analogous results for the single-$\wb$ case can be found
in section~\ref{sectsw}.

{\tt NEXTCALIBUR} does not contain
the exact matrix element for $e^+ e^- \to 4 {\rm}f +\gamma$, therefore
we generate photons always through $p_t$-dependent $ISR$ Structure Functions. 
We used the set of cuts specified in the proposal 
at $\sqrt{s}= 200$ GeV, all diagrams and fermion masses included.
In tables \ref{nexca1} and \ref{nexca2} 
four values of cross section (in pb) are shown.

\begin{table}[hp]\centering
\begin{tabular}{|c|c|}
\hline
&\\
Type & Cross-section \\
&\\
\hline
&\\
$\sigma_{\rm tot}$    &   16.107(9)    \\
$\sigma_{\rm nrad}$   &   15.018(9)    \\
$\sigma_{\rm srad}$   &    1.0697(30)  \\
$\sigma_{\rm drad}$   &    0.0189(4)   \\
\hline
\end{tabular}
\vspace*{3mm}
\caption[]{Cross-sections in fb from {\tt NEXTCALIBUR} for the process
$e^+(1) e^-(2) \to \mu^-(3) \mu^+(4) u(5) \baru(6)$.
$M(34) > 10\,$GeV and $M(56) > 10\,$GeV.
Separation cuts for the photons: $E_{\ph} > 1\,\GeV, 
|\cos\theta_{\ph}| < 0.985$.} 
\label{nexca1}
\end{table}
\begin{table}[hp]\centering
\begin{tabular}{|c|c|}
\hline
&\\
Type & Cross-section \\
&\\
\hline
&\\
$\sigma_{\rm tot}$    &   617.27(59) \\
$\sigma_{\rm nrad}$   &   578.19(58) \\
$\sigma_{\rm srad}$   &   38.54(16)  \\
$\sigma_{\rm drad}$   &    0.54(2)   \\
\hline
\end{tabular}
\vspace*{3mm}
\caption[]{Cross-sections in fb from {\tt NEXTCALIBUR} for the process
$e^+(1) e^-(2) \to \mu^-(3) \barnu_{\mu}(4) u(5) \bard(6)$.
$M(56) > 10\,$GeV.
Separation cuts for the photons: $E_{\ph} > 1\,\GeV, 
|\cos\theta_{\ph}| < 0.985$.} 
\label{nexca2}
\end{table}

The first value, labelled by {\em tot}, is the sum 
of radiative and non radiative events
(within the specified separation cuts for the generated photons).
The second one {\em nrad} corresponds to non-radiative events 
and the third one {\em srad} to single-radiative events, namely events
with only one radiated photon outside the separation cuts.
We also include a fourth entry that represents the small 
fraction of radiative events with $2$ photons ({\em drad}).

To check the sensitivity of the distributions to the chosen form
of Structure Function, we run again the above processes 
with a slightly different implementation of the sub-leading terms, 
without observing
any significant deviation with respect to the previous results.  

\subsubsection*{$4\rmf + \ph$ with {\tt GRACE}}

\subsubsection*{Authors}

\begin{tabular}{l}
Y.~Kurihara, M. Kuroda and Y.~Shimizu \\
\end{tabular}

In this Section we present results from {\tt GRACE} for the $4\rmf+\gamma$ 
processes with $\wb$-pair and single-$\wb$ cuts.
Parameters and cuts used are the same as those of the {\tt WRAP}
and {\tt RacoonWW} collaborations, except that we used $\alpha_{\gf}$ for all 
vertices.
Unfortunately, {\tt GRACE} results cannot be compared directly with those
of {\tt RacoonWW} and {\tt WRAP}; indeed, 
when {\tt GRACE} numbers are be compared with the others one should multiply 
by a factor $\alpha(0)/\alpha_{\gf}$. 
To check the calculations, the following tests have been performed for 
the processes $e^+e^- \to \mu \barnu_{\mu} u \bard \gamma$ at $\sqrt{s} =
200\,$GeV:

\begin{itemize}

\item[--] Gauge parameter independence check;
the amplitude generated by {\tt GRACE} keeps gauge parameters in covariant
gauge. It has been checked numerically that the amplitude is independent of
gauge parameters at several phase-space points. 

\item[--] Ward Identity check;
when the polarization vectors of the external photons are replaced by
their four-momentum, the amplitude must be zero due to Ward-Identity.
We have checked it numerically at several phase-space points.

\item[--] Soft photon check;
the cross-sections with soft-photon emission can be easily calculated
by non-radiation cross-section and the soft-photon emission function.
We have calculated the soft-photon emission cross-section by two methods;
\vspace{2ex}

\begin{enumerate}
\item Using $4\rmf+\gamma$ matrix elements with cuts,
$10^{-4}\,\GeV < E_{\ph} < 10^{-2}\,\GeV$, no angular cut on the photon,
$|\cos\theta_{\mu}| < 0.985, E_{\mu} > 5\,\GeV, M(ud) > 10\,\GeV$, giving
$\sigma = 0.5105 \pm -0.0002\,$pb;

\item Using $4\rmf+\ph$ matrix elements with soft-photon function with cuts,
$10^{-4}\,\GeV < E_{\ph} < 10^{-2}\,\GeV$, no angular cut on the photon,
$|\cos\theta_{\mu}| < 0.985, E_{\mu} > 5\,\GeV, M(ud) > 10\,\GeV$
giving $\sigma = 0.5109 \pm 0.0005\,$pb.
\end{enumerate}
\vspace{2ex}

The two methods, therefore, give consistent results.
We have used exact matrix elements for the calculations of
$4\rmf+\gamma$. For $\wb$-pair processes, we simply used fixed width
for the gauge-boson propagator in the unitary gauge. For single-$\wb$
processes we used a special gauge~\cite{Kurihara} for the $t$-channel photon, 
which shows very small effects from the gauge violation due to the 
gauge-boson width.
     
\end{itemize}
Distributions from {\tt GRACE} are shown in
Fig.~\ref{yr_fig_g12}-\ref{yr_fig_g34sw}.

\clearpage

\bfi
\epsfig{file=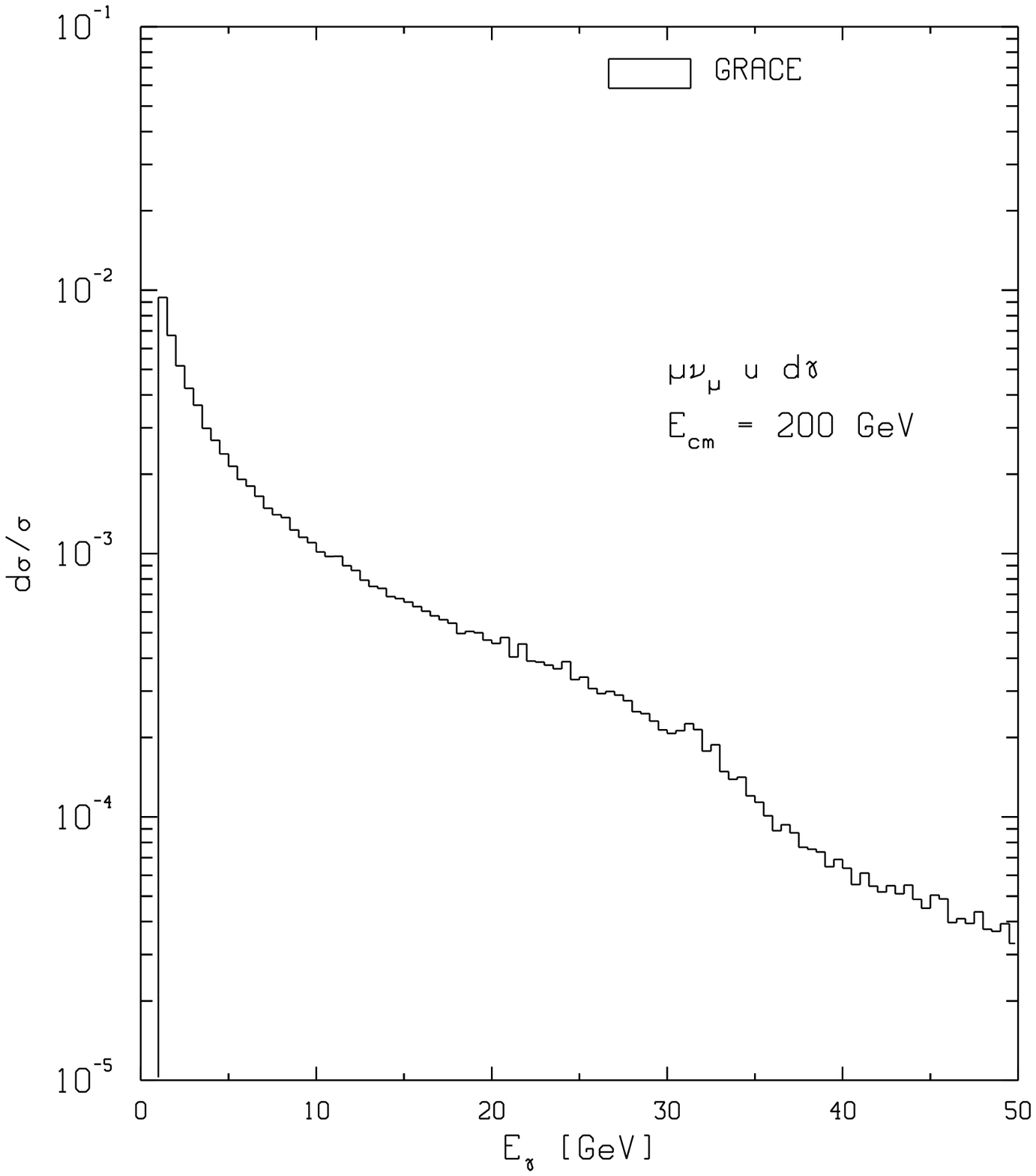,width=0.49\linewidth}
\hfill
\epsfig{file=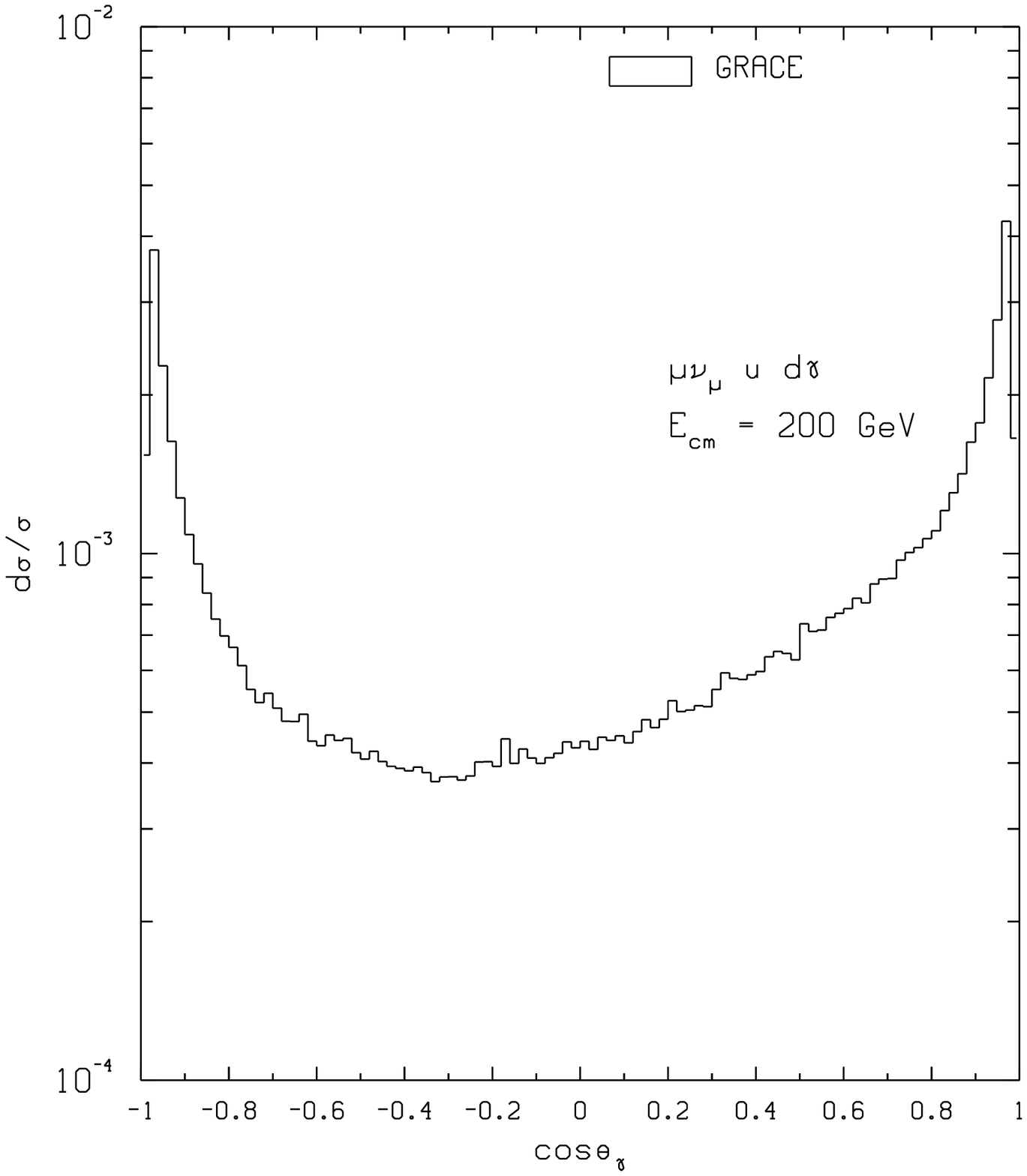,width=0.49\linewidth}
\vskip -1cm
\caption{$E_{\gamma}$ and $\cos \theta_{\ph}$ distributions for the 
process $\mu \fnum u d \gamma$ from {\tt GRACE} with $\wb\wb$-cuts.} 
\label{yr_fig_g12}
\efi
\bfi
\epsfig{file=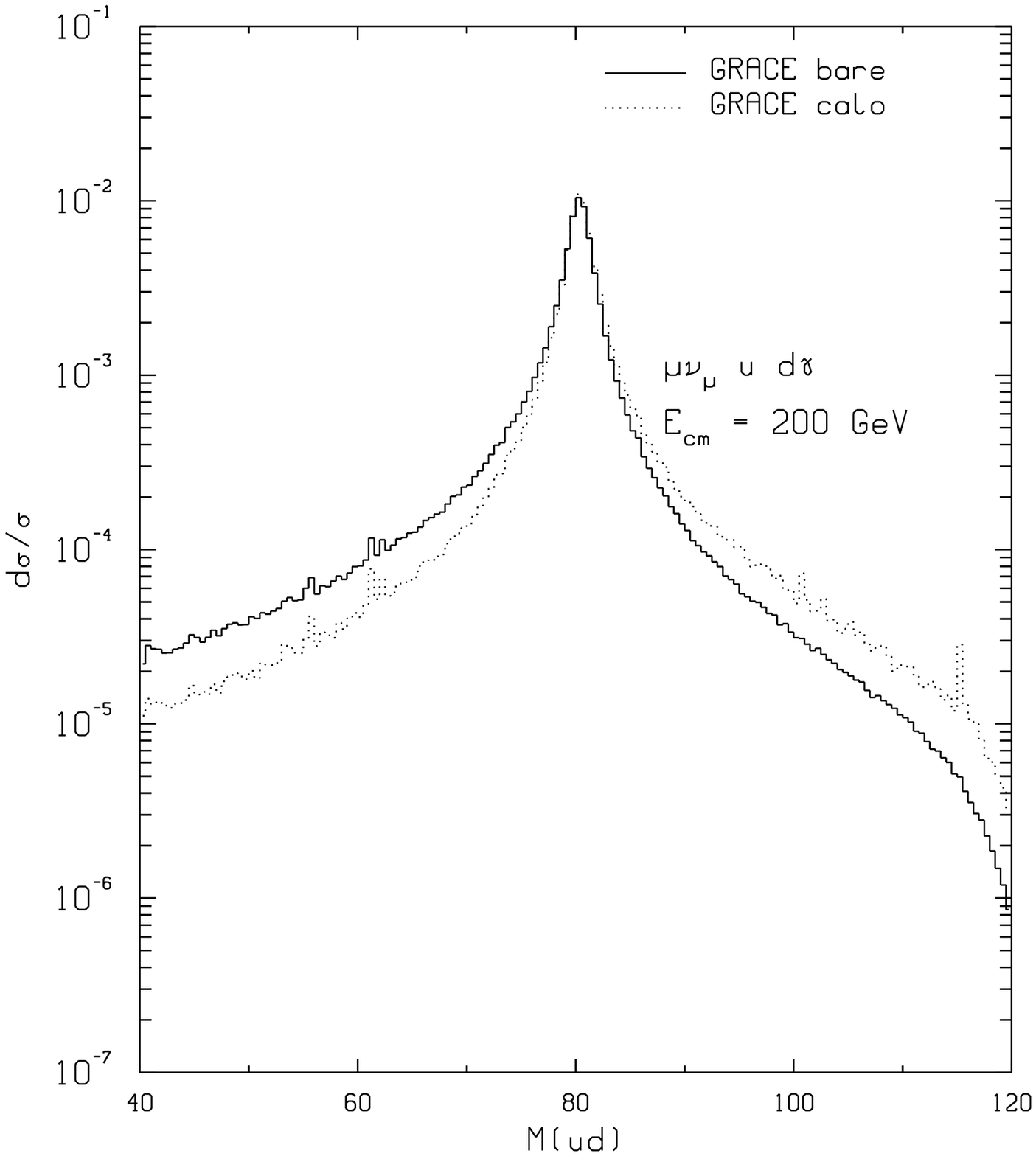,width=0.49\linewidth}
\hfill
\epsfig{file=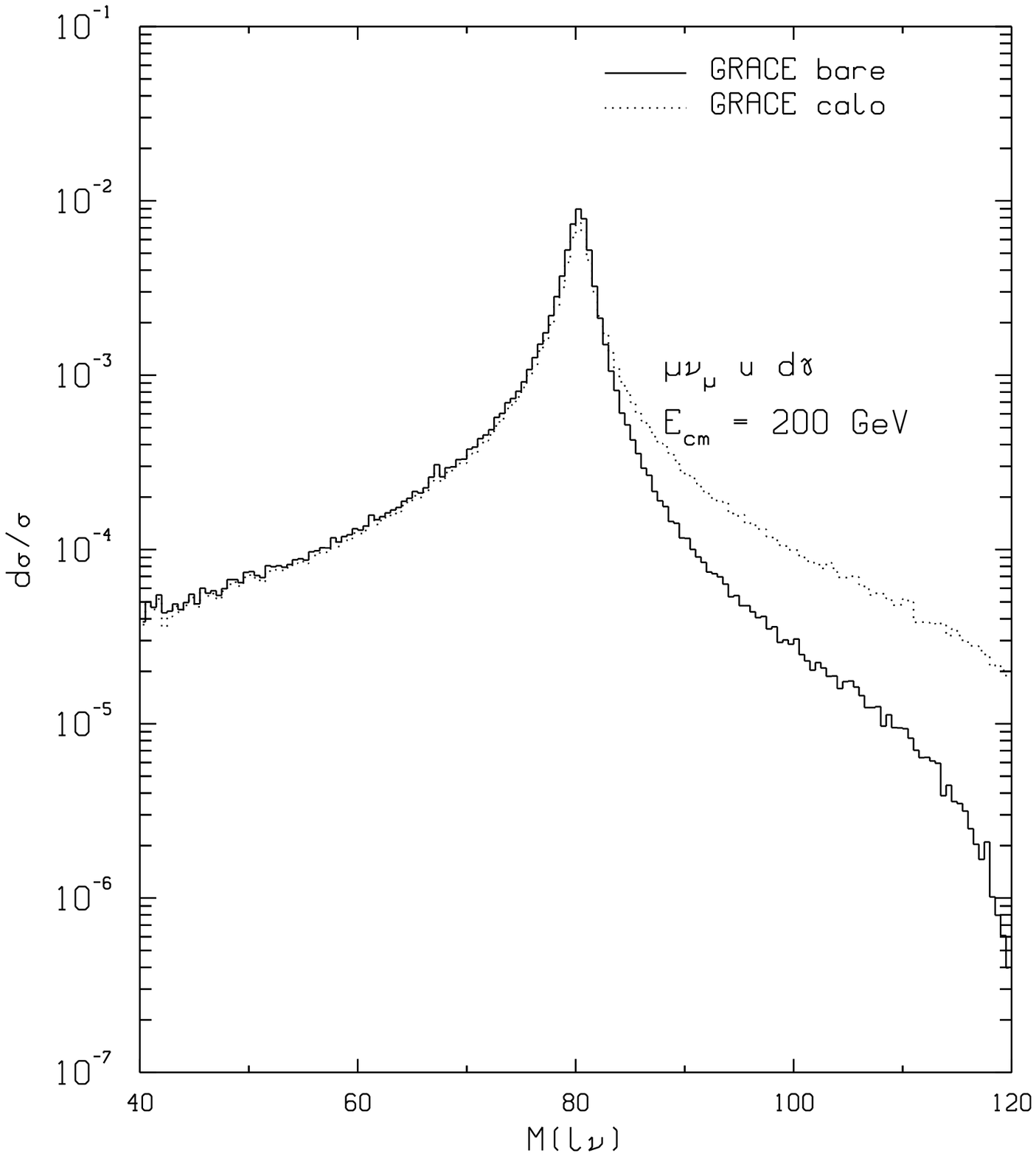,width=0.49\linewidth}
\vskip -1cm
\caption[]{Bare and calo $M(ud)$ distributions for the 
process $\mu \fnum u d \gamma$ from {\tt GRACE} with $\wb\wb$-cuts.} 
\label{yr_fig_g34}
\efi

\clearpage

\bfi
\epsfig{file=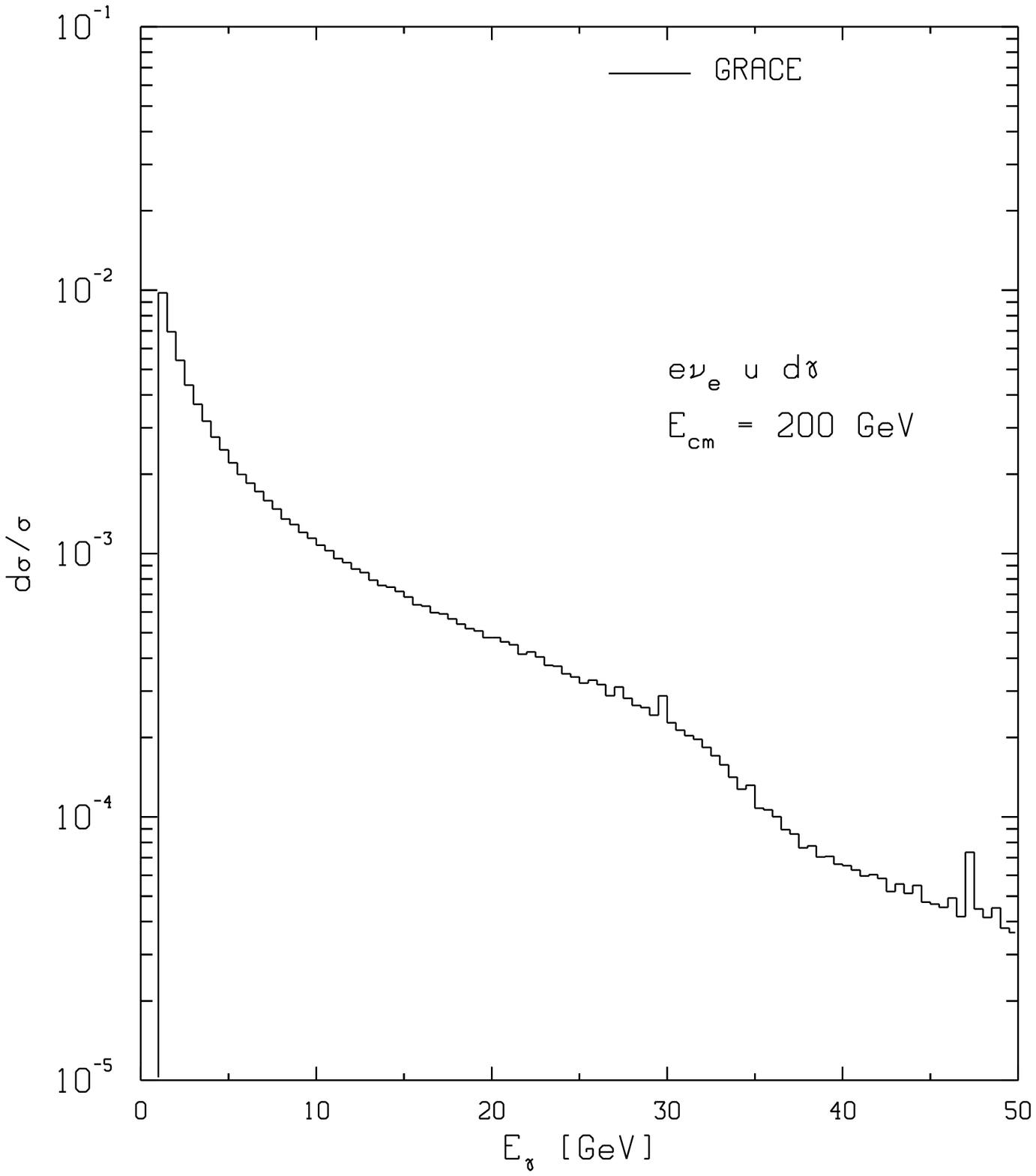,width=0.49\linewidth}
\hfill
\epsfig{file=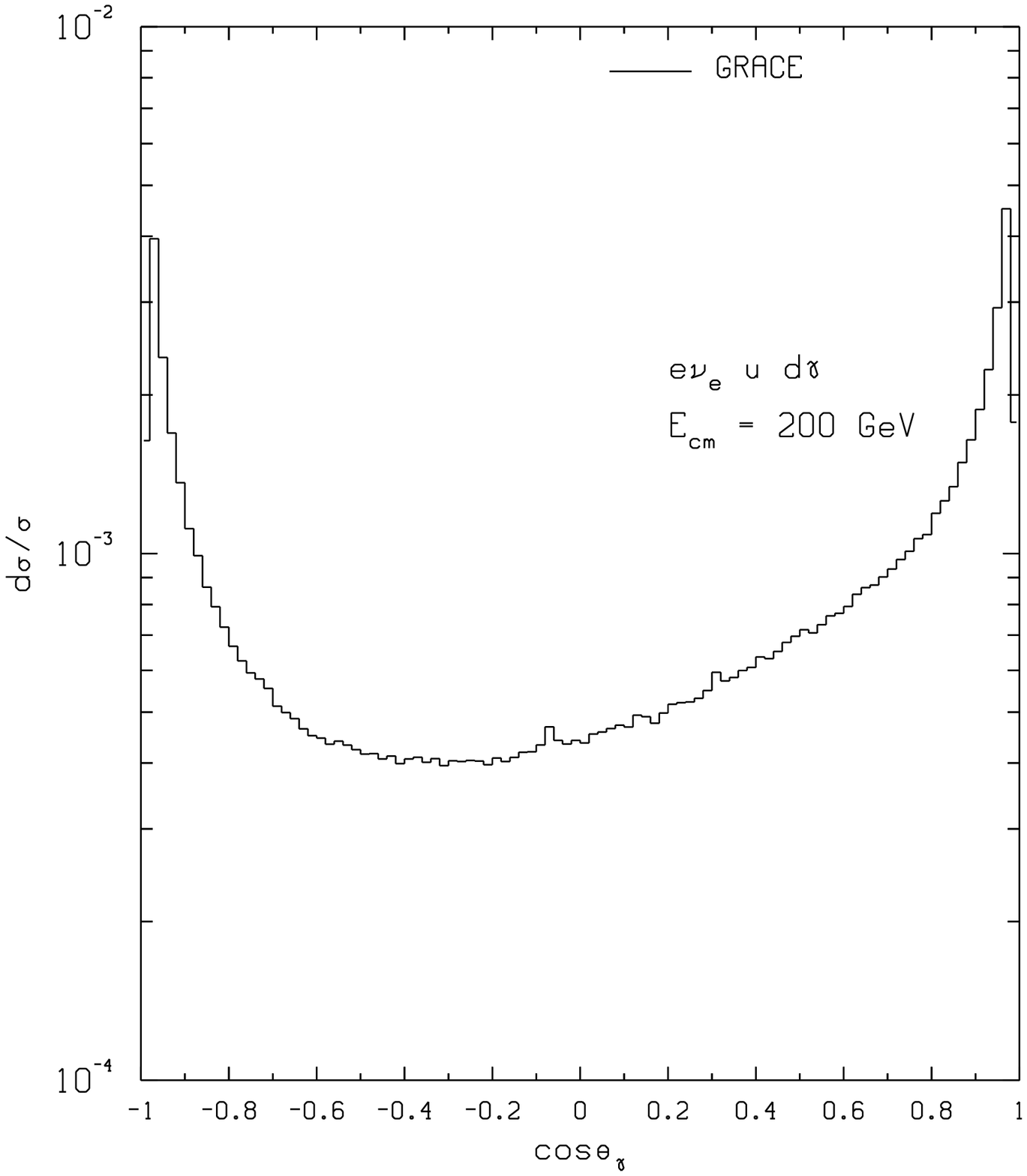,width=0.49\linewidth}
\vskip -1cm
\caption[]{$E_{\gamma}$ and $\cos \theta_{\ph}$ distributions for the 
process $e \nu_e u d \gamma$ from {\tt GRACE} with $\wb\wb$-cuts.} 
\label{yr_fig_g12e}
\efi
\bfi
\epsfig{file=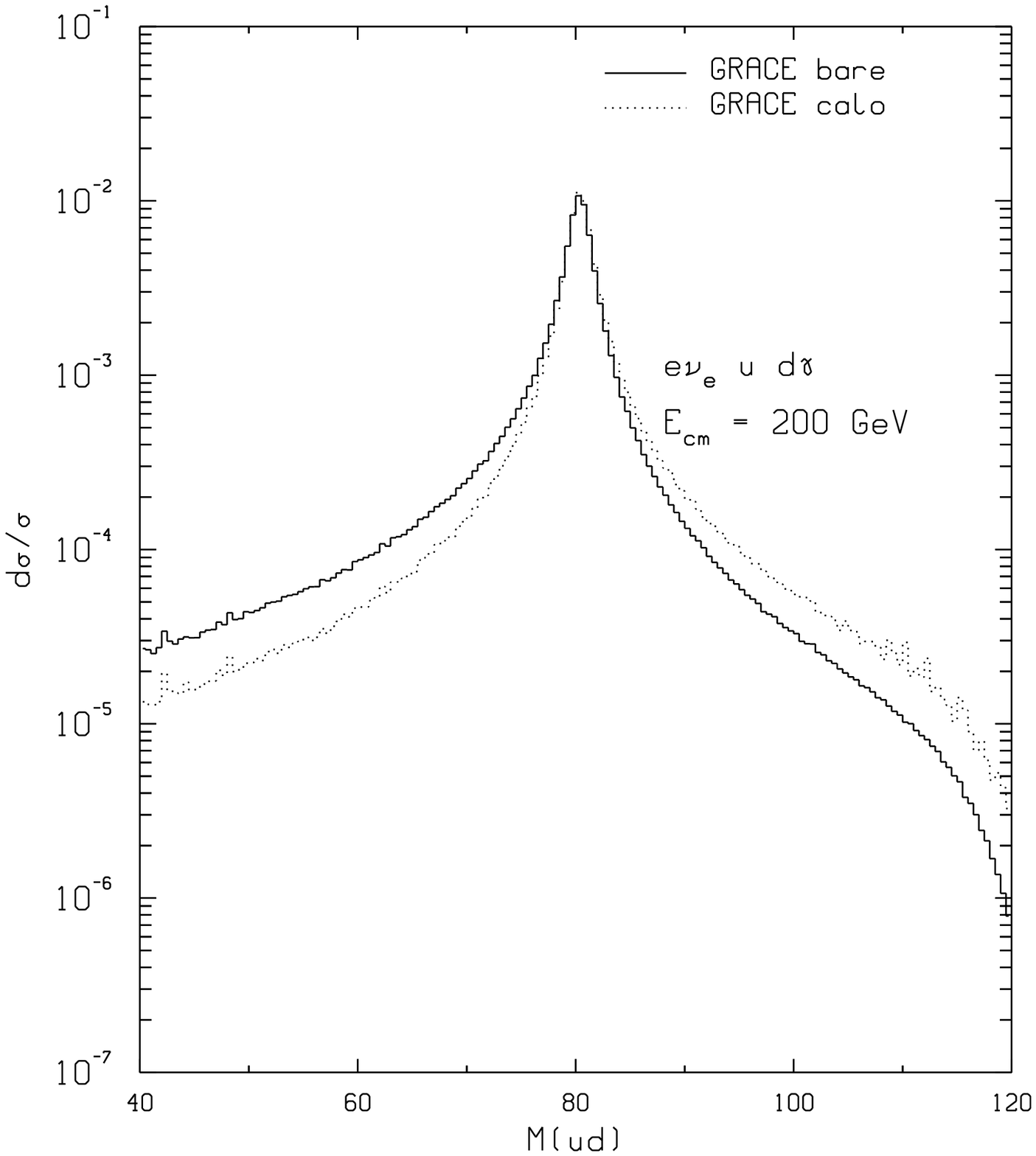,width=0.49\linewidth}
\hfill
\epsfig{file=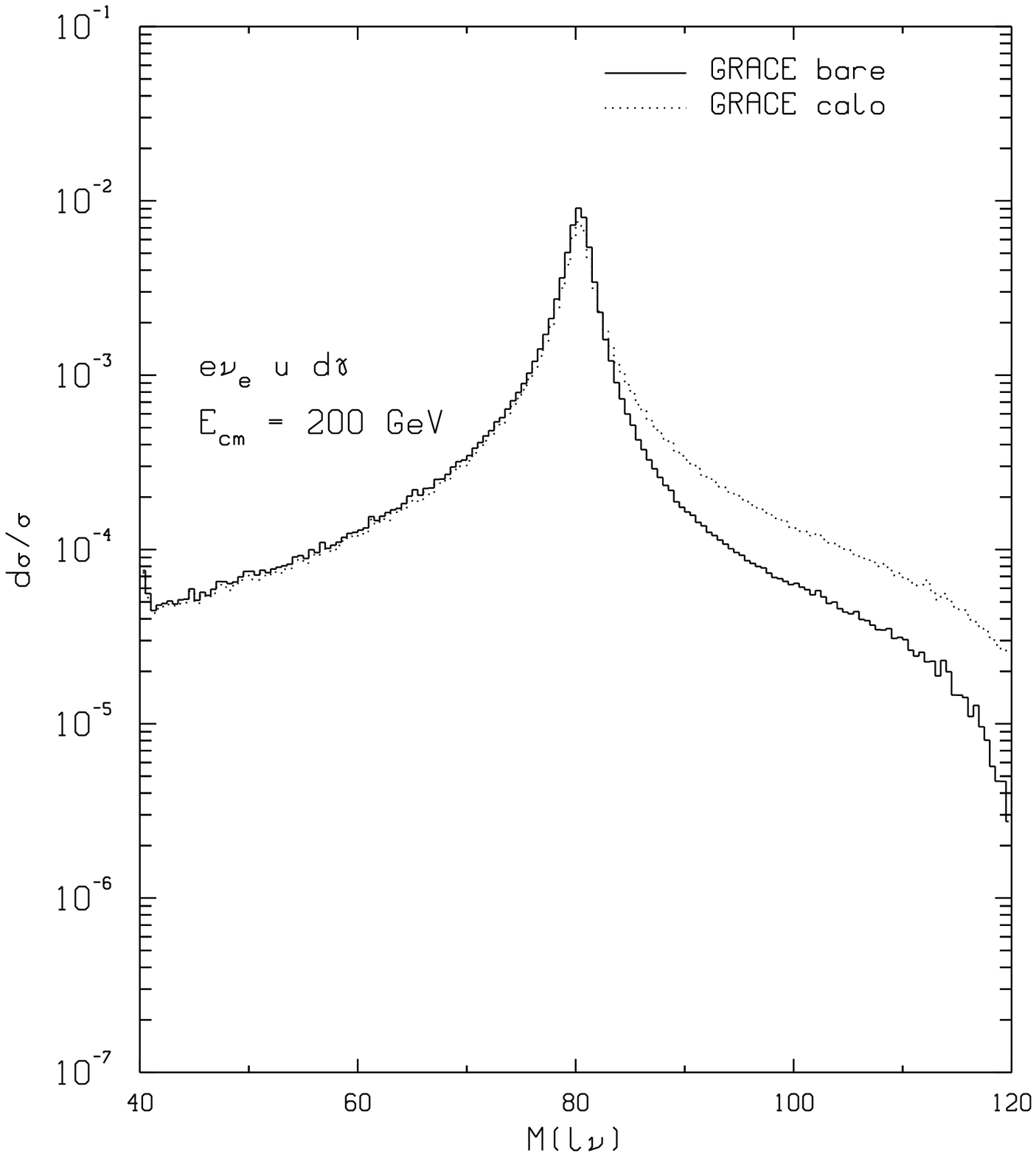,width=0.49\linewidth}
\vskip -1cm
\caption[]{Bare and calo $M(ud)$ distributions for the 
process $e \nu_e u d \gamma$ from {\tt GRACE} with $\wb\wb$-cuts.} 
\label{yr_fig_g34e}
\efi

\clearpage

\bfi
\epsfig{file=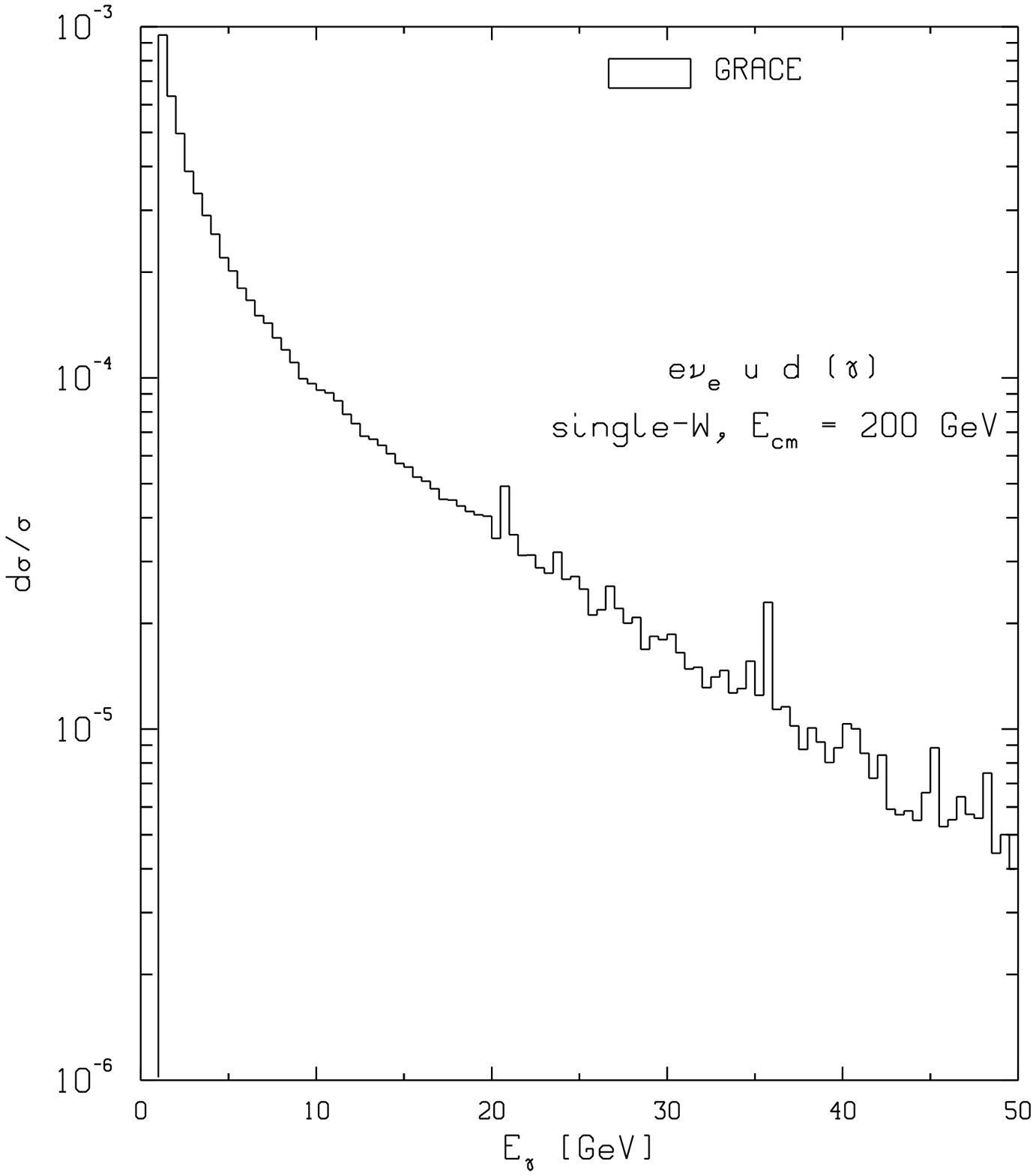,width=0.49\linewidth}
\hfill
\epsfig{file=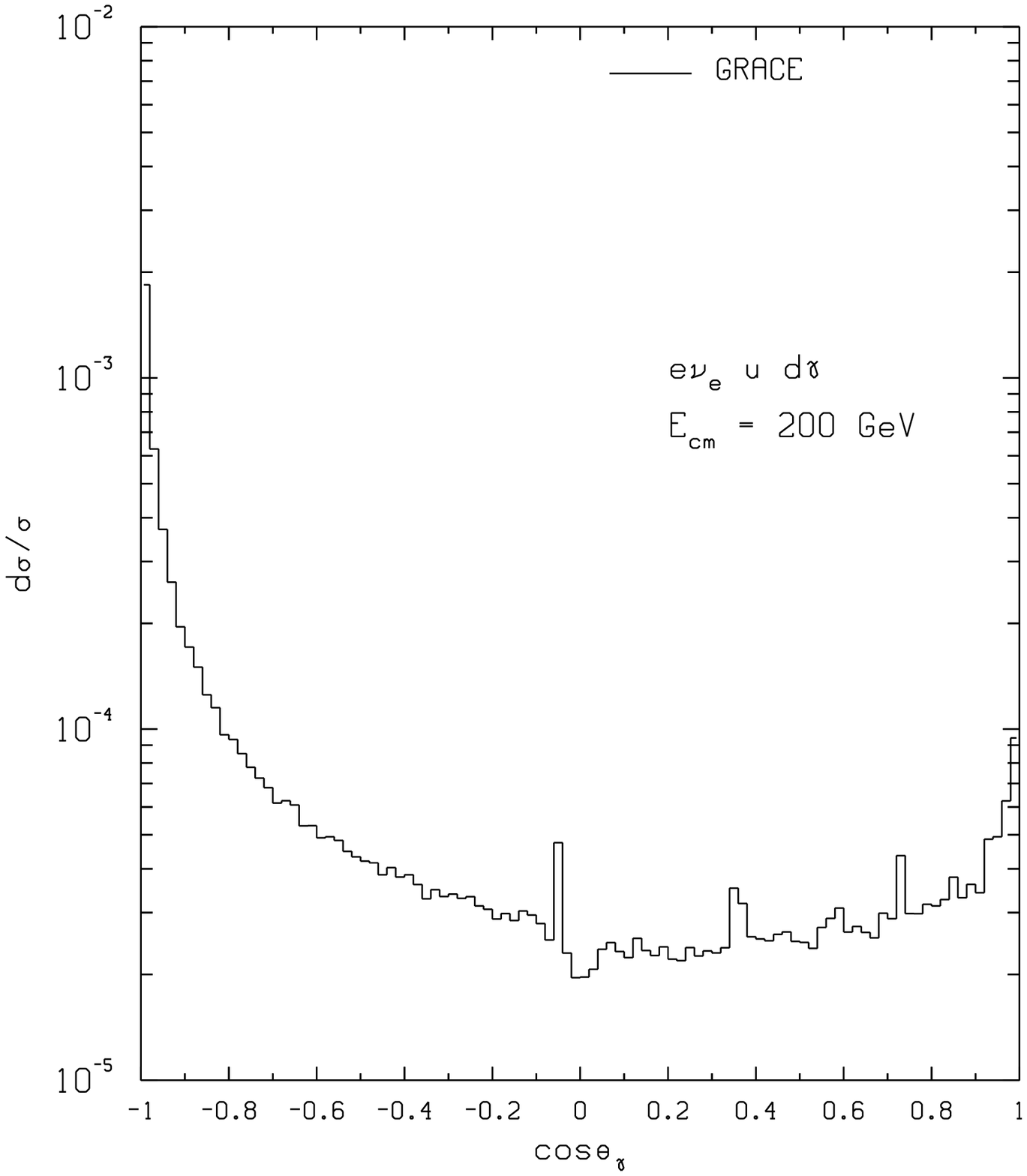,width=0.49\linewidth}
\vskip -1cm
\caption[]{$E_{\gamma}$ and $\cos \theta_{\ph}$ distributions for the 
process $e \nu_e u d \gamma$ from {\tt GRACE} with single-$\wb$ cuts.} 
\label{yr_fig_g12sw}
\efi
\bfi
\epsfig{file=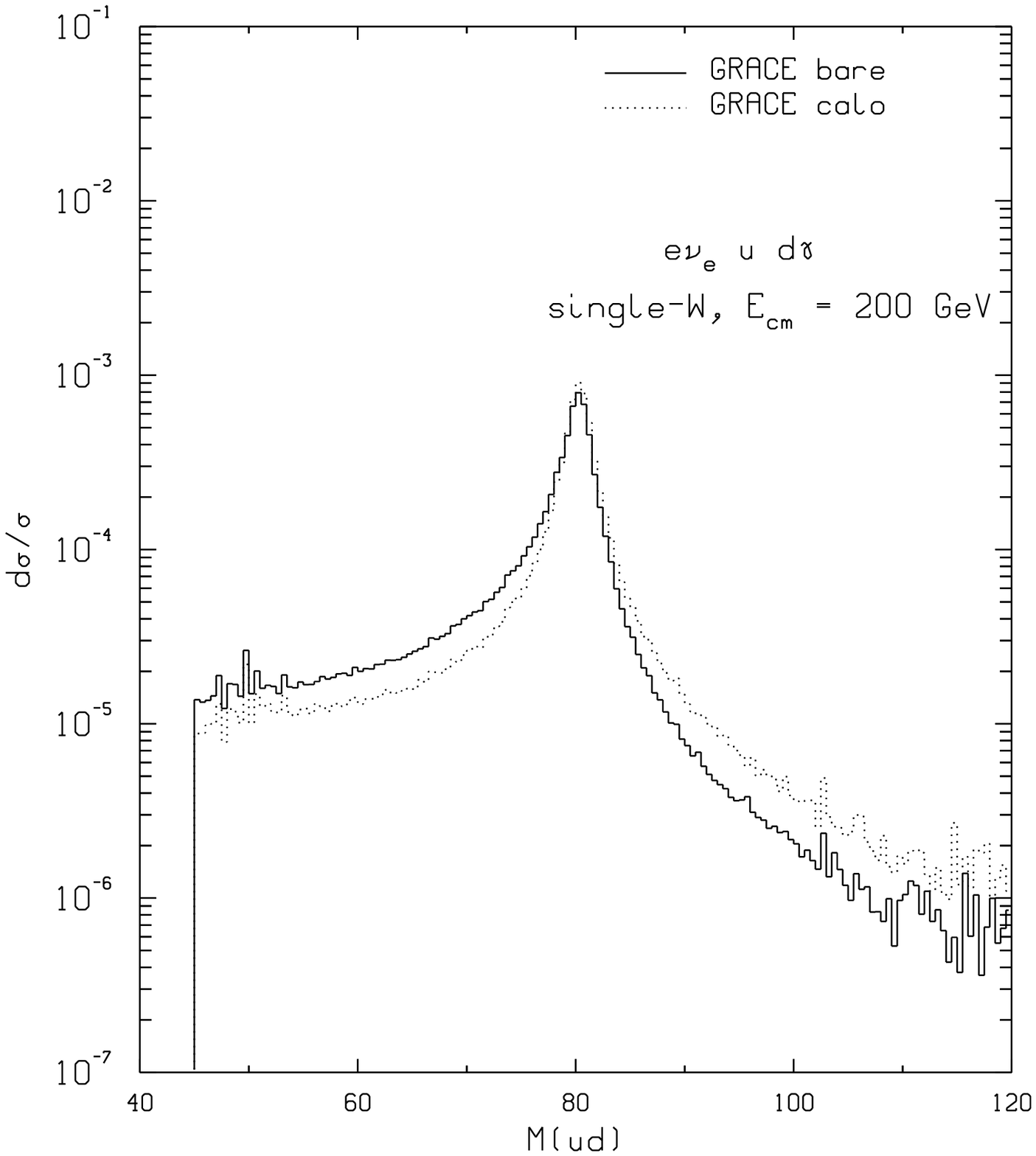,width=0.49\linewidth}
\hfill
\epsfig{file=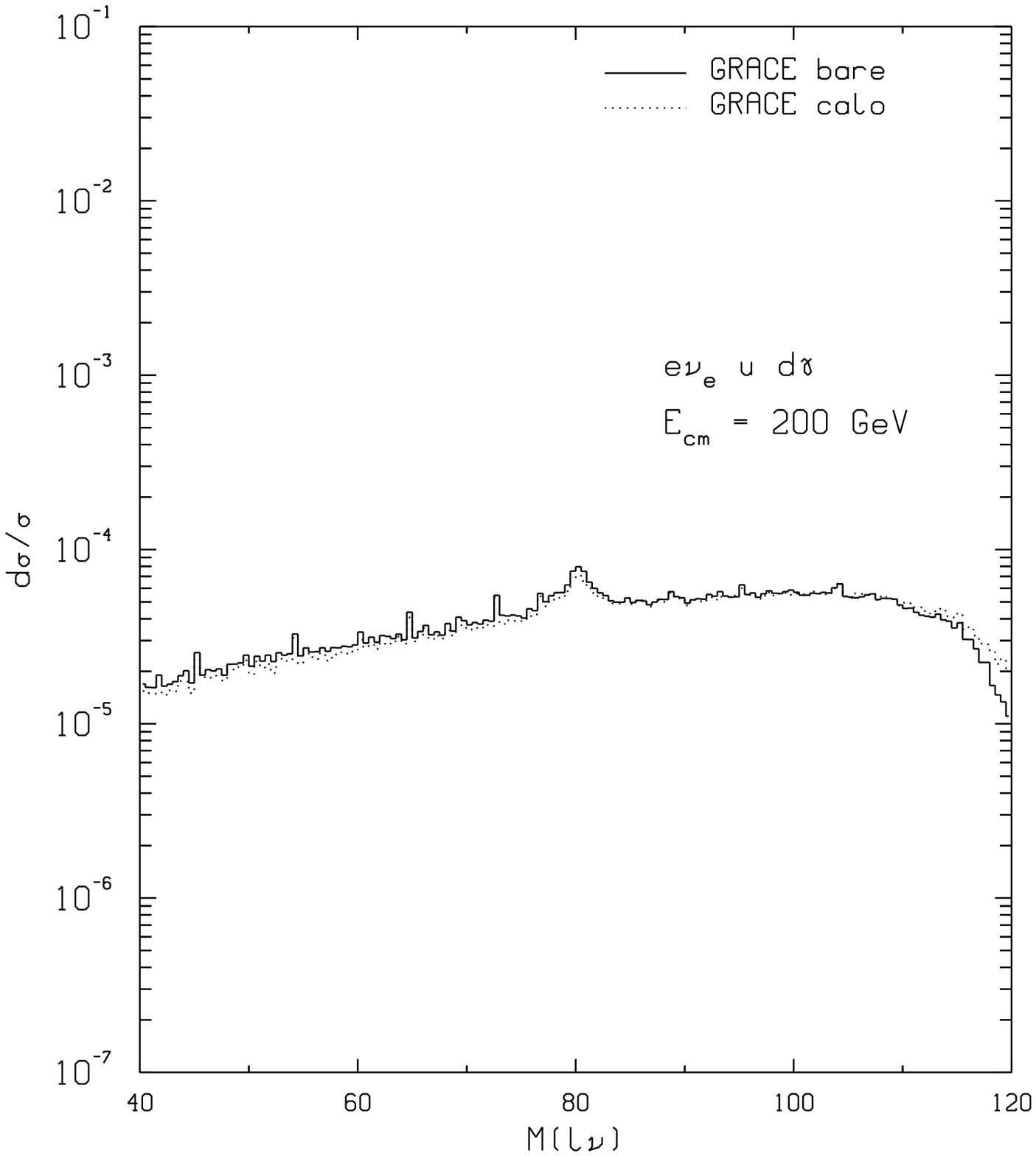,width=0.49\linewidth}
\vskip -1cm
\caption[]{Bare and calo $M(ud)$ distributions for the 
process $e \nu_e u d \gamma$ from {\tt GRACE} with single-$\wb$ cuts.} 
\label{yr_fig_g34sw}
\efi

\clearpage

\subsection{Comparisons for $4\rmf+\ph$}
\label{ac4fg}

A first set of comparisons between the predictions 
of several independent codes, namely {\tt WRAP},
{\tt CompHEP, GRACE} and {\tt RacoonWW} was performed at the beginning of
the workshop.

This comparison covers integrated cross-sections and 
various differential distributions, essentially for 
a CC10 final state. Discrepancies observed at that 
stage are mainly to be ascribed to non-tuned comparisons.
In fact, a detailed tuned comparison between {\tt WRAP},
{\tt RacoonWW} and {\tt PHEGAS/HELAC} presently shows a beautiful agreement 
for several distributions and final states.

Input parameters and cuts used to carry out this tuned comparison 
correspond to those of the $4\rmf$ proposal (in the 
approximation of massless fermions).
In particular, the photon cuts are: $E_{\ph}^{\rm min} = 1$~GeV, 
$|\cos\theta_{\gamma}| <$~0.985, at $\sqrt{s} = 200$~GeV.
In the {\tt PHEGAS} -- {\tt RacoonWW} -- {\tt WRAP} comparison, 
the following final states have been considered: 

\begin{itemize}
\item[--] $\mu \; \barnu_{\mu}\; u\; \bard\; \gamma$
\item[--] $e^- \;\barnu_e\; u\; \bard\; \gamma$
\item[--] $\mu\; \barnu_{\mu}\; \tau^+\; \nu_\tau\; \gamma$
\item[--] $e^-\; \barnu_e\; \tau^+\; \nu_\tau\; \gamma$
\item[--] $s\; {\bar c}\; u\; \bard\; \gamma$
\end{itemize}

The observables studied in the tuned comparison are:

\begin{itemize}
\item[--] integrated cross-sections;
\item[--] $E_{\ph}$ distribution, $d\sigma/dE_{\ph}$ [fb/GeV];
\item[--] distribution in the cosine of the photon angle $\theta_{\ph}$, $d\sigma/d\cos\theta_{\ph}$ [fb];
\item[--] distribution in the opening angle $\theta_{f\gamma}$ between
  the photon and the nearest charged final-state fermion,
  $d\sigma/d\theta_{\ph f}$ [fb];
\item[--] distributions in the bare invariant masses of the 
  $\wb^+$ and $\wb^-$ bosons,
  $M_+ = M_{u \bard}, M_{\tau^+ \nu_\tau}$, $d\sigma/dM_+$ [fb/GeV];
  $M_- = M_{s \barc}, M_{\mu^- \bar\nu_\mu}$, $M_{e^- \bar\nu_e}$,
  $d\sigma/dM_-$ [fb/GeV].
\end{itemize}
All the observables are calculated for $\sqrt{s} = 200\,$GeV 
in the fixed width scheme.
The squared matrix element is 
calculated in the $\gf$ scheme and subsequently multiplied by
$\alpha(0)/\alpha_{\gf}$,
to take exactly into account of the scale of the real photon.

The applied cuts are: 
\begin{itemize}
\item[--] common to all processes: $E_{\ph} > 1\,$GeV,
$|\cos(\theta(\ph,{\rm beam})| < 0.985$,
$\theta(\ph, \rmf) > 5^\circ$, f = charged fermion.

\item[--] for $u d \mu \fnum\gamma$ and $u d e \nu_e\gamma$:
$M(u d) > 10\,$GeV,
$|\cos\theta({\rm l,beam})| < 0.985$
$E_{\rm l} > 5\,$GeV, where l is a charged lepton,

\item[--] for $\tau \fnut \mu \fnum\gamma$ and 
$\tau \fnut e \nu_e\gamma$:
$|\cos\theta({\rm l,beam})| < 0.985$,
$E_{\rm l} > 5\,$GeV,
$M(l^+l^-) > 10\,$GeV,

\item[--] for $u d c s\gamma$:
at least two pairs with $M(q_i q_j) > 10\,$GeV.

\end{itemize}
The generators have produced a huge collection of results and only a small 
sample will be shown here.

The total cross-sections are reported in  Table~\ref{comp4fg_1} 
where the differences between the predictions of
{\tt WRAP}, {\tt RacoonWW} and {\tt PHEGAS/HELAC} are around $0.1\%$,
signalling perfect technical agreement.

In the following we will show few example of predictions.  By
comparing the three different codes with a {\em tuned} comparison we
get a rough estimate of the associated technical uncertainty also for
distributions.  Besides the distributions compared in plots we also be
present ratio-plots, as the distributions themselves are too close to
show a difference between programs in the actual scale.

First we consider the angular distribution, i.e. the
$\cos\theta_{\ph}$ distribution in the range $[-1, 1]$ for various
final states, as shown in Figures~\ref{yr_4fg_12} to~\ref{yr_4fg_34r},
where we also plot the ratios bewteen the predictions.

\clearpage

\begin{table}[p]\centering
\begin{tabular}{|c|c|c|c|}
\hline
Process & {\tt WRAP} & {\tt RacoonWW} & {\tt PHEGAS/HELAC} \\
\hline
&&&\\
$u \bard \mu^- \barnu_{\mu}\gamma$      
& 75.732(22) & 75.647(44) & 75.683(66)\\
$u \bard e^- \barnu_e\gamma$             
& 78.249(43) & 78.224(47) & 78.186(76)\\
$\fnum \mu^+ \tau^- \barnu_{\tau}\gamma$ 
& 28.263(9)  & 28.266(17) & 28.296(22) \\
$\fnum \mu^+ e^- \barnu_e\gamma$     
& 29.304(19) & 29.276(17) & 29.309(25)\\
$u \bard s \barc\gamma$                  
&199.63(10)  &199.60(11) & 199.75(16) \\
&&&\\
\hline
\end{tabular}
\vspace*{3mm}
\caption[]{Comparison between {\tt WRAP}, {\tt RacoonWW} and {\tt PHEGAS/HELAC}
for a sample of total cross-sections (fb).\label{comp4fg_1}}
\label{totxs4fg}
\end{table}
\normalsize

\begin{figure}[p]
\epsfig{file=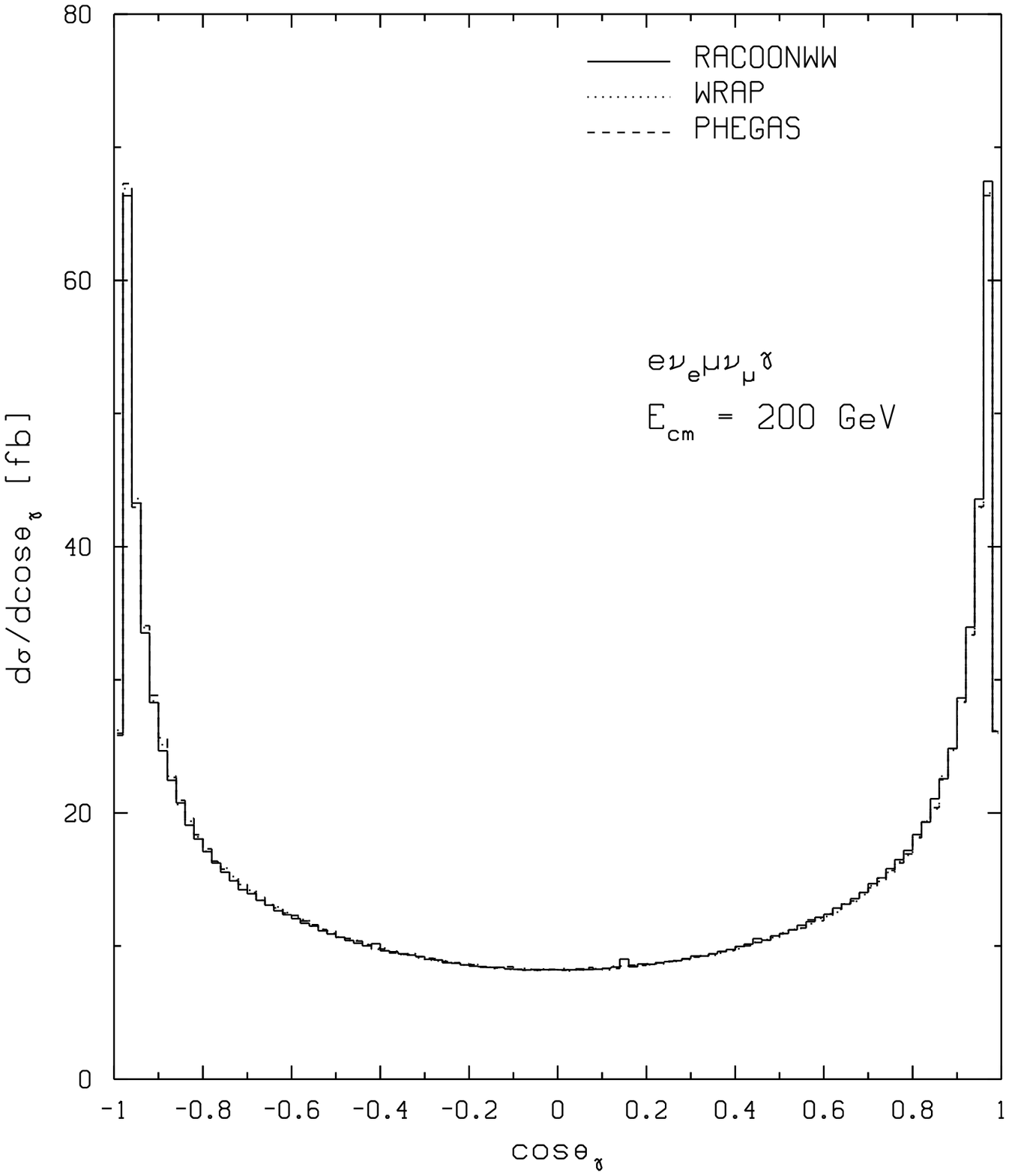,width=0.44\linewidth}
\hfill
\epsfig{file=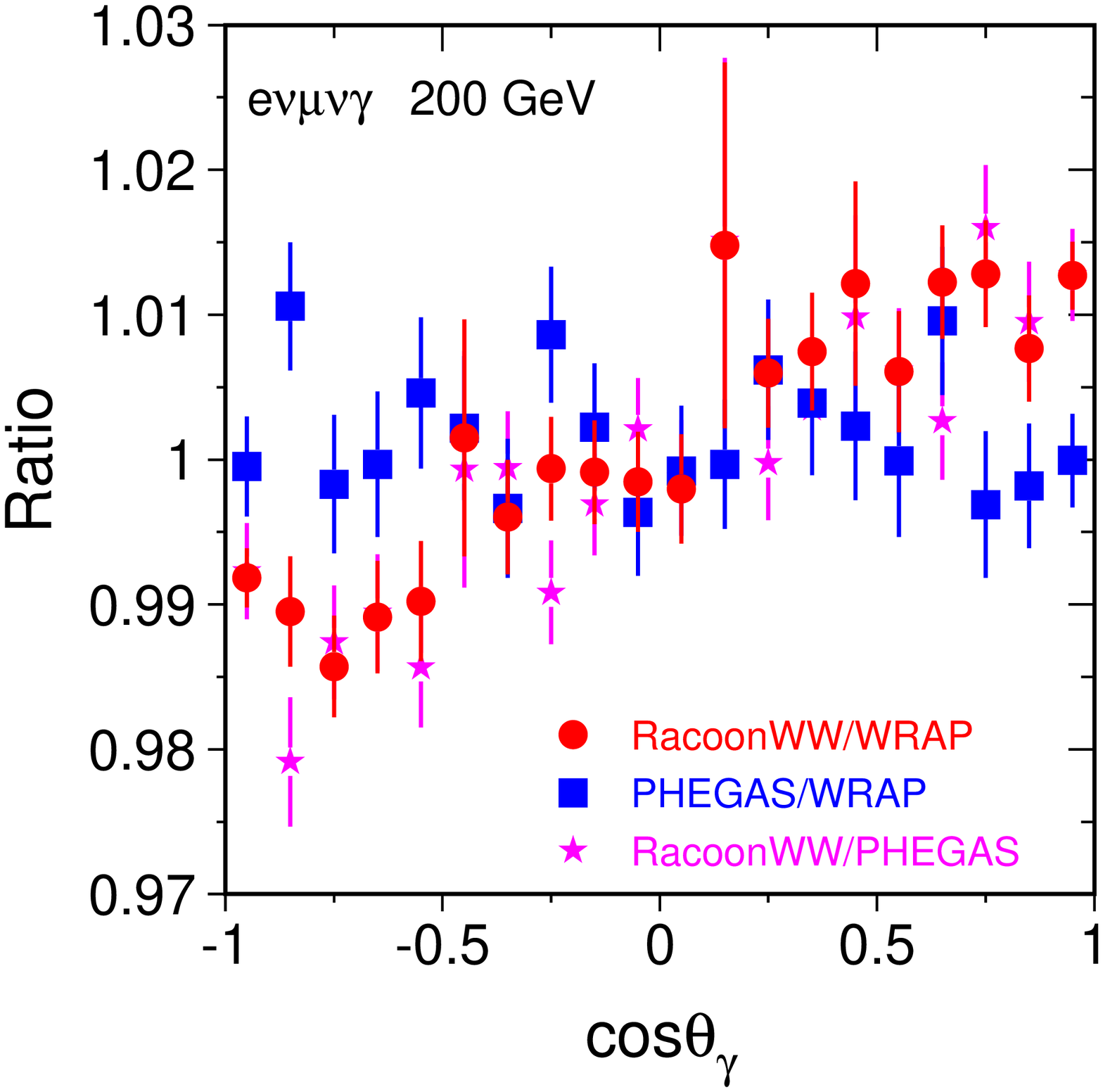,width=0.54\linewidth}
\caption[]{$\cos \theta_{\ph}$ distributions and ratios for the
  processes $\fnum \mu^+ e^- \barnu_e\gamma$.}
\label{yr_4fg_12}
\efi

\clearpage

\begin{figure}[p]
\epsfig{file=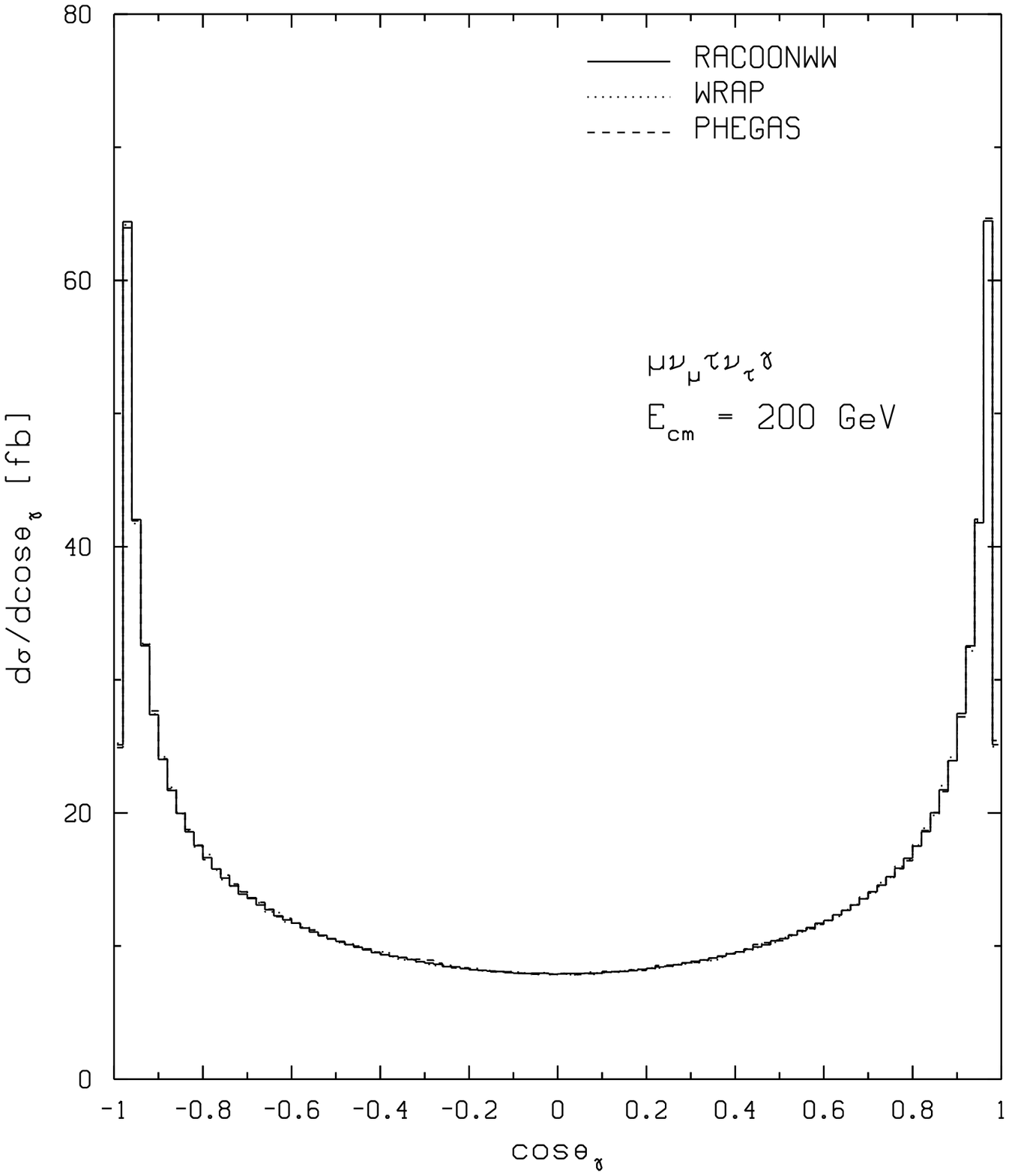,width=0.44\linewidth}
\hfill
\epsfig{file=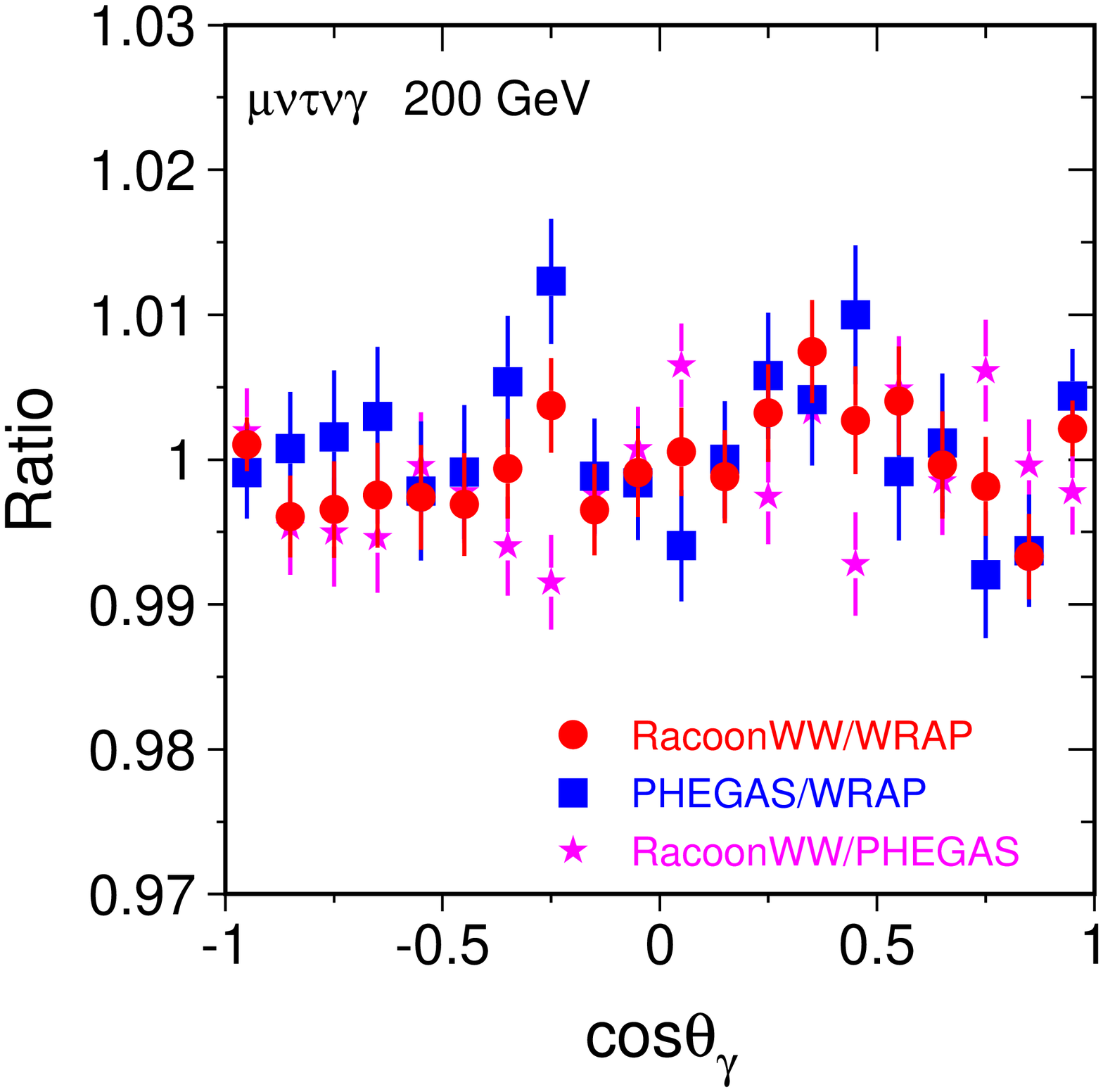,width=0.54\linewidth}
\caption[]{$\cos \theta_{\ph}$ distributions and ratios for the
  processes $\fnum \mu^+ \tau^- \barnu_{\tau}\gamma$.}
\label{yr_4fg_12r}
\efi

\begin{figure}[p]
\epsfig{file=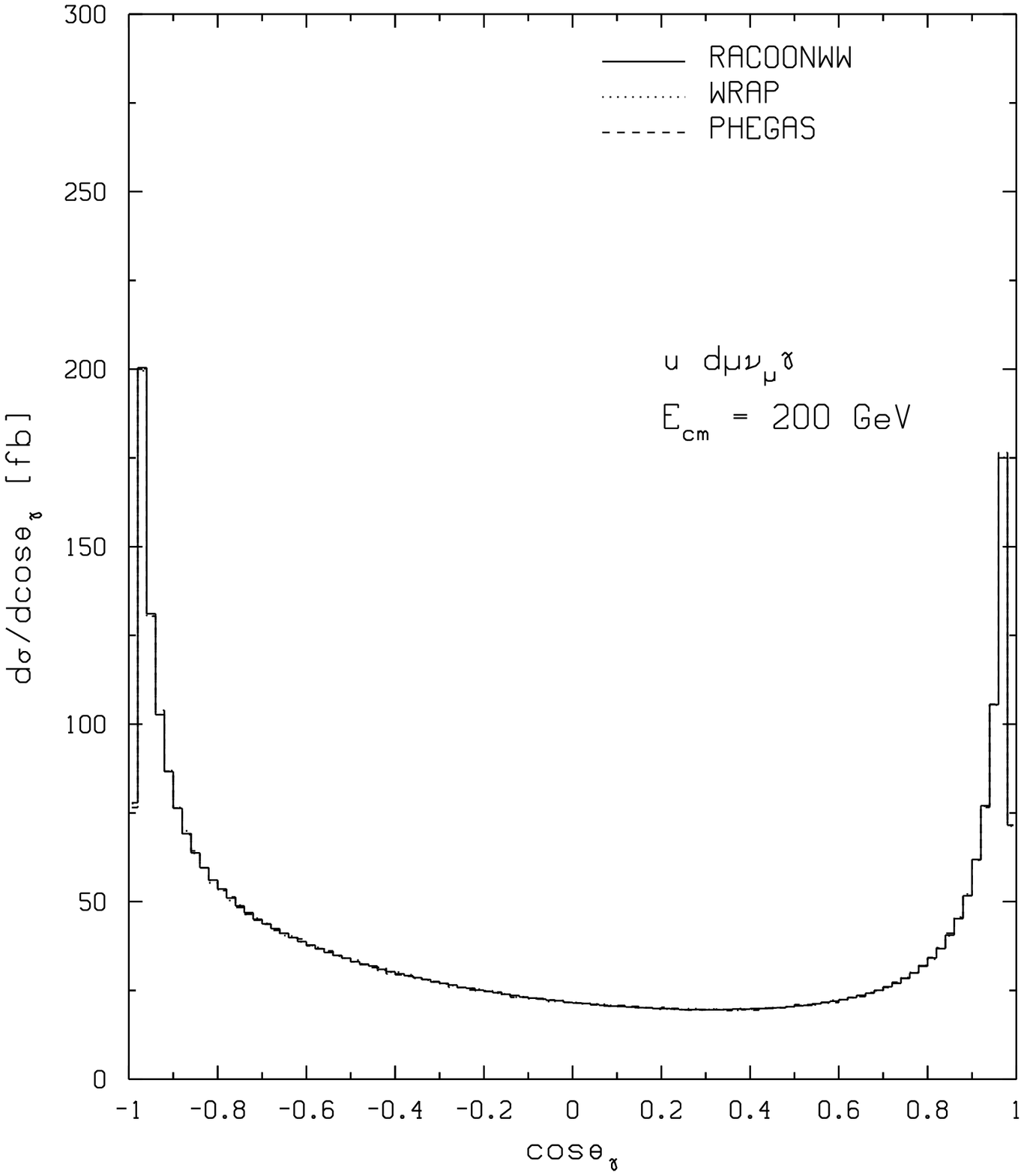,width=0.44\linewidth}
\hfill
\epsfig{file=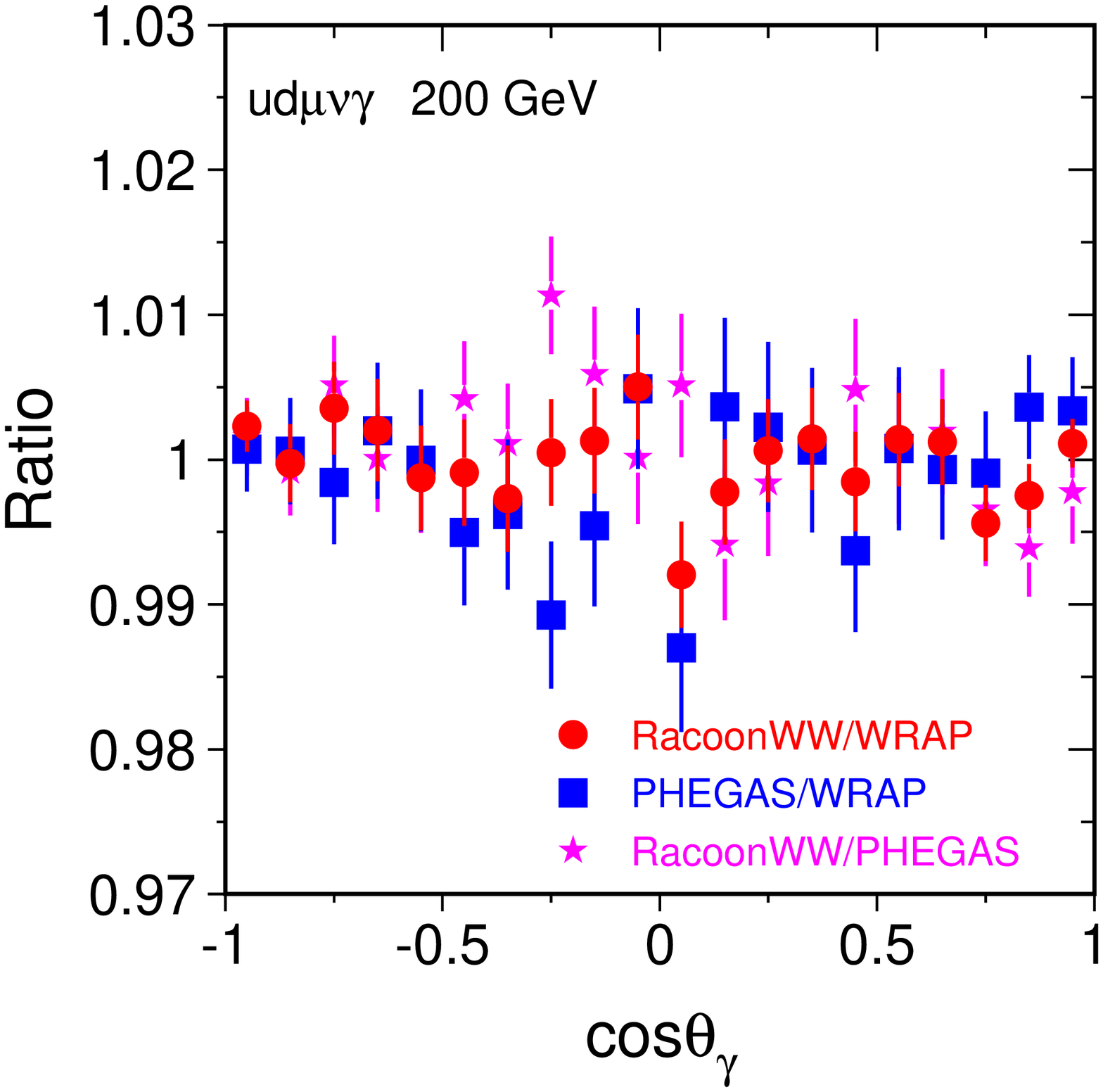,width=0.54\linewidth}
\caption[]{$\cos \theta_{\ph}$ distributions and ratios for the process 
$u \bard \mu^- \barnu_{\mu}\gamma$.}
\label{yr_4fg_5}
\efi

\clearpage

\begin{figure}[p]
\epsfig{file=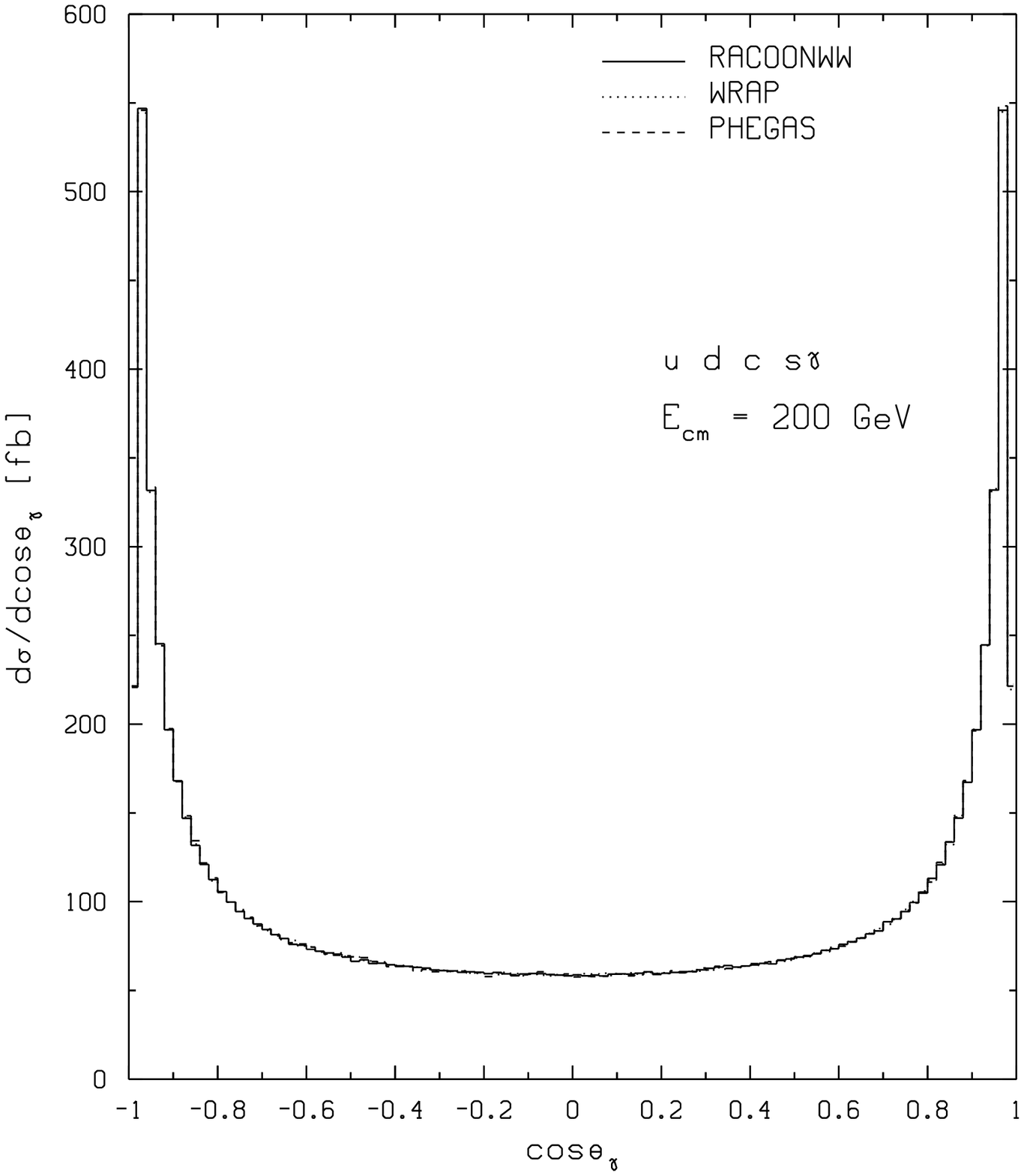,width=0.44\linewidth}
\hfill
\epsfig{file=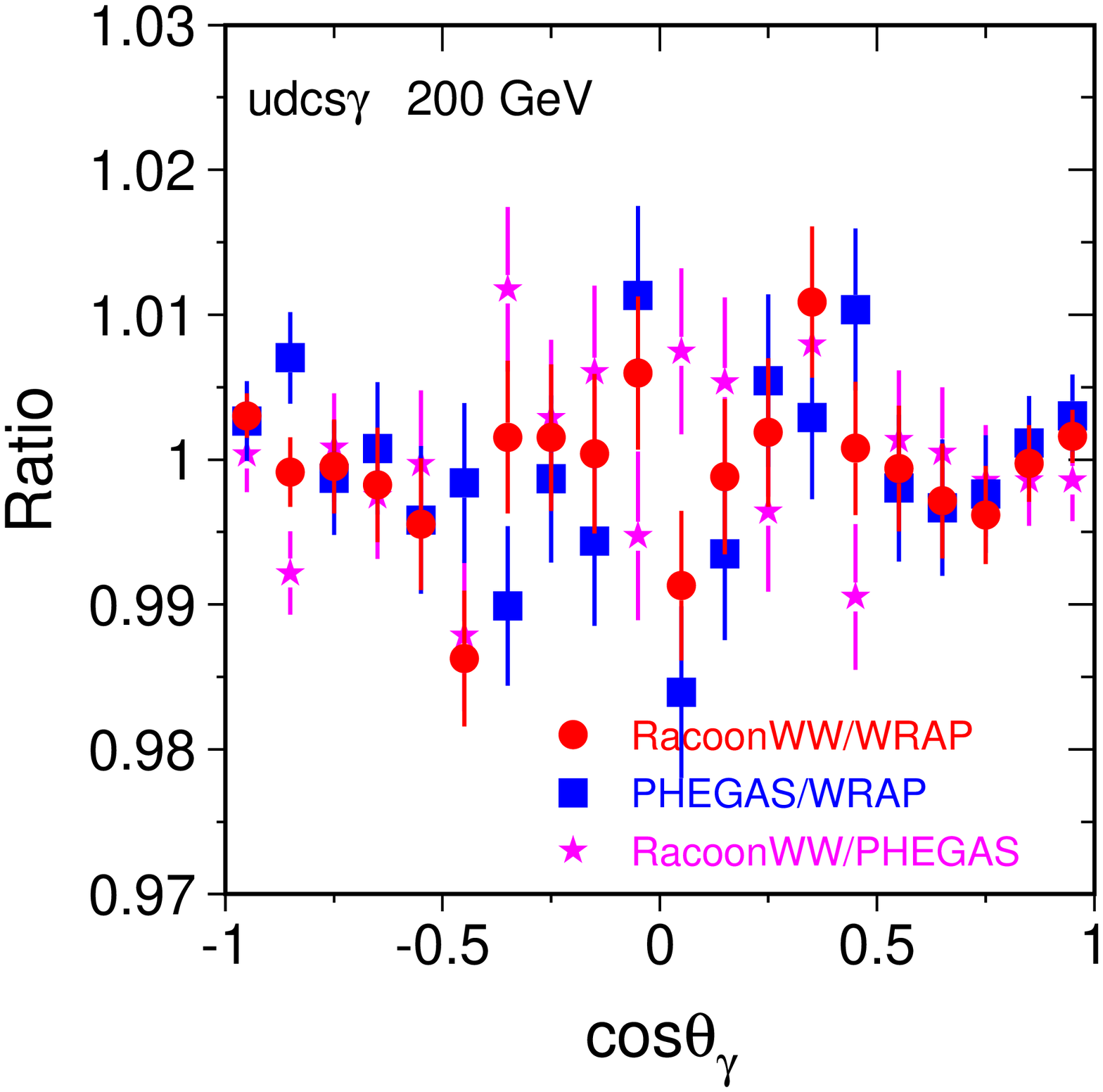,width=0.54\linewidth}
\caption{$\cos \theta_{\ph}$ distribution for the processes 
$u \bard s \barc\gamma$ and $u \bard e^- \barnu_e\gamma$.}
\label{yr_4fg_34}
\efi

\begin{figure}[ht]
\epsfig{file=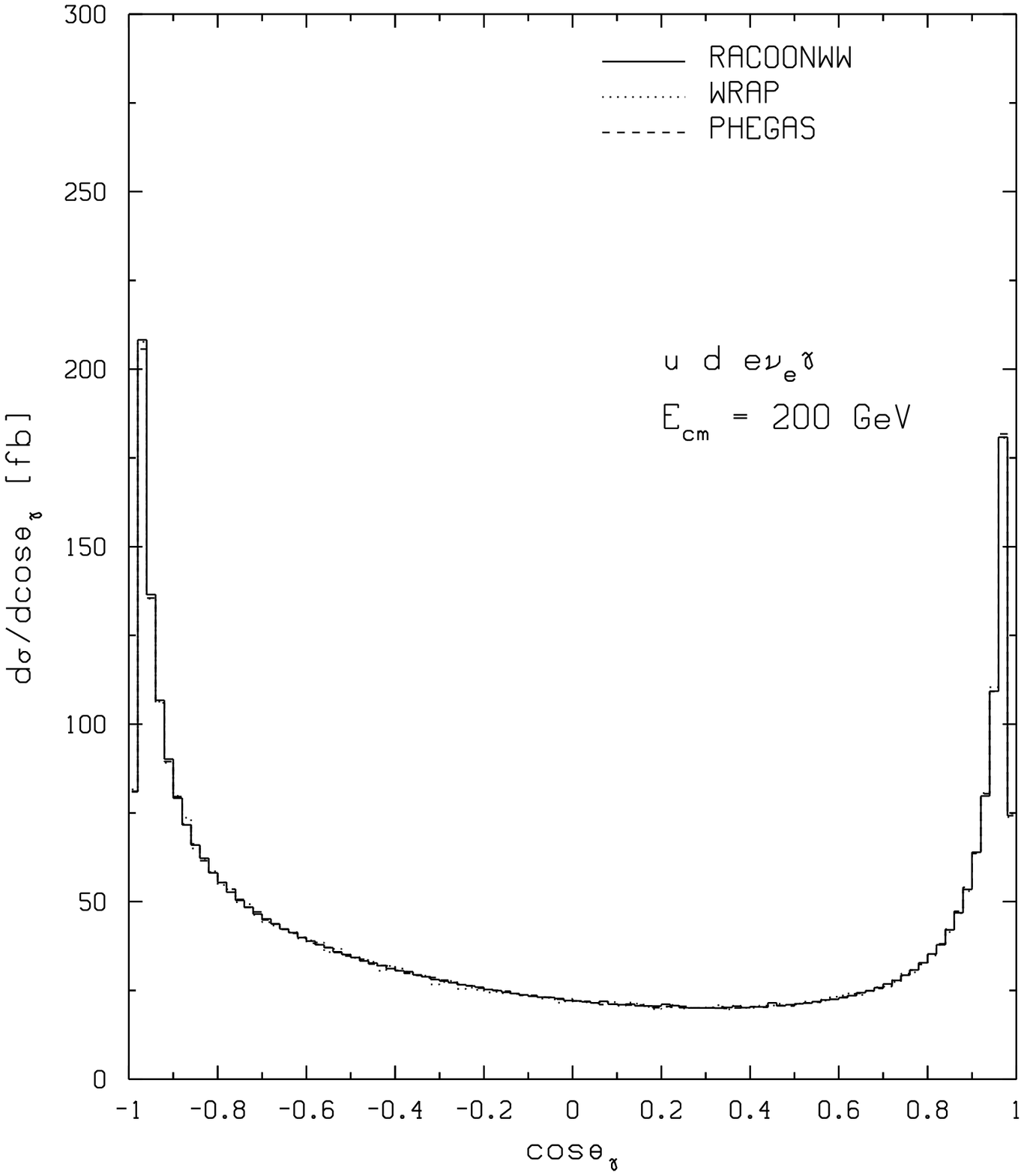,width=0.44\linewidth}
\hfill
\epsfig{file=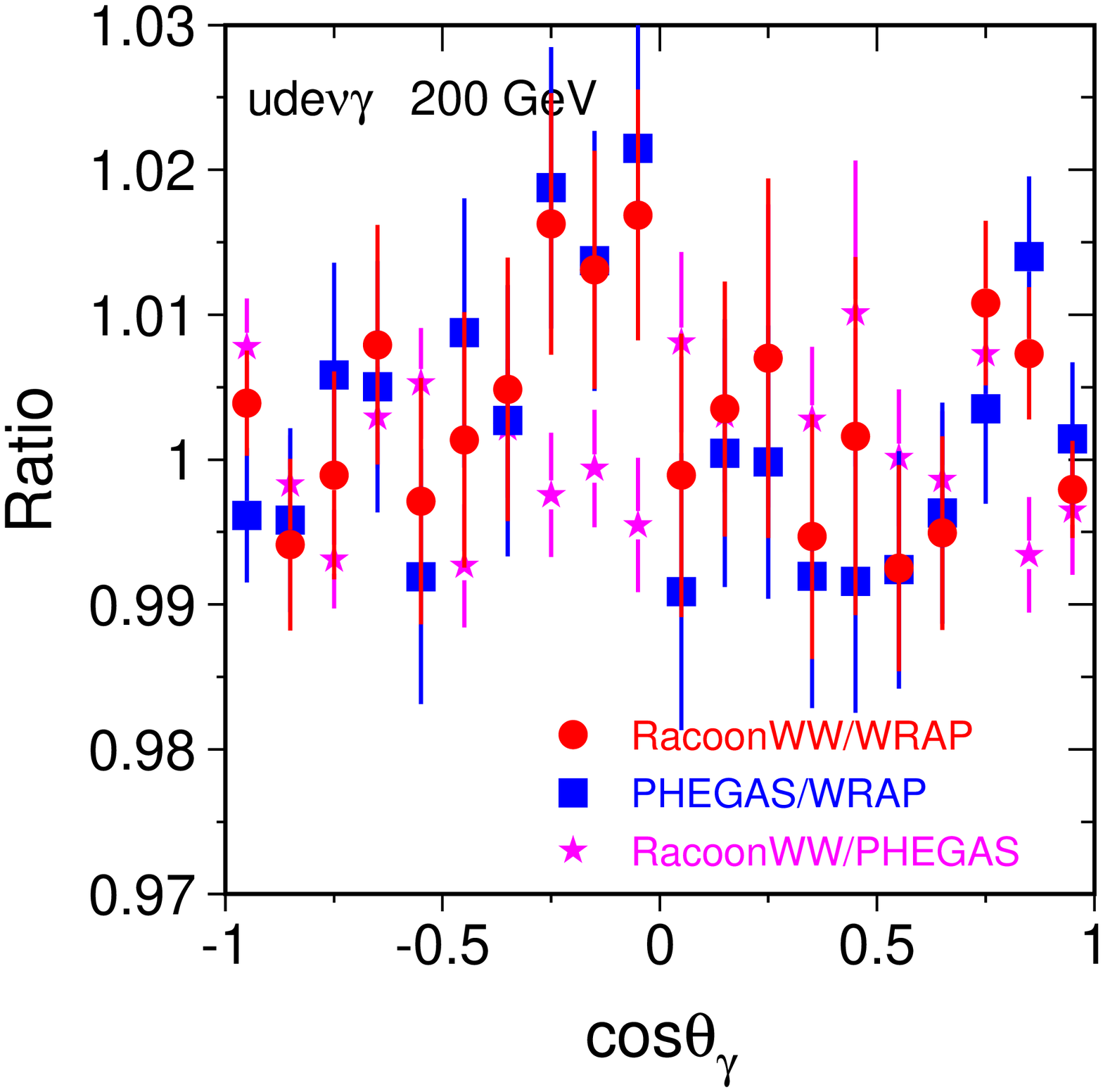,width=0.54\linewidth}
\caption[]{$\cos \theta_{\ph}$ ratios for the processes 
$u \bard s \barc\gamma$ and $u \bard e^- \barnu_e\gamma$.}
\label{yr_4fg_34r}
\efi

\clearpage

Similarly, the $E_{\ph}$ distributions and ratios in the range [1,
50]~GeV are shown for various processes in Figures~\ref{yr_4fg_67}
to~\ref{yr_4fg_89r}.  Note that virtual corrections are not included,
therefore, the photon spectrum starts at some lower boundary of
$1\,$GeV.  Deviations are of the order of $1\%$ for soft photons and
tend to deteriorate for harder ones.  Statistically the deviations are
compatible with zero.  Note that for very hard photons the cross
section and therefore the accuracy of the numerical integration of the
programs becomes poorer.

In \fig{yr_4fg_17} we show the fermion-photon opening angle
$\theta(\ph, \rmf)$ (where f is a charged fermion) distribution. In
the same figure we show the percentage deviation between the three
predictions. The most interesting region occurs for small angles, i.e.
towards the collinear region, where a reasonable agreement is
registered, of the order of a percent.  For the used statistics the
deviations are not yet significant.  The agreement deteriorates for a
larger separation between the photon and the charged particles.
However, in this region the cross-section is an order of magnitude
lower. Note that the peculiar behavior of the distribution towards
$0^\circ$ is only due to the fact that the third bin is between
$3.6^\circ$ and $5.4^\circ$ with a cut at $5^\circ$.

Finally, we compare the distributions in the bare invariant masses.
First the $\wbm$ one as predicted by {\tt WRAP} and {\tt RacoonWW}. Results 
for all considered channels and for the $\wbm$-distribution are shown in 
\fig{yr_4fg_11}(left). Note that the curves for the two purely leptonic 
channels and the two semi-leptonic final states are almost identical. In
\fig{yr_4fg_11} (right) we also present the percentage deviations for
the process $u \bard s \barc\gamma$.  
In \fig{yr_4fg_11b} we give the corresponding $\wbp$ invariant-mass 
distribution including results from the $3$ programs.
In \fig{yr_4fg_15cd} we show the ratio between the $\wbp$ and the $\wbm$
invariant-mass distributions from {\tt WRAP} and {\tt RacoonWW} respectively.

\begin{figure}[hp]
\epsfig{file=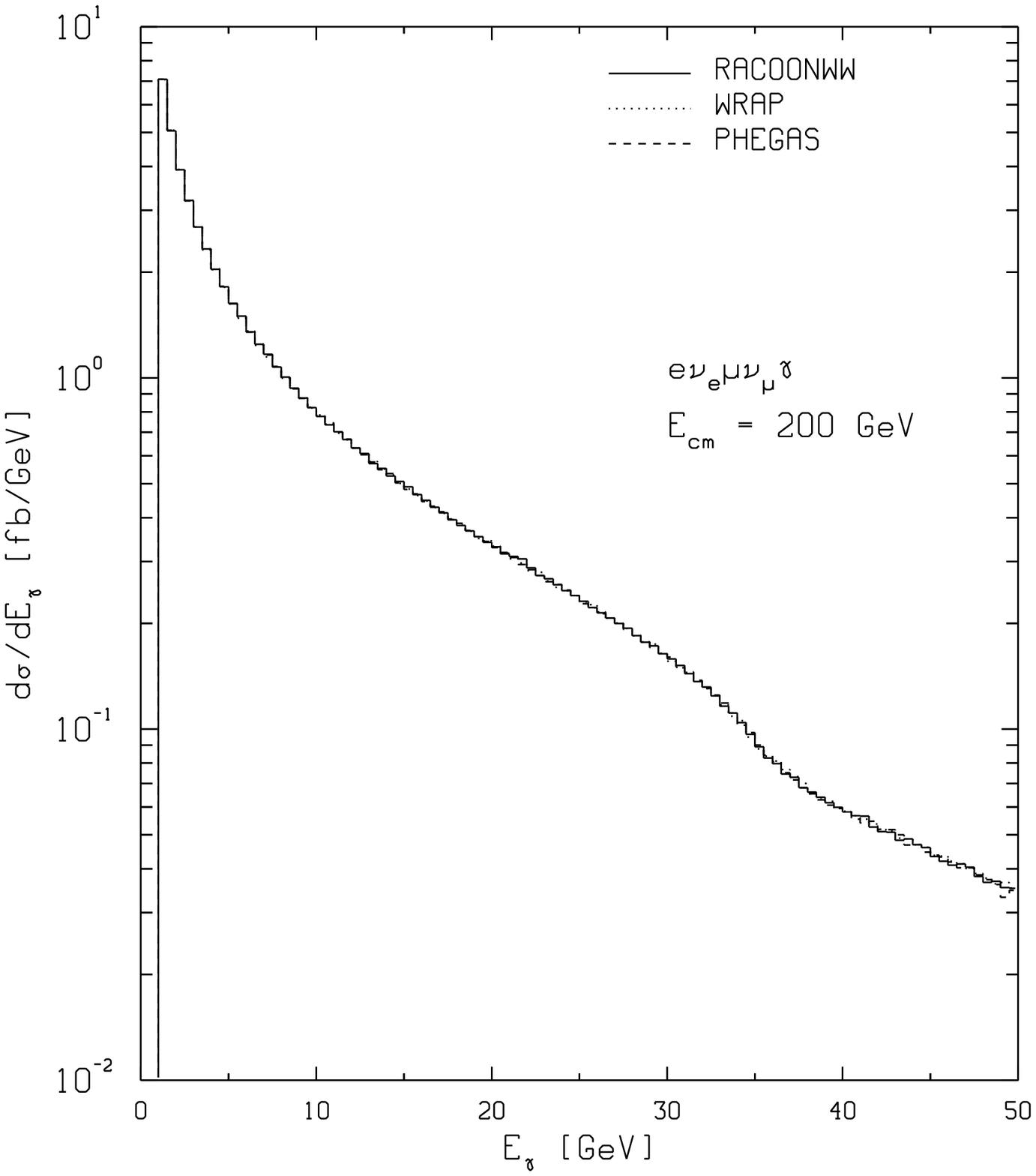,width=0.44\linewidth}
\hfill
\epsfig{file=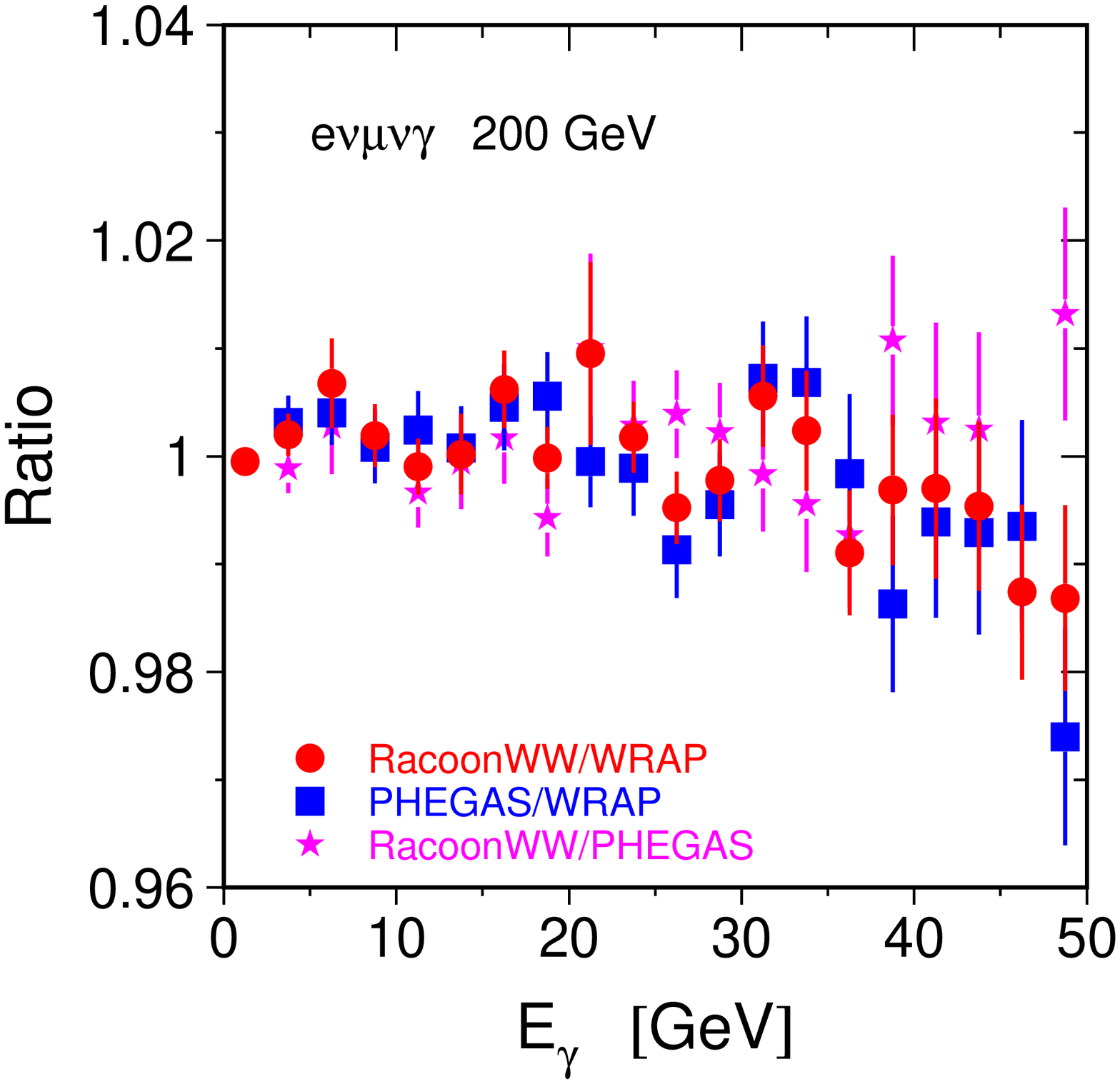,width=0.54\linewidth}
\caption[]{$E_{\ph}$ distributions and ratios for the process
 $\fnum \mu^+ e^- \barnu_e\gamma$.}
\label{yr_4fg_67}
\efi

\clearpage

\begin{figure}[p]
\epsfig{file=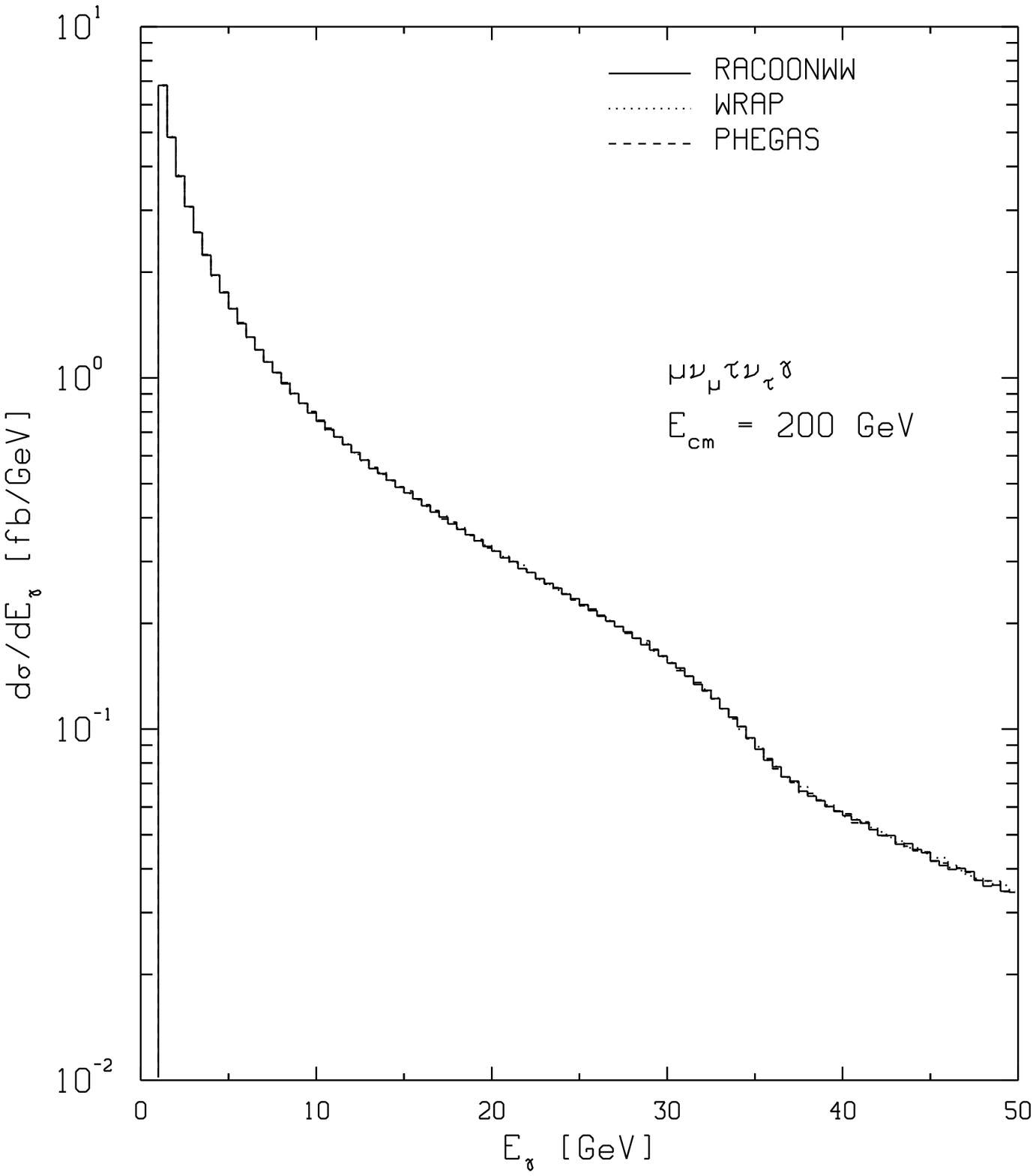,width=0.44\linewidth}
\hfill
\epsfig{file=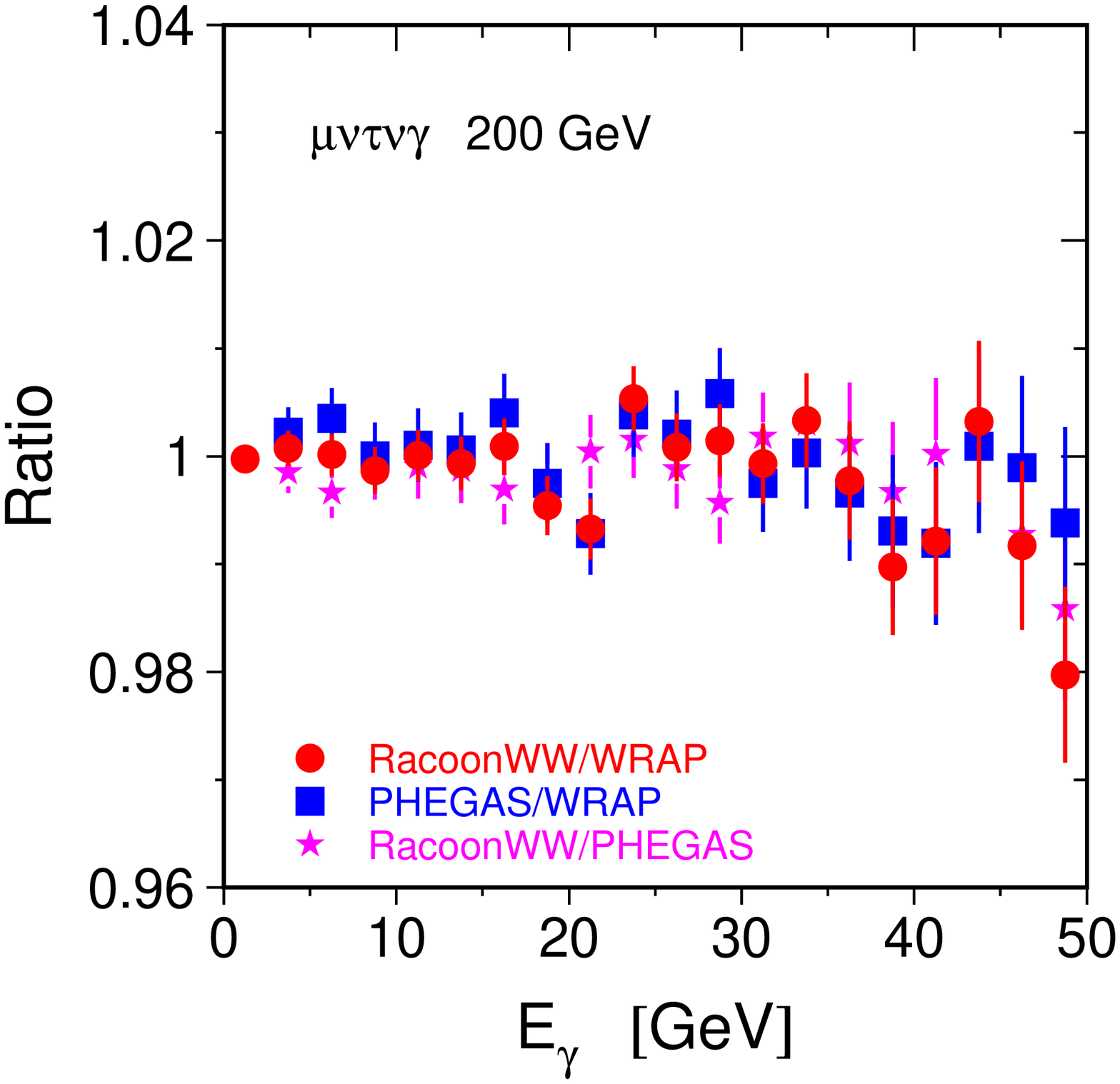,width=0.54\linewidth}
\caption[]{$E_{\ph}$ distributions and ratios for the process
$\fnum \mu^+ \tau^- \barnu_{\tau}\gamma$.}
\label{yr_4fg_67r}
\efi

\begin{figure}[p]
\epsfig{file=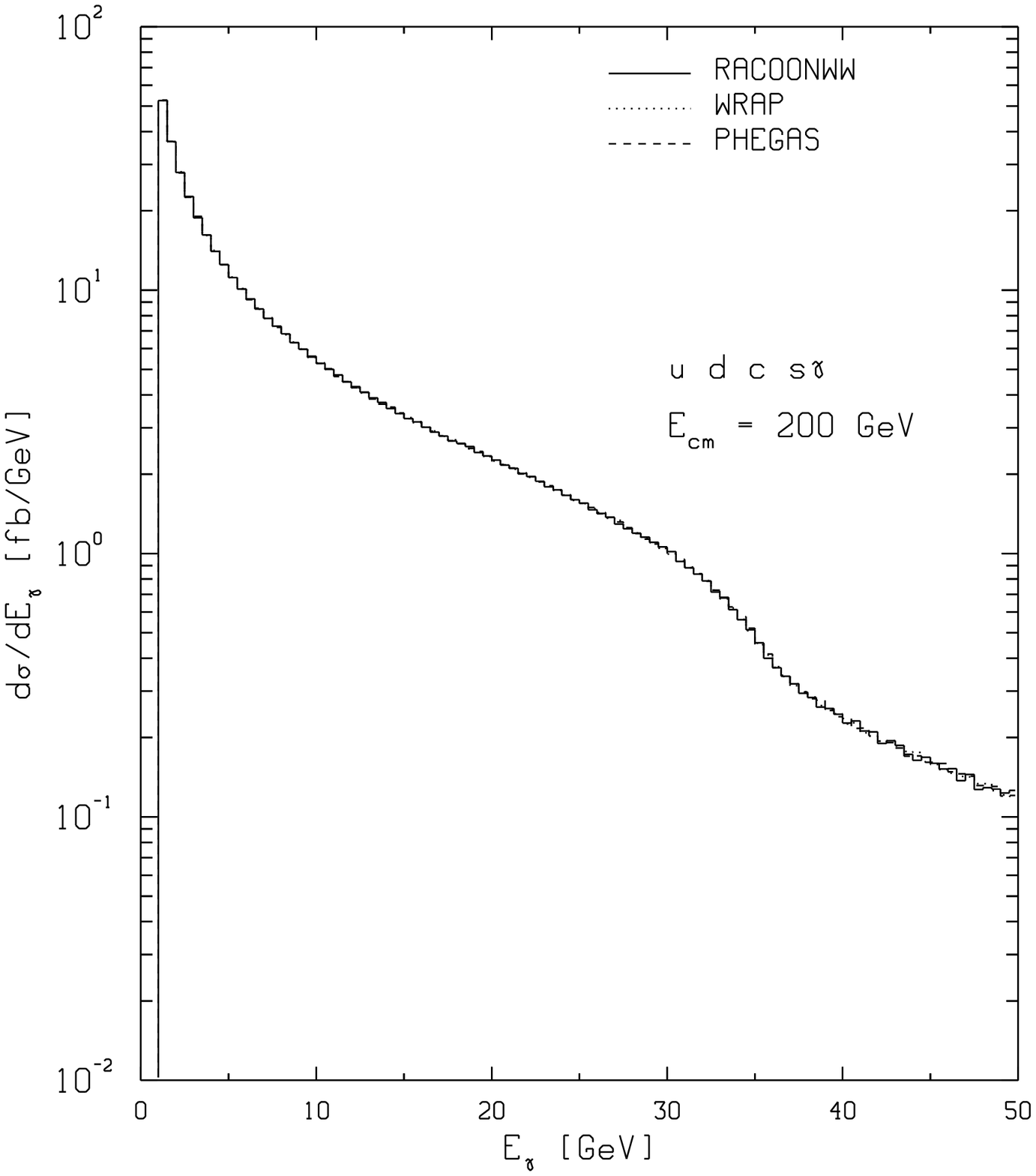,width=0.44\linewidth}
\hfill
\epsfig{file=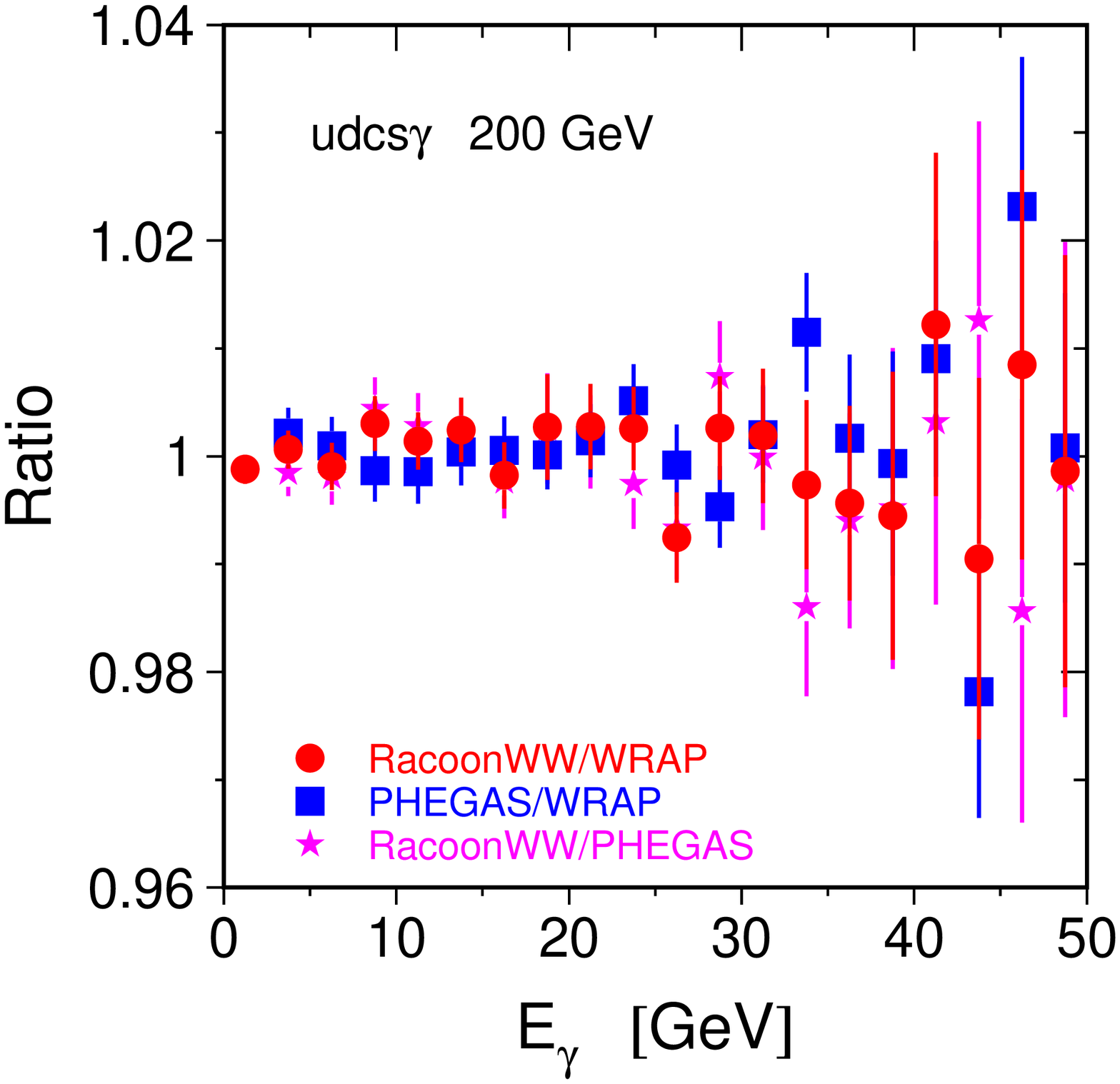,width=0.54\linewidth}
\caption[]{$E_{\ph}$ distributions and ratios for the process 
$u \bard s \barc\gamma$.}
\label{yr_4fg_89}
\efi

\clearpage

\begin{figure}[p]
\epsfig{file=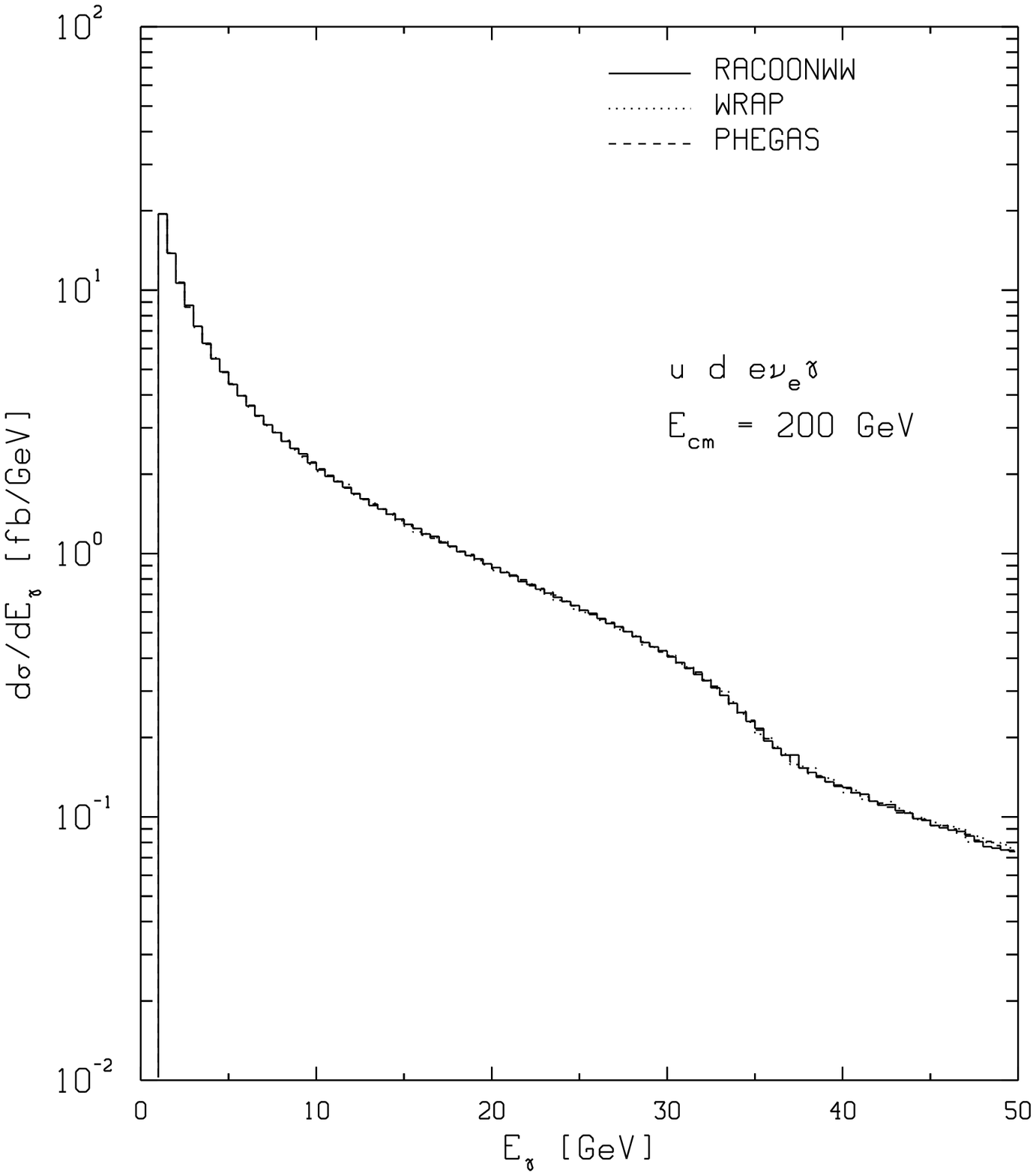,width=0.44\linewidth}
\hfill
\epsfig{file=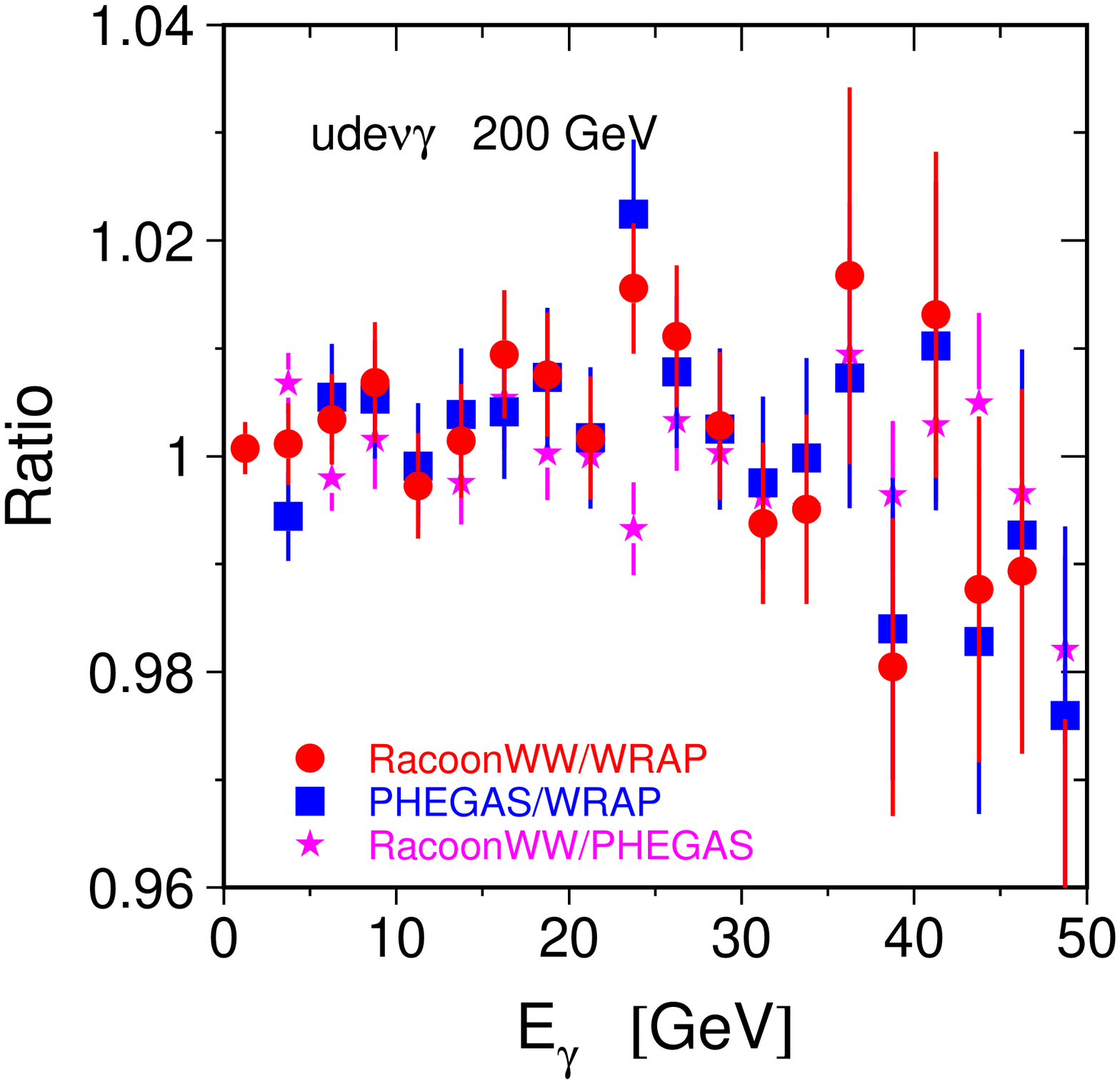,width=0.54\linewidth}
\caption[]{$E_{\ph}$ distributions and ratios for the process
$u \bard e^- \barnu_e\gamma$.}
\label{yr_4fg_89r}
\efi

\bfi
\epsfig{file=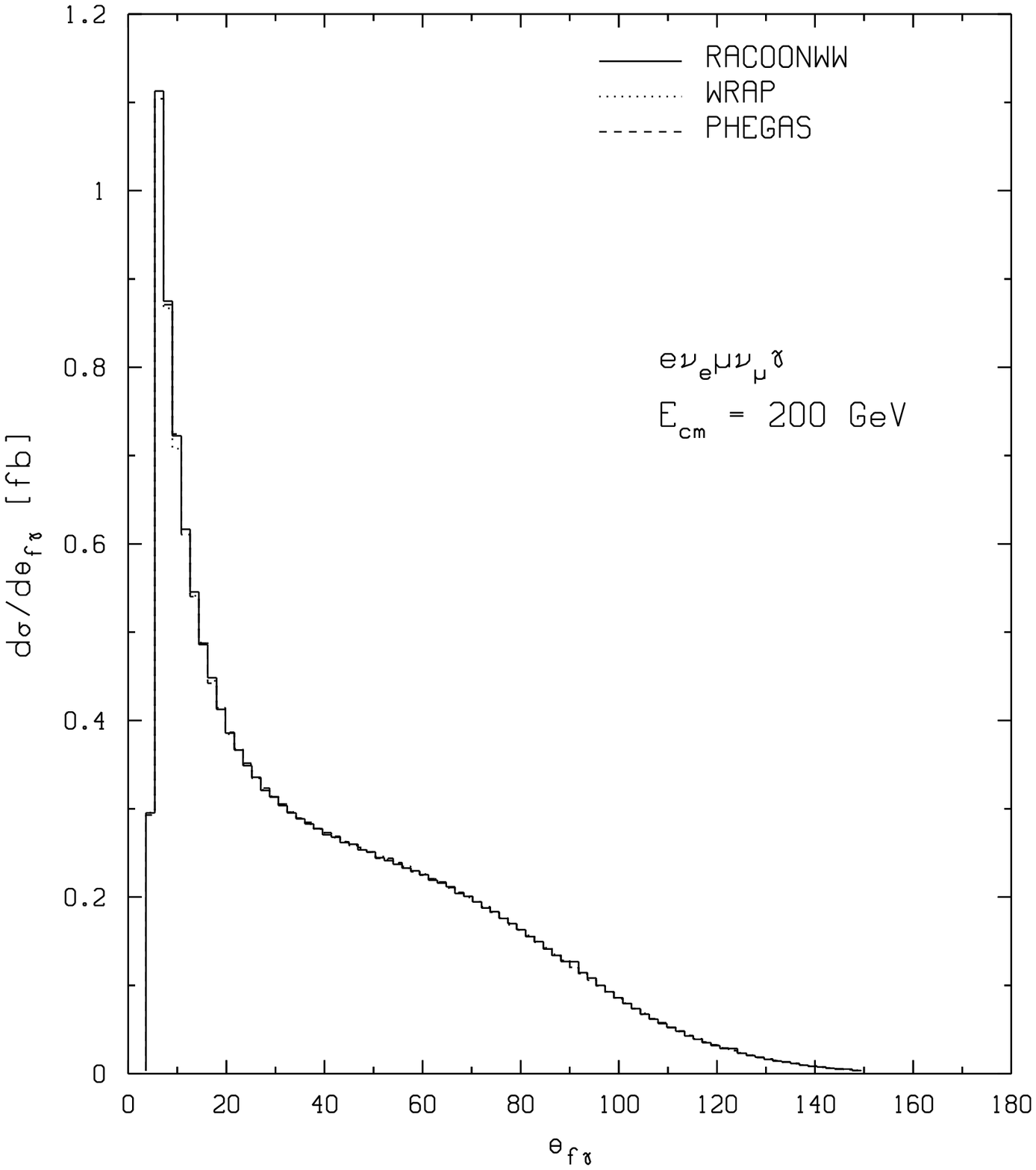,width=0.44\linewidth}
\hfill
\epsfig{file=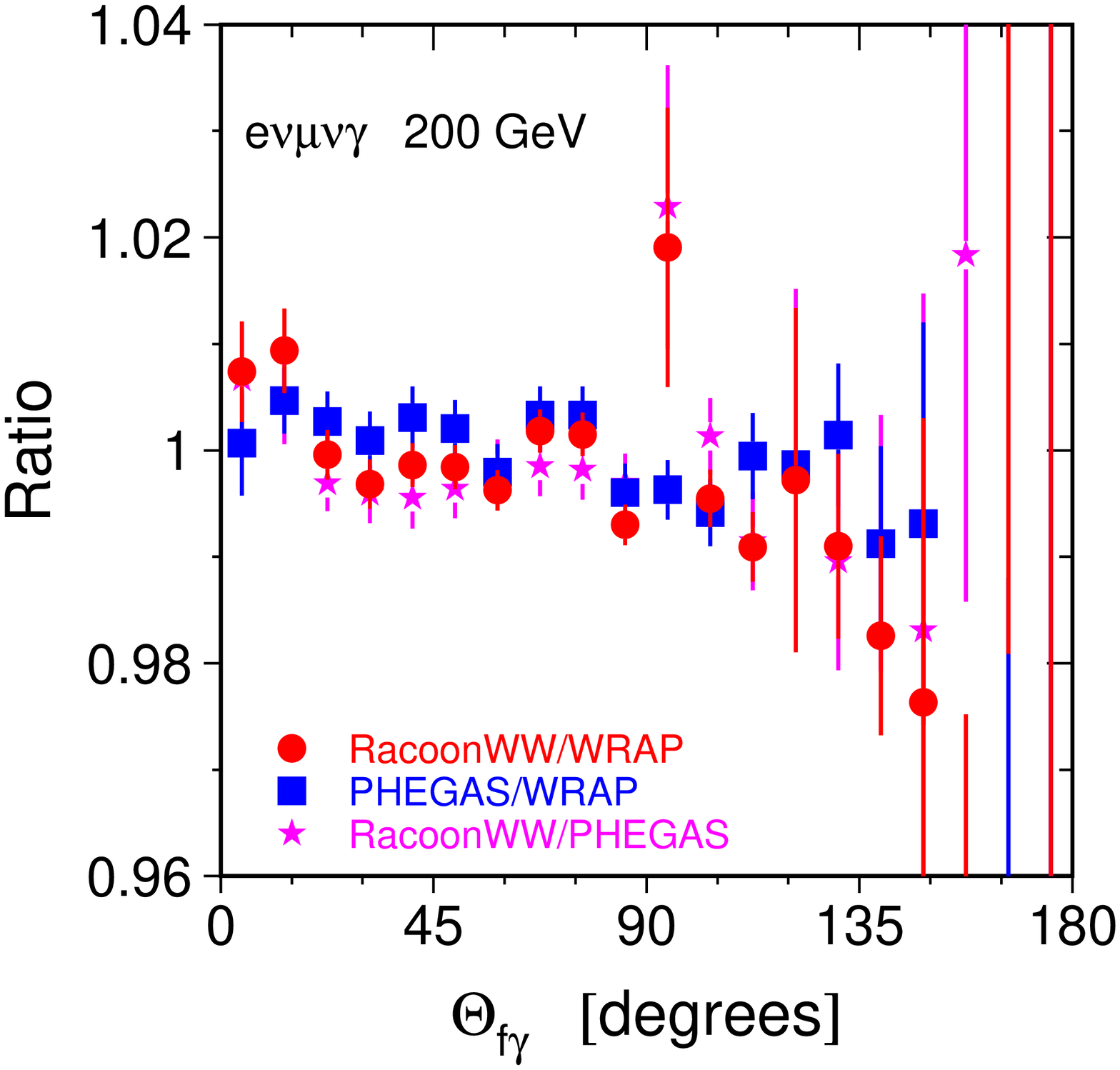,width=0.54\linewidth}
\caption[]{Distribution in the opening angle $\theta(\ph, {\rm
    charged~fermion})$ between the photon and the nearest charged
  final-state fermion in the process $e \nu_e \mu^- \barnu_{\mu}\gamma$ and
  the corresponding ratios.}
\label{yr_4fg_17}
\efi

\clearpage

\begin{figure}[p]
\begin{center}
\vskip -3cm
\epsfig{file=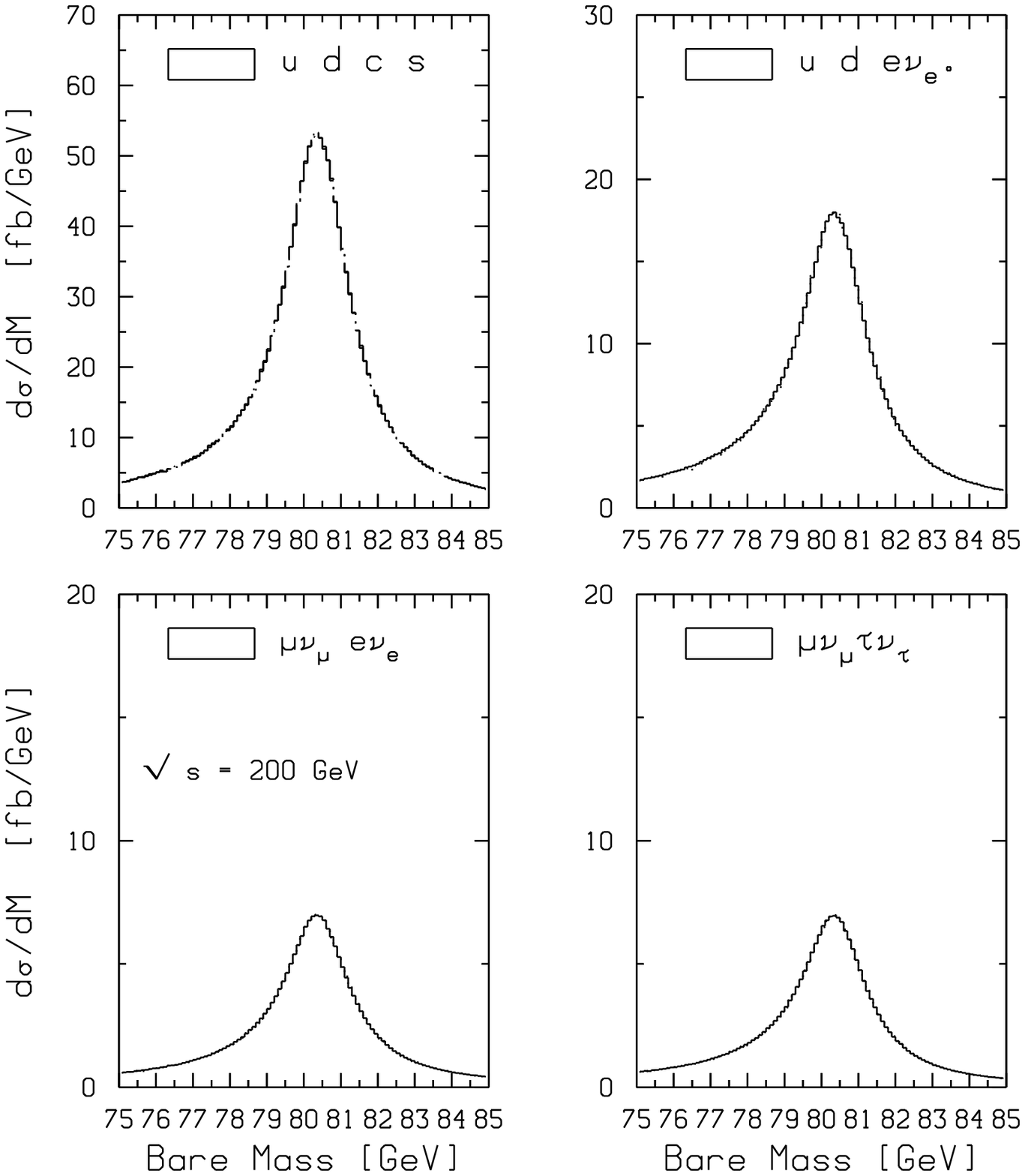,width=0.49\linewidth}
\hfill
\epsfig{file=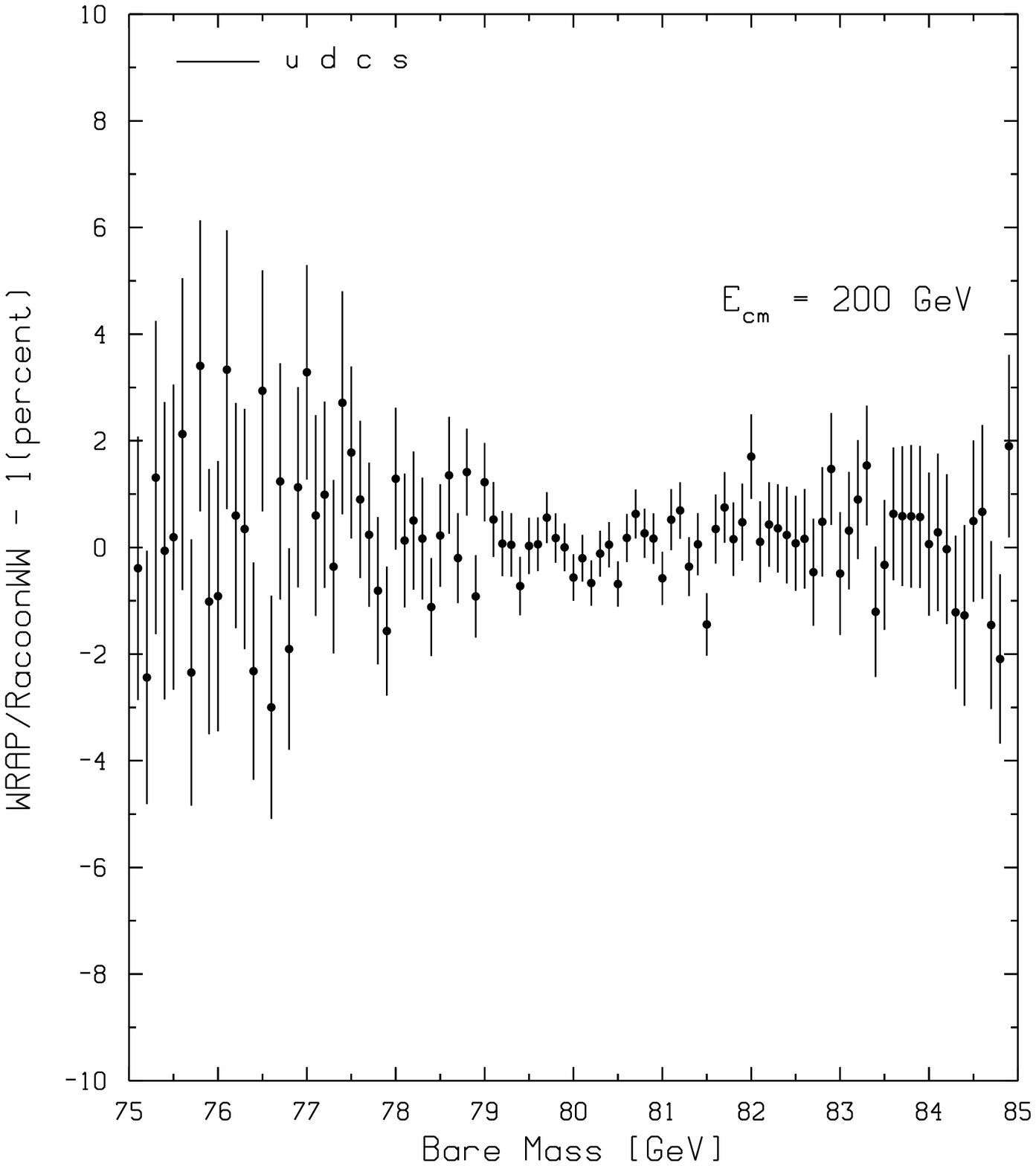,width=0.40\linewidth}
\end{center}
\caption[]{Bare $\wbm$ mass distributions and percentage deviations between
{\tt WRAP} and {\tt RacoonWW} for one specific example, $u \bard s \barc\gamma$.}
\label{yr_4fg_11}
\efi
\begin{figure}[p]
\begin{center}
\vskip -5cm
\epsfig{file=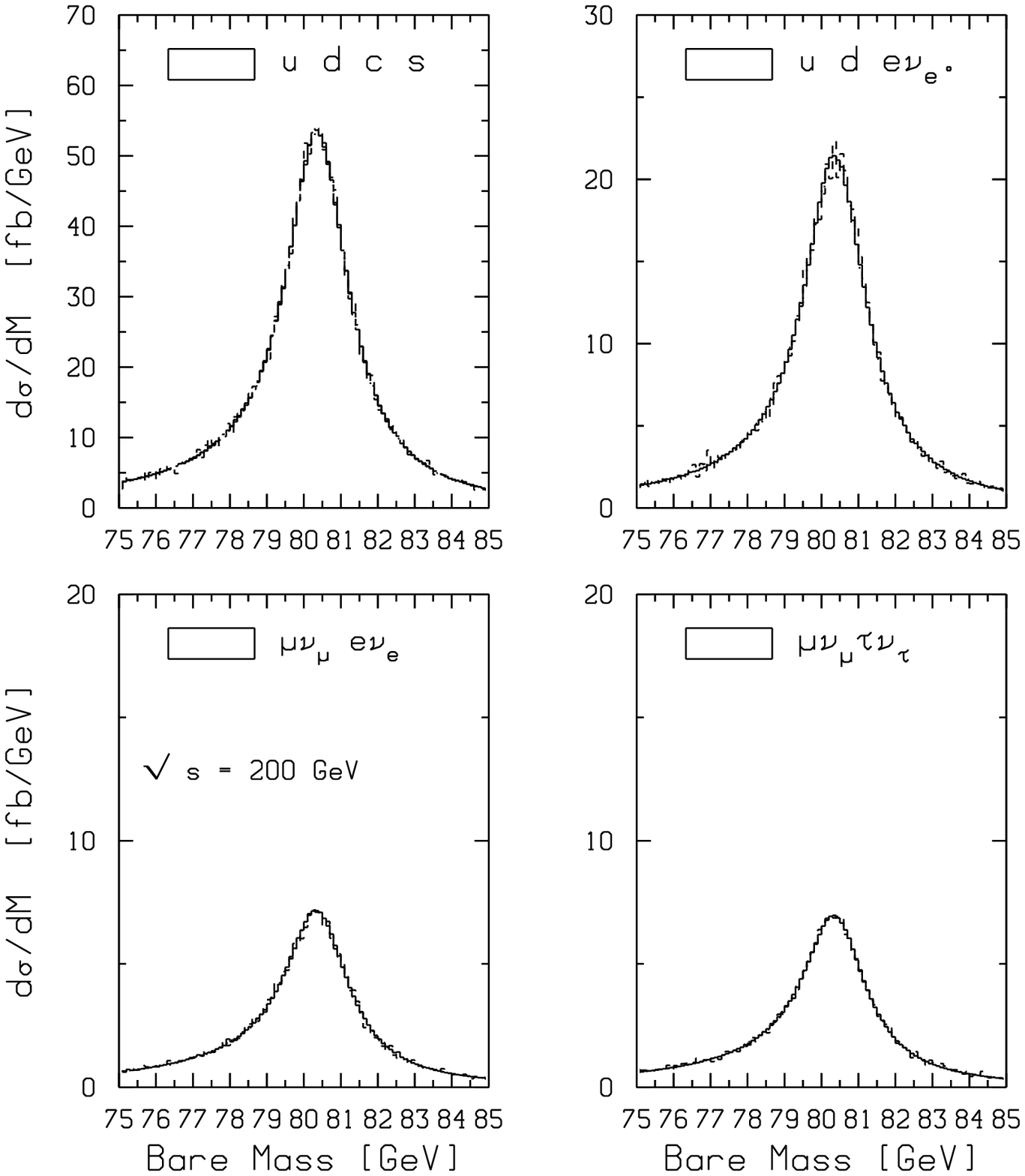,width=0.49\linewidth}
\end{center}
\caption[]{Bare $\wbp$ mass distributions from {\tt WRAP}, {\tt RacoonWW} 
and {\tt PHEGAS}.}
\label{yr_4fg_11b}
\efi
\clearpage
\begin{figure}[ht]
\epsfig{file=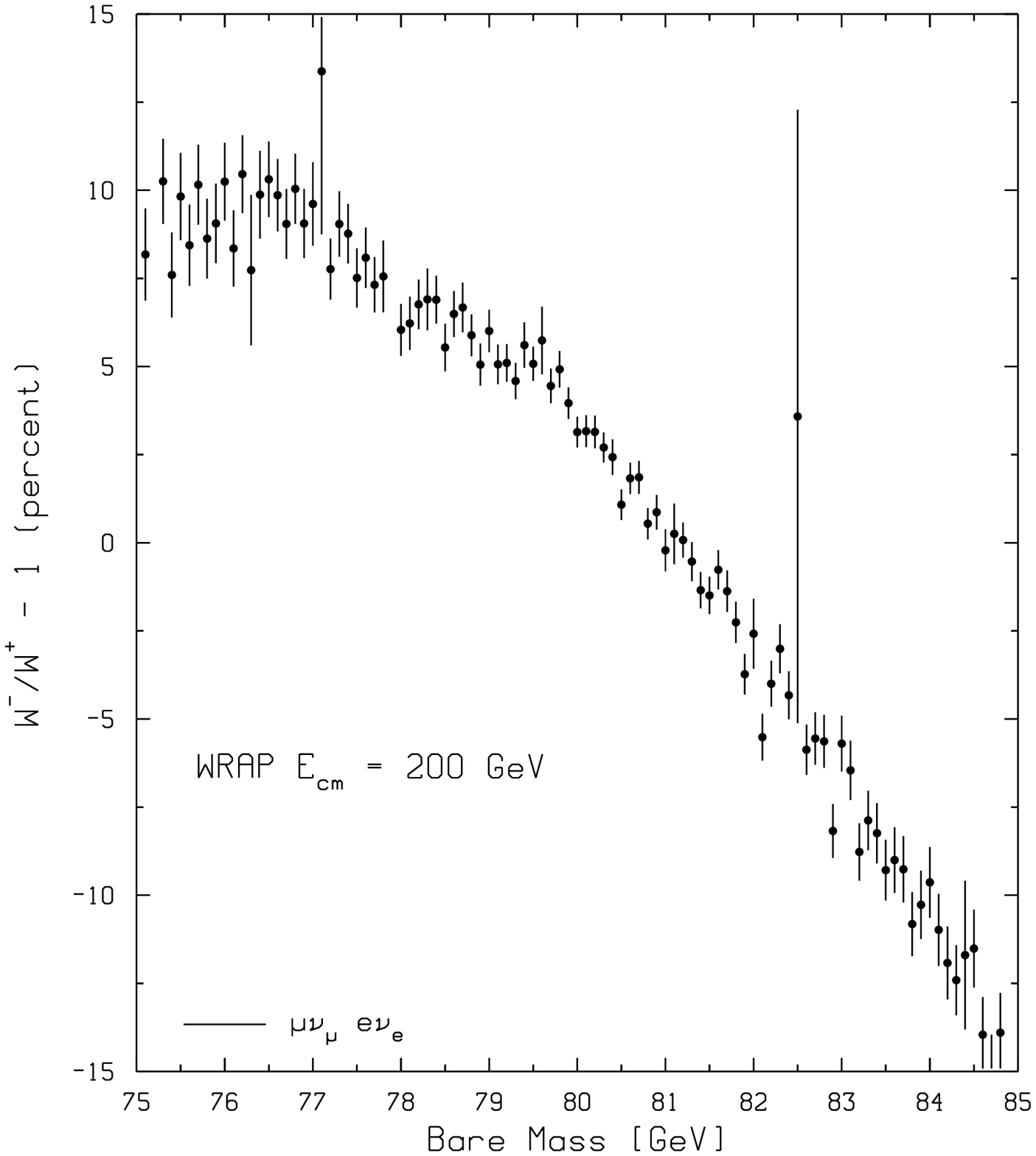,width=0.49\linewidth}
\hfill
\epsfig{file=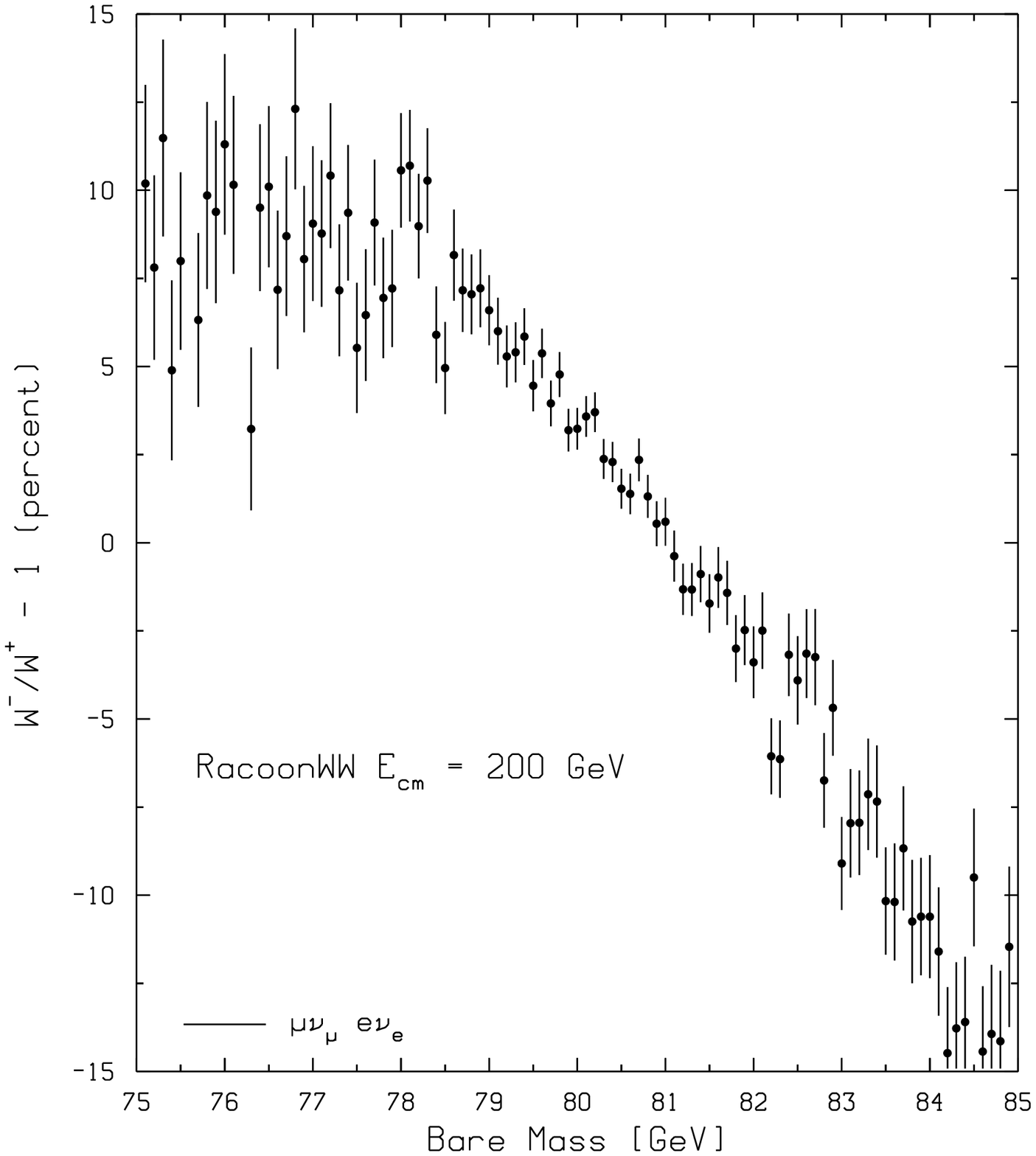,width=0.49\linewidth}
\vskip -1cm
\caption[]{Ratio of invariant mass distributions $\wbm/\wbp$ from
{\tt WRAP} and {\tt RacoonWW} for the process $\mu\fnum e \nu_e$.}
\label{yr_4fg_15cd}
\efi

\subsection{Estimate of theoretical uncertainty}
\label{tu4fg}

No global statement can be given, at the moment, on this issue. The following
programs have agreed to make individual statements:

\subsubsection*{\tt RacoonWW}

Since the program has only tree-level precision for 
$e^+e^- \to 4\rmf+\ph$, a reliable estimate for the theoretical uncertainty 
cannot be given with the present version.
This could be done if leading corrections such as ISR were
included, which is planned in future extensions of the program.

\subsubsection*{\tt WRAP}

{\tt WRAP} has tried to estimate the theoretical uncertainty in 
$4\rmf+\gamma$ processes coming from variations in the 
renormalization scheme. The selected process is 
$e^+ e^- \to u {\bard} \mu^- \bar\fnum \gamma$
with the cuts used in the tuned comparisons. 
The following two schemes have been adopted:
\bqa
I)\quad s_{_{\ssW}}^2 &=& 1 - \frac{\mws}{\mzs}, \quad
      \alpha= \frac{4\,\sqrt{2}\,\gf\mws s_{_{\ssW}}^2}{4\,\pi}, \quad
      g^2 = 4\,\pi\,\frac{\alpha}{s_{_{\ssW}}^2},  \nl             
II)\quad s_{_{\ssW}}^2 &=& \frac{\pi\alpha(2\mw)}{\sqrt{2}\gf\mws}, \quad
      g^2 = 4\sqrt{2}\gf\mws, \quad \mbox{with} \,\,
\alpha(2\mw) = 128.07.
\label{tuiandii}
\eqa
The cross section is always rescaled by the factor 
$\alpha(0)/\alpha$ in order to take into account of the 
scale $\alpha(0)$ for the emitted real photon. 
Here, $\alpha$ is the value computed in the corresponding renormalization 
scheme. The results are shown in \tabn{wraptu}.

\begin{table}[htbp]\centering
\renewcommand{\arraystretch}{1.1}
\begin{tabular}{|c|c|c|}
\hline
$\sqrt{s}\,$[GeV] &  cross section [fb] &   $\delta$ \\
\hline
\hline
200 (I)   &   75.750(29) fb &  \\
200 (II)  &   75.887(29) fb & $0.18\%$ \\
\hline
189   (I) &   71.889(25) fb &  \\
189  (II) &   71.997(25) fb &  $0.15\%$ \\
\hline
183 (I)   &   67.238(22) fb &  \\
183 (II)  &   67.324(22) fb & $0.13\%$ \\
\hline
\end{tabular}
\vspace*{3mm}
\caption[]{Estimate of the contribution to the theoretical uncertainty as due to 
variation of the input parameter set, according to {\tt WRAP}. 
I and II refer to the choices in \eqn{tuiandii} and $\delta$ is the
percentage difference.}
\label{wraptu}
\end{table}
\normalsize

Note that the overall theoretical uncertainty for $4\rmf+\gamma$ production
cannot be below the level of $1 \div 2\%$. In this respect the numbers given 
in \tabn{wraptu} are only a partial indication of possible sources of
uncertainty.
As shown by the previous analysis, 
ISR needs to be taken into account in programs for a 
realistic analysis of $4\rmf+\gamma$ final states. Furthermore, 
in order to avoid double-counting between pre-emission and 
matrix-element radiation, the implementation of QED 
corrections in computational tools for $4\rmf+\gamma$ processes  
should rely upon methods, such parton shower, YFS or $p_t$-dependent 
structure functions, able to keep under control photon $p_t$ effects. 
Effects due to finite fermion masses can become important at some
percent level for small photon-charged fermion separation cuts.

In order to better understand the uncertainty associated 
to the implementation of collinear ISR in $4\rmf + \gamma$ processes, 
a comparison between the effects of 
ISR via collinear SF and $p_t$-dependent SF, respectively,
is shown in \fig{fig3_4fg} for the cross 
section of the the CC10 final state  
$\mu^- \; \barnu_{\mu}\; u\; \bard\; \gamma$, as a 
function of the minimum energy of the observed photon, 
at $\sqrt{s} = 192$~GeV. As can be seen, the two prescriptions
for ISR can differ at $5\%$  level for $E_{\ph}^{\rm min}$ 
close to $1-2\,\GeV$, while the difference becomes smaller and smaller 
as $E_{\ph}^{\rm min}$ increases. In general, the difference between 
collinear and $p_t$-dependent SF is stronger near the soft and 
collinear regions, as a priori expected, and it gives an estimate 
of the size of the double-counting effect at the level of ISR.

\subsubsection*{\tt NEXTCALIBUR}

To check the sensitivity of various distributions to the chosen form
of the Structure Functions, the processes 
$e^+ e^- \to \mu^- \mu^+ u \baru  (\gamma) $ and 
$e^+ e^- \to e^- \barnu_e u \bard (\gamma)$
have been considered with a slightly different implementation 
of the sub-leading terms, without observing
any significant deviation, at the per mille level, with respect to the 
previous results.

\subsection{Summary and conclusions}

While the technical precision in $e^+e^- \to 4{\rm f}+\ph$ does not represent a
problem anymore for all those programs that implement an (exact) matrix 
element, very little effort has been devoted in analyzing the overall
theoretical uncertainty. 
Some of the programs also include the large effect of initial state radiation 
at the leading logarithmic level. When this is done, the bulk of large 
radiative corrections is included. Since however
in general non-logarithmic $\ord{\alpha}$ corrections are not known, the
theoretical accuracy is at the level of $2.5\%$ on integrated 
cross-sections and on inclusive distributions.

\section{Single-$\wb$}
\label{sectsw}

Another interesting process at LEP~2 is the so-called single-$\wb$ production,
$e^+e^- \to \wb e \nu$ which can be seen as a part of the CC20 process,
$e^+e^- \to \barq\,q\,(\mu\,\nu_{\mu},\,\,\tau\,\nu_{\tau})\,e\,\nu_e$, or as
a part of the Mix56 process, $e^+e^- \to e^+\,e^-\,\nu_e\,\barnu_e$. 
For a more detailed theoretical review we refer to~\cite{swc} and to~\cite{bd}.
All processes in the CC20/Mix56 families are usually considered in two regimes,
$|\cos\theta(e^-)| \ge c$ or SA and $|\cos\theta(e^-)| \le c$ or LA.
In the list of observables, the single $\wb$ production is defined by those
events that satisfy $|\cos\theta(e^-)| \ge 0.997$ and therefore is a
SA.

The LA cross-section has been computed by many authors and references can 
be found in \cite{EGWWP}. It represents a contribution to the $e^+e^- \to 
\wbp\wbm$ total cross-section.
From a theoretical point of view the evaluation of a LA cross-section
is free of ambiguity, even in the approximation of massless fermions, as long
as a gauge-preserving scheme is applied and $\theta(e^-)$ is not too small.

For SA instead, one cannot employ the massless approximation anymore.
In other words, in addition to double-resonant $\wb$-pair production with 
one $\wb$ decaying into $e\nu_e$, there are $t$-channel diagrams that give a 
sizeable contribution for small values of the polar scattering angle of the
$t$-channel electron. 
Single-$\wb$ processes are sensitive to the breaking of $U(1)$ gauge 
invariance in the collinear limit, as described in Ref.~\cite{BHF1} (see also
\cite{Kurihara}).
The correct way of handling them is based on the so-called 
Fermion-Loop (FL)scheme~\cite{BHF2}, the gauge-invariant treatment of the 
finite-width effects of $\wb$ and $\zb$ bosons in LEP~2 processes.
However, till very recently, the Fermion-Loop scheme was available only for the
LA-regime.
For $e^+e^- \to e^-\barnu_e f_1\barf_2$, the $U(1)$ gauge 
invariance becomes essential in the region of phase space where the angle 
between the incoming and outgoing electrons is small, see the work of
\cite{BHF1} and also an alternative formulations in \cite{fls-baur}.
In this limit the superficial $1/Q^4$ divergence of the  
propagator structure is reduced to $1/Q^2$ by $U(1)$ gauge  
invariance. In the presence of light fermion masses this gives raise  
to the familiar $\ln(\mes/s)$ large logarithms.
Furthermore, keeping a finite electron mass through the calculation is not 
enough. One of the main results of~\cite{swc} was to show that
there are remaining subtleties in CC20, associated with the zero mass limit
for the remaining fermions.

In~\cite{tfl} a generalization of the Fermion-Loop scheme 
(hereafter EFL) is introduced to account for external, non-conserved, currents.
Another extension has been given in~\cite{abm}
for the imaginary parts of Fermion-Loop contributions, which represents the 
minimal set for preserving gauge invariance.

The most recent numerical results produced for single-$\wb$ production are from
the following codes~\cite{groups}: {\tt CompHEP}, {\tt GRC4F}, 
{\tt NEXTCALIBUR}, {\tt SWAP}, {\tt WPHACT} and {\tt WTO}.

In view of a requested, inclusive cross-section, accuracy of $2\%$ we must
include radiative corrections to the best of our knowledge, at least the
bulk of any large effect.
As we know, the correct scale of the couplings and their differentiation 
between $s$- and $t$-channel is connected to the real part of the corrections,
so that the imaginary FL is not enough, we need a complete FL for 
single-$\wb$, or EFL.
Having all the parts, the tree-level couplings are replaced by running 
couplings at the appropriate momenta and the massive gauge-boson propagators 
are modified accordingly. The vertex coefficients, entering through the 
Yang--Mills vertex, contain the lowest order couplings as well as the 
one-loop fermionic vertex corrections. 

Each calculation aimed to provide some estimate for single-$\wb$ production
is, at least nominally, a tree level calculation. Among other things it will 
require the choice of some Input Parameter Set (IPS) and of certain relations 
among the parameters. 
Thus, different choices of the basic relations among the input parameters can 
lead to different results with deviations which, in some case, can be sizeable
and should be included in the theoretical uncertainty. Here, more work is 
needed.

For instance, a possible choice is to fix the coupling constant $g$ as
\bq
g^2 = {{4\pi\alpha}\over {s_{_{\ssW}}^2}}, \quad
s_{_{\ssW}}^2 = {{\pi\alpha}\over {{\sqrt 2}\gf\mw^2}},
\eq
where $\gf$ is the Fermi coupling constant. Another possibility would be to use
\bq
g^2 = 4{\sqrt 2}\gf\mw^2, 
\eq
but, in both cases, we miss the correct running of the coupling. Ad hoc 
solutions should be avoided, and the running of the parameters must always 
follow from a fully consistent scheme.

Another important issue in dealing with single-$\wb$ production is connected
with the inclusion of QED radiation.
It is well known that universal, $s$-channel structure functions are not 
adequate enough to include the radiation since they generate an excess of
ISR bremsstrahlung. In $t$-channel dominated processes the interference
between incoming fermions becomes very small while the destructive
interference between initial and final states becomes strong.

It is quite a known fact that, among the electroweak corrections, QED radiation
gives the largest contribution and the needed precision requires a
re-summation of the large logarithms.
For annihilation processes, $e^+e^- \to \barf f$, initial state radiation is
a definable, gauge-invariant concept and we have general tools to deal with it;
the structure function approach and also the parton-shower method. However,
when we try to apply the algorithm to four-fermion processes that include
non-annihilation channels we face a problem: it is still possible to
include the large universal logarithms by making use of the standard tools
but an appropriate choice of scale is mandatory.
Such is the case in single-$\wb$.
The problem of the correct scale to be used in QED corrections has been 
tackled by two groups, {\tt GRACE} and {\tt SWAP} and additional results 
will be shown in \subsects{esd}{scalegrace} and in \subsect{rcsw}.

\subsection{Signal definition in single-$\wb$}

The experimental requirements on single-$\wb$ are:

\begin{itemize}

\item[--] CC20 -- Mix56 calculations with some detector acceptance
that are used for a) triple gauge coupling determination, b)
standard model background to searches;

\item[--] the LEP EWWG cross-section definition that is used to
combine the cross-section measurements from the four LEP experiments.

\end{itemize}

During the last WW99 Crete Workshop a proposal has been made to
reach a common {\em signal} definition for the LEP EWWG 
cross-section~\cite{proposal}. 
The persons who participated in the WW99 workshop agreed on some setup to 
define the single-$\wb$ production and now this has been formalized 
in one of the {\tt LEP EWWG} meetings; there, it was decided to have a 
combination of the single-$\wb$ cross-section using the 
signal definitions of \tabn{signalsw} for $e^+e^- \to e^- \barnu_e f' \barf$:
The set of $t$-channel diagrams, all for CC20, are shown in \fig{cc20gz}.
The signal definition uses $10$ diagrams for CC20, $9$ for CC18
and $37$ for Mix56. 

\begin{table}[hp]\centering
\renewcommand{\arraystretch}{1.1}
\begin{tabular}{|c|c|c|}
\hline
Process & diagrams & cuts \\
\hline
$ee\nu\nu$ & $t$-channel only & $E(e^+) > 20\,\GeV,
|\cos\theta(e^+)| < 0.95,  |\cos\theta(e-)| > 0.95$ \\
$e\nu\mu\nu$ & $t$-channel only & $E_(\mu^+)  > 20\,\GeV$ \\
$e\nu\tau\nu$ & $t$-channel only & $E_(\tau^+)  > 20\,\GeV$ \\
$e\nu u d$ & $t$-channel only & $M(ud) > 45 GeV$ \\
$e\nu c s$ & $t$-channel only & $M(cs) > 45 GeV$ \\
\hline
\end{tabular}
\vspace*{3mm}
\caption[]{Signal definition for single-$\wb$ processes.\label{signalsw}}
\end{table}

Note that charge-conjugate state should be taken into account and that
an asymmetric cut has been introduced for $ee\nu\nu$; the latter is due
to the fact that the process itself is CP-even when no cut is applied, but 
an ambiguity remains if one starts to discuss single-$\wb$ with
$e^-$ in the forward direction.
Then we should multiply this process by a factor $2$ as well.
The goal of this common definition is to be able to combine the
different $e\nu\barq q, e\nu\mu\nu, e\nu\tau\nu, e\nu e\nu$ measurements from 
different experiments so that the new theoretical calculations can be
checked with data at a level better than $10\%$. 

Signal definition has a longstanding tradition in LEP physics, the most 
celebrated being the $t$-channel subtraction in Bhabha and the most recent
being the CC03 cross-section.
Here we have a different situation. First of all, nobody has radiative
corrections for single-$\wb$ production, hence the usual argument of the
availability of a sophisticated semi-analytical calculation for the signal
does not apply.
We could avoid a definition of the signal in terms of diagrams and have
recourse to a definition in terms of cuts since, in a very narrow cone around 
the beam axis, the single-$\wb$ family is fully dominated by the $t$-channel 
photons.

\begin{figure}[t]
\vspace{0.2cm}
\bqas
\ba{ccc}
\vcenter{\hbox{
  \SetScale{0.7}
  \begin{picture}(110,100)(0,0)
  \ArrowLine(50,120)(0,140)
  \ArrowLine(100,140)(50,120)
  \ArrowLine(0,0)(50,20)
  \ArrowLine(50,20)(100,0)
  \ArrowLine(100,110)(80,70)
  \ArrowLine(80,70)(100,30)
  \Photon(50,20)(50,70){2}{7}
  \Photon(50,70)(50,120){2}{7}
  \Photon(50,70)(80,70){2}{7}
  \Text(-14,98)[lc]{$e^+$}
  \Text(77,98)[lc]{$\barnu_e$}
  \Text(-14,0)[lc]{$e^-$}
  \Text(77,0)[lc]{$e^-$}
  \Text(77,77)[lc]{$\bard$}
  \Text(77,21)[lc]{$u$}
  \Text(18,60)[lc]{$\wb$}
  \Text(11,26)[lc]{$\gamma,\zb$}
  \end{picture}}}
&\quad+&
\vcenter{\hbox{
  \SetScale{0.7}
  \begin{picture}(110,100)(0,0)
  \ArrowLine(50,120)(0,140)
  \ArrowLine(65,126)(50,120)
  \ArrowLine(100,140)(65,126)
  \ArrowLine(0,0)(50,20)
  \ArrowLine(50,20)(100,0)
  \ArrowLine(100,110)(80,70)
  \ArrowLine(80,70)(100,30)
  \Photon(50,20)(50,120){2}{7}
  \Photon(65,126)(80,70){2}{7}
  \Text(77,77)[lc]{$\bard$}
  \Text(77,21)[lc]{$u$}
  \Text(54,76)[lc]{$\wb$}
  \Text(11,66)[lc]{$\gamma,\zb$}
  \Text(-14,98)[lc]{$e^+$}
  \Text(77,98)[lc]{$\barnu_e$}
  \Text(-14,0)[lc]{$e^-$}
  \Text(77,0)[lc]{$e^-$}
  \end{picture}}}
\ea
\eqas
\bqas
\ba{ccc}
\vcenter{\hbox{
  \SetScale{0.7}
  \begin{picture}(110,100)(0,0)
  \ArrowLine(50,120)(0,140)
  \ArrowLine(100,140)(50,120)
  \ArrowLine(0,0)(50,20)
  \ArrowLine(50,20)(100,0)
  \ArrowLine(100,90)(50,90)
  \ArrowLine(50,90)(50,50)
  \ArrowLine(50,50)(100,50)
  \Photon(50,20)(50,50){2}{7}
  \Photon(50,90)(50,120){2}{7}
  \Text(77,72)[lc]{$\bard$}
  \Text(77,26)[lc]{$u$}
  \Text(22,52)[lc]{$u$}
  \Text(10,20)[lc]{$\gamma,\zb$}
  \Text(17,77)[lc]{$\wb$}
  \Text(-14,98)[lc]{$e^+$}
  \Text(77,98)[lc]{$\barnu_e$}
  \Text(-14,0)[lc]{$e^-$}
  \Text(77,0)[lc]{$e^-$}
  \end{picture}}}
&\quad+&
\vcenter{\hbox{
  \SetScale{0.7}
  \begin{picture}(110,100)(0,0)
  \ArrowLine(50,120)(0,140)
  \ArrowLine(100,140)(50,120)
  \ArrowLine(0,0)(50,20)
  \ArrowLine(50,20)(100,0)
  \ArrowLine(100,90)(50,50)
  \Line(50,90)(65,78)
  \ArrowLine(90,58)(100,50)
  \ArrowLine(50,50)(50,90)
  \Photon(50,20)(50,50){2}{7}
  \Photon(50,90)(50,120){2}{7}
  \Text(77,72)[lc]{$\bard$}
  \Text(77,21)[lc]{$u$}
  \Text(22,52)[lc]{$d$}
  \Text(10,20)[lc]{$\gamma,\zb$}
  \Text(17,77)[lc]{$\wb$}
  \Text(-14,98)[lc]{$e^+$}
  \Text(77,98)[lc]{$\barnu_e$}
  \Text(-14,0)[lc]{$e^-$}
  \Text(77,0)[lc]{$e^-$}
  \end{picture}}}
\ea
\eqas
\bqas
\ba{ccc}
\vcenter{\hbox{
  \SetScale{0.7}
  \begin{picture}(110,100)(0,0)
  \ArrowLine(50,120)(0,140)
  \ArrowLine(100,140)(50,120)
  \ArrowLine(0,0)(50,20)
  \ArrowLine(50,20)(100,0)
  \ArrowLine(100,110)(80,70)
  \ArrowLine(80,70)(100,30)
  \Photon(50,20)(50,120){2}{7}
  \Photon(18,132)(80,70){2}{7}
  \Text(77,77)[lc]{$\bard$}
  \Text(77,21)[lc]{$u$}
  \Text(48,71)[lc]{$\wb$}
  \Text(21,56)[lc]{$\zb$}
  \Text(-14,98)[lc]{$e^+$}
  \Text(77,98)[lc]{$\barnu_e$}
  \Text(-14,0)[lc]{$e^-$}
  \Text(77,0)[lc]{$e^-$}
  \end{picture}}}
&\quad+&
\vcenter{\hbox{
  \SetScale{0.7}
  \begin{picture}(110,100)(0,0)
  \ArrowLine(50,120)(0,140)
  \ArrowLine(100,140)(50,120)
  \ArrowLine(0,0)(50,20)
  \ArrowLine(50,20)(100,0)
  \ArrowLine(100,110)(80,70)
  \ArrowLine(80,70)(100,30)
  \Photon(50,20)(50,120){2}{7}
  \Photon(64,15)(80,70){2}{7}
  \Text(77,77)[lc]{$\bard$}
  \Text(77,21)[lc]{$u$}
  \Text(40,41)[lc]{$\wb$}
  \Text(19,56)[lc]{$\wb$}
  \Text(-14,98)[lc]{$e^+$}
  \Text(77,98)[lc]{$\barnu_e$}
  \Text(-14,0)[lc]{$e^-$}
  \Text(77,0)[lc]{$e^-$}
  \end{picture}}}
\ea
\eqas
\vspace{-2mm}
\caption[]{The $t$-channel component of the CC20 family of diagrams: fusion,
bremsstrahlung and multi-peripheral.}
\label{cc20gz}
\end{figure}

\subsubsection{A study of single-$\wb$ signal definition with {\tt CompHEP}}

\subsubsection*{Authors}

\begin{tabular}{l}
E.~Boos, M.~Dubinin and V.~Ilyin \\
\end{tabular}

\subsubsection*{Single-$\wb$ signal definition in the reaction
             $e^+ e^- \to e^+ e^- \nu_e \barnu_e$}

It is well-known for a long time how the single $\wb$ signal
can be separated with the help of kinematical cuts \cite{Kurihara}.
The typical set of cuts used by ALEPH, DELPHI and L3 collaborations
for the leptonic four fermion states $e^- \barnu_e l^+ \nu_l$
separates the configurations with very forward $e^-$ and a rather
energetic $l^+$ produced at a sufficiently large angle with the
beam. For instance, the L3 cuts to be used in the following
calculations are
$|\cos \theta_{e^-}| \geq 0.997$, $E_l \geq 15 \,\GeV$ and
$|\cos \theta_{l^+}| \leq 0.997$.
In the case of the semi-leptonic states $e^- \barnu_e q \barq^{'}$ 
an additional cut $M(q \barq^{'}) \geq$ 45 GeV have been applied
by OPAL.
In so far as different collaborations are using not exactly the
same cuts (defined by the optimal detector acceptance), the
definition of the $\wb$ signal in terms of angular cuts is not 
universal and some standardization procedure is needed. 
In the recent proposal by LEP experiments~\cite{proposal}
the OPAL collaboration considered the possibility to introduce
the definition of the $\wb$ signal in terms of diagrams.
Angular cuts on the forward electron and the corresponding
anti-lepton are not imposed, so the single $\wb$ cross-section 
depends only on the $E_l$ energy cut and is defined by the gauge 
invariant subset of the $t$-channel single resonant diagrams.
The universality of such definition is satisfactory if the
interferences between the gauge invariant subsets of diagrams in
the channels $e^- \barnu_e l^+ \nu_l$ and $e^- \barnu_e q \barq^{'}$
are always negligible. Then indeed the single $\wb$ cross-section
in terms of diagrams could be meaningful.

We performed a detailed calculation
of the contributions from various diagram sets of the 
Mix56 channel $e^+ e^- \to e^+ e^- \nu_e \barnu_e$ (see Appendix for
\fig{18w}-\fig{2t} referred to in the following). Using
the general approach to the amplitude decomposition into gauge
invariant classes \cite{Boos}, we found
ten gauge invariant subsets of diagrams (see \fig{18w}-\fig{8z}). In 
\tabn{ctab} $18\wb$ denotes two gauge invariant subsets of $9$
diagrams with single $\wb$ (see \fig{18w}), $8\zb$ denotes two gauge
invariant
subsets of $4$ diagrams with single $\zb$ (see \fig{8z}), $9\wbp \wbm $
stands for
the double-resonant subset (\fig{9ww}) and so on. Main contribution 
to the final configurations with forward electron come
from the single $\wb$ and the single $\zb$ production, while various
$\gamma,\zb \to e^+ e^-$ conversion corrections (\fig{4ee}-\fig{2t})
to the $e^+ e^- \to
e^+ e^-, \, \nu_e \barnu_e$ are negligible. For the case of angular
cuts on the forward electron the interference between
the single $\wb$ and single $\zb$ subsets $18\wb$ and $8\zb$ is negative and 
equal
to several fb. However, if the angular cuts are removed, the destructive
interference modulo increases rather considerably (Table~\ref{ctab}). 
This is not an unexpected fact since both single-$\wb$ and single-$\zb$
(NC processes with one lost electron) 
subsets have a similar $t$-channel pole structure. Other interferences are
also not negligible. So in the case of $e^+ e^- \nu_e \barnu_e$ channel
the diagram-based definition of single $\wb$ signal is not completely
satisfactory.

\begin{table}[h]
\begin{center}
\renewcommand{\arraystretch}{1.1}
\begin{tabular}{|c|c|c|c|c|c|c|c|c|c|} \hline
 & 26 $t$-ch. &
18$\wb$ & 8$\zb$ & 9$\wbp \wbm $ & 4$\zb\zb$ 
 & 9$\nu_e \barnu_e$ & 4$e^+ e^-$ & 2$\nu_e \barnu_e$ &
2 $t$-ch
\\ \hline
$\theta_e$,$E_l$  & 49.9& 36.1& 16.4& 0.91& 0.02
 & 8$\cdot$10$^{-3}$        &7$\cdot$10$^{-5}$& 1$\cdot$10$^{-5}$ &  
6$\cdot$10$^{-7}$ 
\\ \hline 
only $E_l$ & 220.5   &106.6 & 153.6 &240.5 &44.9 &15.9 &0.02
&3$\cdot$10$^{-3}$ & 8$\cdot$10$^{-4}$ 
\\ \hline 
\end{tabular}
\caption[]{Contributions of the gauge invariant subsets (fb) at the energy
$\sqrt{s}=$200 GeV. First row - with angular cuts, second row - no
angular cuts for $e^-$, $e^+$. The result for 26 $t$-channel diagrams
(18$\wb$ and 8$\zb$, see Fig.1,2) is indicated in the second column.}
\label{ctab}
\end{center}
\end{table}

\subsection{Description of the programs, results and comparisons}

\subsubsection*{{\tt WTO} and EFL}
\label{se:eflwto}

\subsubsection*{Author}

\begin{tabular}{l}
G. Passarino
\end{tabular}

\subsubsection*{The Fermion-Loop scheme (EFL)}

The EFL scheme for non-conserved currents is described in
Ref.~\cite{tfl} and briefly discussed in \sect{sec:resum/fls}. It consists 
of the re-summation of the fermionic one-loop corrections to the vector-vector,
vector-scalar and scalar-scalar propagators and of the inclusion of all 
remaining fermionic one-loop corrections, in particular those to the 
Yang--Mills vertices. 

In the original formulation, the Fermion-Loop scheme requires that vector
bosons couple to conserved currents, \ie, that the masses of all external
fermions can be neglected. There are several examples where fermion masses must
be kept to obtain a reliable prediction. As already stated, this is the case
for the single-$\wb$ production mechanism, where the outgoing electron is 
collinear, within a small cone, with the incoming electron. Therefore, $\me$ 
cannot be neglected.
Furthermore, among the $20$ Feynman diagrams that contribute (for $e\barnu_e 
u\bard$ final states, up to $56$ for $e^+e^-\nu_e\barnu_e$) there are 
multi-peripheral ones that require a non-vanishing mass also for the other
outgoing fermions.  

As well known in the literature, the Fixed-Width scheme behaves 
properly in the collinear and high-energy regions of phase space, to the 
contrary of the Running-Width scheme, but it completely misses the running 
of the couplings, an effect that is expected to be above the requested 
precision tag of $2\%$.
To be specific the name of Fixed-Width scheme is reserved for the following:
the cross-section is computed using the tree-level amplitude. The massive 
gauge-boson propagators are given by 
$1/(\pmoms-\mlones+\ib\Glone m )$.
This gives an unphysical width in $t$-channel, but retains $U(1)$ gauge 
invariance in the CC20 process.

The correct way of handling this problem is to apply the EFL-scheme and,
by considering the impact of the EFL-scheme on the relevant observables, one
is able to judge on the goodness of naive rescaling procedures or of any
incomplete FL-scheme. One of the problem with the latter is that vertices,
although chosen to respect gauge-invariance, are not uniquely defined.
Furthermore, couplings other than $\alpha_{\rm QED}$ usually do not evolve with
the scale and complex poles, the truly gauge-invariant quantities, are never
introduced or explicitly computed. Finally, programs than cannot split
diagrams and apply an overall rescaling, both in $s$- and $t$-channel,
mistreat single-$\wb$ and/or violates $SU(2)$ invariance.
  
\subsubsection*{Numerical results and recommendations.}

Numerical results for EFL have been shown in Ref.~\cite{nfl}. Here, we limit
the presentation to some useful recommendations:

\begin{itemize}

\item[--] the bulk of the effect is in the running of the e.m. coupling 
constant;

\item[--] one can compute the single-$\wb$ cross-sections for a fixed mass of
the top quark, such as $\mt = 173.8\,$GeV, without finding any significative 
difference with the case where $\mt$ is fixed by a consistency relation.
We are using complex-mass renormalization but we only include fermionic 
corrections. Therefore, we can start with the Fermi coupling constant but also 
with $\mw$ as an input parameter.
Equating the corresponding renormalization conditions yields a relation 
between $\mz$, $\gf$, $\Reb\{\alpha(\mzs)^{-1}\}$, $\mw$, and $\mt$, 
see~\cite{BHF2}. 
This relation can be solved iteratively for $m_t$. For the following input 
parameter set,
$\mw = 80.350\,\GeV$, $\mz = 91.1867\,\GeV$ and $\gf= 1.16639\,
\times\,10^{-5}\,\GeV^{-2}$, we obtain the following solution:
\bq
\mu_{\ssW} = \sqrt{\Reb\lpar \sW\rpar} =  80.324\,\GeV, \quad
\gamma_{\ssW} = - {{\Imb\lpar \sW\rpar}\over {\mu_{\ssW}}} = 2.0581\,\GeV,
\quad \mt = 148.62\,\GeV,
\eq
with $26\,\MeV$ difference between $\mw$ and $\mu_{\ssW}$. See
\sect{swout} for the inclusion of QCD effects.
This type of effect should be included in any incomplete FL-scheme;

\item[--] the main accent in the EFL-scheme is on putting
the correct scale in the running of $\alpha_{\rm QED}$. The latter is 
particularly important for the $t$-channel diagrams, dominated by a scale 
$q^2 \approx 0$ and not $q^2 \approx \mws$. 
However, a consistent implementation of radiative corrections does more than
evolving $\alpha_{\rm QED}$ to the correct scale, other couplings are also 
running, propagators are modified and 
vertices are included; 

\item[--] the effective FW-scheme describes considerably well
the hadronic final state with a cut of $M(u\bard) > 45\,$GeV. However, the 
diminution induced by $\alpha_{\rm QED}(q^2)$ is too large for the leptonic 
final state. The latter is a clear sign that other effects are relevant and 
a naive rescaling does not suffice in reproducing a realistic approximation 
in all situations, at least not within the $2\%$ level of requested 
theoretical accuracy;

\item[--] Modifications induced by the fermionic loops are sensitive
to the relative weight of single-resonant terms and of multi-peripheral
peaks. Furthermore, the effect of radiative corrections inside the 
$\wb$-propagators ($\rho$-factors of Ref.~\cite{nfl}) is far from being 
negligible and tends to compensate the change due to the running of 
$\alpha_{\rm QED}$. 

\end{itemize}
These recommendations are better illustrated by few examples.
At $\sqrt{s} = 183\,$GeV we consider the angular distribution, 
$d\sigma/d\theta_e$ for the $u \bard e^- \barnu_e$ final states. 
The results are shown in \fig{fig:rs183}.
\begin{figure}[p]
\centerline{\epsfig{file=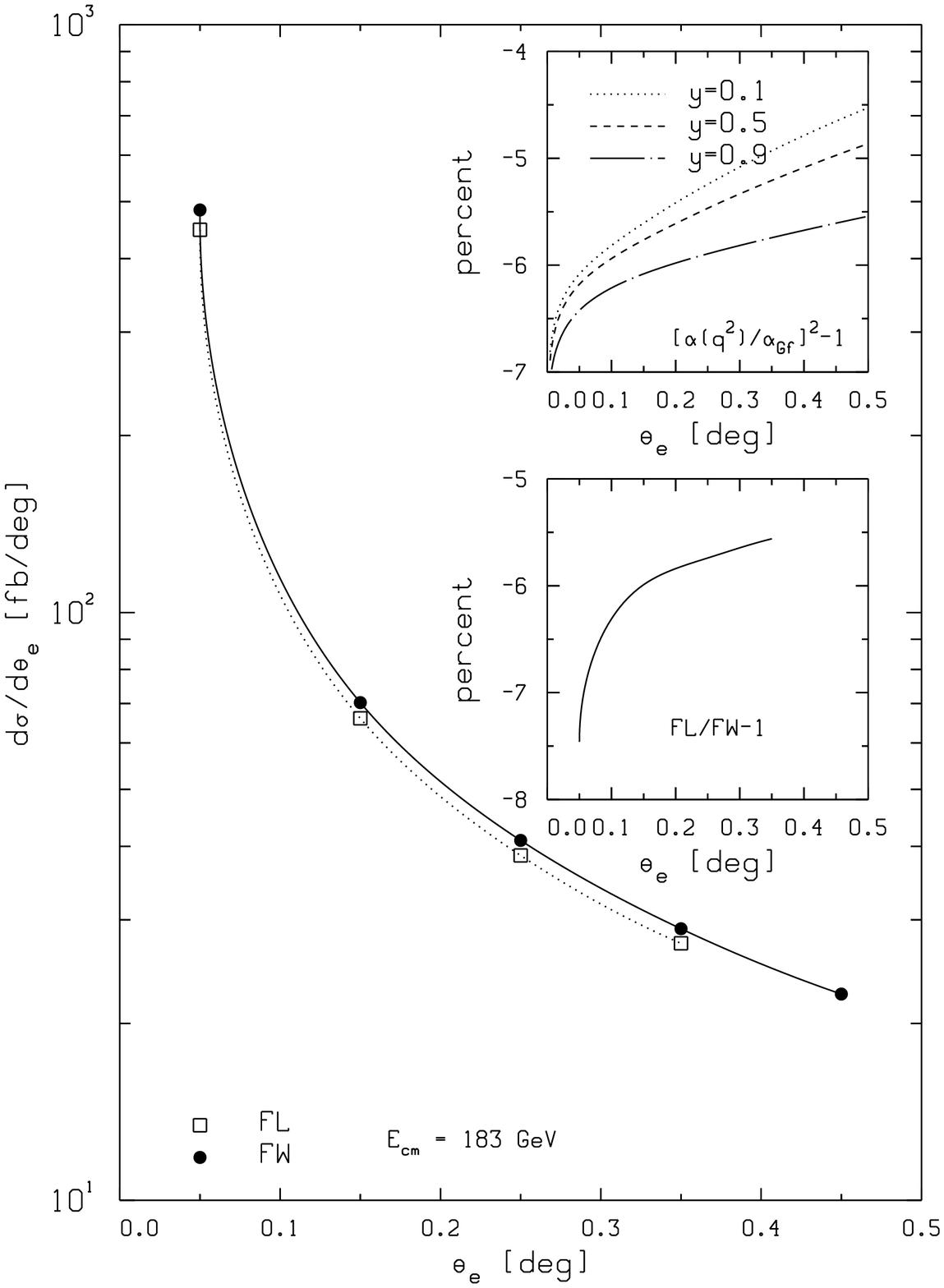,width=\linewidth}}
\caption[]{{\tt WTO} predictions for $d\sigma/d\cos\theta_e$ [fb/degrees] 
for $e^+e^- \to u \bard e^- \barnu_e$ with $M(u\bard) > 45\,$GeV and 
$\sqrt{s} = 183\,\GeV$.}
\label{fig:rs183}
\efi
From \fig{fig:rs183} we see that the EFL prediction is lower than the FW one,
from $-7.46\%$ in the bin $0^\circ - 0.1^\circ$ to $-5.56\%$ in the bin
$0.3^\circ - 0.4^\circ$. Correspondingly, the first bin is $6.78$ higher
than the second one, $11.60(16.37)$ than the third(fourth) one. This is not
a surprise, since the first bin represents $50\%$ of the total single-$\wb$
cross-section. 

Always in the same figure, we have reported the behavior
of $\Bigl[\alpha(q^2)/\alpha_{\gf}-1\Bigr]^2$ as a function of $\theta_e$ for
three values of $y$, using the appropriate relation: 
$q^2 = q^2(\theta_e,y)$, $y$ being the fraction of the 
electron energy carried by the photon. 
The behavior of EFL/FW-1, when we vary $\theta_e$, is very similar to the one
given by the ratio of coupling constants, indicating that the bulk of the 
effect is in the running of the e.m. coupling constant.

For completeness we have reported the numerical results for the three energies
in \tabn{wtotab1}, where the first entry is Fixed-Width distribution and the 
second entry is EFL one. Only the first four bins are shown, owing
to the fact that they are the most significant in the distribution. The third
entry in \tabn{wtotab1} gives EFL/FW-1 in percent.
 \begin{table}[htbp]\centering
 \begin{tabular}{|c||c|c|c|}
 \hline
 $\theta_e\,$[Deg] &  $\sqrt{s} = 183\,\GeV$  & $\sqrt{s} = 189\,\GeV$  & 
$\sqrt{s} = 200\,\GeV$ \\
 \hline
                                &           &           &               \\
   $0.0^\circ \div 0.1^\circ$   &  0.48395  & 0.54721   & 0.67147       \\
                                &  0.44784  & 0.50695   & 0.62357       \\
                                &  -7.46    & -7.36     & -7.13         \\
                                &           &           &               \\
 \hline
                                &           &           &               \\
   $0.1^\circ \div 0.2^\circ$   &  0.07026  & 0.07815   & 0.09323       \\
                                &  0.06605  & 0.07357   & 0.08798       \\
                                &  -5.99    & -5.86     & -5.63         \\
                                &           &           &               \\
 \hline
                                &           &           &               \\
   $0.2^\circ \div 0.3^\circ$   &  0.04095  & 0.04554   & 0.05433       \\
                                &  0.03860  & 0.04298   & 0.05141       \\
                                &  -5.74    & -5.62     & -5.37         \\
                                &           &           &               \\
 \hline
                                &           &           &               \\
   $0.3^\circ \div 0.4^\circ$   &  0.02897  & 0.03223   & 0.03845       \\
                                &  0.02736  & 0.03045   & 0.03646       \\
                                &  -5.56    & -5.52     & -5.18         \\
                                &           &           &               \\
 \hline
 \end{tabular}
\vspace*{3mm}
 \caption[
 ]{
$d\sigma/d\theta_e$ in [pb/degrees], from {\tt WTO}, for the process 
$e^+e^- \to e^- \barnu_e u \bard$, for $M(u\bard) > 45\,$GeV. First entry is 
Fixed-Width distribution, second entry is Fermion-Loop one and third entry 
is EFL/FW-1 in percent.}
 \label{wtotab1}
 \end{table}
 \normalsize

Next we consider $e^+e^- \to e\nu\mu\nu$, with $|\cos\theta_e| > 0.997$, 
$E_{\mu} > 15\,$GeV, and $|\cos\theta_{\mu}| < 0.95$.
In \tabn{wtotab3} we report the comparison between the EFL distribution and the
FW one for $\sqrt{s} = 183\,$GeV. As before, only the most significant bins 
are shown ($0.0^\circ \div 0.4^\circ$).
 \begin{table}[htbp]\centering
 \begin{tabular}{|c||c|c|c|}
 \hline
 $\theta_e\,$[Deg] &  FW  & EFL  & EFL/FW-1 (percent)  \\
 \hline
                                &           &           &               \\
   $0.0^\circ \div 0.1^\circ$   &  0.14154  & 0.13448   &  -4.99      \\
                                &           &           &               \\
 \hline
                                &           &           &               \\
   $0.1^\circ \div 0.2^\circ$   &  0.02113  & 0.02031   &  -3.88      \\
                                &           &           &               \\
 \hline
                                &           &           &               \\
   $0.2^\circ \div 0.3^\circ$   &  0.01238  & 0.01194   &  -3.55      \\
                                &           &           &               \\
 \hline
                                &           &           &               \\
   $0.3^\circ \div 0.4^\circ$   &  0.00880  & 0.00851   &  -3.30      \\
                                &           &           &               \\
 \hline
 \end{tabular}
\vspace*{3mm}
 \caption[
 ]{
$d\sigma/d\theta_e$ in [pb/degrees], from {\tt WTO}, for the process 
$e^+e^- \to e^- \barnu_e \nu_{\mu} \mu^+$, for $|\cos\theta_e| > 0.997$, 
$E_{\mu} > 15\,$GeV, and $|\cos\theta_{\mu}| < 0.95$. Furthermore, 
$\sqrt{s} = 183\,$GeV.}
 \label{wtotab3}
 \end{table}
 \normalsize
As for the hadronic case, the EFL prediction is considerably lower than the
FW one, although the percentage difference between the two is 
approximately $2.2\% \div 2.4\%$ smaller than in the previous case.
Useful comparisons will be presented in the {\tt WPHACT} description
of this Section.

A final comment will be devoted to QED ISR. Very often one can find the 
statement that the choice of the appropriate scale in the structure functions 
is mandatory. This is a jargon for `implementing the correct exponentiation
factor in multi-photon emission'. Note that the usual infrared exponent
$\alpha\,B$ is represented by
\bqa
\alpha\,B &=& \frac{2\,\alpha}{\pi}\,\big[ \frac{1+r^2}{1-r^2}\,
\ln\lpar\frac{1}{r}\rpar - 1\big] \sim \frac{2\,\alpha}{\pi}\,\lpar
\ln\frac{Q^2}{m^2} - 1\rpar, \quad \mbox{for}\quad Q^2 \gg m^2, \nl
\frac{m^2}{Q^2} &=& \frac{r}{(1-r)^2} \sim r,
\label{corrb}
\eqa
where $Q^2$ is the Mandelstam invariant associated with the emitting pair.
For $|t| \gg \mes$ the photon radiation is governed by
$\ln(|t|/\mes)$ rather than by $\ln(s/\mes)$. The difference is again a {\em
large log} and explain the excess of radiation generated by $s$-channel SF.
However, the whole expression for $B$ is known and not only its asymptotic
behavior (the scale). Therefore, for vanishing scattering angles, the correct 
behavior should be read from \eqn{corrb}. In this respect one should
remember that $|t_{\ph}|_{\rm min}$ in single-$\wb$ can be much lower than
$\mes$, being $\mes y^2/(1-y)$ where $y = M^2(\nu_e f_1 \barf_2)/s$.

\subsubsection*{Single-$\wb$ with {\tt WPHACT}}
\label{se:wphact}

\subsubsection*{Authors}

\begin{tabular}{l}
E. Accomando, A. Ballestrero and E. Maina \\
\end{tabular}

A new version of \wph~\cite{wph} is now available. It includes all 
massive matrix elements in addition to the previous ones which accounted for 
$b$-quark masses only. As before, the matrix elements are computed with the 
method of Ref.~\cite{BM}, which has proved to be fast and reliable in 
particular for massive calculations. 
New mappings of the phase space have been added, in order to account in an 
efficient way for the peaking structure of contributions like  single-$\wb$, 
single-$\zb$ and $\gamma \gamma$ contributions.
With the new version one has, therefore, the choice of using 
fully massive or massless calculations. The former are needed in various 
processes which diverge for massless fermions ,
while the latter are faster and give an excellent approximation 
for most cases. 
We start with the introduction of the IFL-scheme showing comparisons
with alternative solutions, designed to deal with gauge-invariance issues.
However, the most important part is contained in the second
Subsection where the effective scaling induced by $\alpha_{\rm QED}$ is 
presented.

\subsubsection*{IFL-scheme}
\label{se:ifl}

The Imaginary Part Fermion-Loop
scheme has been generalized to the fully massive case of non-con\-served weak 
currents in Ref.~\cite{abm}. The results obtained have been compared with 
other gauge restoring schemes used in single-$\wb$ processes computations. 
The following schemes have been considered in the analysis:

\begin{itemize}

\item[--] Imaginary-part FL scheme(IFL): 
    The imaginary part of the fermion-loop corrections, 
    as computed in Ref.~\cite{abm}, are used. Fermion masses  are 
    neglected only in loops but not in the rest of the diagrams. 

\item[--] Fixed width(FW):  The \PW-boson propagators show 
an unphysical width for $p^2<0$, but retains $U(1)$ gauge 
invariance in the CC20 process~\cite{BHF1}.

\item[--] Complex Mass(CM):
All weak boson masses squared $M^2_{_B}\;, B = \wb,\zb$ are changed to
$M^2_{\ssB}-i\,M_{\ssB}\Gamma_{\ssB}$ \cite{racoonww_ee4fa} ($\Gamma_{\ssB}$
is the on-shell $B$ width), including when they appear 
in the definition of the weak mixing angle. This scheme, which again 
gives an unphysical width in some cases, has however the advantage of
preserving both $U(1)$ and $SU(2)$ Ward identities.
    
\item[--] Overall scheme(OA):
    The diagrams for \processcctwenty\ can be split into two sets that are
    separately gauge invariant under $U(1)$. 
    In the actual implementation of the OA-scheme, $t$~channel diagrams
    are computed without any width and are then multiplied
    by $(q^2-M^2)/(q^2-M^2+iM\Gamma)$ where $q$, $M$ and $\Gamma$ are the
    momentum, the mass and  the width of the possibly-resonant \PW-boson.
    This scheme retains $U(1)$ gauge invariance  at the
    expenses of a mistreatment of the non-resonant terms.
    
\end{itemize}

In order to asses the relevance of current non-conservation, 
the imaginary part of the fermion-loop corrections 
have also been implemented with the
assumption that all currents that couple to the fermion-loop are conserved.
In this case the expressions of Ref.~\cite{abm} reduce to those
computed in ~\cite{BHF1}.
Note that the masses of external fermions are nonetheless taken into account
in the calculation of the matrix elements. This scheme violates $U(1)$
gauge-invariance by terms which are proportional to the fermion masses squared,
as already noted in Ref.~\cite{thosetwo}. However they are 
enhanced at high energy by large factors and can be numerically quite relevant.
This scheme will be referred to as the 
imaginary-part FL scheme with conserved currents (hereafter IFLCC).
All schemes described above have been implemented in the new version 
of \wph~\cite{wph} with the fully massive option.

\begin{table}[htb]\centering
\renewcommand{\arraystretch}{1.1}
\begin{tabular}{|c|c|c|c|} 
\hline
            & 190 GeV         & 800 GeV         & 1500 GeV           \\
\hline
\hline
IFL       &   0.11815 (13)   &   1.6978 (15)   &    3.0414 (35)     \\
\hline
FW        &   0.11798 (11)   &   1.6948 (12)   &    3.0453 (41)     \\
\hline
CM        &   0.11791 (12)  &  1.6953 (16)  &  3.0529  (60)   \\
\hline
OA        &   0.11760 (10)   &   1.6953 (13)   &    3.0401 (23)     \\   
\hline
IFLCC     &   0.11813 (12)   &   1.7987 (16)   &    5.0706 (44)     \\
\hline
\end{tabular}
\caption{Cross-sections in pb for the processes $e^+ e^-~\ra\; e^- \bar
\nu_e u\bard$ for various gauge restoring schemes.
No ISR is included and  we apply the following cuts:
$M(u \bard) > 5 \GeV\;, E_u > 3 \GeV, E_{\bard} > 3 \GeV,
\cos (\theta_e ) > .997$}
\label{tabschemes}
\end{table}

In \tabn{tabschemes} the cross-sections for CC20 are given
for the different gauge restoring schemes at LEP~2 and LC energies.
From it, one can immediately deduce that the IFL, FW, CM and the OA schemes
agree within $2\,\sigma$ in almost all cases. The IFLCC scheme agrees with the
other ones at LEP~2 energies but already at 800 $\GeV$ it overestimates the 
total cross-section by about $6\%$. At $1.5 \TeV$ the error is almost a factor 
of two.
On the contrary, even in the presence of non--conserved currents,
\ie of massive external fermions, the FW CM and OA calculations give 
predictions which are in agreement, within a few per mil, with the IFL scheme. 
The agreement with the results of a self-consistent approach justifies, from 
a practical point of view, the ongoing use of the FW, CM and OA schemes.
 
\begin{table}[htb]\centering
\begin{tabular}{|c|c|c|} 
\hline
            & IFL        & FW          \\
\hline
$e^-e^+\to e^-\bar{\nu}_e u\bar{d}$ \rule{3 mm}{0 mm}  $M_{u\bar{d}}> 45$ GeV
      &   0.12043 (10)   &   0.12041 (11)       \\
\hline
$e^-e^+\to e^-\bar{\nu}_e u\bar{d}$ \rule{3 mm}{0 mm} $M_{u\bar{d}}< 45$ GeV
      &   0.028585 (14)   &   0.028564 (14)       \\
\hline
$e^-e^+\to e^-\bar{\nu}_e \mu^+ \nu_\mu $
      &   0.035926 (34)   &   0.035886 (32)   \\
\hline
$e^-e^+\to e^-\bar{\nu}_e e^+ \nu_e $
      &   0.050209 (38)  &  0.050145 (32)   \\
\hline
\end{tabular}
\caption{Comparison of FW and IFL schemes for different single-$\wb$
cross-sections in pb and at $200$ GeV. No ISR is
included. Cuts are defined in the text. }
\label{tabfwifl}
\end{table}

The possible dependence of this agreement on the particular single-$\wb$ process
considered has been examined and we compare in \tabn{tabfwifl} the 
cross-sections obtained in the IFL and FW scheme at $\sqrt{s} = 200$ GeV. 
In this case, as in the following ones, the standard cuts have been applied:
the electron angle is limited in all processes by $|\cos \theta_e^-| > 0.997 $,
the other charged lepton by $|\cos \theta_l| < 0.95 $, its energy has to 
be $E_l > 15$ GeV.
These results confirm that, at LEP~2, there is no dependence of the 
cross-sections on the scheme.  Distribution of several observables have also 
been studied with \wph\ in the IFL and FW schemes. In most variables like 
the electron angle and energy no difference has been found. However, the mass 
spectrum of the $u\bard$ pair shows some scheme dependence, as reported in 
\fig{mud}. The physical motivation for this difference can be traced to the 
fact that the IFL scheme uses, correctly, a running $\wb$-width. 
In fact, comparing IFL mass distribution with a FW calculation in which $\wb$ 
mass and width are properly shifted~\cite{Belep1}, the difference is reduced 
to a small overall factor, as expected, and should not be viewed as a 
theoretical uncertainty.

In any case, in view of possible discrepancies, the use of IFL has to be 
preferred among the schemes analyzed in this section.
 
\begin{figure}[th]
\centerline{\epsfig{file=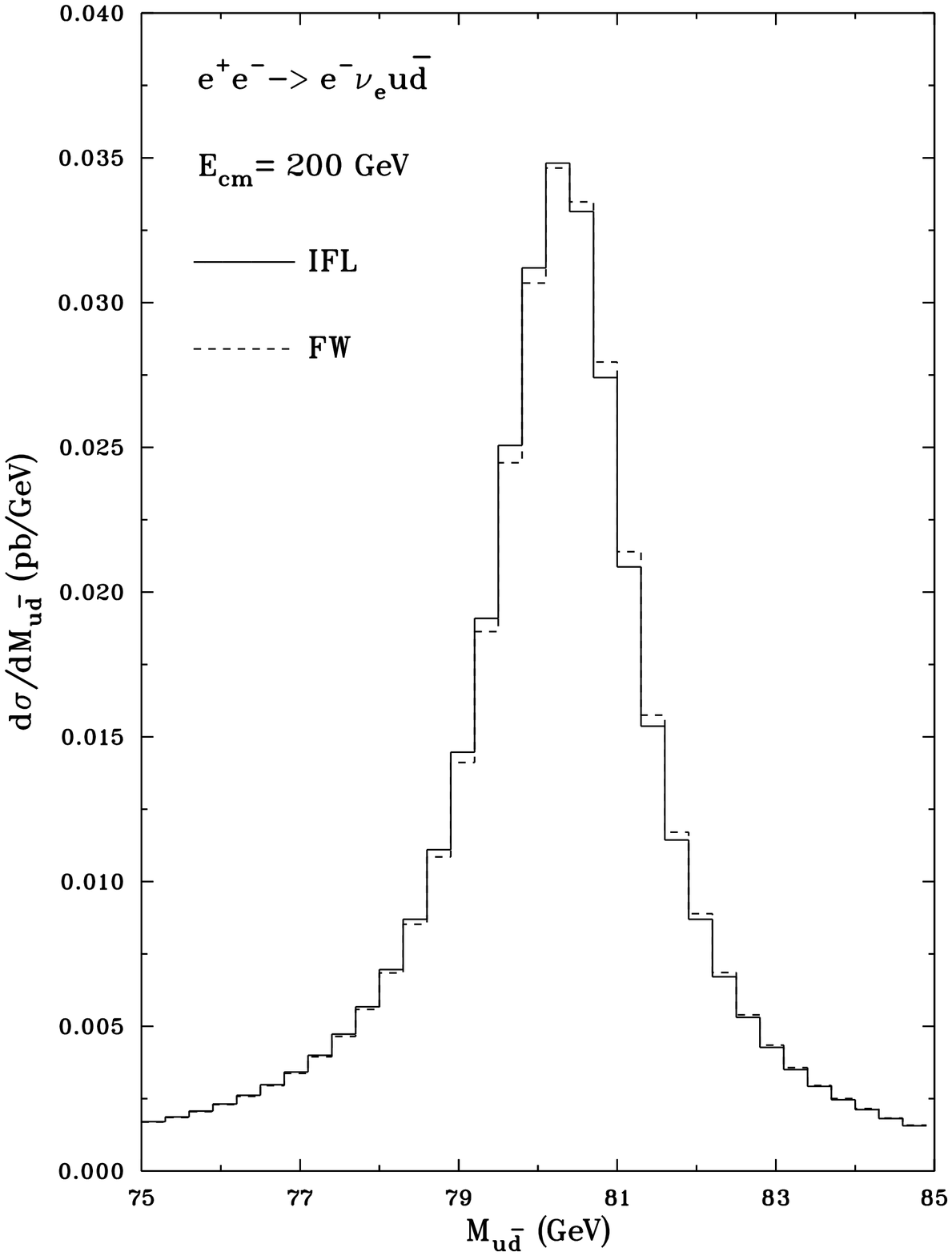,height=10cm,angle=0}}
\caption[]{Mass distribution of the $u\bard$ pair in \processcctwenty\
at $\sqrt{s}$ = 200 GeV in the IFL and FW schemes. No ISR.
$|\cos \theta_e^-| > 0.997 $ }
\label{mud}
\efi

\subsubsection*{Running of $\alpha$, comparisons with EFL.}
\label{se:alpha}

The EFL scheme implemented in \cite{tfl,nfl} for the massive  
(non-conserved currents) case solves the gauge-invariance problems 
exactly, as IFL does, but in addition it computes the real part of 
Fermion-Loop radiative corrections. These terms are known to determine the 
running of the couplings involved in single-$\wb$ processes. One may 
argue, therefore, that considering the running of $\alpha_{\rm QED}$ at an 
appropriate physical scale might account for the most relevant part of EFL 
corrections.
To test the correctness of this argument, a proper $\alpha_{\rm QED}$ 
evolution has been introduced as an option in \wph .
For every set of final momenta, $\alpha_{\rm QED}$ is evaluated at the scale 
$t$, the  virtuality of the photon emitted by the electron  line,   and 
used  for two vertices in the $t$-channel contributions only.  

The separate gauge invariance of $s$- and $t$-channel diagrams makes it 
possible to use a different $\alpha$ for them: $\alpha(t)$ for t-channel and 
$\alpha_{\gf}$ for $s$-channel. Such a separation, which can be
implemented in codes computing Feynman diagrams as \wph, should automatically 
account for the relative weight of $s$ and $t$ contributions for any set of
cuts. 

Computations performed with this choice will be referred to as IFL$_{\alpha}$. 
Several comparisons have been performed between the IFL and IFL$_{\alpha}$  
schemes and with the FW/EFL predictions by \wto \cite{wto}.

The good agreement of the two codes as far as FW and IFL schemes
are concerned is documented in \tabns{tabxsud}{tabangmu} for 
the cross-sections, the electron angular distribution and the quark invariant 
mass distribution. However, this has to be considered as a technical agreement 
more than a physical one.
Whether IFL$_{\alpha}$ can satisfactory reproduce the EFL 
complete calculations seems to depend on the process considered.
Note, in \tabn{tabxsud}, the agreement between IFL$_{\alpha}$ and EFL
for the total cross-section of the process \processcctwenty . 
Only at $200\,$GeV there is a disagreement of less than $0.5\%$ . Moreover, 
the angular distribution studied in \tabn{tabangud}, for the most relevant 
bins, never shows a higher discrepancy.
 
The variation of the cross-section of the process at hand with the invariant
mass $M(u\bard)$ cut is reported 
in \fig{mudmin} from which one deduces that the IFL
and the IFL$_{\alpha}$ schemes practically coincide when the cut reaches the 
mass of the $\wb$-boson.
In \tabn{tabmasud} one sees that, even varying the cuts, the difference 
between FL and IFL$_{\alpha}$ is at most of the order of $1\%$.

The conclusion is, therefore, that at LEP~2 and for \processcctwenty\,
the IFL$_{\alpha}$ scheme is reliable at the percent level.
The same does not apply to \processccemu, as can be verified with the help
of \tabn{tabxsmu} and \tabn{tabangmu}. From these one sees that the 
discrepancy is of the order of $2\%$ or worse.
This confirms that varying the scale of $\alpha_{\rm QED}$,  
on an event by event basis, is not completely satisfactory.
These numerical results point towards an estimate of about $3\%$ theoretical
error for single-$\wb$  predictions via the IFL$_{\alpha}$-scheme.
One can try to apply the running of $\alpha_{\rm QED}$ to only one vertex of 
the $t$-channel diagrams; the agreement obtained 
with this approximation (hereafter IFL$_{\alpha 1}$)
is much better for \processccemu. Of course, it becomes worse 
for \processcctwenty. At $183, 189$ and $200\,$GeV the cross-sections 
for \processccemu\ are respectively $25.65(1), 28.80(2), 34.86(2)$ fb, to be 
compared with the EFL results of \tabn{tabxsmu}.  The first bins of the angular 
distribution are also very close to EFL.  
No physical meaning has to be attributed to this fact: there is no
theoretical reason for using running $\alpha_{\rm QED}$ just at one vertex.
The agreement may be accidental and it is probably due to the fact that 
with the cuts used for
\processccemu\ the contribution of multi-peripheral diagrams is suppressed.

Since the IFL$_{\alpha}$ and IFL$_{\alpha 1}$ schemes are, in turn, in good 
agreement with complete EFL for different processes and cuts, the difference 
between their results will be used as an estimate of the theoretical error 
for  \processmixeevv\ and \processnceevv, where EFL predictions are not 
available. The cross-sections for such processes are presented 
in \tabn{tabxsee} and \tabn{tabangee}. 
The angular distributions for the four processes that we have discussed so 
far are reported in bins of $0.01$ degrees in \fig{ang}.
Note that the relevant part of the cross-section is 
concentrated in the first three or four bins,  also 
for \processmixeevv (Mix) and \processnceevv (NC), as well as for 
the two CC processes.
Finally the comparison between {\tt WPHACT} and {\tt WTO} has been extended to 
cover the LEP~2 signal definition for the hadronic decays of the $\wb$-boson,
but the results will not be presented here.

 \begin{table}[htbp]\centering
 \begin{tabular}{|c||c|c|c|c|c|}
 \hline
            &      &     &    &    &                \\
 $\sqrt{s}$ &  FW  & IFL &IFL$_{\alpha}$ & EFL & EFL/FW-1  \\
            &      &     &    &    &   (percent)             \\
 \hline
            &      &     &          &         &                \\
 $183\,\GeV$ & 88.17(44) & 88.50(4)   & 83.26(5) & 83.28(6)  & -5.5(5)  \\
            &      &     &          &    &                \\
 $189\,\GeV$ & 98.45(25) & 99.26(4)   & 93.60(9) & 93.79(7)  & -4.7(3)  \\
            &      &     &      &    &                \\
 $200\,\GeV$ & 119.77(67)& 120.43(10) & 113.24(8)& 113.67(8) & -5.1(5)  \\
            &      &     &     &    &                \\
 \hline
 \end{tabular}
\vspace*{3mm}
 \caption[
 ]{ 
Total single-$\wb$ cross-section in fb for the process $e^+e^- \to e^- \bar
\nu_e u \bard$, for $M(u\bard) > 45\,$GeV and $|\cos\theta_e| > 0.997$.
FW and EFL are computed by \wto, IFL and IFL$_{\alpha}$ by \wph. No ISR.
The number in parenthesis shows the statistical 
error of the numerical integration on the last digit.}
 \label{tabxsud}
 \end{table}
 \begin{table}[htbp]\centering
 \begin{tabular}{|c||c|c|c|c|c|}
 \hline
            &      &     &    &    &                \\
 $\theta_e\,$[Deg]  &  FW  & IFL &IFL$_{\alpha}$ & EFL & EFL/FW-1  \\
            &      &     &    &    &   (percent)             \\
 \hline
            &      &     &          &         &                \\
$0.0^\circ \div 0.1^\circ$ 
   &0.67147 & 0.67077    & 0.62404   & 0.62357 & -7.13   \\
            &      &     &           &    &                \\
$0.1^\circ \div 0.2^\circ$ 
   &0.09323  & 0.09321     & 0.08753   & 0.08798 & -5.63    \\
            &      &     &      &    &                \\
$0.2^\circ \div 0.3^\circ$ 
   &0.05433 & 0.05455    & 0.05141   & 0.05141 & -5.37     \\
            &      &     &     &     &                \\
$0.3^\circ \div 0.4^\circ$ 
  &0.03845  & 0.03867    & 0.03624   & 0.03646 &-5.18     \\
            &      &     &     &    &                \\
 \hline
 \end{tabular}
\vspace*{3mm}
 \caption[
 ]{
$d\sigma/d\theta_e$ in [pb/degrees] for the process $e^+e^- \to e^- \barnu_e 
u \bard$, for $M(u\bard) > 45\,$GeV, $\sqrt s$=200 GeV. No ISR.
FW and EFL are computed by \wto, IFL and IFL$_{\alpha}$ by \wph.
}
 \label{tabangud}
 \end{table}

 \begin{table}[htbp]\centering
 \begin{tabular}{|c||c|c|c|c|c|}
 \hline
                             &    &    &   &   &                \\
$M_{\rm min}(u\bard)$ & FW & IFL& IFL$_{\alpha}$ &EFL & EFL/FW-1     \\
                             &    &    &    &   &   (percent)    \\
 \hline
                             &    &    &   &   &                \\
45 & 0.04841(3) & 0.04845(3) & 0.04510(4) & 0.04478(3) & -7.5(1) \\
                             &    &    &   &   &                \\
35 & 0.05104(7)  & 0.05107(1) & 0.04754(1) & 0.04711(6) & -7.7(1) \\
                             &    &    &    &   &               \\
25 & 0.0546(1)  & 0.05467(2) & 0.05090(1) & 0.0504(1)   & -7.7(2) \\
                             &    &    &    &   &               \\
15 & 0.0595(1)  & 0.05968(2) & 0.05555(2) & 0.0552(1)  & -7.2(2) \\
                             &    &    &   &   &                \\
10 & 0.0626(1)  & 0.06283(2) & 0.05847(1) & 0.0582(1)  & -7.0(2) \\
                             &    &    &   &   &                \\
5  & 0.0659(1)  & 0.06623(2) & 0.06164(2) & 0.0615(1)  & -6.7(2) \\
                             &    &    &   &   &                \\
1  & 0.0682(1)  & 0.06864(1) & 0.06388(1) & 0.0637(1)  & -6.6(2) \\
                             &    &    &    &   &               \\
 \hline
 \end{tabular}
\vspace*{3mm}
 \caption[
 ]{
Cross-sections for the process $e^+e^- \to e^- \barnu_e 
u \bard$ in pb for
 $0.0^\circ < \theta_e < 0.1^\circ$ and $M(u\bard) \ge 
M_{\rm min}$ (in GeV). $\sqrt{s} = 183\,\GeV$. No ISR.
FW and EFL are computed by \wto, IFL and IFL$_{\alpha}$ by \wph.
The number in parenthesis shows the statistical 
error of the numerical integration on the last digit.}
 \label{tabmasud}
 \end{table}
 \begin{table}[htbp]\centering
 \begin{tabular}{|c||c|c|c|c|c|}
 \hline
            &      &     &    &    &                \\
 $\sqrt{s}$ &  FW  & IFL &IFL$_{\alpha}$ & EFL & EFL/FW-1  \\
            &      &     &    &    &   (percent)             \\
 \hline
            &      &     &          &         &                \\
 $183\,\GeV$ &26.77(14)  &26.45(1)  & 24.90(1)  & 25.53(4)  & -4.6(5)   \\
            &      &     &          &    &                \\
 $189\,\GeV$ & 29.73(14) &29.70(2)  & 27.98(2)  & 28.78(4)  & -3.2(5)     \\
            &      &     &      &    &                \\
 $200\,\GeV$ &36.45(23)  &35.93(4)  & 33.85(4)  & 34.97(6)  & -4.1(6)    \\
            &      &     &     &    &                \\
 \hline
 \end{tabular}
\vspace*{3mm}
 \caption[
 ]{
Total single-$\wb$ cross-section in fb for the process $e^+e^- \to e^- \bar 
\nu_e \mu^+ \nu_{\mu}$, for $|\cos\theta_e| > 0.997$, $E_{\mu} > 15\,$GeV, and 
$|\cos\theta_{\mu}| < 0.95$. No ISR.
FW and EFL are computed by \wto, IFL and IFL$_{\alpha}$ by \wph.
The number in parenthesis shows the statistical 
error of the numerical integration on the last digit.}
 \label{tabxsmu}
 \end{table}
 \begin{table}[htbp]\centering
 \begin{tabular}{|c||c|c|c|c|c|}
 \hline
            &      &     &    &    &                \\
 $\theta_e\,$[Deg]  &  FW  & IFL &IFL$_{\alpha}$ & EFL & EFL/FW-1  \\
            &      &     &    &    &   (percent)             \\
 \hline
            &      &     &          &         &                \\
$0.0^\circ \div 0.1^\circ$ 
            &  0.14154   &0.14170  &0.1319 & 0.13448  &   -4.99                     \\
          &      &     &           &    &                \\
$0.1^\circ \div 0.2^\circ$ 
            &  0.02113   &0.02117  &0.01987 & 0.02031   &  -3.88              \\
            &      &     &      &    &                \\
$0.2^\circ \div 0.3^\circ$ 
            &   0.01238  &0.01240  &0.01166 & 0.01194   &  -3.55                    \\
            &      &     &     &     &                \\
$0.3^\circ \div 0.4^\circ$ 
            &    0.00880 &   0.00879 &0.00830  & 0.00851   &  -3.30           \\
            &      &     &     &    &                \\
 \hline
 \end{tabular}
\vspace*{3mm}
 \caption[
 ]{
$d\sigma/d\theta_e$ in [pb/degrees] for the process $e^+e^- \to e^- \barnu_e 
\nu_{\mu} \mu^+$, for $|\cos\theta_e| > 0.997$, 
$E_{\mu} > 15\,$GeV, and $|\cos\theta_{\mu}| < 0.95$. 
$\sqrt{s} = 183\,$GeV. No ISR. 
FW and EFL are computed by \wto, IFL and IFL$_{\alpha}$ by \wph.
}
 \label{tabangmu}
 \end{table}
 \begin{table}[htbp]\centering
 \begin{tabular}{|c||c|c|c|c|}
 \hline
            &            &      &             &          \\
 $\sqrt{s}$ & final state& IFL  & IFL$_{\alpha}$ & IFL$_{\alpha 1}$   \\
            &            &      &             &                \\
 \hline
            &            &      &             &                \\
 $183\,\GeV$ & $e^+e^-\nu_e\barnu_e$ 
& 38.24(1)  & 35.99(1)   &  37.10(2)    \\
              & $e^+e^-\nu_\mu\barnu_\mu$ 
& 12.81(1)  & 12.05(1) &  12.42(1)    \\
            &            &      &             &                \\
 $189\,\GeV$ & $e^+e^-\nu_e\barnu_e$ 
& 42.38(1)  & 39.86(1) & 41.09(2)     \\
              & $e^+e^-\nu_\mu\barnu_\mu$ 
& 13.74(1)  & 12.92(1) & 13.32(1)     \\
            &            &      &             &                \\
 $200\,\GeV$ & $e^+e^-\nu_e\barnu_e$ 
& 50.20(2)  & 47.25(2) & 48.70(2)     \\
              & $e^+e^-\nu_\mu\barnu_\mu$ 
& 15.37(1)  & 14.46(1) & 14.91(1)     \\
            &            &      &             &                \\
 \hline
 \end{tabular}
\vspace*{3mm}
 \caption[
 ]{
Total single-$\wb$ cross-section in fb by \wph\ for the processes 
$e^+e^- \to e^+ e^- \nu \barnu $ 
 for $|\cos\theta_e^-| > 0.997$, $E_{e^+} > 15\,$GeV, and 
$|\cos\theta_{e^+}| < 0.95$. No ISR.
The number in parenthesis shows the statistical 
error of the numerical integration on the last digit.}
 \label{tabxsee}
 \end{table}
 \begin{table}[htbp]\centering
 \begin{tabular}{|c||c|c|c|c|}
 \hline
            &            &      &             &          \\
 $\theta_e\,$[Deg] & final state& IFL  & IFL$_{\alpha}$ & IFL$_{\alpha 1}$   \\
            &            &      &             &                \\
 \hline
            &            &      &             &                \\
$0.0^\circ \div 0.1^\circ$ & $e^+e^-\nu_e\barnu_e$ 
& 0.27958  & 0.260390 &  0.26990    \\
              & $e^+e^-\nu_\mu\barnu_\mu$ 
& 0.09007  & 0.08386  & 0.08693     \\
            &            &      &             &                \\
$0.1^\circ \div 0.2^\circ$ & $e^+e^-\nu_e\barnu_e$ 
& 0.03890   & 0.03643 & 0.03764      \\
              & $e^+e^-\nu_\mu\barnu_\mu$ 
& 0.01158  & 0.01086  & 0.01123     \\
            &            &      &             &                \\
$0.2^\circ \div 0.3^\circ$  & $e^+e^-\nu_e\barnu_e$ 
& 0.02279   & 0.02146 & 0.02216      \\
              & $e^+e^-\nu_\mu\barnu_\mu$ 
& 0.00680  & 0.00641  & 0.00660     \\
            &            &      &             &                \\
$0.3^\circ \div 0.4^\circ$  & $e^+e^-\nu_e\barnu_e$ 
& 0.01622   & 0.01535 & 0.01573     \\
              & $e^+e^-\nu_\mu\barnu_\mu$ 
& 0.00482  & 0.00456  & 0.00467     \\
            &            &      &             &                \\
 \hline
 \end{tabular}
\vspace*{3mm}
 \caption[
 ]{
$d\sigma/d\theta_e$ in [pb/degrees] by \wph\ for the processes 
$e^+e^- \to e^+ e^- \nu \barnu $ 
 for $|\cos\theta_e^-| > 0.997$, $E_{e^+} > 15\,$GeV, and 
$|\cos\theta_{e^+}| < 0.95$.
$\sqrt{s} = 200\,$GeV. No ISR. 
}
 \label{tabangee}
 \end{table}

\bfi
\centerline{\epsfig{file=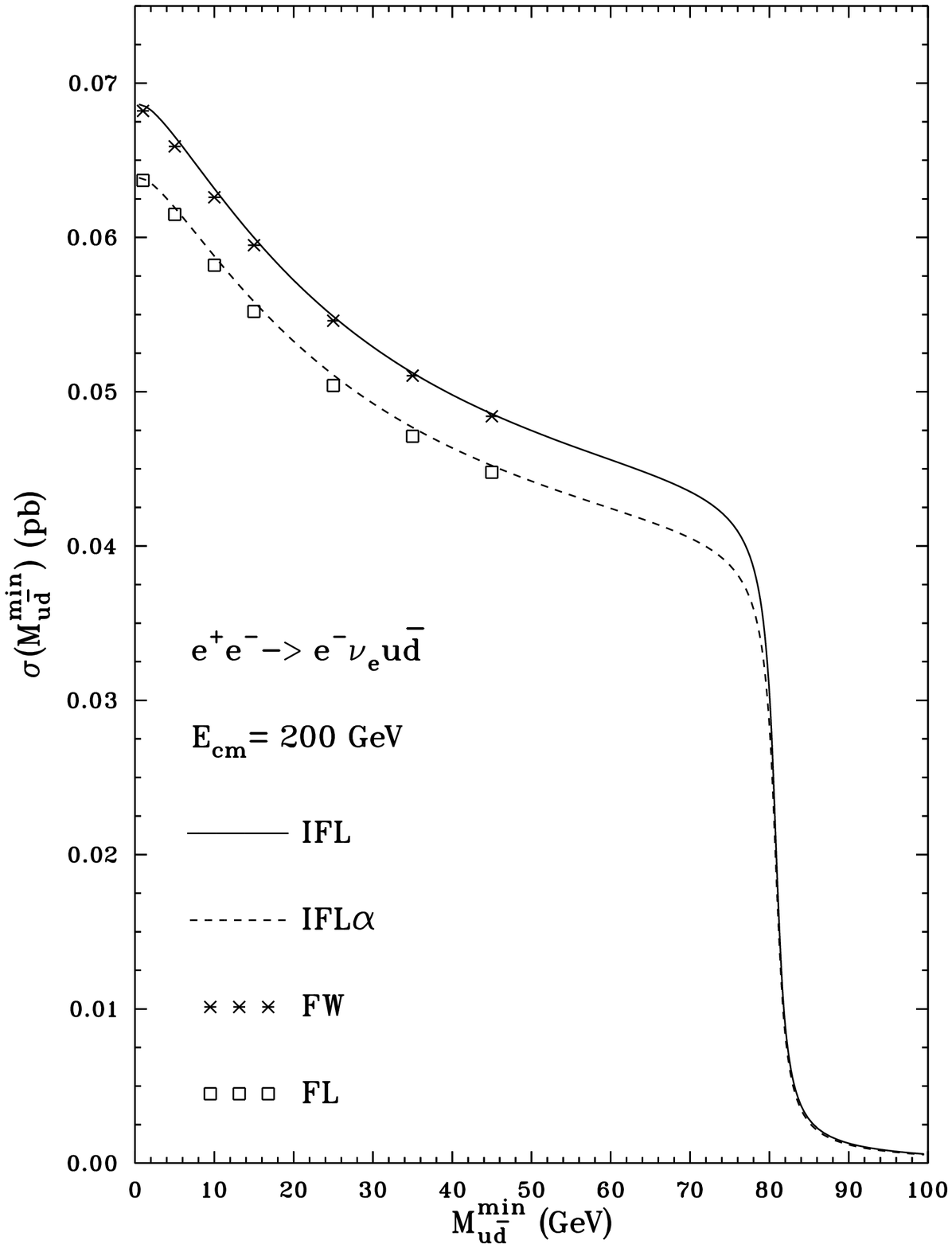,height=12cm,angle=0}}
\caption[]{Total cross-section for \processcctwenty\ at $\sqrt{s}$ = 200 GeV
with $\theta_e\,<\,0.1^\circ$ as a function of the lower 
cut on M$_{u\bard}$ in IFL and IFL$\alpha$ 
schemes. The markers give the results of FW and FL by \wto .  }
\label{mudmin}
\efi


\clearpage

\bfi
\centerline{\epsfig{file=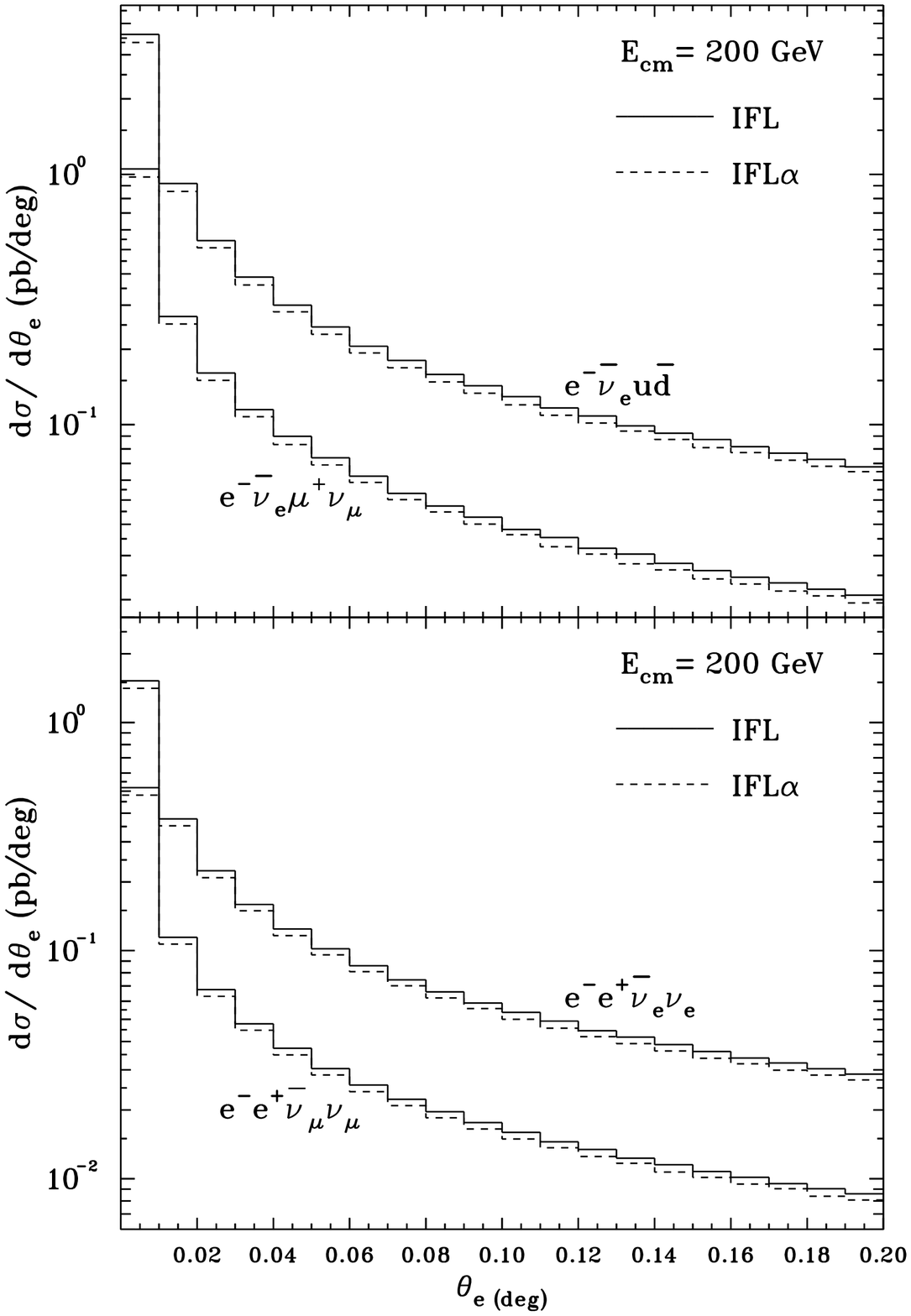,height=12cm,angle=0}}
\caption[]{Angular distributions for different single-$\wb$ processes at
$\sqrt{s}$ = 200 GeV in the IFL and IFL$\alpha$ scheme.}
\label{ang}
\end{figure}

\subsubsection*{Single-$\wb$ and {\tt SWAP}.}

\subsubsection*{Authors}

\begin{tabular}{l}
G. Montagna, M. Moretti, O. Nicrosini, A. Pallavicini and F. Piccinini \\
\end{tabular}

\subsubsection*{Description of the Method}
\label{mesw}
Contributions of the Pavia/{\tt ALPHA} group to the subject of 
single-$\wb$ production are summarized.
The exact matrix elements for single-$\wb$ production 
are computed by means of the {\tt ALPHA} algorithm \cite{alpha} for
the automatic evaluation of the 
Born scattering amplitudes. Fermion masses are exactly 
accounted for in the kinematics and dynamics. 
The contribution of anomalous trilinear gauge couplings
is also taken into account. The anomalous gauge boson couplings 
$\Delta k_{\ph}$, $\lambda_{\ph}$, 
$\delta_{\ssZ}$, $\Delta k_{\ssZ}$ and $\lambda_{\ssZ}$ are 
implemented according to the
parameterization of refs.~\cite{gg,hhpz}. The fixed-width 
scheme is adopted as gauge-restoring approach, as 
motivated in comparison with other gauge-invariance-preserving 
schemes in Ref.~\cite{abm}.

\subsubsection*{Radiative corrections}
\label{rcsw}

Leading-log (LL) QED radiative corrections 
are implemented via the Structure Function (SF) formalism 
in the collinear approximation~\cite{sf}. The $Q^2$-scale entering 
the SF $D(x,Q^2)$ is fixed by comparing the $\ord{\alpha}$ 
expansion of the SF method with the analytic results obtained 
for the $\ord{\alpha}$ double-log photonic corrections as given by soft-photon 
bremsstrahlung from the external legs, its virtual 
counterpart and hard-photon radiation collinear to the final-state particles. 
Notice that, since the goal is to 
determine the scale entering the SF, only the contribution 
of real photons is explicitly calculated, because the 
virtual corrections, in order to preserve the
cancellations of infrared 
singularities, must share the same leading collinear structure of the real
part itself. More details about the derivation in the 
present approach of the soft/collinear 
limit of the $\ord{\alpha}$ correction can be found in \cite{pvscales}.

For example, for the process $e^+e^- \rightarrow e^-{\barnu}u\bard$, 
this comparison translates in the following two $Q^2$-scales: 
(two initial-state (IS) SF are assumed: 
$Q_-^2$  refers to the SF attached to the incoming electron, 
while $Q_+^2$ to the SF attached to the incoming positron)~\cite{pvscales}
\begin{equation}
\label{eq:wscales}
Q_-^2 = 4E^2 \frac{(1-c_-)^2}{\delta^2} \;,\quad
Q_+^2 = 2^{\frac{14}{9}}\,E^2
\frac{\big((1-c_{\bard})(1-c_u)^2\big)^\frac{2}{3}}
{\big((1-c_{u\bard})^2\delta^5\big)^\frac{2}{9}}
\end{equation}
\noindent where $E$ is the beam energy, $c_-$ the cosine
of the electron scattering angle, $c_u$ and $c_{\bard}$ the cosine of 
the quark scattering angles with respect to the initial positron, 
$c_{u\bard}$ the cosine of the relative angle between the quarks, 
$\delta$ the half-opening angle of the electromagnetic jet 
(calorimetric angular resolution). 
It is worth noticing that in the numerical implementation, whenever one of the
two scales is less than a small cut-off 
($\Lambda^2_{\rm cut-off} = 4 \mes$, where 
$\me$ is the electron mass), the 
radiation from the corresponding leg is switched off, in accordance with 
the expected power law behaviour.\footnote{Although this behavior
is exactly known and could be implemented, {\tt SWAP} has evidence for a 
corresponding small effect.}
It was carefully tested that variations of the cut-off do not alter the 
numerical results.

Also a naive ansatz for the 
two scales, as motivated by an analysis of the single-$\wb$ 
process in terms of the Weizs\"acker-Williams approximation, 
can be given~\cite{pvscales} as follows:
\begin{equation}
\label{eq:naive}
  Q_{-,{\rm naive}}^2 = |q^2_{\ph^*}| \;,\quad
  Q_{+,{\rm naive}}^2 = \mws
\end{equation}
where $q^2_{\ph^*}$ is the squared momentum transfer in the $e e \ph^*$ 
vertex and $\mw$ is the mass of the $\wb$ boson. 

The effect of vacuum polarization is also taken into 
account in the calculation, by including the contribution of leptons, 
heavy quarks and light quarks, the latter according to the 
standard parameterization of Ref.~\cite{ej}.

\subsubsection*{Computational tool and obtained results}
\label{rosw}

The theoretical features sketched above have been implemented into a massive 
MonteCarlo (MC) program, named {\tt SWAP} 
({\tt S}ingle {\tt W} process with {\tt A}lpha \& {\tt P}avia). The 
multi-channel importance sampling technique is 
employed to perform the phase-space integration.  
The code supports realistic event selections and can be employed either as
a cross-section calculator or as a true event generator.
The main results obtained in the present study can be summarized as follows:
we have performed a critical analysis of the energy scale for QED 
radiation (see \fig{fig1_sw});
Next, we have evaluated the effect of a running of $\alpha_{\rm QED}$  
(see \fig{fig2_sw});
Finally we have performed a tuned comparisons with other codes.

\begin{figure}
\begin{center}
\includegraphics[width=\linewidth]{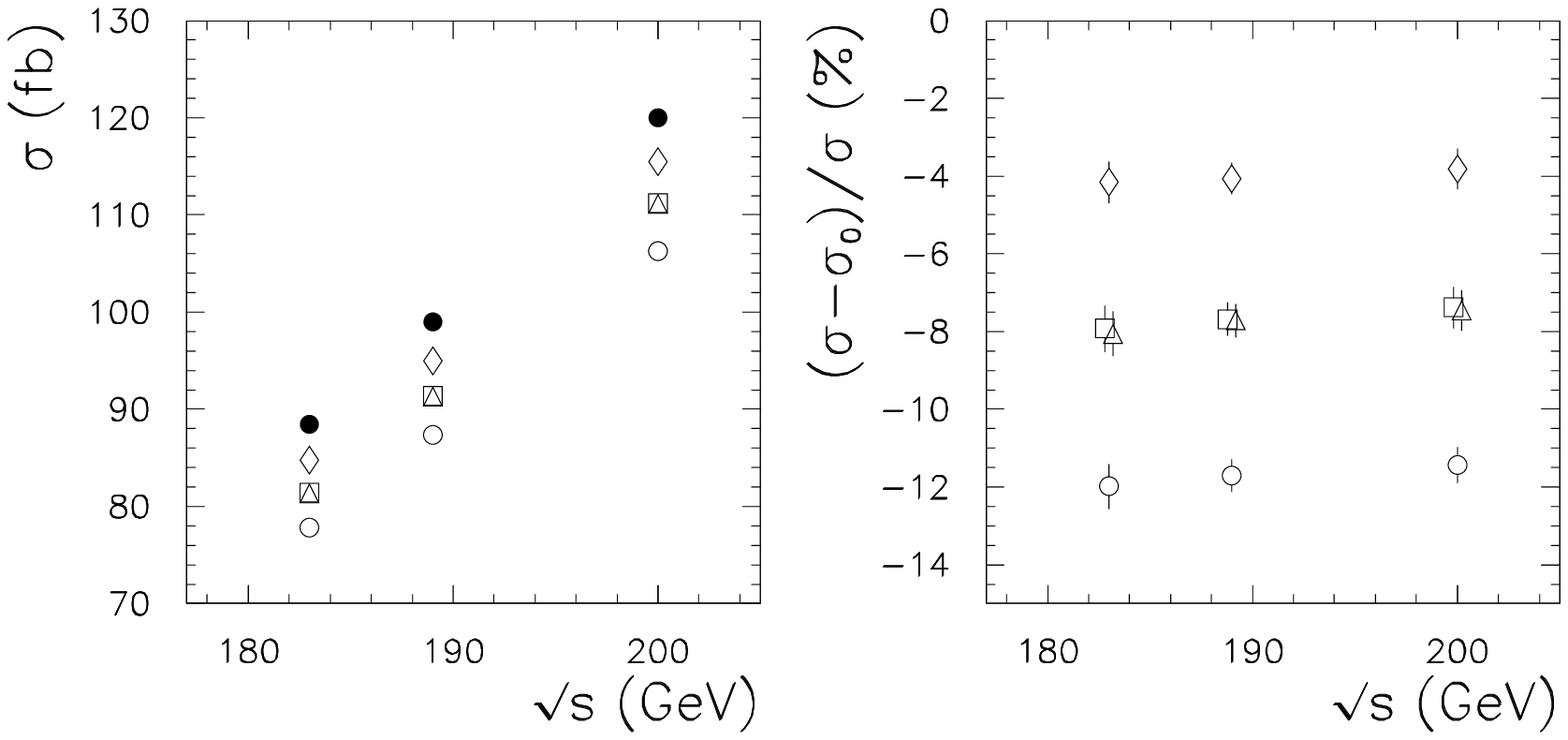}
\caption{The effect of LL QED corrections to the cross 
section of the single-$\wb$ process
$e^+e^-\rightarrow e^-\bar{\nu}u\bar{d}$ for different choices 
of the $Q^2$-scale in the electron/positron SF. 
Left: absolute cross
section values; Right: relative difference 
between QED corrected cross-sections and the Born one. 
The marker
$\bullet$  represents the Born cross-section,
$\scriptstyle\bigcirc$ represents the correction
given by $Q^2_{\pm} = s$ scale, 
$\scriptstyle\diamondsuit$ represents the correction
given by $Q^2_{\pm} = |q^2_{\ph^*}|$ scale,
$\scriptstyle\triangle$ the correction given by the scales 
of eq.~(\ref{eq:wscales}),
the correction given by the naive scales of eq.~(\ref{eq:naive}). 
The entries correspond to 
183, 189, 200 GeV}. 
\label{fig1_sw}
\end{center}
\end{figure}

Input parameters and cuts used to obtain the numerical 
results shown in the following are those of the $4\rmf$ proposal 
for the process $e^+e^-\rightarrow e^-\bar{\nu}u\bar{d}$ 
($|\cos\vartheta_e| > 0.997, M_{u\bard} > 45$~GeV). 
For \fig{fig1_sw} the value of $\delta$ 
parameter entering eq.~(\ref{eq:wscales}) is $\delta = 5^\circ$, 
but it has been checked that the numerical results are very marginally
affected by its actual value.

{\em Scales \& QED radiation}.
In \fig{fig1_sw} the numerical impact of different 
choices of the $Q^2$-scale on the 
cross-section of the single-$\wb$ process
$e^+e^-\rightarrow e^-\bar{\nu}u\bar{d}$ is shown. 
The marker $\bullet$ represents the Born cross-section,
$\scriptstyle\bigcirc$ represents the correction
given by $Q^2_{\pm} = s$ scale for both IS SF(s), 
$\scriptstyle\diamondsuit$ represents the correction
given by $Q^2_{\pm} = |q^2_{\ph^*}|$ scale for both IS SF(s),
$\scriptstyle\triangle$ the correction given by the scales 
of eq.~(\ref{eq:wscales}),
the correction
given by the naive scales of eq.~(\ref{eq:naive}). 
It can be seen that neither the 
$s$ scale, as implemented in computational tools used 
for the analysis of the single-$\wb$ process, nor the 
$|q^2_{\ph^*}|$ 
scale, as recently proposed in Ref.~\cite{bd}, 
are able to reproduce the effects due to the scales 
of eq.~(\ref{eq:wscales}) and eq.~(\ref{eq:naive}). These 
two scales are in good agreement and both predict a 
lowering of the Born cross-section of about $8\%$, almost 
independent of the c.m.s. LEP~2 energy. 
Note the $4\%$ difference between ISR with $s$-scale and the new scale.

{\em Running of $\alpha_{\rm QED}$}.
Because $\gf$, $\mw$ and $\mz$ are the agreed input parameters 
in the $4f$ proposal, the value of the e.m. coupling 
constant $\alpha$ is fixed at tree-level to a high energy value as
specified by the $\gf$-scheme.
On the other hand, the single-$\wb$ process is a 
$q^2_{\ph^*} \simeq 0$ dominated process and therefore the above high-energy 
evaluation of $\alpha$, $\alpha_{\gf}$, needs to be rescaled to its 
correct value at small momentum transfer. In order to take into 
account the effect of the running 
of $\alpha_{\rm QED}$ in a gauge invariant way, a {\em re-weighting}
procedure can be adopted, by simply rescaling the differential cross 
section $d\sigma/dt$ ($t \equiv q^2_{\ph^*}$) in the following way
\bq
\label{eq:running}
{{d\sigma}\over{dt}} 
\rightarrow {{\alpha^2(0)}\over{\alpha^2_{\gf}}}
{{d\sigma}\over{dt}} \;, \qquad 
{{d\sigma}\over{dt}} 
\rightarrow {{\alpha^2(t)}\over{\alpha^2_{\gf}}}
{{d\sigma}\over{dt}} \;,
\eq
where $\alpha(0), \alpha(t)$ is the QED running coupling 
computed at virtuality $q^2_{\ph^*}$ 
equal to $0$ and $t$, respectively. 

\fig{fig2_sw} 
shows the effects of the above re-weighting procedure. 
The $\scriptstyle\triangle$ represent the relative 
difference between the integrated cross-section computed 
in terms of $\alpha_{\gf}$ and the cross-section 
computed in terms of $\alpha(0)$, while 
$\scriptstyle\diamondsuit$ is the relative 
difference between the integrated cross-section computed 
in terms of $\alpha_{\gf}$ and the cross-section 
computed in terms of $\alpha(t)$. 
As can be seen, the rescaling from $\alpha_{\gf}$ to $\alpha(t)$ 
introduces a negative correction of about $5-6\%$ in the LEP~2 
energy range.
The difference between 
$\scriptstyle\triangle$ and $\scriptstyle\diamondsuit$, 
which is about $2-3\%$,  
is a measure of the running of $\alpha_{\rm QED}$ from 
$q^2_{\ph^*} = 0$ to $q^2_{\ph^*} = t$.
A detailed numerical analysis 
of the effect of the running couplings in single-$\wb$ 
production has been very recently 
performed in Ref.~\cite{tfl}, based on the
theoretical results of the massive fermion-loop scheme 
of Ref.~\cite{tfl}. 
The results for the running of $\alpha_{\rm QED}$, 
as shown in \fig{fig2_sw}, are 
in agreement with those of Ref.~\cite{nfl}, as far as 
the effect of $\alpha_{\rm QED}$ is concerned, which is the bulk of the 
EFL contribution, leaving residual differences at the level of $1-2\%$, 
depending on the considered channel and event selection, see also the
discussion in the {\tt WPHACT} part of this Section.

\clearpage

\begin{figure}[ht]
\begin{center}
\includegraphics[width=0.9\linewidth]{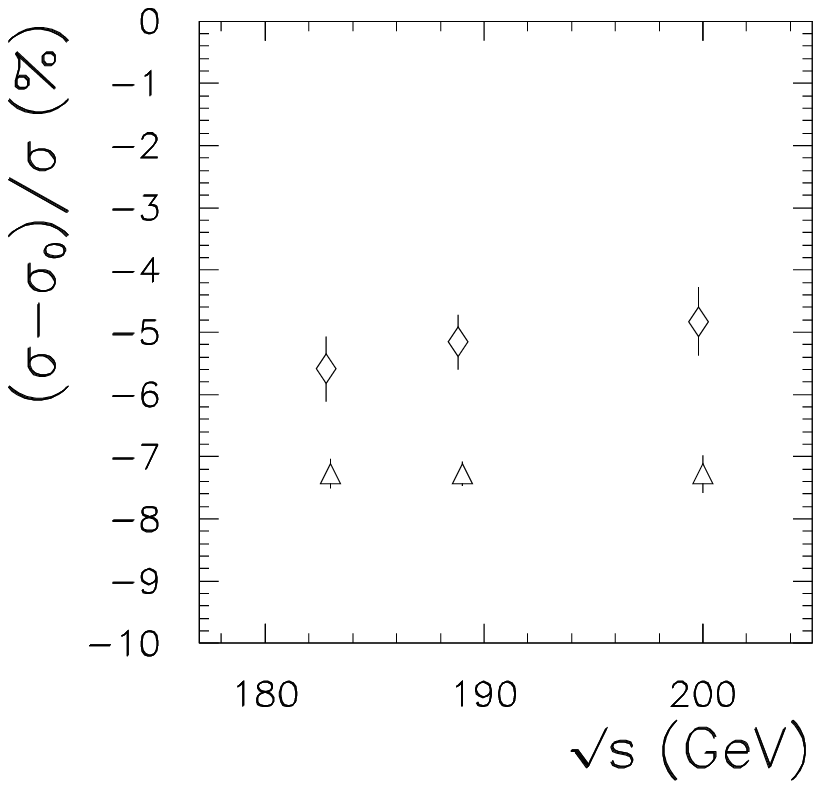}
\caption[]{The effects of the rescaling of 
  $\alpha_{\rm QED}$ from $\alpha_{\gf}$ to $\alpha(q^2_{\ph^*} = 0)$
  ($\scriptstyle\triangle$) and $\alpha(q^2_{\ph^*})$
  ($\scriptstyle\diamondsuit$) on the integrated cross section of the
  single-$\wb$ process $e^+e^-\rightarrow e^-\bar{\nu}u\bar{d}$.
  $\sigma_0$ is the cross-section computed in terms of $\alpha_{\gf}$.
  The entries correspond to 183, 189, 200 GeV. }
\label{fig2_sw}
\end{center}
\end{figure}

\subsubsection*{{\tt NEXTCALIBUR}}
\label{swnexca}

\subsubsection*{Authors}

\begin{tabular}{l}
F.A.Berends, C.~G.~Papadopoulos and R.Pittau \\
\end{tabular}

This section describes the features of a new Monte Carlo program
{\tt NEXTCALIBUR}~\cite{nexca}, which aims at keeping the advantages of
{\tt EXCALIBUR} \cite{exca}, but tries to improve on its
shortcomings.
The advantages, which should be kept are the high speed of the
program and the applicability to all possible 4-fermion final states.
The shortcomings of {\tt EXCALIBUR}, which are partly related
to its assets, are the massless nature of its fermions, the
inclusive treatment of ISR QED corrections (no $p_t$ from
a photon in an event) and the neglect of any running of coupling
constants.

\subsubsection*{The strategy of the code}

To start with, it should be noted that unless stated otherwise
complex gauge boson masses and a complex weak mixing angle are used
to ensure gauge invariant matrix elements \cite{racoonww_ee4fa}. This
procedure has been shown to work well \cite{abm}. The
various wanted improvements will now be successively
discussed.

\subsubsection*{Inclusion of fermion masses}

Inside the program a massive matrix element is needed, for the
calculation of which a recursive method \cite{costas} is used.

This massive matrix element now exists in the whole phase
space, since the singularities of the massless case are 
regularized. Nevertheless serious numerical cancellations
take place in very specific situations. The most dramatic
case is caused by the photonic multi-peripheral diagrams
which blow up for forward scattering. When at the same
time both electron and positron move in the forward direction,
it becomes necessary to perform the calculation in
quadruple precision. When only one is moving in the forward
direction the usual double precision is sufficient.
A version of the program using double precision in
all possible situations is currently under study.

The phase space generation is an extension of the treatment
in  {\tt EXCALIBUR}, \ie a self-adjusting multi-channel
approach, now including the multi-peripheral situation in
an improved form.

With the above mentioned ingredients one indeed has an
event generator for any massive four-fermion final state.
In particular, for the potentially dangerous kinematical
situations events can now be generated, like forward 
single W-production or $\gamma-\gamma$ processes. 
Also all channels, where Higgs exchange can take place now 
indeed contain Higgs exchanges.

To demonstrate the ability of the program to cover all phase-space regions,
without loosing efficiency, we show, in \tabns{pitt1}{pitt2}, the total 
cross-sections for the processes $e^+ e^- \to e^+ e^- \mu^+ \mu^-$ and
$e^+ e^- \to e^+ e^- e^+ e^-$. Where available, we compare our predictions
with the QED numbers published in Ref. \cite{ggtree}.

\begin{table}[hb]\centering
\begin{tabular}{|l||c|c|} \hline 
$\sqrt{s}$       & {\tt BDK}  
                 & {\tt NEXTCALIBUR}              
      \\ \hline \hline
 20      &          98.9 $\pm$ 0.6 &   99.20 $\pm$ 0.98  \\ \hline          
 35      &         131.4 $\pm$ 2.2 &  131.03 $\pm$ 0.88  \\ \hline
 50      &         154.4 $\pm$ 0.9 &  152.33 $\pm$ 0.83  \\ \hline
100      &         205.9 $\pm$ 1.2 &  204.17 $\pm$ 1.73  \\ \hline
200      &            ---          &  263.50 $\pm$ 1.31  \\ \hline 
200 (all)&            ---          &  265.58 $\pm$ 1.44  \\ \hline  
\end{tabular}
\vspace*{3mm}
\caption[]{$\sigma_{tot}$ (in nb) for the process
$e^+ e^- \to e^+ e^- \mu^+ \mu^-$. Only QED diagrams, 
except in the last entry.}
\label{pitt1}
\end{table}

\begin{table}[ht]\centering
\begin{tabular}{|l||c|c|} \hline 
$\sqrt{s}$       & {\tt BDK}  
                 & {\tt NEXTCALIBUR}              
      \\ \hline \hline
 20      & 0.920         $\pm$ .011 &0.905   $\pm$ .011  \\ \hline          
 35      & 1.070         $\pm$ .015 &1.079   $\pm$ .014  \\ \hline
 50      & 1.233         $\pm$ .018 &1.214   $\pm$ .016  \\ \hline
100      & 1.459         $\pm$ .025 &1.485   $\pm$ .020  \\ \hline
200      &              ---         &1.776   $\pm$ .019  \\ \hline 
200 (all)&              ---         &1.787   $\pm$ .030  \\ \hline  
\end{tabular}
\vspace*{3mm}
\caption[]{$\sigma_{tot}$ (in nb $ \times 10^7$) for the process
$e^+ e^- \to e^+ e^- e^+ e^-$. Only QED diagrams, 
except in the last entry.}
\label{pitt2}
\end{table}

\noindent {\tt NEXTCALIBUR} 
contains all electroweak  diagrams, and can therefore 
be used to compute the electroweak background to the above 
$\gamma\,\gamma$ processes.
By looking at the last entry of the tables, the latter is found to be less
than $1\%$ at LEP~2 energies, at least for totally inclusive quantities.

\noindent All numbers have been produced at the Born level, but ISR and running
$\alpha_{\rm QED}$ can be included as described in the next sections.

\subsubsection*{Taking into account the correct scale}
\label{se:excasca}

As mentioned above, the matrix element calculation can
easily be modified. One option would be to take into account
Fermion-Loop corrections, which becomes relevant when
there are different scales in the matrix element, \eg
due to small $t$-channel scales.
A possible solution is the Fermion-Loop approach of Refs. 
\cite{BHF2,nfl}, where all fermion corrections  
are consistently included by introducing 
running couplings $g(s)$ and $e(s)$, together with
the re-summed bosonic propagators.

In presence of the $\wb\wb \gamma$ vertex, the above ingredients 
are not sufficient to ensure gauge invariance, because loop 
mediated vertices have to be consistently included. 
On the contrary, when no $\wb\wb\gamma$ vertex is present, 
the neutral gauge boson vertices, induced by the 
Fermion-Loop contributions, are separately gauge invariant \cite{BHF2}.

\noindent Instead of explicitly including the loop vertices, we follow 
a {\em Modified Fermion-Loop} approach. 
Namely, we neglect the separately 
gauge invariant neutral boson vertices, and include 
only the part of the $\wb\wb\ph$ loop function necessary 
to renormalize the bare $\wb\wb\ph$ vertex and 
to insure the $U(1)$ gauge invariance. 
Our procedure is as follows: besides running couplings,
we use bosonic propagators
\bqa
&&P_w^{\mu\nu}(s)= \left(s-M^2_w(s)\right)^{-1}
 \left(g_{\mu\nu}-\frac{p_\mu p_\nu}{M^2_w(s)}\right)
 \nonumber \\
&&P_z^{\mu\nu}(s)= \left(s-M^2_z(s)\right)^{-1} 
 \left(g_{\mu\nu}-\frac{p_\mu p_\nu}{M^2_z(s)}\right)
\nonumber 
\eqa
with running boson masses defined as
\bqa
&& M^2_w(s)= \mu_w\frac{g^2(s)}{g^2(\mu_w)}
-g^2(s) [T_{\ssW}(s)-T_{\ssW}(\mu_w)]\,\nonumber \\
&& M^2_z(s)= \mu_z \frac{g^2(s)}{c^2_\theta(s)} 
\frac{c^2_\theta(\mu_z)}{g^2(\mu_z)}
-\frac{g^2(s)}{c^2_\theta(s)}[T_{\ssZ}(s)-T_{\ssZ}(\mu_z)]\,. \nonumber
\eqa
$T_{\ssW,\ssZ}(s)$ are contributions due to the top quark,
$\mu_{w,z}$ the complex poles of the propagators (one can
take, for instance, $\mu_{w,z}= M^2_{w,z}-i\Gamma_{w,z} M_{w,z}$) and
$$ s^2_\theta(s)= \frac{e^2(s)}{g^2(s)}\,,~c^2_\theta(s)= 1-s^2_\theta(s)\,.$$
The leading contributions are in the real part of the running couplings
therefore we take only the real part of them. This also means that one can
replace, in the above formulae, $g^2(\mu_{w,z}) \to g^2 (M^2_{w,z})$,
$c^2_\theta(\mu_z) \to c^2_\theta(\mzs)$ and 
also $T_{\ssW,\ssZ}(\mu_{w,z}) \to T_{\ssW,\ssZ}(M^2_{w,z})$.

When the $\wb\wb \gamma$ coupling is present, we 
introduce, in addition, the following effective three gauge boson vertex 
\begin{center}
\begin{picture}(100,100)(50,-50)
\glin{0,0}{30,0}{4}
\glin{30,0}{60,20}{4}
\glin{30,0}{60,-20}{4}
\Vertex(30,0){2.5}
\LongArrow(5,7)(20,7)
\LongArrow(50,25)(40,17)
\LongArrow(50,-25)(40,-17)
\Text(-3,0)[r]{\small $\gamma_\mu$}
\Text(63,20)[bl]{\small $W^+_\nu$}
\Text(63,-20)[tl]{\small$W^-_\rho$}
\Text(7,12)[bl]{\small $p$}
\Text(44,25)[br]{\small  $p_+$}
\Text(44,-23)[tr]{\small $p_-$}
\Text(80,0)[l]{$= i\,e(s)V_{\mu \nu \rho}$}
\end{picture}
\end{center}
with $s = p^2\,,~~s^+= p^2_+\,,~~s^-= p^2_-$ and
\bqa
V_{\mu \nu \rho} &=& 
 g_{\mu\nu}  (p   -p_+)_\rho
+g_{\nu\rho} (p_+ -p_-)_\mu\,(1+\delta_V)
+g_{\rho\mu} (p_- -p  )_\nu  \nonumber \\
&+&\frac{(p_+ -p_-)_\mu}{s^- - s^+}\left[
 \left(\frac{g(s^-)}{g(s^+)}-1 \right)\,p_{+ \nu}p_{+ \rho}
-\left(\frac{g(s^+)}{g(s^-)}-1 \right)\,p_{- \nu}p_{- \rho}
                             \right] \nonumber \\
\delta_V&=&  \frac{1}{g(s^+) g(s^-) (s^- - s^+)} 
  \left[ g^2(s^+) g^2(s^-)\,[ T_{\ssW}(s^-)- T_{\ssW}(s^+) ]  \right. 
\nonumber \\
     &+& \left. [g(s^+)- g(s^-)]\,[ s^- g(s^+) + s^+ g(s^-) ] 
\right]\,. 
\label{effver}
\eqa
It is the easy to see that, with the above choice for $V_{\mu \nu \rho}$,
the $U(1)$ gauge invariance - namely current conservation -
is preserved, even in presence of complex 
masses and running couplings, also with massive final state fermions.

By looking at \eqn{effver}, one can notice at least two effective ways 
to preserve $U(1)$. One can either
compute $g(s)$ at a fixed scale (for example always with $s= \mws$),
while keeping only the running of $e(s)$, or let all the couplings 
run at the proper scale.\footnote{Note, however, that in the complete 
formulation of the EFL-scheme there is no ambiguity and all scales are 
automatically fixed.}

With the first choice the modification of the three gauge boson 
vertex is kept minimal (but the leading running effects included).
With the second choice everything runs at the proper scale, 
but a heavier modification of the Feynman rules is required.
At this point one should not forget that our approach is an effective one, 
the goodness of which can be judged only by comparing with the exact 
calculation of ~Ref. \cite{tfl}.
We found that the second choice gives a better agreement
for leptonic single-$\wb$ final states, while the first one 
is closer to the exact result in the hadronic case, which
is phenomenologically more relevant. Therefore, we adopted this
first option as our default implementation in {\tt NEXTCALIBUR}.
The results of the EFL-scheme are then reproduced 
at $2\%$ accuracy for both leptonic and hadronic 
single-$\wb$ final states.

We want to stress once more that the outlined solution 
is flexible enough to deal with any four-fermion final state, 
whenever small scales dominate.  For example, once the given formulae 
are implemented in the Monte Carlo, the correct running 
of $\alpha_{\rm QED}$ is taken into account also for $s$-channel
processes as $\zb \gamma^\ast$ production.

Also naive QCD corrections can be easily included, without breaking
$U(1)$ gauge invariance, by the usual recipe of rescaling the total
$\wb$-width and the cross-section.

In fact, in our approach, $\Gamma_{_\ssW}$ can be generic, and the
above procedure respects current conservation, provided the same
$\wb$-width is used everywhere\footnote{Note, however, that in a complete
EFL-scheme the relevant objects are the complex poles and QCD corrections
should be computed accordingly, see \sect{swout}}.

\subsubsection*{Improving the treatment of the QED radiation}

Once the matrix element calculation is fixed one can add 
externally the QED leading logarithmic effects in the Structure
Function method \cite{isr}. 
Such a strategy is implemented in most of the programs used for 
the analysis of the LEP~2 data \cite{isr1} and accurately reproduces
the inclusive four-fermion cross-sections, at least for $s$-channel dominated
processes. 
In principle both initial and final state
radiation (ISR and FSR) 
can be treated in this way, as it has been explicitly
done originally for Bhabha scattering \cite{alibaba}. Here
only the implementation of ISR in {\tt NEXTCALIBUR} is
discussed. There are two issues to be discussed. One is the
choice of scale $q^2$ in the leading logarithm
$L= \ln(q^2/\mes)$.
Another is the {\em unfolding} of this leading logarithm
in terms of an emitted photon. For the latter issue a particular
form of $p_t$ dependent 
Structure Functions \cite{mmnp} is implemented. These 
are derived, at the first leading logarithmic
order, for small values of $p_t$. 
\noindent  In practice, we replace the quantity 
$$\ln(\frac{q^2}{\mes})~~~~{\rm by}~~~~  
\frac{1}{1-c_i+2\frac{\mes}{q^2}}$$ in the 
strictly collinear Structure Function for 
the $i^{th}$ incoming particle, by explicitly generating
$c_1$ and $c_2$, the cosines (in the laboratory frame)
of the emitted photons with respect to the incoming particles.
Once $c_{1,2}$ are generated, together with the energy
fractions $x_{1,2}$, 
and the azimuthal angles $\phi_{1,2}$,
the momenta of two ISR photons are known.
The four-fermion event is then generated in the c.m.s.
of the incoming particles {\em after} QED radiation, and then
boosted back to the laboratory frame. 

We also take into account non leading terms 
with the substitution \cite{ptsf1}
$$\ln(\frac{q^2}{\mes})-1~~\to~~~~~ 
\frac{1}{1-c_i+2\frac{\mes}{q^2}} -2 \frac{\mes}{q^2}
\frac{1}{(1-c_i+2\frac{\mes}{q^2})^2}.$$
The above choice ensures that the residue of the 
soft-photon pole gets proportional to $\ln(\frac{q^2}{\mes})-1$,
after integration over $c_i$.

As to the scale $q^2$, $s$ should be taken
for $s$-channel dominated processes, while, when a process is
dominated by small $t$ exchanges and $-t$
is much smaller than $s$, the scale is related to $t$. 
This is \eg. the case in small angle Bhabha scattering \cite{sbhab}
and the proper scale is chosen as the one which reproduces 
roughly the exact first order QED correction, which is known
for Bhabha scattering. A similar procedure now also exists
for the multi-peripheral two photon process \cite{grace:twopho}, since
an exact first order calculation is also available \cite{gge}.
In these $t$-channel dominated processes it is important to know whether
a cross-section with angular cuts is wanted, since then the
$t$-related scale will increase and the QED corrections as well.
When no exact first order calculations are available the
scale occurring in the first order soft corrections is
also used as guideline to guess $q^2$ \cite{grace:twopho,pvscales}.

In {\tt NEXTCALIBUR} the choice of the
scale is performed automatically by the program, event by event, according 
to the selected final state (see \tabn{pitt0}).
\begin{table}[hp]\centering
\begin{tabular}{|l||c|c|} 
\hline 
Final State          & $q^2_-$  & $q^2_+$ \\ \hline \hline
No $e^\pm$ & $s$     & $s$     \\ \hline              
1  $e^-$   & $|t_-|$ & $s$     \\ \hline               
1  $e^+$   & $s$     & $|t_+|$ \\ \hline               
1  $e^-$ and 1  $e^+$  & $|t_-|$    & $|t_+|$ \\ \hline               
2  $e^-$ and 2  $e^+$  & min($|t_-|$)    & min($|t_+|$) \\ \hline
\end{tabular}
\vspace*{3mm}
\caption[] {The choice of the QED scale in {\tt NEXTCALIBUR}.
$q^2_\pm$ are the scales of the incoming $e^\pm$ while
$t_\pm$ represent the $t$-channel invariants obtained
by combining initial and final state $e^\pm$ momenta.
When two combinations are possible, as in the last entry of the table, 
that one with the minimum value of $|t|$ is chosen, event by event.}
\label{pitt0}
\end{table}

\subsubsection*{Numerical results}

In \tabns{pitt3}{pitt4} we show single-$\wb$ numbers 
produced with the Modified Fermion-Loop approach, as discussed in the previous
section.
Comparisons are made with the EFL calculation of Ref.~\cite{tfl}.
The results of EFL are reproduced 
within $2\%$ accuracy for both leptonic and hadronic 
single-$\wb$ final states.

It should also be noted that, 
when neglecting Fermion-Loop corrections, 
one can directly compare {\tt NEXTCALIBUR} with other massive
Monte Carlo's and one finds excellent agreement for
single-$\wb$ production in the whole phase space.
\begin{table}[hp]\centering
\begin{tabular}{|l||c|c|c|} \hline 
$d\sigma/d\theta_e$     & MFL & EFL & MFL/EFL $-$ 1 (percent)  
         \\ \hline \hline
$0.0^\circ \div 0.1^\circ$  & 0.45062(70)  &  0.44784 & +0.62\\ \hline 
$0.1^\circ \div 0.2^\circ$  & 0.06636(28)  &  0.06605 & +0.47\\ \hline 
$0.2^\circ \div 0.3^\circ$  & 0.03848(21)  &  0.03860 & -0.31\\ \hline 
$0.3^\circ \div 0.4^\circ$  & 0.02726(18)  &  0.02736 & -0.37\\ \hline \hline
$\sigma_{tot} $             & 83.26(9)     &  83.28(6)& -0.02\\ \hline
\end{tabular}                 
\vspace*{3mm}
\caption[]{$d\sigma/d\theta_e$ [pb/degrees] and $\sigma_{tot}$ [fb] 
for the process $e^+ e^- \to e^- \barnu_e u \bard$. The first column
is the Modified Fermion-Loop, the second one is the exact Fermion-Loop 
of Ref. \cite{tfl}. $\sqrt{s}= 183$ GeV, $|\cos \theta_e| > 0.997$, 
$M(u\bard) > 45$ GeV. QED radiation not included.
The number in parenthesis is the integration error on the last digits.}
\label{pitt3}
\end{table}
\begin{table}[hp]\centering
\begin{tabular}{|l||c|c|c|} \hline 
 $d\sigma/d\theta_e$        & MFL & EFL & MFL/EFL $-$ 1 (percent)  
         \\ \hline \hline
$0.0^\circ \div 0.1^\circ$  & 0.13218(26) & 0.13448 &-1.7 \\ \hline 
$0.1^\circ \div 0.2^\circ$  & 0.01997(10) & 0.02031 &-1.7 \\ \hline 
$0.2^\circ \div 0.3^\circ$  & 0.01171(8)  & 0.01194 &-1.9 \\ \hline 
$0.3^\circ \div 0.4^\circ$  & 0.00838(6)  & 0.00851 &-1.5 \\ \hline \hline
$\sigma_{tot} $             & 25.01(3)    & 25.53   &-2.0 \\ \hline
\end{tabular}                 
\vspace*{3mm}
\caption[]{$d\sigma/d\theta_e$ [pb/degrees] and $\sigma_{tot}$ [fb] 
for the process $e^+ e^- \to e^- \barnu_e \nu_\mu \mu^+$. The first column
is the Modified Fermion-Loop, the second one is the exact Fermion-Loop 
of Ref. \cite{tfl}. $\sqrt{s}= 183$ GeV, $|\cos \theta_e| > 0.997$,
$|\cos \theta_\mu| < 0.95$ and  $E_\mu > 15$ GeV. QED radiation not included.
The number in parenthesis is the integration error on the last digits.}
\label{pitt4}
\end{table}

\begin{figure}[p]
\begin{center}
\begin{picture}(400,250)(-50,20)
\LinAxis(0,50)(300,50)(10,2,5,0,1.5)
\LinAxis(0,250)(300,250)(10,2,-5,0,1.5)
\LogAxis(0,50)(0,250)(4,-5,0,1.5)
\LogAxis(300,50)(300,250)(4,5,0,1.5)
\Text(0  ,40)[t]{$-1$}
\Text(150,40)[t]{$0$}
\Text(300,40)[t]{$1$}
\Text(-27,100)[l]{$10^{-1}$}
\Text(-27,150)[l]{$10^{0 }$}
\Text(-27,200)[l]{$10^{1 }$}
\Text(-50,227)[l]{$\frac{1}{\sigma} \frac{d\sigma}{d \cos \theta_\gamma}$}
\Text(150,20)[t]{$\cos \theta_\gamma$}
\Line(   0.0, 189.7)(   3.0, 189.7)
\Line(   3.0, 145.6)(   6.0, 145.6)
\Line(   6.0, 135.3)(   9.0, 135.3)
\Line(   9.0, 129.3)(  12.0, 129.3)
\Line(  12.0, 125.4)(  15.0, 125.4)
\Line(  15.0, 119.9)(  18.0, 119.9)
\Line(  18.0, 115.5)(  21.0, 115.5)
\Line(  21.0, 114.6)(  24.0, 114.6)
\Line(  24.0, 110.2)(  27.0, 110.2)
\Line(  27.0, 109.4)(  30.0, 109.4)
\Line(  30.0, 108.5)(  33.0, 108.5)
\Line(  33.0, 107.1)(  36.0, 107.1)
\Line(  36.0, 105.6)(  39.0, 105.6)
\Line(  39.0, 101.8)(  42.0, 101.8)
\Line(  42.0, 102.1)(  45.0, 102.1)
\Line(  45.0, 101.1)(  48.0, 101.1)
\Line(  48.0, 104.9)(  51.0, 104.9)
\Line(  51.0,  98.5)(  54.0,  98.5)
\Line(  54.0,  98.2)(  57.0,  98.2)
\Line(  57.0,  97.5)(  60.0,  97.5)
\Line(  60.0,  94.3)(  63.0,  94.3)
\Line(  63.0,  96.3)(  66.0,  96.3)
\Line(  66.0,  97.8)(  69.0,  97.8)
\Line(  69.0,  98.6)(  72.0,  98.6)
\Line(  72.0,  95.3)(  75.0,  95.3)
\Line(  75.0,  94.9)(  78.0,  94.9)
\Line(  78.0,  94.1)(  81.0,  94.1)
\Line(  81.0,  91.8)(  84.0,  91.8)
\Line(  84.0,  89.8)(  87.0,  89.8)
\Line(  87.0,  91.3)(  90.0,  91.3)
\Line(  90.0,  89.3)(  93.0,  89.3)
\Line(  93.0,  91.4)(  96.0,  91.4)
\Line(  96.0,  87.2)(  99.0,  87.2)
\Line(  99.0,  93.5)( 102.0,  93.5)
\Line( 102.0,  88.1)( 105.0,  88.1)
\Line( 105.0,  88.9)( 108.0,  88.9)
\Line( 108.0,  91.7)( 111.0,  91.7)
\Line( 111.0,  90.7)( 114.0,  90.7)
\Line( 114.0,  90.2)( 117.0,  90.2)
\Line( 117.0,  89.6)( 120.0,  89.6)
\Line( 120.0,  88.0)( 123.0,  88.0)
\Line( 123.0,  85.8)( 126.0,  85.8)
\Line( 126.0,  84.3)( 129.0,  84.3)
\Line( 129.0,  89.3)( 132.0,  89.3)
\Line( 132.0,  89.7)( 135.0,  89.7)
\Line( 135.0,  89.6)( 138.0,  89.6)
\Line( 138.0,  87.5)( 141.0,  87.5)
\Line( 141.0,  91.2)( 144.0,  91.2)
\Line( 144.0,  84.7)( 147.0,  84.7)
\Line( 147.0,  87.8)( 150.0,  87.8)
\Line( 150.0,  89.0)( 153.0,  89.0)
\Line( 153.0,  91.0)( 156.0,  91.0)
\Line( 156.0,  86.8)( 159.0,  86.8)
\Line( 159.0,  88.0)( 162.0,  88.0)
\Line( 162.0,  90.2)( 165.0,  90.2)
\Line( 165.0,  86.9)( 168.0,  86.9)
\Line( 168.0,  87.4)( 171.0,  87.4)
\Line( 171.0,  89.9)( 174.0,  89.9)
\Line( 174.0,  89.1)( 177.0,  89.1)
\Line( 177.0,  89.9)( 180.0,  89.9)
\Line( 180.0,  93.0)( 183.0,  93.0)
\Line( 183.0,  90.1)( 186.0,  90.1)
\Line( 186.0,  90.2)( 189.0,  90.2)
\Line( 189.0,  89.6)( 192.0,  89.6)
\Line( 192.0,  93.0)( 195.0,  93.0)
\Line( 195.0,  88.0)( 198.0,  88.0)
\Line( 198.0,  91.5)( 201.0,  91.5)
\Line( 201.0,  93.5)( 204.0,  93.5)
\Line( 204.0,  93.2)( 207.0,  93.2)
\Line( 207.0,  93.3)( 210.0,  93.3)
\Line( 210.0,  92.9)( 213.0,  92.9)
\Line( 213.0,  92.0)( 216.0,  92.0)
\Line( 216.0,  94.4)( 219.0,  94.4)
\Line( 219.0,  95.6)( 222.0,  95.6)
\Line( 222.0,  96.5)( 225.0,  96.5)
\Line( 225.0,  97.0)( 228.0,  97.0)
\Line( 228.0,  98.5)( 231.0,  98.5)
\Line( 231.0,  94.2)( 234.0,  94.2)
\Line( 234.0,  99.7)( 237.0,  99.7)
\Line( 237.0,  99.7)( 240.0,  99.7)
\Line( 240.0, 100.3)( 243.0, 100.3)
\Line( 243.0, 101.4)( 246.0, 101.4)
\Line( 246.0,  99.9)( 249.0,  99.9)
\Line( 249.0, 102.9)( 252.0, 102.9)
\Line( 252.0, 102.5)( 255.0, 102.5)
\Line( 255.0, 107.2)( 258.0, 107.2)
\Line( 258.0, 106.3)( 261.0, 106.3)
\Line( 261.0, 108.3)( 264.0, 108.3)
\Line( 264.0, 109.3)( 267.0, 109.3)
\Line( 267.0, 111.2)( 270.0, 111.2)
\Line( 270.0, 111.5)( 273.0, 111.5)
\Line( 273.0, 113.4)( 276.0, 113.4)
\Line( 276.0, 115.5)( 279.0, 115.5)
\Line( 279.0, 119.0)( 282.0, 119.0)
\Line( 282.0, 122.7)( 285.0, 122.7)
\Line( 285.0, 126.9)( 288.0, 126.9)
\Line( 288.0, 133.3)( 291.0, 133.3)
\Line( 291.0, 140.6)( 294.0, 140.6)
\Line( 294.0, 150.7)( 297.0, 150.7)
\Line( 297.0, 224.4)( 300.0, 224.4)
\Line(   3.0, 189.7)(   3.0, 145.6)
\Line(   6.0, 145.6)(   6.0, 135.3)
\Line(   9.0, 135.3)(   9.0, 129.3)
\Line(  12.0, 129.3)(  12.0, 125.4)
\Line(  15.0, 125.4)(  15.0, 119.9)
\Line(  18.0, 119.9)(  18.0, 115.5)
\Line(  21.0, 115.5)(  21.0, 114.6)
\Line(  24.0, 114.6)(  24.0, 110.2)
\Line(  27.0, 110.2)(  27.0, 109.4)
\Line(  30.0, 109.4)(  30.0, 108.5)
\Line(  33.0, 108.5)(  33.0, 107.1)
\Line(  36.0, 107.1)(  36.0, 105.6)
\Line(  39.0, 105.6)(  39.0, 101.8)
\Line(  42.0, 101.8)(  42.0, 102.1)
\Line(  45.0, 102.1)(  45.0, 101.1)
\Line(  48.0, 101.1)(  48.0, 104.9)
\Line(  51.0, 104.9)(  51.0,  98.5)
\Line(  54.0,  98.5)(  54.0,  98.2)
\Line(  57.0,  98.2)(  57.0,  97.5)
\Line(  60.0,  97.5)(  60.0,  94.3)
\Line(  63.0,  94.3)(  63.0,  96.3)
\Line(  66.0,  96.3)(  66.0,  97.8)
\Line(  69.0,  97.8)(  69.0,  98.6)
\Line(  72.0,  98.6)(  72.0,  95.3)
\Line(  75.0,  95.3)(  75.0,  94.9)
\Line(  78.0,  94.9)(  78.0,  94.1)
\Line(  81.0,  94.1)(  81.0,  91.8)
\Line(  84.0,  91.8)(  84.0,  89.8)
\Line(  87.0,  89.8)(  87.0,  91.3)
\Line(  90.0,  91.3)(  90.0,  89.3)
\Line(  93.0,  89.3)(  93.0,  91.4)
\Line(  96.0,  91.4)(  96.0,  87.2)
\Line(  99.0,  87.2)(  99.0,  93.5)
\Line( 102.0,  93.5)( 102.0,  88.1)
\Line( 105.0,  88.1)( 105.0,  88.9)
\Line( 108.0,  88.9)( 108.0,  91.7)
\Line( 111.0,  91.7)( 111.0,  90.7)
\Line( 114.0,  90.7)( 114.0,  90.2)
\Line( 117.0,  90.2)( 117.0,  89.6)
\Line( 120.0,  89.6)( 120.0,  88.0)
\Line( 123.0,  88.0)( 123.0,  85.8)
\Line( 126.0,  85.8)( 126.0,  84.3)
\Line( 129.0,  84.3)( 129.0,  89.3)
\Line( 132.0,  89.3)( 132.0,  89.7)
\Line( 135.0,  89.7)( 135.0,  89.6)
\Line( 138.0,  89.6)( 138.0,  87.5)
\Line( 141.0,  87.5)( 141.0,  91.2)
\Line( 144.0,  91.2)( 144.0,  84.7)
\Line( 147.0,  84.7)( 147.0,  87.8)
\Line( 150.0,  87.8)( 150.0,  89.0)
\Line( 153.0,  89.0)( 153.0,  91.0)
\Line( 156.0,  91.0)( 156.0,  86.8)
\Line( 159.0,  86.8)( 159.0,  88.0)
\Line( 162.0,  88.0)( 162.0,  90.2)
\Line( 165.0,  90.2)( 165.0,  86.9)
\Line( 168.0,  86.9)( 168.0,  87.4)
\Line( 171.0,  87.4)( 171.0,  89.9)
\Line( 174.0,  89.9)( 174.0,  89.1)
\Line( 177.0,  89.1)( 177.0,  89.9)
\Line( 180.0,  89.9)( 180.0,  93.0)
\Line( 183.0,  93.0)( 183.0,  90.1)
\Line( 186.0,  90.1)( 186.0,  90.2)
\Line( 189.0,  90.2)( 189.0,  89.6)
\Line( 192.0,  89.6)( 192.0,  93.0)
\Line( 195.0,  93.0)( 195.0,  88.0)
\Line( 198.0,  88.0)( 198.0,  91.5)
\Line( 201.0,  91.5)( 201.0,  93.5)
\Line( 204.0,  93.5)( 204.0,  93.2)
\Line( 207.0,  93.2)( 207.0,  93.3)
\Line( 210.0,  93.3)( 210.0,  92.9)
\Line( 213.0,  92.9)( 213.0,  92.0)
\Line( 216.0,  92.0)( 216.0,  94.4)
\Line( 219.0,  94.4)( 219.0,  95.6)
\Line( 222.0,  95.6)( 222.0,  96.5)
\Line( 225.0,  96.5)( 225.0,  97.0)
\Line( 228.0,  97.0)( 228.0,  98.5)
\Line( 231.0,  98.5)( 231.0,  94.2)
\Line( 234.0,  94.2)( 234.0,  99.7)
\Line( 237.0,  99.7)( 237.0,  99.7)
\Line( 240.0,  99.7)( 240.0, 100.3)
\Line( 243.0, 100.3)( 243.0, 101.4)
\Line( 246.0, 101.4)( 246.0,  99.9)
\Line( 249.0,  99.9)( 249.0, 102.9)
\Line( 252.0, 102.9)( 252.0, 102.5)
\Line( 255.0, 102.5)( 255.0, 107.2)
\Line( 258.0, 107.2)( 258.0, 106.3)
\Line( 261.0, 106.3)( 261.0, 108.3)
\Line( 264.0, 108.3)( 264.0, 109.3)
\Line( 267.0, 109.3)( 267.0, 111.2)
\Line( 270.0, 111.2)( 270.0, 111.5)
\Line( 273.0, 111.5)( 273.0, 113.4)
\Line( 276.0, 113.4)( 276.0, 115.5)
\Line( 279.0, 115.5)( 279.0, 119.0)
\Line( 282.0, 119.0)( 282.0, 122.7)
\Line( 285.0, 122.7)( 285.0, 126.9)
\Line( 288.0, 126.9)( 288.0, 133.3)
\Line( 291.0, 133.3)( 291.0, 140.6)
\Line( 294.0, 140.6)( 294.0, 150.7)
\Line( 297.0, 150.7)( 297.0, 224.4)
\end{picture}
\end{center}
\vskip 0.2cm
\caption[] {$\cos \theta_\gamma$ distribution (with respect 
to the incoming $e^+$), by {\tt NEXTCALIBUR}, for the most energetic photon
in the process $e^+ e^- \to e^- \bar \nu_e u \bar d (\gamma)$.}
\label{coscc20}
\end{figure}
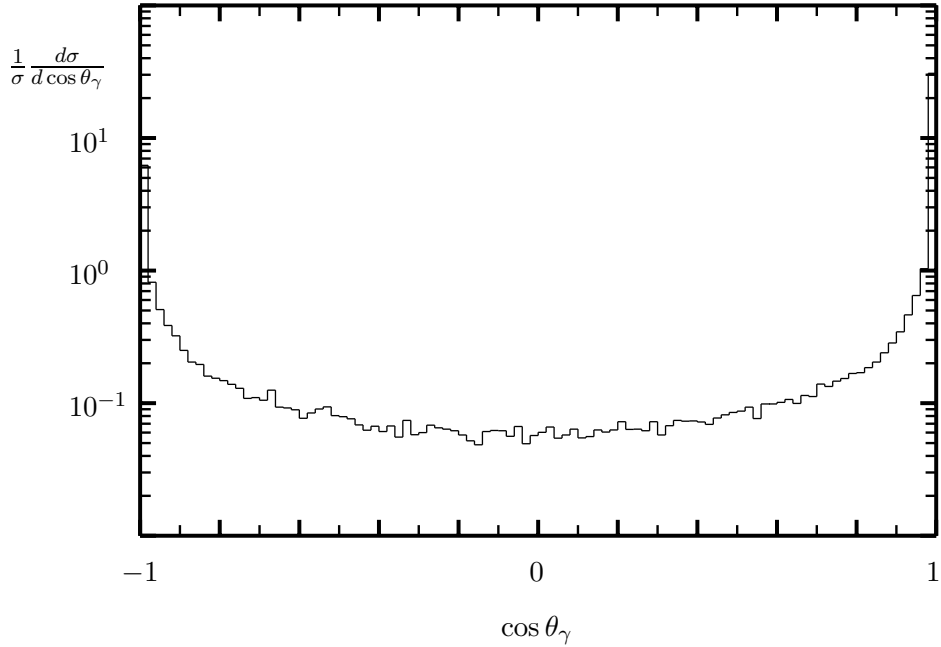

In \figs{coscc20}{enecc20} we show the
$\cos \theta_\gamma$ and $E_\gamma$ distributions for the most 
energetic photon in the process 
$e^+ e^- \to e^- \bar \nu_e u \bar d (\gamma)$.
We used $\sqrt{s}= 200$ GeV, $|\cos \theta_e| > 0.997 $ 
and $M(u \bar d) > 45 $ GeV.
Only ISR photons are taken into account, according to the scheme given 
in \tabn{pitt0}.

\clearpage

\begin{figure}[htbp]
\begin{center}
\vskip 2cm
\begin{picture}(400,300)(-50,-30)
\LinAxis(0,0)(300,0)(5,2,5,0,1.5)
\LinAxis(0,300)(300,300)(5,2,-5,0,1.5)
\LogAxis(0,0)(0,300)(6,-5,0,1.5)
\LogAxis(300,0)(300,300)(6,5,0,1.5)
\Text(60 ,-10)[t]{$10$}
\Text(120,-10)[t]{$20$}
\Text(180,-10)[t]{$30$}
\Text(240,-10)[t]{$40$}
\Text(-27, 50)[l]{$10^{-4}$}
\Text(-27,100)[l]{$10^{-3}$}
\Text(-27,150)[l]{$10^{-2}$}
\Text(-27,200)[l]{$10^{-1}$}
\Text(-27,250)[l]{$10^{0}$}
\Text(-50,277)[l]{$\frac{1}{\sigma} \frac{d\sigma}{d E_\gamma}$}
\Text(150,-30)[t]{$E_\gamma$ {\tt [GeV]}}
\Line(   0.0, 259.1)(   3.0, 259.1)
\Line(   3.0, 196.2)(   6.0, 196.2)
\Line(   6.0, 185.1)(   9.0, 185.1)
\Line(   9.0, 177.6)(  12.0, 177.6)
\Line(  12.0, 172.0)(  15.0, 172.0)
\Line(  15.0, 166.9)(  18.0, 166.9)
\Line(  18.0, 164.9)(  21.0, 164.9)
\Line(  21.0, 160.4)(  24.0, 160.4)
\Line(  24.0, 157.8)(  27.0, 157.8)
\Line(  27.0, 156.8)(  30.0, 156.8)
\Line(  30.0, 153.1)(  33.0, 153.1)
\Line(  33.0, 152.1)(  36.0, 152.1)
\Line(  36.0, 149.3)(  39.0, 149.3)
\Line(  39.0, 148.0)(  42.0, 148.0)
\Line(  42.0, 145.5)(  45.0, 145.5)
\Line(  45.0, 143.3)(  48.0, 143.3)
\Line(  48.0, 141.9)(  51.0, 141.9)
\Line(  51.0, 142.0)(  54.0, 142.0)
\Line(  54.0, 138.8)(  57.0, 138.8)
\Line(  57.0, 137.3)(  60.0, 137.3)
\Line(  60.0, 135.5)(  63.0, 135.5)
\Line(  63.0, 134.4)(  66.0, 134.4)
\Line(  66.0, 132.6)(  69.0, 132.6)
\Line(  69.0, 131.2)(  72.0, 131.2)
\Line(  72.0, 129.4)(  75.0, 129.4)
\Line(  75.0, 129.0)(  78.0, 129.0)
\Line(  78.0, 128.1)(  81.0, 128.1)
\Line(  81.0, 128.5)(  84.0, 128.5)
\Line(  84.0, 126.4)(  87.0, 126.4)
\Line(  87.0, 124.8)(  90.0, 124.8)
\Line(  90.0, 122.9)(  93.0, 122.9)
\Line(  93.0, 123.9)(  96.0, 123.9)
\Line(  96.0, 126.2)(  99.0, 126.2)
\Line(  99.0, 119.2)( 102.0, 119.2)
\Line( 102.0, 119.5)( 105.0, 119.5)
\Line( 105.0, 119.6)( 108.0, 119.6)
\Line( 108.0, 116.5)( 111.0, 116.5)
\Line( 111.0, 116.6)( 114.0, 116.6)
\Line( 114.0, 113.8)( 117.0, 113.8)
\Line( 117.0, 113.6)( 120.0, 113.6)
\Line( 120.0, 115.0)( 123.0, 115.0)
\Line( 123.0, 112.0)( 126.0, 112.0)
\Line( 126.0, 112.3)( 129.0, 112.3)
\Line( 129.0, 109.4)( 132.0, 109.4)
\Line( 132.0, 108.5)( 135.0, 108.5)
\Line( 135.0, 110.4)( 138.0, 110.4)
\Line( 138.0, 111.5)( 141.0, 111.5)
\Line( 141.0, 110.8)( 144.0, 110.8)
\Line( 144.0, 109.3)( 147.0, 109.3)
\Line( 147.0, 106.5)( 150.0, 106.5)
\Line( 150.0, 107.1)( 153.0, 107.1)
\Line( 153.0, 104.9)( 156.0, 104.9)
\Line( 156.0, 102.4)( 159.0, 102.4)
\Line( 159.0, 100.4)( 162.0, 100.4)
\Line( 162.0, 103.5)( 165.0, 103.5)
\Line( 165.0,  99.5)( 168.0,  99.5)
\Line( 168.0,  98.7)( 171.0,  98.7)
\Line( 171.0,  99.4)( 174.0,  99.4)
\Line( 174.0,  98.4)( 177.0,  98.4)
\Line( 177.0,  97.4)( 180.0,  97.4)
\Line( 180.0,  97.9)( 183.0,  97.9)
\Line( 183.0,  96.1)( 186.0,  96.1)
\Line( 186.0,  93.9)( 189.0,  93.9)
\Line( 189.0,  92.9)( 192.0,  92.9)
\Line( 192.0,  97.0)( 195.0,  97.0)
\Line( 195.0,  93.3)( 198.0,  93.3)
\Line( 198.0,  92.3)( 201.0,  92.3)
\Line( 201.0,  89.5)( 204.0,  89.5)
\Line( 204.0,  90.8)( 207.0,  90.8)
\Line( 207.0,  90.6)( 210.0,  90.6)
\Line( 210.0,  89.4)( 213.0,  89.4)
\Line( 213.0,  84.7)( 216.0,  84.7)
\Line( 216.0,  87.1)( 219.0,  87.1)
\Line( 219.0,  87.5)( 222.0,  87.5)
\Line( 222.0,  86.1)( 225.0,  86.1)
\Line( 225.0,  84.0)( 228.0,  84.0)
\Line( 228.0,  82.9)( 231.0,  82.9)
\Line( 231.0,  82.8)( 234.0,  82.8)
\Line( 234.0,  81.2)( 237.0,  81.2)
\Line( 237.0,  78.3)( 240.0,  78.3)
\Line( 240.0,  79.5)( 243.0,  79.5)
\Line( 243.0,  86.4)( 246.0,  86.4)
\Line( 246.0,  76.3)( 249.0,  76.3)
\Line( 249.0,  79.1)( 252.0,  79.1)
\Line( 252.0,  74.8)( 255.0,  74.8)
\Line( 255.0,  79.0)( 258.0,  79.0)
\Line( 258.0,  75.6)( 261.0,  75.6)
\Line( 261.0,  74.4)( 264.0,  74.4)
\Line( 264.0,  82.8)( 267.0,  82.8)
\Line( 267.0,  74.9)( 270.0,  74.9)
\Line( 270.0,  70.5)( 273.0,  70.5)
\Line( 273.0,  75.2)( 276.0,  75.2)
\Line( 276.0,  71.2)( 279.0,  71.2)
\Line( 279.0,  69.3)( 282.0,  69.3)
\Line( 282.0,  74.0)( 285.0,  74.0)
\Line( 285.0,  66.1)( 288.0,  66.1)
\Line( 288.0,  71.6)( 291.0,  71.6)
\Line( 291.0,  67.4)( 294.0,  67.4)
\Line( 294.0,  67.6)( 297.0,  67.6)
\Line( 297.0,  68.4)( 300.0,  68.4)
\Line(   3.0, 259.1)(   3.0, 196.2)
\Line(   6.0, 196.2)(   6.0, 185.1)
\Line(   9.0, 185.1)(   9.0, 177.6)
\Line(  12.0, 177.6)(  12.0, 172.0)
\Line(  15.0, 172.0)(  15.0, 166.9)
\Line(  18.0, 166.9)(  18.0, 164.9)
\Line(  21.0, 164.9)(  21.0, 160.4)
\Line(  24.0, 160.4)(  24.0, 157.8)
\Line(  27.0, 157.8)(  27.0, 156.8)
\Line(  30.0, 156.8)(  30.0, 153.1)
\Line(  33.0, 153.1)(  33.0, 152.1)
\Line(  36.0, 152.1)(  36.0, 149.3)
\Line(  39.0, 149.3)(  39.0, 148.0)
\Line(  42.0, 148.0)(  42.0, 145.5)
\Line(  45.0, 145.5)(  45.0, 143.3)
\Line(  48.0, 143.3)(  48.0, 141.9)
\Line(  51.0, 141.9)(  51.0, 142.0)
\Line(  54.0, 142.0)(  54.0, 138.8)
\Line(  57.0, 138.8)(  57.0, 137.3)
\Line(  60.0, 137.3)(  60.0, 135.5)
\Line(  63.0, 135.5)(  63.0, 134.4)
\Line(  66.0, 134.4)(  66.0, 132.6)
\Line(  69.0, 132.6)(  69.0, 131.2)
\Line(  72.0, 131.2)(  72.0, 129.4)
\Line(  75.0, 129.4)(  75.0, 129.0)
\Line(  78.0, 129.0)(  78.0, 128.1)
\Line(  81.0, 128.1)(  81.0, 128.5)
\Line(  84.0, 128.5)(  84.0, 126.4)
\Line(  87.0, 126.4)(  87.0, 124.8)
\Line(  90.0, 124.8)(  90.0, 122.9)
\Line(  93.0, 122.9)(  93.0, 123.9)
\Line(  96.0, 123.9)(  96.0, 126.2)
\Line(  99.0, 126.2)(  99.0, 119.2)
\Line( 102.0, 119.2)( 102.0, 119.5)
\Line( 105.0, 119.5)( 105.0, 119.6)
\Line( 108.0, 119.6)( 108.0, 116.5)
\Line( 111.0, 116.5)( 111.0, 116.6)
\Line( 114.0, 116.6)( 114.0, 113.8)
\Line( 117.0, 113.8)( 117.0, 113.6)
\Line( 120.0, 113.6)( 120.0, 115.0)
\Line( 123.0, 115.0)( 123.0, 112.0)
\Line( 126.0, 112.0)( 126.0, 112.3)
\Line( 129.0, 112.3)( 129.0, 109.4)
\Line( 132.0, 109.4)( 132.0, 108.5)
\Line( 135.0, 108.5)( 135.0, 110.4)
\Line( 138.0, 110.4)( 138.0, 111.5)
\Line( 141.0, 111.5)( 141.0, 110.8)
\Line( 144.0, 110.8)( 144.0, 109.3)
\Line( 147.0, 109.3)( 147.0, 106.5)
\Line( 150.0, 106.5)( 150.0, 107.1)
\Line( 153.0, 107.1)( 153.0, 104.9)
\Line( 156.0, 104.9)( 156.0, 102.4)
\Line( 159.0, 102.4)( 159.0, 100.4)
\Line( 162.0, 100.4)( 162.0, 103.5)
\Line( 165.0, 103.5)( 165.0,  99.5)
\Line( 168.0,  99.5)( 168.0,  98.7)
\Line( 171.0,  98.7)( 171.0,  99.4)
\Line( 174.0,  99.4)( 174.0,  98.4)
\Line( 177.0,  98.4)( 177.0,  97.4)
\Line( 180.0,  97.4)( 180.0,  97.9)
\Line( 183.0,  97.9)( 183.0,  96.1)
\Line( 186.0,  96.1)( 186.0,  93.9)
\Line( 189.0,  93.9)( 189.0,  92.9)
\Line( 192.0,  92.9)( 192.0,  97.0)
\Line( 195.0,  97.0)( 195.0,  93.3)
\Line( 198.0,  93.3)( 198.0,  92.3)
\Line( 201.0,  92.3)( 201.0,  89.5)
\Line( 204.0,  89.5)( 204.0,  90.8)
\Line( 207.0,  90.8)( 207.0,  90.6)
\Line( 210.0,  90.6)( 210.0,  89.4)
\Line( 213.0,  89.4)( 213.0,  84.7)
\Line( 216.0,  84.7)( 216.0,  87.1)
\Line( 219.0,  87.1)( 219.0,  87.5)
\Line( 222.0,  87.5)( 222.0,  86.1)
\Line( 225.0,  86.1)( 225.0,  84.0)
\Line( 228.0,  84.0)( 228.0,  82.9)
\Line( 231.0,  82.9)( 231.0,  82.8)
\Line( 234.0,  82.8)( 234.0,  81.2)
\Line( 237.0,  81.2)( 237.0,  78.3)
\Line( 240.0,  78.3)( 240.0,  79.5)
\Line( 243.0,  79.5)( 243.0,  86.4)
\Line( 246.0,  86.4)( 246.0,  76.3)
\Line( 249.0,  76.3)( 249.0,  79.1)
\Line( 252.0,  79.1)( 252.0,  74.8)
\Line( 255.0,  74.8)( 255.0,  79.0)
\Line( 258.0,  79.0)( 258.0,  75.6)
\Line( 261.0,  75.6)( 261.0,  74.4)
\Line( 264.0,  74.4)( 264.0,  82.8)
\Line( 267.0,  82.8)( 267.0,  74.9)
\Line( 270.0,  74.9)( 270.0,  70.5)
\Line( 273.0,  70.5)( 273.0,  75.2)
\Line( 276.0,  75.2)( 276.0,  71.2)
\Line( 279.0,  71.2)( 279.0,  69.3)
\Line( 282.0,  69.3)( 282.0,  74.0)
\Line( 285.0,  74.0)( 285.0,  66.1)
\Line( 288.0,  66.1)( 288.0,  71.6)
\Line( 291.0,  71.6)( 291.0,  67.4)
\Line( 294.0,  67.4)( 294.0,  67.6)
\Line( 297.0,  67.6)( 297.0,  68.4)
\end{picture}
\end{center}
\vskip 0.1cm
\caption[] {$E_\gamma$ distribution, by {\tt NEXTCALIBUR},
for the most energetic photon in the process 
$e^+ e^- \to e^- \bar \nu_e u \bar d (\gamma)$.}
\label{enecc20}
\end{figure}
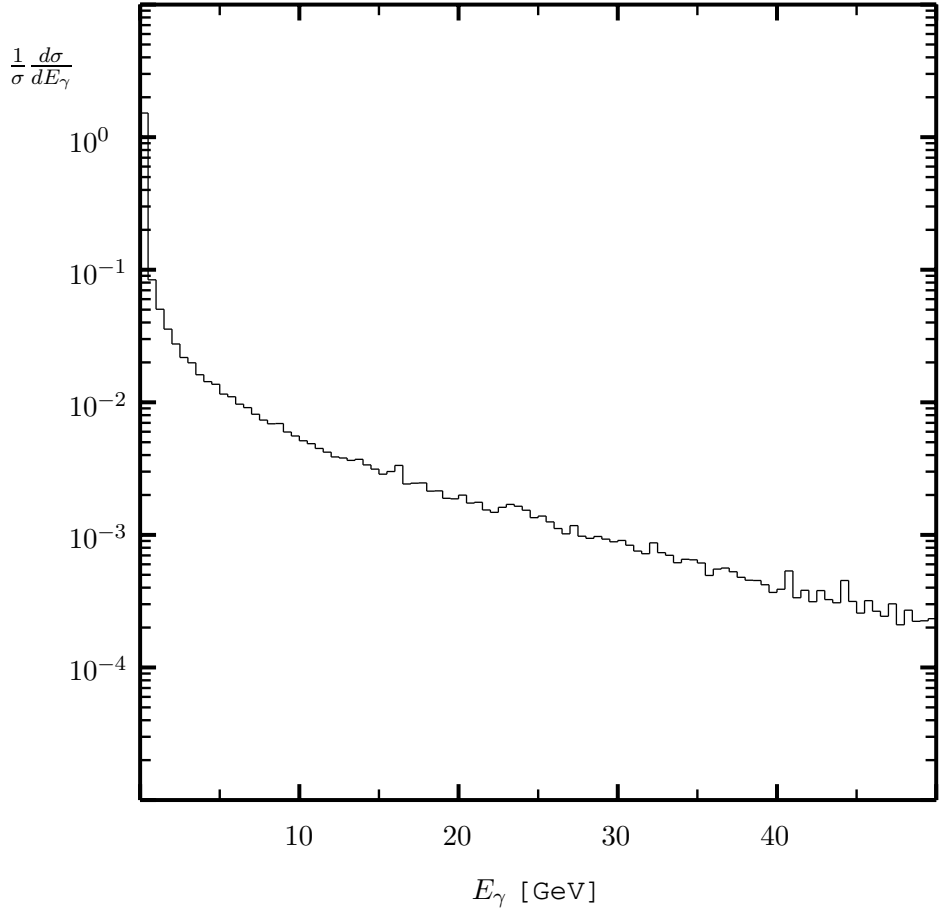

Note, however, that the recipe of using 
Structure Functions with a proper choice of scales is not
enough to determine without ambiguity the pattern of the radiation 
in $t$-channel dominated processes. 
The reason is that, when $|t|= q^2 \sim \mes$, the 
Leading Order Structure Function approach fails and
one has to introduce a minimum value for $|t|$, below which only
non-radiative events from the corresponding leg are generated. 
Since Structure Functions behave like
$\delta$ functions for vanishing $q^2$, this is automatically
achieved by introducing a minimum value
$|t_{min}|$, such that, for events with $|t|< |t_{min}|$, 
the scale in the corresponding Structure Function is always set equal
to $|t_{min}|$\footnote{Note, however, that this behavior
is known and could be implemented. It is enough to consider the standard
YFS infrared emission factor ${\hat{\cal B}}$, see \eg Eq.(3) of
Ref.~\cite{yfsww2:1996}.}.
We observed deviations at the order of $0.5\%$ 
by varying $|t_{min}|$ from $2.71828\,\mes$ to $100\,\mes$. The 
default value of $|t_{min}|$ in {\tt NEXTCALIBUR} is taken to be the latter.

In table (\ref{nexca3}) we also give the cross sections (in pb)
corresponding to the above distributions.  {\em tot} refers to
radiative plus non radiative events (within the specified separation
cuts for the generated photons), {\em nrad} to non-radiative events,
{\em srad} to single-radiative events and {\em drad} to double
radiative events.

Finally, in order to quantify the effects due to ISR scales and
running of $\alpha_{\rm QED}$, we show, in tables (\ref{nexca4}) and
(\ref{nexca5}), the cross sections obtained by using both ISR scales =
$s$ and switching on and off the Modified Fermion loop corrections.

\clearpage

\begin{table}[htb]
\centering
\renewcommand{\arraystretch}{1.1}
\begin{tabular}{|c|c|}
\hline
Type & Cross-section                 \\
\hline
\hline
$\sigma_{\rm tot}$    &   103.68(33) \\
$\sigma_{\rm nrad}$   &    96.54(32) \\
$\sigma_{\rm srad}$   &   7.00(7)    \\
$\sigma_{\rm drad}$   &   0.139(7)   \\
\hline
\end{tabular}
\caption[]{Cross-sections in fb from {\tt NEXTCALIBUR} for the process
$e^+(1) e^-(2) \to e^-(3) \barnu_e(4) u(5) \bard(6)$.
$M(56) > 45\,$GeV, no energy cut.
ISR as in \tabn{pitt0}, Modified Fermion Loop included.
Separation cuts for the photons: 
$E_{\ph} > 1\,\GeV, |\cos\theta_{\ph}| < 0.997$.}
\label{nexca3}
\end{table}
\begin{table}[htb]
\centering
\renewcommand{\arraystretch}{1.1}
\begin{tabular}{|c|c|}
\hline
Type & Cross-section                 \\
\hline
\hline
$\sigma_{\rm tot}$    &   100.73(17) \\
$\sigma_{\rm nrad}$   &    93.39(16) \\
$\sigma_{\rm srad}$   &   7.21(4)    \\
$\sigma_{\rm drad}$   &   0.124(5)   \\
\hline
\end{tabular}
\caption[]{Cross-sections in fb from {\tt NEXTCALIBUR} for the process
$e^+(1) e^-(2) \to e^-(3) \barnu_e(4) u(5) \bard(6)$.
$M(56) > 45\,\GeV$, no energy cut.
Both ISR scales = $s$, Modified Fermion Loop included.
Separation cuts for the photons:
$E_{\ph} > 1\,\GeV, |\cos\theta_{\ph}| < 0.997$.}
\label{nexca4}
\end{table}
\begin{table}[htb]
\centering
\renewcommand{\arraystretch}{1.1}
\begin{tabular}{|c|c|c|}
\hline
Type & Cross-section \\
\hline
\hline
$\sigma_{\rm tot}$       &   106.36(18)  \\
$\sigma_{\rm nrad}$      &     98.62(17) \\
$\sigma_{\rm srad}$      &   7.61(5)     \\
$\sigma_{\rm drad}$      &   0.131(5)    \\
\hline
\end{tabular}
\caption[]{Cross-sections in fb from {\tt NEXTCALIBUR} for the process
$e^+(1) e^-(2) \to e^-(3) \barnu_e(4) u(5) \bard(6)$.
$M(56) > 45\,\GeV$, no energy cut.
Both ISR scales = $s$, Modified Fermion Loop excluded.
Separation cuts for the photons:
$E_{\ph} > 1\,\GeV, |\cos\theta_{\ph}| < 0.997$.}
\label{nexca5}
\end{table}

\subsubsection*{Single-$\wb$ with {\tt GRACE}}

\subsubsection*{Authors}

\begin{tabular}{l}
Y.~Kurihara, M. Kuroda and Y.~Shimizu \\
\end{tabular}

\subsubsection*{Introduction}

The single-$\wb$ production processes present
an opportunity to study the anomalous triple-gauge-couplings (hereafter TGC)
at LEP~2 experiments. In order to proceed
to the precise measurement of TGC, the inclusion of 
an initial state radiative correction (ISR) in any generator is an
inevitable step. 
As a tool for the ISR the structure function (SF)\cite{grace:SF}
and the parton shower\cite{grace:PS}
methods are widely used for the $e^+ e^-$ annihilation processes.
Since the main contribution for the single-$\wb$ production processes comes,
however, from the non-annihilation type diagrams, 
the universal factorization method 
used for the annihilation processes is, obviously, inappropriate.
The main problem lies in the determination of the energy scale of the 
factorization.
According to the study of the two photon process\cite{grace:twopho}, 
SF and QED parton shower (QEDPS) methods were shown to
reproduce the exact $O(\alpha)$ results precisely
even for the non-annihilation processes, when the appropriate energy 
scale is used in those algorithms.

Here, we propose a general method to determine the energy
scale to be used in SF and QEDPS. The numerical
results of testing  SF and QEDPS for 
$e^- e^+ \rightarrow e^- \barnu_e u \bard$ and
$e^- e^+ \rightarrow e^- \barnu_e \mu^+ \nu_\mu$
are given. The systematic errors are also discussed.
 
\subsubsection{Energy Scale Determination in QED corrections}
\label{esd}

Single-$\wb$ is not dominated by annihilation and, therefore, standard methods
as $s$-channel structure functions fail to reproduce the correct result.

The factorization theorem for the QED radiative corrections in the LL 
approximation is valid independently of the structure of the matrix element
of the kernel process. Hence structure functions (hereafter SF) and 
QEDPS must be applicable to 
any $e^+e^-$ scattering processes. However, the choice of the energy 
scale in SF and QEDPS is not a trivial issue.
For simple processes like $e^+e^-$ annihilation into fermion pairs and 
two-photon process (with only the
multi-peripheral diagrams considered so far), the evolution energy scale 
could be determined in terms of the exact perturbative calculations. 
However, for more complicated processes, this is not always possible. 
Hence a way to find a suitable energy scale without knowing the exact 
loop calculations should be established.

First we look at the general consequence of the soft photon
approximation. 

The soft photon cross-section is given, in some approximation, by the Born 
cross-section multiplied by the following correction 
factor~\cite{grace:SW}:
\begin{eqnarray}
\label{softsw}
\frac{d\sigma_{soft}(s)}{d\Omega} &=& 
\frac{d\sigma_0(s)}{d\Omega} 
\left| {\exp}\left[-\frac{\alpha}
{\pi} {\ln}\left( \frac{E}{k_c} \right)  
\sum_{i,j} \frac{e_i e_j \eta_i \eta_j}
{\beta_{ij}} {\ln}\left(\frac{1+\beta_{ij}}{1-\beta_{ij}}\right) 
\right] \right|^2,
\label{ir} \\
\beta_{ij} &=& \left(1-\frac{m_i^2 m_j^2}
{(p_i \cdot p_j)^2}\right)^{\frac{1}{2}},
\end{eqnarray}
where $m_j$ ($p_j$) are the mass(momentum) of $j$-th charged particle,
$k_c$ is the maximum energy of the soft photon (the boundary between soft- and
hard-photons), $E$ is the beam energy, and $e_j$ the electric charge 
in unit of the $e^+$ charge.
The factor $\eta_j$ is $-1$ for the initial particles and $+1$ for the
final particles. The indices ($i,j$) run over all the charged particles 
in the initial and final states. 

The part proportional to $\ln(E/k_c)$ that is shown explicitly in
\eqn{softsw} is exact and not only LL-approximated.
However, the single-logarithmic part is omitted, so that the formula
is not a complete LL-approximation, but it is enough to guess
the energy scale appearing in SF and QEDPS.
For the two-photon process,
$e^-(p_-)+e^+(p_+) \rightarrow e^-(q_-)+e^+(q_+)+\mu^-(k_-)+\mu^+(k_+)$,
it was shown in Ref.\cite{grace:twopho}
that the soft-photon factor in Eq.(\ref{ir}) with a ($p_- \cdot q_-$)-term
gives a good numerical approximation to the exact $\ord{\alpha}$
correction\cite{grace:bdk}.
   
This implies that one is able to make and educated guess about
the possible evolution energy scale in SF from Eq.(\ref{ir}) without an 
explicit loop calculation. 

However, one may 
question why the energy scale $s=(p_-+p_+)^2$ does not appear in 
the soft-photon correction, even if they are included in Eq.(\ref{ir}).
When applying SF to the two-photon process 
we have ignored those terms which come from the photon connecting 
different charged lines. This is because the contributions
from the box diagrams, with photon exchange between the $e^+$ and $e^-$ lines,
is known to be small\cite{grace:vNV}. Fortunately, the infrared part of 
the loop correction is already included in Eq.(\ref{ir}) and there is no need to
know the full form of the loop diagram. 
For the two-photon processes we look at those two terms where, for example, 
($p_- \cdot p_+$)-terms and ($q_- \cdot p_+$)-terms are present; here,
the momentum of $e^-$ is almost 
the same, before and after the scattering($p_- \approx q_-$). 
The difference only appears in $\eta_j \eta_k=+1$ for a ($p_-  p_+$)-term  
and in $\eta_j \eta_k=-1$ for a ($q_- p_+$)-term. Then these terms compensate 
each other after summing them up for the forward scattering, which is 
the dominant kinematical region of this process. This is why the energy 
scale $s=(p_-+p_+)^2$ does not appear in the soft-photon correction,
despite its presence in Eq.(\ref{ir}).

When experimental cuts are imposed, for example the final $e^-$ is
tagged in a large angle, this cancellation is not perfect but only partial 
and the energy scale $s$ must appear in the soft-photon correction. 
In this case the annihilation-type diagrams will also contribute to 
the matrix elements. Then the usual SF and QEDPS formulation for 
the annihilation processes are justified and can be used for ISR with 
the energy scale $s$. One can check which energy scale is dominant 
under the given experimental cuts by numerically integrating 
the soft-photon cross-section given by Eq.(\ref{ir}) over the allowed
kinematical region. Thus, in order to determine the energy scale it is 
sufficient to know the infrared behavior of the radiative process using 
the soft-photon factor. 

Next, we determine the energy-scale of the QED radiative corrections
to the single-$\wb$ production process,
\begin{eqnarray}
e^-(p_-)+e^+(p_+) &\rightarrow& e^-(q_-)+\barnu_e(q_{\nu})
+u(k_u)+\bard(k_d).
\label{pc1}
\end{eqnarray}
The soft-photon correction factor shown in 
Eq.(\ref{ir}) is numerically integrated with the Born matrix element of
the process (\ref{pc1}), with $t$-channel diagrams only and without any cut on
the final fermions. 
The results are shown in Table\ref{grace:TB1}.

\begin{table}[htbp]
\begin{center}
\begin{tabular}{|c|c|c|c|} \hline
all terms & $p_- q_-$ & $p_+ k_u k_d $  & all other combinations \\ \hline
    1     &  $0.38$     &  $0.61$     & $1.9\times10^{-3}$    \\ \hline
\end{tabular}
\end{center}
\caption{\footnotesize
Soft-photon correction factor from different sets of charged particle
combinations in the process of $e^+ e^- \rightarrow e^- \barnu_e 
u \bard$ at the c.m.s. energy of 200 GeV. The total factor 
is normalized to unity. 
}
\label{grace:TB1}
\end{table}

One can see that the main contribution 
comes from an electron-line ($p_- q_-$-term) and 
a positron-line ($p_+ k_u k_d $-term), while all the other
contributions are negligibly small.
As in the case of the two-photon processes, the energy scale  
$s$ does not appear in the soft-photon correction.
Applying SF or QEDPS for the electron and
positron charged-lines individually and with an energy scale given by their 
momentum-transfer squared might be legitimate, according to the
above results.

\subsubsection{Structure Function Method}

The corrected cross-section is given by
\begin{eqnarray}
\sigma_{total}(s) &=&\int dx_{\ssI-}\int dx_{\ssF-}
\int dx_{\ssI+} \int dx_{u} \int dx_{d}\nonumber 
D_{e^-}(x_{\ssI-},-t_-) D_{e^-}(x_{\ssF-},-t_-) \\
&~&D_{e^+}(x_{\ssI+},p_{\ssT ud}^2) D_{u}(x_{u},m_{ud}^2) D_{d}(x_{d},
m_{ud}^2) \sigma_0({\hat s}), 
\end{eqnarray}
using the structure function ($D_f$) with an energy scale
$t_-=(p_- - q_-)^2$, $p_{\ssT ud}^2$ \ie the transverse-momentum squared of
the $u$-${\bar d}$ system and $m_{ud}^2=(k_u+k_d)^2$.

The energy-scale determination for the positron line is rather
ambiguous. The $p_{\ssT u+d}$ is distributing around $\mw/3$, then
the difference between these two energy scales does not give
a significant effect on the correction factor.
After(before) the photon radiation the initial(final) momenta $p_\pm$ 
($q_\pm$) become ${\hat p_\pm}$ (${\hat q_\pm}$) defined by:
\begin{eqnarray}
{\hat p_-} &=& x_{\ssI-}p_-,~~~{\hat q_-}=\frac{1}{x_{\ssF-}}q_-, ...
\end{eqnarray}
Then the c.m.s. energy squared $s$ is scaled as ${\hat s} = 
x_{\ssI-}x_{\ssI+}s$.

\subsubsection{Parton Shower Method}

Instead of the analytic formula of the structure-function approach, a Monte 
Carlo method based on the parton shower algorithm in QED (QEDPS) can be 
used to solve the Altarelli-Parisi equation in 
the LL approximation\cite{grace:AP}. 
The detailed QEDPS-algorithm  can be found in Ref.\cite{grace:QEDPSs}
for the $e^+e^-$ annihilation processes, in Ref.\cite{grace:QEDPSt}
for the Bhabha process, and in Ref.\cite{grace:twopho} for the
two-photon process.
In QEDPS we use the same energy scale as in the SF method. 
One difference between SF and QEDPS is that 
the {\em ad hoc} replacement of the perturbative expansion
coefficient $L(={\ln}({Q^2}/{m_f^2}))$ by $L-1$, which was
realized by hand for SF, does not apply for QEDPS. 
Another significant difference 
between these two methods is that QEDPS can give a correct
treatment of the transverse momentum of emitted photons by imposing the exact 
kinematics at the
$e\rightarrow e\gamma$ splitting. Note that it does not affect the total cross 
sections too much when the final $e^-$ are unconstrained. However, the finite 
recoiling of the final $e^\pm$ may result into a large effect on the tagged 
cross-sections.

As a consequence of the exact kinematics at the $e\rightarrow e\gamma$ 
splitting, the $e^\pm$ are no more on-shell
after photon emission. On the other hand the matrix element of the hard
scattering process must be calculated with the on-shell external particles.
A trick to map the off-shell four-momenta of the initial $e^\pm$ to 
those at on-shell is needed. The following method is used in the calculations:
First ${\hat s}=({\hat p}_- + {\hat p}_+)^2$ is calculated, where
${\hat p}_\pm$ are the four-momenta of the initial $e^\pm$ after the
photon emission by QEDPS. ${\hat s}$ is mainly positive even for the 
off-shell $e^\pm$. (When ${\hat s}$ is negative, that event is discarded.)
Subsequently, all four-momenta are generated in the rest-frame of the initial
$e^\pm$ after the photon emission. Four-momenta
of the hard scattering in their rest-frame are ${\tilde p}_\pm$, where
${\tilde p}_\pm^2=m_e^2$ (on-shell) 
and ${\hat s}=({\tilde p}_- + {\tilde p}_+)^2$.

Finally, all four-momenta are rotated and boosted to match 
the three-momenta of ${\tilde p}_\pm$ with those of ${\hat p}_\pm$.
This method respects the direction of the final $e^\pm$ rather than 
the c.m.s. energy of the collision. The total energy is not conserved 
because of the virtuality of the initial $e^\pm$.
The violation of energy-conservation is of the order of
$10^{-6}$ GeV or less. The probability to violate it by more
than $1$ MeV is $10^{-4}$.

\subsubsection*{Numerical Calculations, the total cross-sections}
\label{scalegrace}

Total and differential cross-sections of the semi-leptonic process 
$e^- e^+ \rightarrow e^- \barnu_e u \bard$ and 
of the leptonic one,
$e^- e^+ \rightarrow e^- \barnu_e \mu^+ \nu_\mu$,
are calculated with the radiative correction by using SF or QEDPS.
Fortran codes to calculate amplitudes of the above processes are 
produced using ${\tt GRACE}$ 
system\cite{grace:grace}. All fermion-masses are kept finite in calculations.
Numerical integrations of the matrix element squared in the four-body
phase space are done using ${\tt BASES}$\cite{grace:bases}.
For the study of the radiative correction for the single-$\wb$ productions,
only $t$-channel diagrams(non-annihilation diagrams) are taken into 
account.

For the total energy of the emitted photons, both methods must give 
the same spectrum, when the same energy scale are used. 
That is confirmed by the results shown in \fig{fig-grace-singlw-1}
at the c.m.s. energy of $200\,$GeV for the semi-leptonic process.

\begin{figure}[htbp]
\begin{center}
\includegraphics[width=8cm]{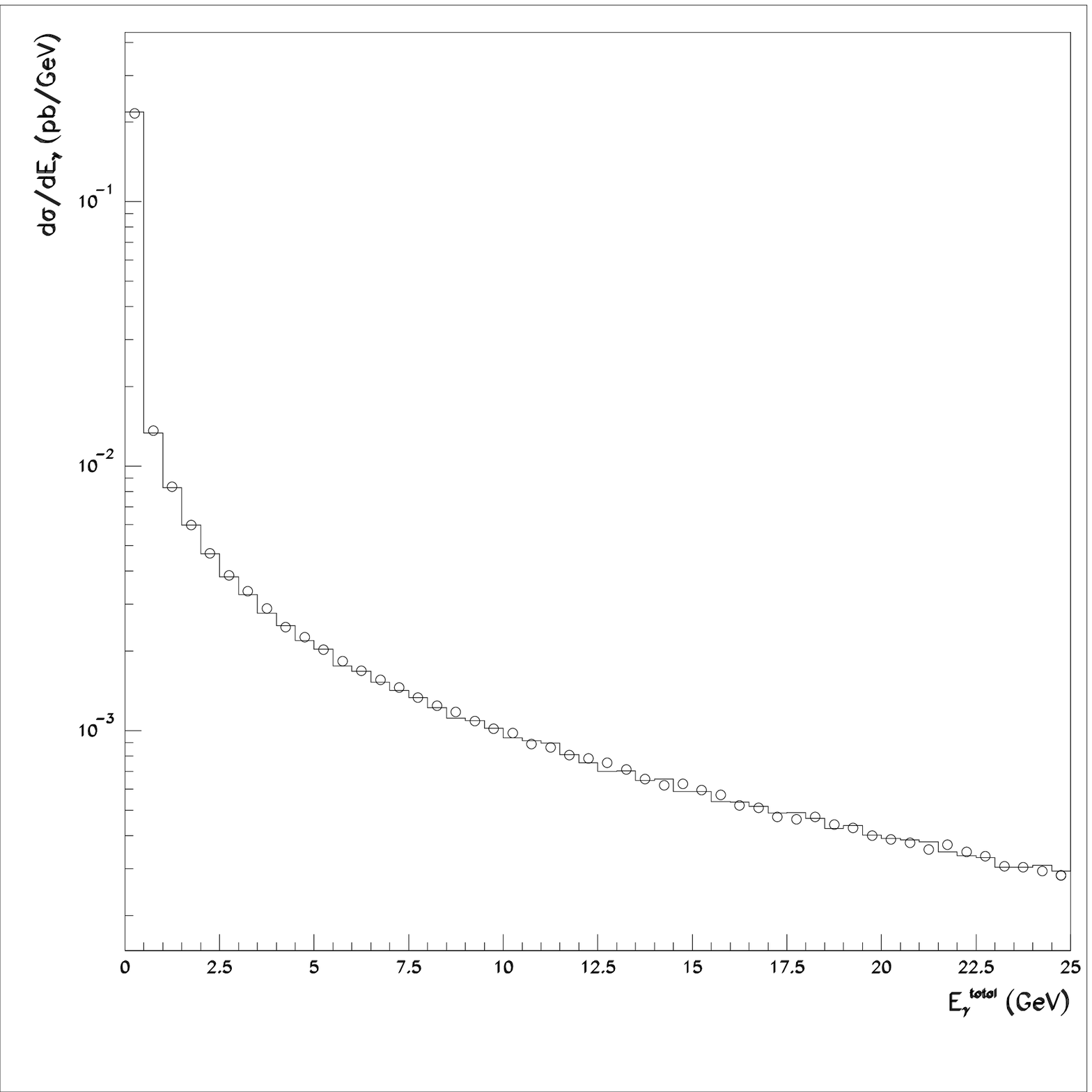}
\caption{
Differential cross-section of the total energy of emitted photon(s)
obtained from QEDPS(histogram) and from SF(circle).
}
\label{fig-grace-singlw-1}
\end{center}
\end{figure}

Total cross-sections as a function of the c.m.s. energies at LEP~2 with
and without experimental cuts are shown in 
\fig{fig-grace-singlw-2}. The experimental cuts applied here are
$M_{q {\barq}}>45 \,\GeV$ and $E_l>20 \,\GeV$.

\begin{figure}[htbp]
\begin{center}
\includegraphics[width=12cm]{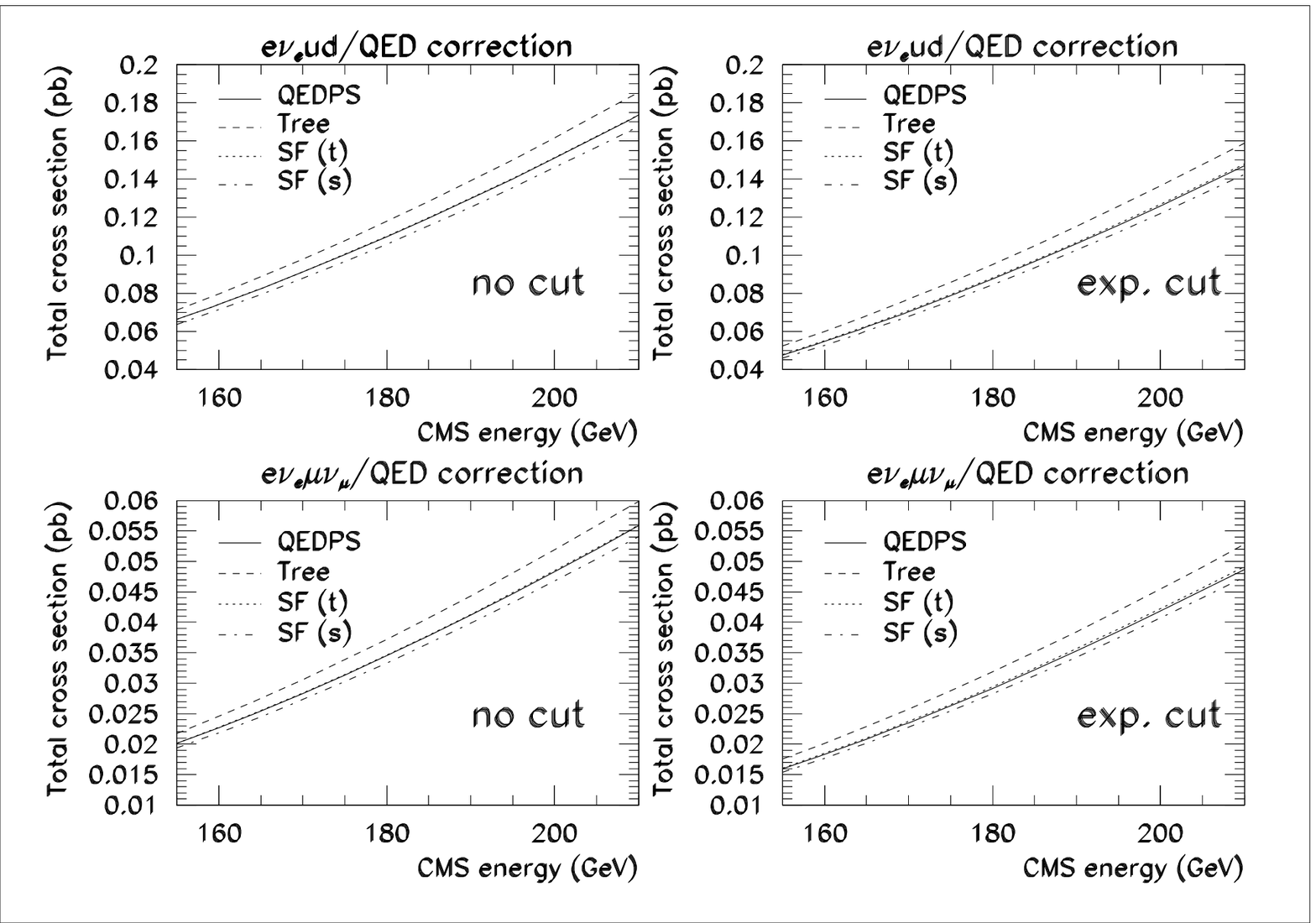}
\caption{
Total cross-sections of $e\nu_e\bar{u}d$ and $e\nu_e\mu\nu_{\nu}$
processes 
without and with experimental cuts.
SF(t) denotes SF with correct energy scale and SF(s) with wrong
energy scale ($s$).
}
\label{fig-grace-singlw-2}
\end{center}
\end{figure}

The effect of the QED
radiative corrections on the total cross-sections are obtained to be
$7$ to $10\%$ on LEP~2 energies. 
If one uses the wrong energy scale $s$ in SF, the ISR effect
is overestimated of about $4\%$ as shown in 
\fig{fig-grace-singlw-3} both with and without cuts.
For the fully extrapolated case the SF-algorithm with a correct energy scale 
is consistent with QEDPS within $0.2\%$. It may reflect
the difference between $L$ and $L-1$, as mentioned in \sect{esd}.
On the other hand, with the experimental cuts the  
SF-method at the correct energy-scale gives a deviation of around $1\%$ 
from QEDPS.

\begin{figure}[htbp]
\begin{center}
\includegraphics[width=12cm]{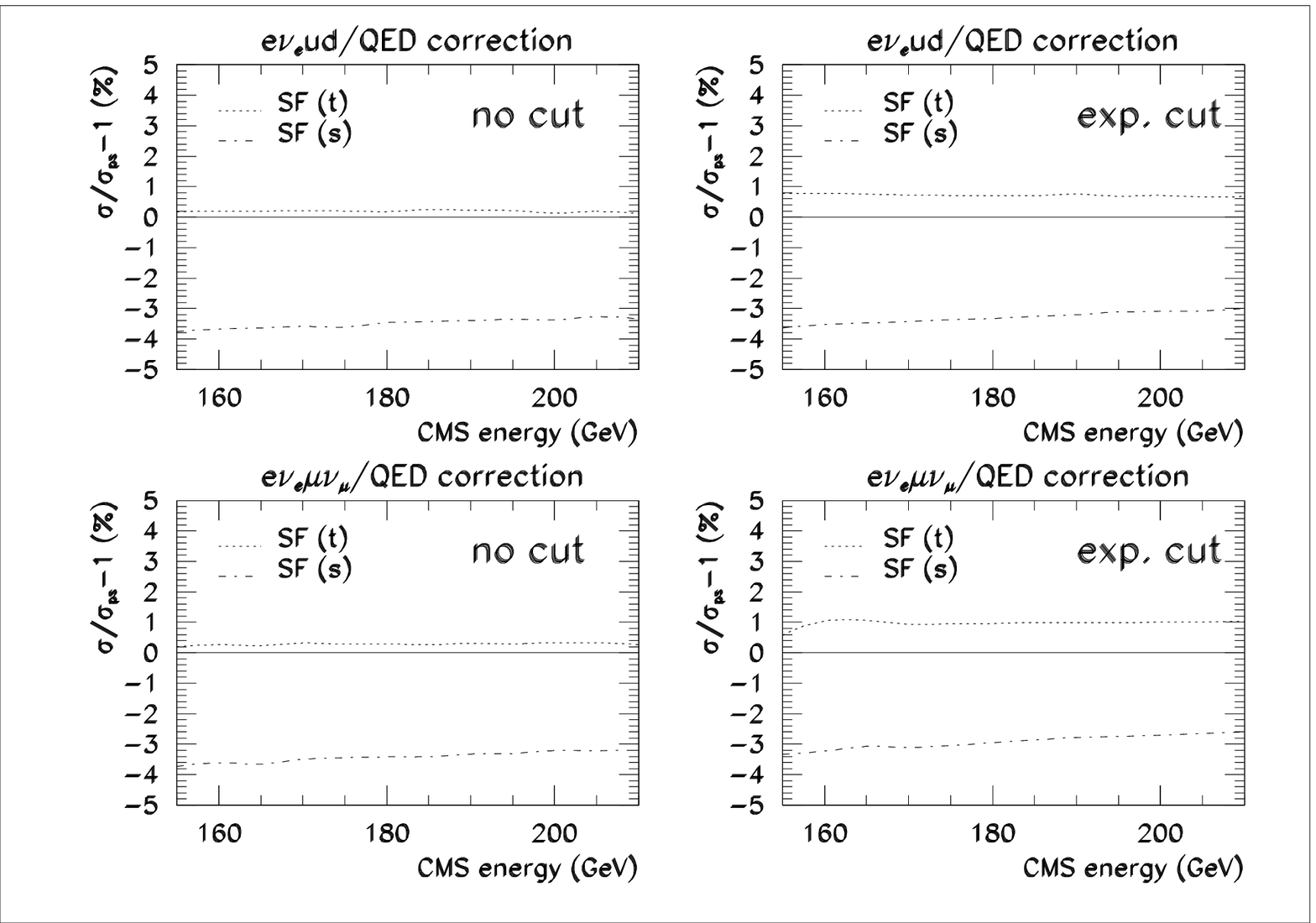}
\caption{
Total cross-sections with SF(s) and SF(t) normalized
to those with QEDPS for $e\nu_e\bar{u}d$ and
$e\nu_e\mu\nu_{\nu}$ processes without and with experimental cuts.
SF(t) denotes SF with correct energy scale and SF(s) with wrong
energy scale ($s$).
}
\label{fig-grace-singlw-3}
\end{center}
\end{figure}

\subsubsection*{Numerical Calculations, the hard photon spectrum}

Energy and angular distributions of the hard photon from QEDPS
are compared with those from the calculations of the exact matrix 
elements. The cross-sections of the process
$e^- e^+ \to e^- \barnu_e u \bard \gamma$
are calculated based on the exact amplitudes generated by ${\tt GRACE}$
and integrated numerically in five-body phase space using ${\tt BASES}$.
To compare the distributions, the soft-photon correction for 
the radiative process must be included. For this purpose
QEDPS is implemented into the calculation of the process
$e^- \barnu_e u \bard \gamma$ with a careful treatment aimed to avoid 
a double-counting of the radiation effect.
The definition of the hard photon is $E_{\gamma} > 1\,\GeV$ with
an opening angle between the photon and the nearest final-state 
charged particles that is greater than $5^\circ$.
The distributions of the hard photons are in good agreement as shown
in \fig{fig-grace-singlw-4}.
The total cross-section of the hard photon emission is consistent 
at the $2\%$ level. On the other hand, if the soft-photon
correction is not implemented on the radiative process, we end up with 
an over-estimate of $30\%$.

\clearpage

\begin{figure}[htbp]
\begin{center}
\includegraphics[width=12cm]{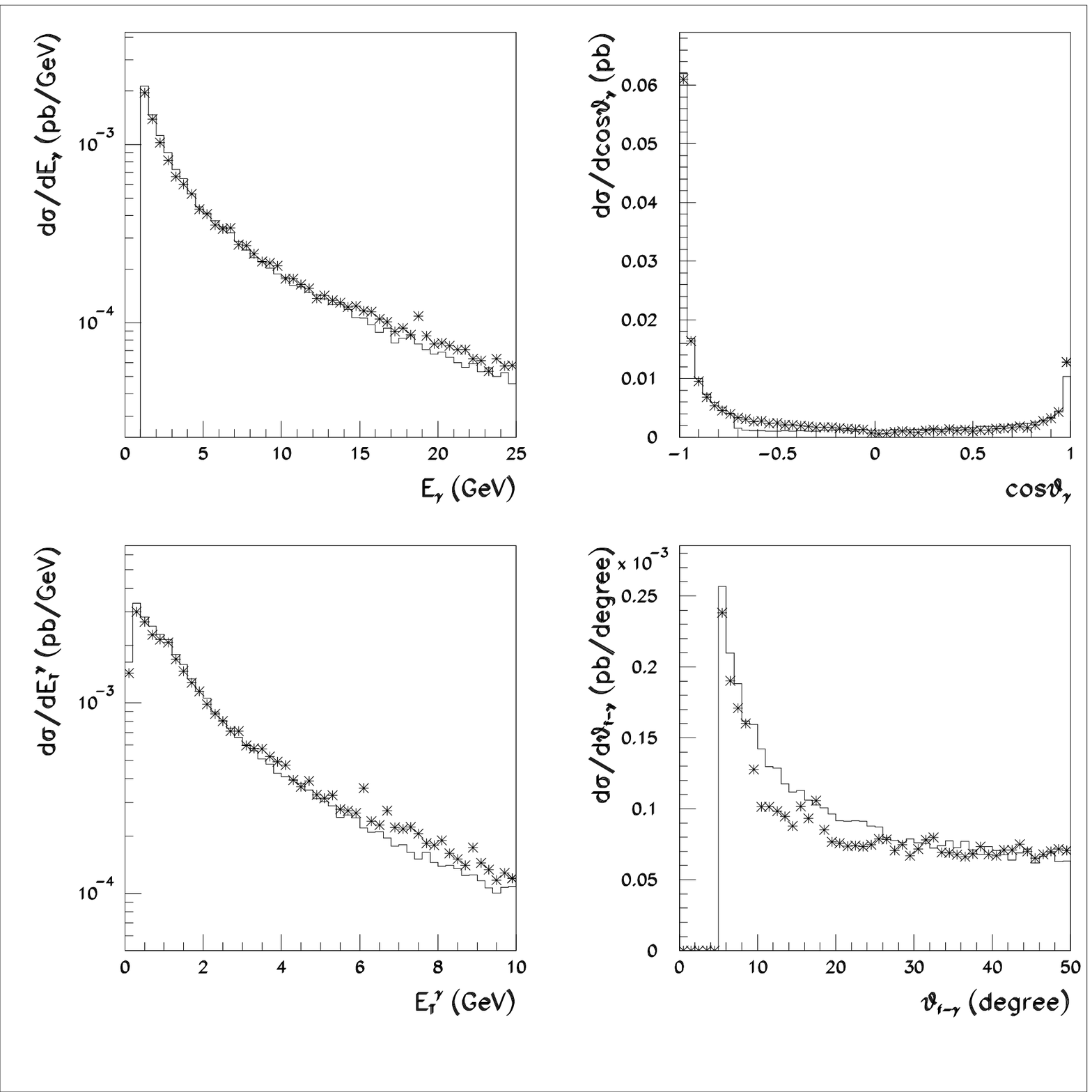}
\caption{
Differential cross-sections of the hard photon;
Energy, transverse energy w.r.t. the beam axis,
cosine of the polar angle, and
opening angle between photon and
nearest charged-fermion.
A histogram shows the QEDPS result and stars from the matrix element
with soft-photon correction.
}
\label{fig-grace-singlw-4}
\end{center}
\end{figure}

\subsection{Technical precision in single-$\wb$}

An old comparison for single-$\wb$ has been extended to cover
\ben
\item $e^+e^- \to q\barq e\nu(\ph)$, $|\cos\theta_e| > 0.997$, either
  $M(q\barq ) > 45\,$GeV or $E_{q_1}, E_{q_2} > 15\,$GeV, inclusive
  cross-section accuracy $2\%$, photon energy and polar angle
  ($|\cos\theta_{\ph}| < 0.997~(0.9995)$) spectrum
\item $e^+e^- \to e\nu e\nu(\ph)$, $|\cos\theta_e| > 0.997$, $E_e >
  15\,$GeV, $|\cos\theta_e| < 0.7~(0.95)$, inclusive cross-section
  accuracy $5\%$, photon energy and polar angle ($|\cos\theta_{\ph}| <
  0.997~(0.9995)$) spectrum.
\item $e^+e^- \to e\nu\mu\nu(\ph)$ and $e^+e^- \to e\nu\tau\nu(\ph)$,
  $|\cos\theta_e| > 0.997$, $E_{\mu/\tau} > 15\,$GeV,
  $|\cos\theta_{\mu/\tau}| < 0.95$, inclusive cross-section accuracy
  $5\%$, photon energy and polar angle ($|\cos\theta_{\ph}| <
  0.997~(0.9995)$) spectrum.
\een
With this comparison we want to check a)
technical precision at the Born level, b) the correct inclusion of QED 
radiation, c) QCD corrections, especially in the low-mass region.

The first answer is that technical precision is not a problem anymore,  
all codes agree on single-$\wb$ cross-sections and distributions, even 
for $\theta_e < 0.1^\circ$, even for leptonic final states.
On $\sigma_{\rm Born}$ the technical accuracy is $0.1\%$, the same for 
$d\sigma/d\theta_e$ for $\theta_e \to 0$.
Not only invariant-mass cuts, but also energy-cuts have been tested as
shown in \tabns{energycut1}{energycut2} and in \fig{teca}.

\clearpage

\begin{table}[hp]\centering
\begin{tabular}{|c|c|c|c|}
\hline
   & & &  \\
   & $\sqrt{s} = 183\,$GeV & $\sqrt{s} = 189\,$GeV & $\sqrt{s} = 200\,$GeV \\
   & & &  \\
\hline
   & & &  \\
{\tt NEXTCALIBUR}  & $26.483 \pm 0.041$ &  $29.679 \pm 0.047$
& $35.893 \pm 0.048$ \\
   & & &  \\
\hline
   & & &  \\
{\tt SWAP}      & $26.47 \pm 0.04$ &  $29.70  \pm 0.04$
& $35.93 \pm 0.05$ \\ 
   & & &  \\
\hline
\end{tabular}
\vspace*{3mm}
\caption[]{Cross-sections [fb] for $e^+e^- \to e^-\barnu_e \mu^+\barnu_{\mu}$.}
\label{energycut1}
\end{table}

\begin{table}[hp]\centering
\begin{tabular}{|c|c|c|c|}
\hline
   & & &  \\
   & $\sqrt{s} = 183\,$GeV & $\sqrt{s} = 189\,$GeV & $\sqrt{s} = 200\,$GeV \\
   & & &  \\
\hline
   & & &  \\
{\tt NEXTCALIBUR}  & $26.422 \pm 0.035$ &  $29.655 \pm 0.046$
& $35.954 \pm 0.052$ \\
   & & &  \\
\hline
   & & &  \\
{\tt SWAP}      & $26.3 \pm 0.2$ &  $29.6  \pm 0.2$
& $35.92 \pm 0.05$ \\ 
   & & &  \\
\hline
\end{tabular}
\vspace*{3mm}
\caption[]{Cross-sections [fb] for $e^+e^- \to e^-\barnu_e \tau^+
\barnu_{\tau}$.}
\label{energycut2}
\end{table}

\begin{figure}[htbp]
\centerline{
\epsfig{file=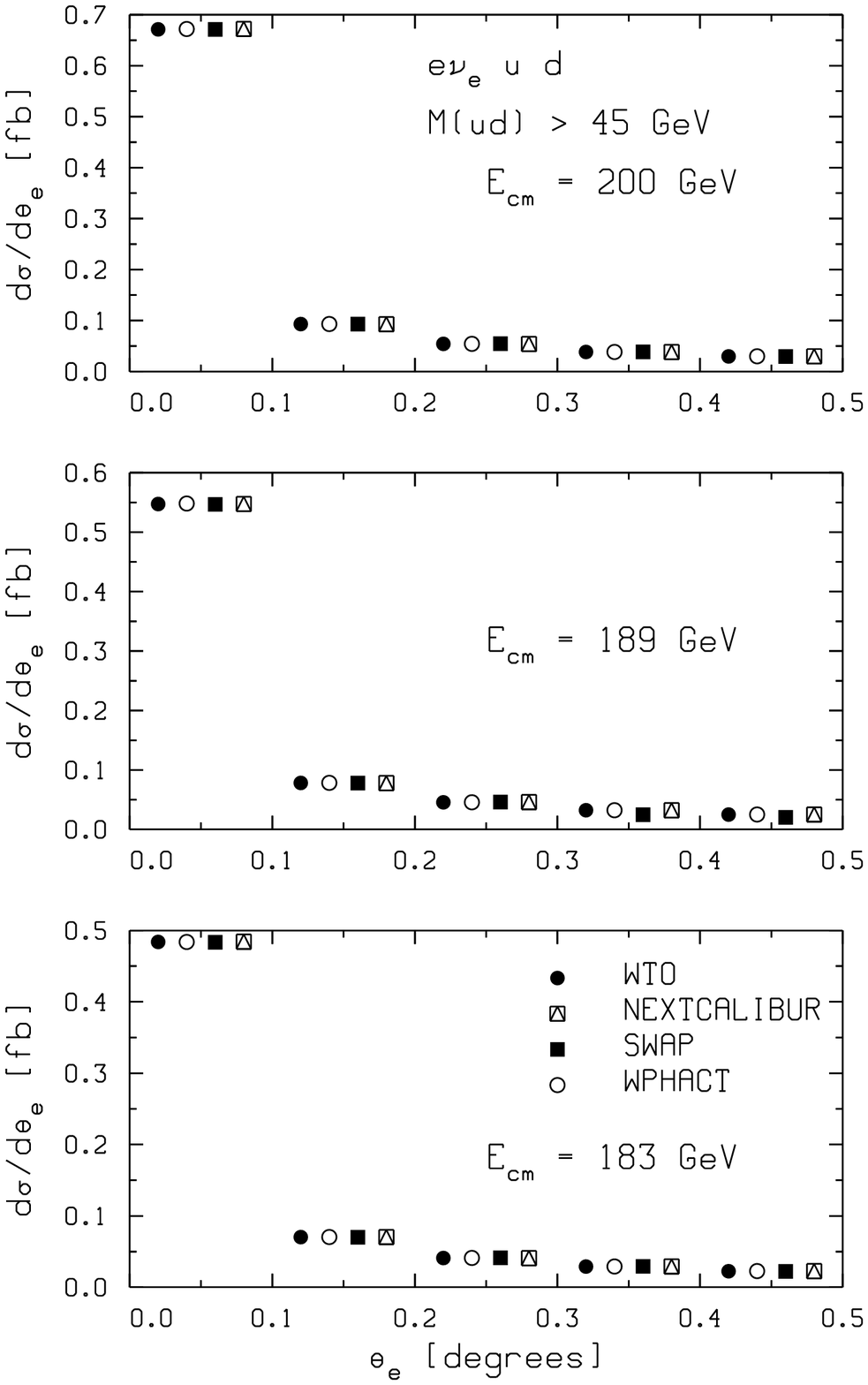,height=14cm,angle=0}
\epsfig{file=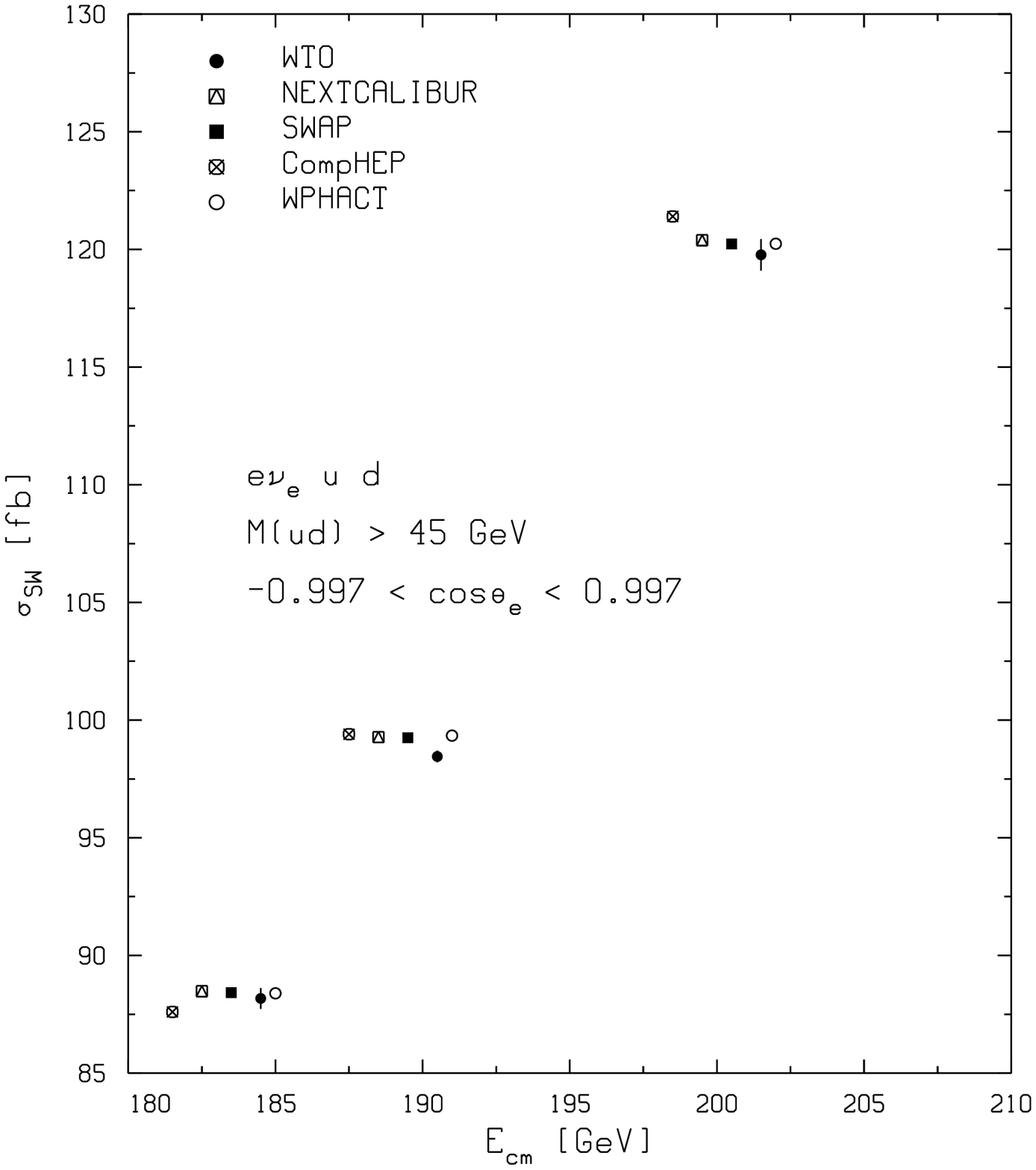,height=10cm,angle=0}}
\caption[]{$\theta_e$ distributions for $u \bard \mu^- \barnu_{\mu}$ and
single-$\wb$ cross-sections for $u \bard \mu^- \barnu_{\mu}$.}
\label{teca}
\efi

\subsubsection{QCD corrections}
\label{swout}

QCD corrections are usually implemented in their {\em naive} form, a
recipe where the total $\wb$-width is corrected by a factor
\bq
\Gamma_{_\ssW} = \Gamma^{\rm EW}_{_\ssW}\,\lpar 1 + \frac{2}{3}\,
{{\alpha_s(\mws)}\over {\pi}}\rpar,
\eq
and the cross-section gets multiplied by $1 + \alpha_s(\mws)/\pi$.
In all those approaches where the Fermion-Loop is included or simulated,
one should pay particular attention to QCD, for instance in {\tt WTO}
QCD corrections are incorporated in the evaluation of the complex poles
by using the $\ord{\alpha\als}$ vector-boson self-energies of Ref.~\cite{ak}
(the location of the poles is gauge-invariant).
Furthermore, the vertices are effectively corrected so that the relevant
Ward identity remains satisfied.
In a similar way {\tt WPHACT} also includes QCD effects in the computation
of the imaginary part of both the re-summed propagators and the vertices,
to preserve gauge invariance.
                                  
To check the effect of QCD corrections we have compared {\tt WPHACT} 
(IFL$_{\alpha}$)
with {\tt WTO} (EFL) for $e\nu_e u d$ final states in LEP~2 configuration
with and without QCD. The comparison is shown in \tabn{swqcd} where the
first error for {\tt WTO} comes from a variation of the scale $\mu$
from $\mu/2$ to $2\,\mu$, where we adopt $\mu = \mw$ as the scale for light
quarks and $\mu = \mt$ for the $b-b, b-t$ and $t-t$ contributions.
Therefore, QCD effects in single-$\wb$ are under control in those programs 
that implement them consistently with Fermion-Loop.
\begin{table}[hp]\centering
\begin{tabular}{|c|c|c|}
\hline
&&  \\
 & without QCD & with QCD  \\
&&  \\
\hline
&&  \\
{\tt WPHACT}       & $107.63 \pm 0.10$ & $109.18 \pm 0.08$  \\
&&  \\
{\tt WTO}          & $108.96 \pm 0.04$ & $110.63^{+0.18}_{-0.04} \pm 0.04$ \\
&&  \\
{\tt WPHACT/WTO} -1 [$\%$] &  -1.22     &  -1.07   \\
&&  \\
\hline
\end{tabular}
\vspace*{3mm}
\caption[]{Cross-sections [fb] at $\sqrt{s} = 182.655\,\GeV$ for 
$e^+e^- \to e^-\barnu_e u \bard$ (LEP~2 configuration)
with and without QCD corrections. The first error in {\tt WTO} comes
from a variation in the scale $\mu$ from $\mu/2$ to $2\,\mu$.}
\label{swqcd}
\end{table}

\clearpage

\subsubsection{Assessing the theoretical uncertainty in single-$\wb$}

If we do not want to use the Fermion-Loop
prediction then, by a careful examination of the most plausible
re-scaling procedure, we end up with approximately $1\%,2\%$ and $3\%$
theoretical uncertainty to be assigned to the energy scale in the
channels $ud$, $\mu\nu_{\mu}$ and $e\nu_e$ respectively.
Therefore, a conservative estimate of the theoretical uncertainty would read 
as follows:
\begin{description}
\item[Energy scale:]  $\pm 2 \div 3\%$ from a tuned comparison among 
{\tt NEXTCALIBUR}, {\tt WPHACT} and {\tt WTO};
\item[ISR for $t$-ch, $p_t$:] $\pm 4\%$: if one uses the wrong energy scale 
$s$ in SF, the ISR effect is, indeed, overestimated by approximately $4\%$ as 
shown in the subsequent analysis.

\end{description}
\noindent 
giving a conservative total upper bound of $\pm 5\%$, see \subsect{summconc} 
for a more complete discussion. 

One should stress that most of the theorists were interested in gauge
invariance issues due to unstable particle for CC20. The experimentalists,
however, were asking from the beginning for ISR $p_t$ effects, comparison 
with QEDPSt, SF and YFS. Unfortunately, only few groups have been working 
on these issues.

In the previous section few recipes have been introduced
to improve upon QED ISR; they are all equivalent insofar as they translate 
into different choices for the scale in the leading-logarithms of the 
structure functions. However this problem has not yet received its
final solution and a full $\ord{\alpha}$ calculation would be needed.

There is, however, an additional complication in the use of QED
structure functions originating from mass effects.  The single-$\wb$
is $s\, \oplus\,t$- channels and the $t$-channel parts look as in
\fig{basic}.

\begin{figure}[htbp]
\vspace{0.2cm}
\bqas
\ba{ccc}
\vcenter{\hbox{
  \begin{picture}(110,100)(0,0)
  \ArrowLine(50,50)(0,100)
  \ArrowLine(100,100)(50,50)
  \ArrowLine(0,0)(50,50)
  \ArrowLine(50,50)(100,0)
  \ArrowLine(100,70)(50,50)
  \ArrowLine(50,50)(100,30)
  \GCirc(50,50){15}{0.5}
  \Text(10,100)[lc]{$e^+$}
  \Text(10,0)[lc]{$e^-$}
  \Text(110,100)[lc]{$\barnu_e$}
  \Text(110,70)[lc]{$\barf_2$}
  \Text(110,30)[lc]{$f_1$}
  \Text(110,0)[lc]{$e^-$}
  \end{picture}}}
&\quad+&
\vcenter{\hbox{
  \begin{picture}(110,100)(0,0)
  \ArrowLine(50,50)(0,100)
  \ArrowLine(100,100)(50,50)
  \ArrowLine(100,70)(50,50)
  \ArrowLine(50,50)(100,30)
  \Photon(50,50)(50,10){2}{7}
  \ArrowLine(0,0)(50,10)
  \ArrowLine(50,10)(100,0)
  \GCirc(50,50){15}{0.5}
  \Text(10,100)[lc]{$e^+$}
  \Text(10,0)[lc]{$e^-$}
  \Text(110,100)[lc]{$\barnu_e$}
  \Text(110,70)[lc]{$\barf_2$}
  \Text(110,30)[lc]{$f_1$}
  \Text(110,0)[lc]{$e^-$}
  \Text(60,20)[lc]{$Q^2$}
  \end{picture}}}
\ea
\eqas
\vspace{-2mm}
\caption[]{The CC20 family of diagrams with the explicit component 
containing a $t$-channel photon.}
\label{basic}
\end{figure}

The corresponding cross-section is proportional to
\bq
\int d\Phi_3 \,\frac{1}{\hq^4}\,{\hat L}_{\mu\nu} {\hat W}_{\mu\nu}, \quad
\hq = {\hat p}_- - q_-, \quad
{\hat L}_{\mu\nu} = \frac{1}{2}\,\hq^2\delta_{\mu\nu} + {\hat p}_{_\mu}
{\hat q}_{_\nu} + {\hat q}_{-\mu}{\hat p}_{-\nu}
\eq
\begin{eqnarray*}
\int d\Phi_3 {\hat W}_{\mu\nu} &=& {\hat W}_1\,( - \delta_{\mu\nu} +
\frac{\hq_{\mu}\hq_{\nu}}{\hq^2}) - \frac{\hq^2}{({\hat p}_+\cdot\hq)^2}\,
{\hat W}_2\,{\cal P}_{\mu}{\cal P}_{\nu}  \nl
{\cal P}^{\mu} &=& {\hat p}_+^{\mu} - \frac{{\hat p}_+\cdot\hq}{\hq^2}\,
\hq^{\mu}  \nl
\end{eqnarray*}
where ${\hat p}$ and ${\hat q}$ denote emission of soft and collinear photons.
Usually, $p_-^2 = {\hat p_-}^2 = 0$ and $q_-^2 = {\hat q_-}^2 = 0$,
and one writes ${\hat p_-} = x_{\rm in}\,p_-$ and ${\hat q_-} = 
q_-/x_{\rm out}$ with the kernel cross-section to be weighted with 
structure functions. Here, however, masses matter should not be neglected and 
the electrons are in a virtual state, \ie off their mass-shell. 
A possible choice is to write
\begin{eqnarray*}
({\hat p_-})^2 &=& - \mes + \frac{1}{2}\,(1-\beta)\,(1-x_{\rm in})\,s 
\sim - x_{\rm in}\,\mes  \nl
\end{eqnarray*}
The important facts are that $\hq_{\mu}{\hat L}_{\mu\nu} = 0$ owing to
gauge invariance. Instead we get 
\bqas
\hq_{\mu}{\hat L}_{\mu\nu} &=& 4\,(1-x_{\rm in})\,\mes\,q_{-\nu} +
4\,(1-\frac{1}{x_{\rm in}})\,\mes\,p_{-\nu}.
\eqas
Even if we insist in putting ${\hat p}_- = x_{\rm in} p_-$ and
$ p_-^2= - \mes$, gauge invariance is violated by terms of 
$\ord{\mes/s}$.
The effect of constant terms on the $Q^2$-integrated photon flux-function
can be as large as $6\%$. Gauge-invariance violation affects this term, 
resulting in some intrinsic theoretical uncertainty, although we expect that 
the effect will be strongly decreased after convolution with SF peaking
at $x_{\rm in/out} = 1$.
Alternatively one may adopt formulations where the electron remains on-shell
after emission but at the price of having collinear photons of non-zero
virtuality, $({\hat p} - p)^2 \not= 0$.

It is worth noticing that, the rescaled 
incoming four-momenta are implemented in {\tt SWAP} as 
$\hat{p}_{\pm} = (x \, E, 0, 0 , \pm \sqrt{x^2 E^2 - m_e^2})$, by interpreting 
$x$ as the energy fraction after photon radiation, as motivated in 
Ref.~\cite{babayaga}. If required, $p_\perp / p_{\ssL}$ effects can be 
implemented in the treatment of ISR, by means of either $p_\perp$-dependent 
SF~\cite{mmnp} or a QED Parton Shower algorithm~\cite{babayaga}.
Therefore, in practice {\tt SWAP} adopts a formulation that preserves 
on-shell incoming electrons.   
Furthermore, in {\tt NEXTCALIBUR} it is possible to have both
on-shell initial state particles and on shell generated photons
but at the price of loosing part of the information on the direction of the 
initial states after radiation.

A final set of comments is needed to quantify the theoretical accuracy of 
single-$\wb$ production. 

\subsubsection*{\tt GRACE}

The method to apply the QED radiative correction on the non-annihilation
processes are established. The conventional method, SF with energy scale $s$
gives about $4\%$ overestimation for the QED radiative effect on the LEP~2
energies. If one wants to look at the hard photon spectrum, the
soft-photon correction on these radiative processes are needed.

\subsubsection*{\tt SWAP}

The difference shown in \fig{fig1_sw} between the predictions given by the two 
set of $Q^2$ scales of \eqn{eq:wscales} and \eqn{eq:naive} is at the per mille 
level, and therefore the simple naive scales of \eqn{eq:naive} are a good 
ansatz for the energy scale of QED radiation, which could be corroborated by 
the comparison with the results of other groups.
QED corrections missing in the present approach are beyond the LL approximation.
The present study shows that the choice 
$Q^2_{\pm} = s$ as scale in the IS QED SF(s)
can lead to over-estimate the effect of LL photonic
corrections by a factor of $1.5$, implying an under-estimate of 
the QED corrected cross-section
of about $4\%$. Also the choice of fixing the scale to 
$Q^2_{\pm} = |q^2_{\ph^*}|$ for both the
IS SF(s), as recently suggested~\cite{bd}, 
leads to an under-estimate of the photon
correction of about $4\%$. Since these effects are not negligible 
in the light of the expected theoretical accuracy, it is 
recommended to adopt the $Q^2$-scales as given in 
eq.~(\ref{eq:wscales}) or eq.~(\ref{eq:naive}), 
which are motivated by the arguments sketched above.
 
Further, the effect of rescaling 
$\alpha_{\rm QED}$ from the high-energy value $\alpha_{\gf}$ 
to $\alpha(q^2_{\ph^*})$ amounts to a negative correction of about $5 
\div 6\%$, to be taken into account carefully.

\subsubsection*{\tt WPHACT}

From the cross-sections of \tabns{tabxsee}{tabangee}
one deduces that the difference between IFL and 
IFL$_{\alpha}$ is of the order of $6\%$, for both \processmixeevv\ and
\processnceevv. The discrepancy between IFL$_{\alpha}$ and IFL$_{\alpha 1}$ 
predictions is always of the order of $3\%$ and one has, therefore, to attribute
an estimate of $3\div 4$ \% error to the IFL$_{\alpha}$ calculations for these 
processes. 
 
Considering all processes together one can conclude that the implementation
of a proper running $\alpha_{\rm QED}$  reduces the theoretical uncertainty by 
about one half with respect to fixed-width or imaginary fermion-loop alone.
In some cases this uncertainty is further reduced to less than one percent, 
but only a comparison with complete EFL calculations, 
as a reference point, may assess whether 
this is the case. If no comparisons are available for the process and cuts at 
hand our study points towards a $3\%$ uncertainty for the calculations using
the running $\alpha_{\rm QED}$, for both single-$\wb$ and single-$\zb$
processes.
Of course, one should add the uncertainty due to the fact that ISR
for annihilation processes is not suited for $t$-channel contributions. Some 
obvious improvements on this point will soon be implemented in \wph, however
a more careful study both for the theoretical uncertainty of these solutions
and for a better treatment of $t$-channel ISR is still needed. 

\subsubsection*{\tt WTO}

Bosonic corrections are still missing and, very often, our 
experience has shown, especially at LEP~1, that bosonic corrections may become 
sizeable~\cite{LEPreport_95}.
A large part of the bosonic corrections, as e.g.\ the leading-logarithmic 
corrections, factorize and can be treated by a convolution. Nevertheless
the remaining bosonic corrections can still be non-negligible, \ie,
of the order of a few percent at LEP~2~\cite{LEP2WWreport}. For the Born 
cross-sections $1 \div 2\%$ 
should, therefore, be understood as the present limit
for the theoretical uncertainty. This will have to be improved, soon or later,
since bosonic corrections are even larger at higher energies 
\cite{ww-review}~\cite{LCreport}
and the single-$\wb$ cross-section will cross over the $\wb\wb$ one at $500\,$
GeV.

\subsection{Summary and conclusions}
\label{summconc}

A fairly large amount of work has been done in the last
years on the topic of single-$\wb$.
In the previous sections we presented the most recent theoretical
developments in single-$\wb$ and their implementation in the generators.
There are common problematic situations with more or less equivalent
solutions. One has to assign an error band to the cross-section 
for our partial knowledge of ISR, with or without $p_t$, and for the
uncertainty in the scale of the running couplings.
As for the energy scale in couplings we have an exact calculation based on 
the EFL-scheme which, at the Born-level (no QED) is known to be at 
the $1\%$ level of accuracy. EFL-scheme, however, is implemented only in one 
generator while the other offer a wide range of approximations based on the 
idea of re-scaling the cross-section. 
Furthermore, no program includes $\ord{\alpha}$ electroweak corrections, not
even in Weizs\"acker-Williams approximation (for the subprocess 
$\Pe\gamma\to\PW\nu_\Pe$), the counterpart of DPA in CC03.

A description of single-$\wb$ processes by means of the EFL-scheme is
mandatory from, at least, two points of view. EFL is the only known 
field-theoretically consistent scheme that preserves gauge invariance in
processes including unstable vector-bosons coupled to e.m. currents.
Furthermore, single-$\wb$ production is a process that depends on several
scales, the single-resonant $s$-channel exchange of $\wb$-bosons, the
exchange of $\wb$-bosons in $t$-channel, the small scattering angle peak
of outgoing electrons. 
A correct treatment of the multi-scale problem can only be achieved when we
include radiative corrections in the calculation, not only one-loop terms
but also the re-summation of leading higher-order terms. Recent months have
shown that this project can be brought to a very satisfactory level by
identifying the correct approximation, process-by-process.
In particular, the $\wb-\wb$ configuration, dominated by double-resonant
terms, can be treated within 
DPA. As a consequence, the theoretical uncertainty associated 
with the determination of the $\wb\wb$ cross-section is sizably decreased. In
principle, the same procedure applies to the determination of the $\zb\zb$
cross-section, where one develops a NC02-DPA  instead of 
the CC03-DPA one.

We have found that all the modifications introduced via the 
EFL-scheme are relevant: running of the couplings, $\rho$-factors 
and vertices, not only the change $\alpha_{\rm QED}({\rm fixed}) \to 
\alpha_{\rm QED}({\rm running})$.
Therefore, a naive rescaling cannot reproduce the EFL answers for
all situations, all kinematical cuts. 
The high-energy Input Parameter Set used in all calculations that are
presently available -- we quote, among the various schemes, the Fixed-Width 
scheme, the Overall scheme and the IFL one -- is
based on $\gf, \mw$ and $\mz$ with $\alpha_{\rm QED}({\rm fixed}) =
1/131.95798$. It allows for the inclusion of part of higher order effects in 
the Born cross-sections but, it fails to give a correct and accurate 
description of the $q^2 \sim 0$ dominated processes. 
A naive, overall, rescaling would lower the single-$\wb$ cross-section of 
about $7\%$. We have found, with the EFL calculation, that this decrease 
is process and cut dependent.
Moreover, the effect is larger in the first bin for $\theta_e$ -- $0.0^\circ 
\div 0.01^\circ$ -- in the distribution $d\sigma/d\theta_e$ and tends to 
become less pronounced for larger scattering angles of the electron. However, 
the first bin represents almost $50\%$ of the total single-$\wb$ cross-section,
so that, in general, the compensations that occur among several effects never 
bring the EFL/FW ratio to one. We obtain a maximal decrease of about $7\%$ in
the result but, on average, the effect is smaller. We have also found that
the effect is rather sensitive to the relative weight of multi-peripheral
contributions.

Finally, the effect of the QED radiative corrections on the total 
cross-sections are between $7\%$ and $10\%$ at LEP~2 energies. 
{\tt grc4f} and {\tt SWAP} have estimated that if one uses the wrong energy 
scale $s$ in the
structure functions, the ISR effect is overestimated of about $4\%$, as shown 
in \figsc{fig1_sw}{fig-grace-singlw-3}, both with and without cuts.
For the no-cut case SF with a correct energy scale is consistent with 
QEDPS around $0.2\%$. 
On the other hand with the experimental cuts, SF with correct energy-scale 
gives around $1\%$ deviation from QEDPS.

At the same time {\tt SWAP} estimates that
the effects due to two different scales (eq.~(\ref{eq:wscales}) and 
eq.~(\ref{eq:naive}) are in good agreement and both predict a 
lowering of the Born cross-section of about $8\%$, almost 
independent of the c.m.s. LEP~2 energy. {\tt SWAP} results show a good
agreement with those of {\tt grc4f} when both are referred to
$s$-channel SF.

Although we register substantial improvements upon the standard treatment
of QED ISR, the problem is not yet fully solved for processes where the
non-annihilation component is relevant. A solution of it should rely on
the complete calculation of the $\ord{\alpha}$ correction, therefore the
the basic YFS approach or any equivalent one augmented by virtual 
corrections.

At the moment, a total upper bound of $\pm 5\%$ theoretical uncertainty should 
still be assigned to the single-$\wb$ cross-section. 
In particular, the difference between annihilation-like QED radiation and
the optimized scales amounts to a $4\%$, which is conservatively used
(by the LEP EWWG) in the global estimate of theoretical uncertainty. 
Alternatively one should use the differences between different implementations 
of ISR in the $t$-channel as a basis for the systematic uncertainty. However, 
we are not yet ready to formulate a strict and definitive statement along 
these lines.
Furthermore, there seems to be and indication of some numerical difference 
arising from different QED treatments in {\tt GRACE} and in {\tt SWAP}.
At present no direct comparison has been attempted to understand the origin.
We could say that QED radiation in single-$\wb$ is understood at a level
better than $4\%$ but we are presently unable to quantify this assertion.

In this sense the current $5\%$ should be considered as a good estimate of 
the global upper limit for theoretical uncertainty.
The origins of this upper bound are as follows. QED effects are bounded
by a $4\%$, saturated only by those programs that do not improve upon the
scale. Effects due to running couplings and vertices are bounded by a 
$2\% \div 3\%$, saturated by those programs that do not implement an
exact massive FL-scheme. To lower this estimate is presently possible only
in a multi-step procedure where program A is used vs. B to assess the
effect of EFL/$\alpha_{\rm QED}$, then A is used vs. C to assess the effect of
QED ISR and finally A is corrected to take into account the missing pieces 
and assign an uncertainty. This procedure should be performed within the
experimental community, using the individual estimates of theoretical 
uncertainties as declared by the programs in this Section.

We expect an improvement upon 
this estimate when more implementation of the Fermion-Loop scheme will be 
available. Presently, the results with a rescaling of $\alpha_{\rm QED}$ 
for the
$t$-channel photon show an agreement with EFL predictions that is between
$1\%$ and $2\%$. Note that in $e\nu e\nu$ EFL is not yet implemented and there
we use the estimate by {\tt WPHACT} of roughly $3\%$.
All programs that implement the correct running of $\alpha_{\rm QED}$
should be able to reach this level of accuracy, but not all of them have
this implementation. 

All program that still implement $s$-channel structure functions saturate 
the $5\%$ level of theoretical accuracy. Further and more complete studies 
are needed for QED corrections and ad hoc solutions, like {\em fudge factors} 
should be avoided.

As stated above, the present level of global theoretical uncertainty, $5\%$, 
comes from different sources and different effects. Some of them have been 
fully understood
from a theoretical point of view but, sometimes, not yet implemented in
most of the programs. There are remaining problems that have not yet
received a satisfactory solution and some of the programs implement educated
guesses. In general we should say that single-$\wb$ remains, to a large
extent, the land of {\em fudge} factors. As for individual programs,
the following collaborations (listed in alphabetic order) have agreed to 
quantify their performances:


{\tt NEXTCALIBUR} tries to include all leading higher order  
effects. At present, by comparing with the Exact-Fermion Loop and
varying the internal parameters of the program, we
can assign a conservative $3\%$ uncertainty coming from our
Modified fermion loop approach. On the other hand, 
our solution to the $t$-channel ISR problem represents the best we 
have so far at the theoretical level. Therefore, the final 
precision of $\pm 5\%$ on single-$\wb$ has to be considered as a
safe estimate of the accuracy reached by the program,
at least in absence of large angle hard photons.

{\tt SWAP} includes exact tree-level matrix 
elements with finite fermion masses and 
anomalous trilinear gauge couplings, the effect of 
vacuum polarization, higher-order leading QED 
corrections according to the treatment for the energy scale 
as given by the (equivalent) choices of eq.~(\ref{eq:wscales}) 
and eq.~(\ref{eq:naive}). Since, apart from the effect of 
the running of $\alpha_{\rm QED}$, other one-loop fermionic and 
bosonic corrections are still missing in {\tt SWAP}, its 
theoretical uncertainty is at the level of $2-3\%$\footnote{to be compared
with the estimated upper bound of $5\%$}, depending on the channel
and/or event selection considered.  

\wph\ can be used for single-$\wb$ in its version IFL$\alpha$. 
This is at present the best choice: all other schemes have been employed for 
studies and comparisons but are not recommended. As already explained, the
theoretical uncertainty due to non implementation of the complete EFL amounts
to $3-4\%$ for CC20/Mix56. This, together with the non correct QED radiation 
for $t$-channel, leads to an estimate of $5-6\%$ accuracy in actual \wph\ 
single-$\wb$ predictions. 

{\tt WTO} can only be used as a benchmark for the determination of
the scales in the coupling constants. In its default {\tt WTO}
saturates the upper bound of $5\%$ of accuracy. Ideally, the difference 
between any program using some approximation and {\tt WTO} should be 
considered as systematic uncertainty for the scale determination (in couplings)
of that program. In practice EFL, the right approach, is only implemented in 
{\tt WTO} and a cross-check is needed before being able to apply the previous 
rule. The correct treatment of QED radiation is still missing, it is a choice 
of the author to avoid ad hoc solutions and a consistent upgrading is 
currently under study. Furthermore, Fermion-Loop (as DPA) implies certain
characteristics and programs that implement a incomplete-FL that does not 
reflect at least a large fraction of them should refrain from using the label 
FL.


The collaborations each take responsibility for the above statements that
range from conservative to more optimistic ones. 

\subsection{Outlook}

A substantial amount of work was done in the last two years on the topic of 
single-$\wb$ production. This has triggered theoretical developments which
can be used also in other areas, \eg massive Fermion-Loop scheme, 
QED radiation in processes dominated by $t$-channel diagrams.
One of the main results of the theoretical activity has been to upgrade
programs that where available prior to the workshop and did not provide a
satisfactory simulation of the process.
They might have given a numerically more or less correct cross section, but 
this was mostly an accident.

Some work has not yet been done, \eg low--invariant-mass $e\nu+\,$hadrons
final states (searches), DPA-equivalent set of radiative corrections
(high-luminosity LC). Finally, we still do not have a complete, 
solution to the ISR problem, although  
there has been considerable progress in the treatment of QED radiation, in
particular in the determination of the radiation scale. Going beyond the
present level of theoretical accuracy would require a complete $\ord{\alpha}$
calculation therefore contributing to improve the present level of 
theoretical accuracy.

\subsection*{Appendix: the diagrams}

\begin{figure}[htbp]
{
\unitlength=1.0 pt
\SetScale{1.0}
\SetWidth{0.7}      
\scriptsize    
{} \qquad\allowbreak
\begin{picture}(95,99)(0,0)
\Text(15.0,80.0)[r]{$e$}
\ArrowLine(16.0,80.0)(37.0,80.0) 
\Text(47.0,84.0)[b]{$e$}
\ArrowLine(37.0,80.0)(58.0,80.0) 
\Text(80.0,90.0)[l]{$\nu_e$}
\ArrowLine(58.0,80.0)(79.0,90.0) 
\Text(54.0,70.0)[r]{$W^+$}
\DashArrowLine(58.0,60.0)(58.0,80.0){3.0} 
\Text(80.0,70.0)[l]{$\bar{\nu}_e$}
\ArrowLine(79.0,70.0)(58.0,60.0) 
\Text(80.0,50.0)[l]{$e$}
\ArrowLine(58.0,60.0)(79.0,50.0) 
\Text(36.0,60.0)[r]{$\gamma$}
\DashLine(37.0,80.0)(37.0,40.0){3.0} 
\Text(15.0,40.0)[r]{$\bar{e}$}
\ArrowLine(37.0,40.0)(16.0,40.0) 
\Line(37.0,40.0)(58.0,40.0) 
\Text(80.0,30.0)[l]{$\bar{e}$}
\ArrowLine(79.0,30.0)(58.0,40.0) 
\Text(47,0)[b] {diagr.1}
\end{picture} \ 
{} \qquad\allowbreak
\begin{picture}(95,99)(0,0)
\Text(15.0,80.0)[r]{$e$}
\ArrowLine(16.0,80.0)(37.0,80.0) 
\Line(37.0,80.0)(58.0,80.0) 
\Text(80.0,90.0)[l]{$\nu_e$}
\ArrowLine(58.0,80.0)(79.0,90.0) 
\Text(33.0,70.0)[r]{$W^+$}
\DashArrowLine(37.0,60.0)(37.0,80.0){3.0} 
\Text(15.0,60.0)[r]{$\bar{e}$}
\ArrowLine(37.0,60.0)(16.0,60.0) 
\Text(47.0,64.0)[b]{$\nu_e$}
\ArrowLine(58.0,60.0)(37.0,60.0) 
\Text(80.0,70.0)[l]{$\bar{e}$}
\ArrowLine(79.0,70.0)(58.0,60.0) 
\Text(54.0,50.0)[r]{$W^+$}
\DashArrowLine(58.0,40.0)(58.0,60.0){3.0} 
\Text(80.0,50.0)[l]{$\bar{\nu}_e$}
\ArrowLine(79.0,50.0)(58.0,40.0) 
\Text(80.0,30.0)[l]{$e$}
\ArrowLine(58.0,40.0)(79.0,30.0) 
\Text(47,0)[b] {diagr.2}
\end{picture} \ 
{} \qquad\allowbreak
\begin{picture}(95,99)(0,0)
\Text(15.0,90.0)[r]{$e$}
\ArrowLine(16.0,90.0)(58.0,90.0) 
\Text(80.0,90.0)[l]{$\nu_e$}
\ArrowLine(58.0,90.0)(79.0,90.0) 
\Text(54.0,80.0)[r]{$W^+$}
\DashArrowLine(58.0,70.0)(58.0,90.0){3.0} 
\Text(80.0,70.0)[l]{$e$}
\ArrowLine(58.0,70.0)(79.0,70.0) 
\Text(54.0,60.0)[r]{$\nu_e$}
\ArrowLine(58.0,50.0)(58.0,70.0) 
\Text(80.0,50.0)[l]{$\bar{\nu}_e$}
\ArrowLine(79.0,50.0)(58.0,50.0) 
\Text(57.0,40.0)[r]{$Z$}
\DashLine(58.0,50.0)(58.0,30.0){3.0} 
\Text(15.0,30.0)[r]{$\bar{e}$}
\ArrowLine(58.0,30.0)(16.0,30.0) 
\Text(80.0,30.0)[l]{$\bar{e}$}
\ArrowLine(79.0,30.0)(58.0,30.0) 
\Text(47,0)[b] {diagr.3}
\end{picture} \ 
{} \qquad\allowbreak
\begin{picture}(95,99)(0,0)
\Text(15.0,90.0)[r]{$e$}
\ArrowLine(16.0,90.0)(58.0,90.0) 
\Text(80.0,90.0)[l]{$\nu_e$}
\ArrowLine(58.0,90.0)(79.0,90.0) 
\Text(54.0,80.0)[r]{$W^+$}
\DashArrowLine(58.0,70.0)(58.0,90.0){3.0} 
\Text(80.0,70.0)[l]{$\bar{\nu}_e$}
\ArrowLine(79.0,70.0)(58.0,70.0) 
\Text(54.0,60.0)[r]{$e$}
\ArrowLine(58.0,70.0)(58.0,50.0) 
\Text(80.0,50.0)[l]{$e$}
\ArrowLine(58.0,50.0)(79.0,50.0) 
\Text(57.0,40.0)[r]{$\gamma$}
\DashLine(58.0,50.0)(58.0,30.0){3.0} 
\Text(15.0,30.0)[r]{$\bar{e}$}
\ArrowLine(58.0,30.0)(16.0,30.0) 
\Text(80.0,30.0)[l]{$\bar{e}$}
\ArrowLine(79.0,30.0)(58.0,30.0) 
\Text(47,0)[b] {diagr.4}
\end{picture} \ 
{} \qquad\allowbreak
\begin{picture}(95,99)(0,0)
\Text(15.0,90.0)[r]{$e$}
\ArrowLine(16.0,90.0)(58.0,90.0) 
\Text(80.0,90.0)[l]{$\nu_e$}
\ArrowLine(58.0,90.0)(79.0,90.0) 
\Text(54.0,80.0)[r]{$W^+$}
\DashArrowLine(58.0,70.0)(58.0,90.0){3.0} 
\Text(80.0,70.0)[l]{$\bar{\nu}_e$}
\ArrowLine(79.0,70.0)(58.0,70.0) 
\Text(54.0,60.0)[r]{$e$}
\ArrowLine(58.0,70.0)(58.0,50.0) 
\Text(80.0,50.0)[l]{$e$}
\ArrowLine(58.0,50.0)(79.0,50.0) 
\Text(57.0,40.0)[r]{$Z$}
\DashLine(58.0,50.0)(58.0,30.0){3.0} 
\Text(15.0,30.0)[r]{$\bar{e}$}
\ArrowLine(58.0,30.0)(16.0,30.0) 
\Text(80.0,30.0)[l]{$\bar{e}$}
\ArrowLine(79.0,30.0)(58.0,30.0) 
\Text(47,0)[b] {diagr.5}
\end{picture} \ 
{} \qquad\allowbreak
\begin{picture}(95,99)(0,0)
\Text(15.0,80.0)[r]{$e$}
\ArrowLine(16.0,80.0)(37.0,80.0) 
\Line(37.0,80.0)(58.0,80.0) 
\Text(80.0,90.0)[l]{$\nu_e$}
\ArrowLine(58.0,80.0)(79.0,90.0) 
\Text(33.0,70.0)[r]{$W^+$}
\DashArrowLine(37.0,60.0)(37.0,80.0){3.0} 
\Text(47.0,64.0)[b]{$W^+$}
\DashArrowLine(58.0,60.0)(37.0,60.0){3.0} 
\Text(80.0,70.0)[l]{$\bar{\nu}_e$}
\ArrowLine(79.0,70.0)(58.0,60.0) 
\Text(80.0,50.0)[l]{$e$}
\ArrowLine(58.0,60.0)(79.0,50.0) 
\Text(36.0,50.0)[r]{$\gamma$}
\DashLine(37.0,60.0)(37.0,40.0){3.0} 
\Text(15.0,40.0)[r]{$\bar{e}$}
\ArrowLine(37.0,40.0)(16.0,40.0) 
\Line(37.0,40.0)(58.0,40.0) 
\Text(80.0,30.0)[l]{$\bar{e}$}
\ArrowLine(79.0,30.0)(58.0,40.0) 
\Text(47,0)[b] {diagr.6}
\end{picture} \ 
{} \qquad\allowbreak
\begin{picture}(95,99)(0,0)
\Text(15.0,80.0)[r]{$e$}
\ArrowLine(16.0,80.0)(37.0,80.0) 
\Line(37.0,80.0)(58.0,80.0) 
\Text(80.0,90.0)[l]{$\nu_e$}
\ArrowLine(58.0,80.0)(79.0,90.0) 
\Text(33.0,70.0)[r]{$W^+$}
\DashArrowLine(37.0,60.0)(37.0,80.0){3.0} 
\Text(47.0,64.0)[b]{$W^+$}
\DashArrowLine(58.0,60.0)(37.0,60.0){3.0} 
\Text(80.0,70.0)[l]{$\bar{\nu}_e$}
\ArrowLine(79.0,70.0)(58.0,60.0) 
\Text(80.0,50.0)[l]{$e$}
\ArrowLine(58.0,60.0)(79.0,50.0) 
\Text(36.0,50.0)[r]{$Z$}
\DashLine(37.0,60.0)(37.0,40.0){3.0} 
\Text(15.0,40.0)[r]{$\bar{e}$}
\ArrowLine(37.0,40.0)(16.0,40.0) 
\Line(37.0,40.0)(58.0,40.0) 
\Text(80.0,30.0)[l]{$\bar{e}$}
\ArrowLine(79.0,30.0)(58.0,40.0) 
\Text(47,0)[b] {diagr.7}
\end{picture} \ 
{} \qquad\allowbreak
\begin{picture}(95,99)(0,0)
\Text(15.0,80.0)[r]{$e$}
\ArrowLine(16.0,80.0)(37.0,80.0) 
\Text(47.0,84.0)[b]{$W^+$}
\DashArrowLine(58.0,80.0)(37.0,80.0){3.0} 
\Text(80.0,90.0)[l]{$\bar{\nu}_e$}
\ArrowLine(79.0,90.0)(58.0,80.0) 
\Text(80.0,70.0)[l]{$e$}
\ArrowLine(58.0,80.0)(79.0,70.0) 
\Text(33.0,70.0)[r]{$\nu_e$}
\ArrowLine(37.0,80.0)(37.0,60.0) 
\Line(37.0,60.0)(58.0,60.0) 
\Text(80.0,50.0)[l]{$\nu_e$}
\ArrowLine(58.0,60.0)(79.0,50.0) 
\Text(36.0,50.0)[r]{$Z$}
\DashLine(37.0,60.0)(37.0,40.0){3.0} 
\Text(15.0,40.0)[r]{$\bar{e}$}
\ArrowLine(37.0,40.0)(16.0,40.0) 
\Line(37.0,40.0)(58.0,40.0) 
\Text(80.0,30.0)[l]{$\bar{e}$}
\ArrowLine(79.0,30.0)(58.0,40.0) 
\Text(47,0)[b] {diagr.8}
\end{picture} \ 
{} \qquad\allowbreak
\begin{picture}(95,99)(0,0)
\Text(15.0,80.0)[r]{$e$}
\ArrowLine(16.0,80.0)(37.0,80.0) 
\Text(47.0,84.0)[b]{$e$}
\ArrowLine(37.0,80.0)(58.0,80.0) 
\Text(80.0,90.0)[l]{$\nu_e$}
\ArrowLine(58.0,80.0)(79.0,90.0) 
\Text(54.0,70.0)[r]{$W^+$}
\DashArrowLine(58.0,60.0)(58.0,80.0){3.0} 
\Text(80.0,70.0)[l]{$\bar{\nu}_e$}
\ArrowLine(79.0,70.0)(58.0,60.0) 
\Text(80.0,50.0)[l]{$e$}
\ArrowLine(58.0,60.0)(79.0,50.0) 
\Text(36.0,60.0)[r]{$Z$}
\DashLine(37.0,80.0)(37.0,40.0){3.0} 
\Text(15.0,40.0)[r]{$\bar{e}$}
\ArrowLine(37.0,40.0)(16.0,40.0) 
\Line(37.0,40.0)(58.0,40.0) 
\Text(80.0,30.0)[l]{$\bar{e}$}
\ArrowLine(79.0,30.0)(58.0,40.0) 
\Text(47,0)[b] {diagr.9}
\end{picture} \ 
}

{
\unitlength=1.0 pt
\SetScale{1.0}
\SetWidth{0.7}      
\scriptsize    
{} \qquad\allowbreak
\begin{picture}(95,99)(0,0)
\Text(15.0,90.0)[r]{$e$}
\ArrowLine(16.0,90.0)(58.0,90.0) 
\Text(80.0,90.0)[l]{$e$}
\ArrowLine(58.0,90.0)(79.0,90.0) 
\Text(57.0,80.0)[r]{$\gamma$}
\DashLine(58.0,90.0)(58.0,70.0){3.0} 
\Text(80.0,70.0)[l]{$\bar{e}$}
\ArrowLine(79.0,70.0)(58.0,70.0) 
\Text(54.0,60.0)[r]{$e$}
\ArrowLine(58.0,70.0)(58.0,50.0) 
\Text(80.0,50.0)[l]{$\nu_e$}
\ArrowLine(58.0,50.0)(79.0,50.0) 
\Text(54.0,40.0)[r]{$W^+$}
\DashArrowLine(58.0,30.0)(58.0,50.0){3.0} 
\Text(15.0,30.0)[r]{$\bar{e}$}
\ArrowLine(58.0,30.0)(16.0,30.0) 
\Text(80.0,30.0)[l]{$\bar{\nu}_e$}
\ArrowLine(79.0,30.0)(58.0,30.0) 
\Text(47,0)[b] {diagr.1}
\end{picture} \ 
{} \qquad\allowbreak
\begin{picture}(95,99)(0,0)
\Text(15.0,80.0)[r]{$e$}
\ArrowLine(16.0,80.0)(37.0,80.0) 
\Line(37.0,80.0)(58.0,80.0) 
\Text(80.0,90.0)[l]{$e$}
\ArrowLine(58.0,80.0)(79.0,90.0) 
\Text(36.0,70.0)[r]{$\gamma$}
\DashLine(37.0,80.0)(37.0,60.0){3.0} 
\Text(15.0,60.0)[r]{$\bar{e}$}
\ArrowLine(37.0,60.0)(16.0,60.0) 
\Text(47.0,64.0)[b]{$e$}
\ArrowLine(58.0,60.0)(37.0,60.0) 
\Text(80.0,70.0)[l]{$\bar{\nu}_e$}
\ArrowLine(79.0,70.0)(58.0,60.0) 
\Text(54.0,50.0)[r]{$W^+$}
\DashArrowLine(58.0,60.0)(58.0,40.0){3.0} 
\Text(80.0,50.0)[l]{$\nu_e$}
\ArrowLine(58.0,40.0)(79.0,50.0) 
\Text(80.0,30.0)[l]{$\bar{e}$}
\ArrowLine(79.0,30.0)(58.0,40.0) 
\Text(47,0)[b] {diagr.2}
\end{picture} \ 
{} \qquad\allowbreak
\begin{picture}(95,99)(0,0)
\Text(15.0,80.0)[r]{$e$}
\ArrowLine(16.0,80.0)(37.0,80.0) 
\Line(37.0,80.0)(58.0,80.0) 
\Text(80.0,90.0)[l]{$e$}
\ArrowLine(58.0,80.0)(79.0,90.0) 
\Text(36.0,70.0)[r]{$\gamma$}
\DashLine(37.0,80.0)(37.0,60.0){3.0} 
\Text(47.0,64.0)[b]{$W^+$}
\DashArrowLine(37.0,60.0)(58.0,60.0){3.0} 
\Text(80.0,70.0)[l]{$\nu_e$}
\ArrowLine(58.0,60.0)(79.0,70.0) 
\Text(80.0,50.0)[l]{$\bar{e}$}
\ArrowLine(79.0,50.0)(58.0,60.0) 
\Text(33.0,50.0)[r]{$W^+$}
\DashArrowLine(37.0,40.0)(37.0,60.0){3.0} 
\Text(15.0,40.0)[r]{$\bar{e}$}
\ArrowLine(37.0,40.0)(16.0,40.0) 
\Line(37.0,40.0)(58.0,40.0) 
\Text(80.0,30.0)[l]{$\bar{\nu}_e$}
\ArrowLine(79.0,30.0)(58.0,40.0) 
\Text(47,0)[b] {diagr.3}
\end{picture} \ 
{} \qquad\allowbreak
\begin{picture}(95,99)(0,0)
\Text(15.0,80.0)[r]{$e$}
\ArrowLine(16.0,80.0)(37.0,80.0) 
\Text(47.0,84.0)[b]{$\nu_e$}
\ArrowLine(37.0,80.0)(58.0,80.0) 
\Text(80.0,90.0)[l]{$e$}
\ArrowLine(58.0,80.0)(79.0,90.0) 
\Text(54.0,70.0)[r]{$W^+$}
\DashArrowLine(58.0,80.0)(58.0,60.0){3.0} 
\Text(80.0,70.0)[l]{$\nu_e$}
\ArrowLine(58.0,60.0)(79.0,70.0) 
\Text(80.0,50.0)[l]{$\bar{e}$}
\ArrowLine(79.0,50.0)(58.0,60.0) 
\Text(33.0,60.0)[r]{$W^+$}
\DashArrowLine(37.0,40.0)(37.0,80.0){3.0} 
\Text(15.0,40.0)[r]{$\bar{e}$}
\ArrowLine(37.0,40.0)(16.0,40.0) 
\Line(37.0,40.0)(58.0,40.0) 
\Text(80.0,30.0)[l]{$\bar{\nu}_e$}
\ArrowLine(79.0,30.0)(58.0,40.0) 
\Text(47,0)[b] {diagr.4}
\end{picture} \ 
{} \qquad\allowbreak
\begin{picture}(95,99)(0,0)
\Text(15.0,90.0)[r]{$e$}
\ArrowLine(16.0,90.0)(58.0,90.0) 
\Text(80.0,90.0)[l]{$e$}
\ArrowLine(58.0,90.0)(79.0,90.0) 
\Text(57.0,80.0)[r]{$Z$}
\DashLine(58.0,90.0)(58.0,70.0){3.0} 
\Text(80.0,70.0)[l]{$\bar{e}$}
\ArrowLine(79.0,70.0)(58.0,70.0) 
\Text(54.0,60.0)[r]{$e$}
\ArrowLine(58.0,70.0)(58.0,50.0) 
\Text(80.0,50.0)[l]{$\nu_e$}
\ArrowLine(58.0,50.0)(79.0,50.0) 
\Text(54.0,40.0)[r]{$W^+$}
\DashArrowLine(58.0,30.0)(58.0,50.0){3.0} 
\Text(15.0,30.0)[r]{$\bar{e}$}
\ArrowLine(58.0,30.0)(16.0,30.0) 
\Text(80.0,30.0)[l]{$\bar{\nu}_e$}
\ArrowLine(79.0,30.0)(58.0,30.0) 
\Text(47,0)[b] {diagr.5}
\end{picture} \ 
{} \qquad\allowbreak
\begin{picture}(95,99)(0,0)
\Text(15.0,80.0)[r]{$e$}
\ArrowLine(16.0,80.0)(37.0,80.0) 
\Line(37.0,80.0)(58.0,80.0) 
\Text(80.0,90.0)[l]{$e$}
\ArrowLine(58.0,80.0)(79.0,90.0) 
\Text(36.0,70.0)[r]{$Z$}
\DashLine(37.0,80.0)(37.0,60.0){3.0} 
\Text(15.0,60.0)[r]{$\bar{e}$}
\ArrowLine(37.0,60.0)(16.0,60.0) 
\Text(47.0,64.0)[b]{$e$}
\ArrowLine(58.0,60.0)(37.0,60.0) 
\Text(80.0,70.0)[l]{$\bar{\nu}_e$}
\ArrowLine(79.0,70.0)(58.0,60.0) 
\Text(54.0,50.0)[r]{$W^+$}
\DashArrowLine(58.0,60.0)(58.0,40.0){3.0} 
\Text(80.0,50.0)[l]{$\nu_e$}
\ArrowLine(58.0,40.0)(79.0,50.0) 
\Text(80.0,30.0)[l]{$\bar{e}$}
\ArrowLine(79.0,30.0)(58.0,40.0) 
\Text(47,0)[b] {diagr.6}
\end{picture} \ 
{} \qquad\allowbreak
\begin{picture}(95,99)(0,0)
\Text(15.0,80.0)[r]{$e$}
\ArrowLine(16.0,80.0)(37.0,80.0) 
\Line(37.0,80.0)(58.0,80.0) 
\Text(80.0,90.0)[l]{$e$}
\ArrowLine(58.0,80.0)(79.0,90.0) 
\Text(36.0,70.0)[r]{$Z$}
\DashLine(37.0,80.0)(37.0,60.0){3.0} 
\Line(37.0,60.0)(58.0,60.0) 
\Text(80.0,70.0)[l]{$\bar{\nu}_e$}
\ArrowLine(79.0,70.0)(58.0,60.0) 
\Text(33.0,50.0)[r]{$\nu_e$}
\ArrowLine(37.0,60.0)(37.0,40.0) 
\Text(15.0,40.0)[r]{$\bar{e}$}
\ArrowLine(37.0,40.0)(16.0,40.0) 
\Text(47.0,44.0)[b]{$W^+$}
\DashArrowLine(37.0,40.0)(58.0,40.0){3.0} 
\Text(80.0,50.0)[l]{$\nu_e$}
\ArrowLine(58.0,40.0)(79.0,50.0) 
\Text(80.0,30.0)[l]{$\bar{e}$}
\ArrowLine(79.0,30.0)(58.0,40.0) 
\Text(47,0)[b] {diagr.7}
\end{picture} \ 
{} \qquad\allowbreak
\begin{picture}(95,99)(0,0)
\Text(15.0,90.0)[r]{$e$}
\ArrowLine(16.0,90.0)(58.0,90.0) 
\Text(80.0,90.0)[l]{$e$}
\ArrowLine(58.0,90.0)(79.0,90.0) 
\Text(57.0,80.0)[r]{$Z$}
\DashLine(58.0,90.0)(58.0,70.0){3.0} 
\Text(80.0,70.0)[l]{$\nu_e$}
\ArrowLine(58.0,70.0)(79.0,70.0) 
\Text(54.0,60.0)[r]{$\nu_e$}
\ArrowLine(58.0,50.0)(58.0,70.0) 
\Text(80.0,50.0)[l]{$\bar{e}$}
\ArrowLine(79.0,50.0)(58.0,50.0) 
\Text(54.0,40.0)[r]{$W^+$}
\DashArrowLine(58.0,30.0)(58.0,50.0){3.0} 
\Text(15.0,30.0)[r]{$\bar{e}$}
\ArrowLine(58.0,30.0)(16.0,30.0) 
\Text(80.0,30.0)[l]{$\bar{\nu}_e$}
\ArrowLine(79.0,30.0)(58.0,30.0) 
\Text(47,0)[b] {diagr.8}
\end{picture} \ 
{} \qquad\allowbreak
\begin{picture}(95,99)(0,0)
\Text(15.0,80.0)[r]{$e$}
\ArrowLine(16.0,80.0)(37.0,80.0) 
\Line(37.0,80.0)(58.0,80.0) 
\Text(80.0,90.0)[l]{$e$}
\ArrowLine(58.0,80.0)(79.0,90.0) 
\Text(36.0,70.0)[r]{$Z$}
\DashLine(37.0,80.0)(37.0,60.0){3.0} 
\Text(47.0,64.0)[b]{$W^+$}
\DashArrowLine(37.0,60.0)(58.0,60.0){3.0} 
\Text(80.0,70.0)[l]{$\nu_e$}
\ArrowLine(58.0,60.0)(79.0,70.0) 
\Text(80.0,50.0)[l]{$\bar{e}$}
\ArrowLine(79.0,50.0)(58.0,60.0) 
\Text(33.0,50.0)[r]{$W^+$}
\DashArrowLine(37.0,40.0)(37.0,60.0){3.0} 
\Text(15.0,40.0)[r]{$\bar{e}$}
\ArrowLine(37.0,40.0)(16.0,40.0) 
\Line(37.0,40.0)(58.0,40.0) 
\Text(80.0,30.0)[l]{$\bar{\nu}_e$}
\ArrowLine(79.0,30.0)(58.0,40.0) 
\Text(47,0)[b] {diagr.9}
\end{picture} \ 
}
\caption{Two gauge invariant subsets for the single $\wb$ production
in the channel $e^+ e^- \to e^+ e^- \nu_e \barnu_e$.
The set of nine diagrams in the first three rows and the set of
diagrams in the last three rows are separately gauge invariant.}
\label{18w}
\end{figure}

\begin{figure}[htbp]
{
\unitlength=1.0 pt
\SetScale{1.0}
\SetWidth{0.7}      
\scriptsize    
{} \qquad\allowbreak
\begin{picture}(95,99)(0,0)
\Text(15.0,80.0)[r]{$e$}
\ArrowLine(16.0,80.0)(37.0,80.0) 
\Text(47.0,84.0)[b]{$e$}
\ArrowLine(37.0,80.0)(58.0,80.0) 
\Text(80.0,90.0)[l]{$e$}
\ArrowLine(58.0,80.0)(79.0,90.0) 
\Text(57.0,70.0)[r]{$Z$}
\DashLine(58.0,80.0)(58.0,60.0){3.0} 
\Text(80.0,70.0)[l]{$\nu_e$}
\ArrowLine(58.0,60.0)(79.0,70.0) 
\Text(80.0,50.0)[l]{$\bar{\nu}_e$}
\ArrowLine(79.0,50.0)(58.0,60.0) 
\Text(36.0,60.0)[r]{$\gamma$}
\DashLine(37.0,80.0)(37.0,40.0){3.0} 
\Text(15.0,40.0)[r]{$\bar{e}$}
\ArrowLine(37.0,40.0)(16.0,40.0) 
\Line(37.0,40.0)(58.0,40.0) 
\Text(80.0,30.0)[l]{$\bar{e}$}
\ArrowLine(79.0,30.0)(58.0,40.0) 
\Text(47,0)[b] {diagr.1}
\end{picture} \ 
{} \qquad\allowbreak
\begin{picture}(95,99)(0,0)
\Text(15.0,80.0)[r]{$e$}
\ArrowLine(16.0,80.0)(37.0,80.0) 
\Text(47.0,81.0)[b]{$Z$}
\DashLine(37.0,80.0)(58.0,80.0){3.0} 
\Text(80.0,90.0)[l]{$\nu_e$}
\ArrowLine(58.0,80.0)(79.0,90.0) 
\Text(80.0,70.0)[l]{$\bar{\nu}_e$}
\ArrowLine(79.0,70.0)(58.0,80.0) 
\Text(33.0,70.0)[r]{$e$}
\ArrowLine(37.0,80.0)(37.0,60.0) 
\Line(37.0,60.0)(58.0,60.0) 
\Text(80.0,50.0)[l]{$e$}
\ArrowLine(58.0,60.0)(79.0,50.0) 
\Text(36.0,50.0)[r]{$\gamma$}
\DashLine(37.0,60.0)(37.0,40.0){3.0} 
\Text(15.0,40.0)[r]{$\bar{e}$}
\ArrowLine(37.0,40.0)(16.0,40.0) 
\Line(37.0,40.0)(58.0,40.0) 
\Text(80.0,30.0)[l]{$\bar{e}$}
\ArrowLine(79.0,30.0)(58.0,40.0) 
\Text(47,0)[b] {diagr.2}
\end{picture} \ 
{} \qquad\allowbreak
\begin{picture}(95,99)(0,0)
\Text(15.0,80.0)[r]{$e$}
\ArrowLine(16.0,80.0)(37.0,80.0) 
\Text(47.0,81.0)[b]{$Z$}
\DashLine(37.0,80.0)(58.0,80.0){3.0} 
\Text(80.0,90.0)[l]{$\nu_e$}
\ArrowLine(58.0,80.0)(79.0,90.0) 
\Text(80.0,70.0)[l]{$\bar{\nu}_e$}
\ArrowLine(79.0,70.0)(58.0,80.0) 
\Text(33.0,70.0)[r]{$e$}
\ArrowLine(37.0,80.0)(37.0,60.0) 
\Line(37.0,60.0)(58.0,60.0) 
\Text(80.0,50.0)[l]{$e$}
\ArrowLine(58.0,60.0)(79.0,50.0) 
\Text(36.0,50.0)[r]{$Z$}
\DashLine(37.0,60.0)(37.0,40.0){3.0} 
\Text(15.0,40.0)[r]{$\bar{e}$}
\ArrowLine(37.0,40.0)(16.0,40.0) 
\Line(37.0,40.0)(58.0,40.0) 
\Text(80.0,30.0)[l]{$\bar{e}$}
\ArrowLine(79.0,30.0)(58.0,40.0) 
\Text(47,0)[b] {diagr.3}
\end{picture} \ 
{} \qquad\allowbreak
\begin{picture}(95,99)(0,0)
\Text(15.0,80.0)[r]{$e$}
\ArrowLine(16.0,80.0)(37.0,80.0) 
\Text(47.0,84.0)[b]{$e$}
\ArrowLine(37.0,80.0)(58.0,80.0) 
\Text(80.0,90.0)[l]{$e$}
\ArrowLine(58.0,80.0)(79.0,90.0) 
\Text(57.0,70.0)[r]{$Z$}
\DashLine(58.0,80.0)(58.0,60.0){3.0} 
\Text(80.0,70.0)[l]{$\nu_e$}
\ArrowLine(58.0,60.0)(79.0,70.0) 
\Text(80.0,50.0)[l]{$\bar{\nu}_e$}
\ArrowLine(79.0,50.0)(58.0,60.0) 
\Text(36.0,60.0)[r]{$Z$}
\DashLine(37.0,80.0)(37.0,40.0){3.0} 
\Text(15.0,40.0)[r]{$\bar{e}$}
\ArrowLine(37.0,40.0)(16.0,40.0) 
\Line(37.0,40.0)(58.0,40.0) 
\Text(80.0,30.0)[l]{$\bar{e}$}
\ArrowLine(79.0,30.0)(58.0,40.0) 
\Text(47,0)[b] {diagr.4}
\end{picture} \ 
}

{
\unitlength=1.0 pt
\SetScale{1.0}
\SetWidth{0.7}      
\scriptsize    
{} \qquad\allowbreak
\begin{picture}(95,99)(0,0)
\Text(15.0,80.0)[r]{$e$}
\ArrowLine(16.0,80.0)(37.0,80.0) 
\Line(37.0,80.0)(58.0,80.0) 
\Text(80.0,90.0)[l]{$e$}
\ArrowLine(58.0,80.0)(79.0,90.0) 
\Text(36.0,70.0)[r]{$\gamma$}
\DashLine(37.0,80.0)(37.0,60.0){3.0} 
\Line(37.0,60.0)(58.0,60.0) 
\Text(80.0,70.0)[l]{$\bar{e}$}
\ArrowLine(79.0,70.0)(58.0,60.0) 
\Text(33.0,50.0)[r]{$e$}
\ArrowLine(37.0,60.0)(37.0,40.0) 
\Text(15.0,40.0)[r]{$\bar{e}$}
\ArrowLine(37.0,40.0)(16.0,40.0) 
\Text(47.0,41.0)[b]{$Z$}
\DashLine(37.0,40.0)(58.0,40.0){3.0} 
\Text(80.0,50.0)[l]{$\nu_e$}
\ArrowLine(58.0,40.0)(79.0,50.0) 
\Text(80.0,30.0)[l]{$\bar{\nu}_e$}
\ArrowLine(79.0,30.0)(58.0,40.0) 
\Text(47,0)[b] {diagr.1}
\end{picture} \ 
{} \qquad\allowbreak
\begin{picture}(95,99)(0,0)
\Text(15.0,80.0)[r]{$e$}
\ArrowLine(16.0,80.0)(37.0,80.0) 
\Line(37.0,80.0)(58.0,80.0) 
\Text(80.0,90.0)[l]{$e$}
\ArrowLine(58.0,80.0)(79.0,90.0) 
\Text(36.0,70.0)[r]{$\gamma$}
\DashLine(37.0,80.0)(37.0,60.0){3.0} 
\Text(15.0,60.0)[r]{$\bar{e}$}
\ArrowLine(37.0,60.0)(16.0,60.0) 
\Text(47.0,64.0)[b]{$e$}
\ArrowLine(58.0,60.0)(37.0,60.0) 
\Text(80.0,70.0)[l]{$\bar{e}$}
\ArrowLine(79.0,70.0)(58.0,60.0) 
\Text(57.0,50.0)[r]{$Z$}
\DashLine(58.0,60.0)(58.0,40.0){3.0} 
\Text(80.0,50.0)[l]{$\nu_e$}
\ArrowLine(58.0,40.0)(79.0,50.0) 
\Text(80.0,30.0)[l]{$\bar{\nu}_e$}
\ArrowLine(79.0,30.0)(58.0,40.0) 
\Text(47,0)[b] {diagr.2}
\end{picture} \ 
{} \qquad\allowbreak
\begin{picture}(95,99)(0,0)
\Text(15.0,80.0)[r]{$e$}
\ArrowLine(16.0,80.0)(37.0,80.0) 
\Line(37.0,80.0)(58.0,80.0) 
\Text(80.0,90.0)[l]{$e$}
\ArrowLine(58.0,80.0)(79.0,90.0) 
\Text(36.0,70.0)[r]{$Z$}
\DashLine(37.0,80.0)(37.0,60.0){3.0} 
\Line(37.0,60.0)(58.0,60.0) 
\Text(80.0,70.0)[l]{$\bar{e}$}
\ArrowLine(79.0,70.0)(58.0,60.0) 
\Text(33.0,50.0)[r]{$e$}
\ArrowLine(37.0,60.0)(37.0,40.0) 
\Text(15.0,40.0)[r]{$\bar{e}$}
\ArrowLine(37.0,40.0)(16.0,40.0) 
\Text(47.0,41.0)[b]{$Z$}
\DashLine(37.0,40.0)(58.0,40.0){3.0} 
\Text(80.0,50.0)[l]{$\nu_e$}
\ArrowLine(58.0,40.0)(79.0,50.0) 
\Text(80.0,30.0)[l]{$\bar{\nu}_e$}
\ArrowLine(79.0,30.0)(58.0,40.0) 
\Text(47,0)[b] {diagr.3}
\end{picture} \ 
{} \qquad\allowbreak
\begin{picture}(95,99)(0,0)
\Text(15.0,80.0)[r]{$e$}
\ArrowLine(16.0,80.0)(37.0,80.0) 
\Line(37.0,80.0)(58.0,80.0) 
\Text(80.0,90.0)[l]{$e$}
\ArrowLine(58.0,80.0)(79.0,90.0) 
\Text(36.0,70.0)[r]{$Z$}
\DashLine(37.0,80.0)(37.0,60.0){3.0} 
\Text(15.0,60.0)[r]{$\bar{e}$}
\ArrowLine(37.0,60.0)(16.0,60.0) 
\Text(47.0,64.0)[b]{$e$}
\ArrowLine(58.0,60.0)(37.0,60.0) 
\Text(80.0,70.0)[l]{$\bar{e}$}
\ArrowLine(79.0,70.0)(58.0,60.0) 
\Text(57.0,50.0)[r]{$Z$}
\DashLine(58.0,60.0)(58.0,40.0){3.0} 
\Text(80.0,50.0)[l]{$\nu_e$}
\ArrowLine(58.0,40.0)(79.0,50.0) 
\Text(80.0,30.0)[l]{$\bar{\nu}_e$}
\ArrowLine(79.0,30.0)(58.0,40.0) 
\Text(47,0)[b] {diagr.4}
\end{picture} \ 
}
\caption{Two gauge invariant subsets for the single $\zb$ production
in the channel $e^+ e^- \to e^+ e^- \nu_e \barnu_e$.
The set of four diagrams in the first two rows and the set of 
diagrams in the last two rows are separately gauge invariant.}
\label{8z}
\end{figure}

\begin{figure}[htbp]
{
\unitlength=1.0 pt
\SetScale{1.0}
\SetWidth{0.7}      
\scriptsize    
{} \qquad\allowbreak
\begin{picture}(95,99)(0,0)
\Text(15.0,90.0)[r]{$e$}
\ArrowLine(16.0,90.0)(37.0,80.0) 
\Text(15.0,70.0)[r]{$\bar{e}$}
\ArrowLine(37.0,80.0)(16.0,70.0) 
\Text(47.0,81.0)[b]{$\gamma$}
\DashLine(37.0,80.0)(58.0,80.0){3.0} 
\Text(80.0,90.0)[l]{$\bar{e}$}
\ArrowLine(79.0,90.0)(58.0,80.0) 
\Text(54.0,70.0)[r]{$e$}
\ArrowLine(58.0,80.0)(58.0,60.0) 
\Text(80.0,70.0)[l]{$\nu_e$}
\ArrowLine(58.0,60.0)(79.0,70.0) 
\Text(54.0,50.0)[r]{$W^+$}
\DashArrowLine(58.0,40.0)(58.0,60.0){3.0} 
\Text(80.0,50.0)[l]{$\bar{\nu}_e$}
\ArrowLine(79.0,50.0)(58.0,40.0) 
\Text(80.0,30.0)[l]{$e$}
\ArrowLine(58.0,40.0)(79.0,30.0) 
\Text(47,0)[b] {diagr.1}
\end{picture} \ 
{} \qquad\allowbreak
\begin{picture}(95,99)(0,0)
\Text(15.0,90.0)[r]{$e$}
\ArrowLine(16.0,90.0)(37.0,80.0) 
\Text(15.0,70.0)[r]{$\bar{e}$}
\ArrowLine(37.0,80.0)(16.0,70.0) 
\Text(47.0,81.0)[b]{$\gamma$}
\DashLine(37.0,80.0)(58.0,80.0){3.0} 
\Text(80.0,90.0)[l]{$e$}
\ArrowLine(58.0,80.0)(79.0,90.0) 
\Text(54.0,70.0)[r]{$e$}
\ArrowLine(58.0,60.0)(58.0,80.0) 
\Text(80.0,70.0)[l]{$\bar{\nu}_e$}
\ArrowLine(79.0,70.0)(58.0,60.0) 
\Text(54.0,50.0)[r]{$W^+$}
\DashArrowLine(58.0,60.0)(58.0,40.0){3.0} 
\Text(80.0,50.0)[l]{$\nu_e$}
\ArrowLine(58.0,40.0)(79.0,50.0) 
\Text(80.0,30.0)[l]{$\bar{e}$}
\ArrowLine(79.0,30.0)(58.0,40.0) 
\Text(47,0)[b] {diagr.2}
\end{picture} \ 
{} \qquad\allowbreak
\begin{picture}(95,99)(0,0)
\Text(15.0,70.0)[r]{$e$}
\ArrowLine(16.0,70.0)(37.0,60.0) 
\Text(15.0,50.0)[r]{$\bar{e}$}
\ArrowLine(37.0,60.0)(16.0,50.0) 
\Text(37.0,62.0)[lb]{$\gamma$}
\DashLine(37.0,60.0)(58.0,60.0){3.0} 
\Text(54.0,70.0)[r]{$W^+$}
\DashArrowLine(58.0,80.0)(58.0,60.0){3.0} 
\Text(80.0,90.0)[l]{$\bar{\nu}_e$}
\ArrowLine(79.0,90.0)(58.0,80.0) 
\Text(80.0,70.0)[l]{$e$}
\ArrowLine(58.0,80.0)(79.0,70.0) 
\Text(54.0,50.0)[r]{$W^+$}
\DashArrowLine(58.0,60.0)(58.0,40.0){3.0} 
\Text(80.0,50.0)[l]{$\nu_e$}
\ArrowLine(58.0,40.0)(79.0,50.0) 
\Text(80.0,30.0)[l]{$\bar{e}$}
\ArrowLine(79.0,30.0)(58.0,40.0) 
\Text(47,0)[b] {diagr.3}
\end{picture} \ 
{} \qquad\allowbreak
\begin{picture}(95,99)(0,0)
\Text(15.0,80.0)[r]{$e$}
\ArrowLine(16.0,80.0)(37.0,80.0) 
\Text(47.0,84.0)[b]{$W^+$}
\DashArrowLine(58.0,80.0)(37.0,80.0){3.0} 
\Text(80.0,90.0)[l]{$\bar{\nu}_e$}
\ArrowLine(79.0,90.0)(58.0,80.0) 
\Text(80.0,70.0)[l]{$e$}
\ArrowLine(58.0,80.0)(79.0,70.0) 
\Text(33.0,60.0)[r]{$\nu_e$}
\ArrowLine(37.0,80.0)(37.0,40.0) 
\Text(15.0,40.0)[r]{$\bar{e}$}
\ArrowLine(37.0,40.0)(16.0,40.0) 
\Text(47.0,44.0)[b]{$W^+$}
\DashArrowLine(37.0,40.0)(58.0,40.0){3.0} 
\Text(80.0,50.0)[l]{$\nu_e$}
\ArrowLine(58.0,40.0)(79.0,50.0) 
\Text(80.0,30.0)[l]{$\bar{e}$}
\ArrowLine(79.0,30.0)(58.0,40.0) 
\Text(47,0)[b] {diagr.4}
\end{picture} \ 
{} \qquad\allowbreak
\begin{picture}(95,99)(0,0)
\Text(15.0,90.0)[r]{$e$}
\ArrowLine(16.0,90.0)(37.0,80.0) 
\Text(15.0,70.0)[r]{$\bar{e}$}
\ArrowLine(37.0,80.0)(16.0,70.0) 
\Text(47.0,81.0)[b]{$Z$}
\DashLine(37.0,80.0)(58.0,80.0){3.0} 
\Text(80.0,90.0)[l]{$\bar{e}$}
\ArrowLine(79.0,90.0)(58.0,80.0) 
\Text(54.0,70.0)[r]{$e$}
\ArrowLine(58.0,80.0)(58.0,60.0) 
\Text(80.0,70.0)[l]{$\nu_e$}
\ArrowLine(58.0,60.0)(79.0,70.0) 
\Text(54.0,50.0)[r]{$W^+$}
\DashArrowLine(58.0,40.0)(58.0,60.0){3.0} 
\Text(80.0,50.0)[l]{$\bar{\nu}_e$}
\ArrowLine(79.0,50.0)(58.0,40.0) 
\Text(80.0,30.0)[l]{$e$}
\ArrowLine(58.0,40.0)(79.0,30.0) 
\Text(47,0)[b] {diagr.5}
\end{picture} \ 
{} \qquad\allowbreak
\begin{picture}(95,99)(0,0)
\Text(15.0,90.0)[r]{$e$}
\ArrowLine(16.0,90.0)(37.0,80.0) 
\Text(15.0,70.0)[r]{$\bar{e}$}
\ArrowLine(37.0,80.0)(16.0,70.0) 
\Text(47.0,81.0)[b]{$Z$}
\DashLine(37.0,80.0)(58.0,80.0){3.0} 
\Text(80.0,90.0)[l]{$e$}
\ArrowLine(58.0,80.0)(79.0,90.0) 
\Text(54.0,70.0)[r]{$e$}
\ArrowLine(58.0,60.0)(58.0,80.0) 
\Text(80.0,70.0)[l]{$\bar{\nu}_e$}
\ArrowLine(79.0,70.0)(58.0,60.0) 
\Text(54.0,50.0)[r]{$W^+$}
\DashArrowLine(58.0,60.0)(58.0,40.0){3.0} 
\Text(80.0,50.0)[l]{$\nu_e$}
\ArrowLine(58.0,40.0)(79.0,50.0) 
\Text(80.0,30.0)[l]{$\bar{e}$}
\ArrowLine(79.0,30.0)(58.0,40.0) 
\Text(47,0)[b] {diagr.6}
\end{picture} \ 
{} \qquad\allowbreak
\begin{picture}(95,99)(0,0)
\Text(15.0,90.0)[r]{$e$}
\ArrowLine(16.0,90.0)(37.0,80.0) 
\Text(15.0,70.0)[r]{$\bar{e}$}
\ArrowLine(37.0,80.0)(16.0,70.0) 
\Text(47.0,81.0)[b]{$Z$}
\DashLine(37.0,80.0)(58.0,80.0){3.0} 
\Text(80.0,90.0)[l]{$\bar{\nu}_e$}
\ArrowLine(79.0,90.0)(58.0,80.0) 
\Text(54.0,70.0)[r]{$\nu_e$}
\ArrowLine(58.0,80.0)(58.0,60.0) 
\Text(80.0,70.0)[l]{$e$}
\ArrowLine(58.0,60.0)(79.0,70.0) 
\Text(54.0,50.0)[r]{$W^+$}
\DashArrowLine(58.0,60.0)(58.0,40.0){3.0} 
\Text(80.0,50.0)[l]{$\nu_e$}
\ArrowLine(58.0,40.0)(79.0,50.0) 
\Text(80.0,30.0)[l]{$\bar{e}$}
\ArrowLine(79.0,30.0)(58.0,40.0) 
\Text(47,0)[b] {diagr.7}
\end{picture} \ 
{} \qquad\allowbreak
\begin{picture}(95,99)(0,0)
\Text(15.0,90.0)[r]{$e$}
\ArrowLine(16.0,90.0)(37.0,80.0) 
\Text(15.0,70.0)[r]{$\bar{e}$}
\ArrowLine(37.0,80.0)(16.0,70.0) 
\Text(47.0,81.0)[b]{$Z$}
\DashLine(37.0,80.0)(58.0,80.0){3.0} 
\Text(80.0,90.0)[l]{$\nu_e$}
\ArrowLine(58.0,80.0)(79.0,90.0) 
\Text(54.0,70.0)[r]{$\nu_e$}
\ArrowLine(58.0,60.0)(58.0,80.0) 
\Text(80.0,70.0)[l]{$\bar{e}$}
\ArrowLine(79.0,70.0)(58.0,60.0) 
\Text(54.0,50.0)[r]{$W^+$}
\DashArrowLine(58.0,40.0)(58.0,60.0){3.0} 
\Text(80.0,50.0)[l]{$\bar{\nu}_e$}
\ArrowLine(79.0,50.0)(58.0,40.0) 
\Text(80.0,30.0)[l]{$e$}
\ArrowLine(58.0,40.0)(79.0,30.0) 
\Text(47,0)[b] {diagr.8}
\end{picture} \ 
{} \qquad\allowbreak
\begin{picture}(95,99)(0,0)
\Text(15.0,70.0)[r]{$e$}
\ArrowLine(16.0,70.0)(37.0,60.0) 
\Text(15.0,50.0)[r]{$\bar{e}$}
\ArrowLine(37.0,60.0)(16.0,50.0) 
\Text(37.0,62.0)[lb]{$Z$}
\DashLine(37.0,60.0)(58.0,60.0){3.0} 
\Text(54.0,70.0)[r]{$W^+$}
\DashArrowLine(58.0,80.0)(58.0,60.0){3.0} 
\Text(80.0,90.0)[l]{$\bar{\nu}_e$}
\ArrowLine(79.0,90.0)(58.0,80.0) 
\Text(80.0,70.0)[l]{$e$}
\ArrowLine(58.0,80.0)(79.0,70.0) 
\Text(54.0,50.0)[r]{$W^+$}
\DashArrowLine(58.0,60.0)(58.0,40.0){3.0} 
\Text(80.0,50.0)[l]{$\nu_e$}
\ArrowLine(58.0,40.0)(79.0,50.0) 
\Text(80.0,30.0)[l]{$\bar{e}$}
\ArrowLine(79.0,30.0)(58.0,40.0) 
\Text(47,0)[b] {diagr.9}
\end{picture} \ 
}
\caption{The $\wbp  \wbm $ double-resonant gauge invariant subset
in the channel $e^+ e^- \to e^+ e^- \nu_e \barnu_e$.}
\label{9ww}
\end{figure}

\begin{figure}[htbp]
{
\unitlength=1.0 pt
\SetScale{1.0}
\SetWidth{0.7}      
\scriptsize    
{} \qquad\allowbreak
\begin{picture}(95,99)(0,0)
\Text(15.0,80.0)[r]{$e$}
\ArrowLine(16.0,80.0)(37.0,80.0) 
\Text(47.0,81.0)[b]{$\gamma$}
\DashLine(37.0,80.0)(58.0,80.0){3.0} 
\Text(80.0,90.0)[l]{$e$}
\ArrowLine(58.0,80.0)(79.0,90.0) 
\Text(80.0,70.0)[l]{$\bar{e}$}
\ArrowLine(79.0,70.0)(58.0,80.0) 
\Text(33.0,60.0)[r]{$e$}
\ArrowLine(37.0,80.0)(37.0,40.0) 
\Text(15.0,40.0)[r]{$\bar{e}$}
\ArrowLine(37.0,40.0)(16.0,40.0) 
\Text(47.0,41.0)[b]{$Z$}
\DashLine(37.0,40.0)(58.0,40.0){3.0} 
\Text(80.0,50.0)[l]{$\nu_e$}
\ArrowLine(58.0,40.0)(79.0,50.0) 
\Text(80.0,30.0)[l]{$\bar{\nu}_e$}
\ArrowLine(79.0,30.0)(58.0,40.0) 
\Text(47,0)[b] {diagr.1}
\end{picture} \ 
{} \qquad\allowbreak
\begin{picture}(95,99)(0,0)
\Text(15.0,80.0)[r]{$e$}
\ArrowLine(16.0,80.0)(37.0,80.0) 
\Text(47.0,81.0)[b]{$Z$}
\DashLine(37.0,80.0)(58.0,80.0){3.0} 
\Text(80.0,90.0)[l]{$\nu_e$}
\ArrowLine(58.0,80.0)(79.0,90.0) 
\Text(80.0,70.0)[l]{$\bar{\nu}_e$}
\ArrowLine(79.0,70.0)(58.0,80.0) 
\Text(33.0,60.0)[r]{$e$}
\ArrowLine(37.0,80.0)(37.0,40.0) 
\Text(15.0,40.0)[r]{$\bar{e}$}
\ArrowLine(37.0,40.0)(16.0,40.0) 
\Text(47.0,41.0)[b]{$\gamma$}
\DashLine(37.0,40.0)(58.0,40.0){3.0} 
\Text(80.0,50.0)[l]{$e$}
\ArrowLine(58.0,40.0)(79.0,50.0) 
\Text(80.0,30.0)[l]{$\bar{e}$}
\ArrowLine(79.0,30.0)(58.0,40.0) 
\Text(47,0)[b] {diagr.2}
\end{picture} \ 
{} \qquad\allowbreak
\begin{picture}(95,99)(0,0)
\Text(15.0,80.0)[r]{$e$}
\ArrowLine(16.0,80.0)(37.0,80.0) 
\Text(47.0,81.0)[b]{$Z$}
\DashLine(37.0,80.0)(58.0,80.0){3.0} 
\Text(80.0,90.0)[l]{$\nu_e$}
\ArrowLine(58.0,80.0)(79.0,90.0) 
\Text(80.0,70.0)[l]{$\bar{\nu}_e$}
\ArrowLine(79.0,70.0)(58.0,80.0) 
\Text(33.0,60.0)[r]{$e$}
\ArrowLine(37.0,80.0)(37.0,40.0) 
\Text(15.0,40.0)[r]{$\bar{e}$}
\ArrowLine(37.0,40.0)(16.0,40.0) 
\Text(47.0,41.0)[b]{$Z$}
\DashLine(37.0,40.0)(58.0,40.0){3.0} 
\Text(80.0,50.0)[l]{$e$}
\ArrowLine(58.0,40.0)(79.0,50.0) 
\Text(80.0,30.0)[l]{$\bar{e}$}
\ArrowLine(79.0,30.0)(58.0,40.0) 
\Text(47,0)[b] {diagr.3}
\end{picture} \ 
{} \qquad\allowbreak
\begin{picture}(95,99)(0,0)
\Text(15.0,80.0)[r]{$e$}
\ArrowLine(16.0,80.0)(37.0,80.0) 
\Text(47.0,81.0)[b]{$Z$}
\DashLine(37.0,80.0)(58.0,80.0){3.0} 
\Text(80.0,90.0)[l]{$e$}
\ArrowLine(58.0,80.0)(79.0,90.0) 
\Text(80.0,70.0)[l]{$\bar{e}$}
\ArrowLine(79.0,70.0)(58.0,80.0) 
\Text(33.0,60.0)[r]{$e$}
\ArrowLine(37.0,80.0)(37.0,40.0) 
\Text(15.0,40.0)[r]{$\bar{e}$}
\ArrowLine(37.0,40.0)(16.0,40.0) 
\Text(47.0,41.0)[b]{$Z$}
\DashLine(37.0,40.0)(58.0,40.0){3.0} 
\Text(80.0,50.0)[l]{$\nu_e$}
\ArrowLine(58.0,40.0)(79.0,50.0) 
\Text(80.0,30.0)[l]{$\bar{\nu}_e$}
\ArrowLine(79.0,30.0)(58.0,40.0) 
\Text(47,0)[b] {diagr.4}
\end{picture} \ 
}
\caption{The $\zb\zb$ double-resonant gauge invariant subset
in the channel $e^+ e^- \to e^+ e^- \nu_e \barnu_e$.}
\label{4zz}
\end{figure}

\begin{figure}[htbp]
{
\unitlength=1.0 pt
\SetScale{1.0}
\SetWidth{0.7}      
\scriptsize    
{} \qquad\allowbreak
\begin{picture}(95,99)(0,0)
\Text(15.0,80.0)[r]{$e$}
\ArrowLine(16.0,80.0)(37.0,80.0) 
\Text(47.0,81.0)[b]{$\gamma$}
\DashLine(37.0,80.0)(58.0,80.0){3.0} 
\Text(80.0,90.0)[l]{$e$}
\ArrowLine(58.0,80.0)(79.0,90.0) 
\Text(80.0,70.0)[l]{$\bar{e}$}
\ArrowLine(79.0,70.0)(58.0,80.0) 
\Text(33.0,70.0)[r]{$e$}
\ArrowLine(37.0,80.0)(37.0,60.0) 
\Line(37.0,60.0)(58.0,60.0) 
\Text(80.0,50.0)[l]{$\nu_e$}
\ArrowLine(58.0,60.0)(79.0,50.0) 
\Text(33.0,50.0)[r]{$W^+$}
\DashArrowLine(37.0,40.0)(37.0,60.0){3.0} 
\Text(15.0,40.0)[r]{$\bar{e}$}
\ArrowLine(37.0,40.0)(16.0,40.0) 
\Line(37.0,40.0)(58.0,40.0) 
\Text(80.0,30.0)[l]{$\bar{\nu}_e$}
\ArrowLine(79.0,30.0)(58.0,40.0) 
\Text(47,0)[b] {diagr.1}
\end{picture} \ 
{} \qquad\allowbreak
\begin{picture}(95,99)(0,0)
\Text(15.0,90.0)[r]{$e$}
\ArrowLine(16.0,90.0)(58.0,90.0) 
\Text(80.0,90.0)[l]{$\nu_e$}
\ArrowLine(58.0,90.0)(79.0,90.0) 
\Text(54.0,80.0)[r]{$W^+$}
\DashArrowLine(58.0,70.0)(58.0,90.0){3.0} 
\Text(80.0,70.0)[l]{$e$}
\ArrowLine(58.0,70.0)(79.0,70.0) 
\Text(54.0,60.0)[r]{$\nu_e$}
\ArrowLine(58.0,50.0)(58.0,70.0) 
\Text(80.0,50.0)[l]{$\bar{e}$}
\ArrowLine(79.0,50.0)(58.0,50.0) 
\Text(54.0,40.0)[r]{$W^+$}
\DashArrowLine(58.0,30.0)(58.0,50.0){3.0} 
\Text(15.0,30.0)[r]{$\bar{e}$}
\ArrowLine(58.0,30.0)(16.0,30.0) 
\Text(80.0,30.0)[l]{$\bar{\nu}_e$}
\ArrowLine(79.0,30.0)(58.0,30.0) 
\Text(47,0)[b] {diagr.2}
\end{picture} \ 
{} \qquad\allowbreak
\begin{picture}(95,99)(0,0)
\Text(15.0,80.0)[r]{$e$}
\ArrowLine(16.0,80.0)(37.0,80.0) 
\Line(37.0,80.0)(58.0,80.0) 
\Text(80.0,90.0)[l]{$\nu_e$}
\ArrowLine(58.0,80.0)(79.0,90.0) 
\Text(33.0,70.0)[r]{$W^+$}
\DashArrowLine(37.0,60.0)(37.0,80.0){3.0} 
\Text(15.0,60.0)[r]{$\bar{e}$}
\ArrowLine(37.0,60.0)(16.0,60.0) 
\Text(47.0,64.0)[b]{$\nu_e$}
\ArrowLine(58.0,60.0)(37.0,60.0) 
\Text(80.0,70.0)[l]{$\bar{\nu}_e$}
\ArrowLine(79.0,70.0)(58.0,60.0) 
\Text(57.0,50.0)[r]{$Z$}
\DashLine(58.0,60.0)(58.0,40.0){3.0} 
\Text(80.0,50.0)[l]{$e$}
\ArrowLine(58.0,40.0)(79.0,50.0) 
\Text(80.0,30.0)[l]{$\bar{e}$}
\ArrowLine(79.0,30.0)(58.0,40.0) 
\Text(47,0)[b] {diagr.3}
\end{picture} \ 
{} \qquad\allowbreak
\begin{picture}(95,99)(0,0)
\Text(15.0,80.0)[r]{$e$}
\ArrowLine(16.0,80.0)(37.0,80.0) 
\Line(37.0,80.0)(58.0,80.0) 
\Text(80.0,90.0)[l]{$\nu_e$}
\ArrowLine(58.0,80.0)(79.0,90.0) 
\Text(33.0,70.0)[r]{$W^+$}
\DashArrowLine(37.0,60.0)(37.0,80.0){3.0} 
\Line(37.0,60.0)(58.0,60.0) 
\Text(80.0,70.0)[l]{$\bar{\nu}_e$}
\ArrowLine(79.0,70.0)(58.0,60.0) 
\Text(33.0,50.0)[r]{$e$}
\ArrowLine(37.0,60.0)(37.0,40.0) 
\Text(15.0,40.0)[r]{$\bar{e}$}
\ArrowLine(37.0,40.0)(16.0,40.0) 
\Text(47.0,41.0)[b]{$\gamma$}
\DashLine(37.0,40.0)(58.0,40.0){3.0} 
\Text(80.0,50.0)[l]{$e$}
\ArrowLine(58.0,40.0)(79.0,50.0) 
\Text(80.0,30.0)[l]{$\bar{e}$}
\ArrowLine(79.0,30.0)(58.0,40.0) 
\Text(47,0)[b] {diagr.4}
\end{picture} \ 
{} \qquad\allowbreak
\begin{picture}(95,99)(0,0)
\Text(15.0,80.0)[r]{$e$}
\ArrowLine(16.0,80.0)(37.0,80.0) 
\Line(37.0,80.0)(58.0,80.0) 
\Text(80.0,90.0)[l]{$\nu_e$}
\ArrowLine(58.0,80.0)(79.0,90.0) 
\Text(33.0,70.0)[r]{$W^+$}
\DashArrowLine(37.0,60.0)(37.0,80.0){3.0} 
\Line(37.0,60.0)(58.0,60.0) 
\Text(80.0,70.0)[l]{$\bar{\nu}_e$}
\ArrowLine(79.0,70.0)(58.0,60.0) 
\Text(33.0,50.0)[r]{$e$}
\ArrowLine(37.0,60.0)(37.0,40.0) 
\Text(15.0,40.0)[r]{$\bar{e}$}
\ArrowLine(37.0,40.0)(16.0,40.0) 
\Text(47.0,41.0)[b]{$Z$}
\DashLine(37.0,40.0)(58.0,40.0){3.0} 
\Text(80.0,50.0)[l]{$e$}
\ArrowLine(58.0,40.0)(79.0,50.0) 
\Text(80.0,30.0)[l]{$\bar{e}$}
\ArrowLine(79.0,30.0)(58.0,40.0) 
\Text(47,0)[b] {diagr.5}
\end{picture} \ 
{} \qquad\allowbreak
\begin{picture}(95,99)(0,0)
\Text(15.0,80.0)[r]{$e$}
\ArrowLine(16.0,80.0)(37.0,80.0) 
\Line(37.0,80.0)(58.0,80.0) 
\Text(80.0,90.0)[l]{$\nu_e$}
\ArrowLine(58.0,80.0)(79.0,90.0) 
\Text(33.0,70.0)[r]{$W^+$}
\DashArrowLine(37.0,60.0)(37.0,80.0){3.0} 
\Text(47.0,61.0)[b]{$\gamma$}
\DashLine(37.0,60.0)(58.0,60.0){3.0} 
\Text(80.0,70.0)[l]{$e$}
\ArrowLine(58.0,60.0)(79.0,70.0) 
\Text(80.0,50.0)[l]{$\bar{e}$}
\ArrowLine(79.0,50.0)(58.0,60.0) 
\Text(33.0,50.0)[r]{$W^+$}
\DashArrowLine(37.0,40.0)(37.0,60.0){3.0} 
\Text(15.0,40.0)[r]{$\bar{e}$}
\ArrowLine(37.0,40.0)(16.0,40.0) 
\Line(37.0,40.0)(58.0,40.0) 
\Text(80.0,30.0)[l]{$\bar{\nu}_e$}
\ArrowLine(79.0,30.0)(58.0,40.0) 
\Text(47,0)[b] {diagr.6}
\end{picture} \ 
{} \qquad\allowbreak
\begin{picture}(95,99)(0,0)
\Text(15.0,80.0)[r]{$e$}
\ArrowLine(16.0,80.0)(37.0,80.0) 
\Line(37.0,80.0)(58.0,80.0) 
\Text(80.0,90.0)[l]{$\nu_e$}
\ArrowLine(58.0,80.0)(79.0,90.0) 
\Text(33.0,70.0)[r]{$W^+$}
\DashArrowLine(37.0,60.0)(37.0,80.0){3.0} 
\Text(47.0,61.0)[b]{$Z$}
\DashLine(37.0,60.0)(58.0,60.0){3.0} 
\Text(80.0,70.0)[l]{$e$}
\ArrowLine(58.0,60.0)(79.0,70.0) 
\Text(80.0,50.0)[l]{$\bar{e}$}
\ArrowLine(79.0,50.0)(58.0,60.0) 
\Text(33.0,50.0)[r]{$W^+$}
\DashArrowLine(37.0,40.0)(37.0,60.0){3.0} 
\Text(15.0,40.0)[r]{$\bar{e}$}
\ArrowLine(37.0,40.0)(16.0,40.0) 
\Line(37.0,40.0)(58.0,40.0) 
\Text(80.0,30.0)[l]{$\bar{\nu}_e$}
\ArrowLine(79.0,30.0)(58.0,40.0) 
\Text(47,0)[b] {diagr.7}
\end{picture} \ 
{} \qquad\allowbreak
\begin{picture}(95,99)(0,0)
\Text(15.0,80.0)[r]{$e$}
\ArrowLine(16.0,80.0)(37.0,80.0) 
\Text(47.0,84.0)[b]{$\nu_e$}
\ArrowLine(37.0,80.0)(58.0,80.0) 
\Text(80.0,90.0)[l]{$\nu_e$}
\ArrowLine(58.0,80.0)(79.0,90.0) 
\Text(57.0,70.0)[r]{$Z$}
\DashLine(58.0,80.0)(58.0,60.0){3.0} 
\Text(80.0,70.0)[l]{$e$}
\ArrowLine(58.0,60.0)(79.0,70.0) 
\Text(80.0,50.0)[l]{$\bar{e}$}
\ArrowLine(79.0,50.0)(58.0,60.0) 
\Text(33.0,60.0)[r]{$W^+$}
\DashArrowLine(37.0,40.0)(37.0,80.0){3.0} 
\Text(15.0,40.0)[r]{$\bar{e}$}
\ArrowLine(37.0,40.0)(16.0,40.0) 
\Line(37.0,40.0)(58.0,40.0) 
\Text(80.0,30.0)[l]{$\bar{\nu}_e$}
\ArrowLine(79.0,30.0)(58.0,40.0) 
\Text(47,0)[b] {diagr.8}
\end{picture} \ 
{} \qquad\allowbreak
\begin{picture}(95,99)(0,0)
\Text(15.0,80.0)[r]{$e$}
\ArrowLine(16.0,80.0)(37.0,80.0) 
\Text(47.0,81.0)[b]{$Z$}
\DashLine(37.0,80.0)(58.0,80.0){3.0} 
\Text(80.0,90.0)[l]{$e$}
\ArrowLine(58.0,80.0)(79.0,90.0) 
\Text(80.0,70.0)[l]{$\bar{e}$}
\ArrowLine(79.0,70.0)(58.0,80.0) 
\Text(33.0,70.0)[r]{$e$}
\ArrowLine(37.0,80.0)(37.0,60.0) 
\Line(37.0,60.0)(58.0,60.0) 
\Text(80.0,50.0)[l]{$\nu_e$}
\ArrowLine(58.0,60.0)(79.0,50.0) 
\Text(33.0,50.0)[r]{$W^+$}
\DashArrowLine(37.0,40.0)(37.0,60.0){3.0} 
\Text(15.0,40.0)[r]{$\bar{e}$}
\ArrowLine(37.0,40.0)(16.0,40.0) 
\Line(37.0,40.0)(58.0,40.0) 
\Text(80.0,30.0)[l]{$\bar{\nu}_e$}
\ArrowLine(79.0,30.0)(58.0,40.0) 
\Text(47,0)[b] {diagr.9}
\end{picture} \ 
}
\caption{The gauge invariant subset with $\gamma,\zb \to e^+ e^-$
conversion corrections to the process $e^+ e^- \to \nu_e \barnu_e$.}
\label{9nunu}
\end{figure}

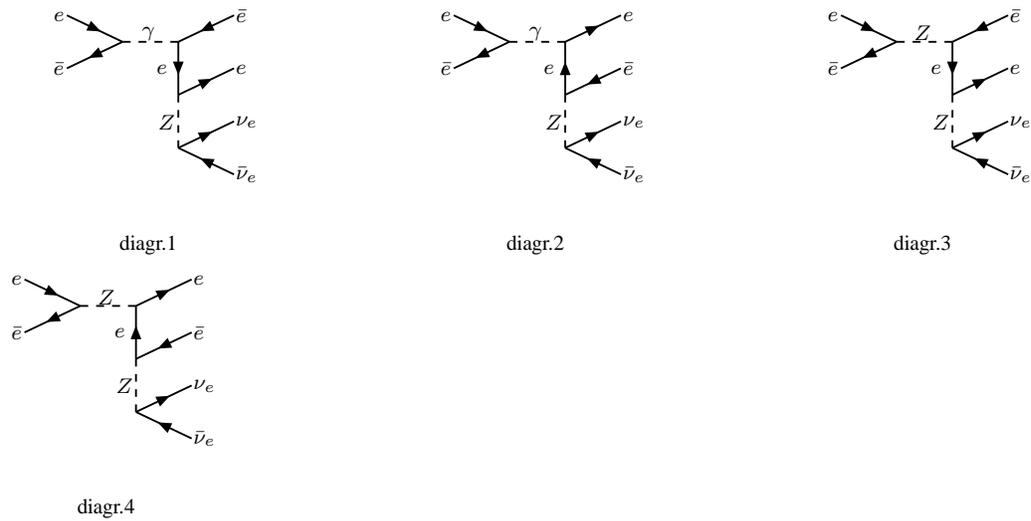
\begin{figure}[htbp]
{
\unitlength=1.0 pt
\SetScale{1.0}
\SetWidth{0.7}      
\scriptsize    
{} \qquad\allowbreak
\begin{picture}(95,99)(0,0)
\Text(15.0,90.0)[r]{$e$}
\ArrowLine(16.0,90.0)(37.0,80.0) 
\Text(15.0,70.0)[r]{$\bar{e}$}
\ArrowLine(37.0,80.0)(16.0,70.0) 
\Text(47.0,81.0)[b]{$\gamma$}
\DashLine(37.0,80.0)(58.0,80.0){3.0} 
\Text(80.0,90.0)[l]{$\bar{e}$}
\ArrowLine(79.0,90.0)(58.0,80.0) 
\Text(54.0,70.0)[r]{$e$}
\ArrowLine(58.0,80.0)(58.0,60.0) 
\Text(80.0,70.0)[l]{$e$}
\ArrowLine(58.0,60.0)(79.0,70.0) 
\Text(57.0,50.0)[r]{$Z$}
\DashLine(58.0,60.0)(58.0,40.0){3.0} 
\Text(80.0,50.0)[l]{$\nu_e$}
\ArrowLine(58.0,40.0)(79.0,50.0) 
\Text(80.0,30.0)[l]{$\bar{\nu}_e$}
\ArrowLine(79.0,30.0)(58.0,40.0) 
\Text(47,0)[b] {diagr.1}
\end{picture} \ 
{} \qquad\allowbreak
\begin{picture}(95,99)(0,0)
\Text(15.0,90.0)[r]{$e$}
\ArrowLine(16.0,90.0)(37.0,80.0) 
\Text(15.0,70.0)[r]{$\bar{e}$}
\ArrowLine(37.0,80.0)(16.0,70.0) 
\Text(47.0,81.0)[b]{$\gamma$}
\DashLine(37.0,80.0)(58.0,80.0){3.0} 
\Text(80.0,90.0)[l]{$e$}
\ArrowLine(58.0,80.0)(79.0,90.0) 
\Text(54.0,70.0)[r]{$e$}
\ArrowLine(58.0,60.0)(58.0,80.0) 
\Text(80.0,70.0)[l]{$\bar{e}$}
\ArrowLine(79.0,70.0)(58.0,60.0) 
\Text(57.0,50.0)[r]{$Z$}
\DashLine(58.0,60.0)(58.0,40.0){3.0} 
\Text(80.0,50.0)[l]{$\nu_e$}
\ArrowLine(58.0,40.0)(79.0,50.0) 
\Text(80.0,30.0)[l]{$\bar{\nu}_e$}
\ArrowLine(79.0,30.0)(58.0,40.0) 
\Text(47,0)[b] {diagr.2}
\end{picture} \ 
{} \qquad\allowbreak
\begin{picture}(95,99)(0,0)
\Text(15.0,90.0)[r]{$e$}
\ArrowLine(16.0,90.0)(37.0,80.0) 
\Text(15.0,70.0)[r]{$\bar{e}$}
\ArrowLine(37.0,80.0)(16.0,70.0) 
\Text(47.0,81.0)[b]{$Z$}
\DashLine(37.0,80.0)(58.0,80.0){3.0} 
\Text(80.0,90.0)[l]{$\bar{e}$}
\ArrowLine(79.0,90.0)(58.0,80.0) 
\Text(54.0,70.0)[r]{$e$}
\ArrowLine(58.0,80.0)(58.0,60.0) 
\Text(80.0,70.0)[l]{$e$}
\ArrowLine(58.0,60.0)(79.0,70.0) 
\Text(57.0,50.0)[r]{$Z$}
\DashLine(58.0,60.0)(58.0,40.0){3.0} 
\Text(80.0,50.0)[l]{$\nu_e$}
\ArrowLine(58.0,40.0)(79.0,50.0) 
\Text(80.0,30.0)[l]{$\bar{\nu}_e$}
\ArrowLine(79.0,30.0)(58.0,40.0) 
\Text(47,0)[b] {diagr.3}
\end{picture} \ 
{} \qquad\allowbreak
\begin{picture}(95,99)(0,0)
\Text(15.0,90.0)[r]{$e$}
\ArrowLine(16.0,90.0)(37.0,80.0) 
\Text(15.0,70.0)[r]{$\bar{e}$}
\ArrowLine(37.0,80.0)(16.0,70.0) 
\Text(47.0,81.0)[b]{$Z$}
\DashLine(37.0,80.0)(58.0,80.0){3.0} 
\Text(80.0,90.0)[l]{$e$}
\ArrowLine(58.0,80.0)(79.0,90.0) 
\Text(54.0,70.0)[r]{$e$}
\ArrowLine(58.0,60.0)(58.0,80.0) 
\Text(80.0,70.0)[l]{$\bar{e}$}
\ArrowLine(79.0,70.0)(58.0,60.0) 
\Text(57.0,50.0)[r]{$Z$}
\DashLine(58.0,60.0)(58.0,40.0){3.0} 
\Text(80.0,50.0)[l]{$\nu_e$}
\ArrowLine(58.0,40.0)(79.0,50.0) 
\Text(80.0,30.0)[l]{$\bar{\nu}_e$}
\ArrowLine(79.0,30.0)(58.0,40.0) 
\Text(47,0)[b] {diagr.4}
\end{picture} \ 
}
\caption{The gauge invariant subset with $\zb \to \nu_e \barnu_e$
conversion corrections to the process $e^+ e^- \to e^+ e^-$. }
\label{4ee}
\end{figure}

\begin{figure}[htbp]
{
\unitlength=1.0 pt
\SetScale{1.0}
\SetWidth{0.7}      
\scriptsize    
{} \qquad\allowbreak
\begin{picture}(95,99)(0,0)
\Text(15.0,90.0)[r]{$e$}
\ArrowLine(16.0,90.0)(37.0,80.0) 
\Text(15.0,70.0)[r]{$\bar{e}$}
\ArrowLine(37.0,80.0)(16.0,70.0) 
\Text(47.0,81.0)[b]{$Z$}
\DashLine(37.0,80.0)(58.0,80.0){3.0} 
\Text(80.0,90.0)[l]{$\bar{\nu}_e$}
\ArrowLine(79.0,90.0)(58.0,80.0) 
\Text(54.0,70.0)[r]{$\nu_e$}
\ArrowLine(58.0,80.0)(58.0,60.0) 
\Text(80.0,70.0)[l]{$\nu_e$}
\ArrowLine(58.0,60.0)(79.0,70.0) 
\Text(57.0,50.0)[r]{$Z$}
\DashLine(58.0,60.0)(58.0,40.0){3.0} 
\Text(80.0,50.0)[l]{$e$}
\ArrowLine(58.0,40.0)(79.0,50.0) 
\Text(80.0,30.0)[l]{$\bar{e}$}
\ArrowLine(79.0,30.0)(58.0,40.0) 
\Text(47,0)[b] {diagr.1}
\end{picture} \ 
{} \qquad\allowbreak
\begin{picture}(95,99)(0,0)
\Text(15.0,90.0)[r]{$e$}
\ArrowLine(16.0,90.0)(37.0,80.0) 
\Text(15.0,70.0)[r]{$\bar{e}$}
\ArrowLine(37.0,80.0)(16.0,70.0) 
\Text(47.0,81.0)[b]{$Z$}
\DashLine(37.0,80.0)(58.0,80.0){3.0} 
\Text(80.0,90.0)[l]{$\nu_e$}
\ArrowLine(58.0,80.0)(79.0,90.0) 
\Text(54.0,70.0)[r]{$\nu_e$}
\ArrowLine(58.0,60.0)(58.0,80.0) 
\Text(80.0,70.0)[l]{$\bar{\nu}_e$}
\ArrowLine(79.0,70.0)(58.0,60.0) 
\Text(57.0,50.0)[r]{$Z$}
\DashLine(58.0,60.0)(58.0,40.0){3.0} 
\Text(80.0,50.0)[l]{$e$}
\ArrowLine(58.0,40.0)(79.0,50.0) 
\Text(80.0,30.0)[l]{$\bar{e}$}
\ArrowLine(79.0,30.0)(58.0,40.0) 
\Text(47,0)[b] {diagr.2}
\end{picture} \ 
}
\caption{The gauge invariant subset with $\zb \to e^+ e^-$
conversion corrections to the process $e^+ e^- \to \nu_e \barnu_e$.}
\label{2nunu}
\end{figure}

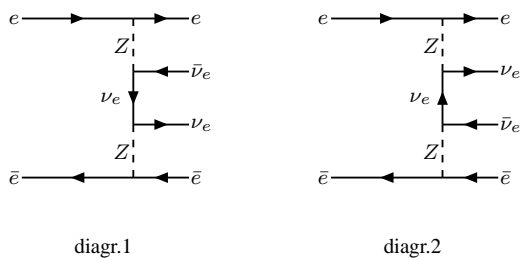
\begin{figure}[htbp]
{
\unitlength=1.0 pt
\SetScale{1.0}
\SetWidth{0.7}      
\scriptsize    
{} \qquad\allowbreak
\begin{picture}(95,99)(0,0)
\Text(15.0,90.0)[r]{$e$}
\ArrowLine(16.0,90.0)(58.0,90.0) 
\Text(80.0,90.0)[l]{$e$}
\ArrowLine(58.0,90.0)(79.0,90.0) 
\Text(57.0,80.0)[r]{$Z$}
\DashLine(58.0,90.0)(58.0,70.0){3.0} 
\Text(80.0,70.0)[l]{$\bar{\nu}_e$}
\ArrowLine(79.0,70.0)(58.0,70.0) 
\Text(54.0,60.0)[r]{$\nu_e$}
\ArrowLine(58.0,70.0)(58.0,50.0) 
\Text(80.0,50.0)[l]{$\nu_e$}
\ArrowLine(58.0,50.0)(79.0,50.0) 
\Text(57.0,40.0)[r]{$Z$}
\DashLine(58.0,50.0)(58.0,30.0){3.0} 
\Text(15.0,30.0)[r]{$\bar{e}$}
\ArrowLine(58.0,30.0)(16.0,30.0) 
\Text(80.0,30.0)[l]{$\bar{e}$}
\ArrowLine(79.0,30.0)(58.0,30.0) 
\Text(47,0)[b] {diagr.1}
\end{picture} \ 
{} \qquad\allowbreak
\begin{picture}(95,99)(0,0)
\Text(15.0,90.0)[r]{$e$}
\ArrowLine(16.0,90.0)(58.0,90.0) 
\Text(80.0,90.0)[l]{$e$}
\ArrowLine(58.0,90.0)(79.0,90.0) 
\Text(57.0,80.0)[r]{$Z$}
\DashLine(58.0,90.0)(58.0,70.0){3.0} 
\Text(80.0,70.0)[l]{$\nu_e$}
\ArrowLine(58.0,70.0)(79.0,70.0) 
\Text(54.0,60.0)[r]{$\nu_e$}
\ArrowLine(58.0,50.0)(58.0,70.0) 
\Text(80.0,50.0)[l]{$\bar{\nu}_e$}
\ArrowLine(79.0,50.0)(58.0,50.0) 
\Text(57.0,40.0)[r]{$Z$}
\DashLine(58.0,50.0)(58.0,30.0){3.0} 
\Text(15.0,30.0)[r]{$\bar{e}$}
\ArrowLine(58.0,30.0)(16.0,30.0) 
\Text(80.0,30.0)[l]{$\bar{e}$}
\ArrowLine(79.0,30.0)(58.0,30.0) 
\Text(47,0)[b] {diagr.2}
\end{picture} \ 
}
\caption{The weak multi-peripheral gauge invariant subset
in the channel $e^+ e^- \to e^+ e^- \nu_e \barnu_e$.}
\label{2t}
\end{figure}

\section{The NC02 cross-section, $\sigma_{ZZ}$}
\label{sectZZ}

The cross-section for $e^+e^- \to \zb\zb$ is defined starting from the NC02
process, very much as $e^+e^- \to \wbp\wbm$ in terms of CC03, hence we
sum over all channels, $e^+e^- \to \zb(\to f_1\barf_1) +
\zb(\to f_2\barf_2)$ including four neutrinos in the final state.
The electroweak corrections to $ee\to \zb\zb$ were calculated in~\cite{zz_corr}.
Actually there the weak corrections have been discussed separately, but 
unfortunately in the $\alpha$ scheme.  

As usual, it is left for the experimenters to evaluate the background, \ie
to define a neutral current observable cross-section as follows:
\bq
\sigma_{\rm NC} = \sigma_{\rm NC02}\,\lpar 1 + \delta^{\rm DPA}_{\rm NC}\rpar + 
\Bigl[ \sigma_{4\rmf} - \sigma_{\rm NC02}\Bigr].
\eq
The theoretical prediction, therefore, should concentrate on
$\sigma_{\rm NC02}$, with or without $\ord{\alpha}$ radiative correction
in DPA-approximation. In particular, the background should account for
the Mixed processes (Mix43).
There is some important remark to be made. When dealing with $u \baru u 
\baru$ etc, \ie with channels containing identical particles, we have
to evaluate the unphysical sum of the two diagrams corresponding to 
$e^+e^- \to \zb(\to u_1 \baru_1) + \zb(\to u_2 \baru_2)$,
tacitly assuming that there are two $u$ quarks, $u$ of type 1 and $u$ of
type 2. 
Since the interferences between the crossings are not double-resonant,
it is customary to consider them as background and to define the $\zb\zb$
signal, \ie $\sigma_{\rm NC02}$, from the absolute squares of the 
double-resonant diagrams only.
This is a matter of definition, \ie, we could define the $\zb\zb$ signal to
contain all crossings in case of four identical flavors in the final
state. That one chooses the first option is largely based on the drawback 
that, with the latter, 
\bq
\sigma( e^+e^- \to \zb\zb)\,\times\,{\rm BR}^2(\zb \to u\baru),
\eq
is no longer $\sigma_{\ssZ\ssZ \to u\baru u\baru}$. It is certainly true that
the cross-section containing all crossings 
would be more physical but, for the time being this is the convention.
Furthermore, one should remember that the 
$\Pep\Pem\to\gamma^*\gamma^*,\gamma^*\PZ$
background is quite large (see \eg Ref.~\cite{racoonww_ee4fa}).

Bearing this in mind, we should stress that the terminology $\sigma^{\rm NC02}$
is, sometimes, unfaithful, simply because this is not what experiments use
in their analysis. A common procedure is to use {\tt EXCALIBUR}
and to restrict it to the complete set of double-resonant diagrams. 
In other words, experiments measure data in some window of invariant mass and 
extrapolate with some coefficient, evaluated by MC, to what one finally
calls the NC02-total cross-section but it represents, instead, the sum of 
all double-resonant $\zb$ diagrams 
(for some channel $4$ instead of $2$).

However, by definition, we select NC02 to be $e^+e^- \to \zb\zb$,
two diagrams ($t$ and $u$ channel), with all $\zb$ decay modes allowed
for both $\zb$-bosons. If one computes everything as production $\,\otimes\,$ 
decay then, as long as one remembers to include factors $1/2$, everything is 
reasonable. The conclusion is based on the following observation. When all
diagrams are taken into account we find
\bqa
\sigma(e^+e^- \to \baru u \barc c) &=& 208.9\,{\rm fb},  \quad
\sigma(e^+e^- \to \baru u \bars s) = 204.4\,{\rm fb},  \nl
\sigma(e^+e^- \to \bard d \bars s) &=& 182.6\,{\rm fb},  \quad
\sigma(e^+e^- \to \baru u \bard d) = 1.980\,{\rm pb},  \nl
\sigma(e^+e^- \to \baru u \baru u) &=& 101.4\,{\rm fb},  \quad
\sigma(e^+e^- \to \bard d \bard d) = 87.88\,{\rm fb},
\label{allnc}
\eqa
and, as a consequence,
\bq
R_{uucc/uuuu} = 2.06, \qquad R_{ddss/dddd} = 2.08.
\eq
In other words, even if we define on-shell and compute off-shell 
the same result, within few percents, is obtained. 

The relative significance of the $\zb\zb$ cross-section is considerably less
than the one attributed to the $\wb\wb$ cross-section. Its is smaller and
with much larger experimental errors, even at the level of projected ones.
As a consequence the NC02 process has received less attention that the CC03
one and, so far, we have no published result on $\ord{\alpha}$ DPA calculations
for it although, in principle, there is no major obstacle to it.

\subsection{Description of programs and results}

\subsubsection*{{\tt YFSZZ}}

\subsubsection*{Authors}

\begin{tabular}{l}
S.~Jadach, W.~Placzek, M.~Skrzypek, B.~Ward and Z.~Was  \\  
\end{tabular}

\subsubsection*{General Description}

The program evaluates the NC02 double resonant process
$e^+e^-\rightarrow \zb\zb\rightarrow 4\rmf$
in the presence of multiple photon radiation using Monte Carlo
event generator techniques. The theoretical formulation is based, 
in the leading pole approximation (LPA), on
$\ord{\alpha^2}$ LL YFS exponentiation
for the production process, with the possibility of
anomalous gauge couplings if the user so desires.
The Monte Carlo algorithm used to realize the YFS exponentiation
is based on the YFS2 algorithm presented in Ref.~\cite{yfs2:1990}
and in Ref.~\cite{koralz3:1991}. In this way, we achieve an event-by-event
realization of our calculation in which arbitrary detector
cuts are possible and in which infrared singularities
are cancelled to all orders in $\alpha$. A detailed description of
our work can be found in Ref.~\cite{yfszz:1997}.

\vspace{2mm}
\noindent

\subsubsection*{Features of the program}

The code is a complete Monte Carlo event generator and gives for each
event the final particle four-momenta for the entire $4f+n\gamma$ final 
state over the entire phase space for each final state particle.
The events may be weighted or unweighted, as it is more or less convenient
for the user accordingly. The code features the realization of the LPA
for the NC02 process that is the analog of that
given in Ref.~\cite{yfsww3:1998} for the CC03 process of
production and decay of $\wb\wb$ pairs. A technical precision
check on the program at the level of 2 per mille for the total cross-section
has been done by comparison with
the results in Ref.~\cite{dimarie}. 
The accuracy of the combined result from {\tt YFSZZ~1.02} and
{\tt KoralW}~1.42, when the combination is taken in analogy
with that presented in Ref.~\cite{00-0101} for {\tt YFSWW3~1.14} and 
{\tt KoralW~1.42},
is expected to be at the level of $2\%$ for the total cross-section, 
due to the missing $\ord{\alpha}$ pure weak corrections in {\tt YFSZZ~1.02}
(we do not expect the other effects missing from our calculation
such as non-universal QED corrections
to enter at this level), 
when all tests are finished. 
These tests are currently in progress.

The operation of the code is entirely analogous to that of the MC
{\tt YFS2} in Refs.~\cite{yfs2:1990}. A crude distribution
based on the primitive Born level distribution and the most dominant
part of the YFS form factors that can be treated analytically is used to
generate a background population of events. The weight for these
events is then computed by standard rejection techniques involving the
ratio of the complete distribution and the crude distribution. As the
user wishes, these weights may be either used directly with the events,
which have the four-momenta of all final state particles available, or they
may be accepted/rejected against a maximal weight WTMAX to produce unweighted
events via again standard MC methods. Standard final statistics
of the run are provided, such as statistical error analysis, 
total cross-sections, etc. The total phase space for the process is always
active in the code.

\vspace{2mm}
\noindent

\subsubsection*{Description of output and availability}

The program prints certain control outputs. The most important
output of the program is the series of Monte Carlo events. The
total cross-section in $fb$ is available for arbitrary cuts 
in the same standard way as it is for {\tt YFS2}, \ie the user
may impose arbitrary detector cuts by the usual rejection methods.

The program is available from the authors via e-mail.
The program is currently posted on {\sl WWW} at
{\sl http://enigma.phys.utk.edu} as well as on {\sl anonymous ftp} at
{\sl enigma.phys.utk.edu} in the form of a {\sl tar.gz} file 
in the {\sl /pub/YFSZZ/} directory
together with all relevant papers and documentation in postscript.

\subsubsection*{{\tt ZZTO}}

\subsubsection*{Author}

\begin{tabular}{l}
G. Passarino \\
\end{tabular}

\subsubsection*{Description.}

{\tt ZZTO} is a newly created code for computing $\sigma^{\rm NC02}$
which, at the moment, has universal Initial State QED, Final State QED, 
Final State QCD, is fully massive with $b$ and $c$ quarks running masses.  
Fermion-Loop is also implemented.
{\tt ZZTO} is missing non-universal QED ISR and purely weak effects (in DPA); 
however, it is under construction with the final goal of including those 
effects.
The code sums over all $\zb\zb$ decay modes, even $\nu\nu\nu\nu$. However,
single channels are available, \ie $qqqq,\, qq\nu\nu,\, qq{\rm ll},\,
{\rm ll}\nu\nu,\, {\rm llll},\, \nu\nu\nu\nu$.
Therefor, inside {\tt ZZTO} we have the exact matrix element for $e^+e^- \to 
\zb\zb \to 4\rmf$ with massive fermions and running masses for the 
$b,c$-quarks.
Cuts are only implemented on the $\zb$ invariant masses, therefore we can 
apply final state QCD correction factors beyond the usual {\em naive} 
correction.
In other words, the total hadronic decay rate of each $\zb$-boson is
split into the sum of the vector current induced rate, $\Gamma^{\ssV}$, and 
of the axial decay rate, $\Gamma^{\ssA}$, which receive different QCD 
corrections evaluated at the scale equal to the virtuality of the 
$q\barq$-pair. Non-factorizable QCD corrections are neglected.
Final state QED corrections are also included, again evaluated at the 
virtuality of the pair, \ie with $\alpha_{\rm QED}(M^2_{\rm pair})$.
Initial state QED corrections include, so far, only the universal part of
the structure functions evaluated at the scale $s$.

To implement the Fermion-Loop scheme we had to incorporate QCD corrections
in the evaluation of the complex pole $\sZ$ and of the $\rho$-parameter 
associated to the $\zb$-propagator. This we have done by taking into account 
also the massive top quark, while the light quarks, including the $b$ one, are 
treated as massless. QCD is exactly implemented by using the $\ord{\alpha\als}$
vector-boson self-energies of Ref.~\cite{ak} with $\mz$ as the scale for light
quarks and $\mt$ for the $b-b, b-t$ and $t-t$ contributions.
For $\mw= 80.350\,\GeV, \mz = 91.1888\,\GeV$ and $\als(\mzs) = 0.120$ we find
a QCD effect illustrated in \tabn{zzqcd}.
\begin{table}[hp]\centering
\begin{tabular}{|c|c|c|}
\hline
&  & \\
& without QCD & with QCD \\
&  & \\
\hline
&  & \\
$\mu_{\ssW} = \sqrt{\Reb\sW}$ [GeV] & 80.324 & 80.322 \\
$-\Imb(\sW)/\mu_{\ssW}$ [GeV] & 2.0581 & 2.1109 \\
$\mu_{\ssZ} = \sqrt{\Reb\sZ}$ [GeV] & 91.155 & 91.153 \\
$-\Imb(\sZ)/\mu_{\ssZ}$ [GeV] & 2.4653 & 2.5315 \\
$\mt$ [GeV] & 148.21 & 156.32 \\
&  & \\
\hline
\end{tabular}
\vspace*{3mm}
\caption[]{Effect of including QCD corrections on the complex $\sW,\sZ$
poles according to {\tt ZZTO}.}
\label{zzqcd}
\end{table}
The program {\tt ZZTO} is currently posted on {\sl WWW} at {\sl 
http://www.to.infn.it/giampier/zzto}.

\subsubsection*{Distributions.}

The $\zb\zb$-signal is basically defined through invariant masses, for 
instance $e^+e^- \to q\barq l^+l^-(\ph)$, $q$-flavour blind or heavy
$q$-flavors, $l=e/\mu/\tau$, $|\cos\theta_{l_1}| < 0.985$, no cut
on the second lepton (only one lepton tagged), $M(q\barq ) > 10~(45)\,$GeV.

Here, we do not discuss invariant mass distributions in terms of the full
processes but only in terms of the signal NC02. The angular cuts are there 
only because of detector holes at the beam pipe. Since for NC02 there
are no poles at edge of phase space, we 
could leave these cuts out for
simplicity. Furthermore, we analyze only $e^+e^- \to q\barq l^+l^-(\ph)$
where the definition of invariant masses is free of ambiguities. 
{\tt ZZTO} includes final state radiations in two different options. In a first
case {\tt ZZTO} implements the exact, factorizable, $\ord{\alpha}$ corrections 
for some extrapolated setup were one can only cut on the $\zb$-virtuality,
see Ref.~\cite{ofsr}.
In the second one, hard and collinear photons are included, within a cone of 
angular resolution $\delta \ll 1$, according to the formalism of 
Ref.~\cite{hcph}. Moreover, soft photons are exponentiated.

Therefore, we can define {\em invariant mass} distributions according to the
following choices: a) $M(l^+l^-\ph)$ or $M(\barq q\ph)$ where $M$
represents the virtuality of the decaying $\zb$-boson and b) $m(l^+l^-)$
where $m$ is the $l^+l^-$ invariant mass and hard photons are included
whenever the angle between the photon and the nearest charged final-state 
fermion is less than $\delta \ll 1$. Above $\delta$ photons are not included
in the mass calculation.
Gluons are always included in $\ord{\als}$ with a fully extrapolated setup, 
\ie the $M$-variable for $\barq q$ final states is always understood as $\zb(M)
\to \barq q + \ph + g$.

In \fig{zz_dist1} we show the $M$-distribution for $e^+e^- +\,$hadrons
and for $\barb b (\barc c) +\,$leptons at one energy, $\sqrt{s} = 188.6\,$GeV.
There is no appreciable difference with $\mu^+\mu^- +\,$hadrons due to
the fact that the FSR correction factor is approximately 
$3/4\,Q^2_f\,\alpha/\pi$ since we cut on the $\zb$-virtuality and not on the
$\barf f$ invariant mass.

\begin{figure}[p]
\vskip -1cm
\epsfig{file=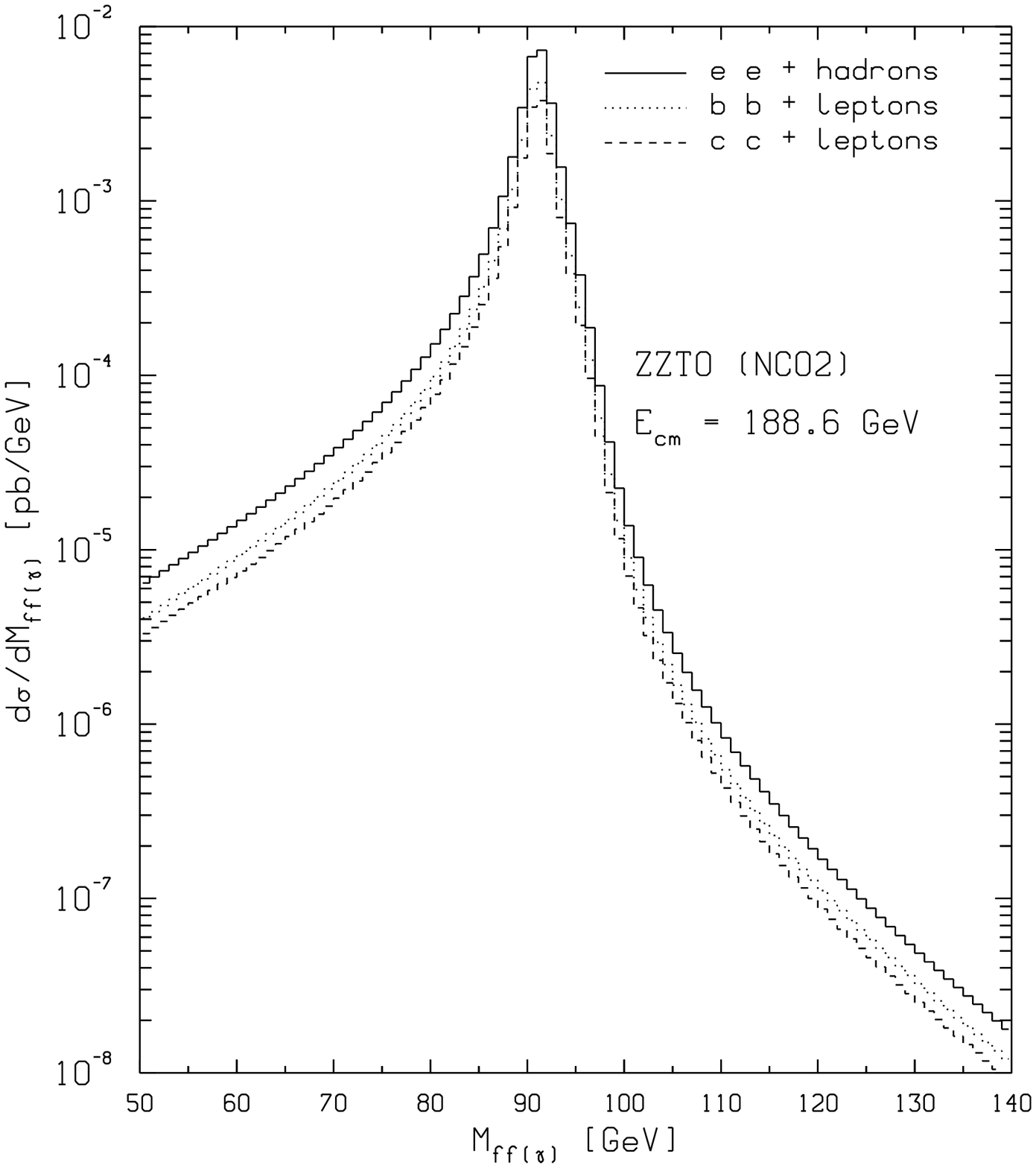,width=0.49\linewidth}
\vskip -1cm
\caption[]{NC02 distributions from {\tt ZZTO}. Here $M(\barf f(\ph,g))$
is the virtuality of the corresponding $\zb$-boson.}
\label{zz_dist1}
\end{figure}

In \fig{zz_dist2} we show $e^+e^- +\,$hadrons and compare $M$ and $m$
distributions for the $e^+e^-(\ph)$ pair. The latter includes collinear
photons within a cone of half-opening angle $\delta = 5^\circ$.
In the same figure we also compare the $m(\barf f)$ distributions for
$e^+e^- +\,$hadrons and $\mu^+\mu^- +\,$hadrons. Since the cut is on the
invariant mass of the pair one starts appreciating differences between different
flavors.

\begin{figure}[p]
\vskip -1cm
\epsfig{file=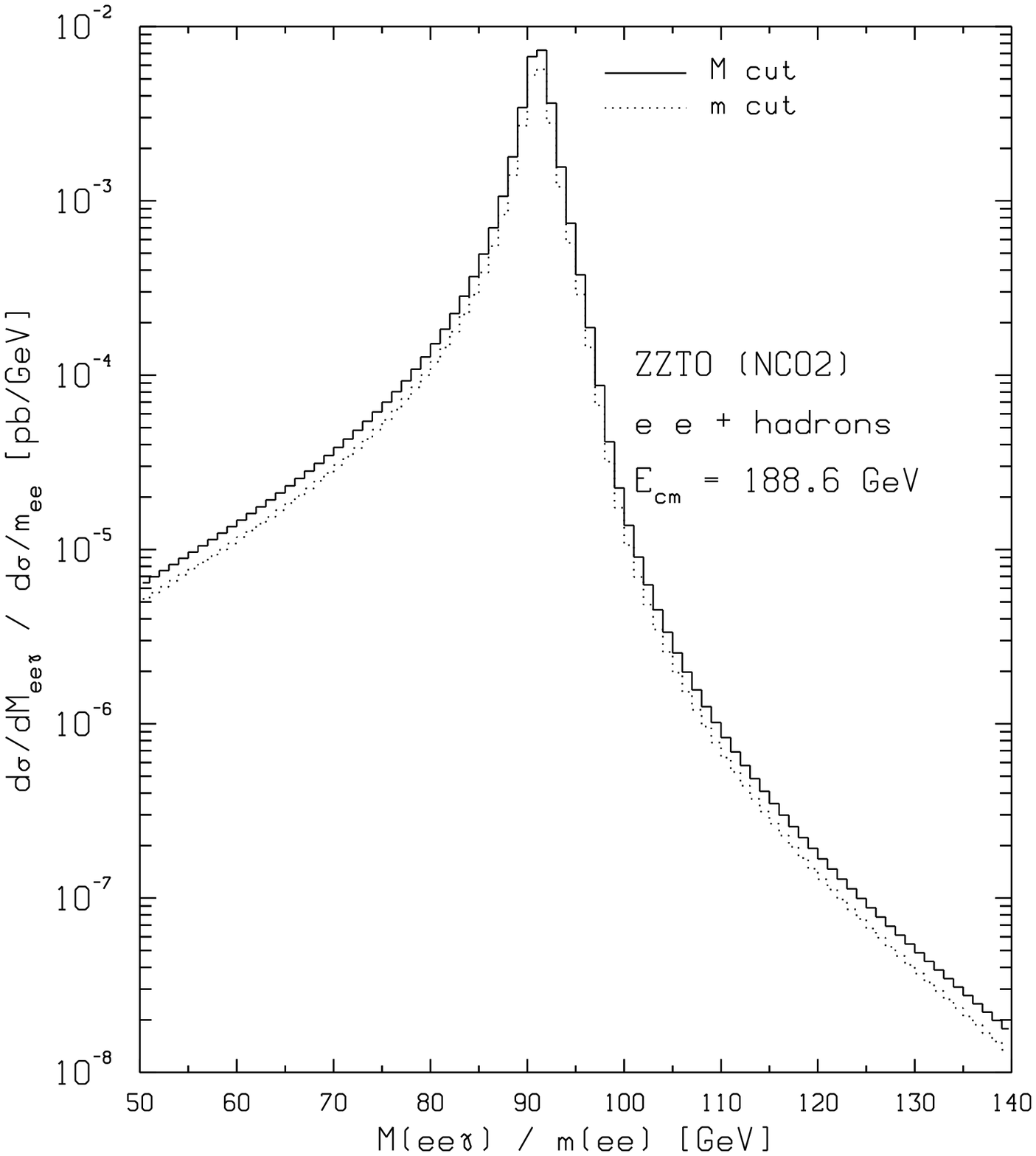,width=0.49\linewidth}
\hfill
\epsfig{file=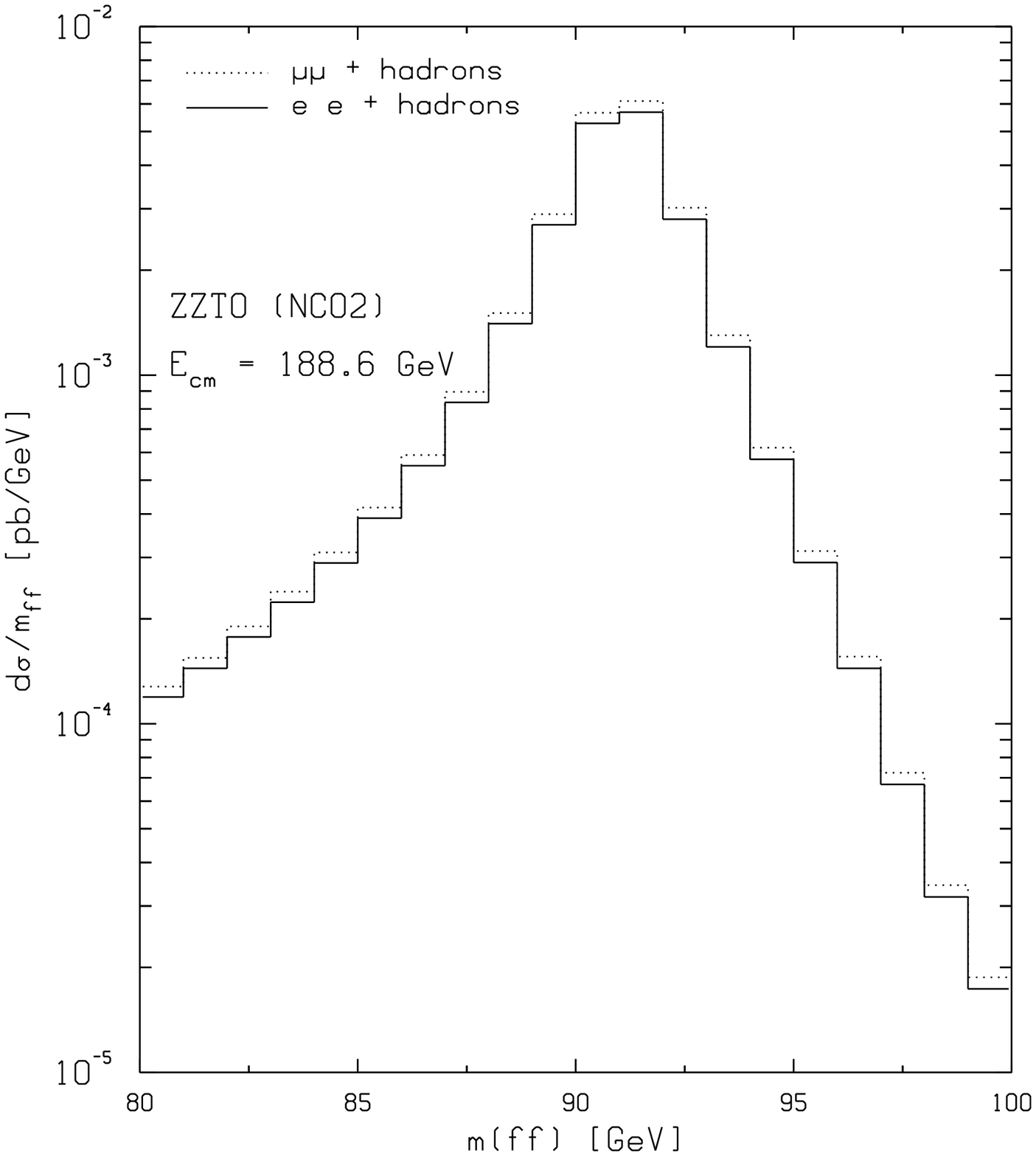,width=0.49\linewidth}
\vskip -1cm
\caption[]{NC02 distributions from {\tt ZZTO}. Here $M(\barf f(\ph,g))$
is the virtuality of the corresponding $\zb$-boson and $m(\barf f)$ is the
$\barf f$ invariant mass with collinear photons that are combined with
the nearest fermion, $\theta(\ph - {\rm nearest}\,f) = 5^\circ$.}
\label{zz_dist2}
\end{figure}

All distributions are computed by {\tt ZZTO} in the Fermion-Loop mode.
The largest effects in the theoretical uncertainty are associated to the fact
that non-factorizable QED corrections are neglected, although one can show that
they vanish in the limit of on-shell $\zb$-bosons, $\mid M^2-\mzs\mid \ll
\Gamma_{\ssZ}\mz$. They also vanish for a fully extrapolated setup, \ie
after integrating over the full range of the two $\zb$ virtualities, which
is not the case for distributions.

\subsubsection*{\tt GENTLE}

\subsubsection*{Authors}

\begin{tabular}{l}
D.~Bardin, A.~Olchevski and T.~Riemann
\end{tabular}

The NC cross-sections in package {\tt 4fan} include now besides the NC32 class 
also the NC02 process (NC08 is unchanged);
also some new options introduced, {\tt ICHNNL}=0,1: switching between NC02 
and NC32 classes (for {\tt IPROC}=2);
Note, that the treatment of {\tt NC08} sub-family
is not changed compared to the version {\tt v.2.10}. It remains accessible
only via {\tt NCqed} branch of the package.

\subsection{Comparisons for the NC02 cross-section}

In this Section we will compare the NC02 cross-section between {\tt
  YFSZZ}, {\tt GENTLE} and the newly created code {\tt ZZTO}.  First,
the comparison between {\tt YFSZZ} and {\tt ZZTO}.  Here, $\sqrt{s} =
188.6\,GeV$ and QCD is not included. The result is shown in
\tabn{zzschemes}.  From \tabn{zzschemes} we see a remarkable
agreement, further quantified in \tabn{zzdiff}. Furthermore, for
$\sigma_{\ssZ\ssZ}$ with Born+ISR+QCD the uncertainty related to the
IPS (Input Parameter Set) is approximately $1\%$.  This does not mean
that the total, true, theoretical uncertainty is $1\%$.  The $\zb\zb$
line-shape, as predicted by {\tt ZZTO} and including QCD corrections
is shown in \tabn{zzls} where the results refer to three schemes,
$\alpha, \gf$ and Fermion-Loop.

Finally, in \fig{zzlineshape} we present the NC02 line-shape for a wide
range of energy, comparing the $\alpha$-scheme with the $\gf$-scheme and
the Fermion-loop one.
Missing an implementation of the Fermion-Loop scheme in other codes, our 
recommendation is to use the $\gf$-scheme since it allows us to
include part of higher order effects in the Born cross-sections.

\clearpage
\begin{table}[p]
\centering
\renewcommand{\arraystretch}{1.1}
\begin{tabular}{|c|c|c|c|}
\hline
channel & {\tt YFSZZ} & {\tt ZZTO} $\gf$-scheme & {\tt ZZTO} $\alpha$-scheme \\
\hline
\hline
$qqqq$            &    294.6794(490) &  298.4411(60)  &    294.5715(59)   \\
$qq\nu\nu$        &    175.4404(302) &  175.5622(35)  &    174.9855(35)   \\
$qq{\rm ll}$      &     88.1805(134) &   88.7146(18)  &     87.9881(18)   \\
${\rm ll}\nu\nu$  &     26.2530(463) &   26.0940(5)   &     26.1342(5)    \\
${\rm llll}$      &      6.5983(15)  &    6.5929(1)   &      6.5706(1)    \\
$\nu\nu\nu\nu$    &     26.1080(71)  &   25.8192(5)   &     25.9868(5)
\\
\hline
total             &    617.2596(755) &  621.2241(124) &    616.2366(123)  \\
\hline
\end{tabular}
\caption[]{Comparison for the NC02 cross-section between {\tt YFSZZ} and
{\tt ZZTO} at $\sqrt{s}= 188.6\,$GeV. The cross-sections are in fb.
\label{zzschemes}}
\end{table}
\begin{table}[p]
\centering
\renewcommand{\arraystretch}{1.1}
\begin{tabular}{|c|c|c|}
\hline
channel & {\tt ZZTO}($\gf$)/{\tt YFSZZ} - 1 & {\tt ZZTO}($\alpha$)/{\tt YFSZZ} 
- 1 \\
\hline
\hline
$qqqq$            &  +1.28  &  -0.04 \\
$qq\nu\nu$        &  +0.07  &  -0.26 \\
$qq{\rm ll}$      &  +0.61  &  -0.22 \\
${\rm ll}\nu\nu$  &  -0.61  &  -0.45 \\
${\rm llll}$      &  -0.08  &  -0.42 \\
$\nu\nu\nu\nu$    &  -1.11  &  -0.46 \\
\hline
total             &  +0.64  &  -0.17 \\
\hline
\end{tabular}
\caption[]{Differences {\tt YFSZZ/ZZTO} for the NC02 cross-section in percent.
\label{zzdiff}}
\end{table}
\begin{table}[p]
\centering
\renewcommand{\arraystretch}{1.1}
\begin{tabular}{|c|c|}
\hline
channel & {\tt ZZTO} $\gf/\alpha - 1$ \\
\hline
\hline
$qqqq$ & +1.31 \\
$qq\nu\nu$ & +0.33 \\
$qq{\rm l}{\rm l}$ & +0.83 \\
${\rm l}{\rm l}\nu\nu$ & -0.15 \\
${\rm l}{\rm l}{\rm l}{\rm l}$ & +0.34 \\
$\nu\nu\nu\nu$ & -0.64 \\
\hline
total& +0.81 \\
\hline
\end{tabular}
\caption[]{Scheme differences in percent for NC02, according to {\tt ZZTO}.
\label{zzperc}}
\end{table}

\clearpage

\begin{table}[hp]\centering
\renewcommand{\arraystretch}{1.3}
\begin{tabular}{|c|c|c|c|c|}
\hline
$\sqrt{s}$ [GeV] & $\sigma^{\ssZ\ssZ}_{\alpha}$ [pb] with QCD & 
$\sigma^{\ssZ\ssZ}_{\gf}$ [pb] with QCD & 
$\sigma^{\ssZ\ssZ}_{\rm FL}$ [pb] with QCD &
$\gf$/FL - 1 [percent] \\
\hline
\hline
180  &          0.12478(1)  & 0.12568(1)  &  0.12669(1) & -0.80 \\
181  &          0.16044(2)  & 0.16160(2)  &  0.16267(2) & -0.66 \\
182  &          0.21135(2)  & 0.21287(2)  &  0.21376(2) & -0.42 \\
183  &          0.27770(2)  & 0.27970(2)  &  0.28009(2) & -0.14 \\
184  &          0.35224(1)  & 0.35477(1)  &  0.35457(1) & -0.03 \\
185  &          0.42644(1)  & 0.42950(1)  &  0.42881(1) & +0.16 \\
186  &          0.49579(1)  & 0.49936(1)  &  0.49833(1) & +0.21 \\
187  &          0.55897(1)  & 0.56299(1)  &  0.56175(1) & +0.22 \\
188  &          0.61596(1)  & 0.62039(1)  &  0.61901(1) & +0.22 \\
189  &          0.66723(1)  & 0.67203(1)  &  0.67057(1) & +0.22 \\
190  &          0.71336(1)  & 0.71848(1)  &  0.71699(1) & +0.21 \\
191  &          0.75487(1)  & 0.76030(1)  &  0.75879(1) & +0.20 \\
192  &          0.79225(1)  & 0.79794(1)  &  0.79643(1) & +0.19 \\
193  &          0.82596(1)  & 0.83190(1)  &  0.83040(1) & +0.18 \\
194  &          0.85643(2)  & 0.86258(2)  &  0.86111(2) & +0.17 \\
195  &          0.88393(2)  & 0.89028(2)  &  0.88884(2) & +0.16 \\
196  &          0.90875(1)  & 0.91528(1)  &  0.91388(1) & +0.15 \\
197  &          0.93118(1)  & 0.93787(1)  &  0.93651(1) & +0.15 \\
198  &          0.95146(1)  & 0.95830(1)  &  0.95698(1) & +0.14 \\
199  &          0.96890(1)  & 0.97677(1)  &  0.97549(1) & +0.13 \\
200  &          0.98635(2)  & 0.99343(2)  &  0.99220(2) & +0.12 \\
201  &          1.00012(2)  & 1.00843(2)  &  1.00724(2) & +0.12 \\
202  &          1.01460(2)  & 1.02189(2)  &  1.02075(2) & +0.11 \\
203  &          1.02660(1)  & 1.03397(1)  &  1.03288(1) & +0.11 \\
204  &          1.03736(1)  & 1.04481(1)  &  1.04376(1) & +0.10 \\
205  &          1.04700(1)  & 1.05452(1)  &  1.05352(1) & +0.09 \\
206  &          1.05561(1)  & 1.06320(1)  &  1.06224(1) & +0.09 \\
207  &          1.06326(1)  & 1.07090(1)  &  1.06998(1) & +0.09 \\
208  &          1.07001(1)  & 1.07770(1)  &  1.07683(1) & +0.08 \\
209  &          1.07594(2)  & 1.08367(2)  &  1.08284(2) & +0.08 \\
210  &          1.08111(2)  & 1.08888(2)  &  1.08809(2) & +0.07 \\
\hline
\end{tabular}
\vspace*{3mm}
\caption[]{NC02 $\zb\zb$ line-shape from {\tt ZZTO}.\label{zzls}}
\end{table}

\clearpage

\begin{figure}[htbp]
\begin{center}
\vskip -1cm
\epsfig{file=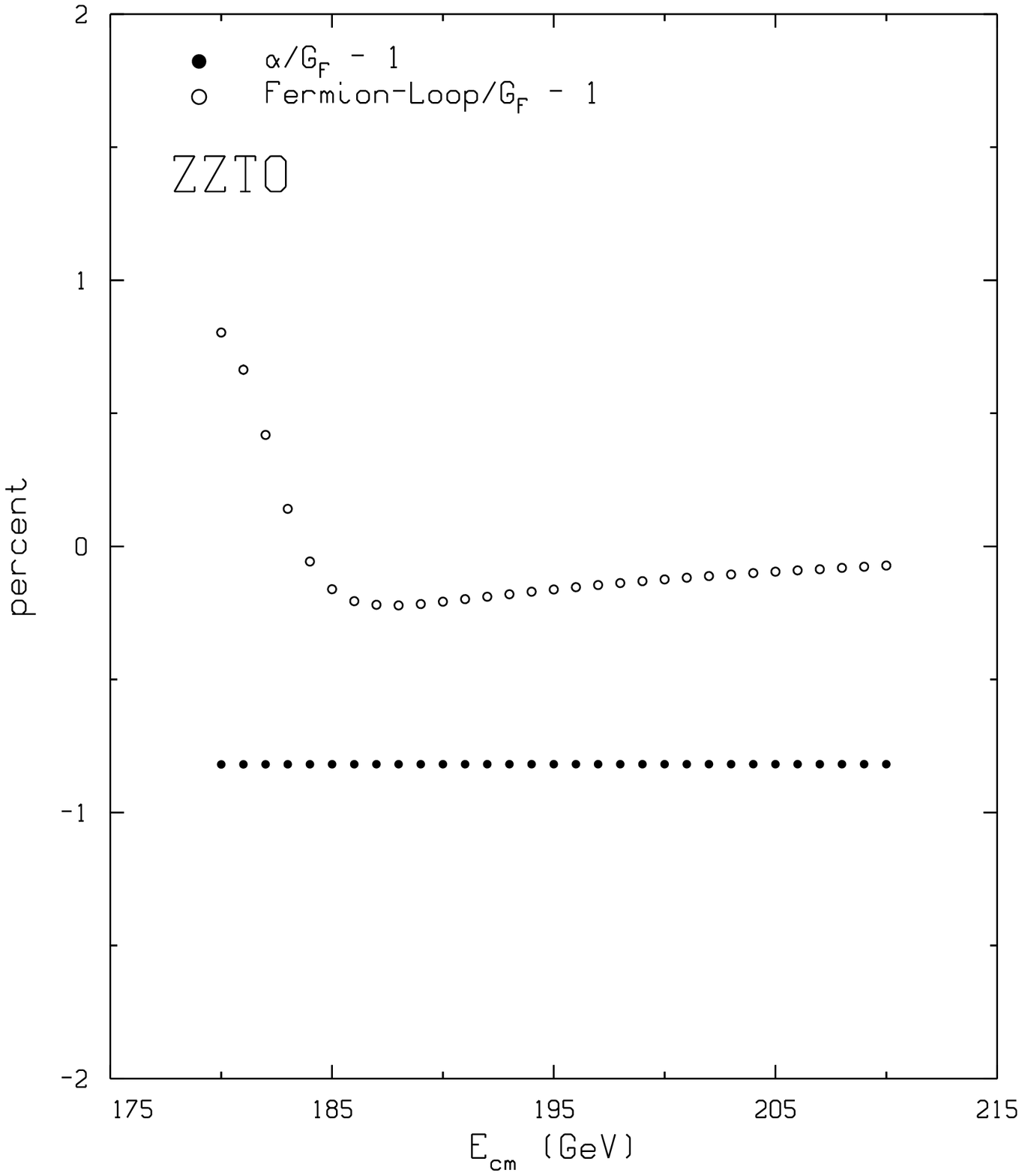,height=12cm,angle=0}
\vskip -2.5cm
\end{center}
\caption{Comparison of different schemes, $\alpha, \gf$ and Fermion-Loop,
for the $\zb\zb$ line-shape from {\tt ZZTO}.}
\label{zzlineshape}
\efi

In \tabn{tab_gentle_nc} we show the $\sigma_{\ssZ\ssZ}$ cross-section as
predicted from {\tt GENTLE}.
\tabn{tab_gentle_nc} is produced with the following {\tt GENTLE/4fan} 
flag settings:
\\
{\tt IPROC,IINPT,IONSHL,IBORNF,IBCKGR,ICHNNL} = 2 2 1 1 0 0
\\
{\tt IGAMZS,IGAMWS,IGAMW,IDCS,IANO,IBIN} = 0 0 0 0 0 0
\\
{\tt ICONVL,IZERO,IQEDHS,ITNONU,IZETTA} = x x 3 0 1
\\
{\tt ICOLMB,IFUDGF,IIFSR,IIQCD} = 0 0 1 0
\\
{\tt IMAP,IRMAX,IRSTP,IMMIN,IMMAX} = 1 0 1 1 1

and with the following NCqed branch settings:
\\
{\tt IPROC ,IINPT ,IONSHL,IBORNF,IBCKGR,ICHNNL}= 3 2 1 1 1 2
\\
{\tt IGAMZS,IGAMWS,IGAMW ,IDCS  ,IANO  ,IBIN}  = 0 0 0 0 0 0
\\
{\tt ICONVL,IZERO ,IQEDHS,ITNONU,IZETTA} = 0 1 x x 1
\\
{\tt ICOLMB,IFUDGF,IIFSR ,IIQCD}         = 2 1 1 0
\\
{\tt IMAP  ,IRMAX ,IRSTP ,IMMIN ,IMMAX}  = 1 0 1 1 1

The Table deserves an extended comment.
Its upper part is obtained with the aid of the standard {\tt GENTLE} approach
to ISR: the band of theoretical uncertainties is produced by choosing
standard structure functions (SF) for the minimum and flux functions (FF) for
the maximum with a reasonable choice in between for the preferred one.
For the maximum, we include LLA second order corrections and 
exclude the lowest order constant term (option {\tt IZERO}=0).
The band has a typical width of about $3 \div 4\%$.  
This approach finds its roots in the treatment of the CC03 cross-section 
where we used the so-called current-splitting technique, the precision of which 
is difficult to evaluate since it takes into account only a part of diagrams. 
We emphasize again that nowadays, after the advent of DPA calculations, 
the theoretical uncertainties in the CC-sector are reduced.

\begin{table}[htbp]\centering
\renewcommand{\arraystretch}{1.1}
\begin{tabular}{|c|r|r|r|}
\hline
channel  & {\tt GENTLE} 2.10  & {\tt GENTLE} $-$ & {\tt GENTLE} $+$ \\
\hline\hline
$qqqq$             & 299.642~~~   & 298.614~~~   & 301.448~~~  \\
$qq\nu\nu$         & 176.076~~~   & 175.410~~~   & 177.137~~~  \\
$qq{\rm ll}$       &  89.187~~~   &  88.851~~~   &  89.720~~~  \\
${\rm ll}\nu\nu$   &  26.204~~~   &  26.103~~~   &  26.362~~~  \\
${\rm llll}$       &   6.637~~~   &   6.612~~~   &   6.677~~~  \\
$\nu\nu\nu\nu$     &  25.857~~~   &  25.766~~~   &  26.013~~~  \\
\hline
total              & 623.602~~~   & 621.356~~~   & 627.356~~~  \\
\hline
\hline
$qqqq$             & 301.448~~~   & 300.418~~~   & 301.522~~~  \\
$qq\nu\nu$         & 177.137~~~   & 176.532~~~   & 177.180~~~  \\
$qq{\rm ll}$       &  89.725~~~   &  89.418~~~   &  89.746~~~  \\
${\rm ll}\nu\nu$   &  26.361~~~   &  26.271~~~   &  26.367~~~  \\
${\rm llll}$       &   6.676~~~   &   6.654~~~   &   6.678~~~  \\
$\nu\nu\nu\nu$     &  26.022~~~   &  25.933~~~   &  26.029~~~  \\
\hline
total              & 627.370~~~   & 625.226~~~   & 627.522~~~  \\
\hline
\end{tabular}
\caption[]{Cross-sections [fb] for $e^+e^- \to \zb\zb \to 4\rmf$ at
$\sqrt{s} = 188.6$ GeV;
first column {\tt GENTLE} 2.10 with preferred flags,
second and third columns estimate variations due 
to theoretical uncertainties. 
The upper part is produced with the 4fan branch and with 
flags: {\tt ICONVL,IZERO}=00,10,01.
The lower part is produced with NCqed branch of {\tt GENTLE/4fan} and with
flags: {\tt IQEDHS,ITNONU}=00,10,11.
\label{tab_gentle_nc}}
\end{table}

For NC-processes, the ISR is well defined and no current-splitting is required.
In paper \cite{Bardin:1996b} we provided the complete lowest order ISR QED 
corrections (option {\tt ITNONU=1}). 
In our complete calculations the constant term
is full reproduced and there are no justifications to exclude it.
This is why in the lower part of the Table we always use {\tt IZERO} =1.
For the theoretical uncertainties, we vary then over three working options
{\tt IQEDHS,ITNONU}=00,10,11 and select preferred, min and max out of them.
As seen from the lower part of the Table, the theoretical uncertainty 
derived in such a way
is about twice as narrow as compared to the upper part. It is important to
emphasize that the two bands overlap, although 
there is a systematic shift towards slightly higher cross-sections.

This shift is due to the constant term. If we had chosen {\tt IZERO}=1
for the upper part, its band would totally contain the band for the lower part.
We tend to consider the lower part to be a more correct treatment of the ISR 
for the case of NC-processes.


\subsection{Summary and conclusions}

Three different programs have produced numbers for the NC02
cross-section showing remarkable agreement over a wide energy range.
{\tt ZZTO} has produced results with two different renormalization
schemes, $\gf$ and $\alpha$, showing differences of the order of a
percent.  {\tt GENTLE} confirms the finding with nearly the same
shifts as {\tt ZZTO} between the two schemes.  It looks plausible to
have a $\pm 2\%$ of theoretical uncertainty assigned to the NC02
cross-section.  There is an indications, coming from the Fermion-Loop
analysis of {\tt ZZTO}, that show smaller deviations with respect to
the $\gf$-scheme and the Fermion-Loop is usually accurate at the $1
\div 2\%$ level.

At the moment the estimated theoretical uncertainty comes from the
comparisons between {\tt GENTLE}, {\tt YFSZZ} and {\tt ZZTO} and it is
roughly about $2\%$.  The size of the uncertainty is confirmed by an
internal estimate of {\tt GENTLE}, as given in \tabn{tab_gentle_nc}.
With the complete lowest order ISR QED included {\tt GENTLE} gives a
total cross-section at $\sqrt{s} = 188.6\,$GeV of
$627.37^{+0.15}_{-2.14}\,$fb where {\tt ZZTO} gives $621.22\,$fb, \ie
{\tt GENTLE} predicts a $0.4\%$ uncertainty with {\tt GENTLE} and {\tt
  ZZTO} differing by roughly $1\%$. Furthermore, {\tt GENTLE} predicts
a $+0.6\%$ shift due to the constant term in ISR and both programs
predict a $-0.8\%$ shift from the $\gf$-scheme to the $\alpha$-scheme.

Given the experimental uncertainty on the cross-section a difference
below $2\%$ is reasonable and, most likely, do not require the
implementation of missing effects which are beyond the reach of the
experiments. Nevertheless, work is in progress for {\tt ZZTO} towards
a complete DPA calculation for NC02.

\section{Conclusions and outlook}
\label{conco}

An extensive collection of theoretical predictions for observables in
$e^+e^-$ interactions at LEP~2 energies had been presented in the 1996
CERN {\it Report of the Workshop on Physics at LEP2}.  However, an
update with improved theoretical prescriptions is needed in order to
match the precision achieved by now in the experimental analyses.

The aim of the four-fermion contribution to this workshop effort is
twofold. We have summarized the most recent theoretical developments
concerning $e^+e^-$ annihilation into four-fermions at LEP~2 energies.
Furthermore, applications to the four most important classes of
processes have been discussed in detail. In decreasing order of
importance they are the $\wb\wb$-signal, the inclusion of an extra
photon in the final state, the single-$\wb$ production and the
$\zb\zb$-signal.

To gauge the priorities of this Report one should remember that the 
experimental situation is rather different for $\wb\wb$ when compared to 
the other processes. For
$\wb$-pairs, LEP (ADLO) is able to test the theory to below $1\%$, \ie, below
the old uncertainty of $\pm 2\%$ established in 1995.  Thus the
CC03-DPA, including non-leading electroweak corrections,
constitutes a very important theoretical development.
However, ADLO cannot test single-$\wb$ or $\zb\zb$-signal to an
equivalent level, since their total cross-section is of the order of
$1\,pb$ or less, $20$ times smaller than that of $\wb$-pair production
\footnote{For $\zb\zb$ with 1997+1998+1999 data, the present 
analyses and global LEP combination method give an average
measurement with $7\%$ accuracy. At the end of LEP, we may reach better 
than $5\%$.}.

The authors of the four-fermion report agree on the following conclusions 
from this study:
\begin{itemize}

\item There is a nice global agreement between the new DPA predictions
  for CC03, which are $2\% \div 3\%$ lower than the old 
approach\footnote{see \sect{sectWW} for a proper definition of the old 
approach.}.

\item The Monte Carlo programs {\tt RacoonWW} and
  {\tt YFSWW3} agree within $0.3\%$ at $\sqrt{s} = 200\,$GeV.
  The present estimated theoretical uncertainty of these programs is $0.4\%$,
  $0.5\%$, and $0.7\%$ for $\sqrt{s} = 200\GeV$, $180\GeV$, and
  $170\GeV$, respectively.

\item There is a general satisfaction with the progress induced by new
  DPA calculations. Nevertheless, the theoretical uncertainty could  
  probably be improved somewhat in the future.

\item More work will be needed to reduce the uncertainty for 
      $4\rmf + \ph$ and of parton shower with $p_t$.

\item In single-$\wb$ production most of the theorists were interested
  in gauge-invariance issues due to unstable particle. The
  experimentalists were asking for ISR and $p_t$ effects, comparisons
  including parton shower, structure functions and exponentiation.
  Unfortunately, only few groups have been working on these issues.
  Their work represents an important result of this Report.

\item In single-$\wb$ production we have a (global) $2\% \div 3\%$ theoretical
  uncertainty associated with the scale of the $t$-channel photon,
  with a projected $1\%$ uncertainty when the implementation of 
  the Fermion-Loop scheme~\cite{tfl} will receive more cross-checks.

\item For simple processes like $e^+e^-$ annihilation and two-photon
  collision, the evolution of the energy scale in the structure
  function or in the parton-shower algorithms can be determined by the
  exact perturbative calculations.  However, this is not available for
  more complicated processes.  When no exact first order calculations
  are available then one resorts to the scale occurring in
  the first order soft corrections. Therefore, at the moment, we may
  apply a very conservative (global) upper bound of $4\%$ theoretical 
  uncertainty  for ISR in single-$\wb$ production.
  Here we repeat one of the conclusions of \sect{sectsw}, we understand
  the implementation of QED radiation in the MC better than before,
  Structure Functions at the scale $s$ are obviously wrong, but 
  we are presently unable to precisely quantify the improvement upon
  the quoted -- global -- upper bound. Single programs may claim to
  have more stringent internal estimates.
  In conclusion, the current upper bound on the global estimate of the
  theoretical uncertainty is $5\%$ for single-$\wb$. A detailed
  explanation of this bound is given in \subsect{summconc}.

\item Compared to the experimental uncertainty on the NC02 $\zb\zb$
  cross-section a difference of about $1\%$ between theoretical
  predictions is acceptable. The global estimate of theoretical uncertainty
  is $2\%$, again acceptable.
  However, it would be nice to improve upon the existing calculations.

\end{itemize}
These points are discussed in more detail in the following.

The new DPA predictions for CC03 are $2\% \div 3\%$ lower than in the
old approach. The new Monte Carlo programs {\tt RacoonWW} and {\tt
  YFSWW3} agree within $0.3\%$ at $\sqrt{s} = 200\GeV$, i.e.\ at a
level that is consistent with the accuracy of the DPA.  The
theoretical uncertainty of these programs for the CC03 $\wb$-pair
cross section, which we estimate to be below $0.4\div 0.5\%$ for
$\sqrt{s}=180$--$210\GeV$, should be compared with the current
experimental precision of $\pm 0.9\%$ with all ADLO data at
$183$--$202\,$GeV combined. It should be mentioned as well that {\tt
  RacoonWW} and the semi-analytical {\tt BBC} calculations agree very
well where they should, i.e.  above $185\,\GeV$.

Turning to distributions, the deviations seem to become somewhat
larger for large $\PWm$ production angles, although compatible with
the statistical accuracy.  The invariant mass distributions agree
within roughly $1\%$ with a distortion of the distributions that is
mainly due to radiation off the final state and the $\wb$~bosons.  We
expect that the present uncertainty of the CC03 $\wb$-pair
cross-section can be reduced somewhat when the sources of the
differences between {\tt RacoonWW} and {\tt YFSWW} and the leading
higher-order corrections will be better analyzed.  To go below the
level of a few per-mille of accuracy would require the complete
calculation of one-loop radiative corrections in four-fermion
production for all 4f final states, a program that does not seem
feasible in a foreseeable future.

The presence of real photons can also change the quantitative
agreement of DPA calculations. For integrated quantities the
differences between alternative approaches are expected to be of the
order of the accuracy of the DPA while for more exclusive observables
larger differences can be expected.  A comparison between {\tt
  RacoonWW} and {\tt YFSWW3} for various distributions in the
semi-leptonic channel $\Pep\Pem\to\Pu\bar\Pd\mu^-\barnu_\mu$ and with
a specified set of separation and recombination cuts reflects,
however, for observables inclusive in the photon the same global
difference as the total cross-section.

The technical precision for $e^+e^- \to 4\rmf+\ph$ has reached high
standards as shown by the comparisons among {\tt PHEGAS/HELAC},
{\tt RacoonWW} and {\tt WRAP}, but at the moment we are unable to present 
any overall
statement on the theoretical uncertainty process by process.  This is
true in particular for the single-$\wb$ configuration.  Furthermore,
no detailed comparison has been performed including parton shower and
hadronization.







In general more work will be needed to establish the uncertainty for 
$4\rmf + \ph$. This should be done process by process, 
with the target of achieving the required accuracy. At the moment we can fix
an upper bound of $2.5\%$ based on missing non-logarithmic 
corrections.

In single-$\wb$ production most of our activity was centered around
gauge-invariance issues due to unstable particle. Although, no coordinate
effort has been performed, at the moment, to study the theoretical
uncertainty induced by ISR $p_t$ effects, comparison with parton
shower, structure functions and exponentiation the
interested reader can find in the Report details on QED corrections as
they stand now. Few programs, noticeably {\tt GRACE} and {\tt SWAP},
have produced a preliminary internal estimate of the uncertainty
associated with the treatment of QED radiation; the net effect of QED
is between $8\%$ and $10\%$ in the LEP~2 energy range, with
$s$-channel structure functions over-estimating the effect by $\approx
4\%$. Furthermore, structure functions with a modified scale seems to
agree with parton shower at the level of $1\%$ when experimental cuts
are included or even better for a fully extrapolated setup.  

As far as the scale of the electromagnetic coupling is concerned we find that
the results with a rescaling of $\alpha_{\rm QED}$ for the $t$-channel
photon that has been implemented in {\tt NEXTCALIBUR, SWAP} and {\tt WPHACT}
show an agreement with {\tt WTO} predictions that is roughly around
$2\%$. 

For single-$\wb$, therefore, we register a conservative,
overall, upper bound of $\pm 5\%$ for the theoretical uncertainty.
Single programs may claim better internal estimates but this does not
transform, yet, into a global one\footnote{
We recall that, at the moment an uncertainty associated to QED ISR is quoted, 
by the LEP EWWG, that follows from taking the average of the Born result with 
the one corrected via $s$-channel structure functions, where 
${\rm SF}(t,p^2_{t\ssW}) > {\rm SF}(s)$ by $+5\%$ at $200\,$GeV and Born 
$\, > {\rm SF}(s)$ by $+12\%$. Note, however, that this is not a real estimate 
of uncertainty but just a pragmatic way of determining the effects of ISR.}.

Implementation of the EFL-scheme in single-$\wb$
(in addition to {\tt WTO}) will give a more solid basis to the estimate
of $1 \div 2\%$ for the uncertainty associated with the 
scale of the e.m. coupling.

The next, obvious, step is represented by the evaluation of
missing $\ord{\alpha}$ electroweak effects, \eg in Weizs\"acker-Williams 
approximation (for the sub-process $\Pe\gamma\to\PW\nu_\Pe$), the analogous 
of DPA for CC03.

A better understanding of QED ISR and of all radiative corrections in 
single-$\wb$ production is certainly needed in order to reduce the 
corresponding uncertainty, hopefully around $1\%$. This, however, requires 
to go beyond the present approximations, not an easy task and with
a considerably large experimental error. Since DPA cannot be applied to
single-$\wb$ production one has to follow some alternative path, like
including radiative corrections in (improved) Weizs\"acker-Williams 
approximation, or WWA. It is expected that already the normal WWA ( \ie 
logarithmic terms only), with a typical Born-accuracy of $5\%$, will yield 
results accurate at the level of $5\%\,\times\,\alpha/\pi$.
For the moment this is not strictly needed but one should consider that
single-$\wb$ will be one of the major processes at LC.

For the NC02 cross-section we have a $1\%$ variation,
obtained by changing the Input Parameter Set in {\tt GENTLE} and in
{\tt ZZTO} and by varying from the standard {\tt GENTLE} approach for ISR to
the complete lowest order corrections. We estimate the real 
uncertainty to be $2\%$.
However, given the experimental uncertainty a theoretical
uncertainty in this order is acceptable and does not seem to require
the implementation of missing effects. Furthermore, {\tt ZZTO} which
is not yet a DPA calculation agrees rather well with {\tt YFSZZ},
roughly below the typical DPA accuracy of $0.5\%$, and the latter
features the realization of the LPA for the NC02 process.
The implementation of a DPA calculation, in more than one code, in the NC02 
$\zb$-pair cross-section will bring the corresponding accuracy at the level of 
$0.5\%$, similar to the CC03 case.

%
%

%
\end{document}